\documentclass[b5paper,11pt,twoside,openright]{book}

\usepackage{geometry}
\usepackage{amsfonts}
\usepackage{amsmath}
\usepackage{mathrsfs}
\usepackage{soul,color}
\usepackage{multirow}
\usepackage{graphicx}
\usepackage{comment}
\usepackage{float}
\usepackage{subcaption}
\usepackage{algorithm}
\usepackage{algpseudocode}
\usepackage{hyperref}
\usepackage{mathpazo}
\usepackage{fancyhdr}

\usepackage{appendix}
\usepackage{chngcntr}
\usepackage{etoolbox}
\usepackage{lipsum}
\AtBeginEnvironment{subappendices}{
\chapter*{Appendix}
\addcontentsline{toc}{chapter}{Appendices}
\counterwithin{figure}{section}
\counterwithin{table}{section}
}

\def\NoNumber#1{{\def\alglinenumber##1{}\State #1}\addtocounter{ALG@line}{-1}}

\newcommand{\eps}{\varepsilon}

\newcommand{\dd}{\mathrm{d}}
\newcommand{\pd}{\partial}
\newcommand{\wtilde}{\widetilde}
\newcommand{\uv}[1]{\hat{\mathbf{#1}}}
\newcommand{\uvg}[1]{\hat{\boldsymbol{#1}}}

\newcommand{\thet}{\vartheta}
\newcommand{\ph}{\varphi}

\def\Xint#1{\mathchoice
{\XXint\displaystyle\textstyle{#1}}
{\XXint\textstyle\scriptstyle{#1}}
{\XXint\scriptstyle\scriptscriptstyle{#1}}
{\XXint\scriptscriptstyle\scriptscriptstyle{#1}}
\!\int}
\def\XXint#1#2#3{{\setbox0=\hbox{$#1{#2#3}{\int}$}
\vcenter{\hbox{$#2#3$}}\kern-.5\wd0}}

\def\dashint{\Xint-}

\usepackage{enumitem}
\usepackage{tikz}
\usetikzlibrary{shapes.geometric, arrows}
\tikzstyle{arrow} = [thick,->,>=stealth]

\begin{document}

\frontmatter

\title{\textbf{{A computational framework for microstructural modelling of polycrystalline materials with damage and failure}}}
\author{ $\;$\\
Ph.D. thesis submitted to the University of Palermo\\
$\;$\\
by\\
$\;$\\
 \textbf{\emph{Vincenzo Gulizzi}}\\
 $\;$\\
Tutor\\
$\;$\\
 \textbf{\emph{Prof. Alberto Milazzo}}\\
 \\
 \textbf{\emph{Prof. Ivano Benedetti}}\\
$
\,
$\\
\\
\\}
\date{Dipartimento di Ing. Civile Ambientale, Aerospaziale, dei Materiali\\ Universit\`a degli Studi di Palermo\\ Scuola Politecnica\\ Viale delle Scienze, Ed. 8 - 90128 Palermo}
\maketitle

\thispagestyle{empty}
{\footnotesize\null \vspace{\stretch{1}}
\noindent\textsc{Vincenzo Gulizzi}\\
Palermo, December 2016\\
\texttt{e-mail:}\verb"vincenzo.gulizzi@unipa.it"\\

\noindent Thesis of the Ph.D. course in \emph{Civil and Environmental Engineering - Materials \\(Ingegneria Civile e Ambientale - Materiali)}\\
Dipartimento di Ingegneria Civile Ambientale, Aerospaziale, dei Materiali\\
Universit\`a degli Studi di Palermo\\
Scuola Politecnica\\
Viale delle Scienze, Ed.8 - 90128 Palermo, ITALY\\

\newpage\thispagestyle{empty}
\normalsize

\chapter*{Preface}
In the present thesis, a computational framework for the analysis of the deformation and damage phenomena occurring at the scale of the constituent grains of polycrystalline materials is presented. The research falls within the area of Computational Micro-mechanics that has been attracting remarkable technological interest due to the capability of explaining the link between the micro-structural details of heterogenous materials and their macroscopic response, and the possibility of fine-tuning the macroscopic properties of engineered components through the manipulation of their micro-structure. However, despite the significant developments in the field of materials characterisation and the increasing availability of High Performance Computing facilities, explicit analyses of materials micro-structures are still hindered by their enormous cost due to the variegate multi-physics mechanisms involved.

Micro-mechanics studies are commonly performed using the Finite Element Method (FEM) for its versatility and robustness. However, finite element formulations usually lead to an extremely high number of degrees of freedom of the considered micro-structures, thus making alternative formulations of great engineering interest. Among the others, the Boundary Element Method (BEM) represents a viable alternative to FEM approaches as it allows to express the problem in terms of boundary values only, thus reducing the total number of degrees of freedom.

The computational framework developed in this thesis is based on a non-linear multi-domain BEM approach for generally anisotropic materials and is devoted to the analysis of three-dimensional polycrystalline microstructures.
Different theoretical and numerical aspects of the polycrystalline problem using the boundary element method are investigated:
first, being the formulation based on a integral representation of the governing equations, a novel and more compact expression of the integration kernels capable of representing the multi-field behaviour of generally anisotropic materials is presented;
second, the sources of the high computational cost of polycrystalline analyses are identified and suitably treated by means of different strategies including an ad-hoc grain boundary meshing technique developed to tackle the large statistical variability of polycrystalline micro-morphologies;
third, non-linear deformation and failure mechanisms that are typical of polycrystalline materials such as inter-granular and trans-granular cracking and generally anisotropic crystal plasticity are studied and the numerical results presented throughout the thesis demonstrate the potential of the developed framework.

\tableofcontents

\mainmatter

\chapter{Grain Boundary formulation for polycrystalline micro-mechanics}\label{ch-intro}
Many materials of technological interest, such as metals, alloys and ceramics, present a polycrystalline micro-structure at a scale usually ranging from nano- to micro-meters (see Figure (\ref{fig-Ch1:micro-scale})). A polycrystalline micro-structure is characterised by the size, shape, random crystallographic orientation and general anisotropic behaviour of each crystal, or \emph{grain}, by the possible presence of randomly distributed flaws and pores and by the physical and chemical properties of the inter-crystalline interfaces, or \emph{grain boundaries}, which play a crucial in polycrystalline micro-mechanics.

\begin{figure}[h]
\centering
\includegraphics[width=\textwidth]{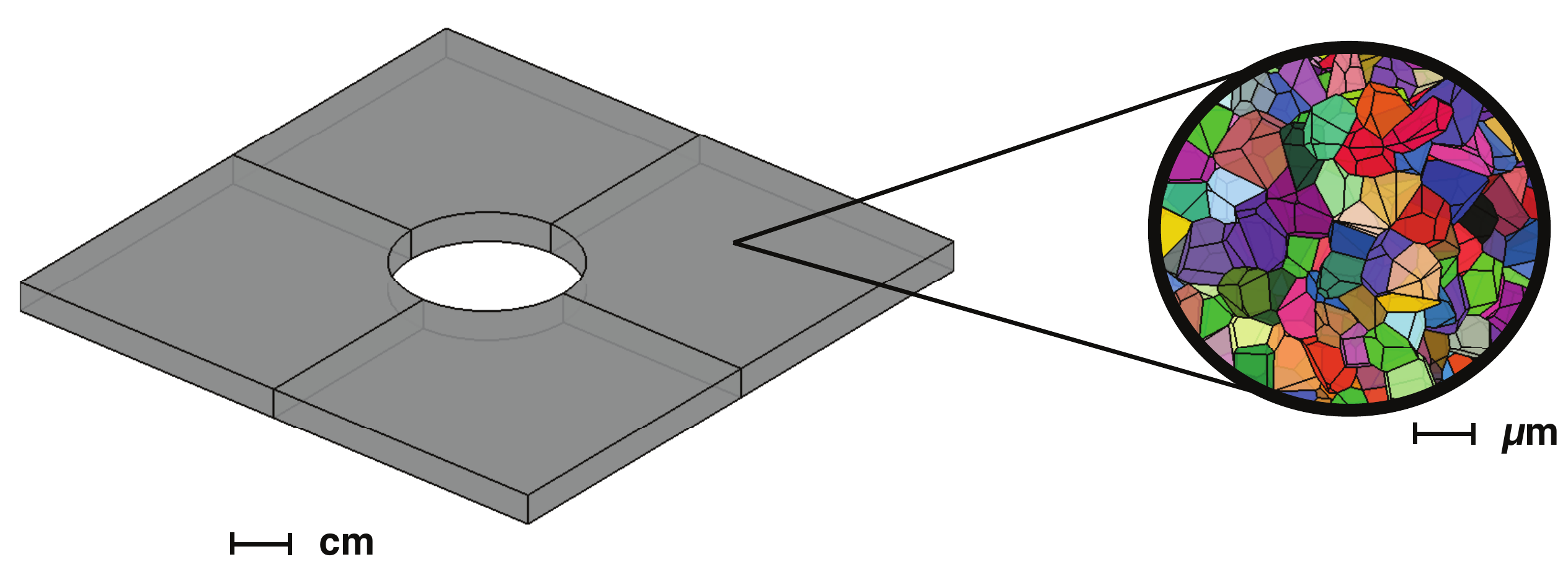}
\caption{Schematic representation of a polycrystalline micro-structure in an engineered structural component.}
\label{fig-Ch1:micro-scale}
\end{figure}

It is nowadays widely recognised that the macroscopic features of a material are strongly influenced by the properties of and the interaction between its micro-constituents, especially when damage initiation and evolution are involved. With current advancements in materials science and technology, there is interest in understanding the link between material micro-structures and macro-properties, i.e.\ the \emph{structure-property relation} \cite{hashin1983,needleman2000,zohdi2008,mura2013,nemat2013}, which also attracts remarkable technological interest due to the possibility of controlling macroscopic properties, such as strength, stiffness, fracture toughness, etc$\dots$, through manipulations of the micro-structure \cite{watanabe1999}.

In this context, experimental measurements and numerical simulations represent two complementary approaches that have been employed in the literature for understanding the micro-macro relation in polycrystalline materials. Experimental techniques based for example on serial sectioning \cite{zhang2004,groeber2006,groeber2008,rowenhorst2010} or x-ray tomography \cite{dobrich2004,ludwig2009} and usually coupled to electron backscattering diffraction (EBSD) measurements have been used to reconstruct polycrystalline micro-morphologies and to determine the crystals' orientation. However, although being essential for the extraction of the statistical features of the micro-morphologies, experimental approaches may be expensive and time-consuming and generally require sophisticated equipment and careful material preparation.

On the other hand, computational approches allow to artificially reproduce the properties of polycrystalline materials and investigate their deformation and damage mechanisms by selectively focus on particular aspects that could be extremely challenging to reproduce in a laboratory. Furthermore, computational micro-mechanics is enormously benefitting from the increasing affordability of High Performance Computing (HPC), which is pushing the boundaries of multi-field micro-mechanics, and from the aforementioned developments in micro-structural materials characterisation, which provide micro-structural details that can be included in high-fidelity computational models, allowing deeper understanding of materials behaviour.

In the literature, several computational approaches have been proposed for the modelling of two- and three-dimensional polycrystalline materials. An important concept in the field of computational micro-mechanics of heterogeneous micro-structures is the concept of the \emph{Representative Volume Element} (RVE). An RVE is usually defined as material sample sufficiently large to statistically represent the heterogeneous nature of the micro-structure, but small enough to be considered as a infinitesimal volume at the macroscopic scale. The resolution of the micro-macro link is known as \emph{material homogenization} \cite{nguyen2011} and is usually performed by computing volume and/or ensemble averages of the fields of interests over multiple RVE realisations subjected to general loading conditions and possibly undergoing internal evolution.

To this date, the most popular computational approach is probably still represented by the Finite Element Method (FEM). Several FEM studies have been proposed to evaluate the effective macroscopic properties of polycrystalline materials \cite{kumar1994,kumar1996} or to study the local behaviour at the grain boundaries \cite{kozaczek1995,zhao2004,kamaya2007}. The concept of the representative volume element has been investigated by Ren and Zheng \cite{ren2002,ren2004} and by Nyg{\aa}rds \cite{nygaards2003}, who studied the dependence of the RVE size on the micro-structural properties of two- and three-dimensional polycrystalline aggregates and on the anisotropy of the crystals. Kanit et al.\ \cite{kanit2003} focused on the concept of the RVE from statistical and numerical points of view, highlighting the possibility of computing the macroscopic properties by means of a sufficiently large number of small RVEs rather than by a large representative volume. The authors also studied the effect of different boundary conditions on the convergence of macroscopic properties of the RVEs. Fritzen et al.\ \cite{fritzen2009} focused on the generation and meshing of geometrically periodic polycrystalline RVEs.

Finite element models have been also employed to study different deformation and damage mechanisms in polycrystalline materials. Due to the anisotropic nature of the constituent crystals, one of the most important as well as studied deformation mechanisms in polycrystalline materials is the mechanism of crystal plasticity \cite{roters2010}, which denotes the plastic slip over specific crystallographic planes defined by the lattice of each crystal. The phenomenon, which has been widely studied within FEM frameworks, has been also addressed in the present thesis and more details are given in Chapter (\ref{ch-CP}). Damage and fracture mechanisms have been usually addressed by means of the \emph{cohesive zone} approach \cite{barenblatt1959,barenblatt1962} in which fracture evolution is represented by a traction-separation law that encloses the complex damage phenomena occurring at a \emph{process zone} of the failing interface into a phenomenological model. Cohesive laws have been derived either by the definition of a cohesive potential \cite{tvergaard1990,xu1993} or by simply assuming a functional form \cite{camacho1996,ortiz1999} and have been used to model different fracture mechanisms \cite{chandra2002,shet2002}.

In the field of computational micro-mechanics, the cohesive zone approach has been employed by Zhai and Zhou \cite{zhai2000,zhai2004} to model static and dynamic micro-mechanical failure of two-phase $\mathrm{Al}_2\mathrm{O}_3$/$\mathrm{Ti}\mathrm{B}_2$ ceramics. Espinosa and Zavattieri \cite{espinosa2003a,espinosa2003b} studied inter-granular fracture of two-dimensional polycrystalline materials subject to static and dynamic loading. The competition between the bulk deformation of grains interior and the fracture at the grain boundaries of three-dimensional columnar polycrystalline structures was investigated by Wei and Anand \cite{wei2004} by coupling a crystal plasticity and an elasto-plastic grain boundary model. Maiti et al.\ \cite{maiti2005} used the cohesive zone approach to model the fragmentation of rapidly expanding polycrystalline ceramics. The cohesive model at the micro-structure scale was also used by Zhou et al.\ \cite{zhou2012}, who investigated the effect of grain size and grain boundary strength on the crack pattern and the fracture toughness of a ceramic polycrystalline material.

The aforementioned works mainly focused on inter-granular fracture of polycrystalline materials, i.e.\ the fracture of the grain boundaries. Another failure mechanism occurring in polycrystalline materials is the fracture of the bulk grains, known as trans-granular fracture. Similarly to the crystal plasticity phenomenon, trans-granular fracture is a highly anisotropic fracture phenomenon mainly controlled by the orientation of the crystals within the aggregate. The problem of trans-granular fracture has been investigated by fewer authors \cite{sukumar2003,musienko2009,clayton2015} within the framework of the finite element method due to its inherent anisotropy and generally higher complexity with respect to the inter-granular fracture mechanism. Trans-granular fracture has been also discussed in the present thesis and further details on the problem are given in Chapter (\ref{ch-TG}).

Fracture in polycrystalline materials is also strongly affected by the operating conditions of the polycrystalline component. Naturally, ductile materials are known to present a brittle fracture behaviour in presence of an aggressive environment, and different studies \cite{kamaya2009,musienko2009,simonovski2011,simonovski2012} focusing on the combined effect of the applied stress and the presence of corrosive species can be found in the literature.

In addition to the finite element approach, the polycrystalline problem has also been addressed within the framework of the Boundary Element Method (BEM) \cite{banerjee1981,wrobel2002,aliabadi2002}. Inter-granular micro-cracking has been developed for two-dimensional \cite{sfantos2007a,sfantos2007b} and three-dimensional \cite{benedetti2013a,benedetti2013b,gulizzi2015,benedetti2015,gulizzi2016,benedetti2016} problems and the formulation has been usually referred to as \emph{grain boundary formulation} for polycrystalline materials. 

More recently, a number of different numerical approaches have been proposed to model deformation and fracture mechanisms in polycrystalline materials. The phase-field framework has been proved to be effective for the modelling of different deformation and fracture mechanisms \cite{aranson2000,miehe2010,steinbach2009,borden2012}. Within the polycrystalline materials framework, the phase field approach has been used to model the failure of both two- and three-dimensional elastic, elasto-plastic and ferroelectric materials \cite{abdollahi2012,abdollahi2014,clayton2015,clayton2016,shanthraj2016}. A peridynamic model for the dynamic fracture of two-dimensional polycrystalline materials was proposed by Ghajari et al.\ \cite{ghajari2014}. The cellular automata approach \cite{das2006,shterenlikht2015,di2016} and the non-local particle method \cite{chen2015} have been also investigated for modelling inter- and trans-granular failure of polycrystalline materials.

In this Chapter, the grain boundary formulation employed for modelling the micro-mechanics of polycrystalline aggregates is described. It is already worth mentioning here that the salient feature of the formulation is the expression of the polycrystalline micro-structural problem in terms of inter-granular displacements and tractions, which are the \emph{primary unknowns} of the problem. In the present context, the term \lq\lq grain-boundary\rq\rq\ refers to this aspect. The formulation has been first proposed by Sfantos and Aliabadi \cite{sfantos2007a,sfantos2007b} for two-dimensional polycrystalline materials and by Benedetti and Aliabadi \cite{benedetti2013a,benedetti2013b,benedetti2015} for three-dimensional polycrystalline problems. The present Chapter is intended to introduce the problem of polycrystalline micro-mechanics, to present the 3D grain-boundary formulation and to set the stage of the subsequent developments of the present thesis.

\section{Notation}
In this Section, the notation used in the present thesis is introduced. This work deals with three-dimensional polycrystalline materials in the 3D space denoted by $\mathbb{R}^3$. Spatial coordinates in $\mathbb{R}^3$ are indicated using bold lower-case letters, i.e.\  $\mathbf{x}$ denotes the set of the three components $\{x_1,x_2,x_3\}$.

Vectorial and tensorial quantities that depend on the spatial variable $\mathbf{x}$ are indicated using the indicial notation; for instance, $u_i(\mathbf{x})$ and $\sigma_{ij}(\mathbf{x})$ are used to denote the set of three components of the vector $\{u_1(\mathbf{x}),u_2(\mathbf{x}),u_3(\mathbf{x})\}$ and the set of the nine components of the tensor $\{\sigma_{11}(\mathbf{x}),\sigma_{12}(\mathbf{x}),\dots,\sigma_{23}(\mathbf{x}),\sigma_{33}(\mathbf{x})\}$, respectively. The same convention also holds for higher-order tensors. Unless otherwise stated, the range of the subscripts is always $1,2,3$.

By means of the indicial notation, the product of two vectorial or tensorial quantities with repeated subscripts implies the summation over the repeated subscripts; as an example, using the implied summation convention, the distance $r$ between two points $\mathbf{x}$ and $\mathbf{y}$ is written as
\begin{equation*}
r=\sqrt{(x_1-y_1)^2+(x_2-y_2)^2+(x_3-y_3)^2}=\sqrt{(x_k-y_k)(x_k-y_k)},
\end{equation*}
where the summation over the subscript $k$ is implied. On the other hand, $N\times1$ vectors and $N\times M$ matrices usually stemming from a discretisation procedure and containing a fixed number of entries are written in bold upper-case letters. For instance, a linear system of $N$ equations is simply written $\mathbf{A}\mathbf{X}=\mathbf{B}$, being $\mathbf{X}$ the $N\times1$ vector containing the unknowns, $\mathbf{A}$ the $N\times N$ coefficients matrix and $\mathbf{B}$ the $N\times 1$ vector representing the right-hand side.

Eventually, superscripts are used to specify a particular dependence of a generic quantity. No summation is employed for repeated superscripts and their range is always specified. As an example, the surface of the generic grain $g$ is indicated as $S^g$.

\begin{figure}[h]
\centering
	\begin{subfigure}{0.49\textwidth}
	\centering
	\includegraphics[width=\textwidth]{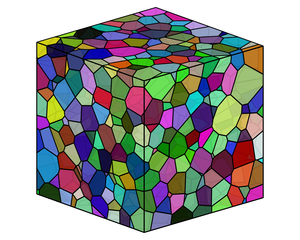}
	\caption{}
	\end{subfigure}
\	
	\begin{subfigure}{0.49\textwidth}
	\centering
	\includegraphics[width=\textwidth]{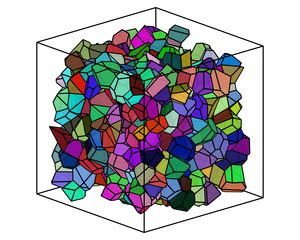}
	\caption{}
	\end{subfigure}
\caption{Example of a polycrystalline aggregate generated using a Voronoi-based scheme: (\emph{a}) 1000-grain polycrystalline aggregate inside a cubic box and (\emph{b}) morphology of the \emph{internal} grains of the aggregate.}
\label{fig-Ch1:3D voronoi tessellation}
\end{figure}

\section{Generation of artificial polycrystalline morphologies}
In this Section, the generation of the micro-morphologies for the study of the deformation and failure of polycrystalline materials is described. The morphology of polycrystalline materials can be reconstructed by experimental observations and measurements or by computer models that are able to reproduce the statistical features of the considered aggregate.
Three-dimensional reconstructions of polycrystalline materials have been performed by using several techniques relying on x-ray micro-tomography \cite{dobrich2004}, diffraction contrast tomography \cite{king2008} and serial sectioning \cite{zhang2004,rowenhorst2010} combined with electron backscatter diffraction (EBSD) maps \cite{groeber2006,groeber2008}. However, such techniques, although necessary to measure the statistical features of polycrystalline micro-morphologies, require generally expensive laboratory equipment and involve time-consuming data processing.
On the other hand, artificial morphologies, generated on the information provided by experimental measurements, offer a valid alternative to experimentally reconstructed aggregates. In the literature, simplified geometrical models (such as cubes or octahedra) of the grains have been employed \cite{zhao2007,ranganathan2008,ritz2008}. Such regular grain shapes have been shown to produce satisfactory results in terms of homogenised properties in linear and crystal plasticity analyses. However, the consequent regular structure of the grain interfaces is not fully appropriate to model inter-granular failure mechanisms, which may require a more accurate representation of the grain boundaries.
Besides regular morphologies, Voronoi tessellation algorithms \cite{voronoi1908} have been extensively used to artificially generate polycrystalline aggregates. Voronoi tessellation are analytically defined given a distribution of points (or \emph{seeds}) scattered inside a domain. In general, the bounded three-dimensional Voronoi tessellation can be defined given a bounded domain $\mathscr{D}\subset \mathbb{R}^3$, a set of generator seeds $\mathscr{S}=\{\mathbf{x}^g\in \mathscr{D}, g=1,\dots,N_g\}$, being $N_g$ the total number of seeds, and a distance function $d$ defined in $\mathbb{R}^3$. For each seed $\mathbf{x}^g$, the corresponding \emph{grain} occupying the volume $V^g$ is defined as
\begin{equation}
V^g=\{\mathbf{y}\in\mathscr{D}|\ d(\mathbf{y},\mathbf{x}^g) < d(\mathbf{y},\mathbf{x}^h)\ \forall h\ne g\}
\end{equation}
and the resulting bounded Voronoi diagram $\mathscr{V}(\mathscr{D},\mathscr{S})$ is obtained as the union of all grains
\begin{equation}
\mathscr{V}(\mathscr{D},\mathscr{S})=\bigcup\limits_{g=1}^{N_g}V^g.
\end{equation}
Approximations of three-dimensional polycrystalline aggregates can be obtained using 3D Voronoi tessellations and the Euclidean distance. Voronoi-based polycrystalline morphologies and their modifications \cite{kumar1992,fan2004,lautensack2008}, described in more details in the next part of this Section, have been widely used within different numerical approaches to model the behaviour of polycrystalline morphologies, see for examples Refs.\ \cite{kumar1994,kumar1996,benedetti2013b,gerard2013} to cite a few. As an example, Figure (\ref{fig-Ch1:3D voronoi tessellation}a) shows a polycrystalline aggregate generated using a Voronoi-based scheme inside a cubic box and Figure (\ref{fig-Ch1:3D voronoi tessellation}b) shows the interior grains of the same aggregate.


\begin{figure}[H]
\centering
	\begin{subfigure}{0.49\textwidth}
	\centering
	\includegraphics[width=\textwidth]{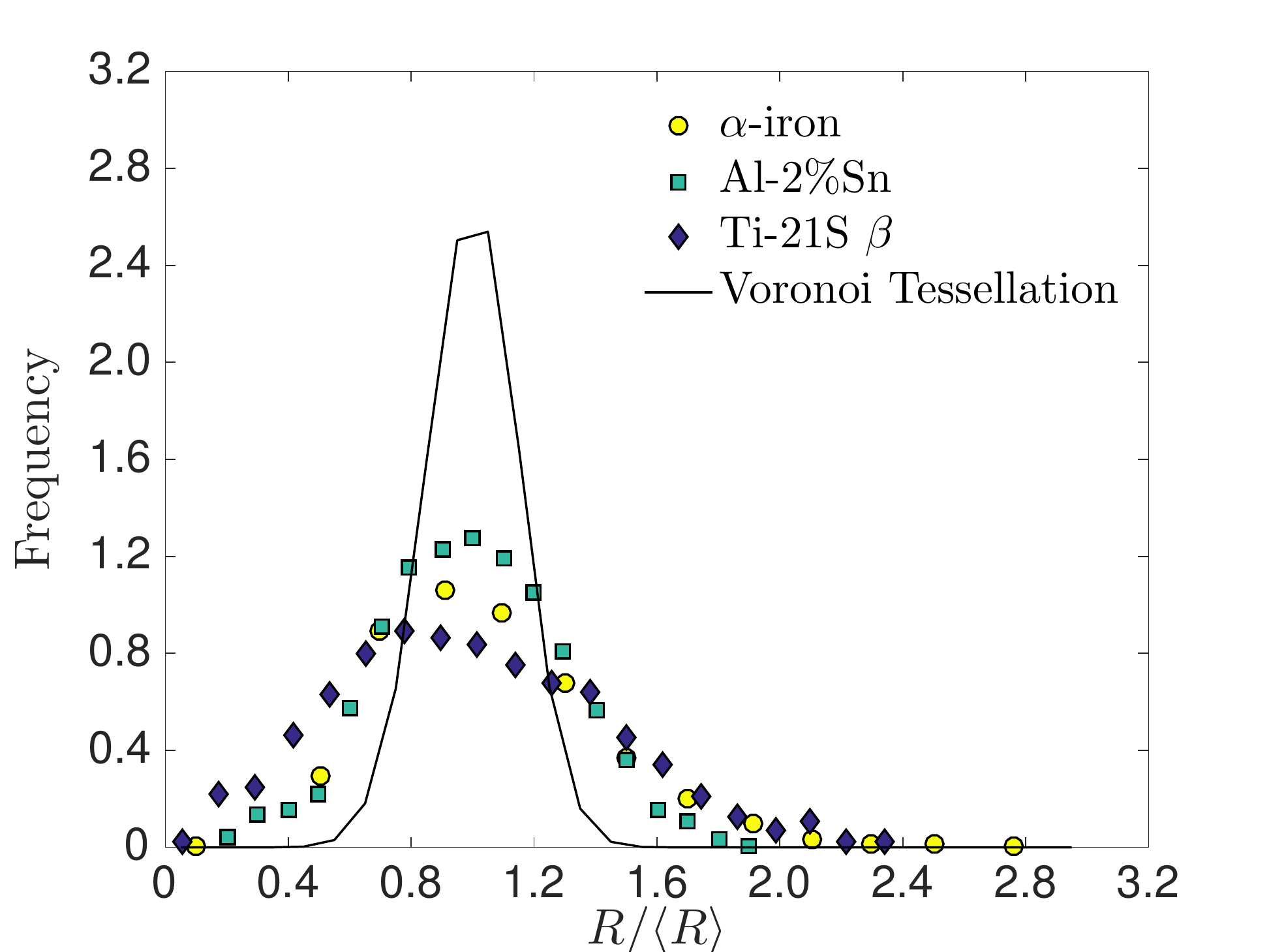}
	\caption{}
	\end{subfigure}
\	
	\begin{subfigure}{0.49\textwidth}
	\centering
	\includegraphics[width=\textwidth]{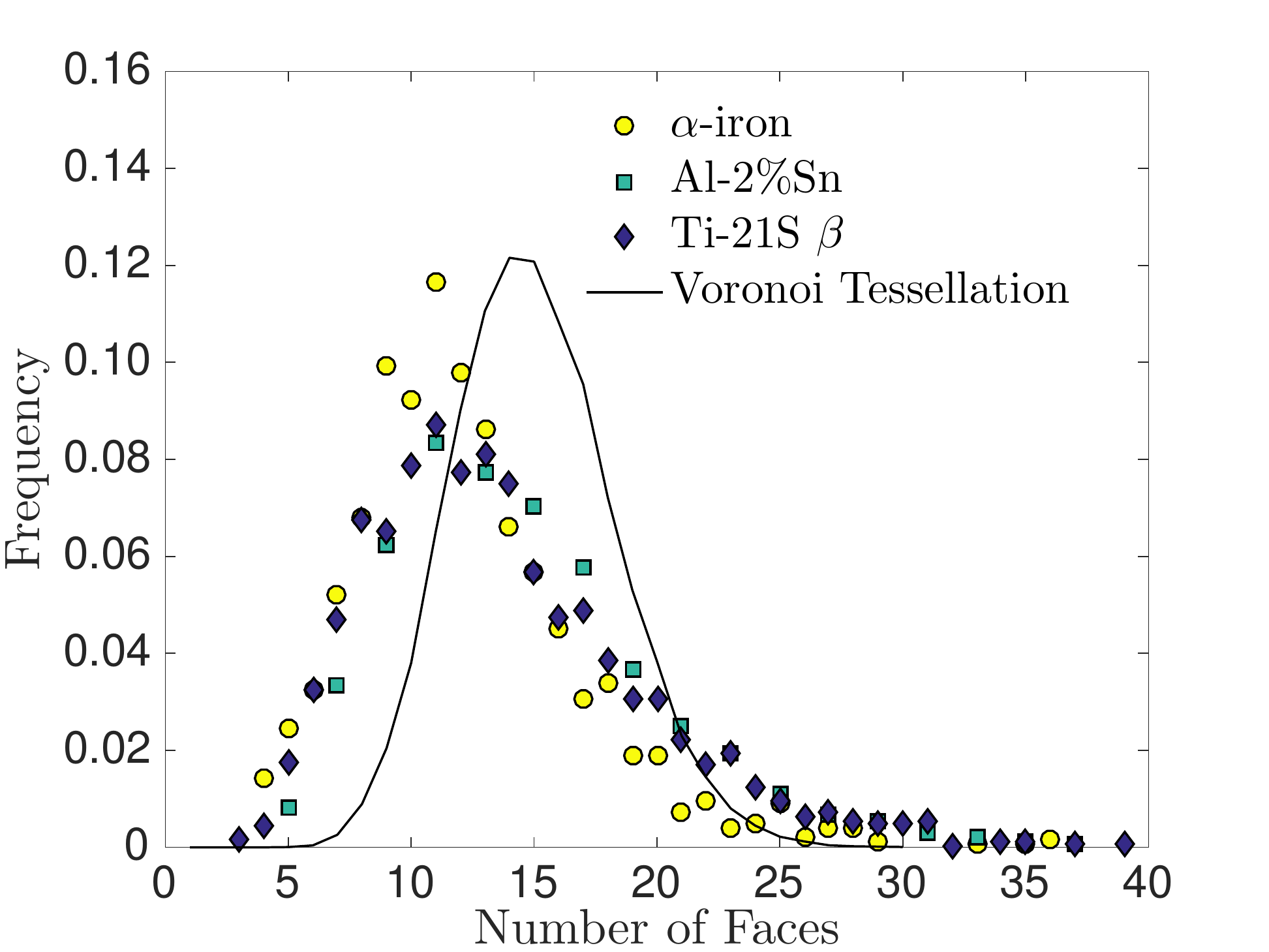}
	\caption{}
	\end{subfigure}
\caption{Statistics of three polycrystalline materials compared to those obtained by using Voronoi tessellations. (\emph{a}) Grain size obtained as the ratio between the spherical equivalent grain radius $R$ (i.e. the radius of a sphere with the same volume of the grain) and the mean radius $\langle R\rangle$. (\emph{b}) Number of grain faces. Experimental data are taken from \cite{zhang2004} (\emph{circles}), \cite{dobrich2004} (\emph{squares}) and \cite{rowenhorst2010} (\emph{diamonds}).}
\label{fig-Ch1:voro stats}
\end{figure}

\begin{figure}[H]
\centering
	\begin{subfigure}{0.49\textwidth}
	\centering
	\includegraphics[width=\textwidth]{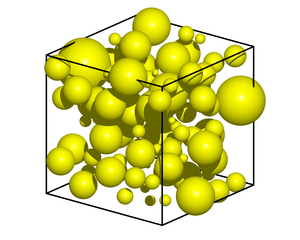}
	\caption{}
	\end{subfigure}
\	
	\begin{subfigure}{0.49\textwidth}
	\centering
	\includegraphics[width=\textwidth]{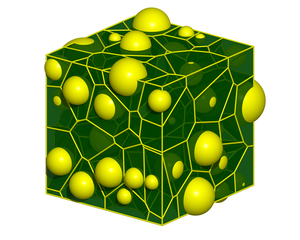}
	\caption{}
	\end{subfigure}
\caption{100-grain hardcore Laguerre-Voronoi tessellation: (\emph{a}) seeds and weights (indicated by the spheres) distribution inside a cubic box and (\emph{b}) resulting tessellation.}
\label{fig-Ch1:HL tessellation}
\end{figure}

Voronoi tessellations can be generated by using open source libraries such as \texttt{TetGen} (\url{http://wias-berlin.de/software/tetgen/}) \cite{si2015}, \texttt{Voro++} (\url{http://math.lbl.gov/voro++/}) \cite{rycroft2009} or \texttt{Neper} (\url{http://neper.sourceforge.net}) \cite{quey2011}. Eventually, artificial polycrystalline morphologies can also be generated by relying on grain growth model in combination with phase field \cite{krill2002} or cellular automata \cite{ding2006} techniques.

In this thesis, the polycrystalline microstructures are generated starting from Voronoi tessellations. Figures (\ref{fig-Ch1:voro stats}a) and (\ref{fig-Ch1:voro stats}b) show two important statistical features of three polycrystalline materials (namely $\alpha$-iron \cite{zhang2004}, Al-$2\%$Sn \cite{dobrich2004} and Ti-21S$\beta$ \cite{rowenhorst2010}), and the corresponding statistical features measured for a 1000-grain Poisson-Voronoi tessellation. Figure (\ref{fig-Ch1:voro stats}a) shows that the artificial morphology underestimates the variability of grain size, whereas Figure (\ref{fig-Ch1:voro stats}b) shows that the Voronoi tessellation slightly overestimates the number of faces per grain.

In order to reduce such differences between artificial Voronoi morphologies and real microstructures, modified Voronoi tessellations have been used in the literature. A first constraint prescribes a minimum distance between the tessellation seeds, to take into account the minimum size of stable nuclei before the solidification process initiates, which leads to the so called \emph{hardcore} Voronoi tessellations. Second, a sphere with random radius from a suitable statistical distribution can be associated to each seed, requiring that the associated grain must contain it, which leads to the so called \emph{Laguerre} tessellations \cite{fan2004,lautensack2008}. Calling $r^g$ the radius associated to the seed $\mathbf{x}^g$, the distance $d_L$ between $\mathbf{x}^g$ and a generic point $\mathbf{y}$ used to compute the Laguerre tessellations is defined as
\begin{equation}
d_L(\mathbf{y},\mathbf{x}^g)=\{[d_E(\mathbf{y},\mathbf{x}^g)]^2-(r^g)^2\}^{1/2}
\end{equation}
where $d_E$ is the Euclidean distance.
Figure (\ref{fig-Ch1:HL tessellation}) shows the Laguerre tessellation obtained combining the two modifications. The weight (radius) of each seed is indicated by a sphere, and the hardcore condition ensures that the spheres do not intersect. Fine-tuning the distribution of grains weights allows to generate micro morphologies that better represent the statistical features of real materials. Figure (\ref{fig-Ch1:alpha voro stats}) compares the statistical features of a 1000-grain Poisson-Voronoi tessellation with those of a 1000-grain hardcore Laguerre tessellation, whose sphere radii distribution was determined from the following Weibull distribution:
\begin{equation}
\psi(x|a,b) = \frac{b}{a}\left(\frac{x}{a}\right)^{b-1}\exp\left[-\left(\frac{x}{a}\right)^b\right]
\end{equation}
where the parameters $a$ and $b$ have been chosen to fit the experimental data reported in Ref.\ \cite{zhang2004}. The latter clearly better approximates the observed real microstructure, in terms of both grain size and number of faces. For such reasons, the hardcore Laguerre tessellations are adopted in the present framework.

\begin{figure}[H]
\centering
	\begin{subfigure}{0.49\textwidth}
	\centering
	\includegraphics[width=\textwidth]{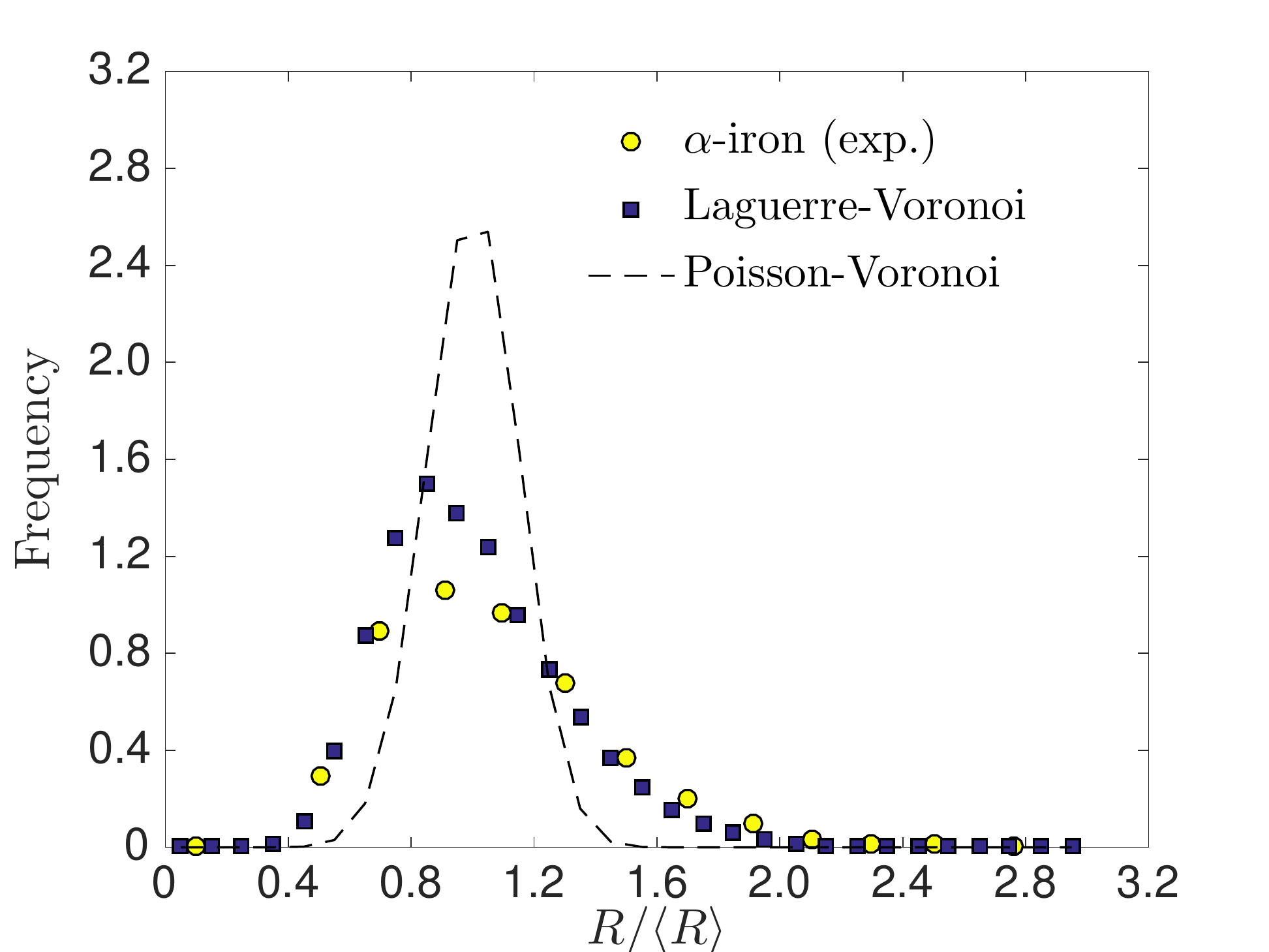}
	\caption{}
	\end{subfigure}
\	
	\begin{subfigure}{0.49\textwidth}
	\centering
	\includegraphics[width=\textwidth]{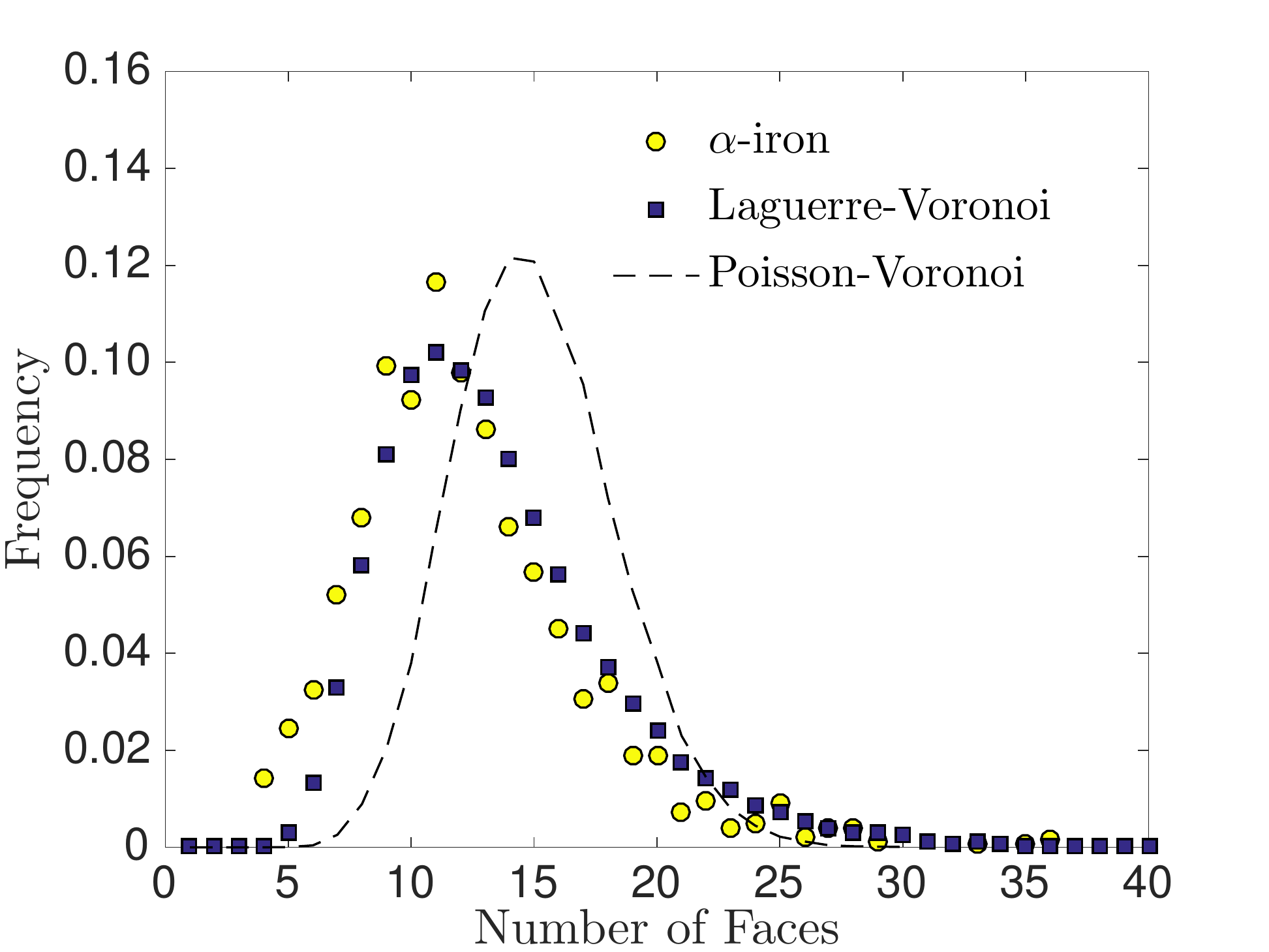}
	\caption{}
	\end{subfigure}
\caption{(\emph{a}) Grain size distribution and (\emph{b}) number of faces statistics generated from standard Poisson-Voronoi tessellations (\emph{dashed line}), experimental measurements \cite{zhang2004} (\emph{circles}) and hardcore Laguerre-Voronoi tessellations based on experimental data (\emph{squares}).}
\label{fig-Ch1:alpha voro stats}
\end{figure}

\begin{figure}[h]
\centering
\includegraphics[width=\textwidth]{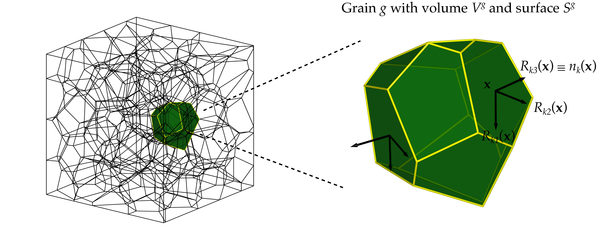}
\caption{Position of the generic grain $g$ inside the polycrystalline aggregate. $V^g$ and $S^g=\pd V^g$ denote the volume occupied by the grain and its boundary, respectively. A local reference system $R_{ki}(\mathbf{x})$ is attached at each point $\mathbf{x}$ of the boundary of the grain to facilitate the expression of the interface equations.}
\label{fig-Ch1:grain RS}
\end{figure}

\section{Grain boundary integral equations}\label{sec-Ch1:grain boundary integral equations}
The polycrystalline morphology is modelled using a multi-domain boundary integral formulation in which each grain $g$ is considered as a generally anisotropic domain with a specific crystallographic orientation in the three-dimensional space.

Let us first consider a generic three-dimensional grain $g$ in a polycrystalline aggregate occupying the volume $V^g\subset \mathbb{R}^3$ and whose boundary is denoted by $S^g=\pd V^g$ (see Figure (\ref{fig-Ch1:grain RS})). In the linear elastic regime, the relation between the second-order stress tensor $\sigma_{ij}^g(\mathbf{x})$ and the second-order strain tensor $\eps_{ij}^g(\mathbf{x})$ can be generally expressed as
\begin{equation}\label{eq-Ch1:constitutive relation}
\sigma_{ij}^g(\mathbf{x})=c_{ijkl}^g\eps_{kl}^g(\mathbf{x})
\end{equation}
where $c_{ijkl}^g$ represents the components of the grain's fourth-order stiffness tensor, which is assumed to be constant for the considered grain. It is recalled here that, in Eq.(\ref{eq-Ch1:constitutive relation}) and throughout the present thesis, repeated subscripts imply summation. In the formulation, the displacements field of the grain is denoted by $u_i^g(\mathbf{x})$ and is related to the strain tensor $\eps_{ij}^g(\mathbf{x})$ whose components are given by
\begin{equation}\label{eq-Ch1:strain-displacement relation}
\eps_{ij}^g(\mathbf{x})=\frac{1}{2}\left[\frac{\pd u_i^g}{\pd x_j}(\mathbf{x})+\frac{\pd u_j^g}{\pd x_i}(\mathbf{x})\right];
\end{equation}
the tractions field defined on a surface with normal $n_i(\mathbf{x})$ is denoted by $t_i^g(\mathbf{x})$ and its components are given by \begin{equation}
t_i^g(\mathbf{x})=\sigma_{ij}^g(\mathbf{x})n_j(\mathbf{x})=n_j(\mathbf{x})c_{ijkl}^g\frac{\pd u_k^g}{\pd x_l}(\mathbf{x}).
\end{equation}

The boundary integral equations for the grain are derived by using the reciprocity relation between the elastic solution of the grain subject to prescribed boundary conditions and the elastic solution in the infinite 3D elastic space of a body force concentrated at the point $\mathbf{y}\in\mathbb{R}^3$.

For any \emph{internal} point $\mathbf{y}$ inside the domain $V^g$, the \emph{displacement boundary integral equations} (DBIE) in absence of body forces are written as follows \cite{banerjee1981,wrobel2002,aliabadi2002}
\begin{equation}\label{eq-Ch1:DBIE}
u_p^g(\mathbf{y})+
\int_{S^g} T_{pi}^g(\mathbf{x},\mathbf{y})u_i^g(\mathbf{x})\dd S(\mathbf{x})=
\int_{S^g} U_{pi}^g(\mathbf{x},\mathbf{y})t_i^g(\mathbf{x})\dd S(\mathbf{x})
\end{equation}
where $\mathbf{x}$ is the \emph{integration point} running over the boundary $S^g$ of the grain $g$. $\mathbf{y}$ is also denoted as the \emph{collocation point} and represents the point at which the boundary integral equations are evaluated. The integration kernels $U_{pi}^g(\mathbf{x},\mathbf{y})$ and $T_{pi}^g(\mathbf{x},\mathbf{y})$ depend on the elastic constants of the considered grain and their expressions are explicitly given in Section (\ref{ssec-Ch1: anisotropic kernels}) in terms of the fundamental solutions of the general anisotropic linear elastic problem. 

The explicit expression of the fundamental solutions is given in Chapter (\ref{ch-FS}). However, it is worth noting here that the kernels in the boundary integral equations considered in this thesis can always be written as the product of a singular function depending on the distance $r=\sqrt{(x_k-y_k)(x_k-y_k)}$ between the collocation point $\mathbf{y}$ and the integration point $\mathbf{x}$, and a regular function depending on the direction $\uv{r}=(\mathbf{x}-\mathbf{y})/r$ and the material properties, i.e.\ each kernel $K(\mathbf{x},\mathbf{y})$ can generally be written as $K(\mathbf{x},\mathbf{y})=r^{-\beta}K(\uv{r})$ being $\beta>0$ an integer representing the order of the singular function. 

For three-dimensional problems, the kernel $U_{pi}^g(\mathbf{x},\mathbf{y})$ has an order of singularity $\beta=1$, whereas the kernel $T_{pi}^g(\mathbf{x},\mathbf{y})$ has an order of singularity $\beta=2$. The order of singularity is of crucial importance when the collocation point is taken at the boundary $S^g$ \cite{banerjee1981,wrobel2002,aliabadi2002}.

In fact, in order to solve the elastic problem given a proper set of displacements and/or tractions boundary conditions over $S^g$, the DBIE (\ref{eq-Ch1:DBIE}) must be evaluated for points belonging to the boundary $S^g$ of the volume $V^g$. However, for a point $\mathbf{y}\in S^g$, given the singular nature of the integration kernels, the integrals appearing in Eq.(\ref{eq-Ch1:DBIE}) must be treated appropriately.

The DBIE for a boundary point are obtained by suitable \emph{limiting processes} that are well described in the BEM literature \cite{banerjee1981,wrobel2002,aliabadi2002}. In order to give the intuition of a boundary limiting process when integrating over the surface $S^g$, one can consider a reference system centered at $\mathbf{y}$. In such a case, the behaviour of the terms $\dd S(\mathbf{x})$, $U_{pi}^g(\mathbf{x},\mathbf{y})$ and $T_{pi}^g(\mathbf{x},\mathbf{y})$, as $\mathbf{x}\rightarrow\mathbf{y}$, are of the order $r^2$, $r^{-1}$ and $r^{-2}$, respectively, being $r$ the distance between $\mathbf{x}$ and $\mathbf{y}$. As a consequence, the integral involving the $U_{pi}^g(\mathbf{x},\mathbf{y})$ is only \emph{weakly singular} and the singularity can be cancelled by the Jacobian of the integration. On the other hand, the integral involving the kernel $T_{pi}^g(\mathbf{x},\mathbf{y})$ is \emph{strongly singular} and the so-called \emph{free terms} remain after the limiting process. For a detailed derivation of the DBIE for a point belonging to the boundary, the reader is referred to \cite{banerjee1981,wrobel2002,aliabadi2002}.

Following considering the limiting process, a more general form of the DBIE can then be written as
\begin{equation}\label{eq-Ch1:DBIE-2}
c_{pi}(\mathbf{y})u_i^g(\mathbf{y})+
\dashint_{S^g} T_{pi}^g(\mathbf{x},\mathbf{y})u_i^g(\mathbf{x})\dd S(\mathbf{x})=
\int_{S^g} U_{pi}^g(\mathbf{x},\mathbf{y})t_i^g(\mathbf{x})\dd S(\mathbf{x})
\end{equation}
where the integration symbol $\dashint$ denotes a Cauchy principal value integral and the terms $c_{pi}^g(\mathbf{y})u_i(\mathbf{y})$ are the free terms that stem from the limiting process. Depending on the position of the point $\mathbf{y}$, the coefficients $c_{pi}(\mathbf{y})$ take the following values
\begin{align}\label{eq-Ch1:DBIE free terms}
c_{pi}(\mathbf{y})=
\begin{cases}
\delta_{pi},& \mathrm{if}\quad \mathbf{y}\ \mathrm{inside}\ V^g\\
\delta_{pi}/2,& \mathrm{if}\quad \mathbf{y}\ \mathrm{on}\ S^g\ \mathrm{smooth}\\
0,&\mathrm{if}\quad \mathbf{y}\ \mathrm{outside}\ V^g
\end{cases},
\end{align}
where $\delta_{pi}$ is the Kronecker delta function.

The strain and the stress fields at any internal point $\mathbf{y}$ inside the grain $g$ can be computed by suitably taking the derivatives of Eq.(\ref{eq-Ch1:DBIE}) with respect to the coordinates $y_i$ of the collocation point $\mathbf{y}$. The \emph{strain boundary integral equations} are then obtained by using the strain-displacement relationships and are given as follows
\begin{equation}\label{eq-Ch1:EBIE}
\eps_{pq}^g(\mathbf{y})+
\int_{S^g} T_{pqi}^{\eps,g}(\mathbf{x},\mathbf{y})u_i^g(\mathbf{x})\dd S(\mathbf{x})=
\int_{S^g} U_{pqi}^{\eps,g}(\mathbf{x},\mathbf{y})t_i^g(\mathbf{x})\dd S(\mathbf{x}),
\end{equation}
where the kernels $U_{pqi}^{\eps,g}(\mathbf{x},\mathbf{y})$ and $T_{pqi}^{\eps,g}(\mathbf{x},\mathbf{y})$ are obtained in terms of the kernels $U_{pi}^g(\mathbf{x},\mathbf{y})$ and $T_{pi}^g(\mathbf{x},\mathbf{y})$, respectively, and their expressions are given in Section (\ref{ssec-Ch1: anisotropic kernels}). By using the grain's constitutive relations, the \emph{stress boundary integral equations} are expressed as
\begin{equation}\label{eq-Ch1:SBIE}
\sigma_{mn}^g(\mathbf{y})+
\int_{S^g} T_{mni}^{\sigma,g}(\mathbf{x},\mathbf{y})u_i^g(\mathbf{x})\dd S(\mathbf{x})=
\int_{S^g} U_{mni}^{\sigma,g}(\mathbf{x},\mathbf{y})t_i^g(\mathbf{x})\dd S(\mathbf{x}),
\end{equation}
where the kernels $U_{mni}^{\sigma,g}(\mathbf{x},\mathbf{y})$ and $T_{mni}^{\sigma,g}(\mathbf{x},\mathbf{y})$ are obtained by applying the constitutive relation to the kernels $U_{pqi}^{\eps,g}(\mathbf{x},\mathbf{y})$ and $T_{pqi}^{\eps,g}(\mathbf{x},\mathbf{y})$, respectively. Eqs.(\ref{eq-Ch1:EBIE}) and (\ref{eq-Ch1:SBIE}) provide a relation to compute the strain and the stress fields, respectively, within the volume $V^g$ of the grain in terms of the values of the displacements and the tractions at the boundary $S^g$ and are usually employed in a post processing stage after the solution of the elastic problem is obtained by means of Eq.(\ref{eq-Ch1:DBIE-2}).

The DBIE given in Eq.(\ref{eq-Ch1:DBIE-2}) represent the starting point of the grain boundary formulation of polycrystalline mechanics. In fact, the DBIE must be written for each crystal of the aggregate and suitable boundary conditions and interface conditions must be enforced to the grain boundaries touching the external domain and to the internal grain boundaries, respectively. To simplify the expression of the interface boundary conditions, the displacements and the tractions at the internal grain boundaries are usually expressed in terms of their \emph{tangential} and \emph{normal} components. It is therefore convenient to define a local reference system over each face (see Figure (\ref{fig-Ch1:grain RS})) of the boundaries of the constituent grains and express displacements and tractions in that reference system. In the local reference system, the displacements and the tractions are written in terms of the local transformation matrix $R_{ij}(\mathbf{x})$, which depends on the position of the considered point on the boundary. The DBIE can then be rewritten in a more convenient form as
\begin{equation}\label{eq-Ch1:DBIE-2 local RS}
\wtilde{c}_{pi}(\mathbf{y})\wtilde{u}_i^g(\mathbf{y})+
\dashint_{S^g} \wtilde{T}_{pi}^g(\mathbf{x},\mathbf{y})\wtilde{u}_i^g(\mathbf{x})\dd S(\mathbf{x})=
\int_{S^g} \wtilde{U}_{pi}^g(\mathbf{x},\mathbf{y})\wtilde{t}_i^g(\mathbf{x})\dd S(\mathbf{x}),
\end{equation}
where
\begin{equation}
\wtilde{c}_{pi}(\mathbf{y})=c_{pk}(\mathbf{y})R_{ki}(\mathbf{y}),
\end{equation}
\begin{equation}
\wtilde{U}_{pi}^g(\mathbf{x},\mathbf{y})=U_{pk}^g(\mathbf{x},\mathbf{y})R_{ki}(\mathbf{x}),\quad
\wtilde{T}_{pi}^g(\mathbf{x},\mathbf{y})=T_{pk}^g(\mathbf{x},\mathbf{y})R_{ki}(\mathbf{x})
\end{equation}
and $R_{ij}(\mathbf{x})$ relate the components of the grain boundary displacements and tractions in the global reference system to the components in the local reference system, i.e.\
\begin{equation}\label{eq-Ch1:local u and t}
u_k^g(\mathbf{x})=R_{ki}(\mathbf{x})\wtilde{u}_i^g(\mathbf{x}),\quad
t_k^g(\mathbf{x})=R_{ki}(\mathbf{x})\wtilde{t}_i^g(\mathbf{x}).
\end{equation}

In Eqs.(\ref{eq-Ch1:DBIE-2 local RS}) to (\ref{eq-Ch1:local u and t}) and throughout the present thesis, the symbol $\wtilde{\cdot}$ denotes a quantity expressed in the local reference system of the grain boundary.

\begin{figure}[h]
\centering
\includegraphics[width=\textwidth]{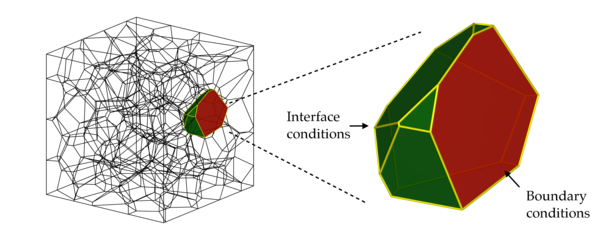}
\caption{Example of a grain for which the external and internal grain boundaries are distinguished. Boundary conditions are enforced on the external boundaries (in red) whereas interface conditions are enforced on internal boundaries (in green).}
\label{fig-Ch1:grain BCs-ICs}
\end{figure}

\begin{figure}[h]
\centering
\includegraphics[width=\textwidth]{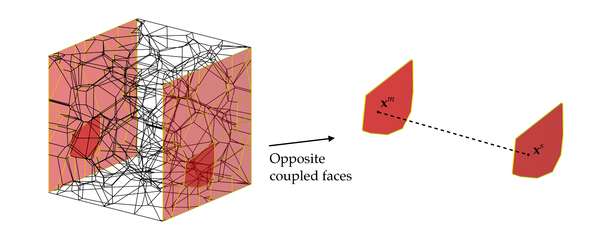}
\caption{Example of a periodic polycrystalline morphology and detail of the pair of opposite master point $\mathbf{x}^m$ and slave point $\mathbf{x}^s$ for the enforcement of periodic boundary conditions.}
\label{fig-Ch1:grain PBCs}
\end{figure}

\subsection{Boundary conditions}\label{ssec-Ch1:BCs}
Boundary conditions (BCs) generally refer to prescribed values of displacements or tractions that are enforced at the boundary of each considered domain. However, in case of the polycrystalline problem external grain boundary and internal grain boundaries must be distinguished. Therefore, throughout the present thesis, boundary conditions will refer to prescribed values of displacements and tractions enforced over the external boundaries of the grains, that is those boundaries of the grains that touch the boundary of the external domain as shown in Figure (\ref{fig-Ch1:grain BCs-ICs}), in which the internal grain boundaries of a representative boundary grain are coloured in green and the external grain boundary is coloured in red.

Different sets of boundary conditions can be assigned to polycrystalline aggregates based on the considered applications. Kinematic or static BCs are usually employed to model the response of a polycrystalline structure to a particular loading condition. On the other hand, to extract homogenised properties of polycrystalline RVEs, periodic boundary conditions (PBCs) have been shown to provide faster convergence to the effective properties with respect to displacement or traction BCs \cite{terada2000}. Considering the external faces of the aggregate, and in particular the pair of opposite \emph{master} point $\mathbf{x}^m$ and \emph{slave} point $\mathbf{x}^s$, PBCs are enforced as follows
\begin{subequations}\label{eq-Ch1:periodic boundary conditions}
\begin{align}
&u_i(\mathbf{x}^s)=u_i(\mathbf{x}^m)+\bar{\Gamma}_{ij}(x_j^s-x_j^m),\\
&t_i(\mathbf{x}^s)+t_i(\mathbf{x}^m)=0,
\end{align}
\end{subequations}
where $\bar{\Gamma}_{ij}$ represents the prescribed strain tensor. Even though it does not represent a requirement, the use of periodic boundary conditions is facilitated by the periodic structure of the aggregate and therefore by the conformity between the meshes of coupled opposite faces. Figure (\ref{fig-Ch1:grain PBCs}) shows a periodic polycrystalline morphology, whose generation is detailed is Section (\ref{sec-Ch3:modified periodic morphology}) of Chapter (\ref{ch-EF}), and the position of the master point $\mathbf{x}^m$ and the slave point $\mathbf{x}^s$ on opposite faces of the aggregate for which the PBCs given in Eq.(\ref{eq-Ch1:periodic boundary conditions}) are enforced.

\subsection{Interface conditions}\label{ssec-Ch1:ICs}
The interface conditions (ICs) express the relationships between the displacements and the traction fields at the interface between two adjacent grains. Consider the two grains in Figure (\ref{fig-Ch1:2 neighbour grains}), where the two adjacent grains $g$ and $h$ share the interface $\mathscr{I}^{gh}=V^g\cap V^h$ that is highlighted by darker green. The interface conditions can be generally expressed, $\forall \mathbf{x} \in \mathscr{I}^{gh}$, as follows
\begin{subequations}\label{eq-Ch1:interface equations}
\begin{align}
&\Psi_i^{gh}[\wtilde{u}_j^g(\mathbf{x}),\wtilde{u}_j^h(\mathbf{x}),\wtilde{t}_j^g(\mathbf{x}),\wtilde{t}_j^h(\mathbf{x})]=0,\\
&\Theta_i^{gh}[\wtilde{t}_j^g(\mathbf{x}),\wtilde{t}_j^h(\mathbf{x})]=0,
\end{align}
\end{subequations}
where $\Theta_i^{gh}$ represents grain boundary equilibrium conditions whereas $\Psi_i^{gh}$ represents generic relations between the grain boundary tractions and displacements that can include from pristine to completely failed interfaces based on the state of the grain boundaries. More details on the interface conditions used for the polycrystalline problems addressed in this thesis are given based on the specific considered application.

As previously mentioned, it is more convenient to express the interface conditions in terms of the normal and tangential components of the grain boundary displacements and tractions and this is the reason why the functional dependence of the ICs in Eq.(\ref{eq-Ch1:interface equations}) is given in terms of the local fields $\wtilde{u}_i$ and $\wtilde{t}_i$. Moreover, in case of a damaged or failing interface, Eq.(\ref{eq-Ch1:interface equations}a) is usually given in terms of the displacements jump at the interface. By defining opposite local reference systems at coupled points of the interface as shown in Figure (\ref{fig-Ch1:2 neighbour grains}), i.e.\ $R_{ki}(\mathbf{x}\in V^g)=-R_{ki}(\mathbf{x}\in V^h)$, the local displacements jump $\delta\wtilde{u}_i^{gh}$ can be written as
\begin{equation}\label{eq-Ch1:local du}
\delta\wtilde{u}_i^{gh}=-(\wtilde{u}_i^g+\wtilde{u}_i^h).
\end{equation}
Similarly, the equilibrium conditions considering opposite local reference systems can be written as
\begin{equation}\label{eq-Ch1:local t}
\wtilde{t}_i^g=\wtilde{t}_i^h=\wtilde{t}_i^{gh}.
\end{equation}
where $\wtilde{t}_i^{gh}$ is used to denote the values of the tractions at the interface $\mathscr{I}^{gh}$.

\begin{figure}[h]
\centering
\includegraphics[width=0.7\textwidth]{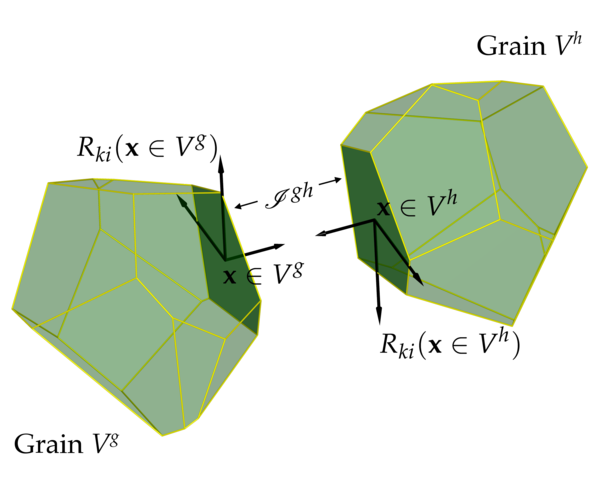}
\caption{Schematic representation of two opposite reference systems at the interface $\mathscr{I}^{gh}$ between the adjacent grains $g$ and $h$.}
\label{fig-Ch1:2 neighbour grains}
\end{figure}

\subsection{Anisotropic Kernels}\label{ssec-Ch1: anisotropic kernels}
The kernels appearing in the boundary integral equations (\ref{eq-Ch1:DBIE}), (\ref{eq-Ch1:DBIE-2}), (\ref{eq-Ch1:EBIE}) and (\ref{eq-Ch1:SBIE}) are computed in terms of the fundamental solutions of the generally anisotropic elastic problem. More specifically, the fundamental solutions $\Phi_{pi}(\mathbf{x},\mathbf{y})$ of general anisotropic elasticity are obtained as solutions of the following problem
\begin{equation}\label{eq-Ch1:fundamental solutions problem}
c_{ijkl}^g\frac{\pd^2 \Phi_{pk}^g}{\pd x_j\pd x_l}(\mathbf{x},\mathbf{y})+\delta_{pi}\delta(\mathbf{x}-\mathbf{y})=0,
\end{equation}
where $\delta(\mathbf{x}-\mathbf{y})$ is the Dirac delta function and the superscript $g$ in $\Phi_{pk}^g$ and in the stiffness tensor $c_{ijkl}^g$ highlights the fact that the fundamental solutions depend on the specific considered material. The explicit solution of Eq.(\ref{eq-Ch1:fundamental solutions problem}) is discussed in a more general fashion in Chapter (\ref{ch-FS}) wherein a novel and unified compact expression of the fundamental solutions of a broader class of partial differential operators, i.e.\ second-order homogeneous elliptic partial differential operators, is obtained in terms of spherical harmonics expansions. Once the fundamental solutions $\Phi_{pi}(\mathbf{x},\mathbf{y})$ and their derivatives are computed, the kernels in Eqs. (\ref{eq-Ch1:DBIE}), (\ref{eq-Ch1:DBIE-2}) are obtained as follows
\begin{subequations}\label{eq-Ch1:DBIE kernels}
\begin{align}
U_{pi}^g(\mathbf{x},\mathbf{y})&=\Phi_{pi}^g(\mathbf{x},\mathbf{y}),\\
\Sigma_{pij}^g(\mathbf{x},\mathbf{y})&=c_{ijkl}^g\frac{\pd \Phi_{pk}^g}{\pd x_l}(\mathbf{x},\mathbf{y}),\\
T_{pi}^g(\mathbf{x},\mathbf{y})&=\Sigma_{pij}^g(\mathbf{x},\mathbf{y})n_j(\mathbf{x}).
\end{align}
\end{subequations}

The kernels appearing in Eq.(\ref{eq-Ch1:EBIE}) are computed by taking the derivatives of the kernels in Eq.(\ref{eq-Ch1:DBIE kernels}) with respect to the collocation point coordinates $y_i$ and are written as
\begin{subequations}\label{eq-Ch1:EBIE kernels}
\begin{align}
U_{pqi}^{g,\eps}(\mathbf{x},\mathbf{y})&=\frac{1}{2}\left[\frac{\pd \Phi_{pi}^g}{\pd y_q}(\mathbf{x},\mathbf{y})+\frac{\pd \Phi_{qi}^g}{\pd y_p}(\mathbf{x},\mathbf{y})\right],\\
\Sigma_{pqij}^{g,\eps}(\mathbf{x},\mathbf{y})&=\frac{1}{2}c_{ijkl}^g\frac{\pd}{\pd x_l}\left[\frac{\pd \Phi_{pk}^g}{\pd y_q}(\mathbf{x},\mathbf{y})+\frac{\pd \Phi_{qk}^g}{\pd y_p}(\mathbf{x},\mathbf{y})\right],\\
T_{pqi}^{g,\eps}(\mathbf{x},\mathbf{y})&=\Sigma_{pqij}^{g,\eps}(\mathbf{x},\mathbf{y})n_j(\mathbf{x}).
\end{align}
\end{subequations}

Finally, the kernels appearing in the stress boundary integral equations (\ref{eq-Ch1:SBIE}) are obtained upon applying the constitutive relations to the kernels given in Eq.(\ref{eq-Ch1:EBIE kernels}) and their expressions are
\begin{subequations}\label{eq-Ch1:SBIE kernels}
\begin{align}
U_{mni}^{g,\sigma}(\mathbf{x},\mathbf{y})&=c_{mnpq}^g\frac{\pd \Phi_{pi}^g}{\pd y_q}(\mathbf{x},\mathbf{y}),\\
\Sigma_{mnij}^{g,\sigma}(\mathbf{x},\mathbf{y})&=c_{mnpq}^gc_{ijkl}^g\frac{\pd^2 \Phi_{pk}^g}{\pd x_l\pd y_q}(\mathbf{x},\mathbf{y}),\\
T_{mni}^{g,\sigma}(\mathbf{x},\mathbf{y})&=\Sigma_{mnij}^{g,\sigma}(\mathbf{x},\mathbf{y})n_j(\mathbf{x}).
\end{align}
\end{subequations}

In Eqs.(\ref{eq-Ch1:DBIE kernels}b), (\ref{eq-Ch1:EBIE kernels}b) and (\ref{eq-Ch1:SBIE kernels}b) the kernels $\Sigma_{pij}^g(\mathbf{x},\mathbf{y})$, $\Sigma_{pqij}^{g,\eps}(\mathbf{x},\mathbf{y})$ and $\Sigma_{mnij}^{g,\sigma}(\mathbf{x},\mathbf{y})$, respectively, have been introduced as their expressions will be used in Chapter (\ref{ch-CP}).

\subsection{Numerical Integration}\label{ssec-Ch1: numerical integration}
To solve numerically the polycrystalline problem, the boundary integral equations (\ref{eq-Ch1:DBIE-2}) are used in the context of the boundary element method for each grain $g$ of the aggregate, according to the following steps:
\begin{itemize}
\item{
The boundary $S^g$ of each grain is subdivided into sets of non-overlapping surface elements following the strategy described in Chapter (\ref{ch-EF}), wherein an enhanced meshing technique for polycrystalline aggregates is developed for computational effectiveness;
}
\item{
The boundary displacements and tractions fields are then expressed in terms of suitable shape functions $\mathbf{N}(\boldsymbol{\xi})$, defined over each element in a local 2D (surface) coordinate system $\boldsymbol{\xi}=\{\xi_1,\xi_2\}$, and nodal values of boundary displacements and tractions expressed in a face local reference system. After this discretisation procedure, a set of nodal points, or simply \emph{nodes}, and boundary elements are identifiable for each grain;
}
\item{
The DBIE (\ref{eq-Ch1:DBIE-2}) are then written for each node of the considered grain and it is \emph{numerically integrated}, considering the explicit approximation of the boundary fields in terms of shape functions and nodal values.
}
\end{itemize}
In this way, a set of $3N_p^g$ equations, where $N_p$ is the number of nodes used in the discretisation of the grain $g$, is written in terms of nodal values of displacements and tractions and the discrete version of the displacement boundary integral equations reads
\begin{equation}\label{eq-Ch1:DBIE discrete}
\mathbf{H}^{g}\mathbf{U}^{g} = \mathbf{G}^{g}\mathbf{T}^{g}
\end{equation}
where the vectors $\mathbf{U}^{g}$ and $\mathbf{T}^{g}$ are $3N_p^g$ vectors collecting the nodal values of boundary displacements and tractions, respectively, and the matrices $\mathbf{H}^{g}$ and $\mathbf{G}^{g}$ are $3N_p^g\times 3N_p^g$ matrices stemming from the integration of the kernels $\wtilde{T}_{ij}^g$ and $\wtilde{U}_{ij}^g$, respectively. In the numerical integration procedure, care must be taken when integrating over the elements that contain, for a given collocation point, the collocation point itself. Such \emph{singular} elements are treated, in the present work, using element subdivision and coordinate transformation for the \emph{weakly singular} kernels $\wtilde{U}_{ij}^{g}$, and rigid body conditions for the \emph{strongly singular} kernels $\wtilde{T}_{ij}^{g}$. The interested readers are referred to \cite{banerjee1981,aliabadi2002} for such boundary element specific aspects.

Enforcing displacement and traction boundary conditions on the faces of the grains lying on the loaded external boundary of the aggregate, Eq.(\ref{eq-Ch1:DBIE discrete}) leads, for \emph{each} grain, to the following system of equations
\begin{equation}\label{eq-Ch1:DBIE discrete + BCs}
\mathbf{A}^{g}\mathbf{X}^{g} = \mathbf{C}^{g}\mathbf{Y}^{g}(\lambda)
\end{equation}
where $\mathbf{X}^g$ collects the unknown nodal values of grain boundary displacements and tractions, $\mathbf{Y}^g$ collects prescribed values of boundary displacements and tractions, and $\mathbf{A}^g$ and $\mathbf{C}^g$ contain suitable combination of the columns of the matrices $\mathbf{H}^{g}$ and $\mathbf{G}^{g}$. In Eq.(\ref{eq-Ch1:DBIE discrete + BCs}), the prescribed values of boundary displacements and tractions are given as a function of the load factor $\lambda$ governing the loading history.

It is clear that, at the internal grain boundaries, neither the displacements nor the tractions are known and therefore the matrix $\mathbf{A}^g$ is not a square matrix; that is, the system of equations (\ref{eq-Ch1:DBIE discrete + BCs}) for a single grain has more unknowns than equations. To solve the polycrystalline problem, Eq.(\ref{eq-Ch1:DBIE discrete + BCs}) must be written for each grain and coupled to suitable interface conditions. The implementation of the polycrystalline system of equations is described next.

\section{Polycrystalline implementation}\label{sec-Ch1: polycrystalline system of equations}
To obtain the system of equations of the entire polycrystalline aggregates, the system of equations given in (\ref{eq-Ch1:DBIE discrete + BCs}) is written for each grain of the aggregate, that is for $g=1,\dots,N_g$; the integrity of the microstructure is retrieved in the model by enforcing suitable inter-granular conditions, which leads to the following system
\begin{equation}\label{eq-Ch1:polycrystalline DBIE discrete}
\mathbf{M}(\mathbf{X},\lambda)=\left\{
\begin{array}{c}
\mathbf{A}\mathbf{X}-\mathbf{B}(\lambda)\\
\mathbf{I}(\mathbf{X})
\end{array}
\right\}=\mathbf{0}
\end{equation}
where
\begin{equation}\label{eq-Ch1:polycrystalline DBIE matrix A b X}
\mathbf{A}=
\left[\begin{array}{cccc}
\mathbf{A}^{1}& \mathbf{0}&\cdots&\mathbf{0}\\
\mathbf{0}& \mathbf{A}^2&\cdots&\mathbf{0}\\
\vdots&\vdots&\ddots&\vdots\\
\mathbf{0}&\mathbf{0}&\cdots&\mathbf{A}^{N_g}
\end{array}\right],\
\mathbf{B}=
\left[\begin{array}{c}
\mathbf{C}^{1}\mathbf{Y}^{1}\\
\mathbf{C}^{2}\mathbf{Y}^{2}\\
\vdots\\
\mathbf{C}^{N_g}\mathbf{Y}^{N_g}
\end{array}\right],\
\mathbf{X}=
\left[\begin{array}{c}
\mathbf{X}^{1}\\
\mathbf{X}^{2}\\
\vdots\\
\mathbf{X}^{N_g}
\end{array}\right],
\end{equation}
being $\mathbf{X}$ the vector collecting the unknown degrees of freedom of the polycrystalline system, i.e.\ the unknown displacements and tractions at the grain boundaries. In Eq.(\ref{eq-Ch1:polycrystalline DBIE discrete}), $\mathbf{I}(\mathbf{X})$ implements the interfaces conditions, which are written for each pair of coupled interface nodes according to Eq.(\ref{eq-Ch1:interface equations}) and therefore are generally function of the unknown nodal displacements and tractions $\mathbf{X}$. 

\subsection{System solution}\label{ssec-Ch1: system solution}
The system of equations given in (\ref{eq-Ch1:polycrystalline DBIE discrete}) must be solved at each step of the loading history. Calling $\lambda_n$ the load factor at the $n$-th load step of the loading history, the corresponding solution $\mathbf{X}_n$ is obtained by solving $\mathbf{M}(\mathbf{X}_n,\lambda_n)=\mathbf{0}$. The solution is computed by employing a Newton-Raphson iterative algorithm, i.e., at the $n$-th load step, the $(k+1)$-th approximation $\mathbf{X}_n^{k+1}$ of the system solution is obtained as
\begin{equation}\label{eq-Ch1:NR solution}
\mathbf{X}_n^{k+1}=\mathbf{X}_n^k-(\mathbf{J}_n^k)^{-1}\mathbf{M}_n^k,
\end{equation}
where $\mathbf{M}_n^k\equiv\mathbf{M}(\mathbf{X}_n^k,\lambda_n)$, the starting guess solution is chosen as the last converged solution, i.e.\ $\mathbf{X}_n^0\equiv\mathbf{X}_{n-1}$, and $\mathbf{J}_n^k\equiv\mathbf{J}(\mathbf{X}_n^k,\lambda_n)$ is the Jacobian of the system defined as
\begin{equation}\label{eq-Ch1:jacobian of the system}
\mathbf{J}(\mathbf{X},\lambda)=\frac{\pd \mathbf{M}}{\pd \mathbf{X}}(\mathbf{X},\lambda).
\end{equation}

Given the structure of the overall system of equations (\ref{eq-Ch1:polycrystalline DBIE discrete}), it is easy to see that the Jacobian defined in Eq.(\ref{eq-Ch1:jacobian of the system}) is a highly sparse matrix, especially for a large number of grains. As a consequence, the package \texttt{PARDISO} (\url{http://www.pardiso-project.org}) \cite{pardiso1,pardiso2,pardiso3} is used as a sparse solver to compute the term $(\mathbf{J}_n^k)^{-1}\mathbf{M}_n^k$ in Eq.(\ref{eq-Ch1:NR solution}). To enhance the convergence and the whole solution calculation, the zero and non-zero elements of the Jacobian are defined at the beginning of the loading history in such a way that its sparsity pattern is retained throughout the analysis. Furthermore, the solver \texttt{PARDISO} is used both as a direct and an iterative solver. More specifically, by considering that during subsequent load steps the changes in the Jacobian matrix are small, the same LU factorisation at a certain load step is used as a preconditioner of the iterative solution at the successive steps, whereas the new LU factorisation is computed only when the iterative approach does not converge. Such an approach drastically decreases the computational cost of the solution. The reader is referred to the \texttt{PARDISO} documentation for more details.

A direction of further computational enhancement could involve the use of specialised high efficiency solvers, based for example on the use of the hierarchical format in conjunction with iterative solution strategies \cite{benedetti2008,benedetti2009}.

\section{Key features of the formulation}
To summarise, the salient features of the method are:
\begin{itemize}
\item{
The polycrystalline microstructural problem is entirely formulated in terms of boundary integral equations, in which the grain-boundary displacements and tractions are \emph{primary unknowns}; in the present context, the term \lq\lq grain-boundary\rq\rq\ refers to this specific aspect;
}
\item{
The grain-boundary nature of the formulation induces a natural simplification of the pre-processing stage, as no volume meshing of the grains' interior volume is required; in a context characterised by an inherent statistical variability of the physical problem morphology, this is an important computational aspect, as it confines potential meshing issues to the boundary of the grains;
}
\item{Since the problem is solved using only nodal points lying on the surface of the grains, the formulation induces a reduction in the order of the solving system, i.e.\ in the number of degrees of freedom; in a multiscale perspective, this represents a computational benefit;
}
\item{
The strain and stress fields at interior points, within the grains, can be computed from grain-boundary variables using a stress boundary integral equation, in a post-processing stage;
}
\item{
In the present form, non-linear constitutive behaviors \emph{within} the grains are not considered; this is not an inherent limitation of the formulation, as such constitutive behaviors can be included introducing suitable volume terms (and volume meshes) in the boundary integral equations, \emph{without introducing additional degrees of freedom}, as it will be shown in Chapter (\ref{ch-CP}) when the problem of crystal plasticity is addressed.
}
\end{itemize}

\section{Content of the thesis}
The following Chapters of this thesis are organised as follows: Chapter (\ref{ch-FS}) presents a novel, unified and compact expression for the fundamental solutions of homogeneous second-order differential operators and their derivatives. An efficient evaluation of the fundamental solutions of partial differential equations is still of great engineering interest as they represent the key ingredient in the boundary integral equations of the boundary element method approach employed in the present thesis, see Eqs.(\ref{eq-Ch1:DBIE},\ref{eq-Ch1:DBIE-2},\ref{eq-Ch1:EBIE},\ref{eq-Ch1:SBIE}). The expression found in this thesis has the advantages of being general for the considered class of differential operators and simpler with respect to the formulas found in the literature, especially when high-order derivatives of the fundamental solutions need to be computed.

In Chapter (\ref{ch-EF}), an enhanced grain boundary framework for the study of polycrystalline materials is presented. The framework is aimed at reducing the high computational cost that hinders the development of three-dimensional non-linear polycrystalline models using the boundary element approach. The use of a regularised tessellation and the development of a novel ad-hoc meshing strategy to tackle the high statistical variability of polycrystalline micro-morphologies are key points of the enhanced framework. The effects of the adopted strategies on the number of degrees of freedom of polycrystalline systems and on the computational costs of micro-cracking analyses are described by several studies.

In Chapter (\ref{ch-TG}), a novel formulation for the competition between inter-granular and trans-granular cracking in polycrystalline materials is developed. The two mechanisms are modelled using the cohesive zone approach. However, unlike inter-granular cracks, which are well defined when the polycrystalline morphologies are generated, trans-granular cracks are introduced on-the-fly during the micro-cracking analysis and the polycrystalline morphologies are remeshed accordingly. 

In Chapter (\ref{ch-CP}), a grain-boundary formulation for the mechanism of small strain crystal plasticity is presented and implemented in polycrystalline materials for the first time. The boundary integral equations involving the presence of general anisotropic plastic strain in polycrystalline materials are extensively described. The iterative incremental algorithm for the solution of the crystal plasticity polycrystalline problems is discussed and numerical tests involving single crystals and polycrystalline materials are presented.

\clearpage

\chapter{Fundamental solutions for multi-field materials based on spherical harmonics expansions}\label{ch-FS}

\section{Introduction}\label{sec-Ch2:intro}
Fundamental solutions are essential to the solution of many boundary value problems in engineering \cite{mura2013} and represent the key ingredient in boundary integral formulations \cite{banerjee1981,wrobel2002,aliabadi2002}. Simple and closed form expressions are available only for simple cases, such as potential problems or isotropic elasticity. Therefore, the development of efficient schemes for computing the fundamental solutions of generic linear systems of partial differential equations (PDEs) still represents a challenge of great scientific interest.

In the field of materials engineering, several works have been devoted to finding the displacements field \cite{fredholm1900,lifshitz1947} and its derivatives \cite{barnett1972} due to a point force in a three-dimensional anisotropic elastic medium. The formal expression of the fundamental solutions of a second order partial differential operator has been classically obtained using either Fourier or Radon transforms, which lead to expressions in terms of a contour integral on a unit circle \cite{synge1957}. By using suitable variable transformations, several researchers replaced the contour integral by an infinite line integral \cite{wang1997}, whose solution has been obtained by means of the Cauchy's residue theorem in terms of the Stroh eigenvalues \cite{wu1998,sales1998,lee2003} or eigenvectors \cite{malen1971,nakamura1997}. However, when the Stroh's eigenvalues or eigenvectors approach is used, the issue of \emph{degeneracy} needs to be robustly addressed, in particular when such an approach is used in a numerical code, e.g.\ in boundary element implementations.  Non-degenerate cases were first studied by Dederichs and Leibfried \cite{dederichs1969} for cubic crystals. Phan et al.\ \cite{phan2004,phan2005} presented a technique to compute the fundamental solutions of a 3D anisotropic elastic solid and their first derivatives in presence of multiple roots and using the residue approach. Shiah et al.\ \cite{shiah2008} used the spherical coordinates differentiation to obtain the explicit expressions of the derivatives of the fundamental solutions for an anisotropic elastic solid up to the second order. Although being exact, the main disadvantage of these approaches is the necessity of using different expressions for each different case of different roots, two coincident roots, and three coincident roots. A unified formulation, valid for degenerate as well as non-degenerate cases, has been first presented by Ting and Lee \cite{ting1997}. Their approach has been further investigated by the recent work of Xie et al.\ \cite{xie2016}, in which the authors developed a unified approach to compute the fundamental solutions of 3D anisotropic solids valid for partially degenerate, fully degenerate and non-degenerate materials. Although not suffering from material's degeneracy, the expressions presented in the work of Xie et al.\ \cite{xie2016} are valid only for anisotropic elastic materials, are given up to second order of differentiation and are rather long and complex, in particular when the derivatives are considered. A 2D Radon transform approach has been also used in the literature as an alternative approach to the problem \cite{buroni2014,xie2016b}.

The aforementioned works have been mainly devoted to the derivation of the fundamental solution of elastic anisotropic materials. Recently, multi-field materials have been receiving an increasing interest for their application in composite multi-functional devices \cite{nan2008,peng2014}. Closed form expressions can be found for transversely isotropic materials showing piezo-electric \cite{dunn1994,dunn1996} and magneto-electro-elastic \cite{wang2002,soh2003,hou2005,ding2005} coupling. Pan and Tonon \cite{pan2000} used the Cauchy's residue theorem to derive the fundamental solutions of non-degenerate anisotropic piezo-electric solids, and a finite difference scheme to obtain their derivatives. Buroni and Saez \cite{buroni2010} used the Cauchy's residue theorem and the Stroh's formalism to obtain the fundamental solutions of degenerate or non-degenerate anisotropic magneto-electro-elastic materials. Their scheme was recently employed in a boundary element code for fracture analysis \cite{munoz2015}.

From a numerical perspective, the fundamental solutions represent the essential ingredient in boundary integral formulations, such as the Boundary Element Method (BEM) \cite{banerjee1981,wrobel2002,aliabadi2002}. In practical BEM analyses of engineering interest, the fundamental solutions and their derivatives are computed in the order of million times and the availability of efficient schemes for their evaluation is thus of great interest, especially for large 3D problems. To accelerate the computations for anisotropic elastic materials, Wilson and Cruse \cite{wilson1978} proposed to pre-compute the values of the fundamental solutions at regularly spaced points of a spatial grid and to use an interpolation scheme with cubic splines to approximate their values in general points during the subsequent BEM analysis. Such an approach and similar interpolation techniques \cite{schclar1994} have been widely employed in the BEM literature \cite{benedetti2009,milazzo2012,benedetti2013a,benedetti2013b,gulizzi2015,benedetti2016}.
Mura and Kinoshita \cite{mura1971} represented the fundamental solutions of a general anisotropic elastic medium in terms of spherical harmonics expansions and used a term-by-term differentiation to obtain the first derivative. Aubrys and Arsenalis \cite{aubry2013} used the spherical harmonics expansions for dislocation dynamics in anisotropic elastic media and pointed out that line integrals and double line integrals could be obtained analytically once the series coefficients were computed.
Recently, Shiah et al.\ \cite{shiah2012} proposed an alternative scheme to compute the fundamental solutions of 3D anisotropic elastic solids based on a double Fourier series representation. The authors expressed the fundamental solutions as given by Ting and Lee \cite{ting1997} in the spherical reference system and then built their Fourier series representation relying on their periodic nature. The authors underlined that the coefficients of the series were computed only once for a given material and employed their method in a BEM code \cite{tan2013}. They also obtained the first and the second derivatives of the fundamental solutions whose complexity increases with the order of differentiation, despite the use of the spherical coordinates to obtain the derivatives. The interested reader is referred to the book by Pan and Chen \cite{pan2015} and to the recent paper by Xie et al.\ \cite{xie2016b} for a comprehensive overview of the available methods to obtain the fundamental solutions.

In the present Chapter, given a generic linear system of PDEs defined by a homogeneous elliptic partial differential operator, the fundamental solutions and their derivatives are computed in a unified fashion in terms of spherical harmonics expansions \cite{gulizzi2016b}. It is here demonstrated that the formula found by Mura and Kinoshita \cite{mura1971} is in fact a particular case of a more general representation of the fundamental solutions and their derivatives, which are not obtained by a term-by-term differentiation and can be computed up to the desired order. The coefficients of the series depend on the material constants and need to be computed only once, thus making the present scheme attractive for efficient boundary integral formulations. Eventually, mathematically degenerate media do not require any specific treatment and the present scheme can be generally employed to cases ranging from simply isotropic to more complex general anisotropic differential operators. 
To the best of the author's knowledge, it is the first time that the fundamental solutions for generally anisotropic multi-field materials and their derivatives up to any order are represented in such compact unified fashion.

The Chapter is organised as follows: Section(\ref{sec-Ch2:problem statement - system of pdes}) introduces the class of systems of partial differential equations and the corresponding fundamental solutions that will be addressed in the present study; Section(\ref{sec-Ch2:fourier transform + plane wave expansion}) illustrates the mathematical steps needed to obtain the expressions of the fundamental solutions and their derivatives in terms of spherical harmonics; Section(\ref{sec-Ch2:results}) presents a few results from the proposed scheme: first it is shown that, in the case of isotropic operators, the proposed representation leads to exact expressions of the fundamental solutions; then a few numerical tests covering isotropic elastic, generally anisotropic elastic, transversely isotropic and generally anisotropic piezo-electric and magneto-electro-elastic media are presented and discussed. 
\section{Problem statement}\label{sec-Ch2:problem statement - system of pdes}
The linear behaviour of different classes of \emph{multi-field} materials, such as Piezo-Electric (PE), Magneto-Electric (ME), or Magneto-Electro-Elastic (MEE) materials, can be represented by a system of generally coupled partial differential equations (PDEs)
\begin{equation}\label{eq-Ch2:system of pdes}
\mathscr{L}_{ij}(\pd_x)\phi_j(\mathbf{x})+f_i(\mathbf{x})=0,
\end{equation}
where $\mathbf{x}=\in \mathbb{R}^3$ is the spatial independent variable, $\phi_{j}(\mathbf{x})$ represent the \emph{unknown} functions of $\mathbf{x}$, $f_i(\mathbf{x})$ represent the known \emph{generalized} volume forces, and $i,j=1,\dots,N$ where $N$ is the number of equations as well as the number of unknown functions. $\mathscr{L}_{ij}(\pd_x)$ is supposed to be a general \emph{homogeneous} \emph{elliptic} partial differential operator involving a linear combination of \emph{second order} derivatives of $\mathbf{x}$, i.e.\ $\mathscr{L}_{ij}(\pd_x)=c_{ijkl}\pd^2\left(\cdot\right)/\pd x_k\pd x_l$, where $c_{ijkl}$ are the material constants.

The system of PDEs (\ref{eq-Ch2:system of pdes}) may be specialised to several specific problems ranging from the classical Laplace equation up to the governing equations for general anisotropic magneto-electro-elastic materials, as shown in Section (\ref{sec-Ch2:results}).

The system of PDEs in Eq.(\ref{eq-Ch2:system of pdes}) is defined $\forall\mathbf{x}\in V\subseteq\mathbb{R}^3$, being $V$ the material domain, and it is mathematically closed by enforcing a suitable set of boundary conditions over the boundary $S=\partial V$ of $V$. The Boundary Element Method \cite{banerjee1981,wrobel2002,aliabadi2002} is based on the integral representation of Eq.(\ref{eq-Ch2:system of pdes}). In particular, using the Green's identities, it is possible to express the values of the functions $\phi_i(\mathbf{y})$ at any interior point $\mathbf{y}\in V$ in terms of the values of $\phi_i(\mathbf{x})$ and their derivatives on the boundary $S$ as follows:
\begin{equation}\label{eq-Ch2:integral representation}
\phi_p(\mathbf{y})=\int_S\left[\Phi_{pi}(\mathbf{x},\mathbf{y})\tau_i(\mathbf{x})-T_{pi}(\mathbf{x},\mathbf{y})\phi_{i}(\mathbf{x})\right]\dd S(\mathbf{x})
+\int_V\Phi_{pi}(\mathbf{x},\mathbf{y})f_i(\mathbf{x})\dd V(\mathbf{x}),
\end{equation}
where
\begin{equation}\label{eq-Ch2:integral representation terms}
\tau_i(\mathbf{x})=n_k(\mathbf{x})c_{ijkl}\frac{\pd \phi_{j}}{\pd x_l}(\mathbf{x}),
\qquad
T_{pi}(\mathbf{x},\mathbf{y})=n_l(\mathbf{x})c_{jikl}\frac{\pd\Phi_{pj}}{\pd x_k}(\mathbf{x},\mathbf{y}).
\end{equation}

$\Phi_{ij}(\mathbf{x},\mathbf{y})$ are the fundamental solutions of the system of PDEs (\ref{eq-Ch2:system of pdes}) that, for a given material specified by the constants $c_{ijkl}$, depend only on the relative position between the \emph{collocation} point $\mathbf{y}$ and the \emph{observation} point $\mathbf{x}$, i.e.\ $\Phi_{ij}(\mathbf{x},\mathbf{y})\equiv\Phi_{ij}(\mathbf{x}-\mathbf{y})$. Defining the distance vector $\mathbf{r}\equiv\mathbf{x}-\mathbf{y}$, $\Phi_{ij}(\mathbf{r})$ is obtained as the solution of the adjoint problem:
\begin{equation}\label{eq-Ch2:system of pdes adjoint}
\mathscr{L}_{ij}^*(\pd_r)\Phi_{jp}(\mathbf{r})+\delta_{pi}\delta(\mathbf{r})=0,
\end{equation}
where $\mathscr{L}_{ij}^*(\pd_r)=\mathscr{L}_{ji}(\pd_r)$, $\delta_{pi}$ is the Kronecker delta function and $\delta(\mathbf{r})$ is the Dirac delta function: $\Phi_{jp}(\mathbf{r})$ is then the value of the field component function $\phi_{j}(\mathbf{r})$ at the point $\mathbf{r}=\mathbf{x}-\mathbf{y}$ due to a generalised unit point force $\delta_{pi}$ at $\mathbf{r}=\mathbf{0}$.

Upon applying the Fourier transform to Eq.(\ref{eq-Ch2:system of pdes adjoint}), the fundamental solutions $\Phi_{ij}(\mathbf{r})$ can be written as
\begin{equation}\label{eq-Ch2:system of pdes fundamental solutions Fourier transform representation}
\Phi_{ij}(\mathbf{r})=\frac{1}{8\pi^3}\int_{\mathbb{R}^3}\wtilde{\Phi}_{ij}(\boldsymbol{\xi})\exp(\mathrm{i}\,\xi_kr_k)\dd \boldsymbol{\xi}
\end{equation}
where
\begin{equation}\label{eq-Ch2:system of pdes fundamental solution fourier}
\wtilde{\Phi}_{ij}(\boldsymbol{\xi})=\left[\mathscr{L}_{ij}^*(\boldsymbol{\xi})\right]^{-1}.
\end{equation}

In Eqs.(\ref{eq-Ch2:system of pdes fundamental solutions Fourier transform representation}-\ref{eq-Ch2:system of pdes fundamental solution fourier}), $\boldsymbol{\xi}\in \mathbb{R}^3$ is the variable spanning the Fourier transform domain and $\mathrm{i}=\sqrt{-1}$ denotes the imaginary unit; $\mathscr{L}_{ij}^*(\boldsymbol{\xi})=c_{jikl}\xi_k\xi_l$ is usually referred to as the \emph{symbol} of the partial differential operator $\mathscr{L}_{ij}^*(\pd_x)$ \cite{wells2007}. Here and in the following, it is assumed that $\mathscr{L}_{ij}^*(\boldsymbol{\xi})$ can be inverted $\forall\boldsymbol{\xi}\in\mathbb{R}^3$. It is worth noting that the symbol $\mathscr{L}_{ij}^*(\boldsymbol{\xi})$ is homogeneous of order $n=2$ and, therefore, $\wtilde{\Phi}_{ij}(\boldsymbol{\xi})$ is homogeneous of order $n=-2$.

In Eq.(\ref{eq-Ch2:system of pdes fundamental solutions Fourier transform representation}), $\Phi_{ij}(\mathbf{r})$ are given in integral form and an explicit closed form expression can be obtained for particular cases only, e.g. for isotropic  materials (Kelvin solution); for general anisotropic materials, the integral in Eq.(\ref{eq-Ch2:system of pdes fundamental solutions Fourier transform representation}) can be explicitly given in terms of the Stroh's eigenvalues \cite{ting1997,lee2009,buroni2010,xie2016}, which however have to be computed for each couple of collocation and observation points, thus inducing high computational costs when employed in numerical codes. In this work, the fundamental solutions $\Phi_{ij}(\mathbf{r})$ and their derivatives are obtained using the spherical harmonics expansion. The method is illustrated in the next Section.

\section{Fourier transform and Rayleigh expansion}\label{sec-Ch2:fourier transform + plane wave expansion}
Let us consider a function $F(\mathbf{r})$, $\mathbf{r}\in\mathbb{R}^3$, and its Fourier transform representation
\begin{equation}\label{eq-Ch2:fourier representation}
F(\mathbf{r})=\frac{1}{8\pi^3}\int_{\mathbb{R}^3}\wtilde{F}(\boldsymbol{\xi})\exp(\mathrm{i}\,\xi_kr_k)\dd\boldsymbol{\xi}
\end{equation}
where $\wtilde{F}(\boldsymbol{\xi})$ is the Fourier transform of $F(\mathbf{r})$, i.e.\
\begin{equation}\label{eq-Ch2:inverse fourier representation}
\wtilde{F}(\boldsymbol{\xi})=\int_{\mathbb{R}^3}F(\mathbf{r})\exp(-\mathrm{i}\,\xi_kr_k)\dd\mathbf{r}.
\end{equation}

Following Adkins \cite{adkins2013} and using the Rayleigh expansion (see Eq.(\ref{app-Ch2:plane wave expansion})), in combination with the spherical harmonic addition theorem (see Eq.(\ref{eq-Ch2:spherical harmonics addition theorem})), the quantity $\exp\left(\mathrm{i}\,\xi_k x_k\right)$ in Eq.(\ref{eq-Ch2:fourier representation}) can be expressed as
\begin{equation}\label{eq-Ch2:expansion}
\exp\left(\mathrm{i}\,\xi_k r_k\right)=4\pi\sum_{\ell=0}^{\infty}\sum_{m=-\ell}^{\ell}\mathrm{i}^\ell J_\ell(\xi\, r)Y_\ell^m(\uv{r})\bar{Y}_\ell^m(\uvg{\xi})
\end{equation}
where $r=\sqrt{r_kr_k}$, $\xi=\sqrt{\xi_k\xi_k}$, $\uv{r}=\mathbf{r}/r$, $\uvg{\xi}=\boldsymbol{\xi}/\xi$, $J_\ell(\cdot)$ are the spherical Bessel functions, $Y_\ell^m(\cdot)$ are the spherical harmonics and $\bar{(\cdot)}$ indicates complex conjugation. Upon substituting Eq.(\ref{eq-Ch2:expansion}) into Eq.(\ref{eq-Ch2:fourier representation}), one obtains
\begin{equation}\label{eq-Ch2:fourier representation + expansion}
F(\mathbf{r})=\frac{1}{2\pi^2}\sum_{\ell=0}^{\infty}\sum_{m=-\ell}^{\ell}\mathrm{i}^\ell Y_\ell^m(\uv{r})\int_{\mathbb{R}^3}\wtilde{F}(\boldsymbol{\xi})J_\ell(\xi\, r)\bar{Y}_\ell^m(\uvg{\xi})\dd\boldsymbol{\xi}.
\end{equation}

If we let $\wtilde{F}(\boldsymbol{\xi})$ be homogeneous of order $n$, then $\wtilde{F}(\xi\uvg{\xi})=\xi^n \wtilde{F}(\uvg{\xi})$, and the integral over $\mathbb{R}^3$ in Eq.(\ref{eq-Ch2:fourier representation + expansion}) can be separated into the integral over the unit sphere and the integral over $\xi\in[0,\infty)$, i.e.\
\begin{multline}\label{eq-Ch2:fourier representation + expansion + F homogeneous}
F(\mathbf{r})=
\frac{1}{2\pi^2}\sum_{\ell=0}^{\infty}\sum_{m=-\ell}^{\ell}\mathrm{i}^\ell Y_\ell^m(\uv{r})\int_{\mathbb{R}^3}\xi^n\wtilde{F}(\uvg{\xi})J_\ell(\xi \,r)\bar{Y}_\ell^m(\uvg{\xi})\dd\boldsymbol{\xi}=\\
=\frac{1}{2\pi^2}\sum_{\ell=0}^{\infty}\sum_{m=-\ell}^{\ell}\mathrm{i}^\ell Y_\ell^m(\uv{r})\int_{0}^{\infty}\xi^{n+2} J_\ell(\xi \,r)\dd \xi\int_{S_1}\wtilde{F}(\uvg{\xi})\bar{Y}_\ell^m(\uvg{\xi})\dd S(\uvg{\xi}),
\end{multline}
where $S_1$ the surface of the unit sphere. In Eq.(\ref{eq-Ch2:fourier representation + expansion + F homogeneous}), the integral with respect to $\xi$ is evaluated as
\begin{equation}\label{eq-Ch2:xi integral}
\int_{0}^{\infty}\xi^{n+2} J_\ell(\xi\, r)\dd \xi=\sqrt{\pi}\frac{2^{n+1}}{r^{n+3}}\frac{\Gamma[(\ell+n+3)/2]}{\Gamma[(\ell-n)/2]},
\end{equation}
where $\Gamma(\cdot)$ is the Gamma function. The result in Eq.(\ref{eq-Ch2:xi integral}) is in general valid for $-3<n+\ell<-1+\ell$; however, it can be extended to the range $n+\ell>-3$, as showed by Adkins \cite{adkins2013}. Using the result obtained in Eq.(\ref{eq-Ch2:xi integral}), Eq.(\ref{eq-Ch2:fourier representation + expansion + F homogeneous}) becomes
\begin{equation}\label{eq-Ch2:fourier representation 2}
F(\mathbf{r})=\frac{1}{\pi^{2}r^{n+3}}\sum_{\ell=0}^{\infty}a_\ell(n)\sum_{m=-\ell}^{\ell}\wtilde{F}^{\ell m}Y_\ell^m(\uv{r})
\end{equation}
where
\begin{multline}\label{eq-Ch2:a_l}
a_\ell(n)=\sqrt{\pi}\ \mathrm{i}^\ell 2^n\frac{\Gamma[(\ell+n+3)/2]}{\Gamma[(\ell-n)/2]}=\\
\sqrt{\pi}\left[\cos\left(\frac{\pi}{2}\ell\right)+\mathrm{i}\,\cdot\sin\left(\frac{\pi}{2}\ell\right)\right]2^n\frac{\Gamma[(\ell+n+3)/2]}{\Gamma[(\ell-n)/2]}
\end{multline}
and
\begin{equation}\label{eq-Ch2:F_l^m}
\wtilde{F}^{\ell m}=\int_{S_1}\wtilde{F}(\uvg{\xi})\bar{Y}_\ell^m(\uvg{\xi})\dd S(\uvg{\xi}).
\end{equation}
From the second equality in Eq.(\ref{eq-Ch2:a_l}), it is possible to show that the coefficients $a_\ell(n)$ are linked to the associated Legendre polynomials $P_\ell^m(t)$ and $Q_\ell^m(t)$ evaluated at $t=0$. Indeed, using Eq.(\ref{app-Ch2:associated legendreP value at 0}), the following identities hold
\begin{subequations}\label{eq-Ch2:al explicit expressions}
\begin{align}
a_\ell(n)&=-\pi\frac{(-1)^{n/2}}{4}P_\ell^{n+2}(0)+\mathrm{i}\,\frac{(-1)^{n/2}}{2}Q_\ell^{n+2}(0),\quad n\ \mathrm{even},\\
a_\ell(n)&=\frac{(-1)^{(n-1)/2}}{2}Q_\ell^{n+2}(0)+\mathrm{i}\,\pi\frac{(-1)^{(n-1)/2}}{4}P_\ell^{n+2}(0),\quad n\ \mathrm{odd}.
\end{align}
\end{subequations}
As an example, Figure (\ref{fig-Ch2:a_l}) shows the behaviour of the coefficients $a_\ell(n)$ for $n=-2$ and $n=0$, which, as it will be demonstrated next, are related to the fundamental solutions and their second derivatives, respectively.

\begin{figure}[ht]
\centering
	\begin{subfigure}{0.49\textwidth}
	\centering
	\includegraphics[width=\textwidth]{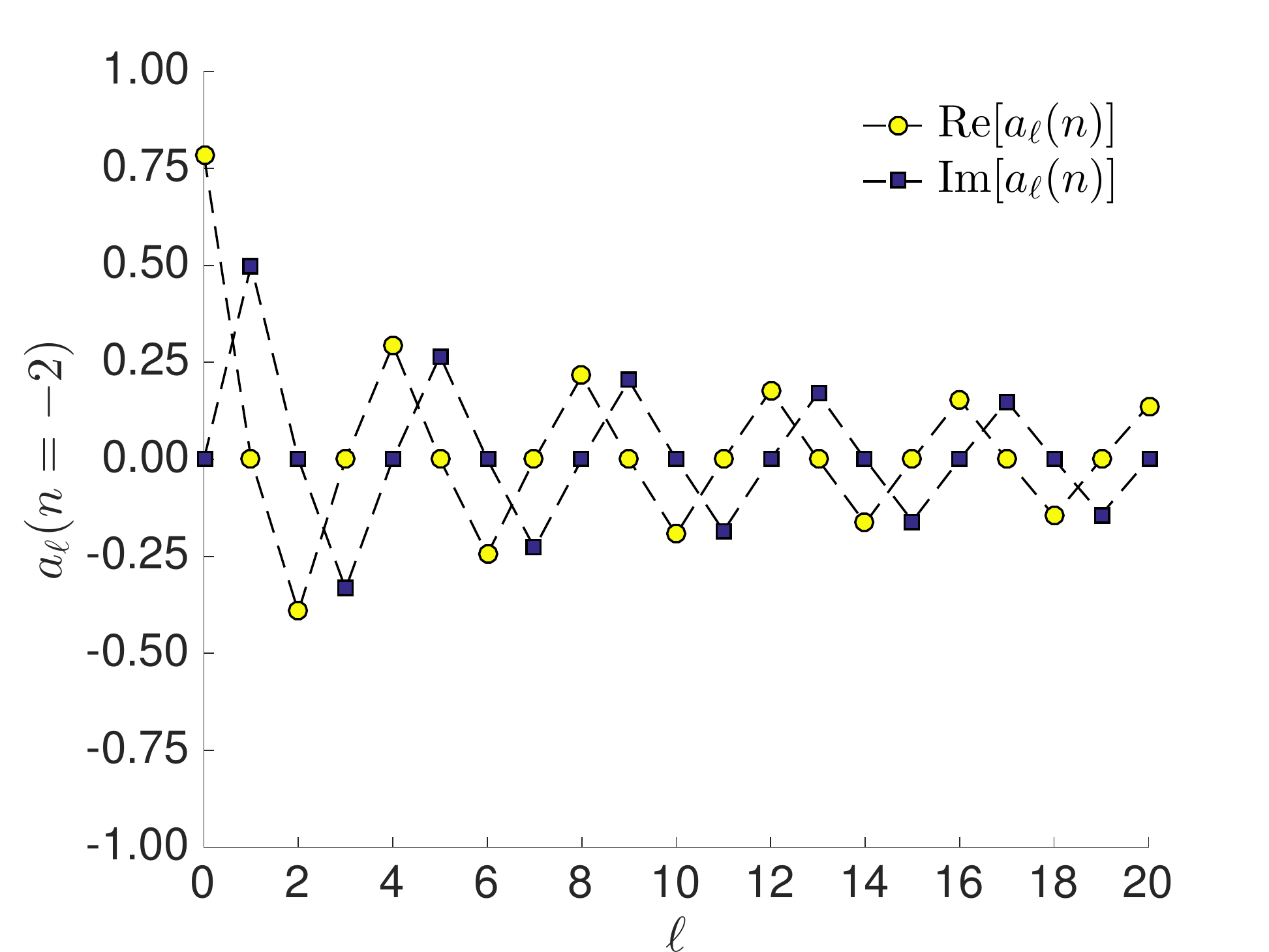}
	\caption{}
	\end{subfigure}
\
	\begin{subfigure}{0.49\textwidth}
	\centering
	\includegraphics[width=\textwidth]{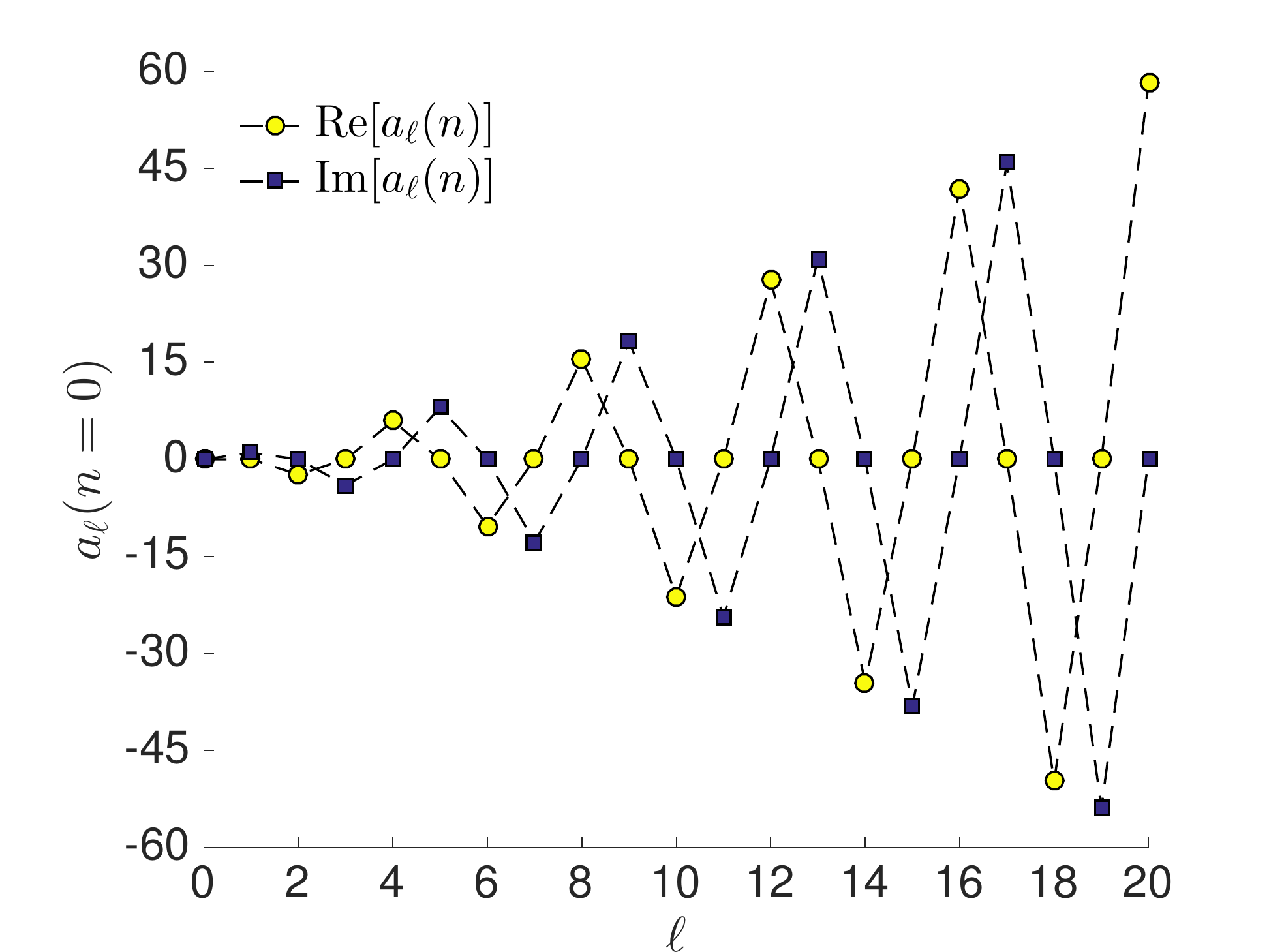}
	\caption{}
	\end{subfigure}
\caption{Behavior of the coefficients $a_\ell(n)$ for (\emph{a}) $n=-2$, (\emph{b}) $n=0$.}
\label{fig-Ch2:a_l}
\end{figure}

\subsection{Fundamental solutions of general anisotropic materials}\label{ssec-Ch2: fundamental solutions with spherical harmonics}
Starting from Eq.(\ref{eq-Ch2:system of pdes fundamental solutions Fourier transform representation}) and recalling that $\wtilde{\Phi}_{ij}(\boldsymbol{\xi})=\left[\mathscr{L}_{ij}^*(\boldsymbol{\xi})\right]^{-1}$ is homogeneous of order $n=-2$, the fundamental solutions of the system of PDEs in (\ref{eq-Ch2:system of pdes}) can be written using Eq.(\ref{eq-Ch2:fourier representation 2}) and Eqs.(\ref{eq-Ch2:al explicit expressions}). Before writing the series expression of the fundamental solutions, it is worth noting that, since $\wtilde{\Phi}_{ij}(\boldsymbol{\xi})$ is a real quantity, the sum over $m$ in Eq.(\ref{eq-Ch2:fourier representation 2}) is also real and, therefore, in order to obtain a real quantity with respect to the sum over $\ell$, only the real part of the coefficients $a_\ell(-2)$ is retained. The following expression is then obtained
\begin{equation}\label{eq-Ch2:system of pdes fundamental solutions spherical harmonics}
\Phi_{ij}(\mathbf{r})=
\frac{1}{4\pi r}\sum_{\substack{\ell=0\\ \ell\, \mathrm{even}}}^{\infty}P_{\ell}(0)\sum_{m=-\ell}^{\ell}\wtilde{\Phi}_{ij}^{\ell m}Y_\ell^{m}(\uv{r})
\end{equation}
where
\begin{equation}\label{eq-Ch2:system of pdes fundamental solutions spherical harmonics coefficients}
\wtilde{\Phi}_{ij}^{\ell m}=\int_{S_1}\wtilde{\Phi}_{ij}(\uvg{\xi})\bar{Y}_\ell^m(\uvg{\xi})\dd S(\uvg{\xi}).
\end{equation}
In Eq.(\ref{eq-Ch2:system of pdes fundamental solutions spherical harmonics}), the sum over $\ell$ ranges over even values since $P_\ell(0)=0$ for odd $\ell$.

The first derivatives of the fundamental solutions can be computed using Eq.(\ref{eq-Ch2:system of pdes fundamental solution fourier}) and the properties of the Fourier transform of a function derivative. In the Fourier transform domain, the first derivative of the fundamental solutions with respect to $r_\alpha$ can be written as $\mathrm{i}\,\xi_\alpha\wtilde{\Phi}_{ij}(\boldsymbol{\xi})$, which is homogeneous of order $n=-1$. In this case, $n$ is odd and the quantity $\mathrm{i}\,\xi_\alpha\wtilde{\Phi}_{ij}(\boldsymbol{\xi})$ is purely imaginary. Therefore, only the imaginary part of the coefficients $a_\ell(-1)$ is retained. The following expression is obtained 
\begin{equation}\label{eq-Ch2:system of pdes fundamental solutions 1st der spherical harmonics}
\frac{\pd\Phi_{ij}}{\pd r_\alpha}(\mathbf{r})=
\frac{1}{4\pi r^2}\sum_{\substack{\ell=1\\ \ell\, \mathrm{odd}}}^{\infty}P_\ell^{1}(0)\sum_{m=-\ell}^{\ell}\wtilde{\Phi}_{ij,\alpha}^{\ell m}Y_\ell^{m}(\uv{r})
\end{equation}
where
\begin{equation}\label{eq-Ch2:system of pdes fundamental solutions 1st der spherical harmonics coefficients}
\wtilde{\Phi}_{ij,\alpha}^{\ell m}=\int_{S_1}\hat{\xi}_\alpha\wtilde{\Phi}_{ij}(\uvg{\xi})\bar{Y}_\ell^m(\uvg{\xi})\dd S(\uvg{\xi})
\end{equation}
In Eq.(\ref{eq-Ch2:system of pdes fundamental solutions 1st der spherical harmonics}), the sum over $\ell$ ranges over odd $\ell$ values as $P_\ell^1(0)=0$ for even $\ell$.

Similarly, the spherical harmonics expansion fot the second derivative with respect to $r_\alpha$ and $r_\beta$ is obtained as
\begin{equation}\label{eq-Ch2:system of pdes fundamental solutions 2nd der spherical harmonics}
\frac{\pd^2\Phi_{ij}}{\pd r_\alpha\pd r_\beta}(\mathbf{r})=
\frac{1}{4\pi r^3}\sum_{\substack{\ell=0\\ \ell\, \mathrm{even}}}^{\infty}P_\ell^2(0)\sum_{m=-\ell}^{\ell}\wtilde{\Phi}_{ij,\alpha\beta}^{\ell m}Y_\ell^{m}(\uv{r})
\end{equation}
where
\begin{equation}\label{eq:system of pdes fundamental solutions 2nd der spherical harmonics coefficients}
\wtilde{\Phi}_{ij,\alpha\beta}^{\ell m}=\int_{S_1}\hat{\xi}_\alpha\hat{\xi}_\beta\wtilde{\Phi}_{ij}(\uvg{\xi})\bar{Y}_\ell^m(\uvg{\xi})\dd S(\uvg{\xi}).
\end{equation}

The results given in Eqs.(\ref{eq-Ch2:system of pdes fundamental solutions spherical harmonics}), (\ref{eq-Ch2:system of pdes fundamental solutions 1st der spherical harmonics}) and (\ref{eq-Ch2:system of pdes fundamental solutions 2nd der spherical harmonics}) can be generalised to higher-order derivatives of the fundamental solutions and written using the following unified form
\begin{equation}\label{eq-Ch2:system of pdes fundamental solutions nth der spherical harmonics}
\frac{\pd^{(I)}\Phi_{ij}}{\pd r_1^{(i_1)}\pd r_2^{(i_2)}\pd r_3^{(i_3)}}(\mathbf{r})=
\frac{1}{4\pi r^{I+1}}\sum_{\ell\in\mathcal{L}}^{\infty}P_\ell^{I}(0)\sum_{m=-\ell}^{\ell}\wtilde{\Phi}_{ij,(i_1,i_2,i_3)}^{\ell m}Y_\ell^{m}(\uv{r}),
\end{equation}
where 
\begin{equation}\label{eq-Ch2:system of pdes fundamental solutions nth der spherical harmonics coefficients}
\wtilde{\Phi}_{ij,(i_1,i_2,i_3)}^{\ell m}=\int_{S_1}(\hat{\xi}_1)^{i_1}(\hat{\xi}_2)^{i_2}(\hat{\xi}_3)^{i_3}\wtilde{\Phi}_{ij}(\uvg{\xi})\bar{Y}_\ell^m(\uvg{\xi})\dd S(\uvg{\xi}).
\end{equation}

In Eq.(\ref{eq-Ch2:system of pdes fundamental solutions nth der spherical harmonics}), $I=i_1+i_2+i_3$, $\mathcal{L}$ is the set of positive even (odd) integers when $I$ is even (odd). Mura and Kinoshita \cite{mura1971} pointed out that the $\ell$-th coefficients of the spherical harmonics expansion of the fundamental solutions is the value of the Legendre polynomial of degree $\ell$ evaluated at zero. Here, it is proved that such a relationship is in fact a particular case of the more general expression given in Eq.(\ref{eq-Ch2:system of pdes fundamental solutions nth der spherical harmonics}). Furthermore, Eq.(\ref{eq-Ch2:system of pdes fundamental solutions nth der spherical harmonics}) shows that the fundamental solutions as well as their derivatives can be written as the product of a \emph{regular} part, expressed as a spherical harmonics expansion, by a \emph{singular} part depending only on powers of $r$.

In the following, the compact notation $\wtilde{\Phi}_{ij,(i_1,i_2,i_3)}^{\ell m}$ will be used to denote the coefficients of degree $\ell$ and order $m$ of the fundamental solutions $\Phi_{ij}(\mathbf{r})$, whose derivative is taken $i_1$ times with respect to $r_1$, $i_2$ times with respect to $r_2$ and $i_3$ times with respect to $r_3$. The terms $\wtilde{\Phi}_{ij,(0,0,0)}^{\ell m}$ will then represent the coefficients of the fundamental solutions' expansion.

From the expression given by Eq.(\ref{eq-Ch2:system of pdes fundamental solutions nth der spherical harmonics coefficients}), two properties of the expansion coefficients $\wtilde{\Phi}_{ij,(i_1,i_2,i_3)}^{\ell m}$ are worth mentioning. First, the coefficients $\wtilde{\Phi}_{ij,(i_1,i_2,i_3)}^{\ell m}$ depend only on the material properties and \emph{can be computed only once}, in advance in any numerical implementation; additionally, using Eq.(\ref{eq-Ch2:spherical harmonics definition negative m}), it is possible to show that
\begin{equation}\label{eq-Ch2:system of pdes fundamental solutions nth der spherical harmonics coefficients negative coefficients}
\wtilde{\Phi}_{ij,(i_1,i_2,i_3)}^{\ell(-m)}=(-1)^m\bar{\wtilde{\Phi}}_{ij,(i_1,i_2,i_3)}^{\ell m}.
\end{equation}
Second, it is interesting to note that when the variable $\uvg{\xi}$ is expressed in spherical coordinates $\{\thet,\ph\}$ as $\uvg{\xi}=\{\sin\thet\cos\ph,\sin\thet\sin\ph,\cos\thet\}$, the following identities hold for its components
\begin{subequations}\label{eq-Ch2:spherical coordinates with spherical harmonics}
\begin{equation}
\hat{\xi}_1=\sqrt{\frac{2\pi}{3}}\left[Y_1^{-1}(\uvg{\xi})-Y_1^{1}(\uvg{\xi})\right],
\end{equation}
\begin{equation}
\hat{\xi}_2=\mathrm{i}\sqrt{\frac{2\pi}{3}}\left[Y_1^{-1}(\uvg{\xi})+Y_1^{1}(\uvg{\xi})\right],
\end{equation}
\begin{equation}
\hat{\xi}_3=\sqrt{\frac{4\pi}{3}}Y_1^0(\uvg{\xi}).
\end{equation}
\end{subequations}
Then, using the expression of the Clebsch-Gordon series in Eq.(\ref{eq-Ch2:spherical harmonics Clebsch-Gordon series}), it is possible to compute the expansion coefficients $\wtilde{\Phi}_{ij,(i_1+1,i_2,i_3)}^{\ell m}$, $\wtilde{\Phi}_{ij,(i_1,i_2+1,i_3)}^{\ell m}$ and $\wtilde{\Phi}_{ij,(i_1,i_2,i_3+1)}^{\ell m}$ in terms of the coefficients $\wtilde{\Phi}_{ij,(i_1,i_2,i_3)}^{\ell m}$, thus avoiding further integration. In fact, it is sufficient to compute the coefficients of the fundamental solutions (including the coefficients with odd values of $\ell$), and the use the following recurrence relations for the coefficients of higher-order derivatives:
\begin{subequations}\label{eq-Ch2:system of pdes fundamental solutions nth der spherical harmonics coefficients 2}
\begin{equation}
\wtilde{\Phi}_{ij,(i_1+1,i_2,i_3)}^{\ell m}=
\sqrt{\frac{2\ell+1}{2}}\left[
-\sum_{k=k_{m}^{\ell}(-1)}^{\ell+1}\frac{\tilde{c}^{\ell k}_{m}(-1)}{\sqrt{2k+1}}\wtilde{\Phi}_{ij,(i_1,i_2,i_3)}^{k(m+1)}+
\sum_{k=k_{m}^{\ell}(1)}^{\ell+1}\frac{\tilde{c}^{\ell k}_{m}(1)}{\sqrt{2k+1}}\wtilde{\Phi}_{ij,(i_1,i_2,i_3)}^{k(m-1)}
\right]
\end{equation}
\begin{equation}
\wtilde{\Phi}_{ij,(i_1,i_2+1,i_3)}^{\ell m}=
\mathrm{i}\sqrt{\frac{2\ell+1}{2}}\left[
\sum_{k=k_{m}^{\ell}(-1)}^{\ell+1}\frac{\tilde{c}^{\ell k}_{m}(-1)}{\sqrt{2k+1}}\wtilde{\Phi}_{ij,(i_1,i_2,i_3)}^{k(m+1)}+
\sum_{k=k_{m}^{\ell}(1)}^{\ell+1}\frac{\tilde{c}^{\ell k}_{m}(1)}{\sqrt{2k+1}}\wtilde{\Phi}_{ij,(i_1,i_2,i_3)}^{k(m-1)}
\right]
\end{equation}
\begin{equation}
\wtilde{\Phi}_{ij,(i_1,i_2,i_3+1)}^{\ell m}=\sqrt{2\ell+1}\sum_{k=k_m^{\ell}(0)}^{\ell+1}\frac{\tilde{c}^{\ell k}_{m}(0)}{\sqrt{2k+1}}\wtilde{\Phi}_{ij,(i_1,i_2,i_3)}^{km}
\end{equation}
\end{subequations}
where $k_{m}^{\ell}(m')\equiv\max\{|1-\ell|,|m+m'|\}$ and $\tilde{c}^{\ell k}_{m}(m')\equiv c_{000}^{1\ell k}c_{m'(-m)(-m+m')}^{1\ell k}$.

As a last remark on the series expansion given in Eq.(\ref{eq-Ch2:system of pdes fundamental solutions nth der spherical harmonics}), it is interesting to note that such a representation allows to obtain some integral properties of the fundamental solutions, which may be useful for numerical implementation. As an example, the technique proposed by Gao and Davis \cite{gao2000} for the evaluation of strongly singular volume integrals, arising e.g.\ in isotropic plasticity problems, required that the integral over the unit sphere $S_1$ of the regular part of the strain kernel, involving the second derivatives of the fundamental solution, vanished. Such a requirement has been proved by Gao and Davis \cite{gao2000} for isotropic elasto-plasticity; however, it was only numerically verified by Benedetti et al.\ \cite{benedetti2016} for anisotropic crystal plasticity of copper. Using Eq.(\ref{eq-Ch2:system of pdes fundamental solutions nth der spherical harmonics}) and considering that $\int_{S_1}Y_\ell^m(\uv{r})\dd S(\uv{r})=0$ $\forall\ell,m\ne0$ and $P_0^I(0)=0$ $\forall I>0$, it is possible to show that, in fact, the integral over $S_1$ of the regular part of \emph{all} the derivatives of the fundamental solutions of \emph{any} second-order homogeneous elliptic operator vanishes. This constitutes a relevant byproduct of the present unified treatment. 


\subsection{Convergence}\label{ssec-Ch2:convergence}
In this Section, it is shown that \emph{the spherical harmonics representation in Eq.}(\ref{eq-Ch2:system of pdes fundamental solutions nth der spherical harmonics}) \emph{in fact coincides with the representation of the fundamental solution in terms of unit circle integration}.

Considering the fundamental solutions and using Eq.(\ref{eq-Ch2:system of pdes fundamental solutions nth der spherical harmonics}), one has
\begin{multline}
\Phi_{ij}(\mathbf{r})=\frac{1}{4\pi r}\sum_{\ell=0}^{\infty}P_\ell(0)\sum_{m=-\ell}^{\ell}\wtilde{\Phi}_{ij}^{\ell m}Y_\ell^{m}(\uv{r})=\\
\frac{1}{8\pi^2r}\int_{S_1}\sum_{\ell=0}^{\infty}\frac{2\ell+1}{2}P_\ell(0)P_\ell(\hat{\xi}_k\hat{r}_k)\wtilde{\Phi}_{ij}(\uvg{\xi})\dd S(\uvg{\xi})
\end{multline}
where the second equality is obtained using the addition theorem (\ref{eq-Ch2:spherical harmonics addition theorem}). Using the completeness property of the Legendre polynomials reported in Eq.(\ref{app-Ch2:completeness legendreP}), the above expression simplifies to
\begin{equation}
\Phi_{ij}(\mathbf{r})=\frac{1}{8\pi^2r}\int_{S_1}\wtilde{\Phi}_{ij}(\uvg{\xi})\delta(\hat{\xi}_k\hat{r}_k)\dd S(\uvg{\xi})=\frac{1}{8\pi^2r}\oint_{C}\wtilde{\Phi}_{ij}(\uvg{\xi})\dd c
\end{equation}
where the last integral is taken over the unit circle $C$ defined over the plane $\hat{\xi}_k\hat{r}_k=0$, which is \emph{exactly} the classical unit circle integral.

Consider now the derivative with respect to $r_\alpha$ of the fundamental solutions, which will be indicated as $\Phi_{ij,\alpha}\equiv\pd\Phi_{ij}/\pd r_\alpha$. Using the addition theorem (\ref{eq-Ch2:spherical harmonics addition theorem}), the spherical harmonics expansion reads
\begin{equation}\label{eq-Ch2:spherical harmonics decomposition 1st der proof}
\Phi_{ij,\alpha}(\mathbf{r})=\frac{1}{8\pi^2r^2}\int_{S_1}\sum_{\ell=0}^{\infty}\frac{2\ell+1}{2}P_\ell^1(0)P_\ell(\hat{\xi}_k\hat{r}_k)\wtilde{\Phi}_{ij}(\uvg{\xi})\hat{\xi}_\alpha\dd S(\uvg{\xi}).
\end{equation}
Using now the completeness relation given in Eq.(\ref{app-Ch2:completeness legendreP}) and applying the operator $(-1)(1-t^2)^{1/2}\pd(\cdot)/\pd t$ to both sides, one obtains
\begin{equation}
\sum_{\ell=0}^{\infty}\frac{2\ell+1}{2}P_\ell^1(t)P_\ell(s)=(1-t^2)^{1/2}\delta'(s-t).
\end{equation}
where the right-hand side is intended in a distributional sense. Using the above relation with $t=0$, the definition of the derivative of the Dirac delta function and Eq.(\ref{eq-Ch2:spherical harmonics decomposition 1st der proof}), one obtains
\begin{equation}\label{eq-Ch2:spherical harmonics decomposition 1st der proof 2}
\Phi_{ij,\alpha}(\mathbf{r})=\frac{1}{8\pi^2r^2}\int_{S_1}\wtilde{\Phi}_{ij}(\uvg{\xi})\hat{\xi}_\alpha\delta'(\hat{\xi}_k\hat{r}_k)\dd S(\uvg{\xi})=\frac{1}{8\pi^2r^2}\oint_{C}\left.\frac{\pd[\wtilde{\Phi}_{ij}(\uvg{\xi})\hat{\xi}_\alpha]}{\pd(\hat{\xi_k}\hat{r}_k)}\right|_{\hat{\xi}_k\hat{r}_k=0}\dd c.
\end{equation}
Upon considering a reference system aligned with the direction $\hat{r}$, the above expression coincides exactly with that used by Schclar \cite{schclar1994} to derive the first derivative of the fundamental solutions of an anisotropic elastic material.

The above technique can be generalised to higher-order derivatives and confirms that the spherical harmonics expansion in Eq.(\ref{eq-Ch2:system of pdes fundamental solutions nth der spherical harmonics}) is in fact an alternative representation of the fundamental solutions and their derivatives.

\subsection{Pseudo-Algorithm}
To conclude the present Section, the following steps are presented to show the compactness of representing the fundamental solutions and their derivatives in terms of spherical harmonics, given a generic elliptic system of PDEs:
\begin{enumerate}
\item{Consider an elliptic system of PDEs defined by the operator $\mathscr{L}_{ij}(\pd_x)$ as given in Eq.(\ref{eq-Ch2:system of pdes})}
\item{Define the \emph{symbol} $\mathscr{L}_{ij}^*(\boldsymbol{\xi})$ of the adjoint differential operator $\mathscr{L}_{ij}^*(\pd_x)\equiv\mathscr{L}_{ji}(\pd_x)$;}
\item{Compute the coefficients $\wtilde{\Phi}_{ij,(i_1,i_2,i_3)}^{\ell m}$ for the desired orders $i_1$, $i_2$, $i_3$ using Eqs.(\ref{eq-Ch2:system of pdes fundamental solutions nth der spherical harmonics coefficients}) and (\ref{eq-Ch2:system of pdes fundamental solutions nth der spherical harmonics coefficients 2});}
\item{Compute the fundamental solutions $\wtilde{\Phi}_{ij}(\mathbf{r})$ and their derivatives using Eq.(\ref{eq-Ch2:system of pdes fundamental solutions nth der spherical harmonics}).}
\end{enumerate}

\section{Results}\label{sec-Ch2:results}
In the present Section, the proposed technique is employed to compute the fundamental solutions of different classes of systems of PDEs (\ref{eq-Ch2:system of pdes}). The present scheme is first used to exactly retrieve the fundamental solutions for two isotropic cases, namely the classic Laplace equation and the equations of isotropic elasticity. Then, the scheme is used to compute the fundamental solutions and their derivatives for transversely isotropic and generally anisotropic materials, covering from the elastic to the magneto-electro-elastic cases. In such cases, the series in Eq.(\ref{eq-Ch2:system of pdes fundamental solutions nth der spherical harmonics}) are truncated and the coefficients $\wtilde{\Phi}_{ij,(i_1,i_2,i_3)}^{\ell m}$ are computed for $\ell=0,1,\dots,L$, where $L$ will be referred to as the \emph{series truncation number}. To demonstrate the accuracy of the present scheme for any combination of source point $\mathbf{y}$ and observation point $\mathbf{x}$, the following error is defined over the unit sphere $S_1$ centered at $\mathbf{y}$
\begin{equation}\label{eq-Ch2:difference over S_1}
\Delta\left[\Phi_{ij}(\uv{r})\right]=\frac{\Phi_{ij}(\uv{r})-\Phi_{ij}^R(\uv{r})}{||\Phi_{ij}^R(\uv{r})||_{S_1}}.
\end{equation}
Similarly, the norm of the error over $S_1$ is defined as
\begin{equation}\label{eq-Ch2:error over S_1}
e_{S_1}\left[\Phi_{ij}(\uv{r})\right]=\frac{||\Phi_{ij}(\uv{r})-\Phi_{ij}^R(\uv{r})||_{S_1}}{||\Phi_{ij}^R(\uv{r})||_{S_1}}.
\end{equation}
In Eqs.(\ref{eq-Ch2:difference over S_1}) and (\ref{eq-Ch2:error over S_1}), $||\cdot||_{S_1}=\int_{S_1}|\cdot|\dd S(\uv{r})$, $\Phi_{ij}(\uv{r})$ represents the fundamental solutions using the present scheme, and $\Phi_{ij}^R(\uv{r})$ represents a reference value for the fundamental solutions computed using a reference technique.

\subsection{Isotropic materials examples}\label{ssec-Ch2:isotropic materials}
In this Section, two classic isotropic cases are considered. In particular, the well-known fundamental solutions of the Laplace equation and the isotropic elasticity are retrieved showing that, in both cases, the solutions are obtained in exact form as the series expansions involve only a finite number of terms.

\subsubsection{Fundamental solution of the isotropic Laplace equation}\label{sssec-Ch2:fundamental solution laplace equation}
In the present case, referring to Eq.(\ref{eq-Ch2:system of pdes}), the unknown function is denoted by $\phi(\mathbf{x})$ and the differential operator $\mathscr{L}_{ij}(\pd_x)$ is the Laplace operator, i.e.\ $\mathscr{L}_{ij}\equiv\nabla^2\equiv\pd^2\left(\cdot\right)/\pd x_k\pd x_k$. The adjoint operator $\mathscr{L}_{ij}^*\equiv\mathscr{L}^*$ coincides with $\nabla^2$ (self-adjoint operator) and its symbol is simply $\mathscr{L}^*(\boldsymbol{\xi})\equiv\xi_k\xi_k\equiv\xi^2$. Then, it follows that $[\mathscr{L}^*(\uvg{\xi})]^{-1}=1$. In the previous expressions, the indices $i,j$ have been dropped being $\phi(\mathbf{x})$ a scalar function.

Denoting the fundamental solution by $\Phi(\mathbf{r})$ and considering Eq.(\ref{eq-Ch2:system of pdes fundamental solutions nth der spherical harmonics coefficients}), the coefficients of the spherical harmonics expansions are computed as follows
\begin{equation}\label{eq-Ch2:laplace equation fundamental solution nth der coefficients}
\wtilde{\Phi}_{,(i_1,i_2,i_3)}^{\ell m}=\int_{S_1}(\hat{\xi}_1)^{i_1}(\hat{\xi}_2)^{i_2}(\hat{\xi}_3)^{i_3}\bar{Y}_\ell^m(\uvg{\xi})\dd S(\uvg{\xi}),
\end{equation}
The integrals in Eqs.(\ref{eq-Ch2:laplace equation fundamental solution nth der coefficients}) can be analytically evaluated using Eqs.(\ref{eq-Ch2:spherical coordinates with spherical harmonics}) and the orthogonality property of the spherical harmonics over the unit sphere, see Eq.(\ref{eq-Ch2:spherical harmonics orthogonality}); it is possible to show that the coefficients $\Phi_{,(i_1,i_2,i_3)}^{\ell m}$ are identically zero for $\ell>i_1+i_2+i_3$. As an example, the expressions of the fundamental solution $\Phi(\mathbf{r})$ and its derivatives $\Phi_{,1}(\mathbf{r})$, $\Phi_{,12}(\mathbf{r})$, $\Phi_{,112}(\mathbf{r})$ are explicitly computed in the following. The directional dependence of the fundamental solution is indicated by means of the unit vector $\uv{r}=\{\hat{r}_1,\hat{r}_2,\hat{r}_3\}=\{\sin\thet\cos\ph,\sin\thet\sin\ph,\cos\thet\}$. Moreover, only the coefficients with $m\ge0$ will be given since the coefficients with negative $m$ can be obtained using Eq.(\ref{eq-Ch2:system of pdes fundamental solutions nth der spherical harmonics coefficients negative coefficients}).

Consider the fundamental solution $\Phi(\mathbf{r})$ first. From Eq.(\ref{eq-Ch2:laplace equation fundamental solution nth der coefficients}), the coefficients $\Phi_{,(0,0,0)}^{\ell m}$ represent the integrals of the spherical harmonics over $S_1$, which are different from zero only for $\ell=m=0$. The integral of $\bar{Y}_0^0(\uv{r})$ over $S_1$ equals $2\sqrt{\pi}$ and, using Eq.(\ref{eq-Ch2:system of pdes fundamental solutions nth der spherical harmonics}), one has
\begin{equation}
\Phi(\mathbf{r})=\frac{1}{4\pi r}.
\end{equation}

Next, consider the first derivative $\Phi_{,1}(\mathbf{r})$ of the fundamental solution. Using Eq.(\ref{eq-Ch2:laplace equation fundamental solution nth der coefficients}), the coefficients $\Phi_{,(1,0,0)}^{\ell m}$ are computed as
\begin{equation}\label{eq-Ch2:laplace equation fundamental solution x1 derivative coefficients}
\Phi_{,(1,0,0)}^{\ell m}=\int_{S_1}\hat{\xi}_1\bar{Y}_\ell^m(\uvg{\xi})\dd S(\uvg{\xi}).
\end{equation}
The only non-zero coefficient is in fact
\begin{equation}
\wtilde{\Phi}_{,(1,0,0)}^{11}=-\sqrt{2\pi/3},
\end{equation}
which, using Eq.(\ref{eq-Ch2:system of pdes fundamental solutions nth der spherical harmonics}), provides
\begin{equation}
\Phi_{,1}(\mathbf{r})=-\frac{\hat{r}_1}{4\pi r^2}.
\end{equation}

Then, consider the second derivative $\Phi_{,12}(\mathbf{r})$. The coefficients $\Phi_{,(1,1,0)}^{\ell m}$ are computed as
\begin{equation}\label{eq-Ch2:laplace equation fundamental solution x1x2 derivative coefficients}
\wtilde{\Phi}_{,(1,1,0)}^{\ell m}=\int_{S_1}\hat{\xi}_1\hat{\xi}_2\bar{Y}_\ell^m(\uvg{\xi})\dd S(\uvg{\xi}).
\end{equation}
In this case, the only non-zero coefficient is
\begin{equation}
\wtilde{\Phi}_{,(1,1,0)}^{22}=-\mathrm{i}\sqrt{2\pi/15},
\end{equation}
which, using Eq.(\ref{eq-Ch2:system of pdes fundamental solutions nth der spherical harmonics}), provides
\begin{equation}
\Phi_{,12}(\mathbf{r})=\frac{3\hat{r}_1\hat{r}_2}{4\pi r^3}.
\end{equation}

Eventually, the third derivative $\Phi_{,112}(\mathbf{r})$ is considered. 
The coefficients $\Phi_{,(2,1,0)}^{\ell m}$ are computed as
\begin{equation}\label{eq-Ch2:laplace equation fundamental solution x1x1x2 derivative coefficients}
\wtilde{\Phi}_{,(2,1,0)}^{\ell m}=\int_{S_1}(\hat{\xi}_1)^2\hat{\xi}_2\bar{Y}_\ell^m(\uvg{\xi})\dd S(\uvg{\xi})
\end{equation}
and the only non-zero coefficients are
\begin{equation}
\wtilde{\Phi}_{,(2,1,0)}^{11}=\mathrm{i}\frac{1}{5}\sqrt{2\pi/3},\quad
\wtilde{\Phi}_{,(2,1,0)}^{31}=-\mathrm{i}\frac{1}{5}\sqrt{\pi/21},\quad
\wtilde{\Phi}_{,(2,1,0)}^{33}=\mathrm{i}\sqrt{\pi/35},
\end{equation}
which, using Eq.(\ref{eq-Ch2:system of pdes fundamental solutions nth der spherical harmonics}), provide
\begin{equation}
\Phi_{,112}(\mathbf{r})=\frac{3\hat{r}_2(1-5\hat{r}_1^2)}{4\pi r^4}.
\end{equation}

\subsubsection{Fundamental solution of isotropic elasticity}\label{sssec-Ch2:fundamental solution isotropic elasticity}
The fundamental solutions for isotropic elasticity are here retrieved using the proposed unified formulation. In the present case, the unknown functions are denoted by $u_j(\mathbf{x})$, $j=1,2,3$ and represent the components of the displacement field. The operator $\mathscr{L}_{ij}(\pd_x)$ is
\begin{equation}\label{eq-Ch2:Lij operator for isotropic elasticity}
\mathscr{L}_{ij}=(\lambda+\mu)\frac{\pd^2}{\pd x_i\pd x_j}+\mu\frac{\pd^2}{\pd x_k\pd x_k}\delta_{ij}
\end{equation}
where $\lambda$ and $\mu$ are the Lam\'e constants. As $\mathscr{L}_{ij}$ is symmetric, the adjoint operator $\mathscr{L}_{ij}^*$ coincides with $\mathscr{L}_{ij}$ and its symbol is $\mathscr{L}_{ij}^*(\boldsymbol{\xi})=(\lambda+\mu)\xi_i\xi_j+\mu\xi_k\xi_k\delta_{ij}$. Then, it follows that the inverse of $\mathscr{L}_{ij}^*(\uvg{\xi})$ is
\begin{equation}\label{eq-Ch2:inverse of adjoint of Lij for isotropic elasticity}
\left[\mathscr{L}_{ij}^*(\uvg{\xi})\right]^{-1}=\frac{1}{\mu}\delta_{ij}-\frac{1}{2\mu(1-\nu)}\hat{\xi}_i\hat{\xi}_j
\end{equation}
where $\nu$ is the Poisson coefficient.

Denoting the fundamental solutions of isotropic elasticity by $\Phi_{ij}(\mathbf{r})$, the coefficients $\wtilde{\Phi}_{ij,(i_1,i_2,i_3)}^{\ell m}$ are computed as
\begin{equation}\label{eq-Ch2:isotropic elasticity fundamental solution nth der coefficients}
\wtilde{\Phi}_{ij,(i_1,i_2,i_3)}^{\ell m}=\int_{S_1}(\hat{\xi}_1)^{i_1}(\hat{\xi}_2)^{i_2}(\hat{\xi}_3)^{i_3}\left(\frac{1}{\mu}\delta_{ij}-\frac{1}{2\mu(1-\nu)}\hat{\xi}_i\hat{\xi}_j\right)\bar{Y}_\ell^m(\uvg{\xi})\dd S(\uvg{\xi}).
\end{equation}
The integrals in Eqs.(\ref{eq-Ch2:isotropic elasticity fundamental solution nth der coefficients}) can be once again evaluated in closed form using Eqs.(\ref{eq-Ch2:spherical coordinates with spherical harmonics}) and the orthogonality property Eq.(\ref{eq-Ch2:spherical harmonics orthogonality}) of the spherical harmonics and, in particular, it is possible to show that $\wtilde{\Phi}_{ij,(i_1,i_2,i_3)}^{\ell m}$ are identically zero for $\ell>i_1+i_2+i_3+2$. As an example, the fundamental solution $\Phi_{11}(\mathbf{r})$ and its fourth derivative $\Phi_{11,1123}(\mathbf{r})$ will be computed. Also in this case, only the coefficients with $m\ge0$ will be reported as the coefficients with negative $m$ can be obtained using Eq.(\ref{eq-Ch2:system of pdes fundamental solutions nth der spherical harmonics coefficients negative coefficients}).

Let us consider the fundamental solution $\Phi_{11}(\mathbf{r})$. Using Eq.(\ref{eq-Ch2:isotropic elasticity fundamental solution nth der coefficients}), the coefficients $\Phi_{11,(0,0,0)}^{\ell m}$ are computed as
\begin{equation}\label{eq-Ch2:isotropic elasticity fundamental solution coefficients}
\wtilde{\Phi}_{11,(0,0,0)}^{\ell m}=\int_{S_1}\left(\frac{1}{\mu}\delta_{ij}-\frac{1}{2\mu(1-\nu)}\hat{\xi}_i\hat{\xi}_j\right)\bar{Y}_\ell^m(\uvg{\xi})\dd S(\uvg{\xi}).
\end{equation}
The only non-zero coefficients are
\begin{equation}
\wtilde{\Phi}_{11,(0,0,0)}^{00}=\frac{\sqrt{\pi}(5-6\nu)}{3\mu(1-\nu)},\quad
\wtilde{\Phi}_{11,(0,0,0)}^{20}=\frac{\sqrt{\pi/5}}{3\mu(1-\nu)},\quad
\wtilde{\Phi}_{11,(0,0,0)}^{22}=-\frac{\sqrt{\pi/30}}{\mu(1-\nu)},
\end{equation}
which, using Eq.(\ref{eq-Ch2:system of pdes fundamental solutions nth der spherical harmonics}), provide
\begin{equation}
\Phi_{11}(\mathbf{r})=\frac{3-4\nu+\hat{r}_1^2}{16\pi\mu(1-\nu)r}.
\end{equation}

Consider now the fourth derivative $\Phi_{11,1123}(\mathbf{r})$ of the fundamental solution $\Phi_{11}(\mathbf{r})$. Using Eq.(\ref{eq-Ch2:isotropic elasticity fundamental solution nth der coefficients}), the coefficients $\Phi_{11,(2,1,1)}^{\ell m}$ are computed as
\begin{equation}\label{eq-Ch2:isotropic elasticity fundamental solution 4th der coefficients}
\wtilde{\Phi}_{11,(2,1,1)}^{\ell m}=\int_{S_1}(\hat{\xi}_1)^2\hat{\xi}_2\hat{\xi}_3\left(\frac{1}{\mu}\delta_{ij}-\frac{1}{2\mu(1-\nu)}\hat{\xi}_i\hat{\xi}_j\right)\bar{Y}_\ell^m(\uvg{\xi})\dd S(\uvg{\xi}).
\end{equation}
In this case, the non-zero coefficients are
\begin{equation*}
\wtilde{\Phi}_{11,(2,1,1)}^{21}=\mathrm{i}\frac{\sqrt{\pi/30}(5-6\nu)}{21\mu(1-\nu)},\quad
\wtilde{\Phi}_{11,(2,1,1)}^{41}=-\mathrm{i}\frac{\sqrt{\pi/5}(8-11\nu)}{231\mu(1-\nu)},
\end{equation*}
\begin{equation}
\wtilde{\Phi}_{11,(2,1,1)}^{43}=\mathrm{i}\frac{\sqrt{\pi/35}(8-11\nu)}{33\mu(1-\nu)},\quad
\wtilde{\Phi}_{11,(2,1,1)}^{61}=\mathrm{i}\frac{\sqrt{\pi/546}}{33\mu(1-\nu)},
\end{equation}
\begin{equation*}
\wtilde{\Phi}_{11,(2,1,1)}^{63}=\mathrm{i}\frac{\sqrt{3\pi/455}}{22\mu(1-\nu)},\quad
\wtilde{\Phi}_{11,(2,1,1)}^{65}=-\mathrm{i}\frac{\sqrt{3\pi/1001}}{6\mu(1-\nu)},
\end{equation*}
which provide
\begin{equation}
\Phi_{11,1123}(\mathbf{r})=\frac{15[1-4\nu+14(1+2\nu)\hat{r}_1^2-63\hat{r}_1^4]\hat{r}_2\hat{r}_3}{16\pi\mu(1-\nu)r^5}.
\end{equation}

\subsection{Anisotropic materials examples}
In the present Section, the proposed scheme is used to compute the fundamental solutions of anisotropic elastic, piezoelectric and magneto-electro-elastic materials. The governing equations are given for a general magneto-electro-elastic material and then particularised to the elastic and piezo-electric cases. The system of PDEs for an anisotropic MEE material can be written as follows \cite{pan2002,buroni2010,munoz2015}:
\begin{equation}\label{eq-Ch2:system of pdes MME}
c_{iJKl}\frac{\pd^2 \phi_K}{\pd x_l\pd x_i}(\mathbf{x})+f_J(\mathbf{x})=0
\end{equation}
where the unknown functions $\phi_K(\mathbf{x})$, $K=1,2,\dots,5$ are given by $\phi_K(\mathbf{x})=u_K(\mathbf{x})$ if $K\leq3$, $\phi_K(\mathbf{x})=\ph(\mathbf{x})$ if $K=4$ and $\phi_K(\mathbf{x})=\psi(\mathbf{x})$ if $K=5$, being $u_k(\mathbf{x})$, $k=1,2,3$ the components of the displacement field, $\ph(\mathbf{x})$ the electric potential and $\psi(\mathbf{x})$ the magnetic potential. The constants $c_{iJKl}$ are the multi-field coupling coefficients of the MEE material and are defined by
\begin{equation}\label{eq-Ch2:MME constants}
c_{iJKl}=\left\{
\begin{array}{ll}
c_{ijkl},&J,K\le3\\
e_{lij},&J\le3,K=4\\
e_{ikl},&J=4,K\le3\\
q_{lij},&J\le3,K=5\\
q_{ikl},&J=5,K\le3\\
-\lambda_{il},&J=4,K=5\,\mathrm{or}\,J=5,K=4\\
-\eps_{il},&J=K=4\\
-\kappa_{il},&J=K=5
\end{array}
\right.
\end{equation}
where $c_{ijkl}$, $\eps_{il}$ and $\kappa_{il}$ are the elastic stiffness tensor, the dielectric permittivity tensor and the magnetic permeability tensor, respectively, whereas $e_{lij}$, $q_{lij}$ and $\lambda_{il}$ are the piezoelectric, piezomagnetic and magneto-electric coupling tensors. The aforementioned tensors satisfy the following symmetries
\begin{align}
&c_{ijkl}=c_{jikl}=c_{klij},\quad e_{lij}=e_{lji},\quad q_{kij}=q_{kji}\\
&\eps_{il}=\eps_{li},\quad \lambda_{il}=\lambda_{li},\quad \kappa_{il}=\kappa_{li},\nonumber
\end{align}
which ensure that the symbol $\mathscr{L}_{JK}(\boldsymbol{\xi})\equiv C_{iJKl}\xi_l\xi_i$ of the system of PDEs (\ref{eq-Ch2:system of pdes MME}) and its adjoint $\mathscr{L}_{JK}^*(\boldsymbol{\xi})$ are symmetric and coincident. Eventually, the generalised volume forces are $\left\{f_J(\mathbf{x})\right\}=\{f_1(\mathbf{x}),f_2(\mathbf{x}),f_3(\mathbf{x}),-f^e(\mathbf{x}),-f^m(\mathbf{x})\}$, being $f_j(\mathbf{x})$, $j=1,2,3$ the mechanical body forces, $f^e(\mathbf{x})$ the electric charge density and $f^m(\mathbf{x})$ electric current density.

The fundamental solutions of a generally anisotropic magneto-electro-elastic material are indicated by $\Phi_{JK}(\mathbf{r})$ and are computed using the general form of Eq.(\ref{eq-Ch2:system of pdes fundamental solutions nth der spherical harmonics}), where the coefficients of the expansions are computed using Eq.(\ref{eq-Ch2:system of pdes fundamental solutions nth der spherical harmonics coefficients}) and where $\wtilde{\Phi}_{JK}(\uvg{\xi})$ is defined as $\wtilde{\Phi}_{JK}(\uvg{\xi})=[\mathscr{L}_{JK}^*(\uvg{\xi})]^{-1}=[C_{iJKl}\hat{\xi}_l\hat{\xi}_i]^{-1}$.

The fundamental solutions of elastic, piezoelectric and magneto-electro-elastic materials are presented next.

\clearpage

\subsubsection{Elastic materials}
First, the fundamental solutions of anisotropic elastic materials are presented. In this case, the coupling tensors $e_{lij}$, $q_{lij}$ and $\lambda_{il}$ are set to zero and only the elastic response is considered. Three FCC crystals are considered, namely Nickel (Ni), Gold (Au) and Copper (Cu). The non-zero elastic constants are reported in Table (\ref{tab-Ch2:mat properties Cu, Au, and Ni}) of Section (\ref{app-Ch2:materials properties}).
\begin{figure}[ht]
\centering
	\begin{subfigure}{0.49\textwidth}
	\centering
	\includegraphics[width=\textwidth]{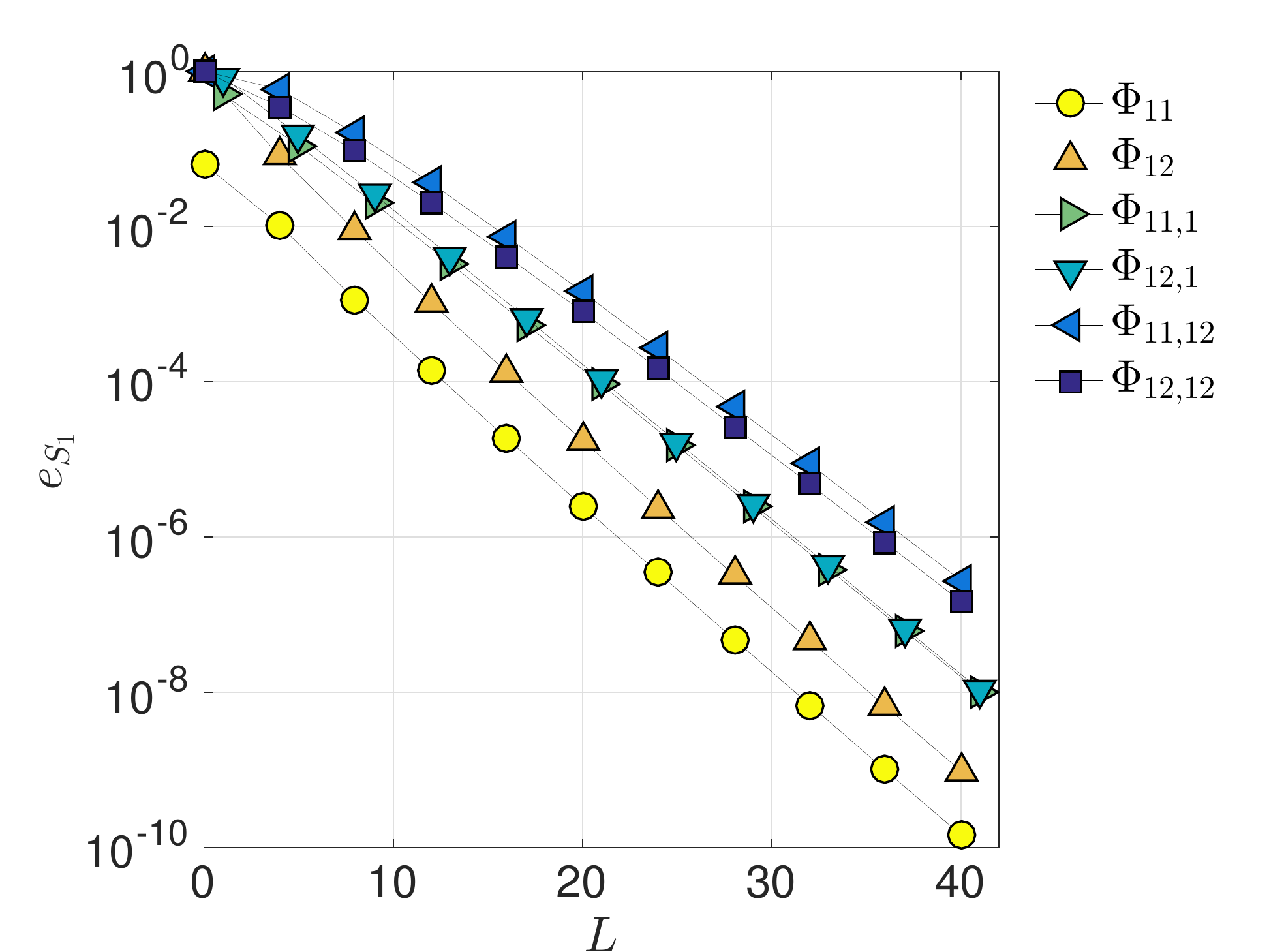}
	\caption{Ni}
	\end{subfigure}
\	
	\begin{subfigure}{0.49\textwidth}
	\centering
	\includegraphics[width=\textwidth]{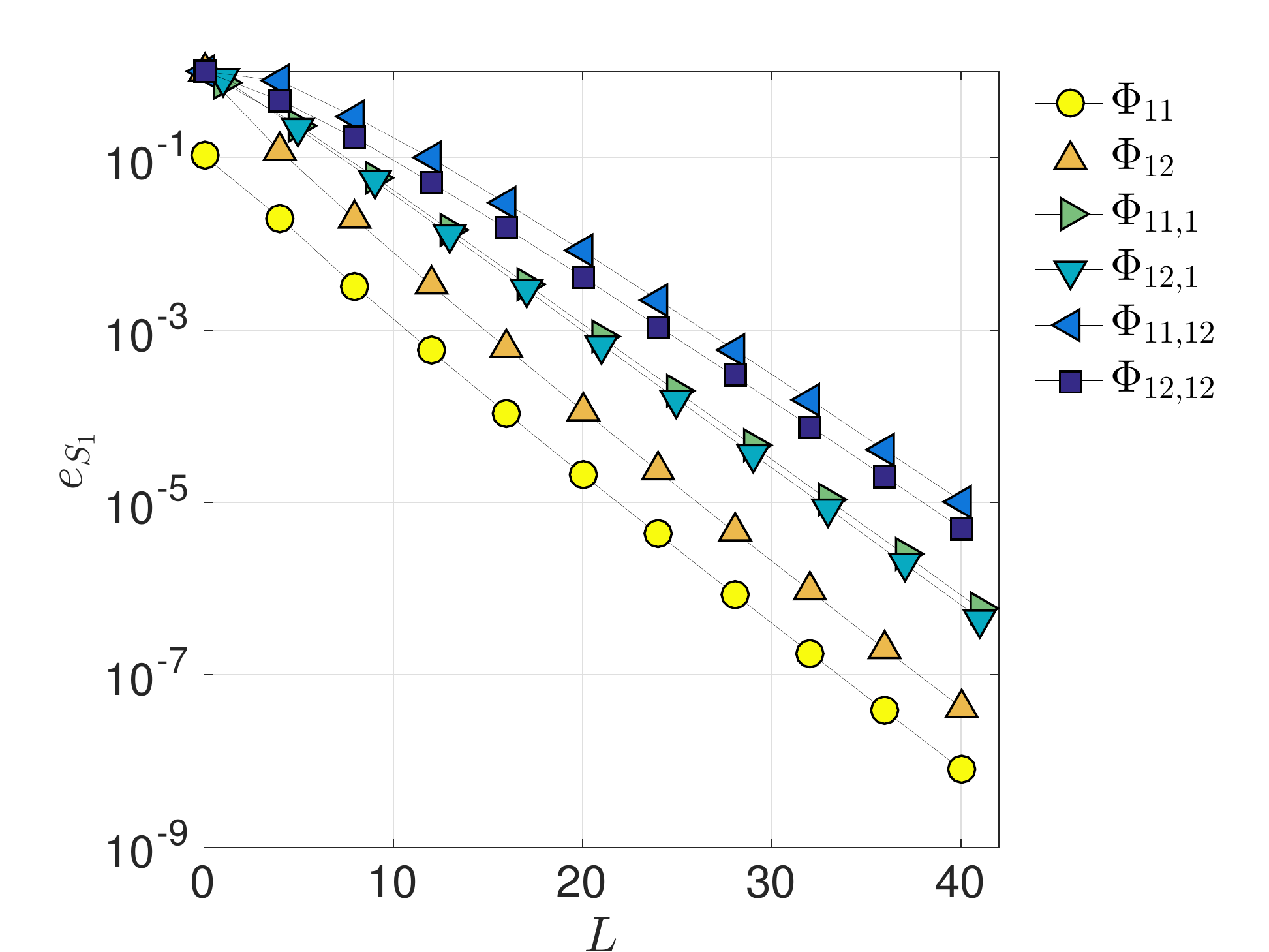}
	\caption{Au}
	\end{subfigure}
\	
	\begin{subfigure}{0.49\textwidth}
	\centering
	\includegraphics[width=\textwidth]{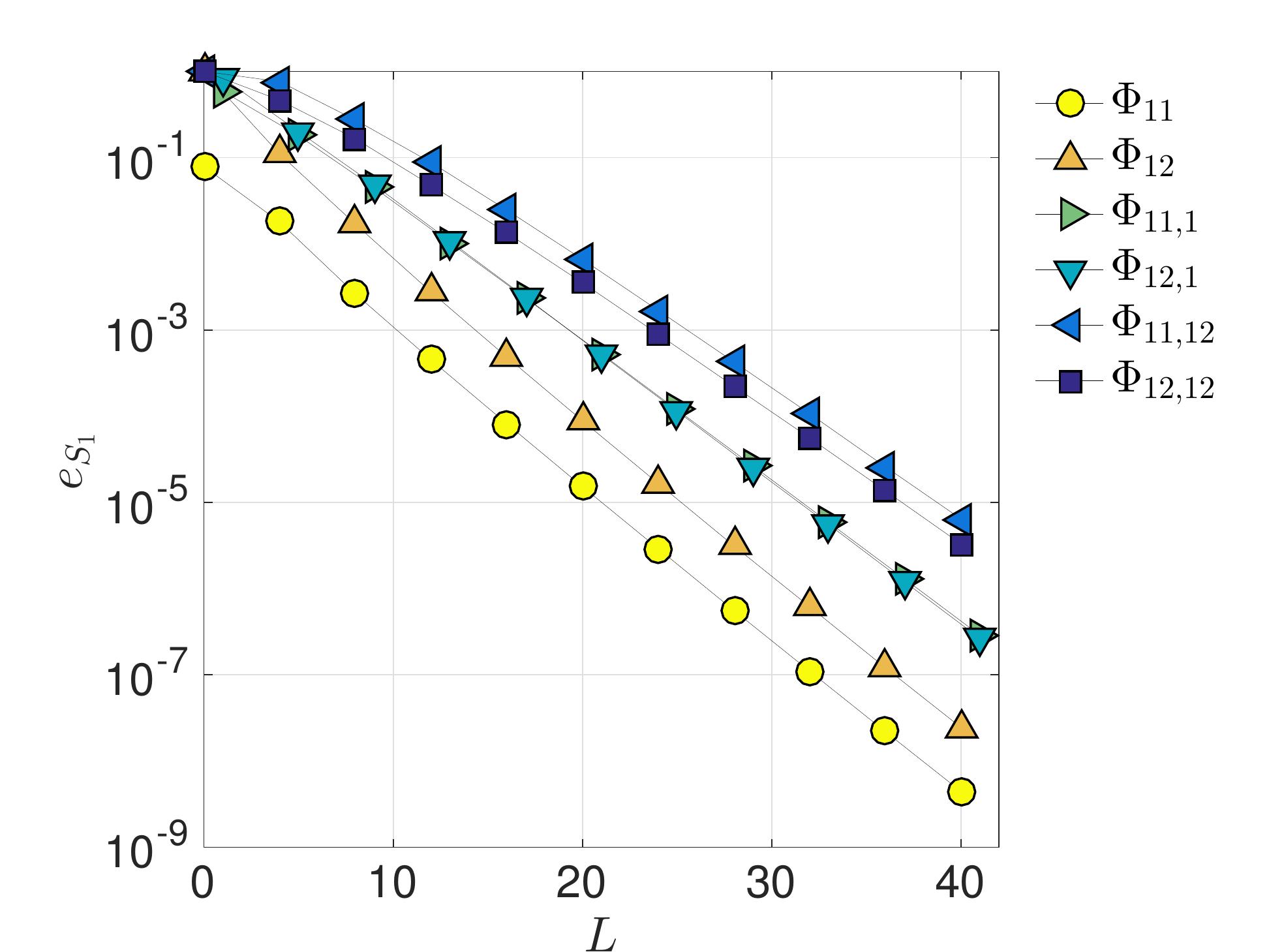}
	\caption{Cu}
	\end{subfigure}
\	
	\begin{subfigure}{0.49\textwidth}
	\centering
	\includegraphics[width=\textwidth]{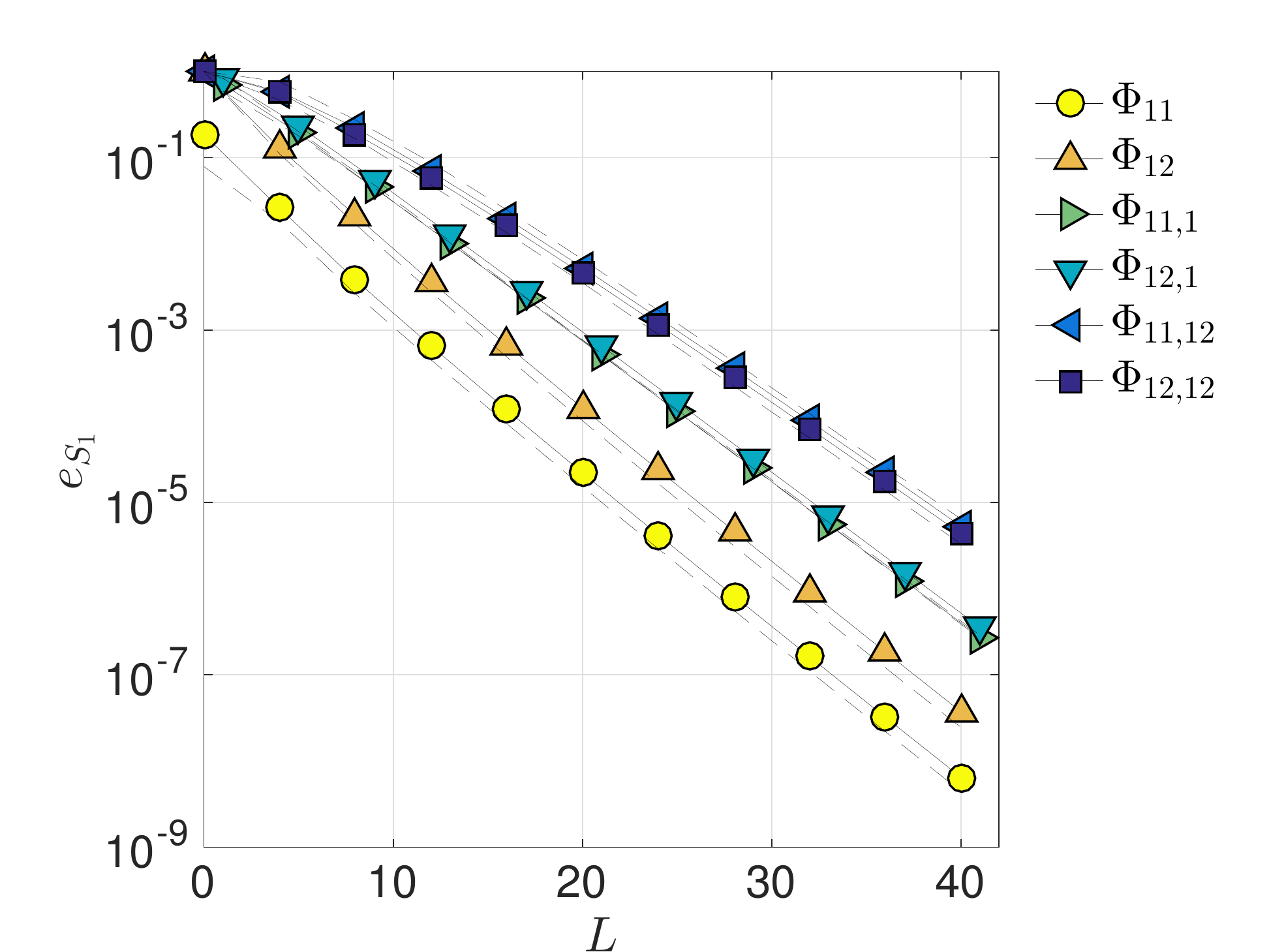}
	\caption{Cu rotated}
	\end{subfigure}
\caption{Error over the unit sphere for the fundamental solutions up to the second derivative as a function of the series truncation number $L$ for three FCC crystals: (\emph{a}) Nickel, (\emph{b}) Gold, (\emph{c}) Copper in principal reference system and (\emph{d}) Copper in a rotated reference system. The reference values are computed using the unit circle integration.}
\label{fig-Ch2:FCC crystals error over S1}
\end{figure}
Figure (\ref{fig-Ch2:FCC crystals error over S1}) reports the error $e_{S_1}$ over the unit sphere, as defined in Eq.(\ref{eq-Ch2:error over S_1}), as a function of the series truncation number $L$. The unit circle integration with an accuracy of 14 digits (decimal places) is used as reference value. The figure reports the fundamental solutions up to the second derivative and shows that the higher the order of derivation the higher is the number of terms to be retained to obtain the same level of accuracy.

Figure (\ref{fig-Ch2:FCC crystals error over S1}) also shows that the rate of convergence is linear in a logarithmic diagram with respect to the series truncation number. Furthermore, it is worth underlining that rotating the material reference system does not substantially affect the efficiency of the scheme. In particular, Figure (\ref{fig-Ch2:FCC crystals error over S1}d) shows the error over the unit sphere for the spherical harmonics expansion of the fundamental solutions of the FCC Copper whose reference system has been inclined by 60 degrees with respect to the $x_1$-$x_2$ plane and rotated by 30 degrees in the $x_1$-$x_2$ plane. The dashed lines represent the convergence of the spherical harmonics expansion when the material properties matrix is expressed in the principal reference system. Several combinations of rotation angles have been tested, but the convergence of the spherical harmonics expansions remained almost unaffected.

The effect of the level of anisotropy on the accuracy of the spherical harmonics expansion is assessed considering an FCC elastic crystal with different values of the Zener anisotropy ratio.
\begin{figure}[ht]
\centering
\includegraphics[width=0.5\textwidth]{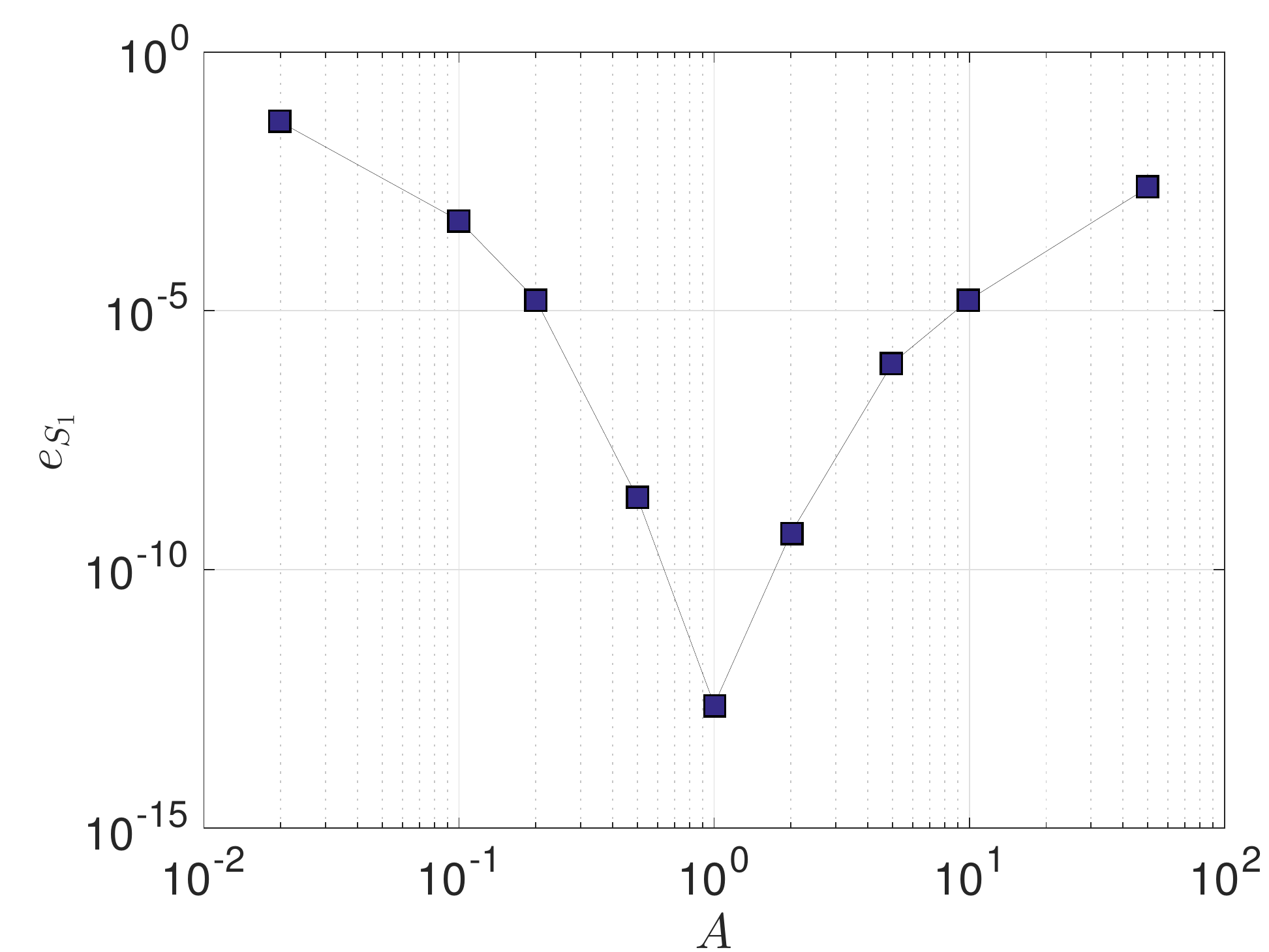}
\caption{Error over the unit sphere for the fundamental solutions $\Phi_{ij}(\uv{r})$ of FCC crystals as a function of the Zener anisotropic ratio for fixed series truncation number $L=40$.}
\label{fig-Ch2:Cu crystals error zener}
\end{figure}
Figure (\ref{fig-Ch2:Cu crystals error zener}) shows the error over the unit sphere for the fundamental solutions of an FCC elastic crystal whose elastic coefficients $c_{11}$ and $c_{12}$ are taken from crystalline Copper, and the coefficient $c_{44}$ is chosen as $c_{44} = A(c_{11}-c_{12})/2$, where the Zener anisotropy ratio $A$ takes values from 1/50 to 50. From the figure it is possible to assess the effect of the degree of anisotropy on the error. In particular, as the value of $A$ moves away from 1, i.e. the isotropic case, the error increases.

\begin{figure}[ht]
\centering
\includegraphics[width=0.5\textwidth]{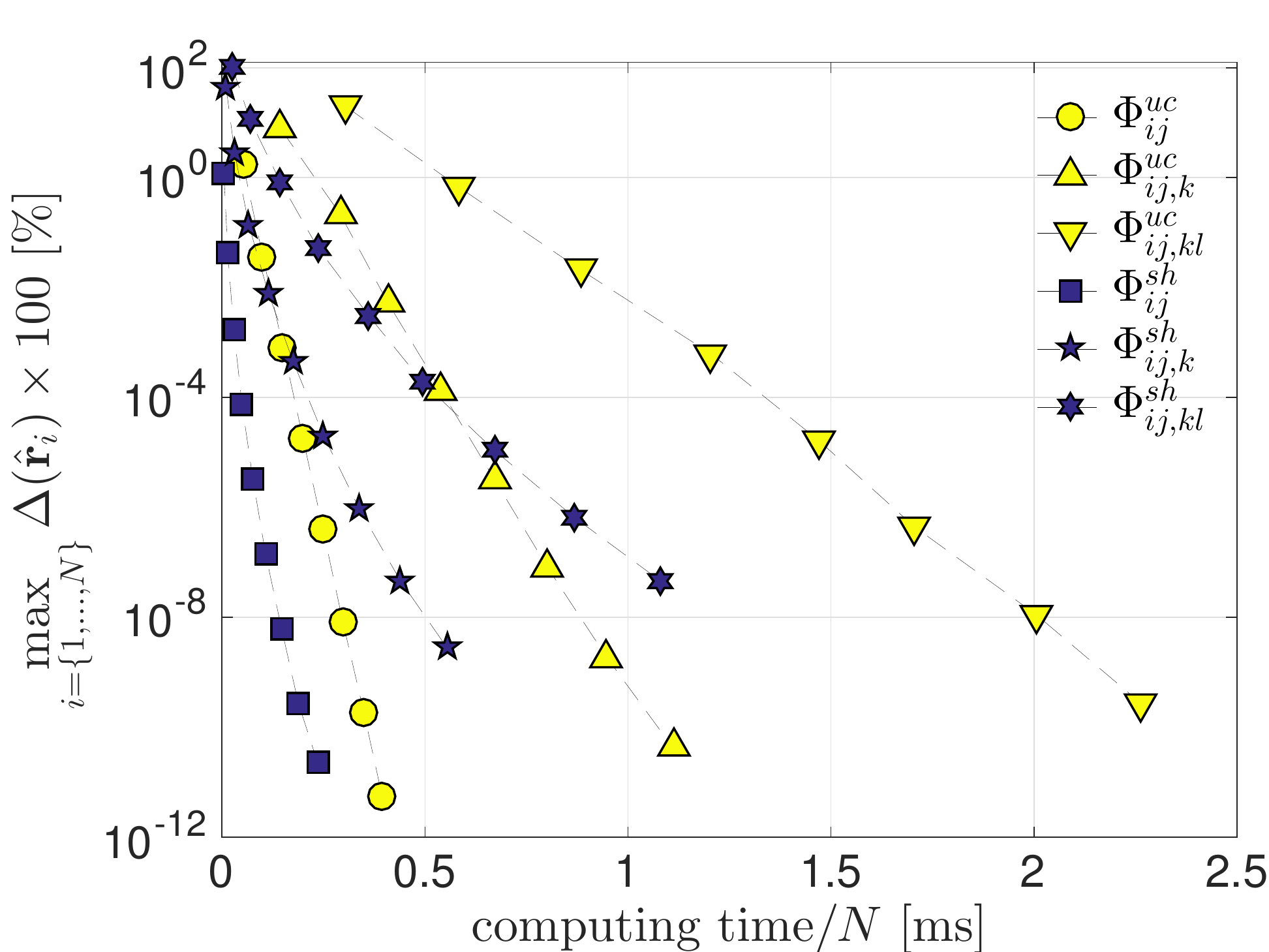}
\caption{Fundamental solutions for crystalline Cu: computing time versus maximum error for the fundamental solutions $\Phi_{ij}$ and their first and second derivatives obtained using the unit circle integration and the spherical harmonics expansion. $N$ denotes the number of source and observation points. In the legend, the superscript $uc$ stands for \emph{unit circle} whereas the superscript $sh$ for \emph{spherical harmonics}.}
\label{fig-Ch2:Cu crystals efficiency comparison}
\end{figure}

To show the efficiency of the spherical harmonics technique, Figure (\ref{fig-Ch2:Cu crystals efficiency comparison}) plots the average computing time per collocation/observation couple versus the maximum error of the considered fundamental solutions with respect to the selected reference value. The figure has been obtained by computing the fundamental solutions of an FCC Cu crystal and their derivatives up to the second order for $N$ couples of collocation and observation points, using both the unit circle integration and spherical harmonics expansion.
The points in the figure have been obtained by varying the number of Gauss points of the unit circle integration and the series truncation number $L$. It is clear that the higher the number of Gauss points (as well as the series truncation number), the higher the computing time and the smaller the error. From the figure, it is possible to see that the use of the spherical harmonics technique is more efficient than the unit circle integration since, in order to achieve the same accuracy in the fundamental solutions, it requires less computing time. However, a few considerations should be made:
\begin{itemize}
\item{The figure does not include the time needed to compute the coefficients of the spherical harmonics series since such time is fixed for any number of evaluation points and can be predominant only if a small number of computation points is considered. Such coefficients would be computed ad stored in advance in any effective implementation;}
\item{The error of the unit circle integration is linear with respect to the required computing time, whereas the error of the spherical harmonics expansion has a quadratic behaviour with respect to the required computing time (since number of coefficients to be computed scales with $L^2$); as a consequence, there is a level of accuracy at which the unit circle integration results more advantageous than the spherical harmonics expansion. However, reaching such a level of accuracy may be not always necessary in a numerical code or beyond the machine precision;}
\item{The presented results depend on the degree of anisotropy of the considered material as well as on the number of unknown functions of the considered system of PDEs. A more exhaustive understanding of the efficiency of the spherical harmonics expansion would definitely benefit from an a priori knowledge of the number of series coefficients required to obtain a certain level of accuracy, once the material properties are specified. However, the task is not trivial and is left open to further investigation.}
\end{itemize}

\begin{figure}[H]
\centering
\includegraphics[width=0.5\textwidth]{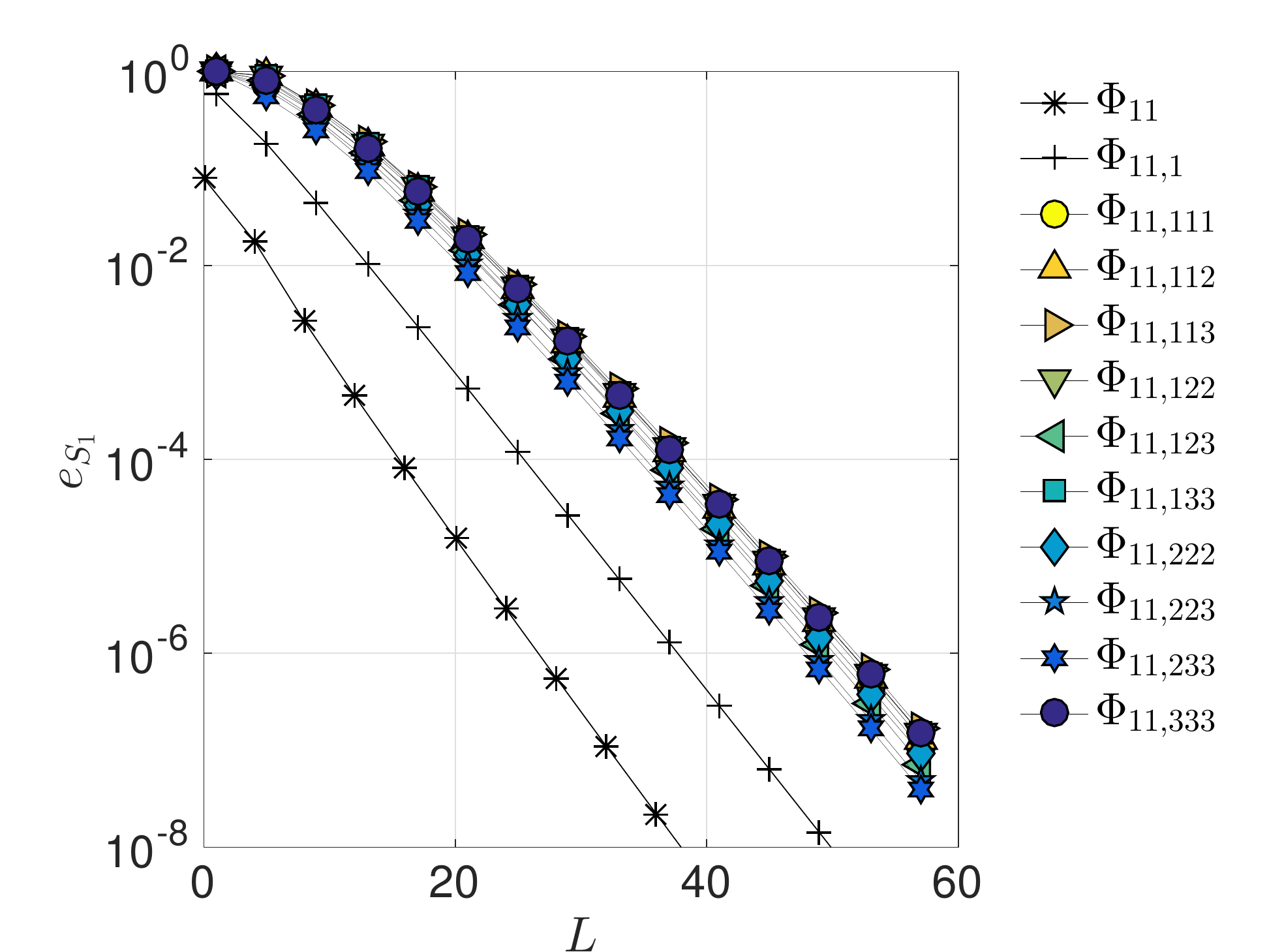}
\caption{Fundamental solutions for crystalline Cu: error, $e_{S_1}[\Phi_{11,ijk}(\uv{r})]$, over the unit sphere for the third derivatives of the fundamental solutions $\Phi_{11}$ as a function of the series truncation number $L$. The error for $\Phi_{11}(\uv{r})$ and $\Phi_{11,1}(\uv{r})$ are reported for comparison purposes.}
\label{fig-Ch2:Cu crystals Phi_ijk error over S1}
\end{figure}

Figure (\ref{fig-Ch2:Cu crystals Phi_ijk error over S1}) shows the error $e_{S_1}$ for the third derivatives of the fundamental solution $\Phi_{11}$ of FCC Cu crystals as a function of the series truncation number $L$. In this case, the fundamental solutions computed with $L=60$ are used as reference. The value $L = 60$ has been selected as it has been verified that the addition of further terms would only affect after the 14th decimal place, thus the found values would practically coincide with those provided by the unit circle integral. For comparison purposes, the figure also reports the error $e_{S_1}$ for the fundamental solution $\Phi_{11}(\uv{r})$ and its first derivative $\Phi_{11,1}(\uv{r})$.

Figure (\ref{fig-Ch2:Cu ball plots}) presents a few selected fundamental solutions for Cu crystals. In order to appreciate their directional dependence, Figures (\ref{fig-Ch2:Cu ball plots}a,c,e) plot the fundamental solutions $\Phi_{11}(\uv{r})$, $\Phi_{11,1}(\uv{r})$, $\Phi_{11,12}(\uv{r})$ respectively, using a spherical representation defined as
\begin{equation}\label{eq-Ch2:spherical plot definition}
\left\{\begin{array}{l}
x(\thet,\ph)=\varrho(\thet,\ph)\sin\thet\cos\ph\\
y(\thet,\ph)=\varrho(\thet,\ph)\sin\thet\sin\ph\\
z(\thet,\ph)=\varrho(\thet,\ph)\cos\thet
\end{array}\right.,
\end{equation}
where the amplitude $\varrho(\thet,\ph)$ is given by the absolute value of the considered fundamental solution. Figures (\ref{fig-Ch2:Cu ball plots}b,d,f) show the relative difference between the fundamental solutions computed using the spherical harmonics expansion and the fundamental solutions computed using the unit circle integration. The difference is plotted over the unit sphere $S_1$.

In Figure (\ref{fig-Ch2:Cu ball plots higher-order derivatives}), some selected third derivatives are plotted using the spherical representation (\ref{eq-Ch2:spherical plot definition}).

\newpage
\vfill

\begin{figure}[H]
\centering
	\begin{subfigure}{0.49\textwidth}
	\centering
	\includegraphics[width=\textwidth]{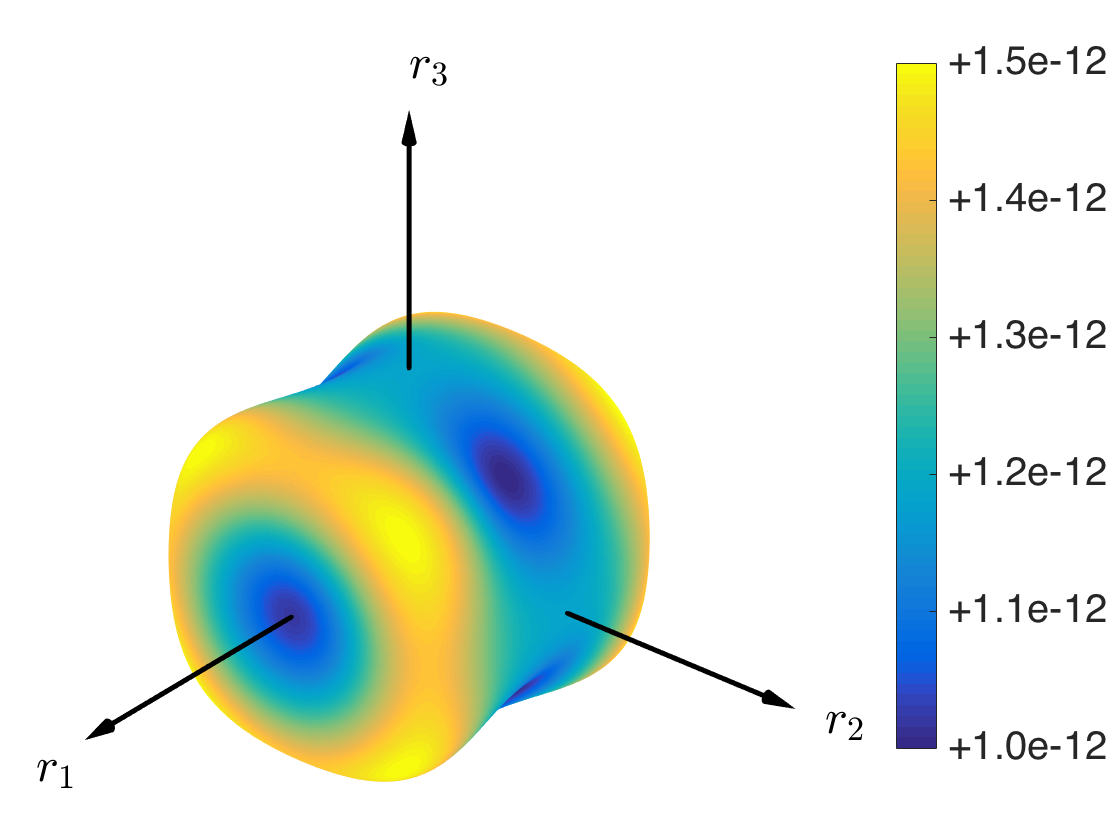}
	\caption{$\Phi_{11}(\uv{r})$ [m]}
	\end{subfigure}
\	
	\begin{subfigure}{0.49\textwidth}
	\centering
	\includegraphics[width=\textwidth]{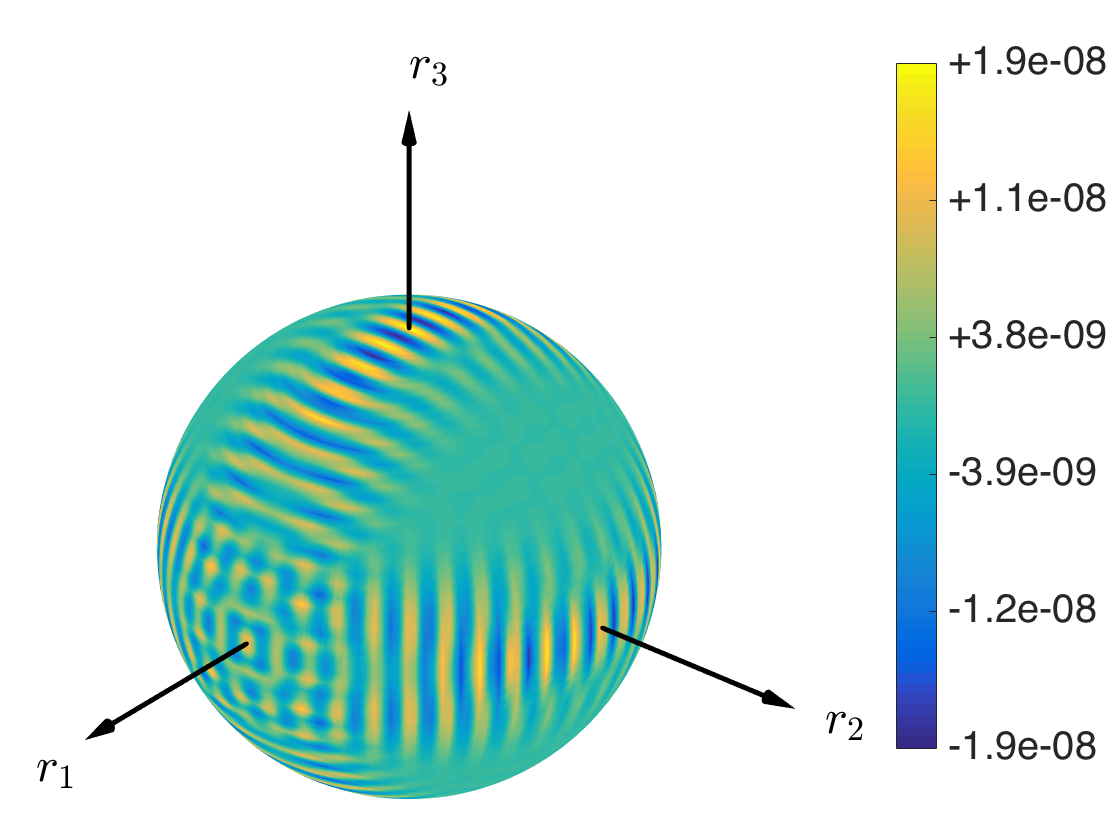}
	\caption{$\Delta[\Phi_{11}(\uv{r})]$}
	\end{subfigure}
\
	\begin{subfigure}{0.49\textwidth}
	\centering
	\includegraphics[width=\textwidth]{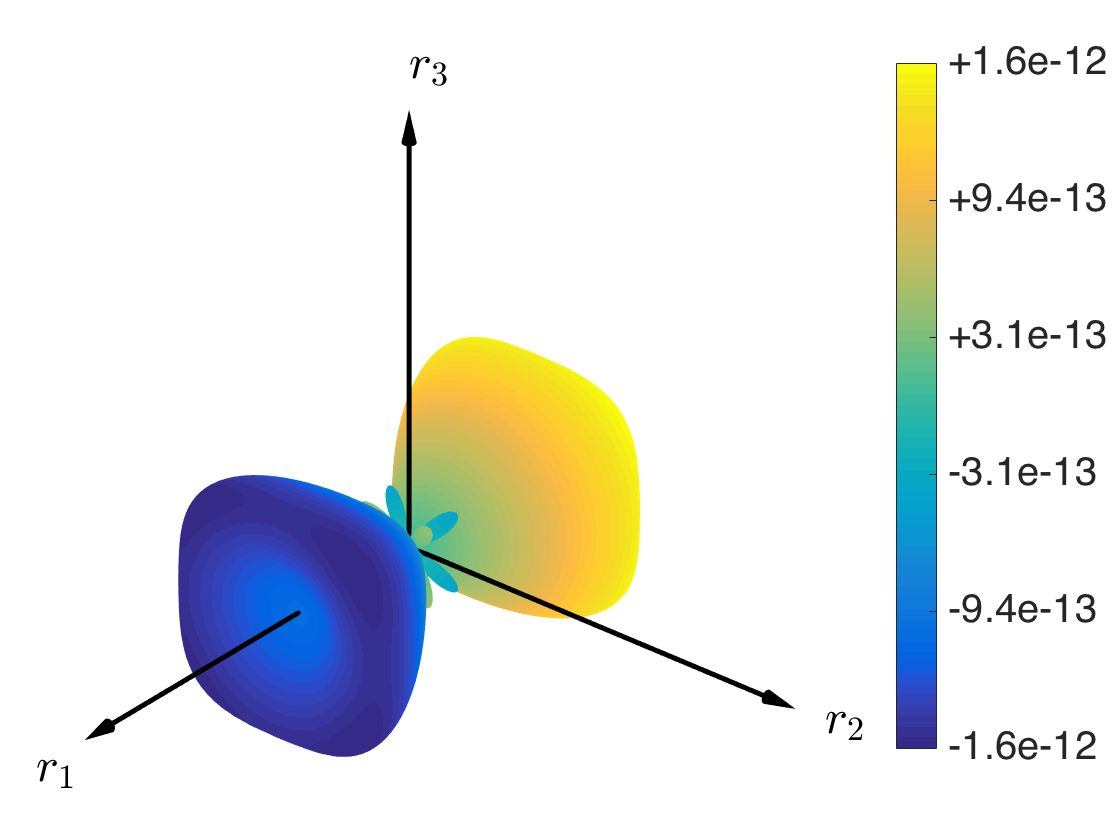}
	\caption{$\Phi_{11,1}(\uv{r})$ [-]}
	\end{subfigure}
\	
	\begin{subfigure}{0.49\textwidth}
	\centering
	\includegraphics[width=\textwidth]{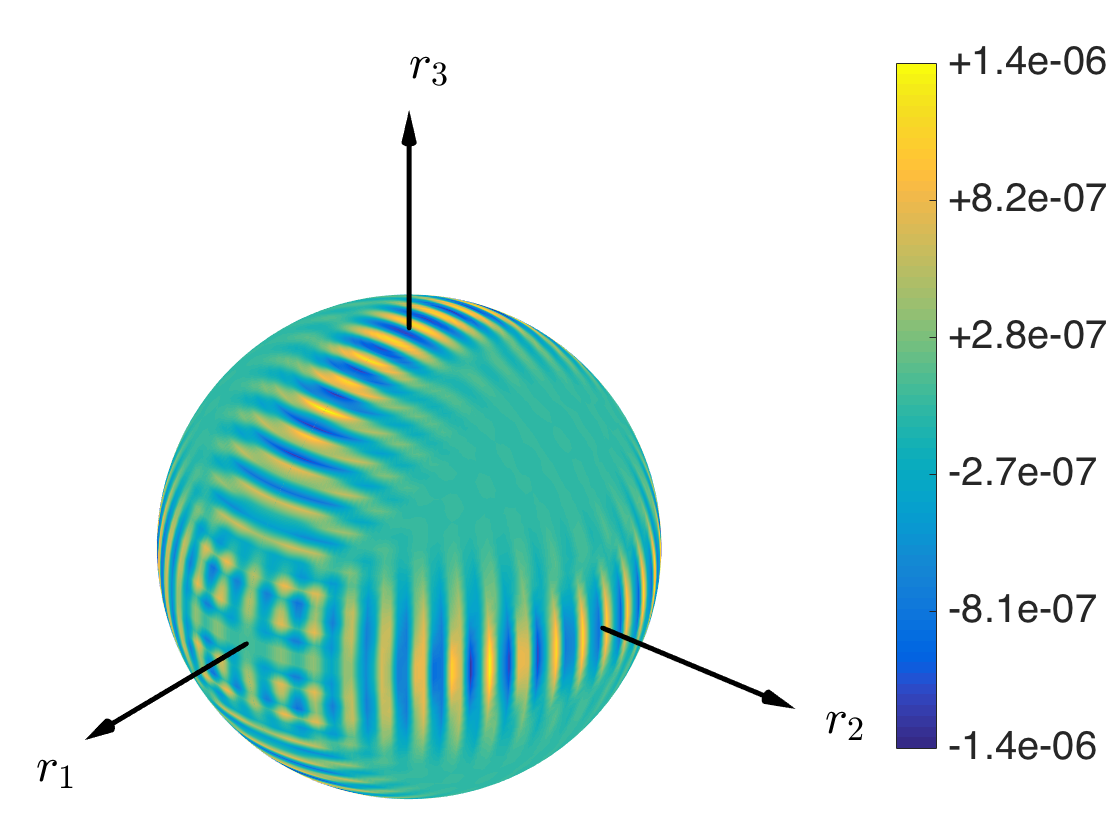}
	\caption{$\Delta[\Phi_{11,1}(\uv{r})]$}
	\end{subfigure}
\
	\begin{subfigure}{0.49\textwidth}
	\centering
	\includegraphics[width=\textwidth]{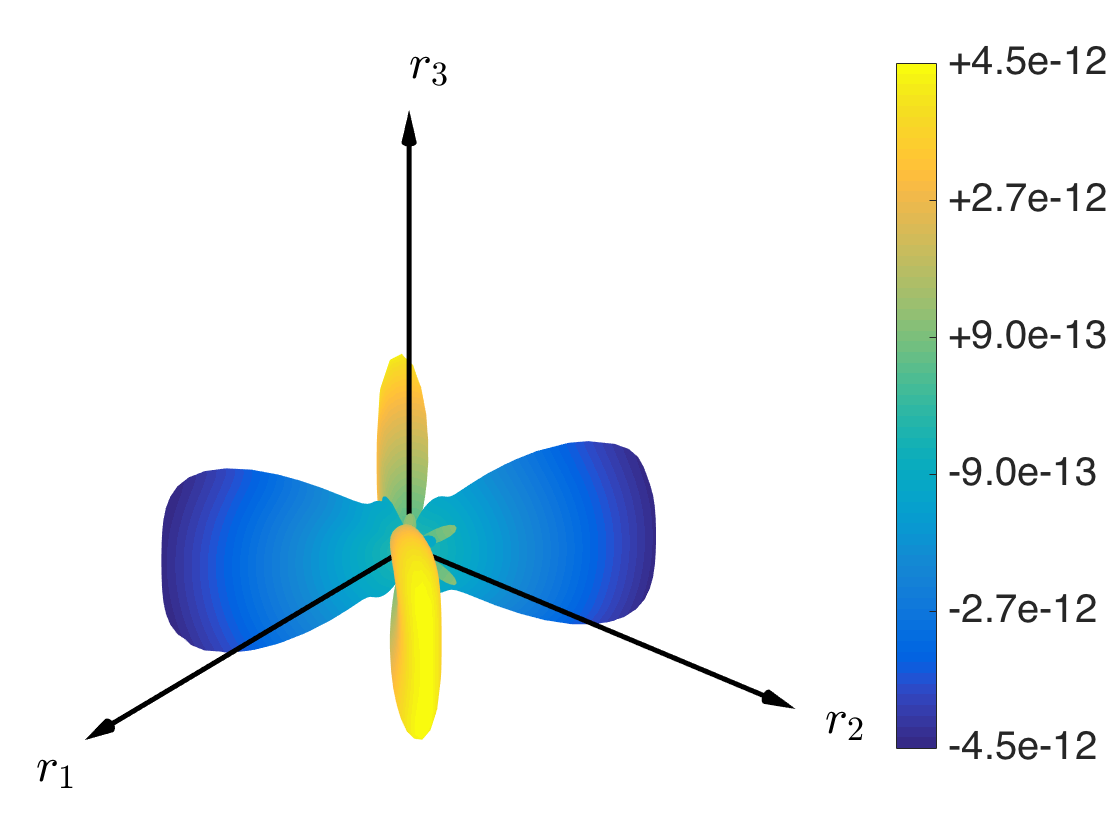}
	\caption{$\Phi_{11,12}(\uv{r})$ [$\mathrm{m}^{-1}$]}
	\end{subfigure}
\	
	\begin{subfigure}{0.49\textwidth}
	\centering
	\includegraphics[width=\textwidth]{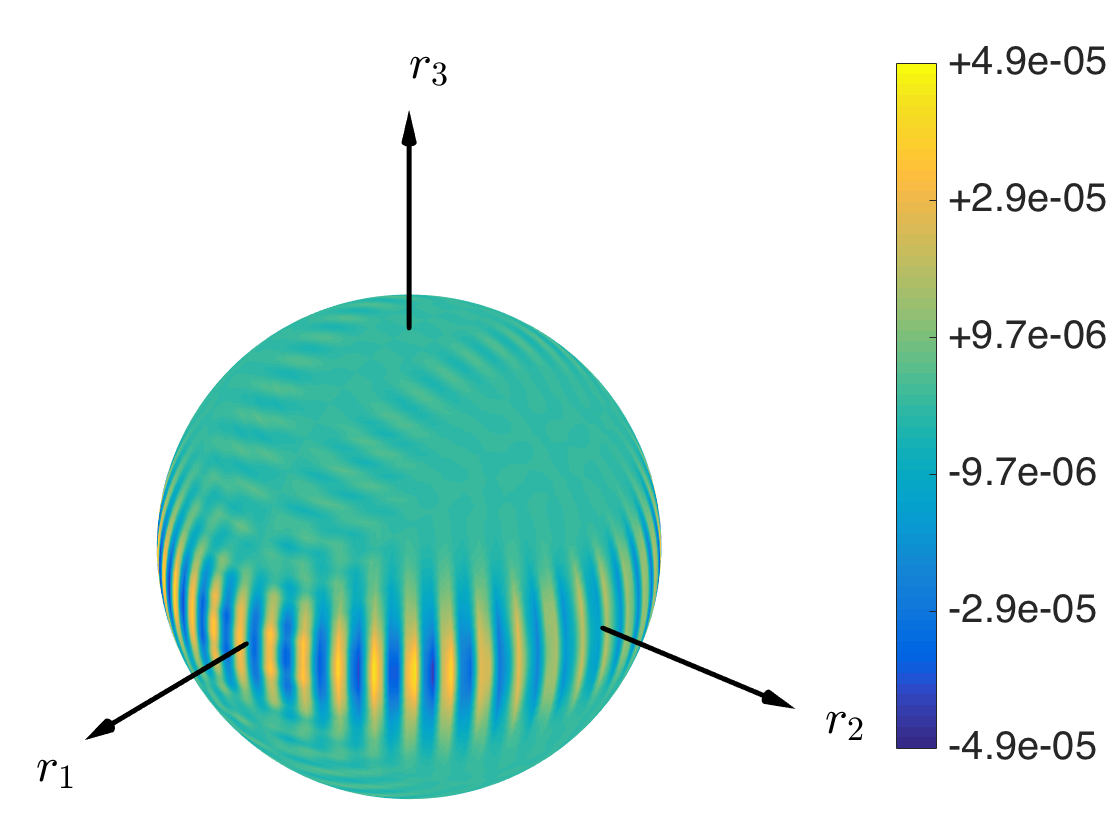}
	\caption{$\Delta[\Phi_{11,12}(\uv{r})]$}
	\end{subfigure}
\caption{Fundamental solutions of $Cu$ crystal: (\emph{a},\emph{c},\emph{e}) spherical plots of a few selected fundamental solutions; (\emph{b},\emph{d},\emph{f}) plot over the unit sphere of the difference between the fundamental solutions computed using the spherical harmonics expansions and the unit circle integration.}
\label{fig-Ch2:Cu ball plots}
\end{figure}

\begin{figure}[H]
\centering
	\begin{subfigure}{0.49\textwidth}
	\centering
	\includegraphics[width=\textwidth]{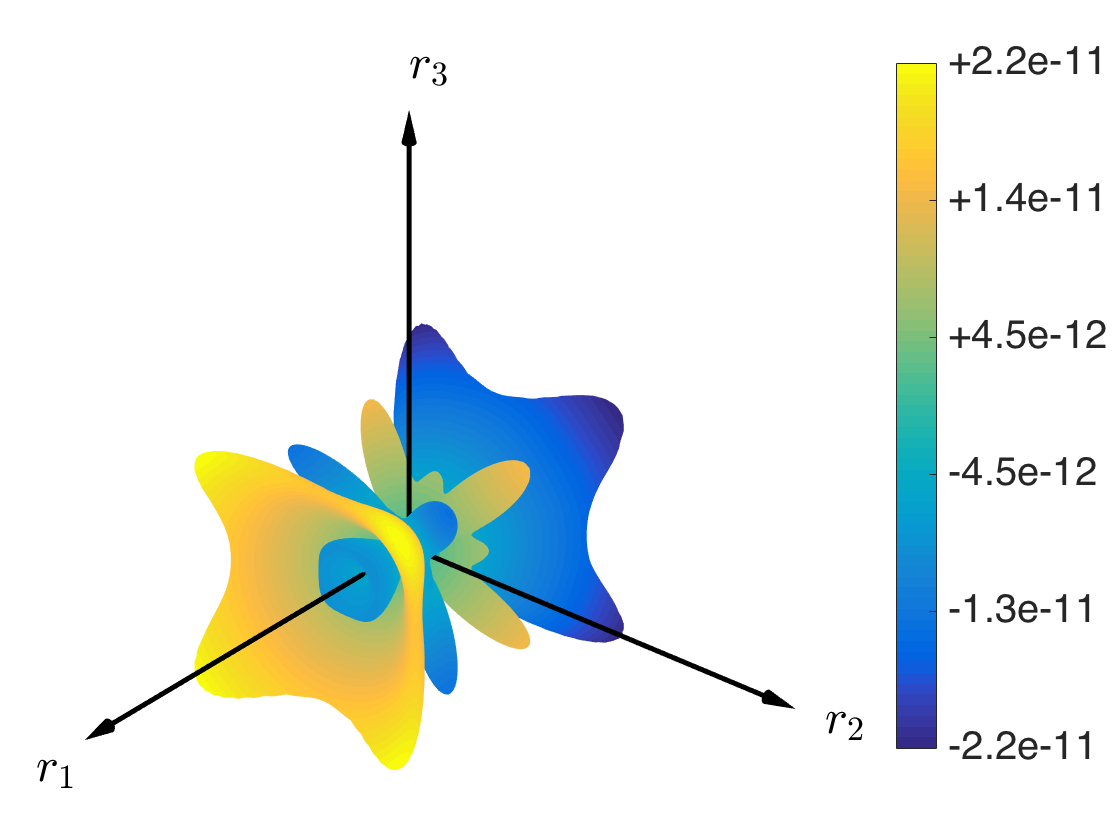}
	\caption{$\Phi_{11,111}(\uv{r})$ [$\mathrm{m}^{-2}$]}
	\end{subfigure}
\
	\begin{subfigure}{0.49\textwidth}
	\centering
	\includegraphics[width=\textwidth]{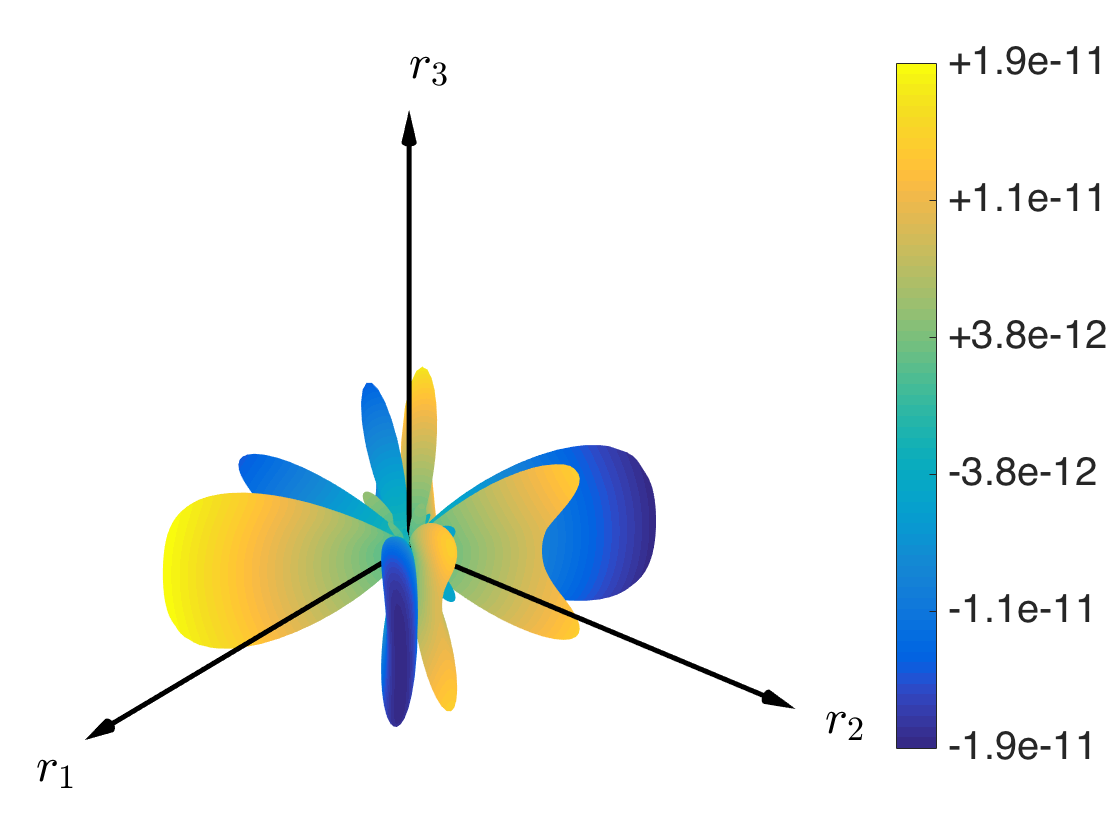}
	\caption{$\Phi_{11,112}(\uv{r})$ [$\mathrm{m}^{-2}$]}
	\end{subfigure}
\	
	\begin{subfigure}{0.49\textwidth}
	\centering
	\includegraphics[width=\textwidth]{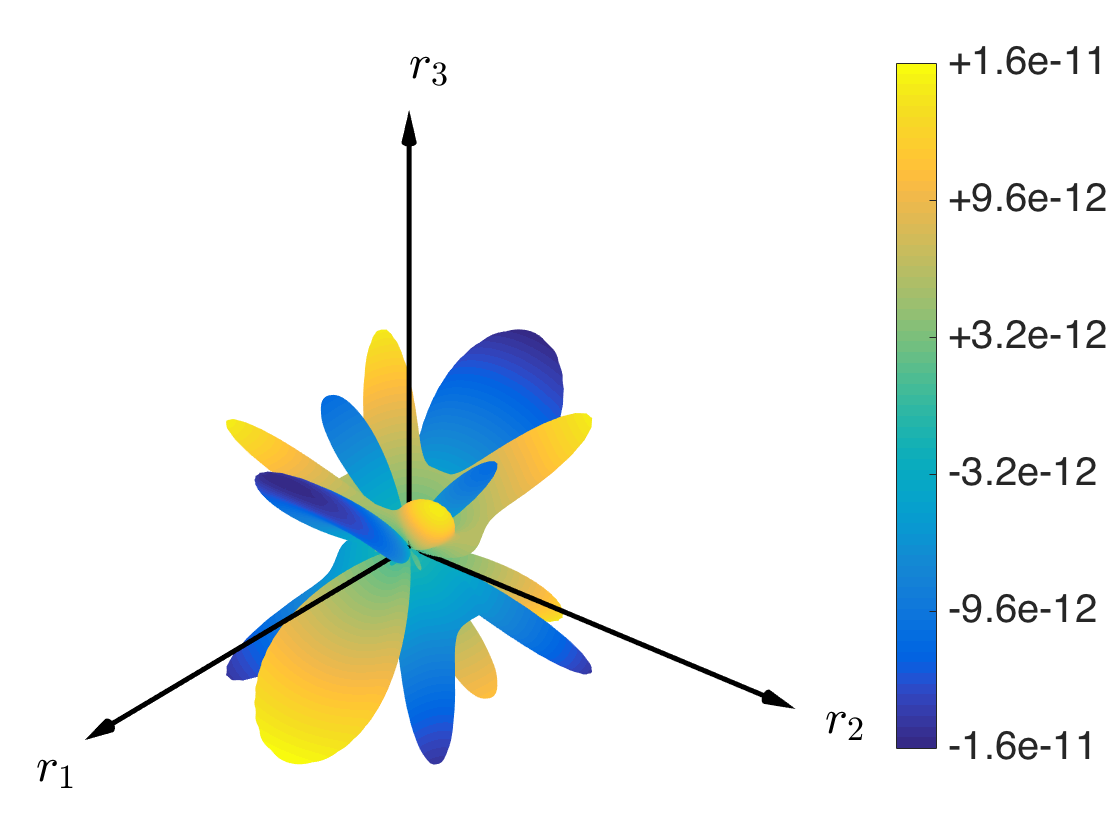}
	\caption{$\Phi_{11,113}(\uv{r})$ [$\mathrm{m}^{-2}$]}
	\end{subfigure}
\
	\begin{subfigure}{0.49\textwidth}
	\centering
	\includegraphics[width=\textwidth]{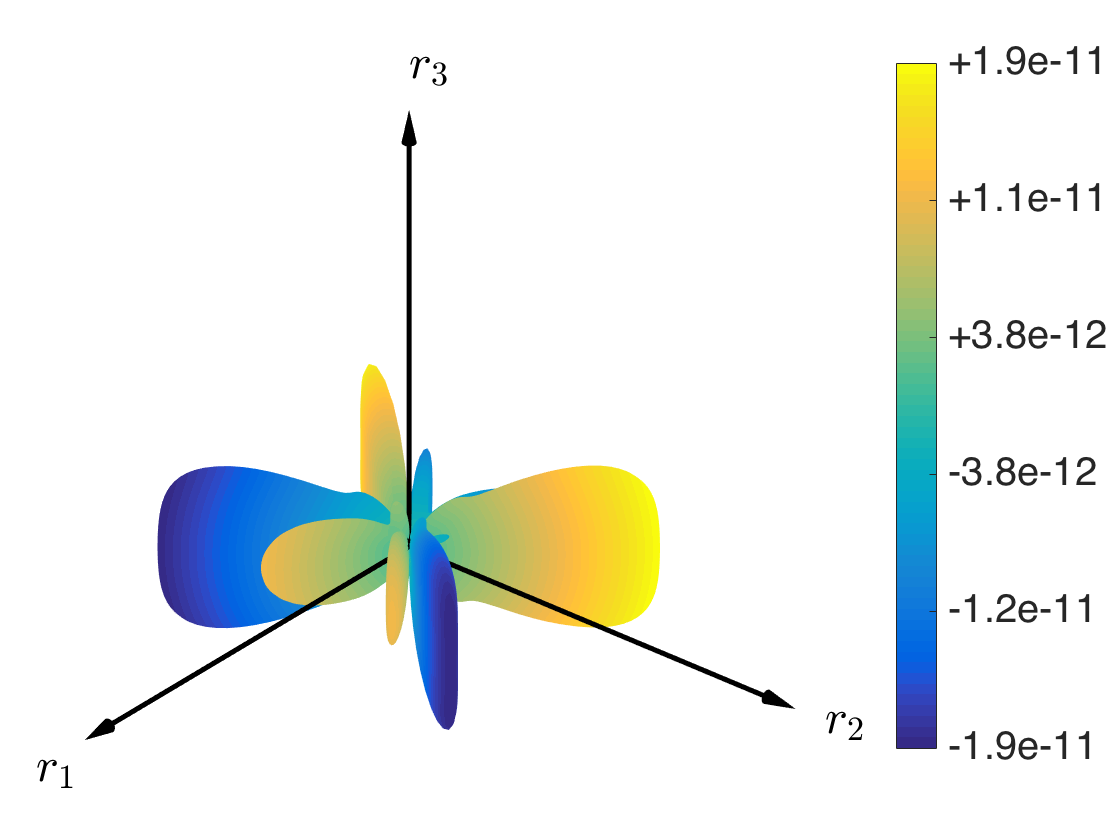}
	\caption{$\Phi_{11,122}(\uv{r})$ [$\mathrm{m}^{-2}$]}
	\end{subfigure}
\
	\begin{subfigure}{0.49\textwidth}
	\centering
	\includegraphics[width=\textwidth]{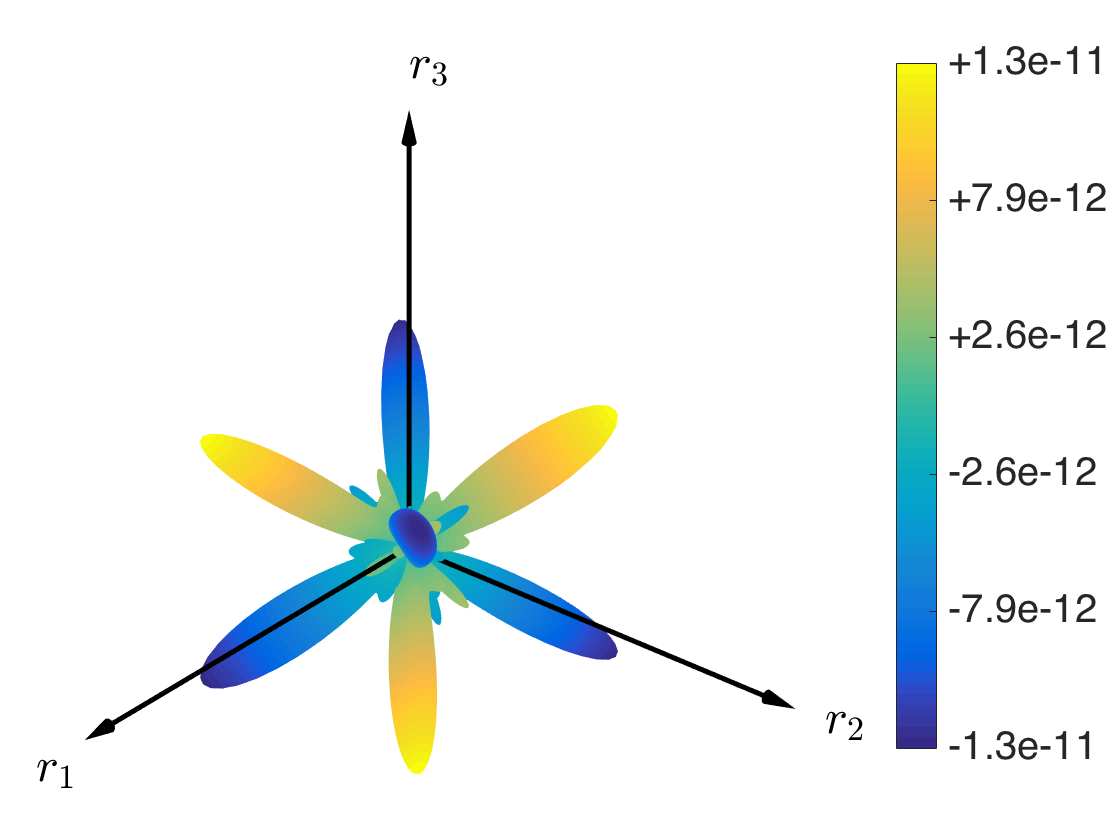}
	\caption{$\Phi_{11,123}(\uv{r})$ [$\mathrm{m}^{-2}$]}
	\end{subfigure}
\
	\begin{subfigure}{0.49\textwidth}
	\centering
	\includegraphics[width=\textwidth]{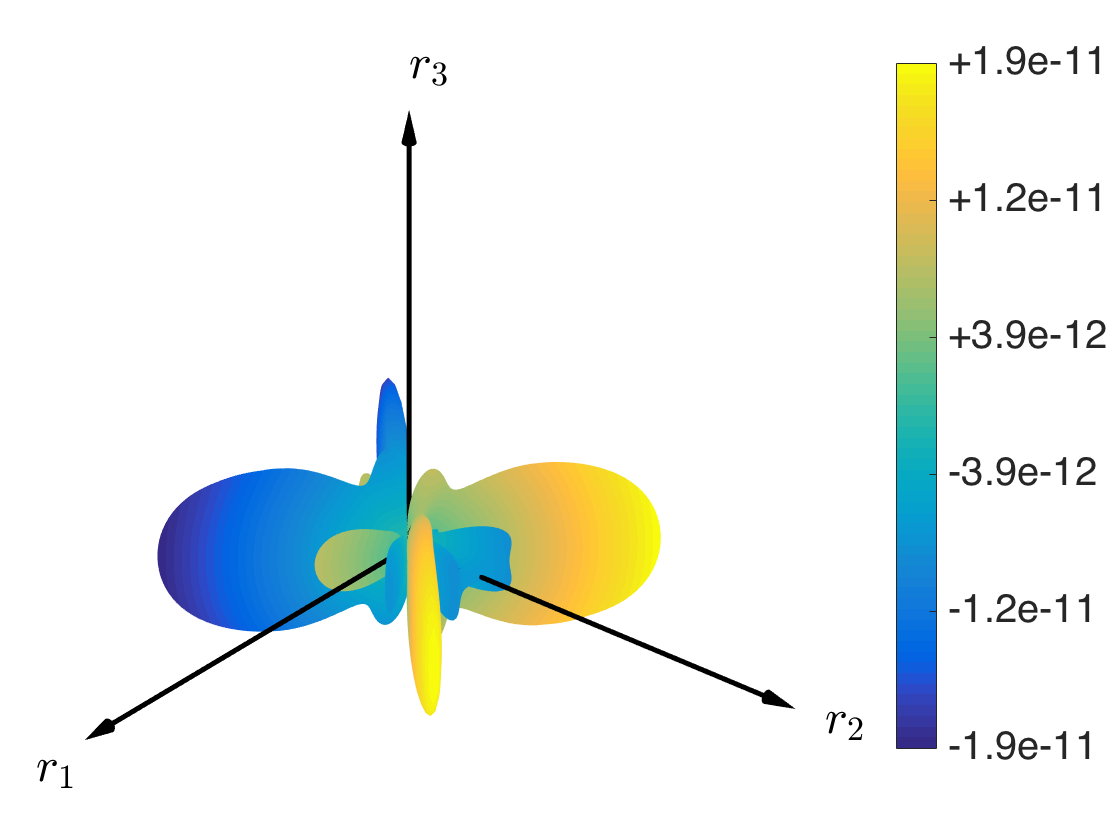}
	\caption{$\Phi_{11,222}(\uv{r})$ [$\mathrm{m}^{-2}$]}
	\end{subfigure}
\caption{Fundamental solutions for crystalline $Cu$: spherical plots of some selected third derivatives of the fundamental solution $\Phi_{11}$.}
\label{fig-Ch2:Cu ball plots higher-order derivatives}
\end{figure}

\clearpage

\subsubsection{Piezo-electric materials}
Next, piezo-electric materials are studied. In this case, the coupling tensors $q_{lij}$ and $\lambda_{il}$ are set to zero, whereas the piezoelectric coupling tensor $e_{lij}$ is retained. Two piezoelectric materials are considered, namely a transversely isotropic lead zirconate titanate (PZT-4) ceramic and an orthotropic piezoelectric polyvinylidene fluoride (PVDF), whose non-zero elastic, piezoelectric and dielectric constants are reported in Table (\ref{tab-Ch2:mat properties PE trav iso}) and Table (\ref{tab-Ch2:mat properties PE ortho}), respectively. The plane of isotropy of PZT-4 is the $x_1$-$x_2$ plane.

Figure (\ref{fig-Ch2:PZT-4 error over S1}a) shows the convergence of the spherical harmonics expansions of the first derivative $\Phi_{ij,2}(\mathbf{r})$ of PZT-4 computed at $\mathbf{r}=\{1,1,1\}$. As reference, the results reported in Ref.\ \cite{buroni2010} and computed using the explicit expressions of the fundamental solutions are used. Figure (\ref{fig-Ch2:PZT-4 error over S1}b) shows the error $e_{S_1}$ for the same derivative.

\begin{figure}[ht]
\centering
	\begin{subfigure}{0.49\textwidth}
	\centering
	\includegraphics[width=\textwidth]{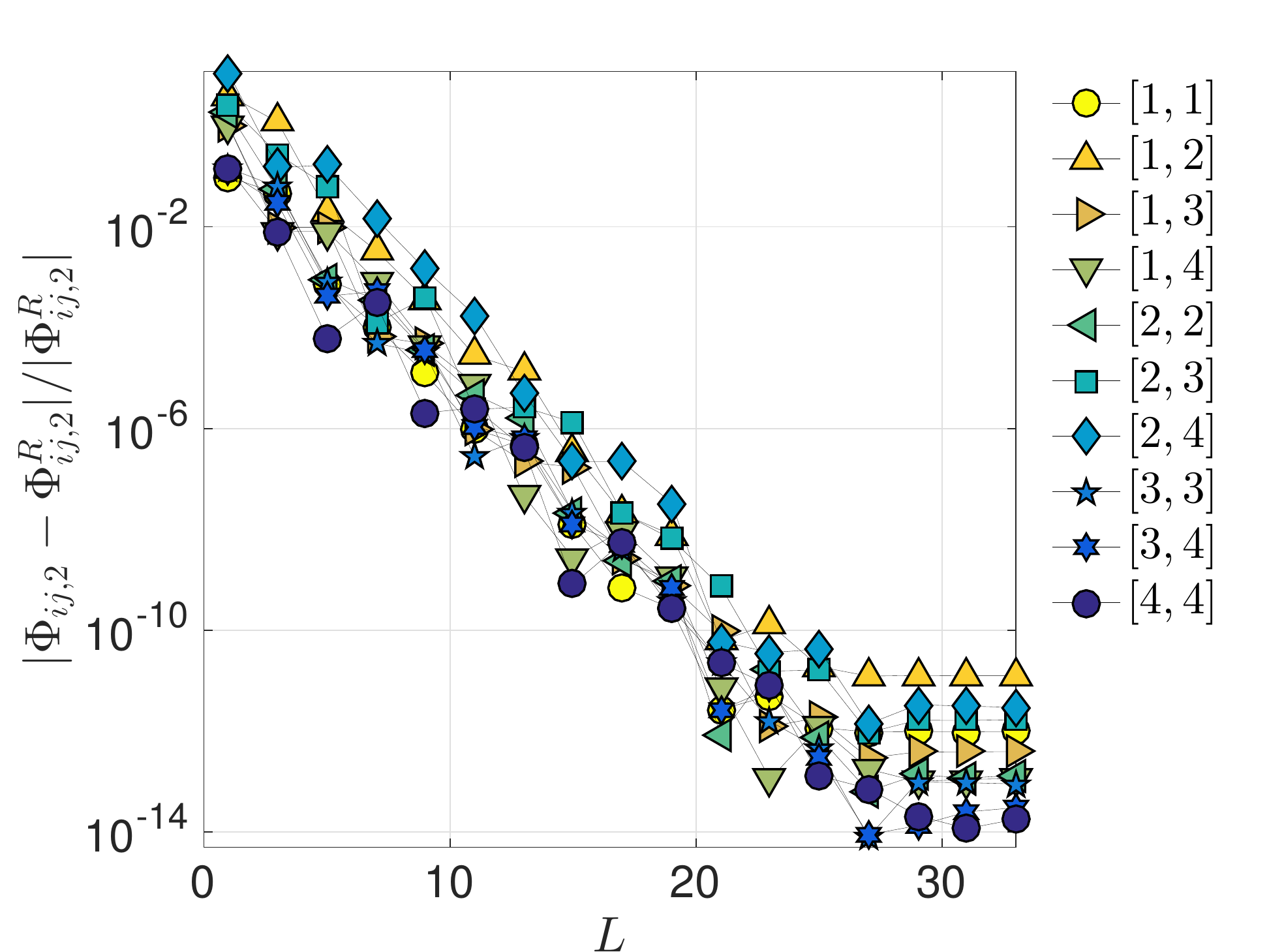}
	\caption{}
	\end{subfigure}
\	
	\begin{subfigure}{0.49\textwidth}
	\centering
	\includegraphics[width=\textwidth]{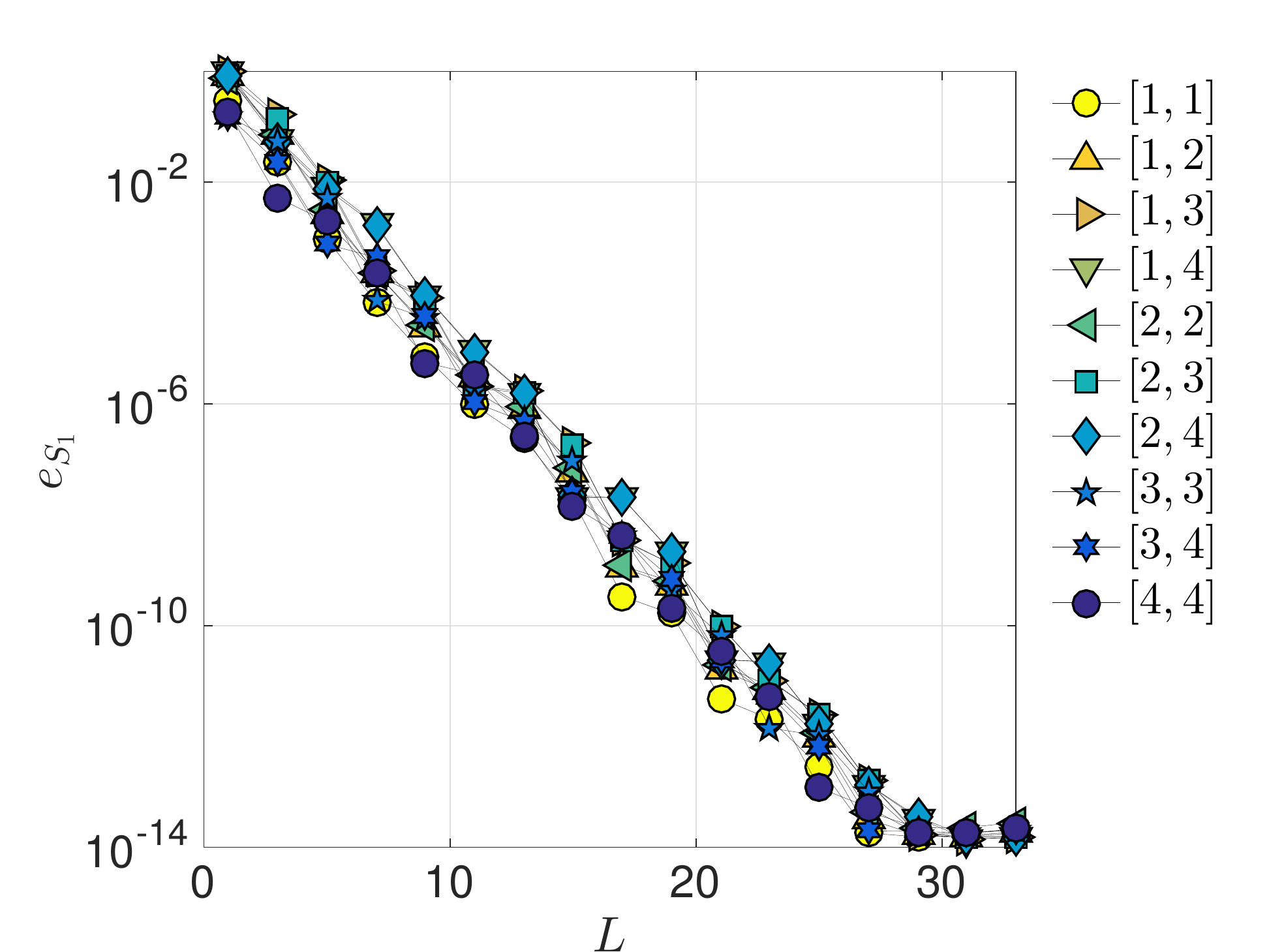}
	\caption{$e_{S_1}[\Phi_{ij,2}(\uv{r})]$}
	\end{subfigure}
\caption{Fundamental solutions for transversely isotropic PZT-4: \emph{a}) error for the first derivative $\Phi_{ij,2}(\mathbf{r})$ computed at $\mathbf{r}=\{1,1,1\}$ as a function of the series truncation number $L$ (the reference values are computed using the explicit expressions of the fundamental solutions as in Ref.\ \cite{buroni2010}); \emph{b}) error over the unit sphere for the first derivatives $\Phi_{ij,2}(\uv{r})$ for the fundamental solutions as a function of the series truncation number $L$ (the reference values are computed using the unit circle integration).}
\label{fig-Ch2:PZT-4 error over S1}
\end{figure}

Figure (\ref{fig-Ch2:PVDF error over S1}) reports the error $e_{S_1}$ of the fundamental solutions and their derivatives up to the second order for the orthotropic piezoelectric material PVDF. The unit circle integration is used as reference value. Figure (\ref{fig-Ch2:PVDF ball plots}) presents a few selected fundamental solutions for PVDF. Figures (\ref{fig-Ch2:PVDF ball plots}a,c,e) plot the fundamental solutions $\Phi_{34}(\uv{r})$, $\Phi_{34,1}(\uv{r})$, $\Phi_{34,12}(\uv{r})$ respectively using the spherical representation (\ref{eq-Ch2:spherical plot definition}), whereas Figures (\ref{fig-Ch2:PVDF ball plots}b,d,f) show the error with respect to the unit circle integration.

\begin{figure}[ht]
\centering
	\begin{subfigure}{0.49\textwidth}
	\centering
	\includegraphics[width=\textwidth]{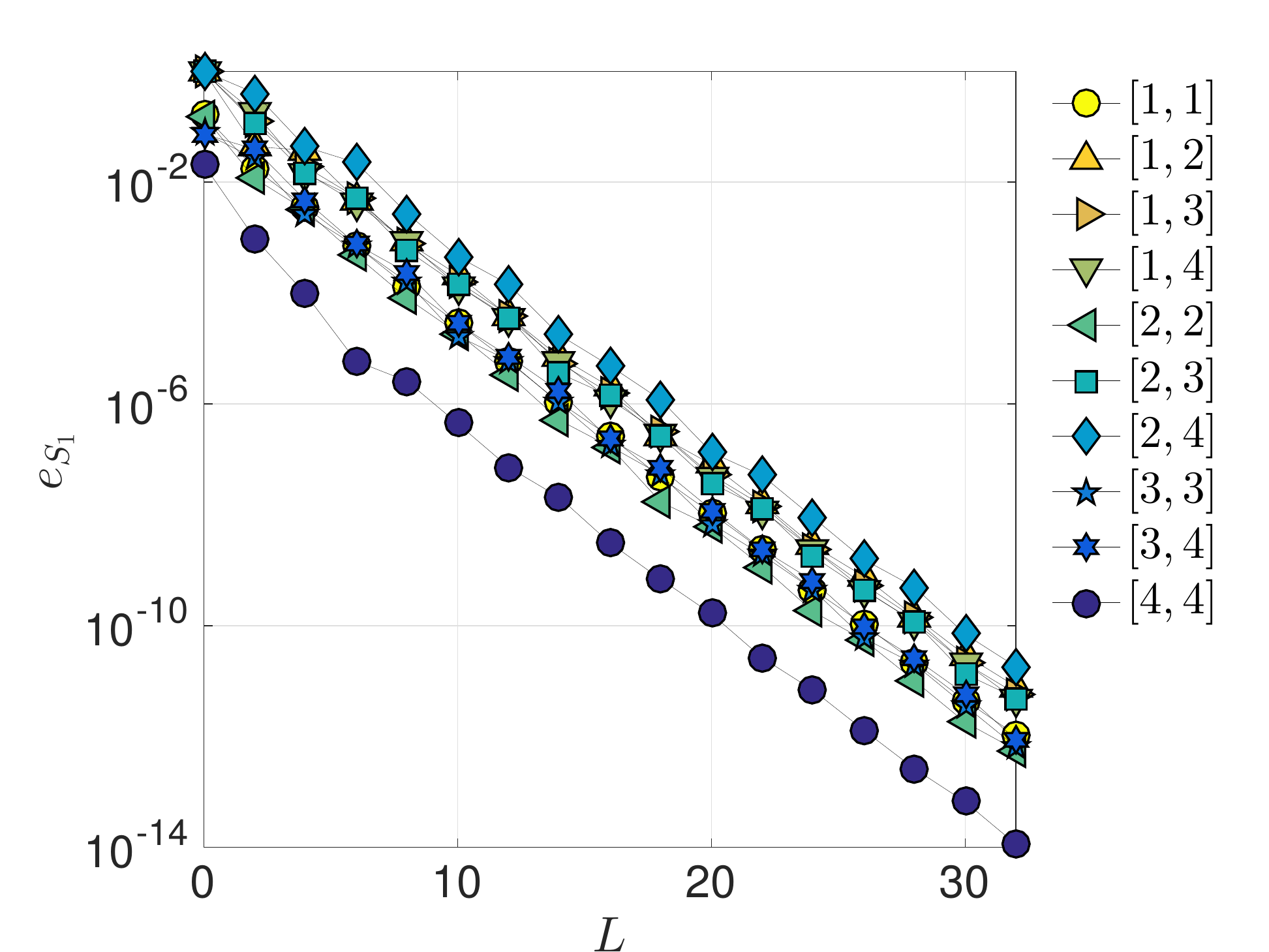}
	\caption{$e_{S_1}[\Phi_{ij}(\uv{r})]$}
	\end{subfigure}
\	
	\begin{subfigure}{0.49\textwidth}
	\centering
	\includegraphics[width=\textwidth]{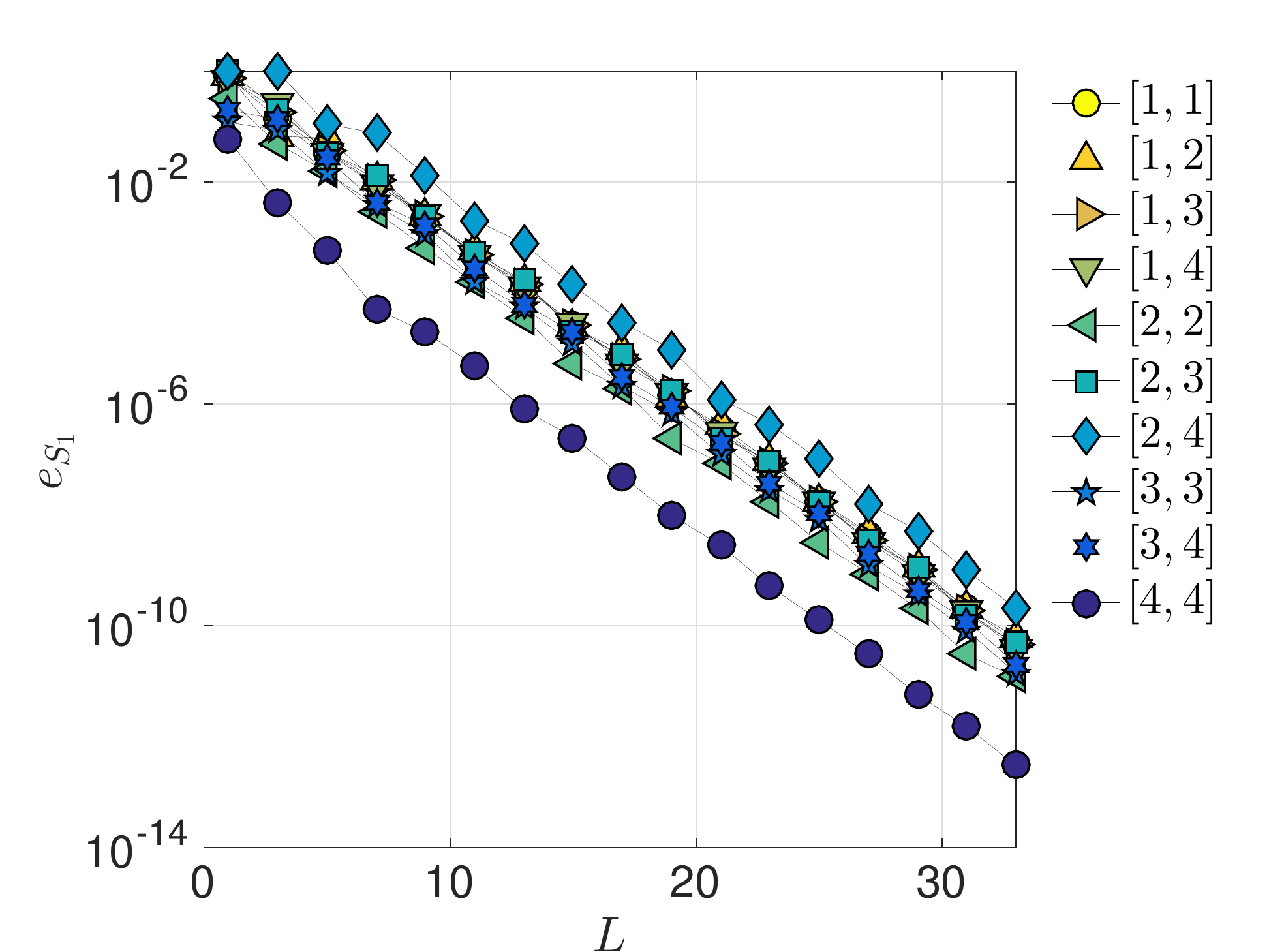}
	\caption{$e_{S_1}[\Phi_{ij,1}(\uv{r})]$}
	\end{subfigure}
\	
	\begin{subfigure}{0.49\textwidth}
	\centering
	\includegraphics[width=\textwidth]{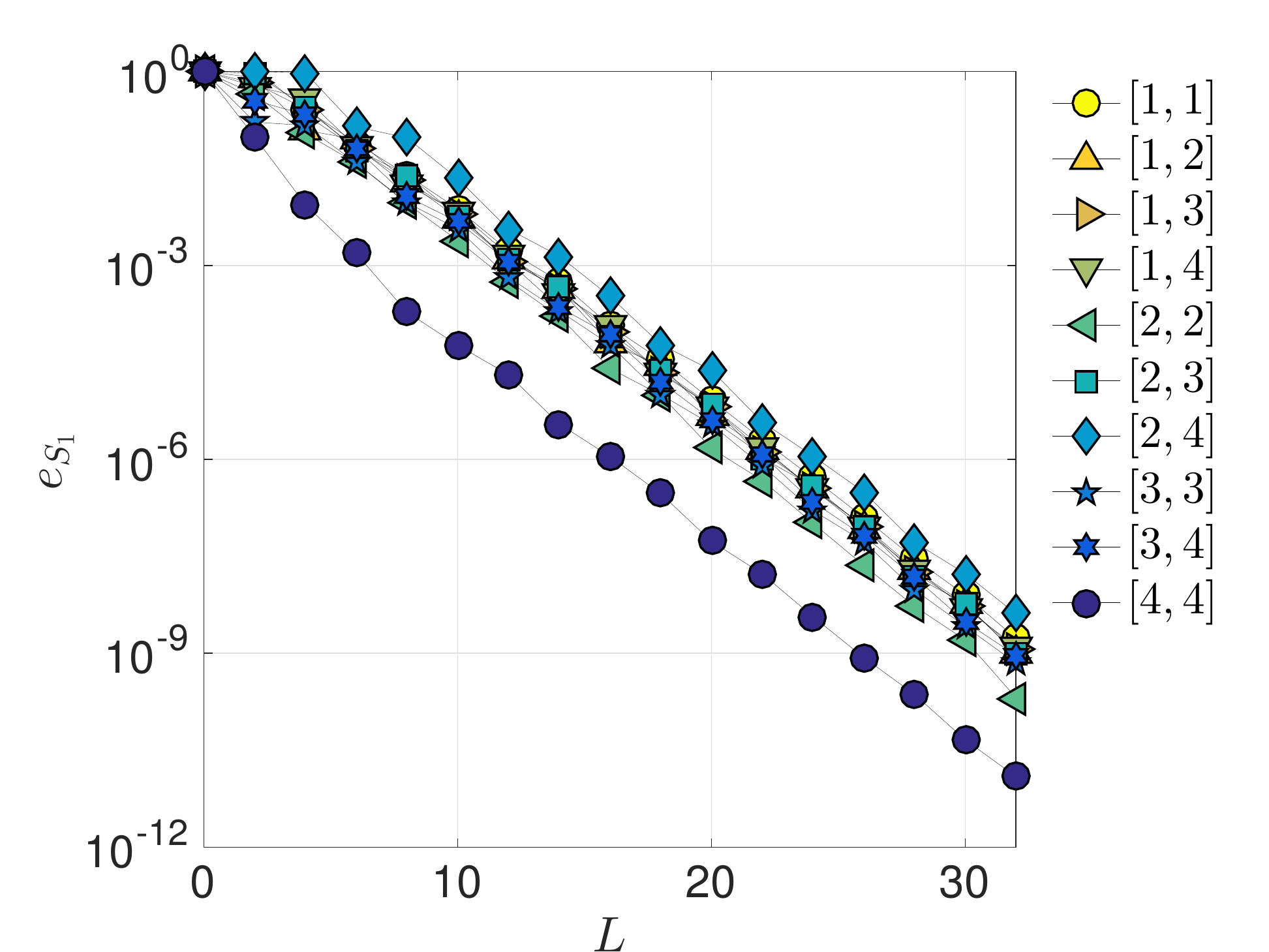}
	\caption{$e_{S_1}[\Phi_{ij,11}(\uv{r})]$}
	\end{subfigure}
\	
	\begin{subfigure}{0.49\textwidth}
	\centering
	\includegraphics[width=\textwidth]{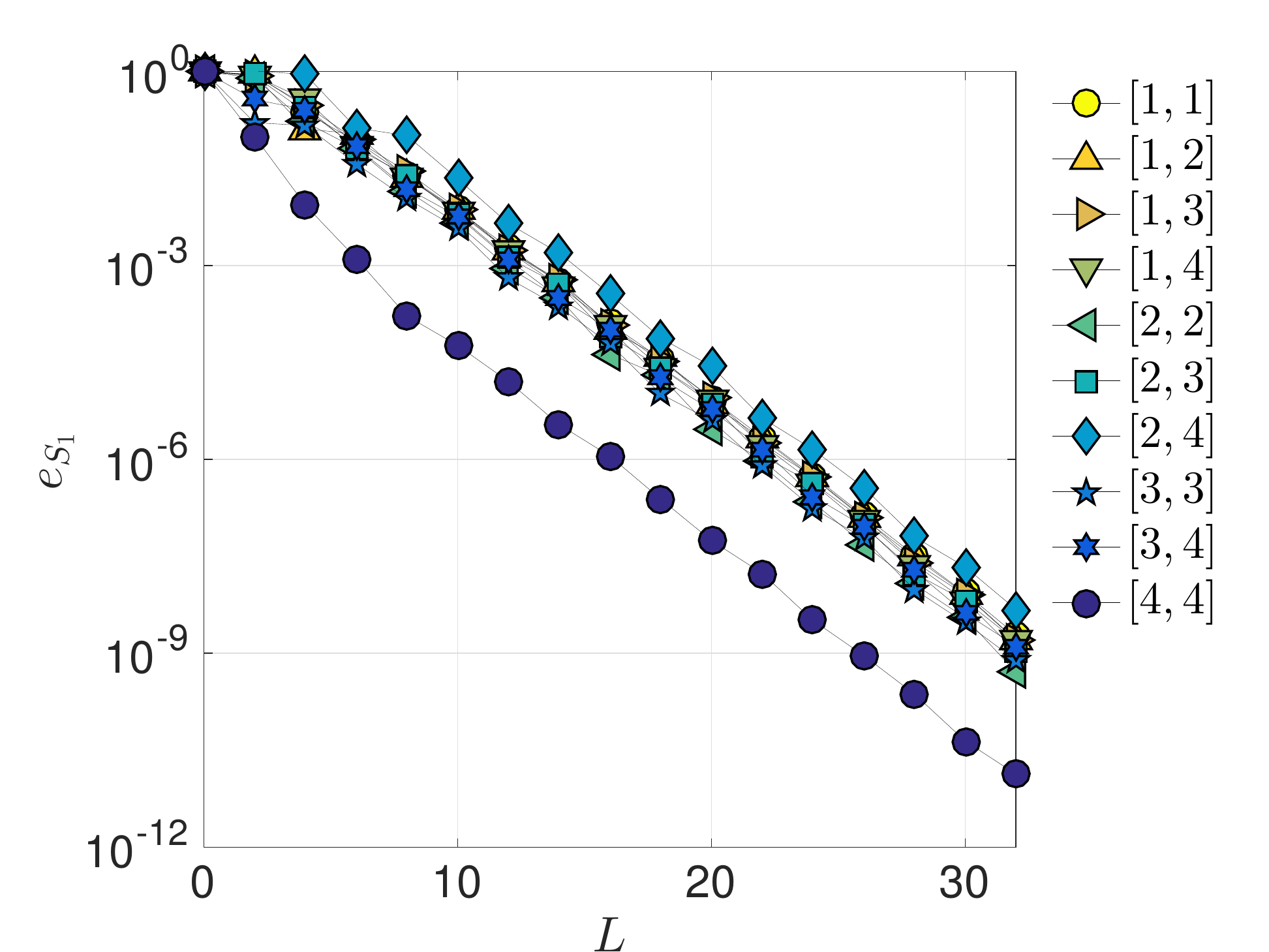}
	\caption{$e_{S_1}[\Phi_{ij,12}(\uv{r})]$}
	\end{subfigure}
\caption{Fundamental solutions for orthotropic piezoelectric PVDF: error for \emph{a}) the fundamental solutions $\Phi_{ij}(\uv{r})$, \emph{b}) the first derivative $\Phi_{ij,1}(\uv{r})$, \emph{c}) the second derivative $\Phi_{ij,11}(\uv{r})$, and \emph{d}) the second derivative $\Phi_{ij,12}(\uv{r})$ as a function of the series truncation number $L$. The reference values are computed using the unit circle integration.}
\label{fig-Ch2:PVDF error over S1}
\end{figure}

\newpage
\vfill

\begin{figure}[H]
\centering
	\begin{subfigure}{0.49\textwidth}
	\centering
	\includegraphics[width=\textwidth]{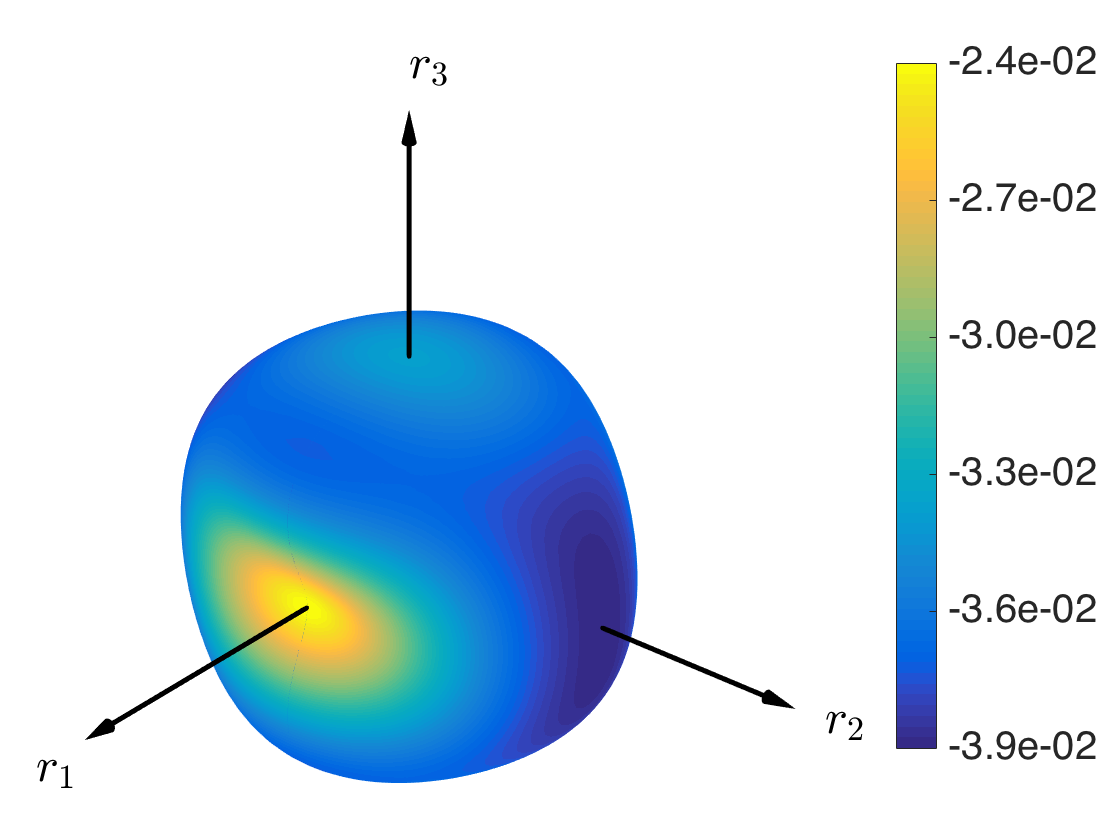}
	\caption{$\Phi_{34}(\uv{r})$ [m]}
	\end{subfigure}
\	
	\begin{subfigure}{0.49\textwidth}
	\centering
	\includegraphics[width=\textwidth]{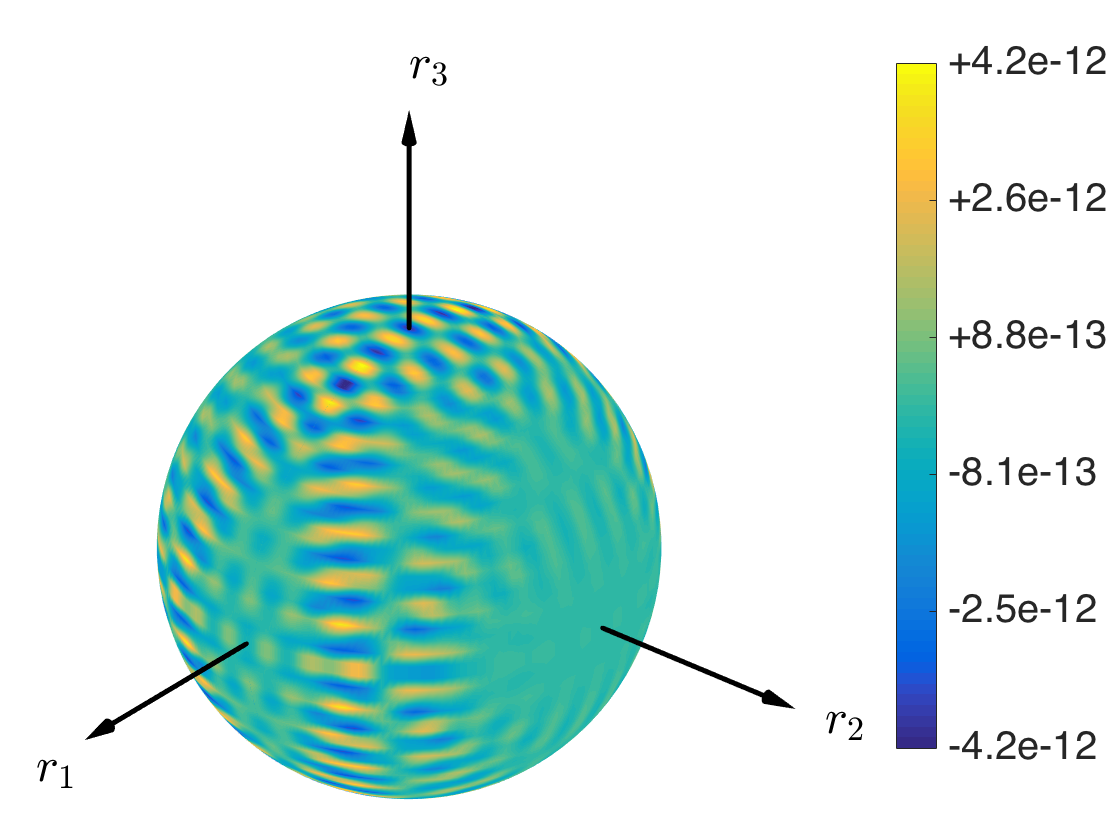}
	\caption{$\Delta_{S_1}[\Phi_{34}(\uv{r})]$}
	\end{subfigure}
\	
	\begin{subfigure}{0.49\textwidth}
	\centering
	\includegraphics[width=\textwidth]{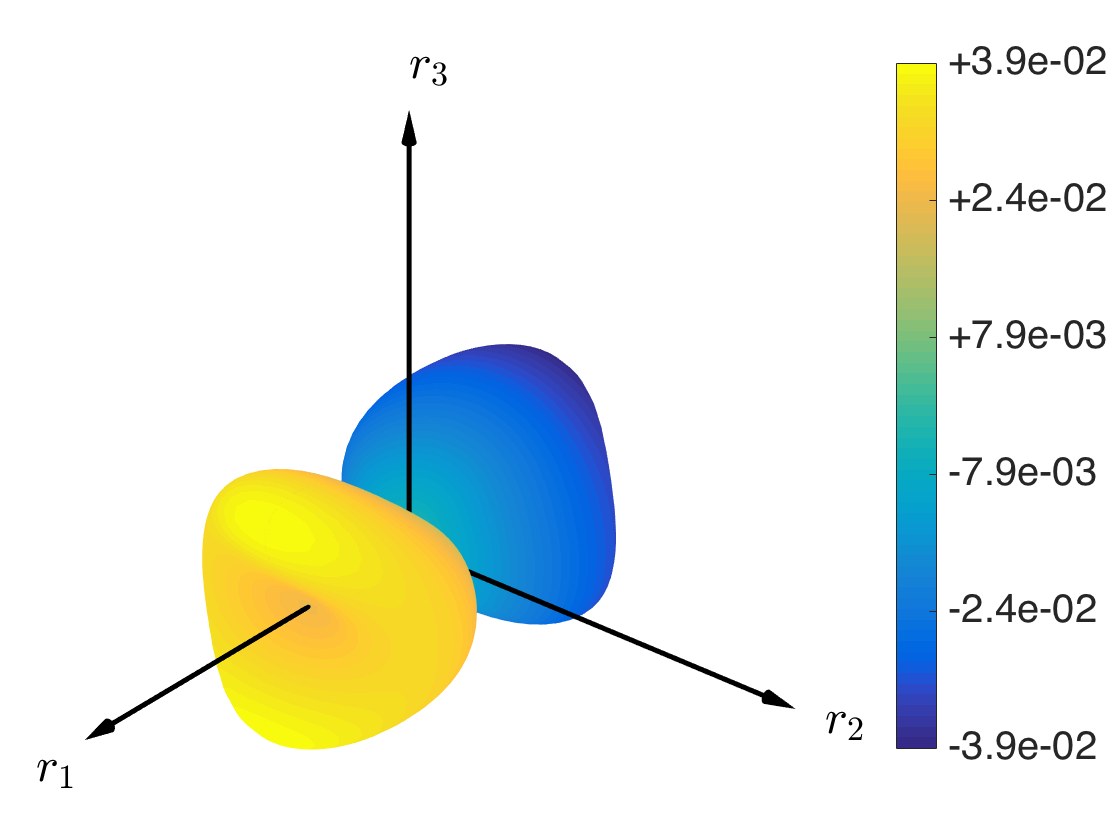}
	\caption{$\Phi_{34,1}(\uv{r})$ [-]}
	\end{subfigure}
\	
	\begin{subfigure}{0.49\textwidth}
	\centering
	\includegraphics[width=\textwidth]{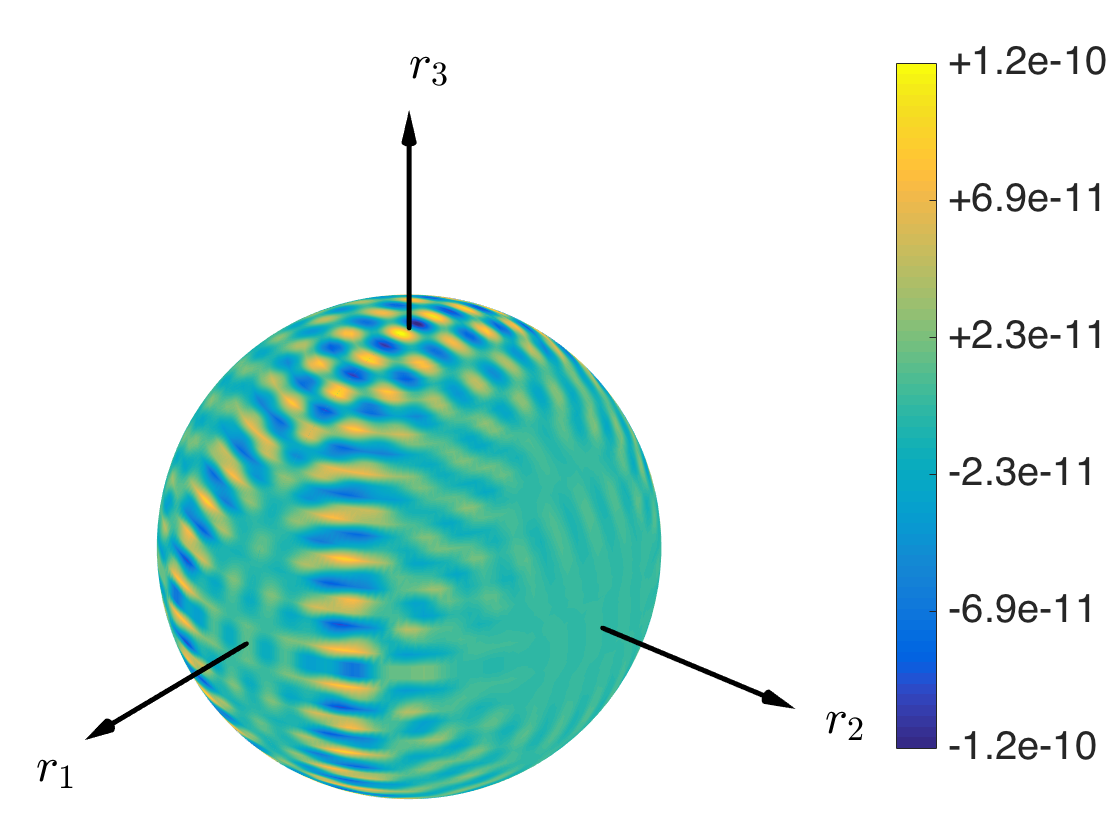}
	\caption{$\Delta_{S_1}[\Phi_{34,1}(\uv{r})]$}
	\end{subfigure}
\	
	\begin{subfigure}{0.49\textwidth}
	\centering
	\includegraphics[width=\textwidth]{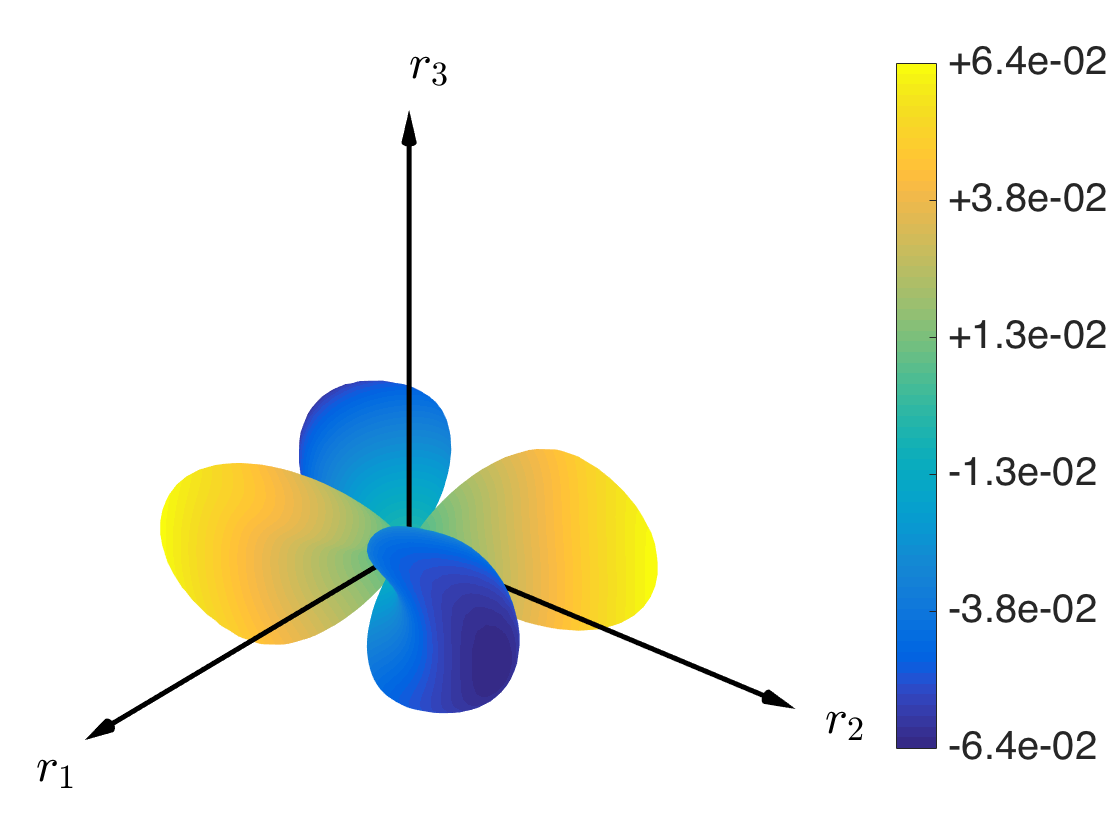}
	\caption{$\Phi_{34,12}(\uv{r})$ [$\mathrm{m}^{-1}$]}
	\end{subfigure}
\	
	\begin{subfigure}{0.49\textwidth}
	\centering
	\includegraphics[width=\textwidth]{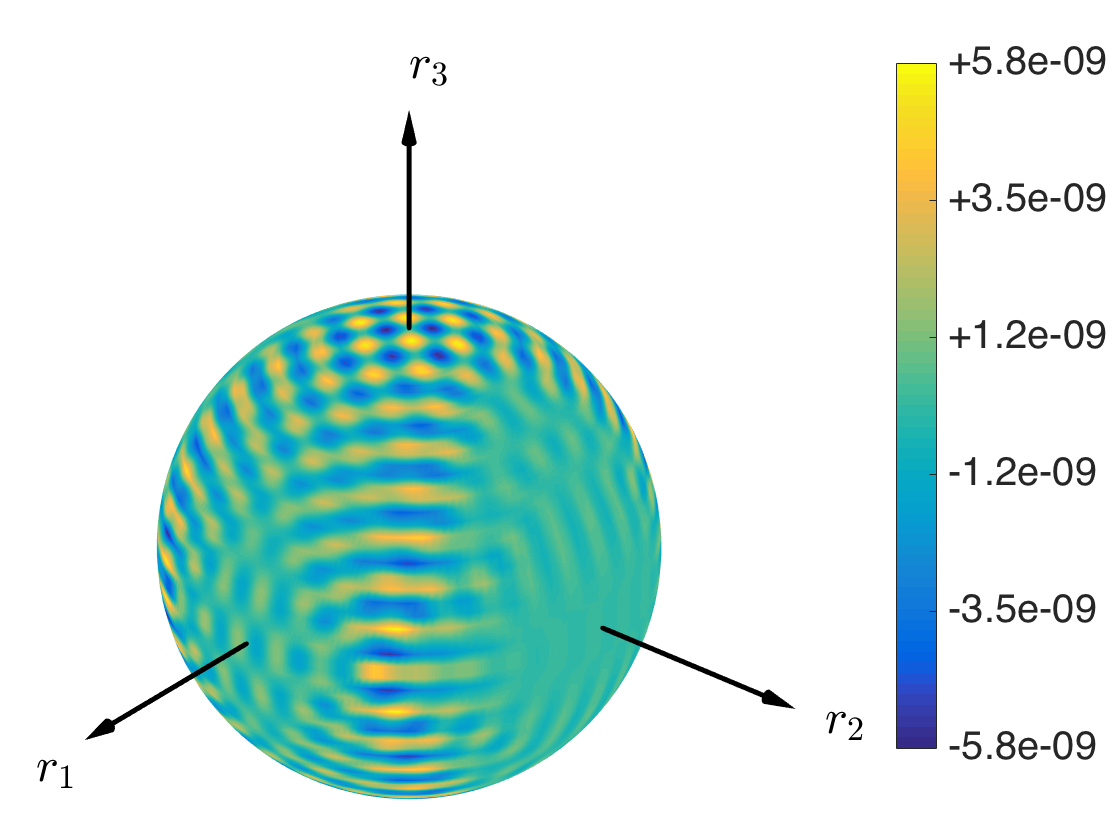}
	\caption{$\Delta_{S_1}[\Phi_{34,12}(\uv{r})]$}
	\end{subfigure}
\caption{Fundamental solutions for orthotropic piezoelectric PVDF: (\emph{a},\emph{c},\emph{e}) spherical plots of a few selected fundamental solutions; (\emph{b},\emph{d},\emph{f}) plot over the unit sphere of the difference between the fundamental solutions computed using the spherical harmonics expansions and the unit circle integration.}
\label{fig-Ch2:PVDF ball plots}
\end{figure}

\clearpage

\subsubsection{Magneto-electro-elastic materials}
The fundamental solutions of magneto-electro-elastic materials are finally computed. Two MEE materials are considered: a transversely isotropic MEE material indicated by $M_1$ and an orthotropic MEE material indicated by $M_2$, whose properties are reported in Tables (\ref{tab-Ch2:mat properties M1}) and (\ref{tab-Ch2:mat properties M2}), respectively.

\begin{figure}[ht]
\centering
	\begin{subfigure}{0.49\textwidth}
	\centering
	\includegraphics[width=\textwidth]{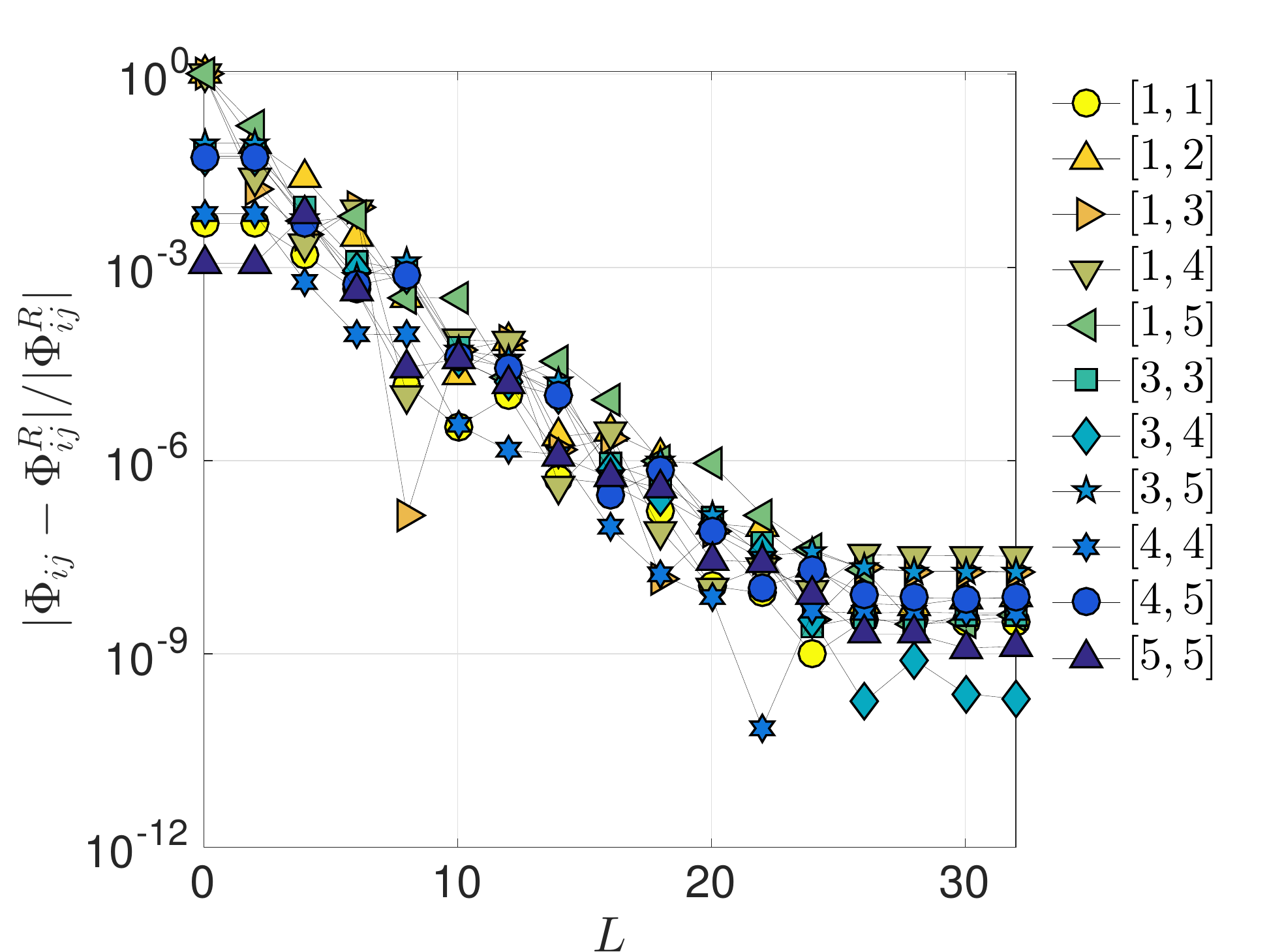}
	\caption{}
	\end{subfigure}
\	
	\begin{subfigure}{0.49\textwidth}
	\centering
	\includegraphics[width=\textwidth]{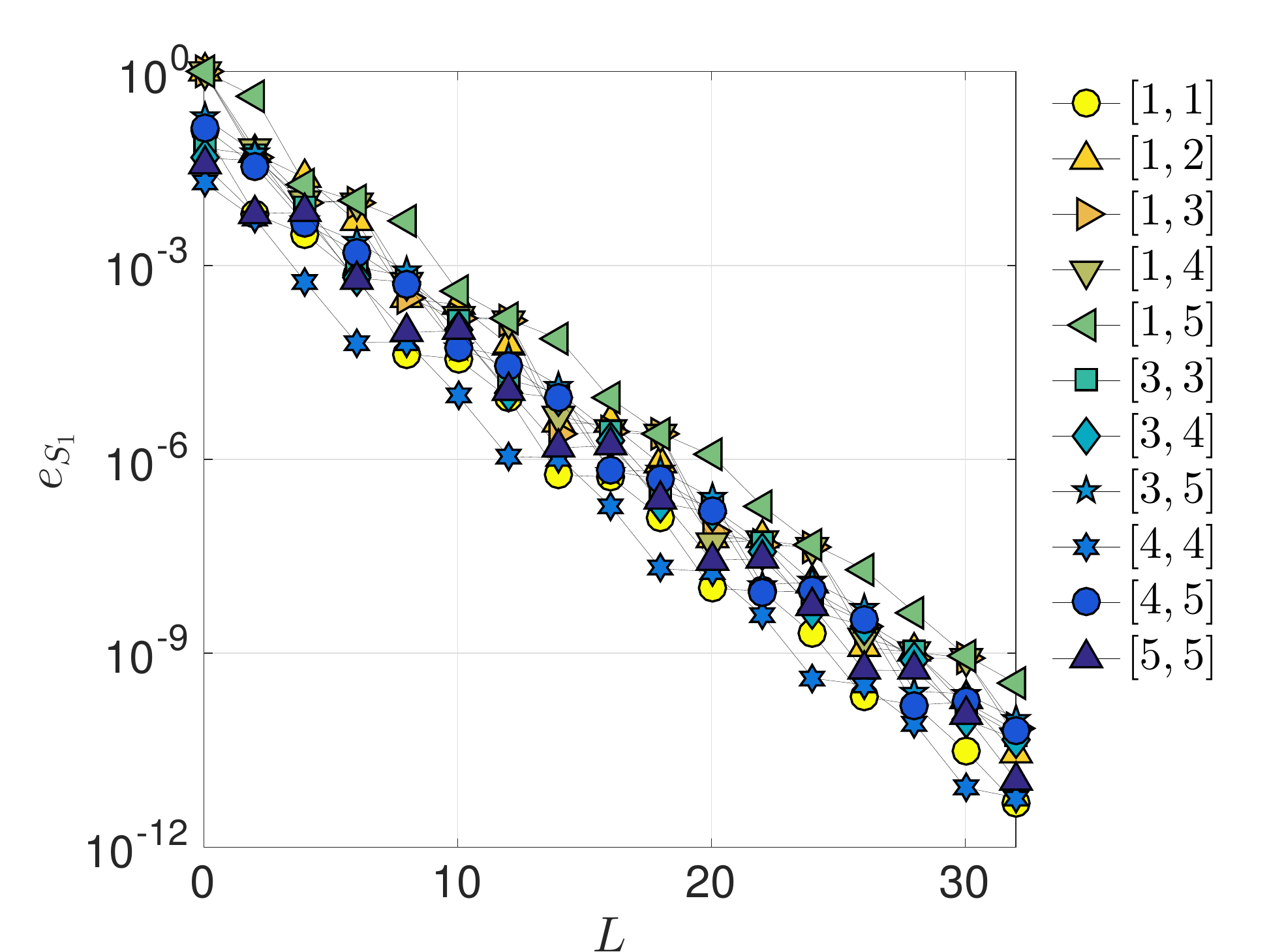}
	\caption{$e_{S_1}[\Phi_{ij}(\uv{r})]$}
	\end{subfigure}
\caption{Fundamental solutions for transversely isotropic magneto-electro-elastic material $M_1$: (\emph{a}) error for the fundamental solutions $\Phi_{ij}(\mathbf{r})$ computed at $\mathbf{r}=\{1,1,-1\}$ as a function of the series truncation number $L$ (the reference values are computed using the explicit expressions of the fundamental solutions as in Ref.\ \cite{pan2002}); (\emph{b}) error over the unit sphere for the fundamental solution $\Phi_{ij}(\uv{r})$ as a function of the series truncation number $L$ (the reference values are computed using the unit circle integration).}
\label{fig-Ch2:M1 error over S1}
\end{figure}

Figure (\ref{fig-Ch2:M1 error over S1}a) shows the convergence of the spherical harmonics expansions of the fundamental solutions $\Phi_{ij}(\mathbf{r})$ of the material $M_1$ computed at $\mathbf{r}=\{1,1,-1\}$. As reference, the results reported in Ref.\ \cite{pan2002} and computed using the explicit expressions of the fundamental solutions are used. Figure (\ref{fig-Ch2:M1 error over S1}b) shows the error $e_{S_1}$ for the same fundamental solutions.

Figure (\ref{fig-Ch2:M2 error over S1}) shows the error $e_{S_1}$ for the fundamental solutions and their derivatives up to the second order of the orthotropic MEE material $M_2$. The unit circle integration is used as reference value. Figure (\ref{fig-Ch2:M2 ball plots}) displays a few selected fundamental solutions for the considered material: Figures (\ref{fig-Ch2:M2 ball plots}a,c,e) plot the fundamental solutions $\Phi_{11}(\uv{r})$, $\Phi_{11,1}(\uv{r})$, $\Phi_{11,12}(\uv{r})$ respectively using the spherical representation (\ref{eq-Ch2:spherical plot definition}), whereas Figures (\ref{fig-Ch2:M2 ball plots}b,d,f) show the error with respect to the unit circle integration.

Figure (\ref{fig-Ch2:M2 Phi_ijkl error over S1}) shows the error $e_{S_1}$ for the fourth derivatives of the fundamental solution $\Phi_{11}$ of the MEE material $M_2$ as a function of the series truncation number $L$. In this case, the fundamental solutions computed with $L=40$ are used as reference. For comparison purposes, the figure also reports the error $e_{S_1}$ for the fundamental solution $\Phi_{11}(\uv{r})$ and its first, second and third derivative $\Phi_{11,1}(\uv{r})$, $\Phi_{11,11}(\uv{r})$ and $\Phi_{11,111}(\uv{r})$. In Figure (\ref{fig-Ch2:M2 ball plots higher-order derivatives}) some selected fourth derivatives are plotted using the spherical representation (\ref{eq-Ch2:spherical plot definition}).

\begin{figure}[H]
\centering
	\begin{subfigure}{0.49\textwidth}
	\centering
	\includegraphics[width=\textwidth]{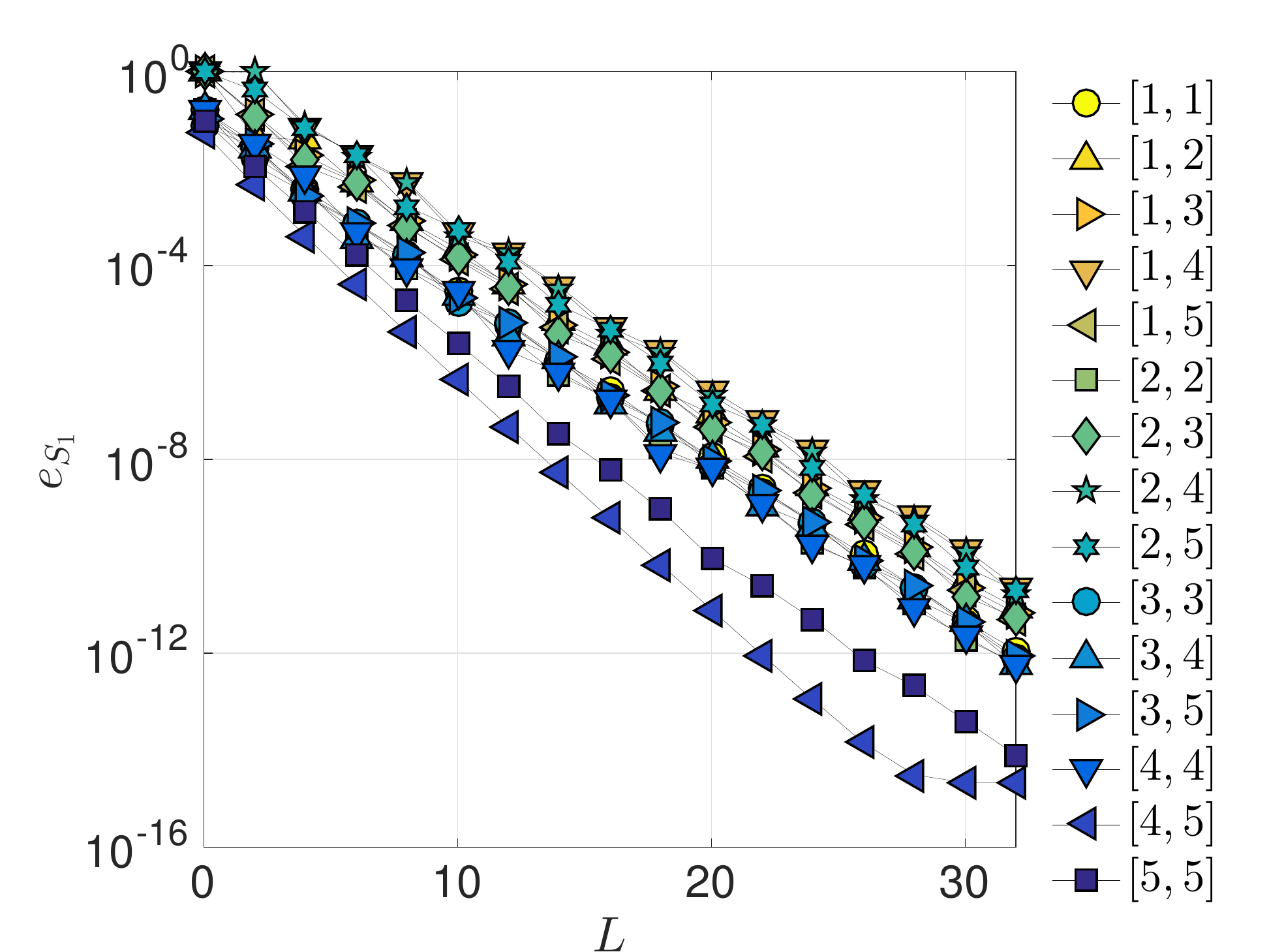}
	\caption{$e_{S_1}[\Phi_{ij}(\uv{r})]$}
	\end{subfigure}
\
	\begin{subfigure}{0.49\textwidth}
	\centering
	\includegraphics[width=\textwidth]{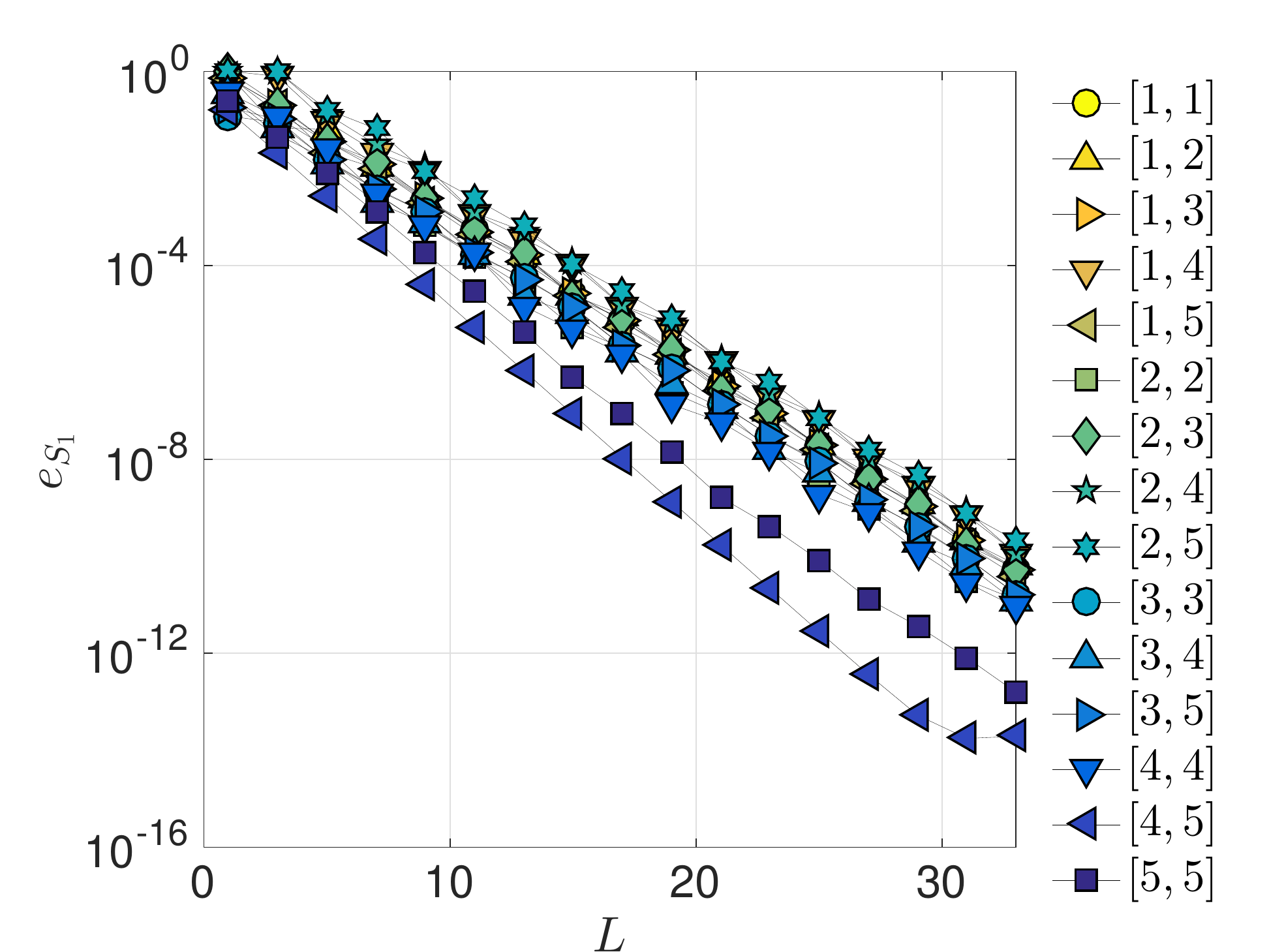}
	\caption{$e_{S_1}[\Phi_{ij,1}(\uv{r})]$}
	\end{subfigure}
\
	\begin{subfigure}{0.49\textwidth}
	\centering
	\includegraphics[width=\textwidth]{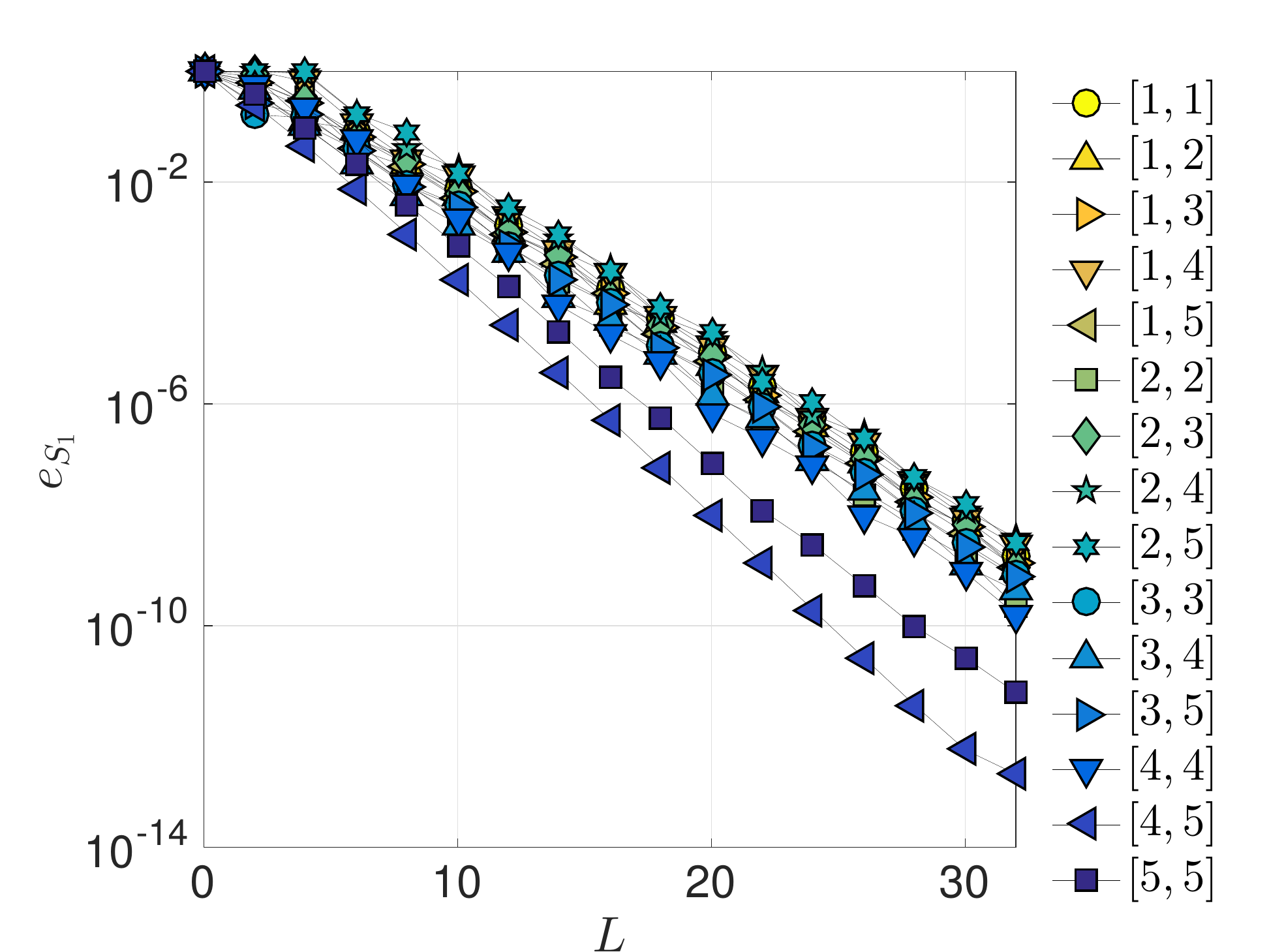}
	\caption{$e_{S_1}[\Phi_{ij,11}(\uv{r})]$}
	\end{subfigure}
\
	\begin{subfigure}{0.49\textwidth}
	\centering
	\includegraphics[width=\textwidth]{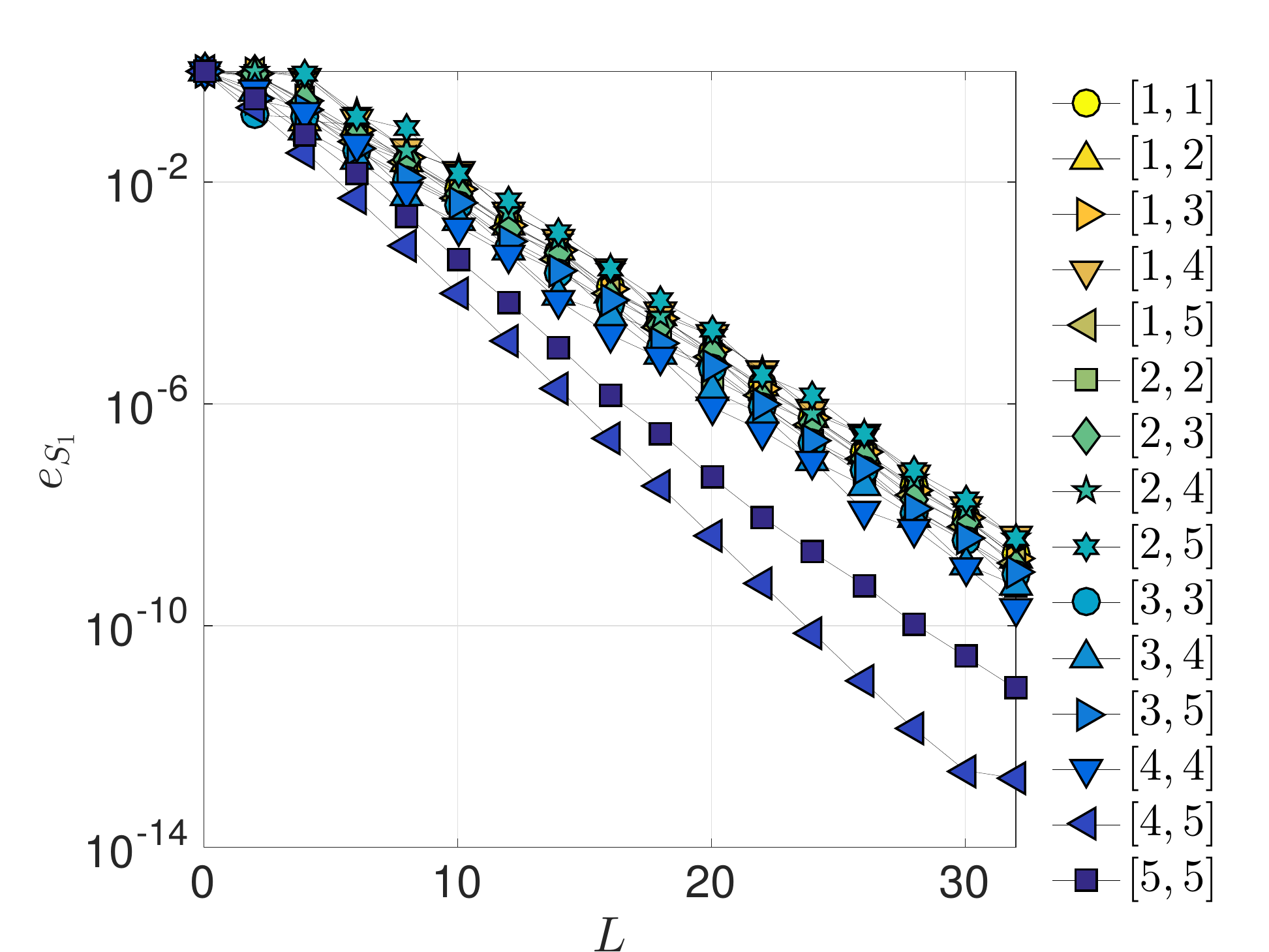}
	\caption{$e_{S_1}[\Phi_{ij,12}(\uv{r})]$}
	\end{subfigure}
\caption{Fundamental solutions for orthotropic magneto-electro-elastic material $M_2$: error for (\emph{a}) the fundamental solutions $\Phi_{ij}(\uv{r})$, (\emph{b}) the first derivative $\Phi_{ij,1}(\uv{r})$, (\emph{c}) the second derivative $\Phi_{ij,11}(\uv{r})$, and (\emph{d}) the second derivative $\Phi_{ij,12}(\uv{r})$ as a function of the series truncation number $L$. The reference values are computed using the unit circle integration.}
\label{fig-Ch2:M2 error over S1}
\end{figure}

\begin{figure}[H]
\centering
\includegraphics[width=0.5\textwidth]{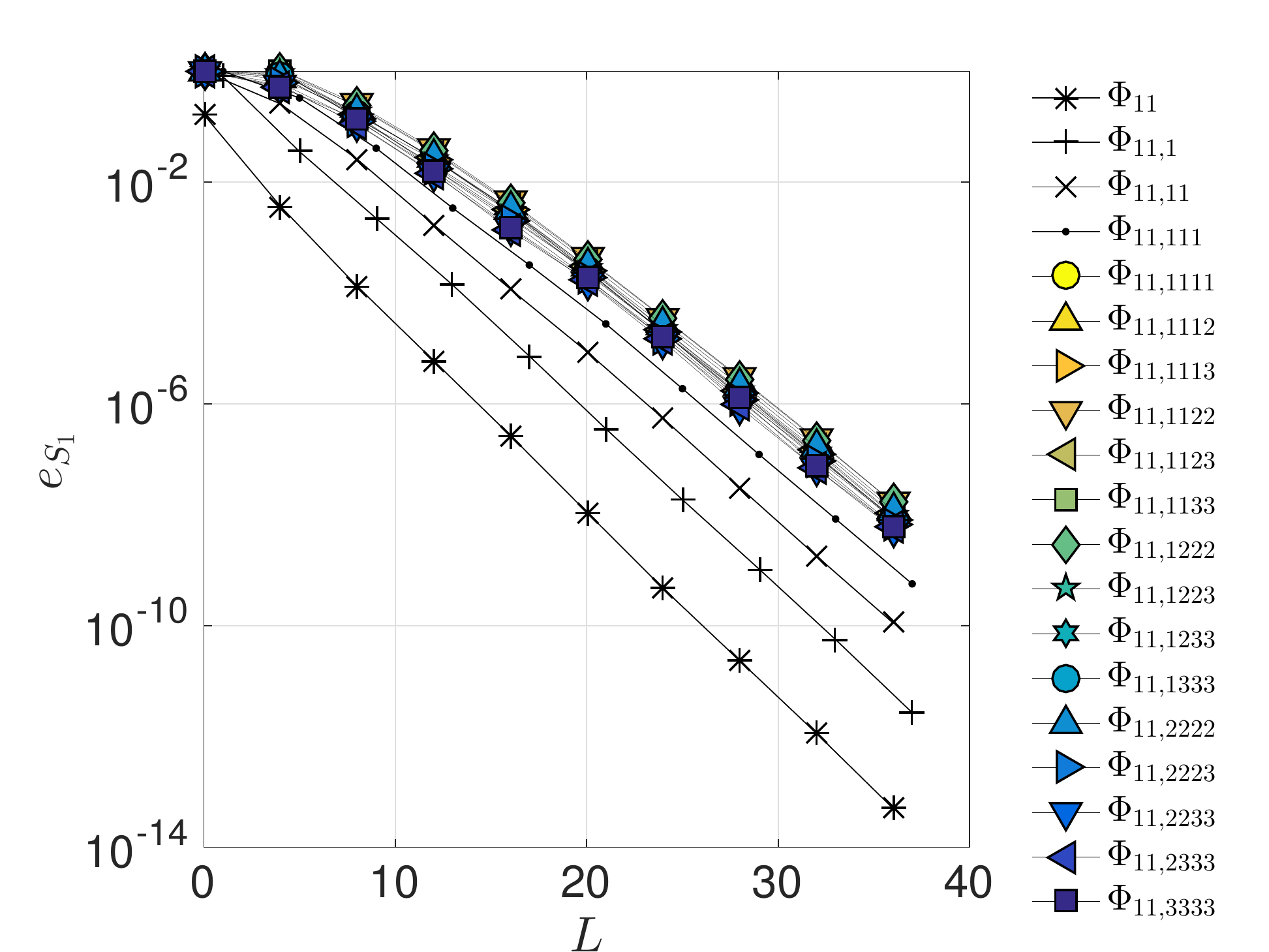}
\caption{Fundamental solutions for orthotropic magneto-electro-elastic material $M_2$: error, $e_{S_1}[\Phi_{11,ijkl}(\uv{r})]$, over the unit sphere for the fourth derivatives of the fundamental solutions $\Phi_{11}$ as a function of the series truncation number $L$. The reference values are fundamental solutions computed for $L=40$. The error for $\Phi_{11}(\uv{r})$, $\Phi_{11,11}(\uv{r})$, and $\Phi_{11,111}(\uv{r})$ are reported for comparison purposes.}
\label{fig-Ch2:M2 Phi_ijkl error over S1}
\end{figure}

\newpage
\vfill

\begin{figure}[H]
\centering
	\begin{subfigure}{0.49\textwidth}
	\centering
	\includegraphics[width=\textwidth]{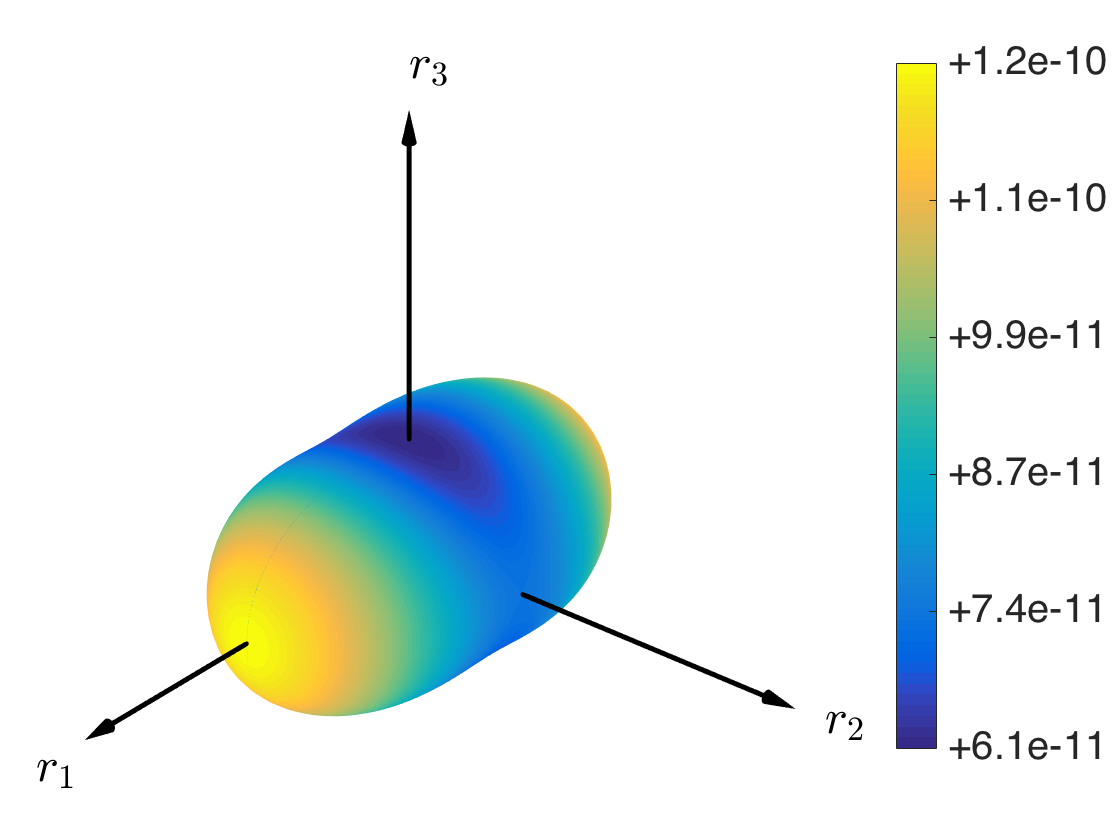}
	\caption{$\Phi_{11}(\uv{r})$ [m]}
	\end{subfigure}
\
	\begin{subfigure}{0.49\textwidth}
	\centering
	\includegraphics[width=\textwidth]{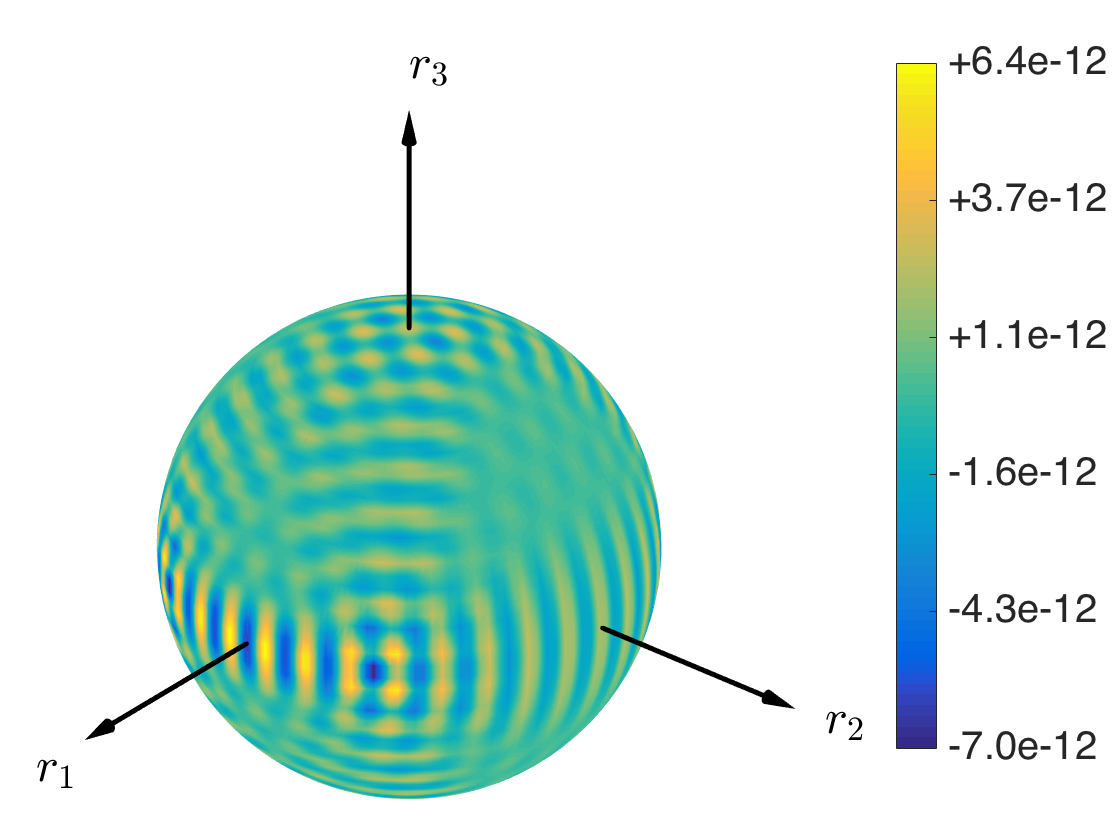}
	\caption{$\Delta_{S_1}[\Phi_{11}(\uv{r})]$}
	\end{subfigure}
\
	\begin{subfigure}{0.49\textwidth}
	\centering
	\includegraphics[width=\textwidth]{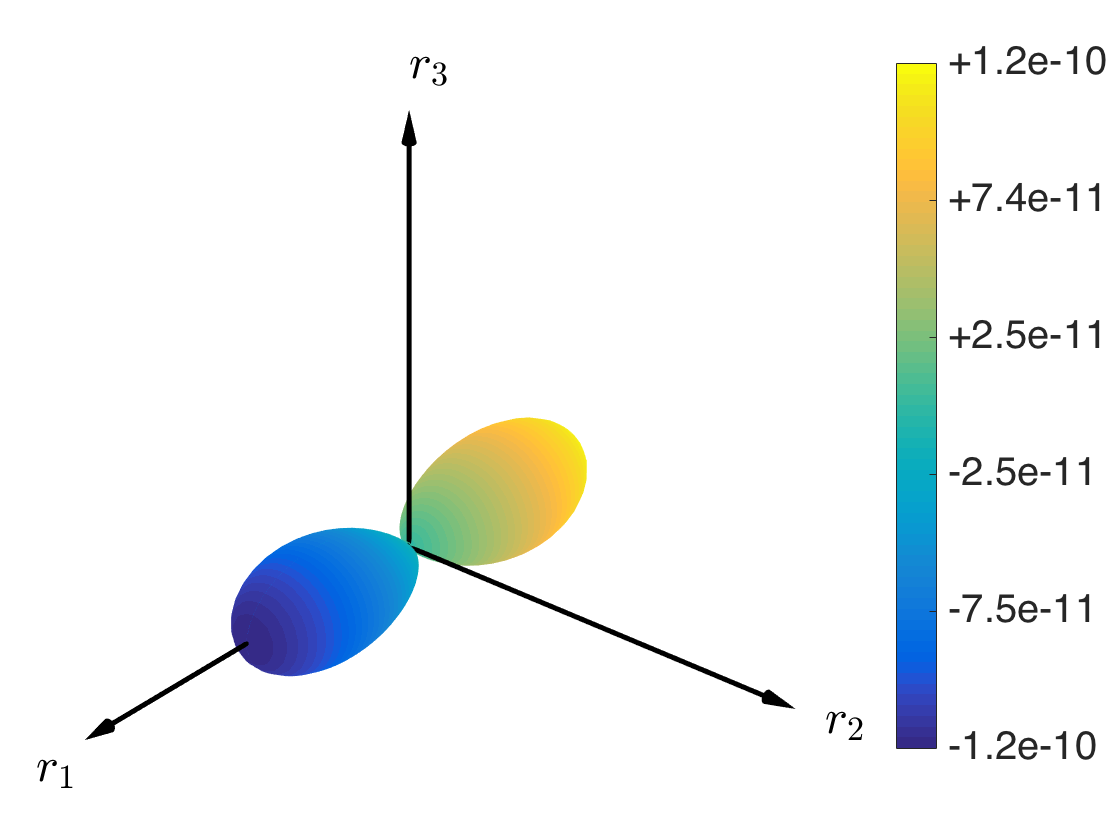}
	\caption{$\Phi_{11,1}(\uv{r})$ [-]}
	\end{subfigure}
\
	\begin{subfigure}{0.49\textwidth}
	\centering
	\includegraphics[width=\textwidth]{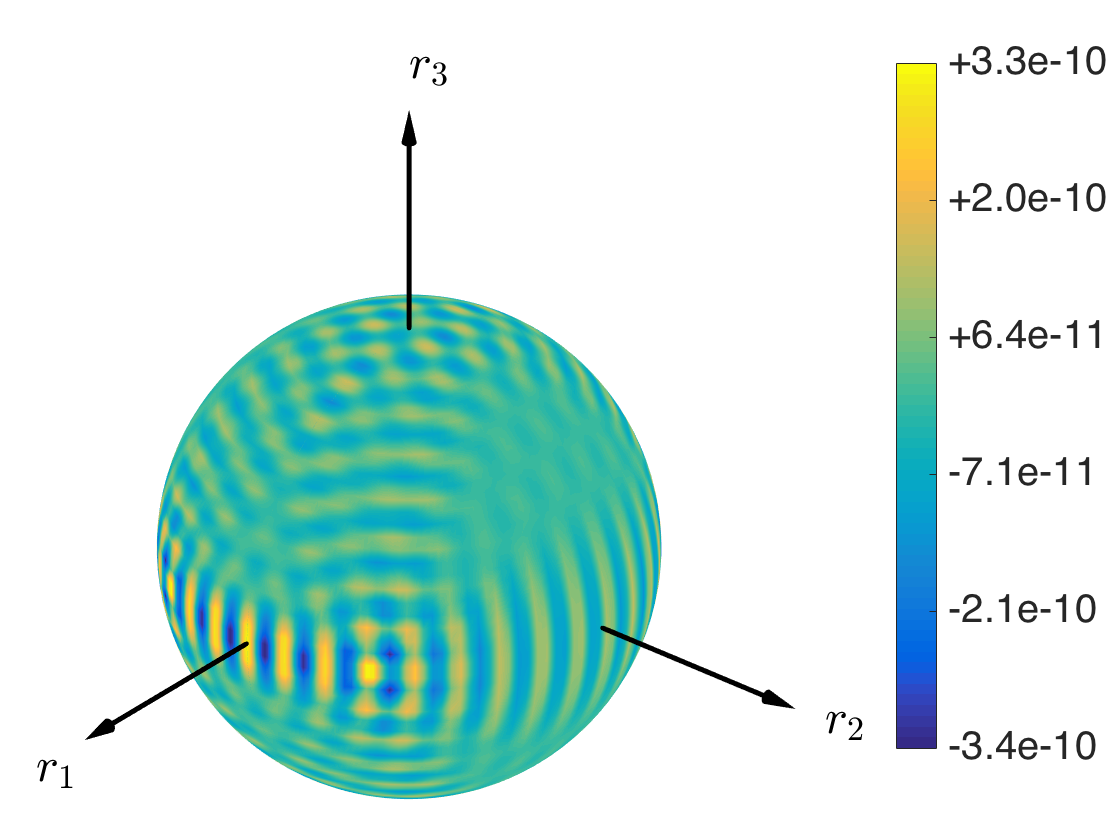}
	\caption{$\Delta_{S_1}[\Phi_{11,1}(\uv{r})]$}
	\end{subfigure}
\
	\begin{subfigure}{0.49\textwidth}
	\centering
	\includegraphics[width=\textwidth]{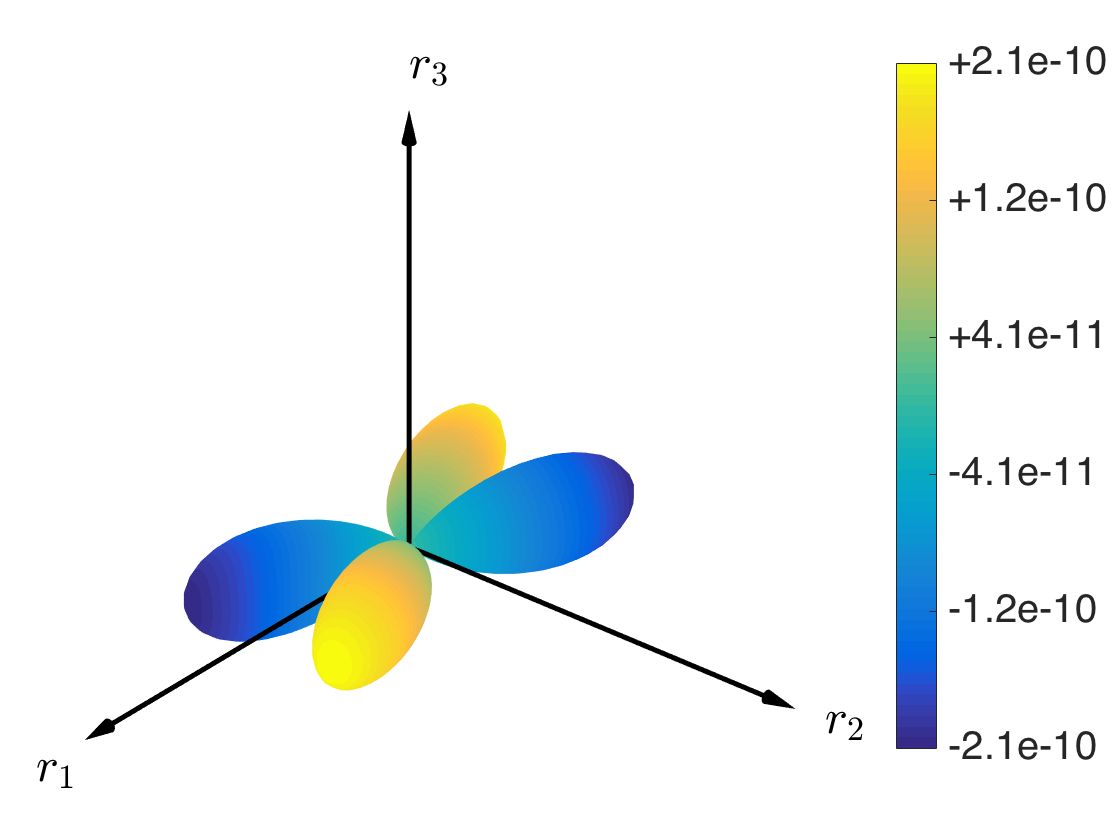}
	\caption{$\Phi_{11,12}(\uv{r})$ [$\mathrm{m}^{-1}$]}
	\end{subfigure}
\
	\begin{subfigure}{0.49\textwidth}
	\centering
	\includegraphics[width=\textwidth]{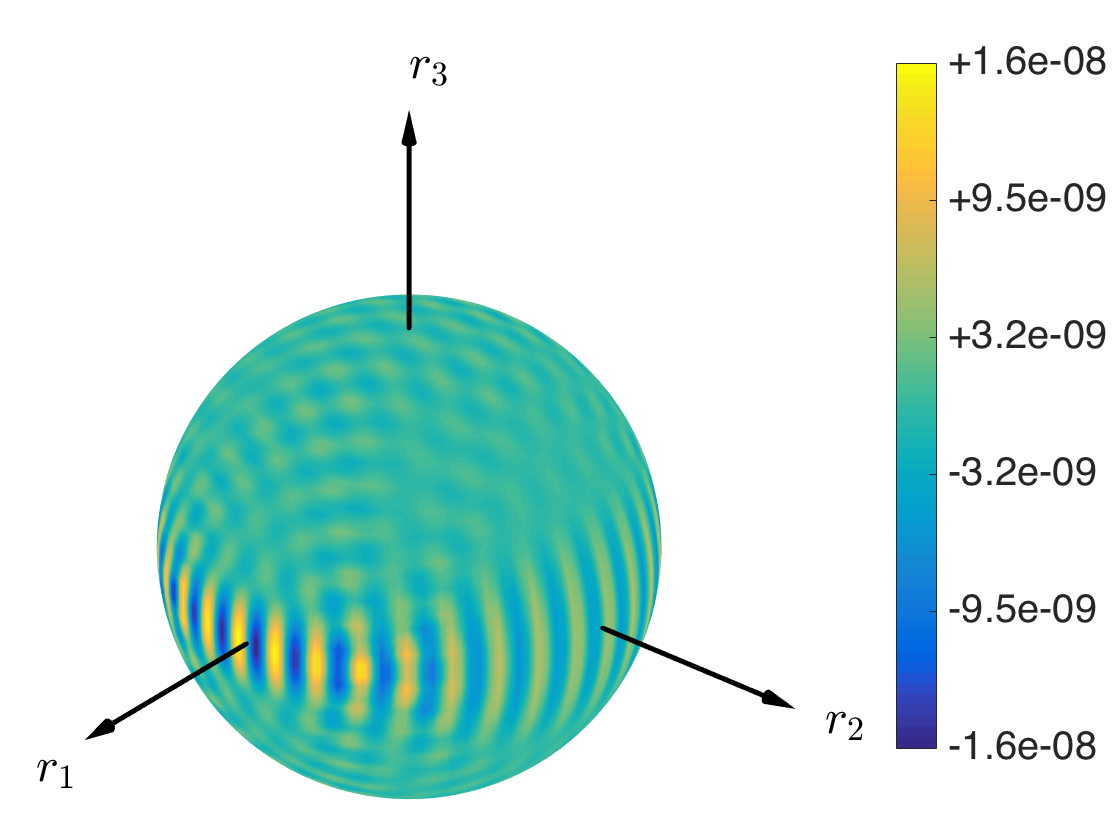}
	\caption{$\Delta_{S_1}[\Phi_{11,12}(\uv{r})]$}
	\end{subfigure}
\caption{Fundamental solutions for orthotropic magneto-electro-elastic material $M_2$: (\emph{a},\emph{c},\emph{e}) spherical plots of a few selected fundamental solutions; (\emph{b},\emph{d},\emph{f}) plot over the unit sphere of the difference between the fundamental solutions computed using the spherical harmonics expansions and the unit circle integration.}
\label{fig-Ch2:M2 ball plots}
\end{figure}

\begin{figure}[H]
\centering
	\begin{subfigure}{0.49\textwidth}
	\centering
	\includegraphics[width=\textwidth]{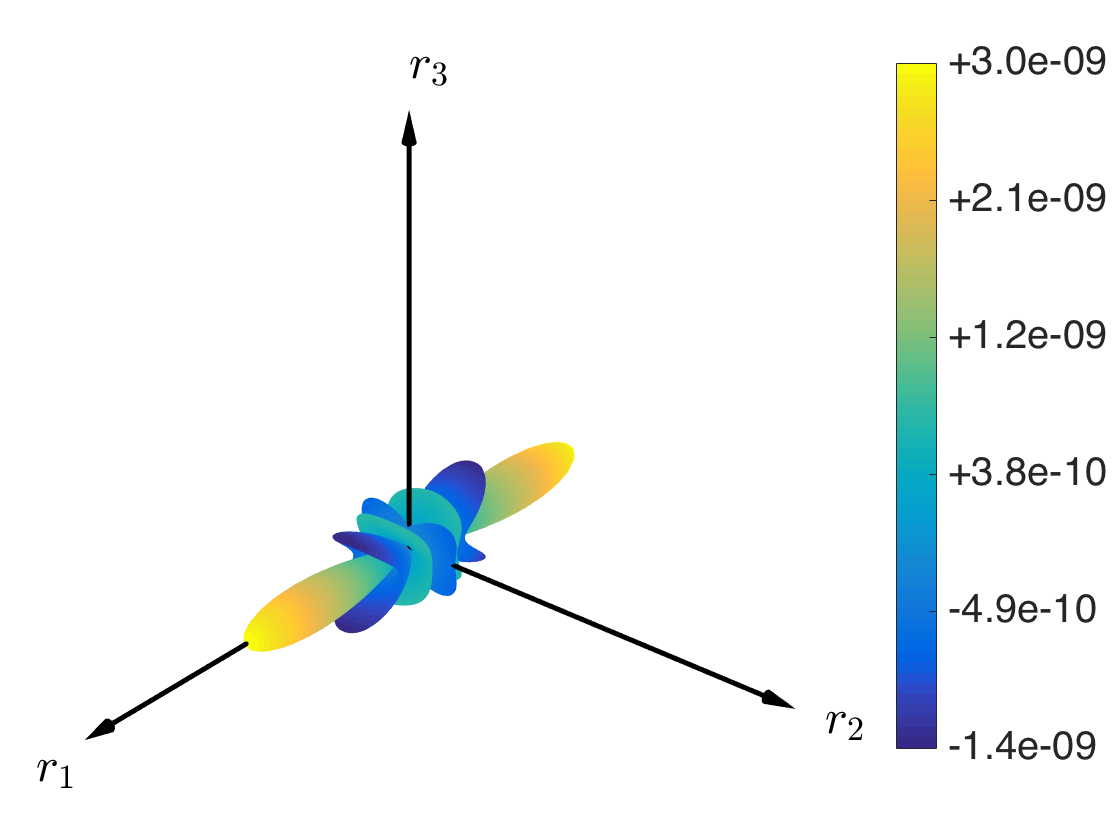}
	\caption{$\Phi_{11,1111}(\uv{r})$ [$\mathrm{m}^{-3}$]}
	\end{subfigure}
\
	\begin{subfigure}{0.49\textwidth}
	\centering
	\includegraphics[width=\textwidth]{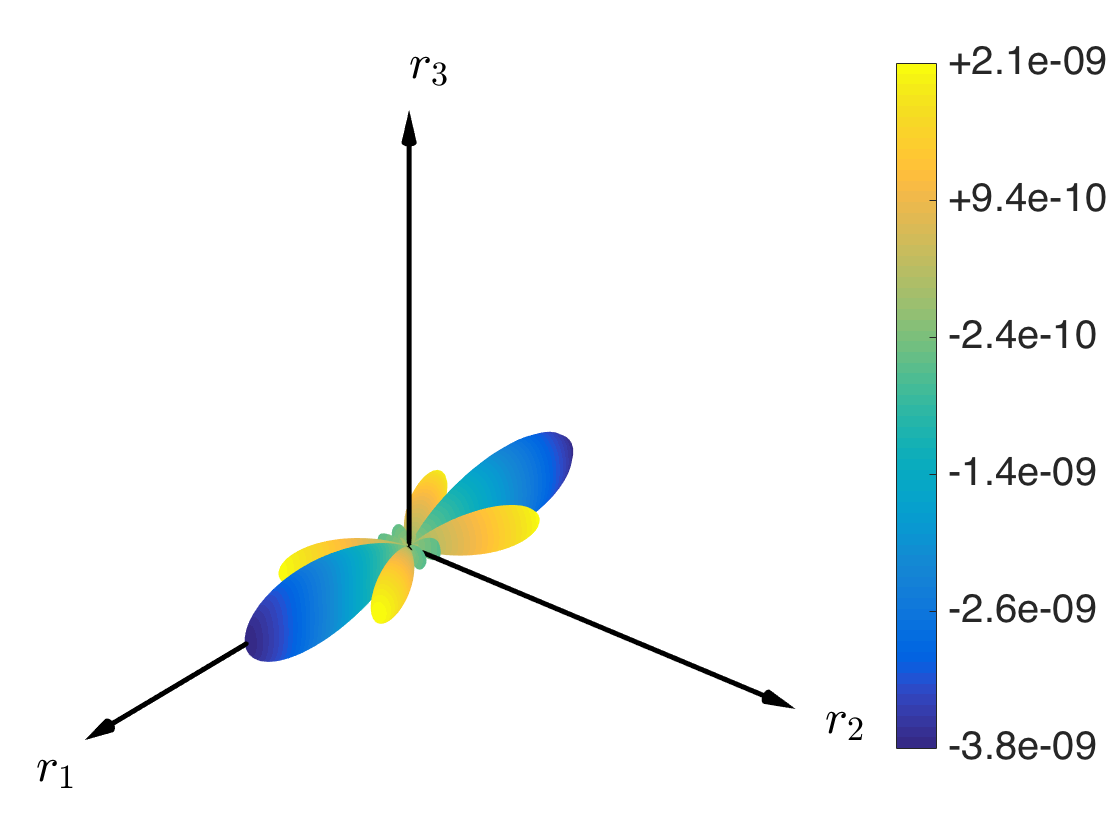}
	\caption{$\Phi_{11,1122}(\uv{r})$ [$\mathrm{m}^{-3}$]}
	\end{subfigure}
\	
	\begin{subfigure}{0.49\textwidth}
	\centering
	\includegraphics[width=\textwidth]{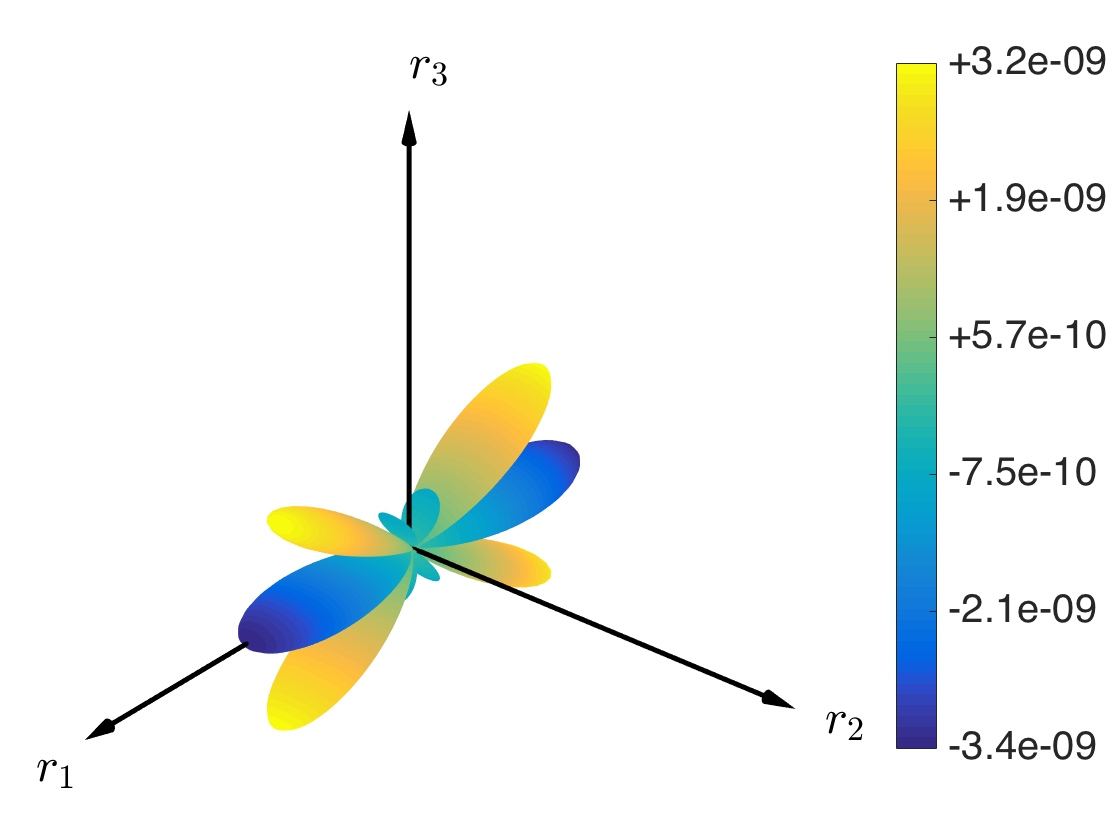}
	\caption{$\Phi_{11,1133}(\uv{r})$ [$\mathrm{m}^{-3}$]}
	\end{subfigure}
\	
	\begin{subfigure}{0.49\textwidth}
	\centering
	\includegraphics[width=\textwidth]{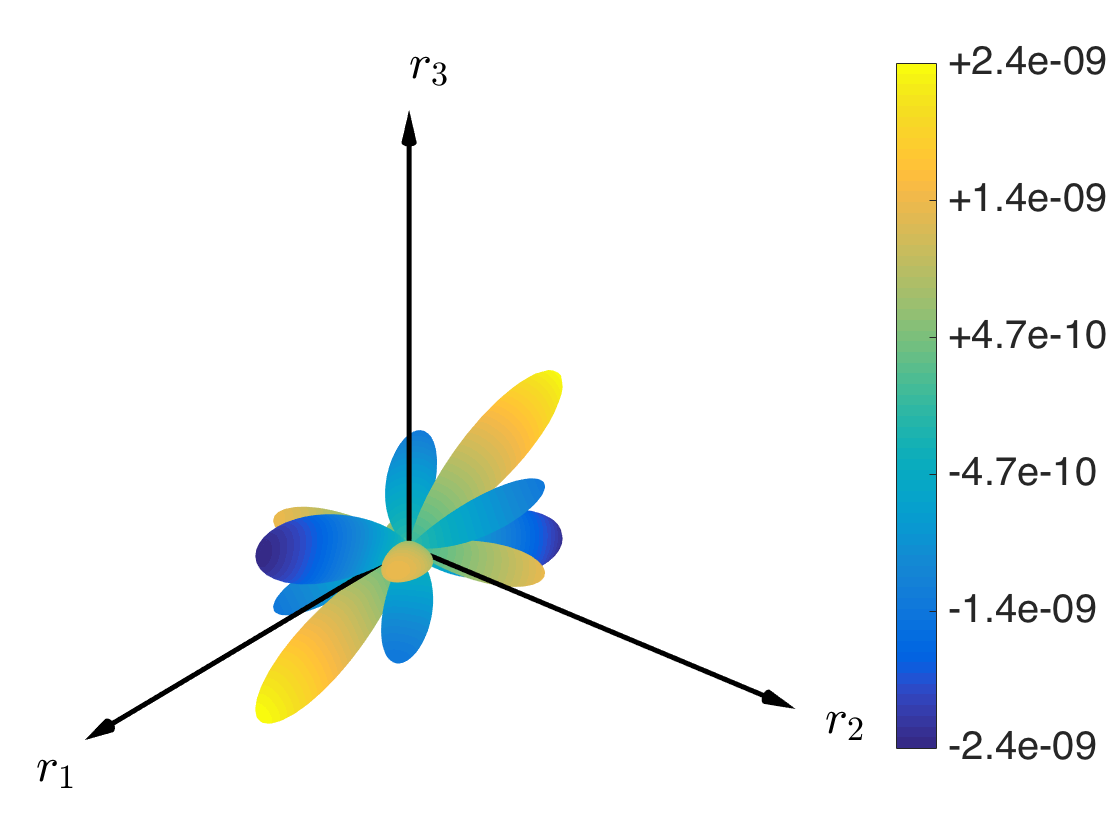}
	\caption{$\Phi_{11,1223}(\uv{r})$ [$\mathrm{m}^{-3}$]}
	\end{subfigure}
\	
	\begin{subfigure}{0.49\textwidth}
	\centering
	\includegraphics[width=\textwidth]{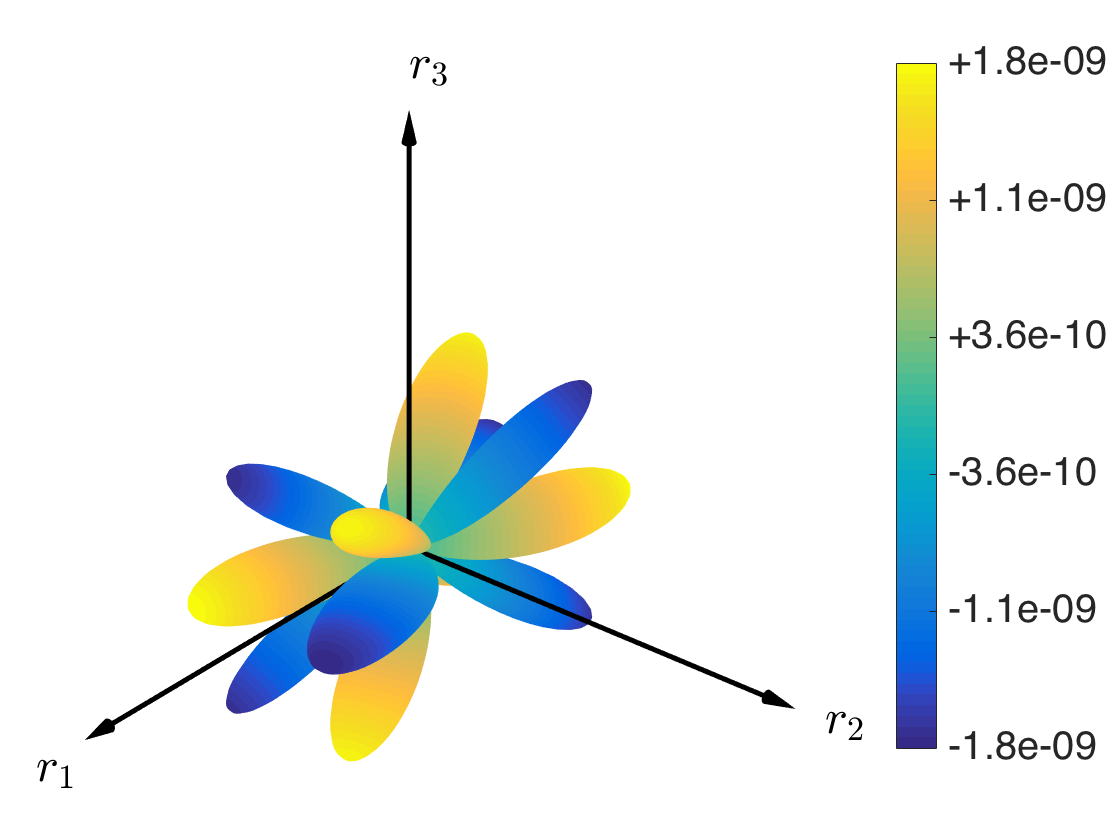}
	\caption{$\Phi_{11,1233}(\uv{r})$ [$\mathrm{m}^{-3}$]}
	\end{subfigure}
\	
	\begin{subfigure}{0.49\textwidth}
	\centering
	\includegraphics[width=\textwidth]{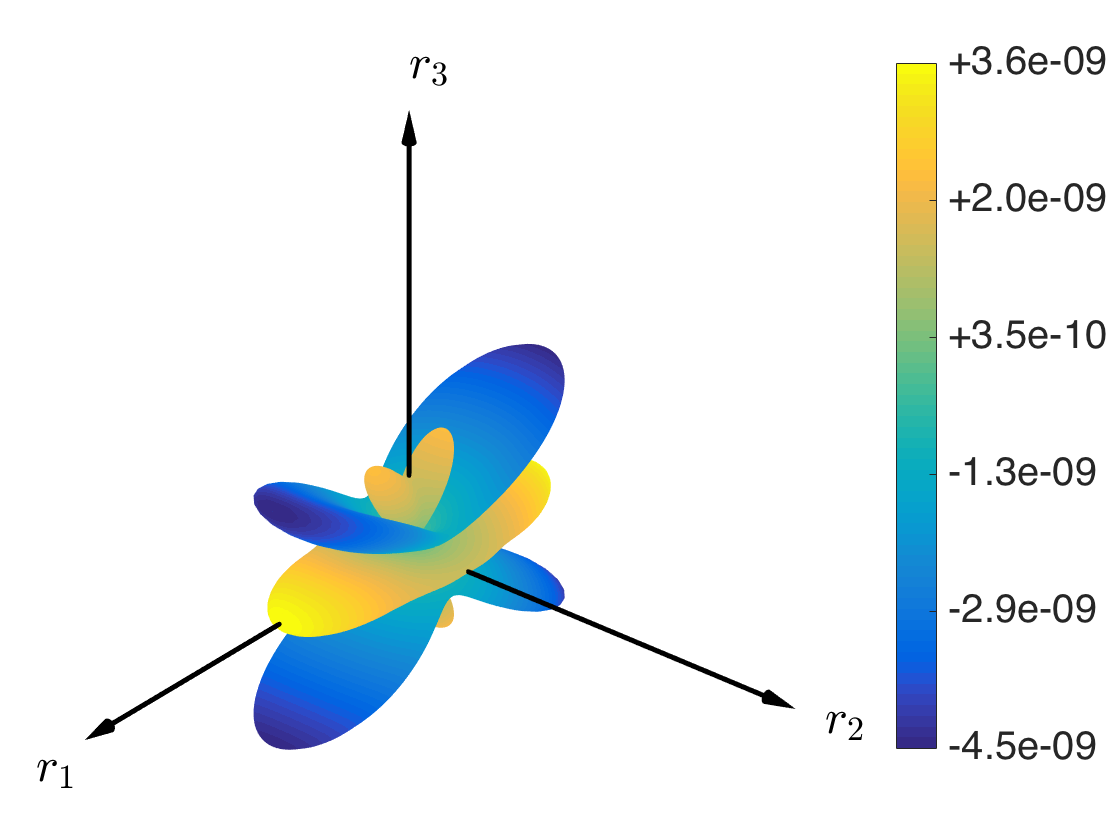}
	\caption{$\Phi_{11,3333}(\uv{r})$ [$\mathrm{m}^{-3}$]}
	\end{subfigure}
\caption{Fundamental solutions for orthotropic magneto-electro-elastic material $M_2$: spherical plots of a few selected fourth derivatives of the fundamental solution $\Phi_{11}$.}
\label{fig-Ch2:M2 ball plots higher-order derivatives}
\end{figure}

\clearpage

\section{Associated Legendre polynomials}\label{app-Ch2:legendreP}
This appendix summarises some of the properties of the associated Legendre polynomials used in this Chapter. Further details about orthogonal polynomials and more comprehensive discussions on their properties can be found in Ref.\ \cite{abramowitz1964}. The associated Legendre polynomials are the particularisation to integer values of $\ell$ of the associated Legendre functions, which are solutions of the associated Legendre differential equation
\begin{equation}\label{eq-Ch2:associated legendre differential equation}
\frac{\dd}{\dd t}\left[(1-t^2)\frac{\dd}{\dd t}y_\ell^m(t)\right]+\left[\ell(\ell+1)-\frac{m^2}{1-t^2}\right]y_\ell^m(t)=0.
\end{equation}
The general solution of Eq.(\ref{eq-Ch2:associated legendre differential equation}) is written as $y_\ell^m(t)=\alpha P_\ell^m(t)+\beta Q_\ell^m(t)$, where $\alpha$ and $\beta$ are constants and $P_\ell^m(t)$ and $Q_\ell^m(t)$ are referred to as the associated Legendre functions of the first and second kind, respectively. Since the associated Legendre polynomials $P_\ell^m(t)$ are mainly employed for obtaining the fundamental solutions in terms of spherical harmonics expansions, some of their properties are listed in the present appendix. 

The associated Legendre polynomials of degree $\ell$ and order $m$, $P_\ell^m(t)$, can be expressed in terms of the Legendre polynomials as
\begin{equation}\label{eq-Ch2:associated legendre polynomial}
P_\ell^m(t)=(-1)^m(1-t^2)^{m/2}\frac{\dd^m}{\dd t^m}P_\ell(t)=\frac{(-1)^m}{2^\ell \ell!}(1-t^2)^{m/2}\frac{\dd^{\ell+m}}{\dd t^{\ell+m}}(t^2-1)^\ell
\end{equation}
where $(\cdot)!$ denotes the factorial. From their definition, the associated Legendre polynomials with negative $m$, $P_\ell^{-m}$, can be written as
\begin{equation}\label{eq-Ch2:associated legendre polynomial negative m}
P_\ell^{-m}(t)=(-1)^m\frac{(\ell-m)!}{(\ell+m)!}P_\ell^m(t).
\end{equation}

The associated Legendre polynomials satisfy several identities and recurrence relations that can be found in handbooks \cite{abramowitz1964}. One of the identities, which is of interest for the developments presented here, is the value of $P_\ell^m(t)$ and $Q_\ell^m(t)$ at $t=0$:
\begin{subequations}\label{app-Ch2:associated legendreP value at 0}
\begin{align}
P_\ell^m(0)&=\frac{2^m}{\sqrt{\pi}}\cos\left[\frac{\pi}{2}(\ell+m)\right]\frac{\Gamma[(\ell+m+1)/2]}{\Gamma[(\ell-m+2)/2]};\\
Q_\ell^m(0)&=-2^{m-1}\sqrt{\pi}\sin\left[\frac{\pi}{2}(\ell+m)\right]\frac{\Gamma[(\ell+m+1)/2]}{\Gamma[(\ell-m+2)/2]}.
\end{align}
\end{subequations}

For $m=0$, the associated Legendre polynomials coincide with the Legendre polynomials that satisfy the following completeness relation
\begin{equation}\label{app-Ch2:completeness legendreP}
\sum_{\ell=0}^{\infty}\frac{2\ell+1}{2}P_\ell(t)P_\ell(s)=\delta(t-s)
\end{equation}
where $\delta(\cdot)$ is the Dirac delta function and $t,s\in(-1,1)$. Furthermore, it is worth recalling here the Rayleigh expansion, also known as \emph{plane wave expansion}, which expresses a plane wave as the infinite sum of spherical waves \cite{arfken2011} and is written as follows
\begin{equation}\label{app-Ch2:plane wave expansion}
\exp(\mathrm{i} \xi_k r_k)=\sum_{\ell=0}^{\infty}(2\ell+1)\mathrm{i}^\ell J_\ell(\xi\ r)P_\ell(\hat{\xi}_k \hat{r}_k)
\end{equation}
where $J_\ell(\cdot)$ are the spherical Bessel functions, $\mathbf{r}=\{r_k\}$ and $\boldsymbol{\xi}=\{\xi_k\}$, $k=1,2,3$, are three-dimensional vectors and $r=\sqrt{r_k r_k}$, $\xi=\sqrt{\xi_k \xi_k}$, $\hat{r}_k=r_k/r$ and $\hat{\xi}_k=\xi_k/\xi$.

\section{Spherical harmonics}\label{app-Ch2:spherical harmonics}
This appendix summarises some of the properties of the spherical harmonics $Y_\ell^m(\uvg{\xi})$ used in the present Chapter. Further details about the spherical harmonics and more comprehensive discussions on their properties can be found in Ref.\ \cite{arfken2011}. Let $\uvg{\xi}=\uvg{\xi}(\thet,\ph)$, $\thet\in[0,\pi]$, $\ph\in[0,2\pi]$, a unit vector spanning the unit sphere $S_1$ in $\mathbb{R}^3$, the spherical harmonic $Y_\ell^m(\uvg{\xi})\equiv Y_\ell^m(\thet,\ph)$ of degree $\ell$ and order $m$ is defined as follows:
\begin{equation}\label{eq-Ch2:spherical harmonics definition}
Y_\ell^m(\uvg{\xi})=Y_\ell^m(\thet,\ph)=c_{\ell}^mP_\ell^m(\cos\thet)\exp(\mathrm{i} m\ph),
\end{equation}
with
\begin{equation}\label{eq-Ch2:spherical harmonics clm}
c_\ell^m=\sqrt{\frac{2\ell+1}{4\pi}\frac{(\ell-m)!}{(\ell+m)!}},
\end{equation}
and where $\mathrm{i}=\sqrt{-1}$ is the imaginary unit, $(\cdot)!$ denotes the factorial and $P_\ell^m(\cdot)$ are the associate Legendre polynomials. The normalisation coefficient $c_\ell^m$ is chosen such that two spherical harmonics $Y_{\ell_1}^{m_1}(\thet,\ph)$ and $Y_{\ell_2}^{m_2}(\thet,\ph)$ are orthogonal over $S_1$ as follows
\begin{equation}\label{eq-Ch2:spherical harmonics orthogonality}
\int_{S_1}Y_{\ell_1}^{m_1}(\thet,\ph)\bar{Y}_{\ell_2}^{m_2}(\thet,\ph)\dd S(\thet,\ph)=\delta_{m_1m_2}\delta_{\ell_1\ell_2},
\end{equation}
where $\bar{(\cdot)}$ denotes the complex conjugate and $\delta_{ij}$ is the Kronecker delta function. The spherical harmonic with negative $m$, $Y_\ell^{-m}$, can be written as
\begin{equation}\label{eq-Ch2:spherical harmonics definition negative m}
Y_\ell^{-m}(\thet,\ph)=(-1)^m\bar{Y}_\ell^m(\thet,\ph).
\end{equation}

The product of two spherical harmonics can be obtained using the Clebsch-Gordon series
\begin{equation}\label{eq-Ch2:spherical harmonics Clebsch-Gordon series}
Y_{\ell_1}^{m_1}(\thet,\ph)Y_{\ell_2}^{m_2}(\thet,\ph)=
\sum_{k=k_i}^{k_f}\sqrt{\frac{(2\ell_1+1)(2\ell_2+1)}{4\pi(2k+1)}}c_{000}^{\ell_1\ell_2k}c_{m_1m_2(m_1+m_2)}^{\ell_1\ell_2k} Y_k^{m_1+m_2}(\thet,\ph)
\end{equation}
where $k_i=\max\{|\ell_1-\ell_2|,|m_1+m_2|\}$, $k_f=\ell_1+\ell_2$ and $c_{m_1 m_2 m}^{\ell_1\ell_2\ell}$ denotes the Clebsch-Gordon coefficients, that is $c_{m_1m_2m}^{\ell_1\ell_2\ell}\equiv\langle \ell_1\ \ell_2\ m_1\ m_2|\ell_1\ \ell_2\ \ell\ m\rangle$.

The spherical harmonics are related to the Legendre polynomials through the following addition theorem
\begin{equation}\label{eq-Ch2:spherical harmonics addition theorem}
\frac{2\ell+1}{2}P_\ell(\hat{\xi}_k\hat{r}_k)=2\pi\sum_{m=-\ell}^{\ell}Y_\ell^m(\uvg{\xi})Y_\ell^m(\uv{r})
\end{equation}
where $\uv{r}$ and $\uvg{\xi}$ are two unit vectors, and $\hat{\xi}_k\hat{r}_k=\hat{\xi}_1\hat{r}_1+\hat{\xi}_2\hat{r}_2+\hat{\xi}_3\hat{r}_3$.

\clearpage

\section{Materials properties}\label{app-Ch2:materials properties}
\begin{table}[ht]
\small
\begin{center}
\caption{Elastic properties for Copper (Cu), Gold (Au) and Nickel (Ni) from Ref.\ \cite{benedetti2013a}.}
\label{tab-Ch2:mat properties Cu, Au, and Ni}
\begin{tabular}{llll}
\hline
\hline
material&property&component&value\\
\hline
Cu&elastic constants [$10^{9}\,\mathrm{N}/\mathrm{m}^{2}$]&$c_{1111}$, $c_{2222}$, $c_{3333}$&168\\
&&$c_{1122}$, $c_{1133}$, $c_{2233}$&121\\
&&$c_{1212}$, $c_{1313}$, $c_{2323}$&75\\
\hline
Au&elastic constants [$10^{9}\,\mathrm{N}/\mathrm{m}^{2}$]&$c_{1111}$, $c_{2222}$, $c_{3333}$&185\\
&&$c_{1122}$, $c_{1133}$, $c_{2233}$&158\\
&&$c_{1212}$, $c_{1313}$, $c_{2323}$&39.7\\
\hline
Ni&elastic constants [$10^{9}\,\mathrm{N}/\mathrm{m}^{2}$]&$c_{1111}$, $c_{2222}$, $c_{3333}$&251\\
&&$c_{1122}$, $c_{1133}$, $c_{2233}$&150\\
&&$c_{1212}$, $c_{1313}$, $c_{2323}$&124\\
\hline
\hline
\end{tabular}
\end{center}
\end{table}

\begin{table}[ht]
\small
\begin{center}
\caption{Piezoelectric properties for transversely isotropic lead zirconate titanate (PZT-4) ceramic from Ref.\ \cite{pan2000}.}
\label{tab-Ch2:mat properties PE trav iso}
\begin{tabular}{lll}
\hline
\hline
property&component&value\\
\hline
elastic constants [$10^{9}\,\mathrm{N}/\mathrm{m}^{2}$]&$c_{1111}$, $c_{2222}$&139\\
&$c_{3333}$&115\\
&$c_{1122}$&77.8\\
&$c_{1133}$, $c_{2233}$&74.3\\
&$c_{2323}$, $c_{1313}$&25.6\\
&$c_{1212}$&$(c_{1111}-c_{1122})/2$\\
piezoelectric constants [$\mathrm{C}/\mathrm{m}^{2}$]&$e_{113}$, $e_{223}$&12.7\\
&$e_{333}$&15.1\\
&$e_{322}$, $e_{311}$&-5.2\\
dielectric permeability&$\eps_{11}$, $\eps_{22}$&6.461\\
constants [$10^{-9}\,\mathrm{C}/(\mathrm{V}\cdot\mathrm{m})$]&$\eps_{33}$&5.620\\
\hline
\hline
\end{tabular}
\end{center}
\end{table}

\begin{table}[ht]
\begin{center}
\caption{Piezoelectric properties for orthotropic piezoelectric polyvinylidene fluoride (PVDF) from Ref.\ \cite{munoz2015}.}
\label{tab-Ch2:mat properties PE ortho}
\begin{tabular}{lll}
\hline
\hline
property&component&value\\
\hline
elastic constants [$10^{9}\,\mathrm{N}/\mathrm{m}^{2}$]&$c_{1111}$&3.61\\
&$c_{1122}$&1.61\\
&$c_{1133}$&1.42\\
&$c_{2222}$&3.13\\
&$c_{2233}$&1.31\\
&$c_{3333}$&1.63\\
&$c_{2323}$&0.55\\
&$c_{1313}$&0.59\\
&$c_{1212}$&0.69\\
piezoelectric constants [$\mathrm{C}/\mathrm{m}^{2}$]&$e_{113}$&-0.016\\
&$e_{223}$&-0.013\\
&$e_{333}$&-0.021\\
&$e_{311}$&0.032\\
&$e_{322}$&-0.004\\
dielectric permeability&$\eps_{11}$&0.054\\
constants [$10^{-9}\,\mathrm{C}/(\mathrm{V}\cdot\mathrm{m})$]&$\eps_{22}$&0.066\\
&$\eps_{33}$&0.059\\
\hline
\hline
\end{tabular}
\end{center}
\end{table}

\begin{table}[ht]
\begin{center}
\caption{Magneto-electro-elastic properties for the transversely isotropic material $M_1$ from Ref.\ \cite{pan2002}.}
\label{tab-Ch2:mat properties M1}
\begin{tabular}{lll}
\hline
\hline
property&component&value\\
\hline
elastic constants [$10^{9}\,\mathrm{N}/\mathrm{m}^{2}$]&$c_{1111}$, $c_{2222}$&166\\
&$c_{3333}$&162\\
&$c_{1122}$&77\\
&$c_{1133}$, $c_{2233}$&78\\
&$c_{1313}$, $c_{2323}$&43\\
&$c_{1212}$&$(c_{1111}-c_{1122})/2$\\
piezoelectric constants [$\mathrm{C}/\mathrm{m}^{2}$]&$e_{113}$, $e_{223}$&11.6\\
&$e_{333}$&18.6\\
&$e_{322}$, $e_{311}$&-4.4\\
piezomagnetic constants [$\mathrm{N}/(\mathrm{A}\cdot\mathrm{m})$]&$q_{113}$, $q_{223}$&550.0\\
&$q_{333}$&699.7\\
&$q_{322}$, $q_{311}$&580.3\\
magneto-electric&$\lambda_{11}$, $\lambda_{22}$&0\\
coefficients [$\mathrm{N}\cdot\mathrm{s}/(\mathrm{A}\cdot\mathrm{m})$]&$\lambda_{33}$&0\\
dielectric permeability&$\epsilon_{11}$, $\epsilon_{22}$&11.2\\
coefficients [$10^{-9}\mathrm{C}/(\mathrm{V}\cdot\mathrm{m})$]&$\epsilon_{33}$&12.6\\
magnetic permeability&$\kappa_{11}$, $\kappa_{22}$&5\\
coefficients [$10^{-6}\mathrm{N}\cdot\mathrm{s}^{2}/\mathrm{C}^2$]&$\kappa_{33}$&10\\
\hline
\hline
\end{tabular}
\end{center}
\end{table}

\begin{table}[ht]
\begin{center}
\caption{Magneto-electro-elastic properties for the orthotropic material $M_2$ from Ref.\ \cite{munoz2015}.}
\label{tab-Ch2:mat properties M2}
\begin{tabular}{lll}
\hline
\hline
property&component&value\\
\hline
elastic constants [$10^{9}\,\mathrm{N}/\mathrm{m}^{2}$]&$c_{1111}$&3.61\\
&$c_{1122}$&1.61\\
&$c_{1133}$&1.42\\
&$c_{2222}$&3.13\\
&$c_{2233}$&1.31\\
&$c_{3333}$&1.63\\
&$c_{2323}$&0.55\\
&$c_{1313}$&0.59\\
&$c_{1212}$&0.69\\
piezoelectric constants [$\mathrm{C}/\mathrm{m}^{2}$]&$e_{113}$&-0.016\\
&$e_{223}$&-0.013\\
&$e_{333}$&-0.021\\
&$e_{311}$&0.032\\
&$e_{322}$&-0.004\\
piezomagnetic constants [$\mathrm{N}/(\mathrm{A}\cdot\mathrm{m})$]&$q_{113}$&550.0\\
&$q_{223}$&570.0\\
&$q_{333}$&699.7\\
&$q_{311}$&580.3\\
&$q_{322}$&590.0\\
magneto-electric&$\lambda_{11}$&0.6\\
coefficients [$\mathrm{N}\cdot\mathrm{s}/(\mathrm{A}\cdot\mathrm{m})$]&$\lambda_{22}$&0.8\\
&$\lambda_{33}$&1.0\\
dielectric permeability&$\eps_{11}$&0.054\\
constants [$10^{-9}\,\mathrm{C}/(\mathrm{V}\cdot\mathrm{m})$]&$\eps_{22}$&0.066\\
&$\eps_{33}$&0.059\\
magnetic permeability&$\kappa_{11}$&5.0\\
coefficients [$10^{-6}\mathrm{N}\cdot\mathrm{s}^{2}/\mathrm{C}^2$]&$\kappa_{22}$&7.0\\
&$\kappa_{33}$&10.0\\
\hline
\hline
\end{tabular}
\end{center}
\end{table}

\clearpage

\chapter{An enhanced grain-boundary framework}\label{ch-EF}
In this Chapter, an enhanced three-dimensional framework for computational homogenisation and intergranular cracking of polycrystalline materials is presented. The framework is aimed at reducing the computational cost of polycrystalline micro simulations. The scheme is based on the grain-boundary formulation described in Chapter (\ref{ch-intro}). A regularisation scheme is used to avoid excessive mesh refinements often induced by the presence of small edges and surfaces in mathematically exact 3D Voronoi or Laguerre morphologies. For homogenisation purposes, periodic boundary conditions are enforced on \emph{non-prismatic} periodic micro Representative Volume Elements ($\mu$RVEs), eliminating pathological grains generally induced by the procedures used to generate \emph{prismatic} periodic $\mu$RVEs. An original meshing strategy is adopted to retain mesh effectiveness without inducing numerical complexities at grain edges and vertices. The proposed methodology offers remarkable \emph{computational savings} and \emph{high robustness}, both highly desirable in a multiscale perspective. The determination of the effective properties of several polycrystalline materials demonstrate the accuracy of the technique. Several microcracking simulations complete the study and confirm the performance of the method.

\section{Introduction}
In Chapter (\ref{ch-intro}), it was underlined how the high and less expensive affordability of HPC facilities has been pushing the boundaries of more and more realistic micro-mechanical models taking into account multiple aspects of the micro-structures of interest.
However, despite such remarkable advancements, any analysis of macro components or structures including all the details of the micro-structure, although conceivable in principle, is \emph{practically unfeasible}, due to its enormous computational cost. As a consequence, there is interest in developing techniques able to extract information from the finer scales that can be profitably used at the scale relevant to the technological application. Upon the premise of spatially separated length-scales, \emph{multi-scale modelling}, responds to such demand, being in fact an effective way to exchange information between the scales, avoiding the excessive cost of fine-scale modelling of the whole problem domain. In this framework, computational homogenisation plays a key role: a $\mu$RVE is associated to each relevant point of a macro-continuum and provides, through computational homogenisation, the macro-continuum constitutive behaviour. The collection of $\mu$RVEs identify the micro-scale underlying the application macro-scale.

Although multiscale modelling should in principle provide huge savings with respect to fine scale analyses of the whole domain, the need of resolving many micro Boundary Value Problems ($\mu$BVPs), nested within the macroscale, still generates huge computational costs, especially when 3D modelling is pursued, when fine details are included in the $\mu$RVEs morphologies, and when highly non-linear phenomena are investigated, e.g.\ initiation and propagation of damage. For such reasons, in recent years, remarkable efforts have been focused on the development of techniques for reducing the computational costs of $\mu$RVEs computations \cite{yvonnet2007,lamari2010}.

Due to the increased affordability of HPC, three-dimensional modelling of polycrystalline RVEs has recently received much attention \cite{musienko2009,kamaya2009,simonovski2012,benedetti2013b,benedetti2015}. However, the computational cost of 3D $\mu$RVEs simulations is consistently reported as a limiting factor for the scalability of the proposed methods and, as a consequence, only few 3D \emph{multiscale} frameworks have been developed. A two-scale finite element approach, successfully demonstrated on two 3D applications, has been developed by Han and Dawson \cite{han2007}. Nakamachi et al.\ \cite{nakamachi2007} proposed a multi-scale finite element procedure employing realistic 3D micro polycrystalline structures obtained by scanning electron microscopy and electron backscattering diffraction (SEM-EBSD) measurements. The technique was further refined for formability tests \cite{kuramae2010} and accelerated with HPC and domain partitioning techniques. Benedetti and Aliabadi \cite{benedetti2015} have recently proposed a 3D, homogenisation based, grain-boundary two-scale formulation for degradation and failure in polycrystalline materials: although the potential of the technique has been demonstrated, the two-scale 3D non-linear simulations have been hindered by the computational costs of the $\mu$RVEs simulations, pointing out the strong need for model order reduction.

In this Chapter, the polycrystalline problem is tackled adopting a comprehensive strategy aimed at reducing the order of the $\mu$RVEs simulations. The methodology is based on the synergistic use of state-of-the-art techniques for the analysis of polycrystalline microstructures. First, regularised Laguerre tessellations \cite{quey2011} are employed to remove the majority of those mathematically exact, but physically inessential, small entities (edges and faces) that generally challenge the meshing algorithms, inducing unduly large mesh refinements. The regularised tessellations are then employed within the grain-boundary formulation developed in Chapter (\ref{ch-intro}), which employs only grain-boundary variables, thus avoiding the grains internal meshing.

The grain-boundary meshing is enhanced employing continuous and semi-discontinuous elements for the faces of the grains, allowing a smoother and computationally more effective representation of the intergranular fields with respect to the basic discontinuous scheme. Additionally, a modified version for the polycrystalline unit cell is adopted: usually, Voronoi or Laguerre unit cells are obtained by cutting the unbounded tessellation, generated by some initial seeds, through some \emph{cutting walls} used to produce a prismatic RVE. The \emph{cutting} operation, however, often re-introduces unnecessary mesh refinements, probably due to the impossibility of avoiding small or pathological residual entities produced by the cutting operation itself. This can be simply avoided by retaining, in the final morphology, only those grains generated by the original seeds. This simple strategy produces polycrystalline RVEs with more regular and affordable overall meshes, at the expense of irregular external shapes and a slightly more complex implementation of the RVEs' boundary conditions.
Although this is a relatively simple expedient, it has not been used often in the literature, see e.g.\ \cite{nygaards2002,hlilou2009,gerard2013}. The above enhancements synergistically contribute to outstanding savings in terms of overall DoFs count. Their practical implementation and effectiveness is challenged by the high statistical variability of Voronoi or Laguerre aggregates, in terms of topology and morphology, so that many different cases can be met. These aspects are thoroughly discussed in the paper, which is organised as follows. The generation and regularisation of polycrystalline morphologies is discussed in Sections (\ref{sec-Ch3:modified morphology}) and (\ref{sec-Ch3:modified periodic morphology}), while the adopted meshing strategy is described in Section (\ref{sec-Ch3:meshing}). The tests in Section (\ref{sec-Ch3: computational tests}) complete the study.
\section{Laguerre tessellations regularisation}\label{sec-Ch3:modified morphology}
The adoption of the hardcore and Laguerre constraints, described in Chapter (\ref{ch-intro}), produces more regular morphologies with respect to the basic Poisson–Voronoi tessellation, namely it reduces the occurrence of excessively small grains edges and faces; however, it does not ensure a minimum smallest edge length in the tessellation. In other words, very small entities are still mathematically possible in the tessellation and, if they occur, they usually induce pathological meshing behaviours, i.e.\ excessive mesh refinements.

Quey et al.\ \cite{quey2011} have recently  developed a regularisation scheme that iteratively removes the small edges whose length is below a predetermined threshold value. The removal is accomplished by merging the two vertices defining the small edges. If the algorithm encounters a face formed by three edges and one of them must be removed, the whole face is deleted. The technique naturally leads to topological and morphological modifications of the tessellation that, strictly speaking, \emph{it is not a Voronoi or a Laguerre tessellation anymore}: the resulting tessellation may have more than four grains intersecting at one vertex and more than three faces intersecting at one edge. Moreover, while the grain faces lying on the external boundary of the aggregate remain flat, a regularised intergranular interface may result not planar, then inducing a loss of convexity for the two grains in contact through the considered interface itself. However, the regularised morphology results more affordable in terms of generated mesh, as the number of small geometrical entities (edges and faces) is drastically reduced. Further information on the effects of the regularization on the statistical features of the artificial polycrystalline topology and morphology can be found in Ref.\ \cite{quey2011}.

In the present work, regularised tessellations, produced in accordance with the referred procedure, are used for the first time in conjunction with the grain-boundary formulation presented in Chapter (\ref{ch-intro}). This is expected to produce noticeable computational advantages, which will be tested in the next sections. The regularisation is performed using the open source software \texttt{Neper} (\texttt{http://neper.sourceforge.net/}) \cite{quey2011}.

\section{Removal of \emph{cut} boundary grains}\label{sec-Ch3:modified periodic morphology}
Besides the adoption of regularised tessellations, another modification is proposed here for the generation of computationally effective polycrystalline $\mu$RVEs.

In a material homogenisation framework, different boundary conditions can be enforced on the $\mu$RVEs to retrieve the macro-scale properties \cite{kanit2003}.
\emph{Periodic} boundary conditions (PBCs) have been shown to provide faster convergence to the overall macro properties with respect to displacement or traction boundary conditions \cite{terada2000}. Even though it does not represent a requirement, the use of PBCs is facilitated by the conformity between the meshes of coupled opposite faces, and therefore by a periodic structure of the $\mu$RVE.

Fritzen et al.\ \cite{fritzen2009} suggested the following procedure to obtain periodic microstructures:
\begin{enumerate}
\item{given the initial seeds distribution inside the \emph{original} domain, the $N_g$ seeds are copied into the 26 boxes surrounding the original one;}
\item{the obtained \emph{extended} bounded domain, containing $27\times N_g$ seeds, is partitioned into convex polyhedra by means of a suitable tessellation algorithm;}
\item{the periodic RVE is extracted cutting out the original domain from the extended one.}
\end{enumerate}
Although this procedure leads to an apparently simple cubic periodic micro-structure, the use of the cutting planes introduces additional small geometrical entities, e.g.\ very small grains, which introduces additional sources of undesired mesh refinement.

Periodic microstructures may also be obtained by repeating steps 1 and 2 above (see Figure (\ref{fig-Ch4:repeated domain})) and, instead of cutting out the extended tessellation, by \emph{considering only the cells originated by original seeds, that is, those seeds that have been copied}. Given, the same initial distribution of seeds within a unit cube, two periodic morphologies generated using the above mentioned procedures are shown in Figure (\ref{fig-Ch4:CPvsP}).

Two couples of opposite boundary faces of the cut morphology are highlighted in Figure (\ref{fig-Ch3:CP50Gcf}). The adopted construction algorithm enforces opposite faces of the domain to be of the same shape (they can be perfectly superimposed through simple translation). This feature facilitates the generation of conforming meshes between opposite faces of the domain external boundary, and thus the application of PBCs on the $\mu$RVEs. Although this procedure has been successfully employed in homogenisation and multi-scale analyses, see e.g.\ \cite{benedetti2015}, the cutting process generally leads to the generation of very small grains. These small entities cannot be avoided during the analysis of the RVE and cannot be generally regularised, using for example \texttt{Neper}, and as a consequence they represent a source of overrefinement in the $\mu$RVE's mesh.

Figure (\ref{fig-Ch3:P50Gcf}) shows instead some of the coupled faces from the periodic microstructure, obtained by considering only the grains corresponding to the initial seeds, those contained within the original domain. In this case, the presence of small cells depends only on the seeds distribution and weights. Nevertheless, the microstructure retains the consistency between opposite faces of the tessellation. In fact, it can be shown that for each external grain face, i.e.\ each face that is not an intergranular interface, there exists an opposite periodic face of the same shape: PBCs can then be applied as in the case of cubic unit cells, the only difference resulting from a slightly more involved detection of coupled opposite periodic faces.

The effects of the regularisation and of the non-prismatic periodic morphology on the number of degrees of freedom for polycrystalline RVEs are investigated and discussed in Section (\ref{ssec-Ch3: computational savings}).

\begin{figure}
\centering
\includegraphics[width=\textwidth]{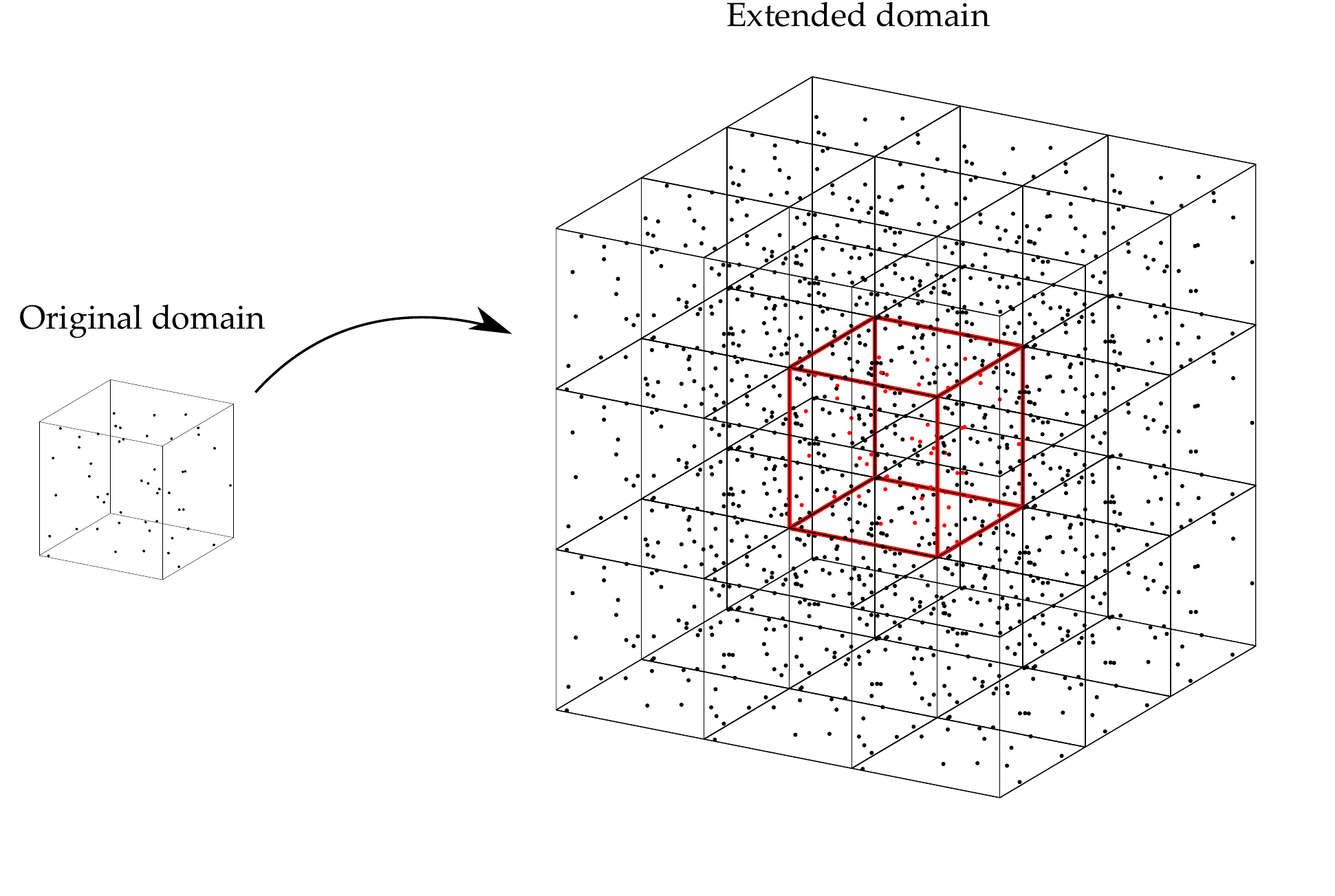}
\caption{Schematic generation of periodic morphologies: original domain containing $N_g$ scattered seeds and extended volume surrounding the original box (marked in red) and containing additional $26\times N_g$ points.}\label{fig-Ch4:repeated domain}
\end{figure}

\begin{figure}
\centering
	\begin{subfigure}{0.49\textwidth}
	\centering
	\includegraphics[width=\textwidth]{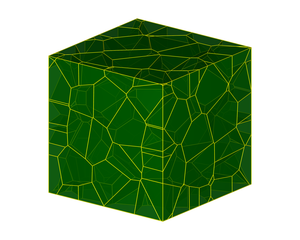}
	\caption{}
	\end{subfigure}
	\
	\begin{subfigure}{0.49\textwidth}
	\centering
	\includegraphics[width=\textwidth]{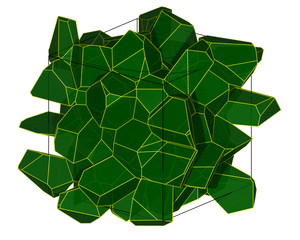}
	\caption{}
	\end{subfigure}
\caption{(\textit{a}) periodic morphology as obtained after the cutting process; (\textit{b}) periodic morphology obtained considering only the cells whose generator seeds fall within the original domain (the original box is sketched in black lines).}\label{fig-Ch4:CPvsP}
\end{figure}

\begin{figure}
\centering
	\begin{subfigure}{0.49\textwidth}
	\centering
	\includegraphics[width=\textwidth]{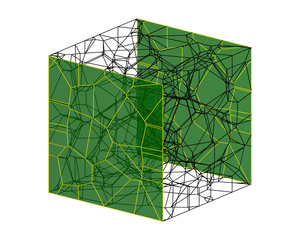}
	\caption{}
	\end{subfigure}
	\
	\begin{subfigure}{0.49\textwidth}
	\centering
	\includegraphics[width=\textwidth]{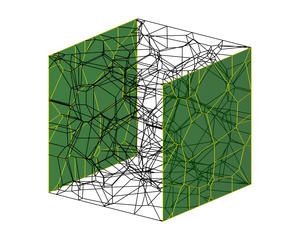}
	\caption{}
	\end{subfigure}
\caption{(\textit{a}-\textit{b}) Two examples of opposite coupled faces obtained after the cutting process.}
\label{fig-Ch3:CP50Gcf}
\end{figure}

\begin{figure}
\centering
	\begin{subfigure}{0.49\textwidth}
	\centering
	\includegraphics[width=\textwidth]{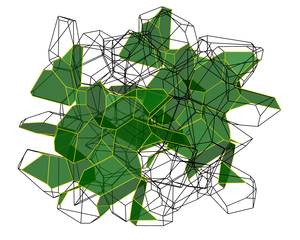}
	\caption{}
	\end{subfigure}
	\
	\begin{subfigure}{0.49\textwidth}
	\centering
	\includegraphics[width=\textwidth]{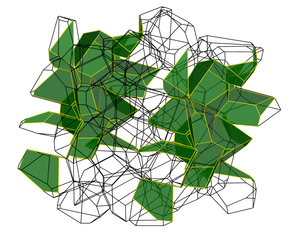}
	\caption{}
	\end{subfigure}
\caption{(\textit{a}-\textit{b}) Two examples of opposite coupled faces from the non-prismatic periodic microstructure.}
\label{fig-Ch3:P50Gcf}
\end{figure}
\section{Grain-boundary meshing}\label{sec-Ch3:meshing}
The generation of appropriate meshes for general polycrystalline microstructures is a demanding task, due to the need of preserving accuracy while maintaining affordable system order.
Apart the case of regular morphologies (cubes, octahedra, dodecahedra, etc.), for which the inherent regularity enables regular meshes, the difficulty of meshing randomly generated morphologies originates from their \emph{high statistical variability}.

In the literature either \emph{structured} or \emph{unstructured} meshes have been employed for polycrystalline aggregates \cite{fritzen2009,bohlke2010,barbe2011}. Structured meshes use elements of regular shape, such as cubes, to discretise the entire domain and attribute to each element the crystallographic properties of the grain they belong to. Such meshes are easy to generate and present high regularity.

However, the interfaces between the grains are poorly resolved, even if in case of very refined meshes, so that they do not appear adequate for the modelling of intergranular fields. Unstructured meshes, on the other hand, are built on the morphology itself and are thus naturally suitable to represent grains boundaries. Unstructured meshes are usually generated starting from 1D meshes of the tessellation edges. The nodes of this 1D mesh are then used as input to mesh the grains faces (2D mesh), which in turn provide the nodes for meshing the grains interior volumes (3D mesh). However, the presence of small entities causes either the presence of very low quality element or overly resolved regions.

Several meshing algorithms have been proposed in the literature to generate high quality polycrystalline meshes \cite{shewchuk1996,schoberl1997,geuzaine2009,quey2011}. In this study, the algorithm proposed by Persson and Strang \cite{persson2004} has been suitably modified for the generation of regular meshes of the grain faces. The difference with respect to the original scheme is that the mesh nodes of the grains edges were used as input for the grains faces meshing, to ensure consistency between the mesh of contiguous grain faces. Mesh homogeneity was enforced as in Ref.\ \cite{benedetti2013a}, and a mesh density parameter $d_m$ to control mesh refinement. Figures (\ref{fig-Ch3:dmeffect}a), (\ref{fig-Ch3:dmeffect}b) and (\ref{fig-Ch3:dmeffect}c) show the surface mesh of a Laguerre cell for three different values of $d_m$, respectively $d_m=1.0$, $d_m=2.0$, and $d_m=3.0$.

\begin{figure}[H]
\centering
	\begin{subfigure}{0.32\textwidth}
	\centering
	\includegraphics[width=\textwidth]{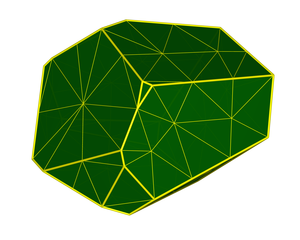}
	\caption{}
	\end{subfigure}
	\
	\begin{subfigure}{0.32\textwidth}
	\centering
	\includegraphics[width=\textwidth]{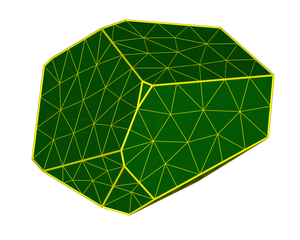}
	\caption{}
	\end{subfigure}
	\
	\begin{subfigure}{0.32\textwidth}
	\centering
	\includegraphics[width=\textwidth]{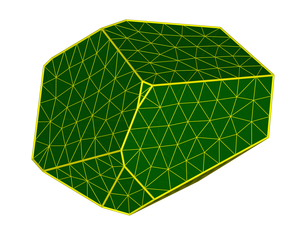}
	\caption{}
	\end{subfigure}
\caption{Surface mesh of a Laguerre cell for different values of the discretisation parameter $d_m$: (\emph{a}) $d_m=1.0$; (\emph{b}) $d_m=2.0$; (\emph{c}) $d_m=3.0.$}
\label{fig-Ch3:dmeffect}
\end{figure}

\subsection{Geometrical nodes and collocation nodes}\label{ssec-Ch3:gnodes and fnodes}
As described in Section (\ref{sec-Ch1:grain boundary integral equations}) of Chapter (\ref{ch-intro}), the displacement boundary integral equations (\ref{eq-Ch1:DBIE-2 local RS}), which are reported here for the sake of readability
\begin{equation}\label{eq-Ch2:DBIE}
\wtilde{c}_{pi}^g(\mathbf{y})\wtilde{u}_i(\mathbf{y})+\dashint_{S^g}\wtilde{T}_{pi}^g(\mathbf{x},\mathbf{y})\wtilde{u}_i(\mathbf{x})\dd S(\mathbf{x})=
\int_{S^g}\wtilde{U}_{pi}^g(\mathbf{x},\mathbf{y})\wtilde{t}_i(\mathbf{x})\dd S(\mathbf{x}),
\end{equation}
are numerically evaluated at the \emph{collocation} points $\mathbf{y}$ of the grain $g$ in order to obtain the discretised version $\mathbf{H}^g\mathbf{U}^g=\mathbf{G}^g\mathbf{T}^g$ (see Eq.(\ref{eq-Ch1:DBIE discrete})). The procedure is straightforward if such points lie on a smooth surface, as the normal unit outward vectors are uniquely defined; however, if the collocation points belong to edges or vertices of the tessellation, the unit normal vector is not unique. This is a well known issue in the boundary element literature: \emph{corner points}, as they are usually referred to, induce more unknowns than equations when they are external points with prescribed displacements and when they are interface points \cite{gray1990}. In a polycrystalline $\mu$RVE, such issue is encountered at \emph{each} grain edge and vertex, so that the approach adopted to tackle it has a noticeable effect on the overall effectiveness of the methodology.

In \cite{benedetti2013a,benedetti2013b}, for simplicity of implementation, discontinuous triangular elements were employed, so to ensure that the collocation points would always lie on smooth surfaces. However, this has been identified as a source of computational ineffectiveness, leading to an unnecessary increase of number of DoFs, particularly critical in the case of intergranular degradation modelling. In polycrystalline modelling, and in general in the framework of multi-scale modelling, the average number of DoFs used in the $\mu$BVP is of crucial importance when memory and solution time performances are of concern. For this reason and in this sense, model reduction with accuracy preservation are undoubtedly of great interest.

Different strategies can be used to address corner points in boundary elements \cite{gray1990,deng2013}. In this work, to reduce the computational inefficiency of discontinuous meshing while keeping a relatively easy implementation, continuous and semi-discontinuous elements are simultaneously used for the representation of the grain boundary fields, while a continuous representation of the geometry is naturally retained.

Consider a sample face as shown in Figure (\ref{fig-Ch3:tri-quad mesh}a), where the edge nodes are indicated with big dots. The 1D mesh of the edges of the entire tessellation is performed first and then used as input for the 2D mesh of the tessellation faces. The face is subdivided into non-overlapping triangles, Figure (\ref{fig-Ch3:tri-quad mesh}b) by means of the procedure described in Section (\ref{sec-Ch3:meshing}). The use of discontinuous elements for the approximation of the field variables leads to $3\times N_t$ collocation nodes, where $N_t$ denotes the number of discretisation triangles. Figure (\ref{fig-Ch3:tri-quad mesh}c) shows the position of the geometrical nodes, which are indicated with big dots, and the position of the collocation points, indicated with crosses.

The use of continuous and semi-discontinuous elements considerably reduces the number of collocation points. Two different distributions of the collocation nodes can be obtained, depending on the algorithm selected. The first distribution, as shown in Figure (\ref{fig-Ch3:tri-quad mesh}d), is obtained by moving the nodes lying on a face's edge \emph{inside} the triangles. In this case each edge node generates as many collocation points as the number of triangles sharing that specific edge node. Internal geometrical nodes must be added to such \emph{edge} collocation nodes to provide the overall number of collocation points associated with a given grain face.
\begin{figure}[H]
\centering
	\begin{subfigure}{0.49\textwidth}
	\centering
	\includegraphics[width=\textwidth]{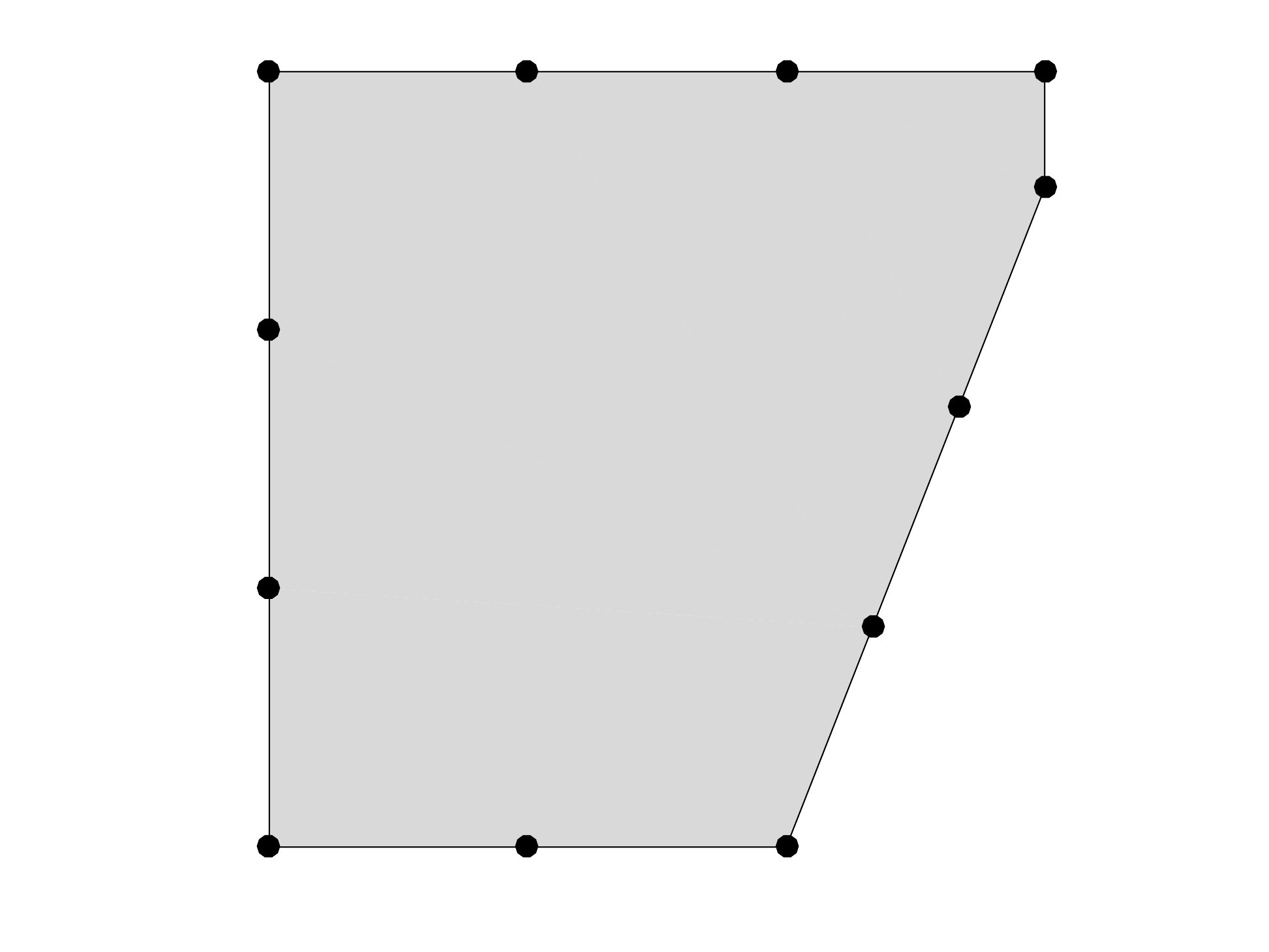}
	\caption{}
	\end{subfigure}
\	
	\begin{subfigure}{0.49\textwidth}
	\centering
	\includegraphics[width=\textwidth]{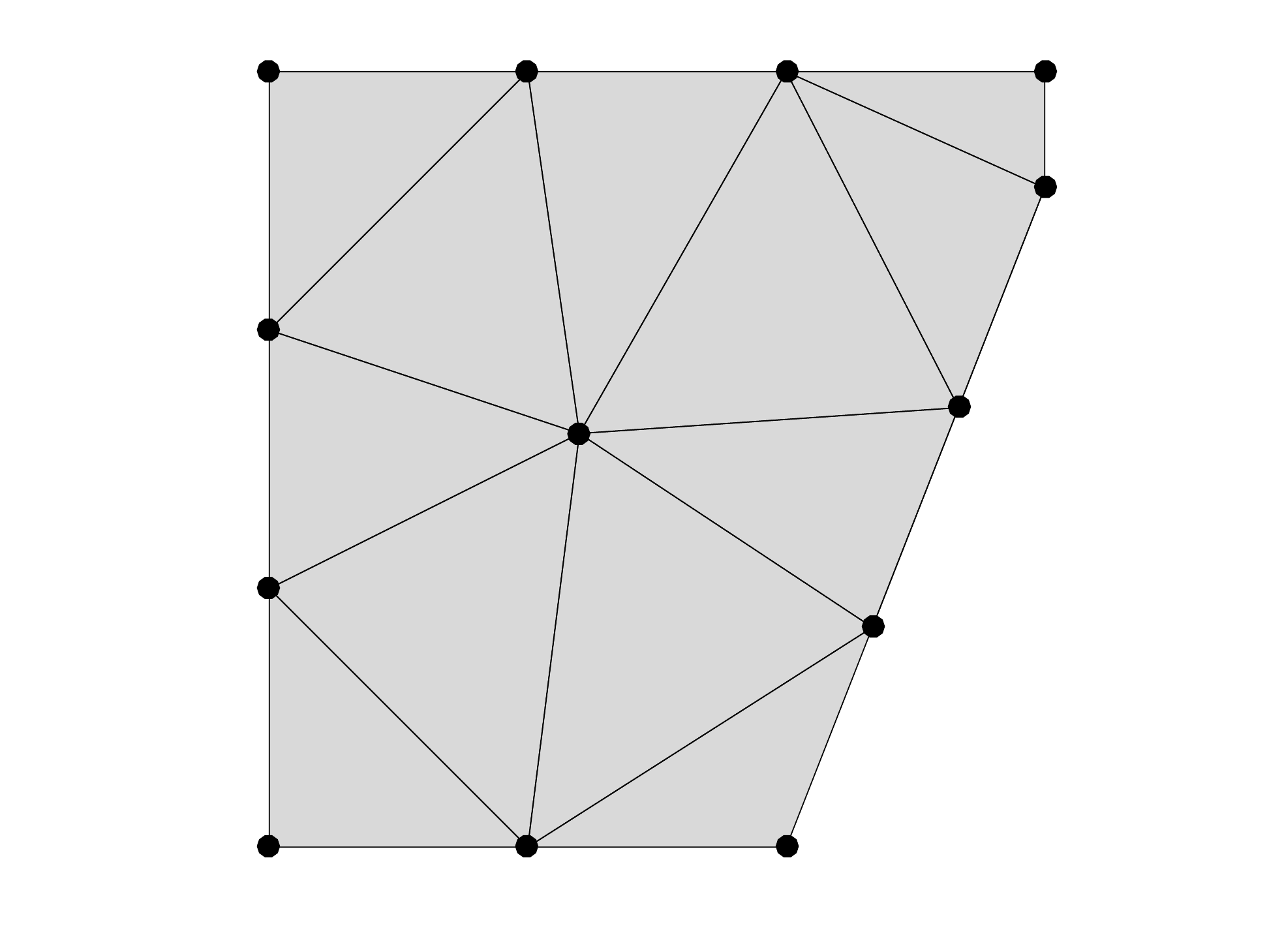}
	\caption{}
	\end{subfigure}
\	
	\begin{subfigure}{0.49\textwidth}
	\centering
	\includegraphics[width=\textwidth]{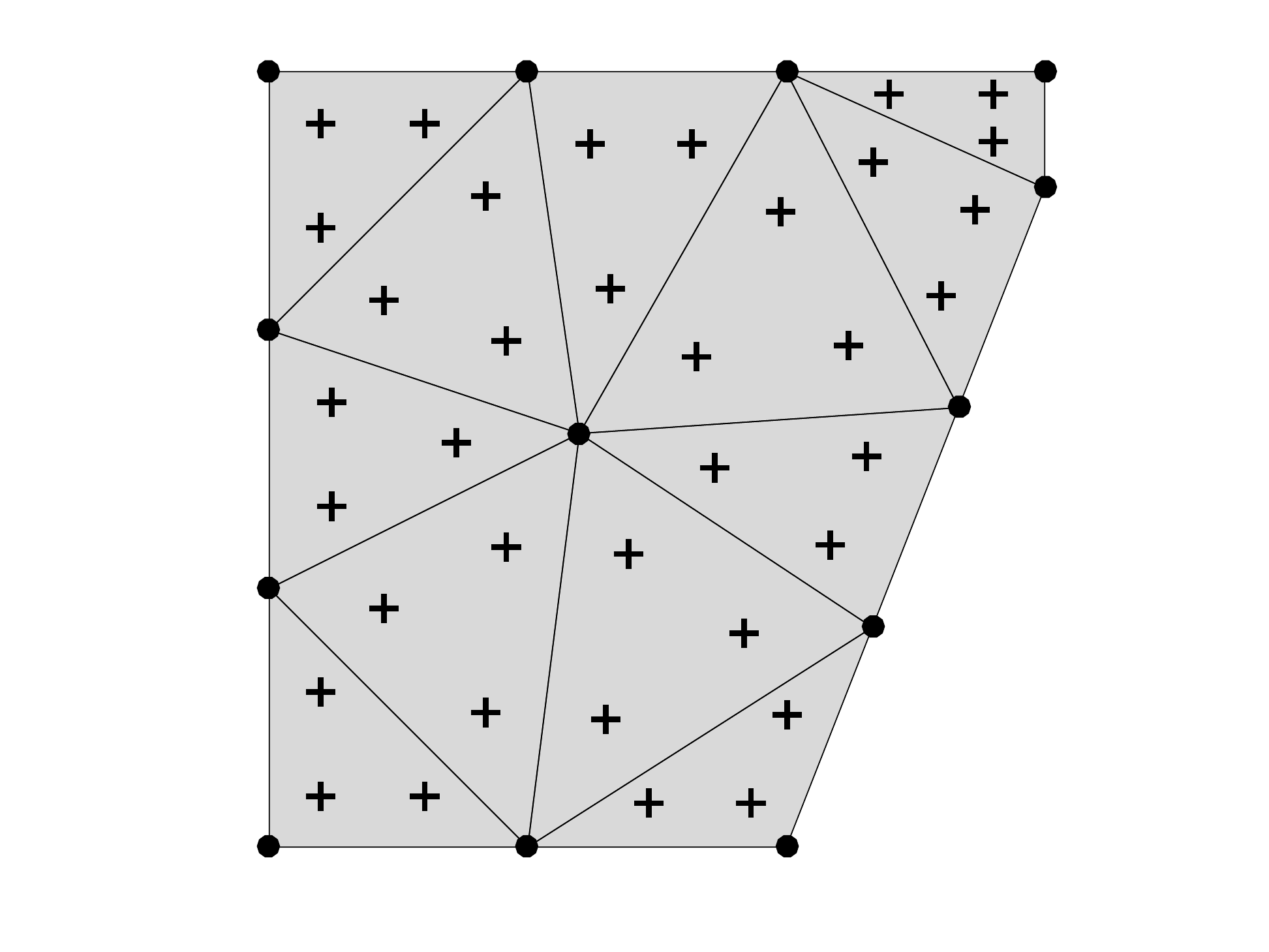}
	\caption{}
	\end{subfigure}
\	
	\begin{subfigure}{0.49\textwidth}
	\centering
	\includegraphics[width=\textwidth]{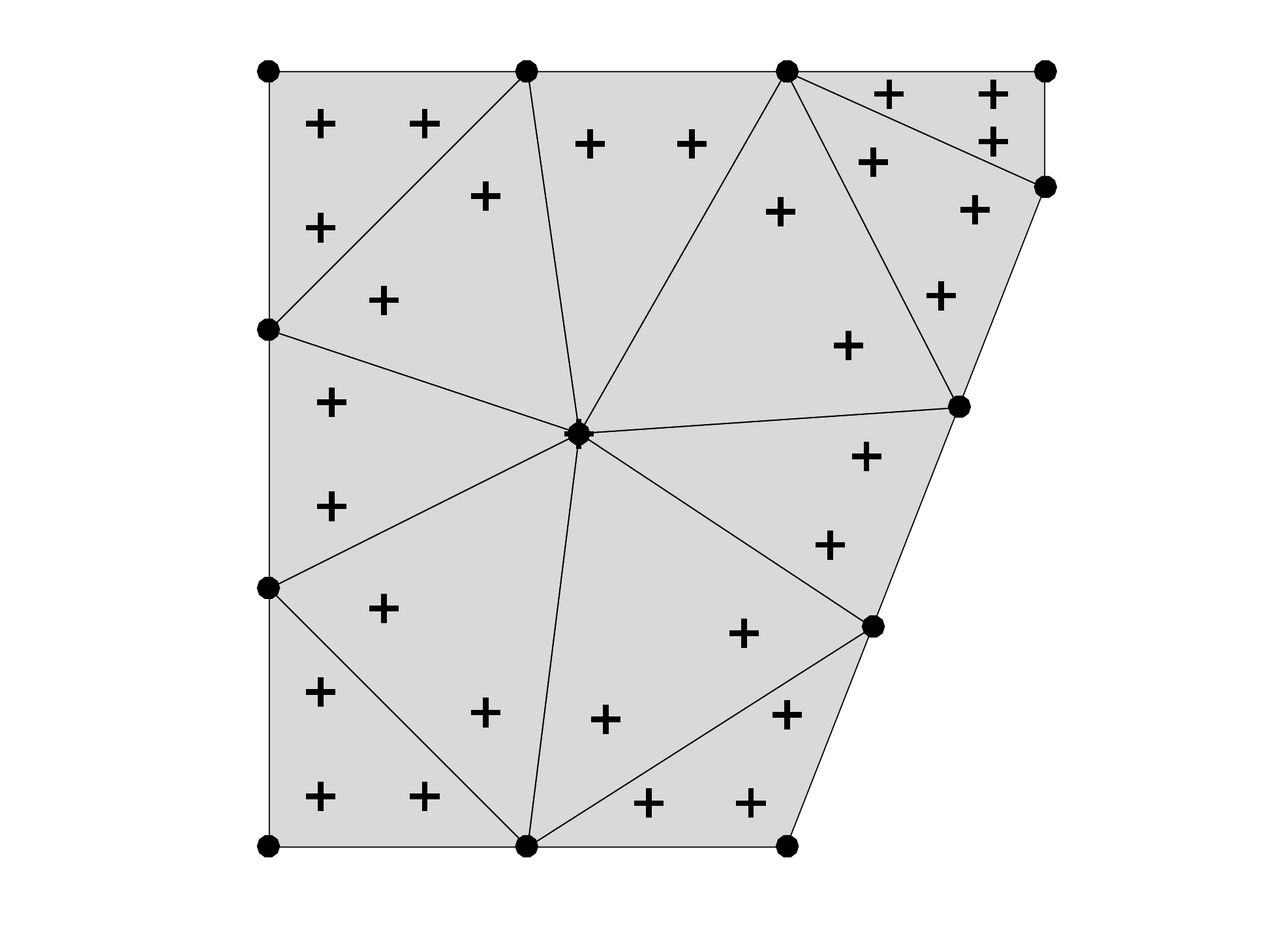}
	\caption{}
	\end{subfigure}
\	
	\begin{subfigure}{0.49\textwidth}
	\centering
	\includegraphics[width=\textwidth]{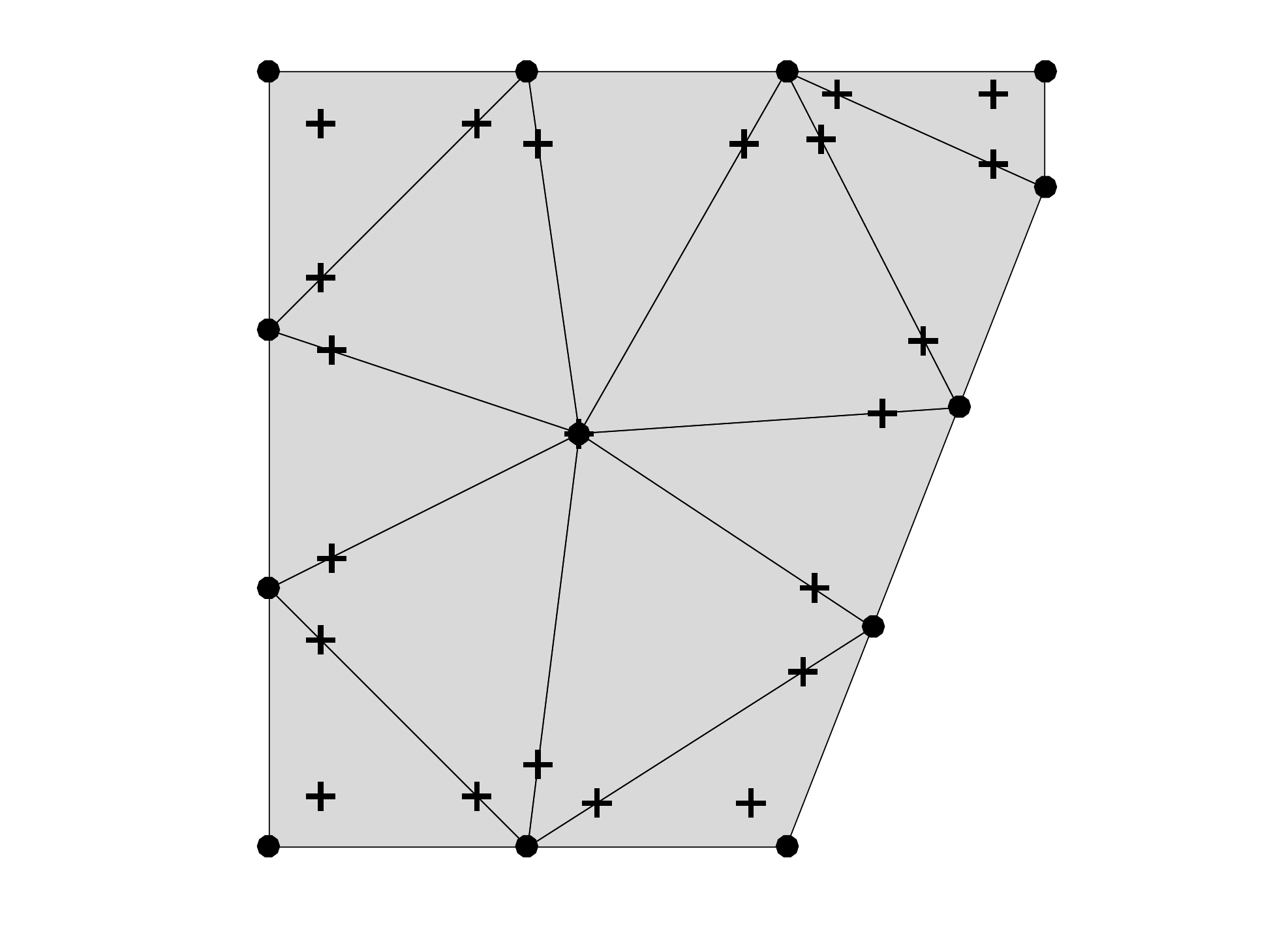}
	\caption{}
	\end{subfigure}
\caption{(\emph{a}) Sample face with marked edge nodes; (\emph{b}) 2D triangulation of the face; (\emph{c}) Position of the collocation nodes when a discontinuous mesh is adopted; (\emph{d},\emph{e}) Position of the collocation nodes when two continuous/semi-discontinuous approximations are adopted. In the figures, the small circles represent the geometrical nodes whereas the small crosses represent the collocation nodes.}
\label{fig-Ch3:tri-quad mesh}
\end{figure}
The second distribution, as shown in Figure (\ref{fig-Ch3:tri-quad mesh}e), is instead obtained by moving the edge nodes \emph{along} the edges of the mesh triangles. In this case, each edge node generates as many collocation points as the number of triangles' edges meeting at the considered node. The overall number of collocation nodes is obtained by adding to these the number of internal geometrical nodes. As it can be inferred from Figures (\ref{fig-Ch3:tri-quad mesh}d) and (\ref{fig-Ch3:tri-quad mesh}e), the former distribution produces a higher number of collocation nodes with respect to the latter. However, each triangle may always contain only three collocation nodes. The latter distribution, on the contrary, produces a smaller number of collocation nodes, but the resulting triangles may contain from three to six collocation nodes. \emph{The latter distribution therefore requires a suitable specialisation of the shape functions for each possible case}, i.e.\ number of collocation nodes per triangle.

In the last part of this Section, it is shown that a surface mesh combining triangular and quadrangular elements leads to a further reduction of collocation nodes with respect to the approach shown in Figure (\ref{fig-Ch3:tri-quad mesh}e), while retaining simplicity in the shape functions representation. Figure (\ref{fig-Ch3:tri-quad mesh 2}a) shows some triangles containing more than three collocation nodes after the semi-discontinuous approximation described in Figure (\ref{fig-Ch3:tri-quad mesh}e). In particular, the four dark-grey triangles contain five collocation nodes, whereas the mid-grey triangle contains four collocation nodes. By looking at the neighbouring triangles, it can be noted that, by appropriately merging two triangles, a quadrangle with four collocation node can be obtained.
This is shown in Figure (\ref{fig-Ch3:tri-quad mesh 2}b), where the mid-grey triangles represent the couples of triangles that, after the merging procedure, would produce a quadrangle with four collocation nodes. Figure (\ref{fig-Ch3:tri-quad mesh 2}b) also shows two triangles (in dark-grey) whose merging would not produce a quadrangle with four collocation nodes. To avoid the generation of this unwanted element, an additional internal node is added to the face, which is triangulated again as shown in Figure (\ref{fig-Ch3:tri-quad mesh 2}c). Additional internal nodes are added until the presence of either triangles with three nodes or quadrangles with four nodes is ensured (usually very few additional nodes are needed). Triangles with four collocation nodes are allowed, as they are treated as collapsed quadrangles. Finally, the merging algorithm is employed and a continuous/semi-discontinuous mesh is obtained as shown in Figure (\ref{fig-Ch3:tri-quad mesh 2}d). The resulting mesh has the advantage of drastically reducing the number of collocation nodes with respect to a completely  discontinuous approach, as the one previously adopted in \cite{benedetti2013a,benedetti2013b}. Furthermore, as it combines triangular and quadrangular elements, the shape functions are straightforward to deal with.

\begin{figure}[H]
\centering
	\begin{subfigure}{0.49\textwidth}
	\centering
	\includegraphics[width=\textwidth]{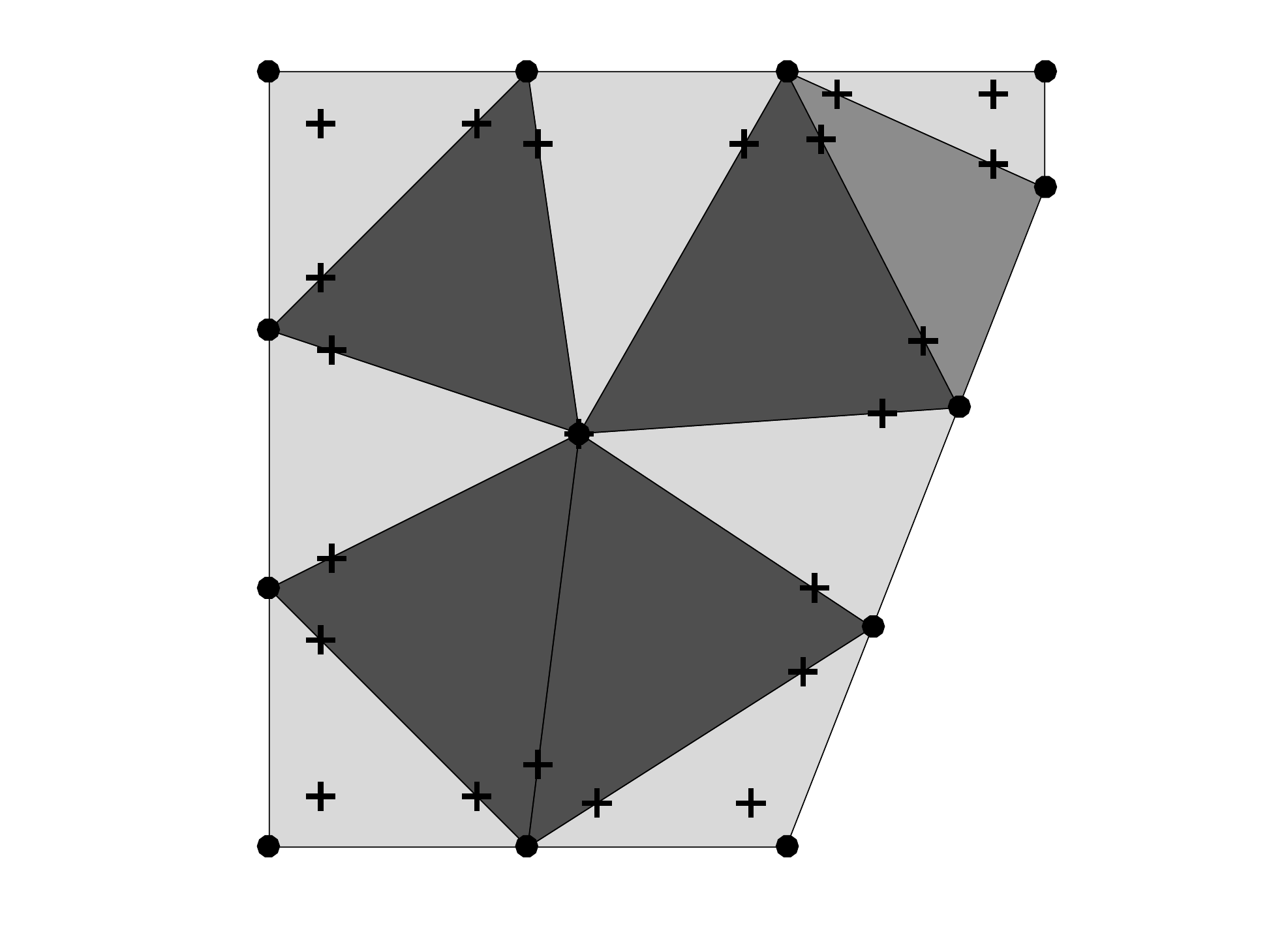}
	\caption{}
	\end{subfigure}
\	
	\begin{subfigure}{0.49\textwidth}
	\centering
	\includegraphics[width=\textwidth]{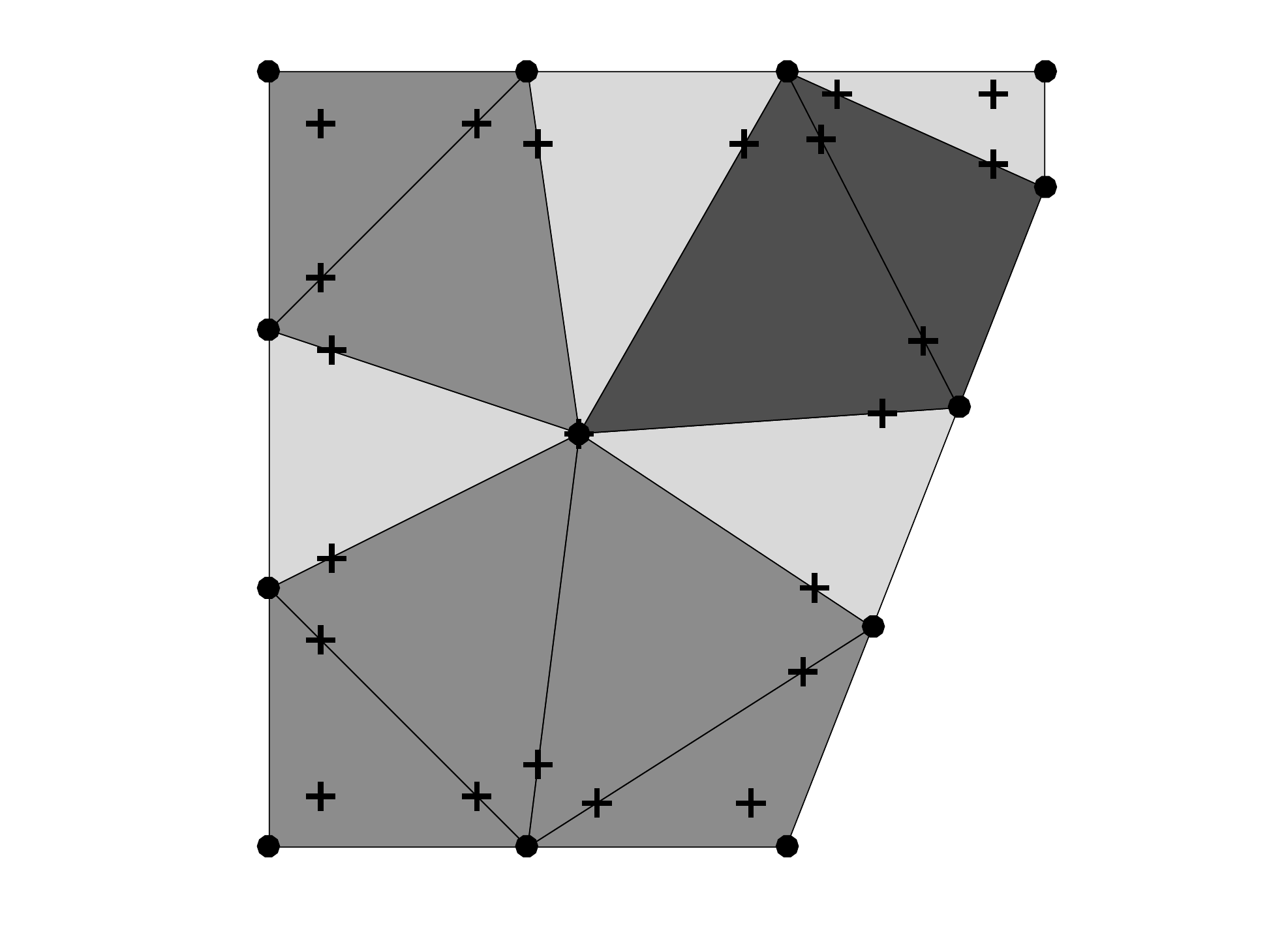}
	\caption{}
	\end{subfigure}
\	
	\begin{subfigure}{0.49\textwidth}
	\centering
	\includegraphics[width=\textwidth]{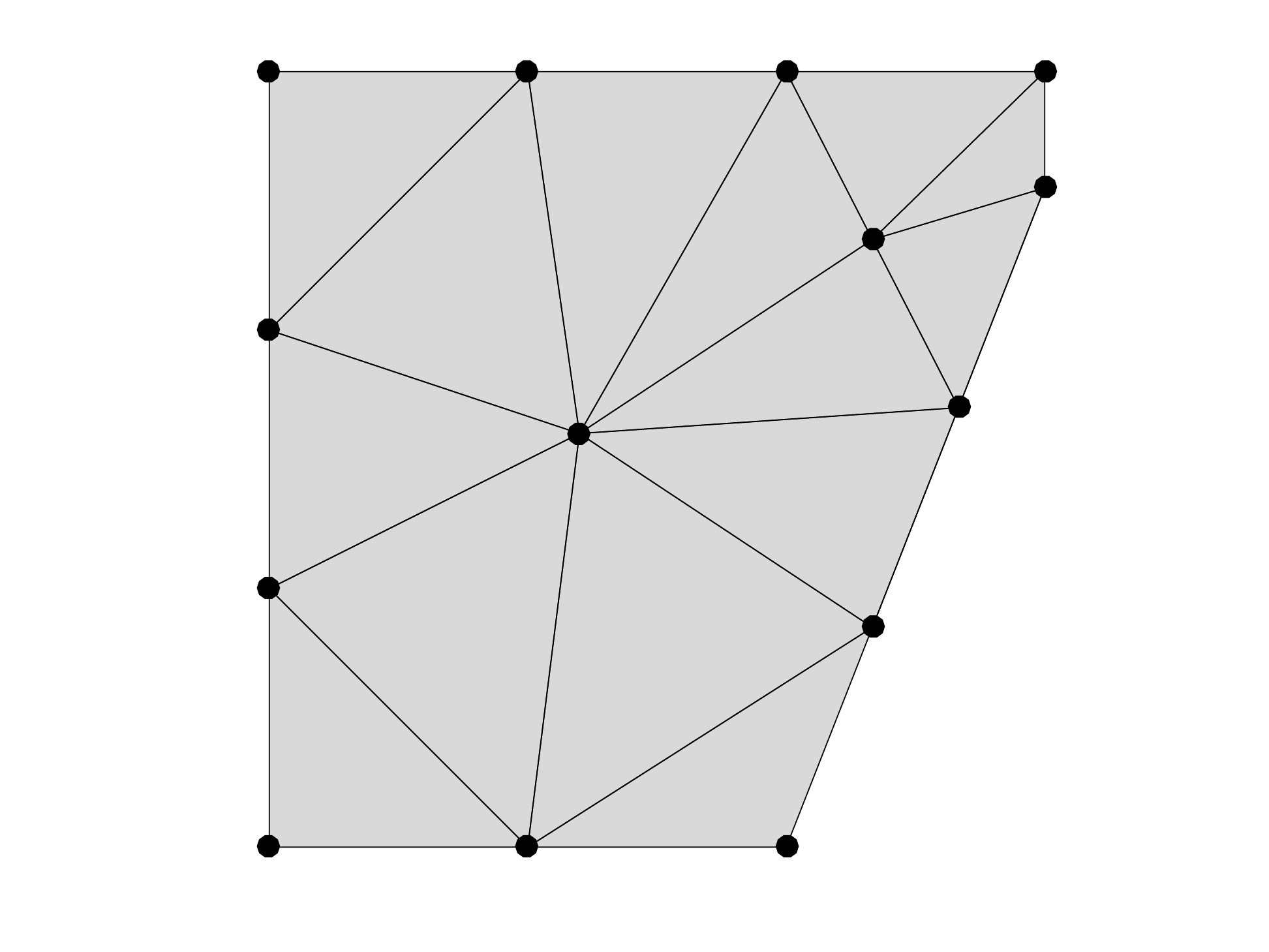}
	\caption{}
	\end{subfigure}
\	
	\begin{subfigure}{0.49\textwidth}
	\centering
	\includegraphics[width=\textwidth]{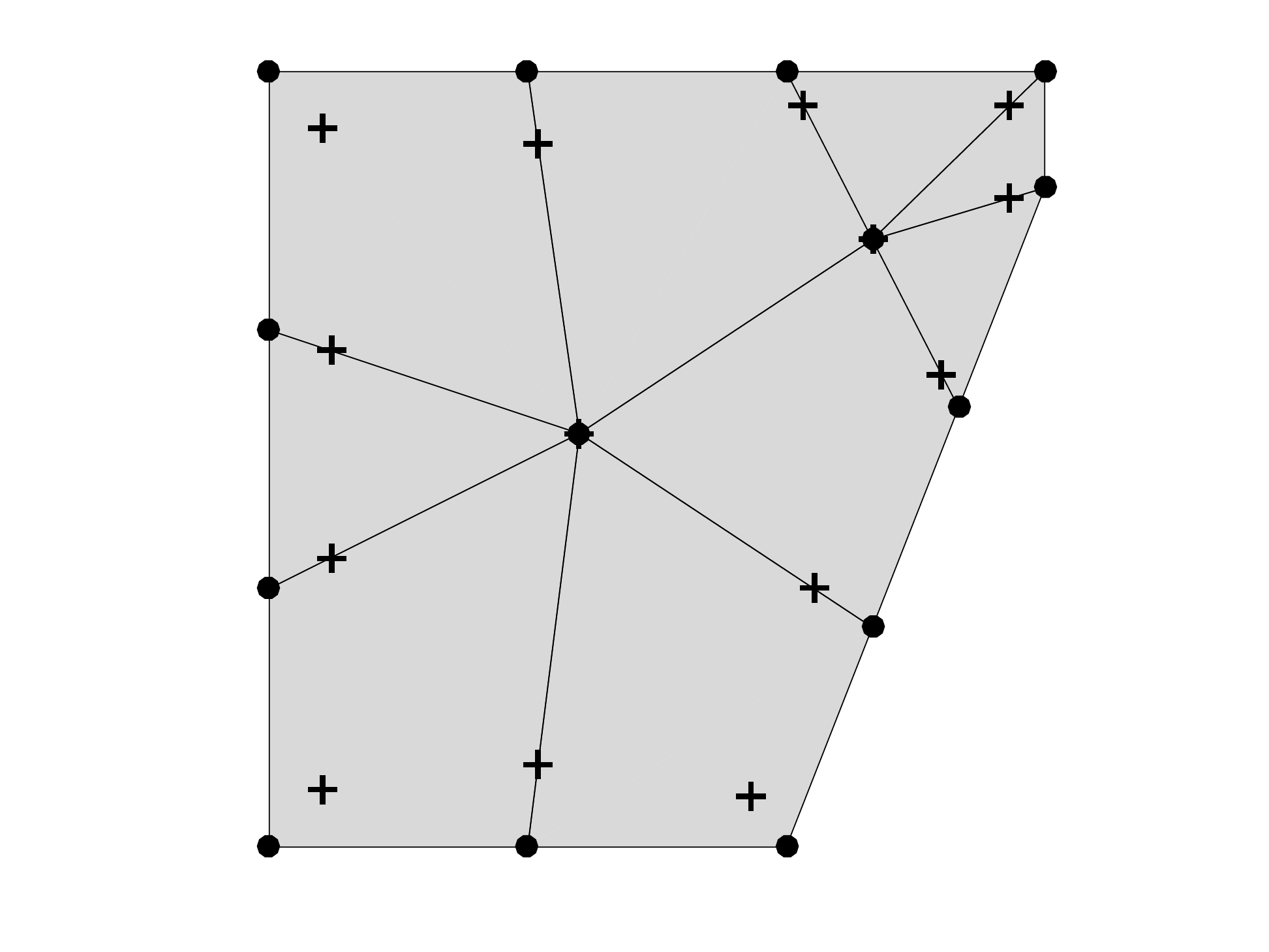}
	\caption{}
	\end{subfigure}
\caption{(\emph{a}) Sample grain face showing semi-discontinuous triangles containing more than three collocation nodes; (\emph{b}) Couples of matched triangles whose merging produces quadrangular elements with four (\emph{mid-grey}) or five (\emph{dark-grey}) collocation nodes; (\emph{c}) New triangulation obtained by adding supplementary internal nodes to avoid generation of quadrangular elements with more than four collocation points; (\emph{d}) Resulting mixed triangular-quadrangular continuous-semidiscontinuous grain face's mesh. In the figures, the small circles represent the geometrical nodes whereas the small crosses represent the collocation nodes.}
\label{fig-Ch3:tri-quad mesh 2}
\end{figure}

Figure (\ref{fig-Ch3:dmeffect tq}) shows the surface mesh of the same Laguerre cell as in Figure (\ref{fig-Ch3:dmeffect}) when the described merging algorithm is employed. It is worth noting that the internal structure of the mesh is retained as the triangles are very well-shaped thanks to the meshing algorithm of Section (\ref{sec-Ch3:meshing}). Moreover, merging two \emph{internal} triangles into a quadrangle would not reduce the number of collocation nodes. On the other hand, on the faces edges, merging triangles into quadrangles leads to a reduction of the DoFs associated to the Laguerre cell. Figs (\ref{fig-Ch3:dmeffect tq}a), (\ref{fig-Ch3:dmeffect tq}b) and (\ref{fig-Ch3:dmeffect tq}c) show the surface mesh for the three different values of $d_m$, respectively $d_m=1.0$, $d_m=2.0$, and $d_m=3.0$, produced by the optimized meshing algorithm. In this thesis, the last described meshing algorithm, that is the one combining triangular and quadrangular semi-discontinuous elements, is adopted.

\begin{figure}[h]
\centering
	\begin{subfigure}{0.32\textwidth}
	\centering
	\includegraphics[width=\textwidth]{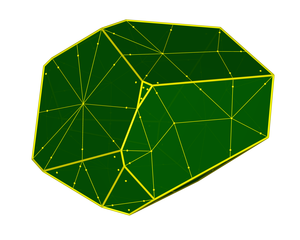}
	\caption{}
	\end{subfigure}
	\
	\begin{subfigure}{0.32\textwidth}
	\centering
	\includegraphics[width=\textwidth]{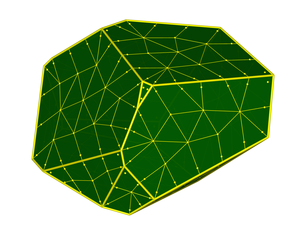}
	\caption{}
	\end{subfigure}
	\
	\begin{subfigure}{0.32\textwidth}
	\centering
	\includegraphics[width=\textwidth]{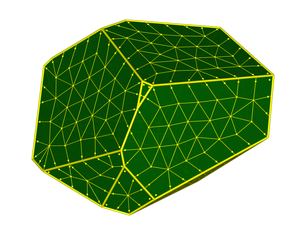}
	\caption{}
	\end{subfigure}
\caption{Surface mesh of a Laguerre cell obtained by using the optimised meshing algorithm with triangular and quadrangular elements, for different values of the discretisation parameter $d_m$: (\emph{a}) $d_m=1.0$; (\emph{b}) $d_m=2.0$; (\emph{c}) $d_m=3.0$. The small dots represent the collocation points of the mesh.}
\label{fig-Ch3:dmeffect tq}
\end{figure}

\subsection{Continuous/semi-discontinuous elements}
In this Section, the triangular and quadrangular elements used for the grain boundary mesh are discussed with more details.

Similarly to a finite element discretisation, the surface of a grain is discretised into non-overlapping elements. For each element, the geometrical position of a point $\mathbf{x}=\{x_1,x_2,x_3\}$ in the global coordinate system can be represented in terms of the coordinates $\boldsymbol{\xi}=\{\xi_1,\xi_2\}$ of the element's local coordinate system by means of a transformation $\mathbf{x}=\mathbf{x}(\boldsymbol{\xi})$ that, for the triangular and quadrangular elements used in this thesis, can be expressed as follows
\begin{equation}\label{eq-Ch3:geo apprx}
\mathbf{x} = \sum_{\alpha=1}^{\mu}M^{(\alpha)}(\boldsymbol{\xi})\mathbf{x}^{(\alpha)},
\end{equation}
where $M^{(\alpha)}(\boldsymbol{\xi})$ represents the $\alpha$-th shape function, $\mathbf{x}^{(\alpha)}$ represents the coordinates of the $\alpha$-th geometrical node of the element and $\mu$ represent the number of geometrical nodes of the elements. In this thesis, linear triangular and quadrangular boundary elements have been employed. The two types of elements are reported in Figure (\ref{fig-Ch3:disc elements}) where the black dots indicate the geometrical nodes.

Unlike the geometry that is represented by continuous shape functions $M^{(\alpha)}(\boldsymbol{\xi})$, a generic unknown boundary field $\phi$ is approximated by means of continuous/semi-discontinuous shape functions that are suitably specialised to avoid corner problems discussed in Section (\ref{ssec-Ch3:gnodes and fnodes}). The field $\phi$ is approximated as follows
\begin{equation}\label{eq-Ch3:field apprx}
\phi = \sum_{\beta=1}^{\nu}N^{(\beta)}(\boldsymbol{\xi},\mathbf{d})\phi^{(\beta)},
\end{equation}
where $N^{(\beta)}(\boldsymbol{\xi},\mathbf{d})$ represents the $\beta$-th shape function, $\mathbf{d}$ is a set of discontinuity parameters that can be adjusted to obtain continuous to completely discontinuous elements, $\phi^{(\beta)}$ is the value of the field at the collocation node $\mathbf{y}^{(\beta)}$ of the element and $\nu$ is the number of nodal values of the field as well as the number of collocation nodes of the element. Figures (\ref{fig-Ch3:disc elements}a) and (\ref{fig-Ch3:disc elements}b) show the position of the collocation nodes of a discontinuous triangular and a discontinuous quadrangular element, respectively, indicated by small crosses. The position of the collocation nodes and the expressions of the shape functions in terms of the discontinuity parameters $\mathbf{d}$ are discussed next.

\begin{figure}[h]
\centering
	\begin{subfigure}{0.49\textwidth}
	\centering
	\includegraphics[width=\textwidth]{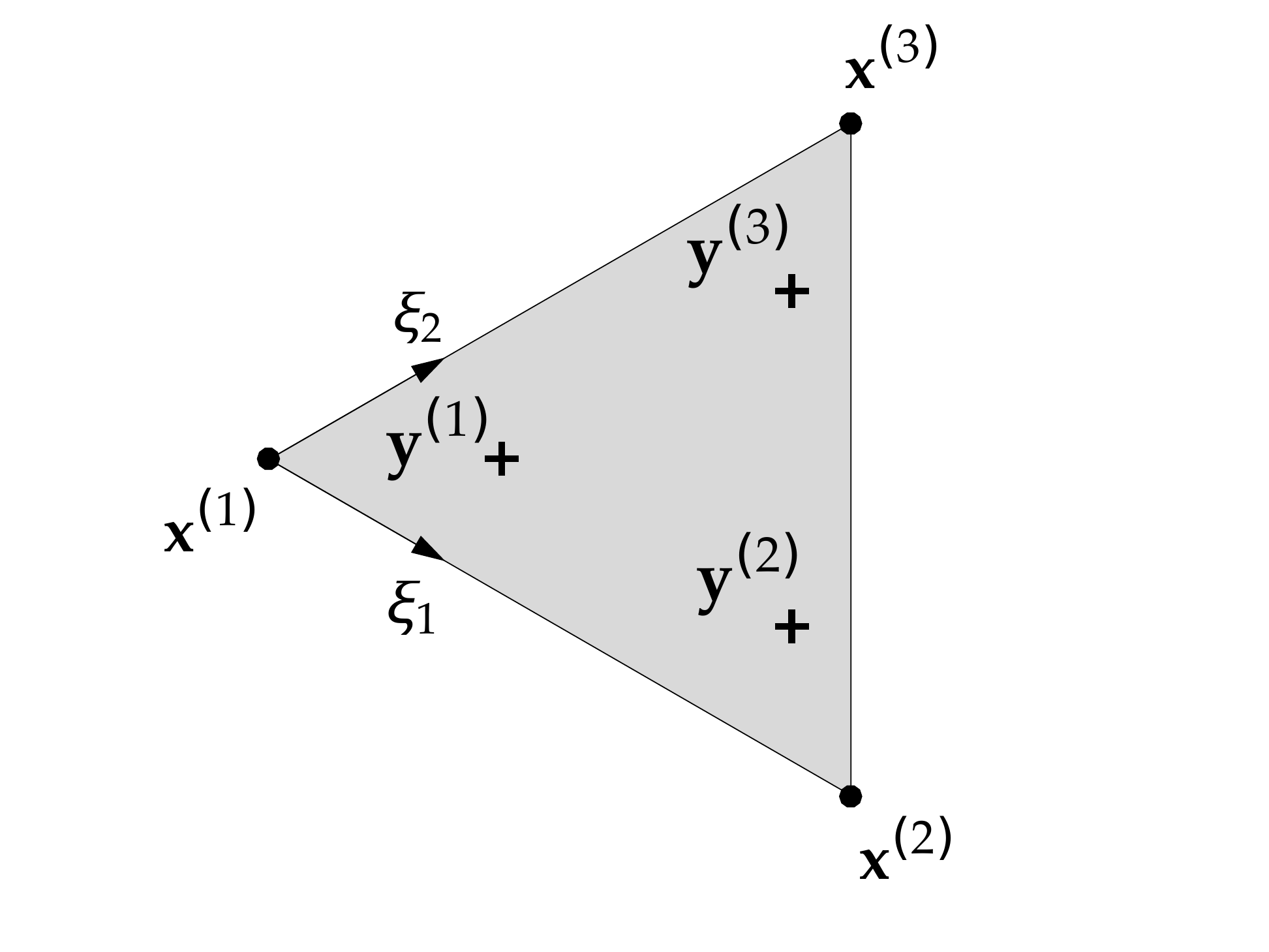}
	\caption{}
	\end{subfigure}
\	
	\begin{subfigure}{0.49\textwidth}
	\centering
	\includegraphics[width=\textwidth]{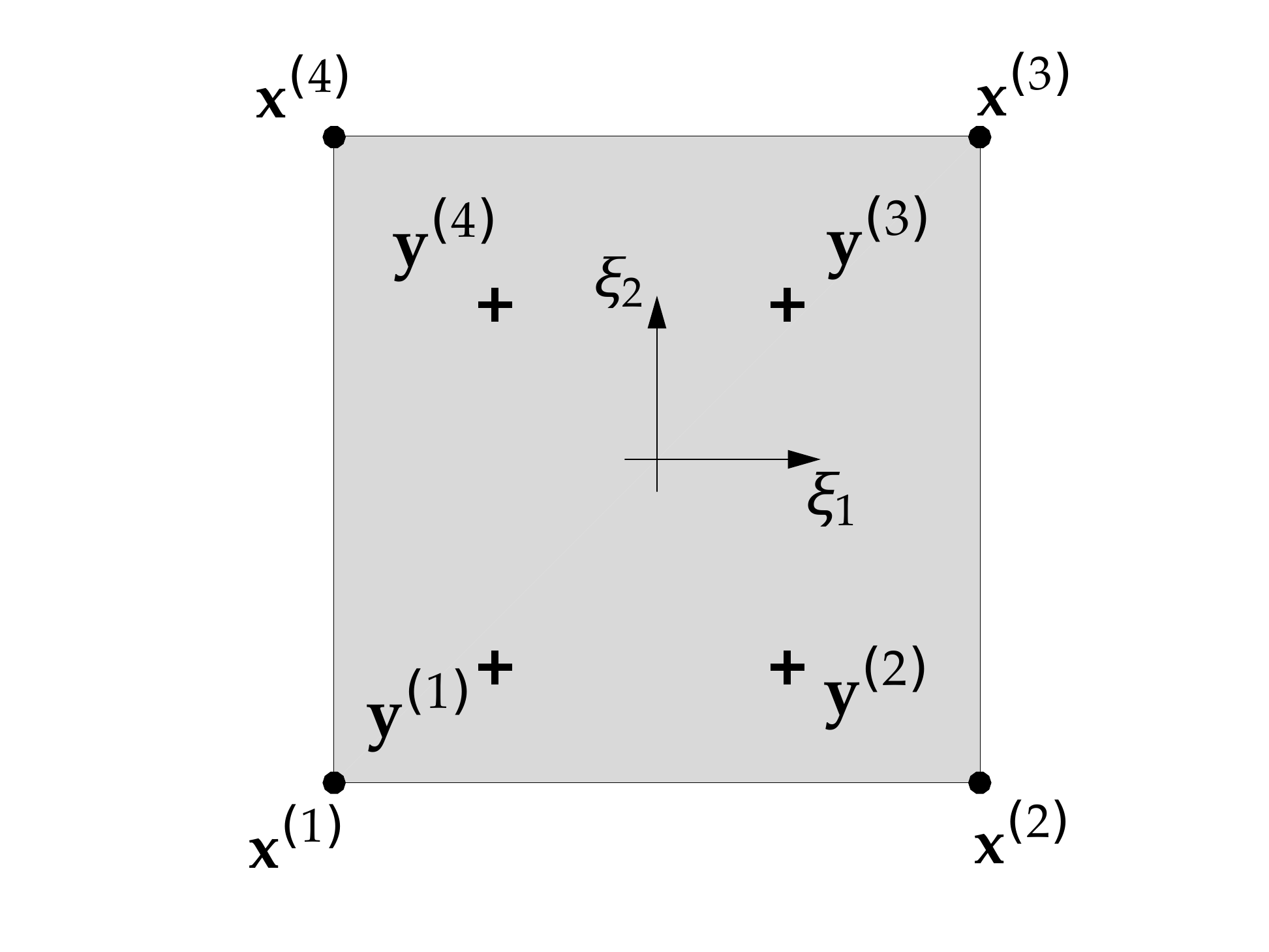}
	\caption{}
	\end{subfigure}
\caption{(\emph{a}) discontinuous triangle; (\emph{b}) discontinuous quadrangle. In the figures, $\mathbf{x}^{(\alpha)}$, $\alpha=1,\dots,\mu$ are the geometrical nodes whereas $\mathbf{y}^{(\beta)}$, $\beta=1,\dots,\nu$ are the collocation nodes.}
\label{fig-Ch3:disc elements}
\end{figure}

\subsubsection{Continuous/semi-discontinuous triangles}
For continuous/semi-discontinuous triangular elements, the number of geometrical nodes $\mu$ as well as the number of collocation nodes $\nu$ is 3; the classical local triangular reference system is centred at $\mathbf{x}^{(1)}$, as shown in Figure (\ref{fig-Ch3:disc elements}a). The position of the collocation nodes $\mathbf{y}^{(\beta)}$ and the expressions of the shape functions $N^{(\beta)}(\boldsymbol{\xi},\mathbf{d})$ are given by \emph{two} discontinuity parameters, i.e.\ $\mathbf{d}\equiv\{\lambda_1,\lambda_2\}$. Referring to Figure (\ref{fig-Ch3:disc elements}a), the local coordinates $\boldsymbol{\xi}^{(\beta)}$ of the collocation nodes are given by
\begin{subequations}\label{eq-Ch3:ds3ptri fnodes pos}
\begin{align}
\boldsymbol{\xi}^{(1)}&=\{\lambda_2,\lambda_2\}\\
\boldsymbol{\xi}^{(2)}&=\{1-\lambda_1-\lambda_2,\lambda_2\}\\
\boldsymbol{\xi}^{(3)}&=\{\lambda_2,1-\lambda_1-\lambda_2\}.
\end{align}
\end{subequations}
The geometrical position of the collocation node in the global reference system is then obtained as $\mathbf{y}^{(\beta)}=\sum_\alpha M^{(\alpha)}[\boldsymbol{\xi}^{(\beta)}]\mathbf{x}^{(\alpha)}$. From Eqs.\ (\ref{eq-Ch3:ds3ptri fnodes pos}), it is clear that, if $\lambda_1=\lambda_2=0$, then the collocation nodes coincide with the geometrical nodes, i.e.\ $\mathbf{y}^{(\beta)}\equiv \mathbf{x}^{(\beta)}$, $\beta=1,2,3$, and the element is continuous.

The shape functions $N^{(\beta)}(\boldsymbol{\xi},\mathbf{d})$ of a continuous/semi-discontinuous triangular element can be expressed as follows
\begin{subequations}\label{eq-Ch3:ds3ptri shape fncs}
\begin{align}
N^{(1)}(\boldsymbol{\xi},\mathbf{d})&=\frac{1}{\Delta}(1-\lambda_1-\xi_1-\xi_2)\\
N^{(2)}(\boldsymbol{\xi},\mathbf{d})&=\frac{1}{\Delta}(\xi_1-\lambda_2)\\
N^{(3)}(\boldsymbol{\xi},\mathbf{d})&=\frac{1}{\Delta}(\xi_2-\lambda_2)
\end{align}
\end{subequations}
where $\Delta=1-\lambda_1-2\lambda_2$. It is straightforward to see that the shape functions defined in Eqs.\ (\ref{eq-Ch3:ds3ptri shape fncs}) satisfy the condition that $N^{(\beta)}(\boldsymbol{\xi},\mathbf{d})$ has unit value at $\boldsymbol{\xi}=\boldsymbol{\xi}^{(\beta)}$ and that $\sum_{\beta}N^{(\beta)}(\boldsymbol{\xi},\mathbf{d})=1$.

\subsubsection{Continuous/semi-discontinuous quadrangles}
For continuous/semi-discontinuous quadrangular elements, the number of geometrical nodes $\mu$ as well as the number of collocation nodes $\nu$ is 4; the local reference system is centred at $\boldsymbol{\xi}=\{0,0\}$, as shown in Figure (\ref{fig-Ch3:disc elements}b). The position of the collocation nodes $\mathbf{y}^{(\beta)}$ and the expressions of the shape functions $N^{(\beta)}(\boldsymbol{\xi},\mathbf{d})$ are given by \emph{four} discontinuity parameters, i.e.\ $\mathbf{d}\equiv\{\lambda_1,\lambda_2,\kappa_1,\kappa_2\}$. Referring to Figure (\ref{fig-Ch3:disc elements}b), the local coordinates $\boldsymbol{\xi}^{(\beta)}$ of the collocation nodes are given by
\begin{subequations}\label{eq-Ch3:ds4pquad fnodes pos}
\begin{align}
\boldsymbol{\xi}^{(1)}&=\{-1+\lambda_1,-1+\kappa_1\}\\
\boldsymbol{\xi}^{(2)}&=\{+1-\lambda_2,-1+\kappa_1\}\\
\boldsymbol{\xi}^{(3)}&=\{+1-\lambda_2,+1-\kappa_2\}\\
\boldsymbol{\xi}^{(4)}&=\{-1+\lambda_1,+1-\kappa_2\}
\end{align}
\end{subequations}
The geometrical position of the collocation node in the global reference system is then obtained as $\mathbf{y}^{(\beta)}=\sum_\alpha M^{(\alpha)}[\boldsymbol{\xi}^{(\beta)}]\mathbf{x}^{(\alpha)}$. From Eqs.\ (\ref{eq-Ch3:ds4pquad fnodes pos}), it is clear that, if $\lambda_1=\lambda_2=\kappa_1=\kappa_2=0$, then the collocation nodes coincide with the geometrical nodes, i.e.\ $\mathbf{y}^{(\beta)}\equiv \mathbf{x}^{(\beta)}$, $\beta=1,2,3,4$, and the element is continuous.

The shape functions $N^{(\beta)}(\boldsymbol{\xi},\mathbf{d})$ of a continuous/semi-discontinuous quadrangular element can be expressed as follows
\begin{subequations}\label{eq-Ch3:ds4pquad shape fncs}
\begin{align}
N^{(1)}(\boldsymbol{\xi},\mathbf{d})&=\frac{1}{\Delta}(1-\lambda_2-\xi_1)(1-\kappa_2-\xi_2)\\
N^{(2)}(\boldsymbol{\xi},\mathbf{d})&=\frac{1}{\Delta}(1-\lambda_1+\xi_1)(1-\kappa_2-\xi_2)\\
N^{(3)}(\boldsymbol{\xi},\mathbf{d})&=\frac{1}{\Delta}(1-\lambda_1+\xi_1)(1-\kappa_1+\xi_2)\\
N^{(4)}(\boldsymbol{\xi},\mathbf{d})&=\frac{1}{\Delta}(1-\lambda_2-\xi_1)(1-\kappa_1+\xi_2)
\end{align}
\end{subequations}
where $\Delta=(2-\lambda_1-\lambda_2)(2-\kappa_1-\kappa_2)$. Similarly to the triangular element, the shape functions defined in Eqs.\ (\ref{eq-Ch3:ds4pquad shape fncs}) satisfy the condition that $N^{(\beta)}(\boldsymbol{\xi},\mathbf{d})$ has unit value at $\boldsymbol{\xi}=\boldsymbol{\xi}^{(\beta)}$ and that $\sum_{\beta}N^{(\beta)}(\boldsymbol{\xi},\mathbf{d})=1$.

\clearpage

\section{Computational tests}\label{sec-Ch3: computational tests}
In this Section, the performance and accuracy of the developed framework are assessed: the overall computational savings in terms of DoFs count, the computational homogenisation accuracy and the performance of the micro-cracking algorithm are considered. The homogenisation and micro-cracking computations have been performed on single 8-core processors of the Galileo super-computing facility of the CINECA Super Computing Applications and Innovation Department in Bologna, Italy (\url{http://www.hpc.cineca.it/hardware/galileo}).

\subsection{Overall computational savings}\label{ssec-Ch3: computational savings}
The computational savings obtained with the proposed methodology are presented and discussed in terms of overall number of DoFs. First, the effect of the adoption of the continuous and semi-discontinuous mesh elements is discussed, as it is valid for both non-periodic and periodic morphologies. Then, the number of DoFs of the non-prismatic periodic morphologies and cubic periodic morphologies are compared.

Figure (\ref{fig-Ch3:mesh dof red}a) shows the number of DoFs, obtained using the different meshing approaches discussed in Section (\ref{sec-Ch3:meshing}), as a function of the number of tessellation grains $N_g$. The represented data is the average value over $100$ different realisations for each selected number of grains. The dashed lines with smaller markers refer to non-regularised tessellations, whereas the solid lines with bigger markers refer to regularised tessellations. It can be noted that, in all cases, the overall number of DoFs varies linearly with respect to the number of grains, as the DoFs of each grain add linearly over the DoFs of the aggregate, but the adoption of the continuous/semi-discontinuous approach, with simultaneous use of quadrangular and triangular elements, guarantees the lowest rate after that of the totally continuous approach, which would however introduce complexities in terms of corner nodes in the formulation.

Considering, as an example, tessellations with $N_g=200$, it might be of interest to observe that the use of only discontinuous elements increases the number of DoFs of approximately 210$\%$ with respect to the use of continuous elements only, whereas the use of the continuous/semi-discontinuous mesh increases the number of DoFs of only 60$\%$ with respect to the continuous mesh, see Table (\ref{tab-Ch3:dof red}). The same savings ratios among different meshing schemes, for the considered number of grains, are maintained if regularised tessellations are considered, as shown in the second row of Table (\ref{tab-Ch3:dof red}). On the other hand, Table (\ref{tab-Ch3:dof red reg}) exemplifies the effect of the morphology regularisation on the number of DoFs for the different meshing approaches, showing that the regularisation algorithm provides in average an additional 10$\%$ reduction in the number of DoFs with respect to a non-regularised morphology for tessellations with $N_g=200$. 

Figure (\ref{fig-Ch3:mesh dof red}b) shows the average number of DoFs as a function of the mesh density parameter $d_m$, for morphologies with $N_g=200$ over $100$ realisations. For discontinuous meshes, the number of DoFs grows faster with $d_m$ with respect to semi-discontinuous and continuous meshes.

Figure (\ref{fig-Ch3:mesh dof red PvsNP}) shows the number of DoFs as a function of the number of grains $N_g$, for the periodic morphologies discussed in Section (\ref{sec-Ch3:meshing}). The non-prismatic periodic morphology allows to further reduce the number of DoFs in average from $45.3\%$ for $N_g=10$ to $27.8\%$ for $N_g=200$ with respect to the cubic periodic morphology. The fact that the higher the number of grains, the smaller the reduction is consistent with the fact less grains are affected by the finite size of the cube when $N_g$ increases.

Finally, as an example, considering $N_g=200$, it can be interesting to note that by combining the continuous/semi-discontinuous mesh, the regularisation and the non-prismatic morphology, the overall reduction in terms of DoFs is around $70\%$.

\begin{table}[H]
\small
\center
\caption{Average increase in the number of DoFs with respect to the (ideal) continuous mesh for the different analysed meshing schemes, for tessellations with $N_g=200$. The second row refers to regularised morphologies.}
\begin{tabular}{lcccc}
\hline
\hline
&disc.&semidisc.\ A&semidisc.\ B&cont./semidisc.\\
\hline \\ [-1.5ex]
$\frac{\mathrm{DoF}-\mathrm{DoF}_\mathrm{cont}}{\mathrm{DoF}_\mathrm{cont}}[\%]$&$216.8\%$&$201.8\%$&$111.2\%$&$59.9\%$\\
\\ [-1.5ex]
\hline \\ [-1.5ex]
$\left.\frac{\mathrm{DoF}-\mathrm{DoF}_\mathrm{cont}}{\mathrm{DoF}_\mathrm{cont}}\right|_\mathrm{reg}[\%]$&$211.1\%$&$200.6\%$&$110.7\%$&$58.7\%$\\
\\ [-1.5ex]
\hline
\hline
\end{tabular}
\label{tab-Ch3:dof red}
\end{table}

\begin{table}[H]
\small
\center
\caption{Average effect of the morphology regularisation on the number of DoFs for tessellations with $N_g=200$.}
\begin{tabular}{lcccc}
\hline
\hline
&disc.&semidisc.\ A&semidisc.\ B&cont./semidisc.\\
\hline\\ [-1.5ex]
$\frac{\mathrm{DoF}-\mathrm{DoF}_\mathrm{reg}}{\mathrm{DoF}}[\%]$&$11.8\%$&$10.5\%$&$10.4\%$&$9.7\%$\\
\\ [-1.5ex]
\hline
\hline
\end{tabular}
\label{tab-Ch3:dof red reg}
\end{table}

\begin{figure}[H]
\centering
	\begin{subfigure}{0.49\textwidth}
	\centering
	\includegraphics[width=\textwidth]{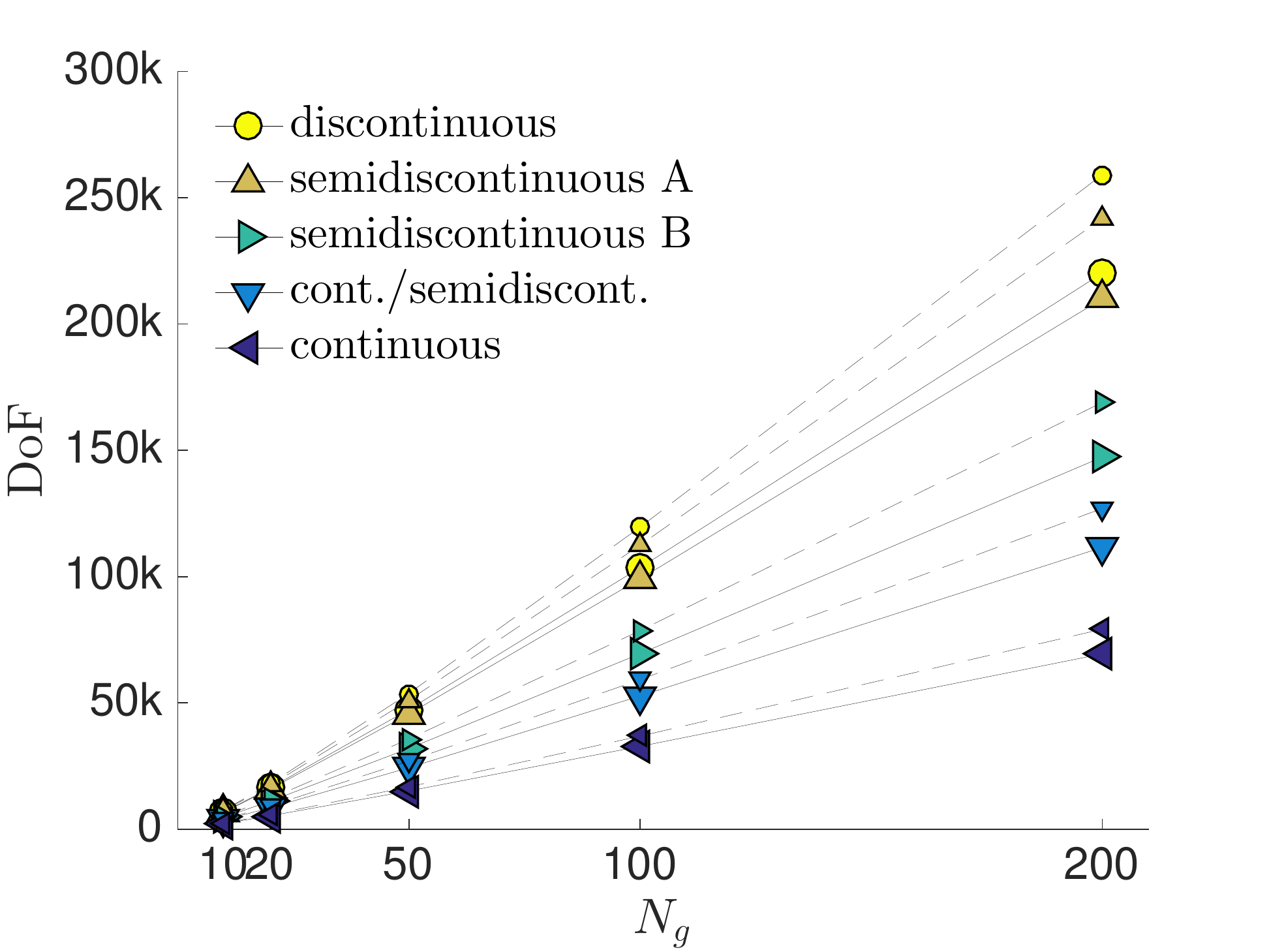}
	\caption{}
	\end{subfigure}
	\begin{subfigure}{0.49\textwidth}
	\centering
	\includegraphics[width=\textwidth]{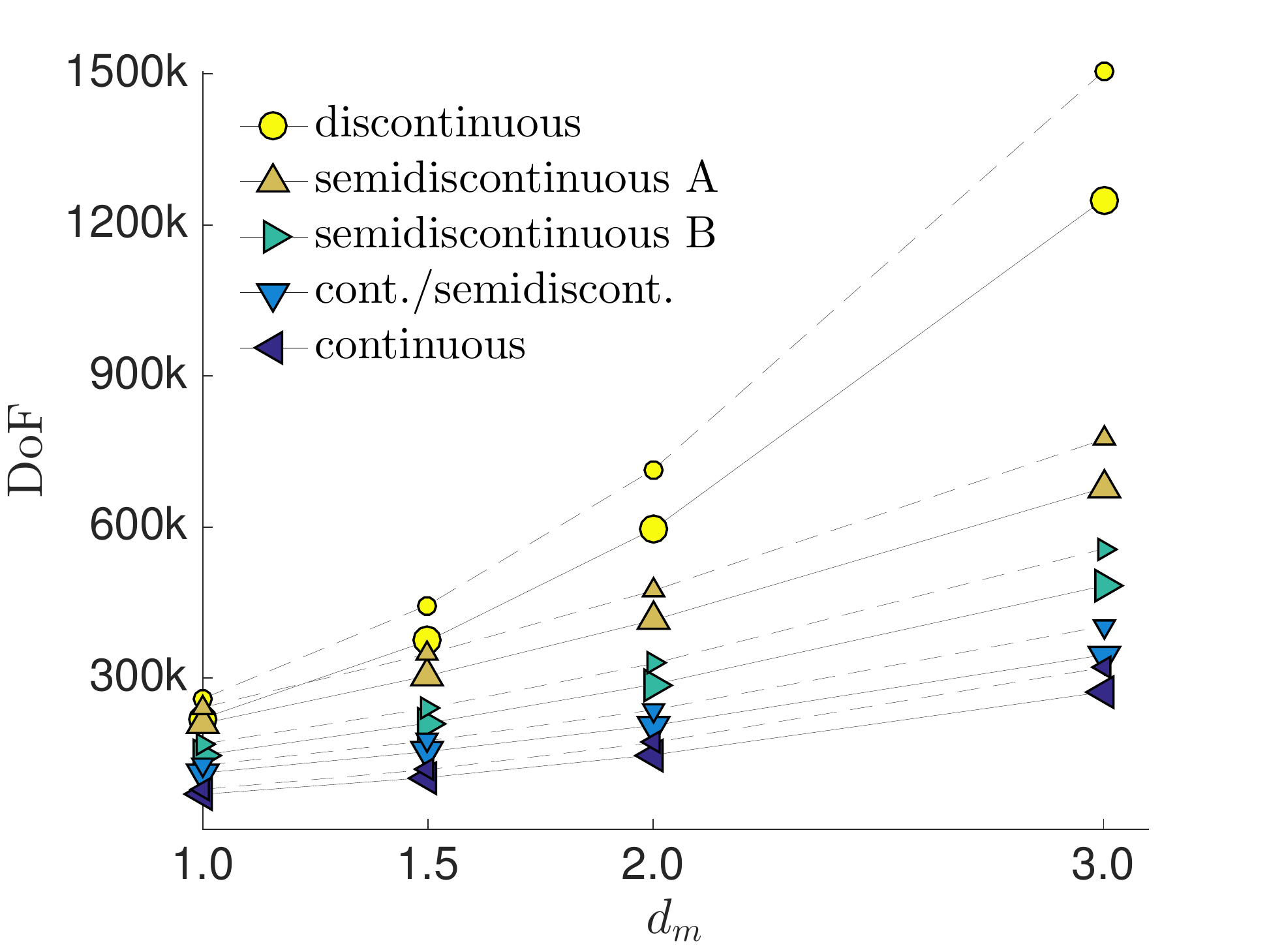}
	\caption{}
	\end{subfigure}
\caption{(\emph{a}) Comparison of the number of DoFs obtained using different meshing approaches. The dashed lines with smaller markers refer to non-regularised morphologies; the solid lines with bigger markers refer to regularised morphologies. Data collected on 100 realisations using $d_m=1.0$; (\emph{b}) Number of DoFs as a function of the mesh density parameter $d_m$. The dashed lines with smaller markers refer to non-regularised morphologies; the solid lines with bigger markers refer to regularised morphologies. Data collected on 100 realisations with $N_g=200$.}
\label{fig-Ch3:mesh dof red}
\end{figure}

\begin{figure}[H]
\centering
\includegraphics[width=0.5\textwidth]{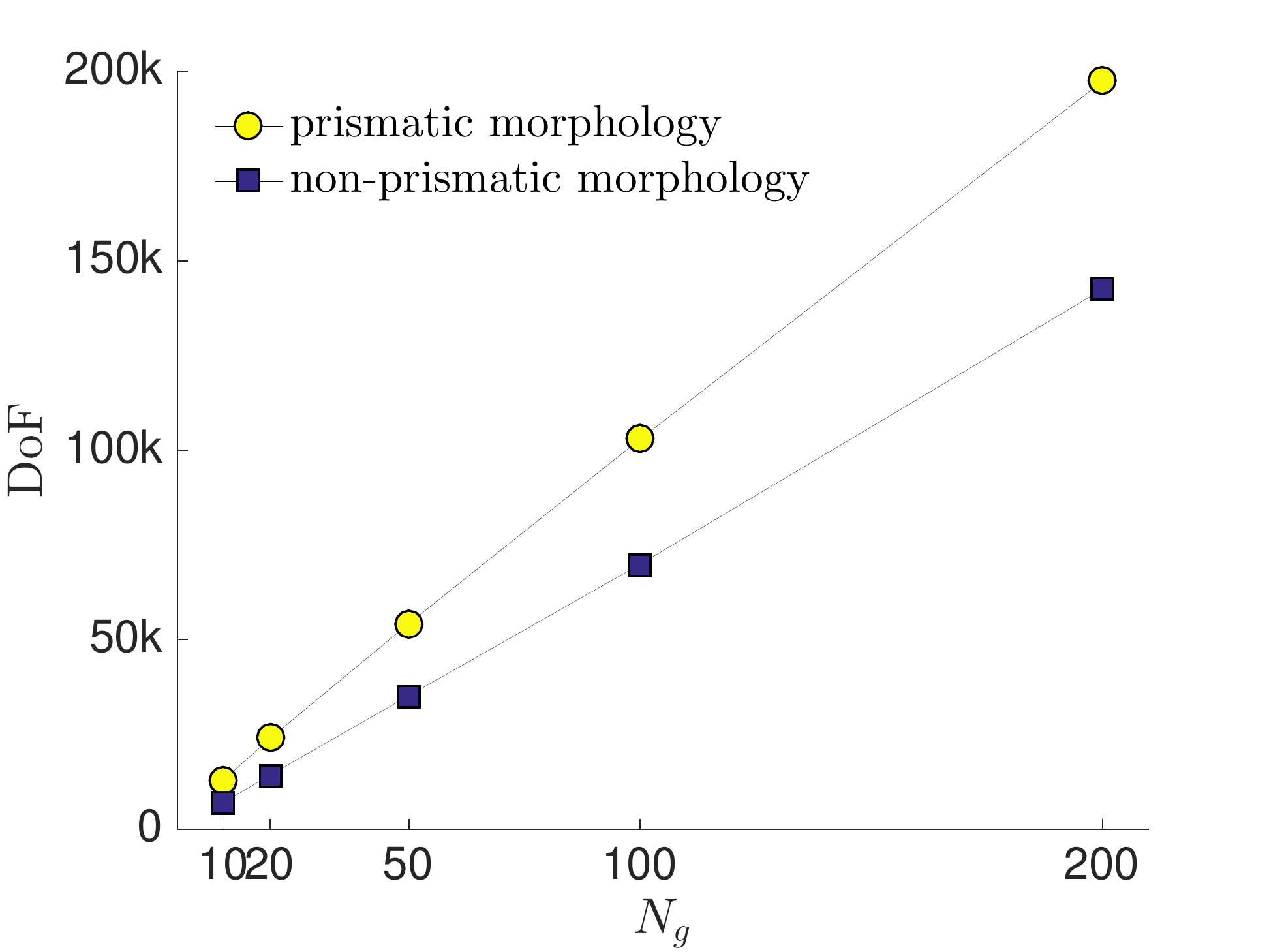}
\caption{Number of DoFs vs $N_g$ for non-prismatic and prismatic periodic morphologies using the continuous/semi-discontinuous mesh. Data collected over 100 realisations.}
\label{fig-Ch3:mesh dof red PvsNP}
\end{figure}

\clearpage

\subsection{Computational homogenisation}\label{ssec-Ch3: computational homogenisation}
In this Section, the developed model is employed to determine the elastic properties of polycrystalline aggregates, using the non-prismatic periodic morphologies presented in Section (\ref{sec-Ch3:modified morphology}) and the periodic BCs introduced in Section (\ref{ssec-Ch1:BCs}) of Chapter (\ref{ch-intro}). For these analyses, the grain boundaries are considered as perfect and displacements continuity and tractions equilibrium are enforced, i.e.\ the interface equations (\ref{eq-Ch1:interface equations}) of Section (\ref{ssec-Ch1:ICs}) are particularised as follows
\begin{subequations}\label{eq-Ch3:interface equations - pristine}
\begin{align}
&\Psi_i^{gh}=\wtilde{u}_i^g(\mathbf{x})+\wtilde{u}_i^h(\mathbf{x})=-\delta\wtilde{u}_i^{gh}(\mathbf{x})=0,\\
&\Theta_i^{gh}=\wtilde{t}_i^g(\mathbf{x})-\wtilde{t}_i^h(\mathbf{x})=0,
\end{align}
\end{subequations}
where it is recalled that the symbol $\wtilde{\cdot}$ denotes a quantity expressed in the boundary local reference system, $u_i$ and $t_i$ denote the grain boundary displacements and tractions, respectively, $\delta u_i$ denotes the displacements jumps and $g$ and $h$ are the two generic adjacent grains sharing an interface. Similarly, in order to account for the general orientation of the faces in the non-prismatic periodic morphology, the periodic boundary conditions given in Eq.(\ref{eq-Ch1:periodic boundary conditions}) are expressed in the local reference system of the periodic faces, i.e.\
\begin{subequations}\label{eq-Ch3:periodic boundary conditions}
\begin{align}
&\wtilde{u}_i(\mathbf{x}^s)+\wtilde{u}_i(\mathbf{x}^m)=R_{ki}(\mathbf{x}^s)\bar{\Gamma}_{ki}(x_j^s-x_j^m),\\
&\wtilde{t}_i(\mathbf{x}^s)-\wtilde{t}_i(\mathbf{x}^m)=0,
\end{align}
\end{subequations}
where the change of signs is obtained by choosing the rotation matrix $R_{ki}(\mathbf{x}^s)$ at the slave point as opposite to the rotation matrix $R_{ki}(\mathbf{x}^m)$ at the master point.

The macro-stress and macro-strain averages are calculated for the $\mu$RVE using the following boundary integrals
\begin{equation*}
\Sigma_{ij}=\frac{1}{V}\int_S t_i(\mathbf{x})x_j\dd S(\mathbf{x}), \qquad
\Gamma_{ij}=\frac{1}{2V}\int_S(u_i(\mathbf{x})n_j(\mathbf{x})+u_j(\mathbf{x})n_i(\mathbf{x}))\dd S(\mathbf{x}),
\end{equation*}
which eventually allow to compute the \emph{apparent} properties $C_{ijkl}$ entering the constitutive relationships $\Sigma_{ij}=C_{ijkl}\Gamma_{kl}$, see e.g.\ \cite{benedetti2013a}.

As first test, the isotropy of the macroscopic elastic properties of an aggregate of Copper (Cu) crystals is assessed. Similar tests, for the basic grain-boundary formulation, have been carried out in \cite{benedetti2013a}. However, it necessary to assess that the new regularisation and meshing strategies, and the non-prismatic morphology, do not introduce any artefact in the homogenisation procedure, for example altering the isotropy of the apparent constants $C_{ijkl}$. For this purpose, the complete apparent stiffness matrix $\mathbf{C}$, ($6\times6$, in Voigt notation) of a single 200-grain aggregate of Cu crystals has been computed obtaining:
\begin{equation*}
\mathbf{C} = \left[\begin{array}{cccccc}
200.1&105.9&102.1& -0.6& -4.9&  0.5\\
105.1&199.3&105.2&  0.6&  1.7&  2.4\\
102.9&105.0&201.1& -0.5&  0.3& -0.8\\
 -0.3&  0.5&  0.6& 49.1&  0.0&  0.3\\
 -1.9&  1.0&  0.4& -1.4& 46.4& -0.3\\
 -1.5&  2.8& -1.4&  1.0&  0.7& 49.6\\
\end{array}\right].
\end{equation*}
The matrix has been obtained from six linearly independent mechanical tests, conducted in displacement control (three independent tensile and three independent shear tests). As it appears, the aggregate, containing $200$ \textit{randomly oriented} crystals, tends to behave macroscopically as an isotropic material. It is underlined that only a single microstructural realisation has been considered, and increased accuracy is to be expected if ensemble averages over several microstructural realisations are carried out. However, the previous result confirms an almost isotropic behaviour and is completely satisfactory. For further details about the effect of the \emph{number of grains} and \emph{number of realisations} on the macroscopic averages, the interested reader is referred to Ref.\ \cite{benedetti2013a}.

After the previous assessment, other homogenisation tests are carried out and the results are presented in terms of effective Young's and shear moduli, computed as ensemble averages of the apparent properties over $100$ realisations, for aggregates with $N_g =\{10,\:20,\:50,\:100,\:200\}$ grains. In this study, crystalline Cu, crystalline SiC and a two-phase steel have been considered as test materials.
Figure (\ref{fig-Ch3:homoCu}) reports the apparent Young's and shear moduli versus $N_g$ for copper crystal aggregates. For comparison, also values obtained with kinematic boundary conditions \cite{benedetti2013a} are displayed. As expected, it is apparent that when periodic boundary conditions are enforced, the apparent properties approach the effective ones much faster with respect to those obtained using displacement boundary conditions. The error bars represent the standard deviation of the homogenisation results. The dashed line, corresponding to the reference effective property, has been taken as the mean value of the fifth order bounds reported in Ref.\ \cite{fritzen2009}.
In Figure (\ref{fig-Ch3:homoCu}), the theoretical bounds for the elastic properties of copper polycrystalline aggregates are reported \cite{berryman2005}. More specifically, the light grey region corresponds to the upper and lower 1st-order bounds, which are the Voigt and Reuss bounds, respectively; the dark grey region corresponds to the Hashin and Shtrikman bounds \cite{hashin1962a,hashin1962b}.

Figure (\ref{fig-Ch3:homoSiC}) shows the apparent Young's and shear moduli for polycrystalline SiC. The error bars represent the standard deviation of the homogenisation results, whereas the reference value of the effective macroscopic properties has been taken from Ref.\ \cite{benedetti2013b}. Also in this case, the theoretical bounds are reported. Again, the light grey region corresponds to the upper and lower 1st-order bounds, whereas the dark grey region corresponds to the Hashin and Shtrikman bounds.

Finally, Figure (\ref{fig-Ch3:homoSAF}) shows the apparent Young's and shear moduli for a two-phase steel alloy, namely two-phase stainless steel SAF $2507$. The properties of the micro-constituents have been taken from Ref.\ \cite{jia2008}, and consist of $58\%$ austenite ($\gamma$) and $42\%$ ferrite ($\alpha$) in volume. The figures show the apparent properties of ferrite aggregates, of austenite aggregates, and of the steel alloy. The theoretical bounds for the alloy constituents are also reported. In this case, only the Hashin and Shtrikman bounds are reported. It can be noted that the alloy's apparent properties approach the known values for SAF $2507$. The slight misalignment of the curves has been intentionally added for the sake of clarity in the figure.

\begin{figure}[ht]
\centering
	\begin{subfigure}{0.49\textwidth}
	\centering
	\includegraphics[width=\textwidth]{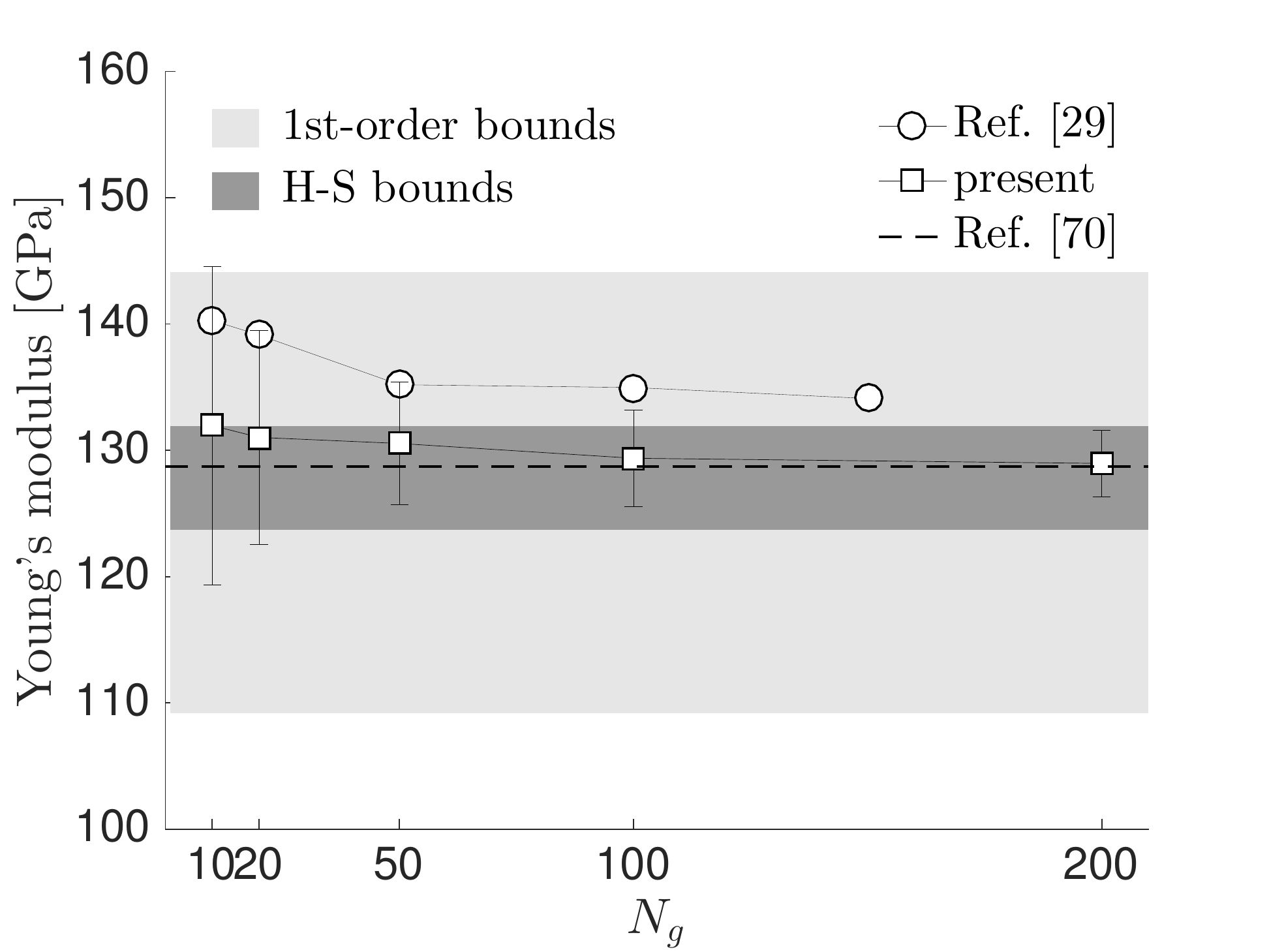}
	\caption{}
	\end{subfigure}
	\
	\begin{subfigure}{0.49\textwidth}
	\centering
	\includegraphics[width=\textwidth]{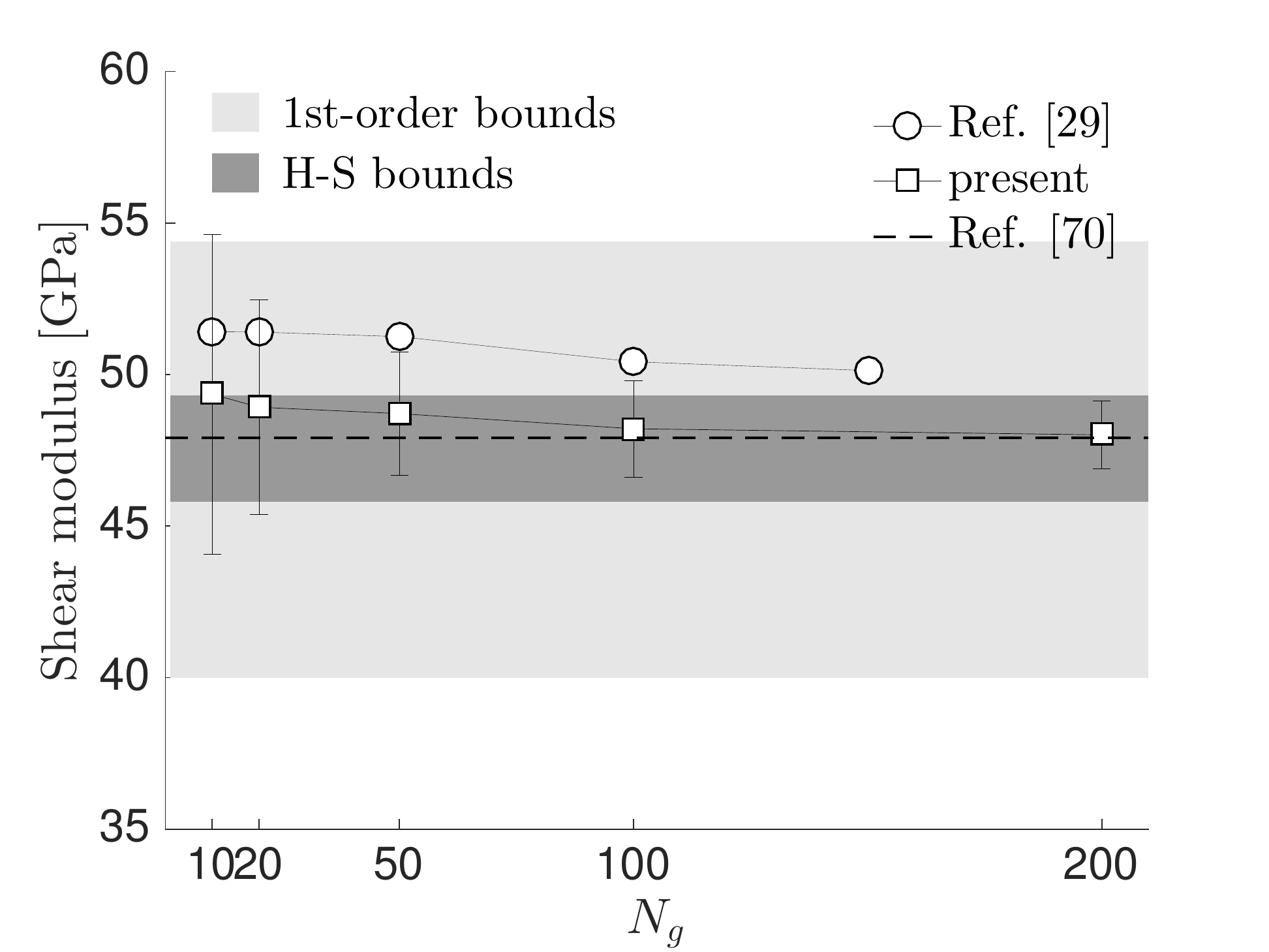}
	\caption{}
	\end{subfigure}
\caption{Apparent (\emph{a}) Young's and (\emph{b}) shear moduli for polycrystalline copper. Data collected on $100$ realisations. The light grey region represents the 1st order bounds, whereas the dark grey region represents the Hashin and Shtrikman bounds.}
\label{fig-Ch3:homoCu}
\end{figure}

\begin{figure}[ht]
\centering
	\begin{subfigure}{0.49\textwidth}
	\centering
	\includegraphics[width=\textwidth]{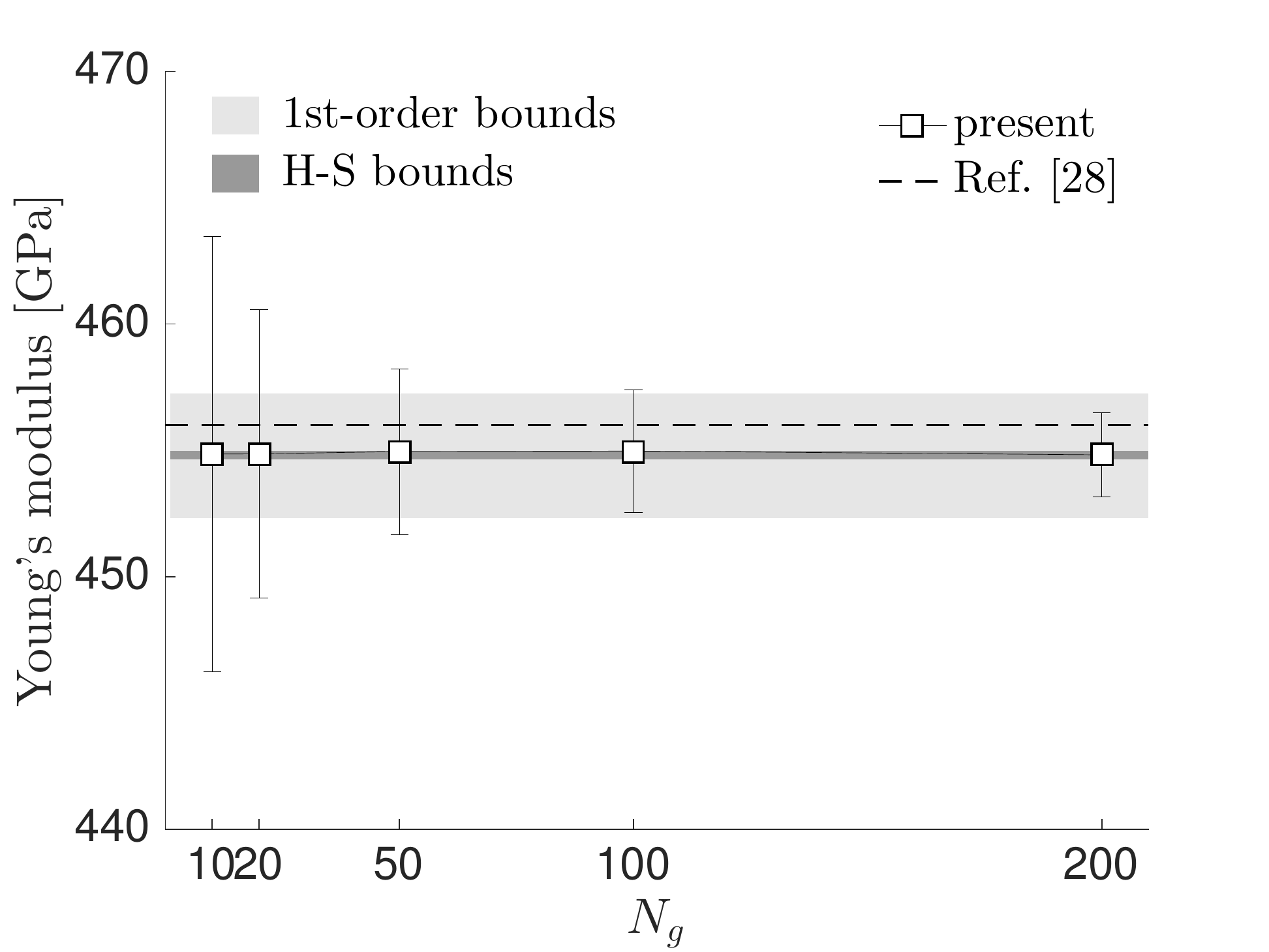}
	\caption{}
	\end{subfigure}
	\
	\begin{subfigure}{0.49\textwidth}
	\centering
	\includegraphics[width=\textwidth]{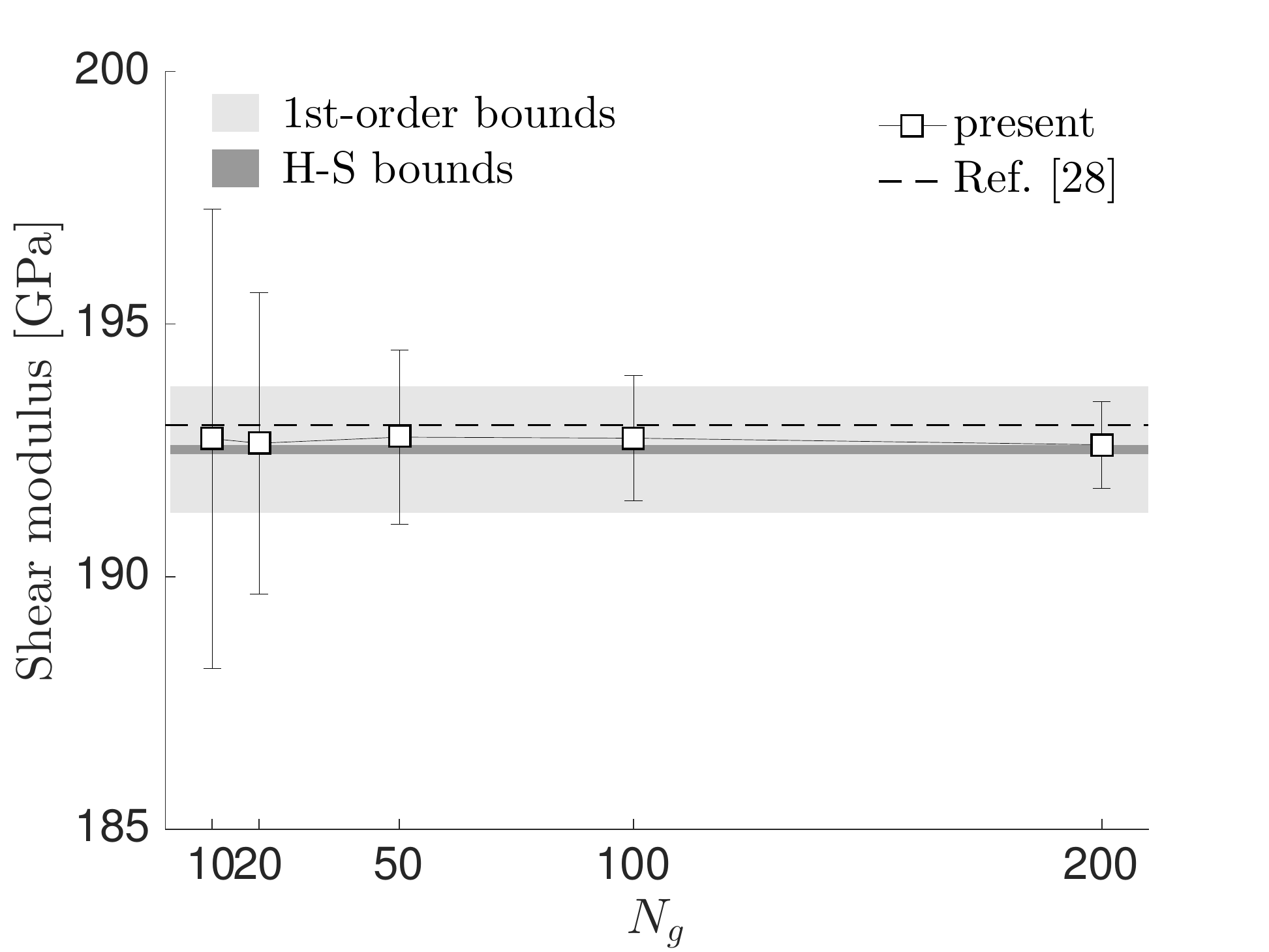}
	\caption{}
	\end{subfigure}
\caption{Apparent (\emph{a}) Young's and (\emph{b}) shear moduli for polycrystalline SiC. Data collected on $100$ realisations. The light grey region represents the 1st order bounds, whereas the dark grey region represents the Hashin and Shtrikman bounds.}
\label{fig-Ch3:homoSiC}
\end{figure}

\begin{figure}[ht]
\centering
	\begin{subfigure}{0.49\textwidth}
	\centering
	\includegraphics[width=\textwidth]{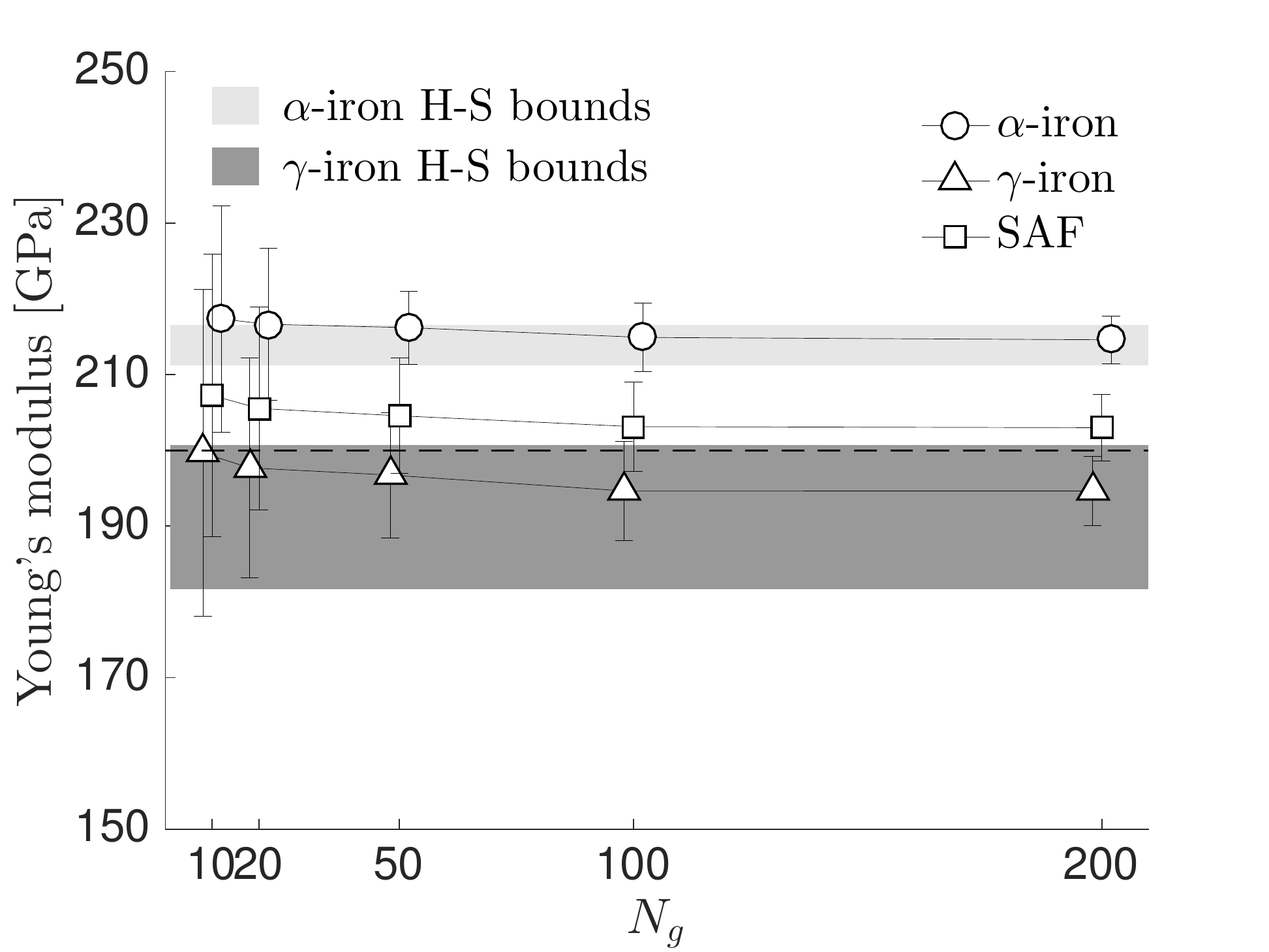}
	\caption{}
	\end{subfigure}
	\
	\begin{subfigure}{0.49\textwidth}
	\centering
	\includegraphics[width=\textwidth]{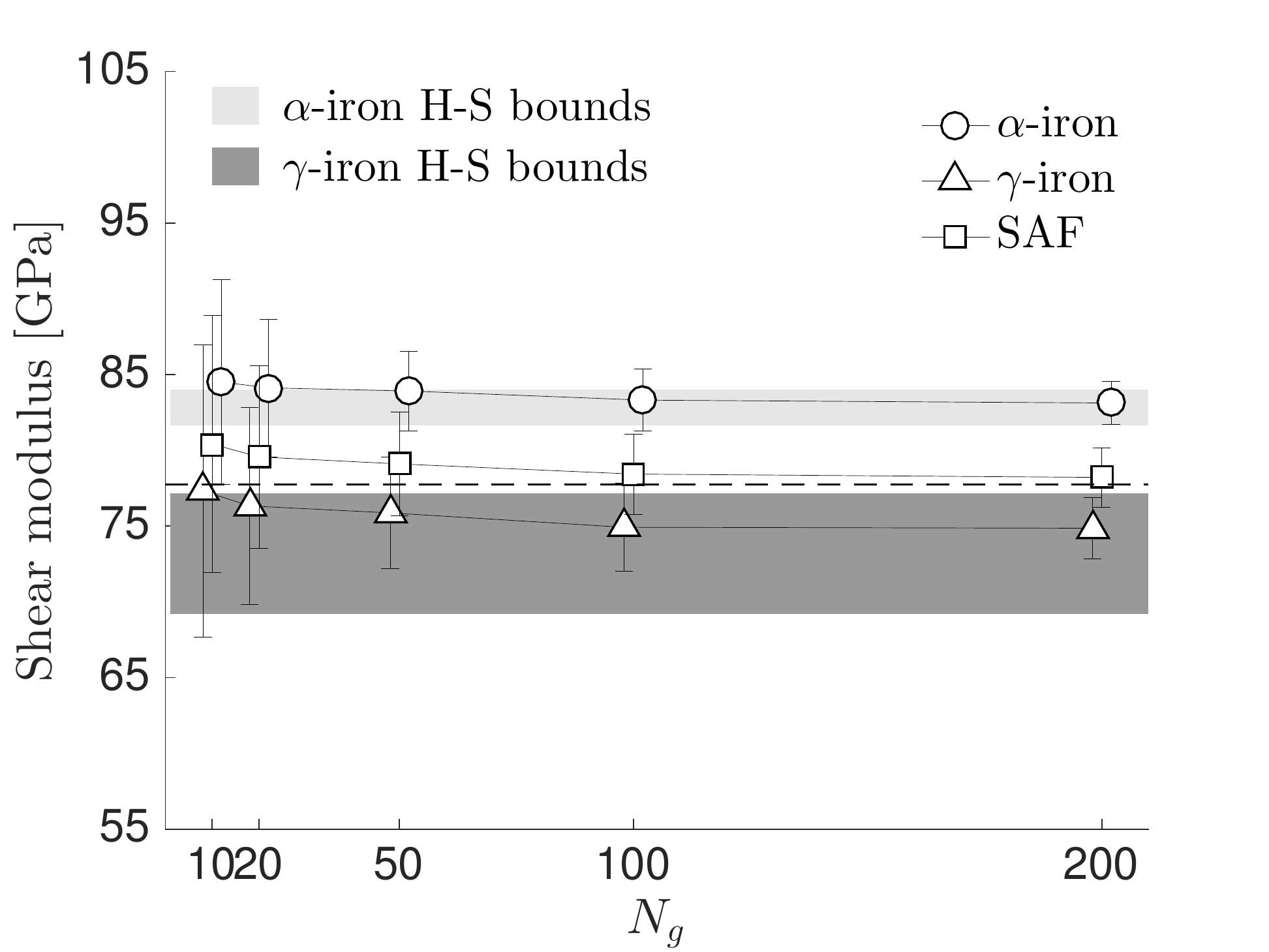}
	\caption{}
	\end{subfigure}
\caption{Apparent (\emph{a}) Young's and (\emph{b}) shear moduli for polycrystalline aggregates of a two-phase steel. Data collected on $100$ realisations. The grey regions represent the Hashin and Shtrikman bounds.}
\label{fig-Ch3:homoSAF}
\end{figure}

\clearpage

\subsection{Interface modelling for micro-cracking}\label{ssec-Ch3: cohesive law}
Prior to presenting the results of micro-cracking simulations of polycrystalline aggregates, the explicit expression of the interface equations introduced in Section (\ref{ssec-Ch1:ICs}) and herein used for modelling polycrystalline micro-cracking is recalled for completeness. The ICs given in Eqs.(\ref{eq-Ch1:interface equations}) are particularised to be the cohesive laws used in Ref.\ \cite{benedetti2013b} for modeling polycrystalline inter-granular micro-cracking.

As detailed in \cite{benedetti2013b}, during the loading history the generic grain boundary interface $\mathscr{I}^{gh}$, shared by the grains $g$ and $h$, undergoes different states, which are subsequently modelled using the appropriate interface equations. At the beginning of the analysis all the interfaces are assumed to be in a \emph{pristine state} and, likewise the homogenisation analyses, displacements continuity and tractions equilibrium are enforced as in Eqs.(\ref{eq-Ch3:interface equations - pristine}).

The pristine state of $\mathscr{I}^{gh}$ holds as long as the following inter-granular threshold condition
\begin{equation}\label{eq-Ch3:te less than Tmax}
\tau_e < T_\mathrm{max}
\end{equation}
is satisfied. In Eq.(\ref{eq-Ch3:te less than Tmax}), $T_\mathrm{max}$ is the \emph{interface cohesive strength} and $\tau_e$ is an \emph{effective} traction expressed as a function of the local grain boundary traction $\wtilde{t}_i^{gh}$ as follows
\begin{equation}\label{eq-Ch3:effective traction}
\tau_e = \sqrt{\langle\tau_n\rangle^2+\left(\frac{\beta}{\alpha}\tau_s\right)^2},
\end{equation}
where $\tau_n$ and $\tau_s$ are the \emph{normal} and \emph{tangential} tractions components defined as $\tau_n=\wtilde{t}_3^{gh}$ and $\tau_s=\sqrt{(\wtilde{t}_1^{gh})^2+(\wtilde{t}_2^{gh})^2}$, respectively, $\alpha$ and $\beta$ are suitable constants weighing the sliding and opening failure modes of the interface and $\langle\cdot\rangle$ are the Macaulay brackets, defined as $\langle \tau\rangle=\max(0,\tau)$. It is recalled that $\wtilde{t}_i^{gh}$ denotes the tractions at the interface $\mathscr{I}^{gh}$, which, due to the equilibrium conditions at the interface (see Eq.(\ref{eq-Ch1:local t})), is equal to the traction $\wtilde{t}_i^g$ or the traction $\wtilde{t}_i^h$, i.e.\ $\wtilde{t}_i^{gh}=\wtilde{t}_i^g=\wtilde{t}_i^{h}$.

As soon as the effective traction overcomes the interface strength, the interface enters the \emph{damage state} and micro-cracking initiation and evolution is modelled by replacing the displacement continuity equations (\ref{eq-Ch3:interface equations - pristine}a) with a cohesive law embodying suitably defined parameters representing the initiation and evolution of damage. The cohesive law can be derived from the definition of a potential \cite{ortiz1999} or its expression can assumed \emph{a priori}. The reader is referred to Ref.\cite{benedetti2013b} for the derivation of the cohesive law used in the present thesis. Here, the following \emph{irreversible extrinsic cohesive traction-separation} law is introduced in place of Eq.(\ref{eq-Ch3:interface equations - pristine}a)
\begin{subequations}\label{eq-Ch3:interface equations - cohesive law}
\begin{align}
&\Psi_i^{gh}=C_s(d^*)\delta\wtilde{u}_i^{gh}-\wtilde{t}_i^{gh}=0,\\
&\Psi_3^{gh}=C_n(d^*)\delta\wtilde{u}_3^{gh}-\wtilde{t}_3^{gh}=0,
\end{align}
\end{subequations}
where Eq.(\ref{eq-Ch3:interface equations - cohesive law}a) is valid for $i=1,2$. The term \emph{irreversible} refers to the fact the damage of the interface is allowed only to accumulate during the loading history and is represented by the monotonically increasing parameter $d^*$; the term \emph{extrinsic} refers to the fact that such cohesive law is introduced only after the threshold condition ceases to hold; finally, the term \emph{traction-separation} refers to the fact the the cohesive law expresses a relation between the tractions $\wtilde{t}_i^{gh}$ and the displacements jump $\delta\wtilde{u}_i^{gh}$ at the interface between the grain $g$ and $h$.

In Eqs.(\ref{eq-Ch3:interface equations - cohesive law}), $C_s$ and $C_n$ are the constitutive constants of the cohesive law given as
\begin{subequations}\label{eq-Ch3:interface equations - cohesive law constants}
\begin{align}
&C_s(d^*)=\frac{\alpha T_\mathrm{max}}{\delta u_s^{cr}}\frac{1-d^*}{d^*},\\
&C_n(d^*)=\frac{T_\mathrm{max}}{\delta u_n^{cr}}\frac{1-d^*}{d^*},
\end{align}
\end{subequations}
where $\delta u_s^{cr}$ and $\delta u_n^{cr}$ denote the critical values of the normal and sliding displacement jumps at which the interface fails in the case of pure opening mode or pure sliding mode, respectively. In Eqs.(\ref{eq-Ch3:interface equations - cohesive law}) and (\ref{eq-Ch3:interface equations - cohesive law constants}), $d^*$ denotes the irreversible damage parameter, which is given by the maximum value reached by an \emph{effective opening} $d$ during the loading history $\mathscr{H}_d$, i.e.
\begin{equation}\label{eq-Ch3:irreversible damage}
d^*=\max_{\mathscr{H}_d}d\in[0,1],
\end{equation}
where $d$ is defined in terms of the local displacements jumps $\delta\wtilde{u}_i^{gh}$ at the interface $\mathscr{I}^{gh}$ as
\begin{equation}\label{eq-Ch3:effective opening}
d = \sqrt{\left\langle\frac{\delta u_n}{\delta u_n^{cr}}\right\rangle^2+\left(\beta\frac{\delta u_s}{\delta u_s^{cr}}\right)^2}
\end{equation}
being $\delta u_s=\sqrt{(\delta\wtilde{u}_1^{gh})^2+(\delta\wtilde{u}_2^{gh})^2}$ and $\delta u_n=\delta\wtilde{u}_3^{gh}$ the tangential and normal components of the displacements jump, respectively.

It is worth noting that Eqs.(\ref{eq-Ch3:interface equations - cohesive law}), in the case of very low levels of damage $d^*$, still provide consistent interface conditions; in fact, it is straightforward to see that, by multiplying $\Psi_i^{gh}$ by $d^*\approx0$, Eqs.(\ref{eq-Ch3:interface equations - cohesive law}) corresponds to enforce $\delta \wtilde{u}_i^{gh}=0$, i.e.\ pristine interface conditions. Furthermore, in case the two grains $g$ and $h$ come in contact, the continuity of the displacements along the local normal direction is enforced, i.e.\ Eq.(\ref{eq-Ch3:interface equations - cohesive law}b) is replaced by $\delta\wtilde{u}_3^{gh}=0$.

Given the definition of $\tau_s$ and $\tau_n$, it is easy to see that Eqs.(\ref{eq-Ch3:interface equations - cohesive law}) can also be written in terms of the tangential and normal components as
\begin{subequations}\label{eq-Ch3:interface equations - cohesive law 2}
\begin{align}
&\tau_s=C_s(d^*)\delta{u}_s,\\
&\tau_n=C_n(d^*)\delta{u}_n,
\end{align}
\end{subequations}
and, therefore, it is possible to show that the work of separation $G_I$ in case of pure normal mode and the work of separation $G_{II}$ in pure sliding mode can be written as
\begin{subequations}\label{eq-Ch3:interface equations - work of separation}
\begin{align}
&G_I=\int_{0}^{\delta u_n^{cr}}\tau_n\dd \delta u_n = \frac{1}{2}T_\mathrm{max}\delta u_n^{cr},\\
&G_{II}=\int_{0}^{\delta u_s^{cr}}\tau_s\dd \delta u_s = \frac{\alpha}{2\beta^2}T_\mathrm{max}\delta u_s^{cr},
\end{align}
\end{subequations}
respectively. Eqs.(\ref{eq-Ch3:interface equations - work of separation}) will be used to compute the critical displacement $\delta u_s^{cr}$ in sliding mode in terms of the interface strength $T_\mathrm{max}$, the critical displacement $\delta u_n^{cr}$ in normal mode, the parameters $\alpha$ and $\beta$ and the ratio $G_{II}/G_I$.

Eventually, the interface enters the \emph{failed state} when $d^*$ reaches the critical values $d^*=1$. In such a case, the cohesive law given in Eqs.(\ref{eq-Ch3:interface equations - cohesive law}) is replaced by the condition of zero tractions, i.e.\ by the equations $\wtilde{t}_i^{gh}=0$, $i=1,2,3$. However, also in this case, if the two grains come in contact, $\delta\wtilde{u}_3^{gh}=0$ is enforced.

For a clearer representation of cohesive law used in this thesis, Figures (\ref{fig-Ch3:cohesive law}a) and (\ref{fig-Ch3:cohesive law}b) show the behaviour of the tangential and normal traction-separation law given in Eqs.(\ref{eq-Ch3:interface equations - cohesive law 2}a) and (\ref{eq-Ch3:interface equations - cohesive law 2}b), respectively. In the figures, the shaded region of the plane $(\beta\delta u_s/\delta u_s^{cr})$-$(\delta u_n/\delta u_n^{cr})$ corresponds to the region $d\ge 1$, i.e.\ the region where the interface is in the failed stated. For further details about the derivation of the cohesive law described in this Section and used in this thesis, the reader is referred to Ref.\cite{benedetti2013b} and the references therein.

\begin{figure}
\centering
	\begin{subfigure}{0.49\textwidth}
	\centering
	\includegraphics[width=\textwidth]{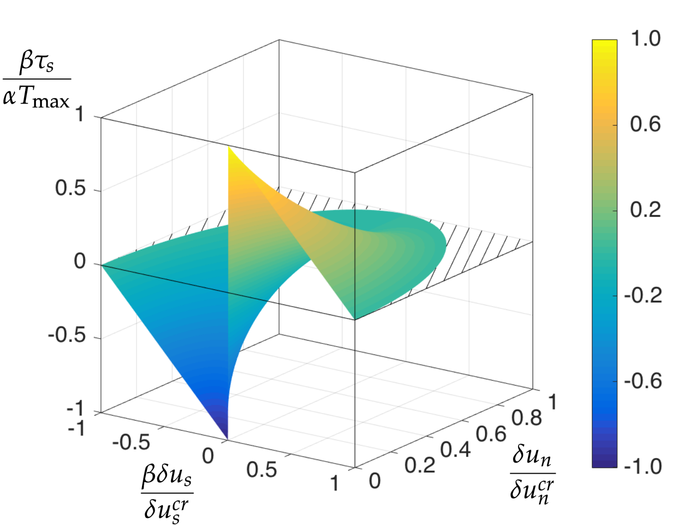}
	\caption{}
	\end{subfigure}
	\
	\begin{subfigure}{0.49\textwidth}
	\centering
	\includegraphics[width=\textwidth]{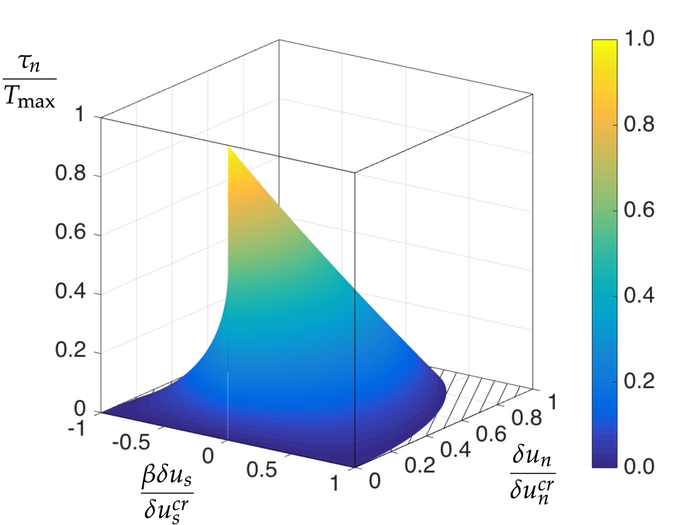}
	\caption{}
	\end{subfigure}
\caption{Schematic representation of the (\emph{a}) tangential component $\tau_s$ and (\emph{b}) normal component $\tau_n$ of the tractions of the cohesive law used in this thesis given in Eqs.(\ref{eq-Ch3:interface equations - cohesive law 2}a) and (\ref{eq-Ch3:interface equations - cohesive law 2}b), respectively.}
\label{fig-Ch3:cohesive law}
\end{figure}

\subsection{Microcracking simulations}\label{ssec-Ch3: micro-cracking}
In this Section, the results of some micro-cracking simulations for SiC aggregates are reported. The SiC grains bulk elastic constants are taken from Ref.\ \cite{benedetti2013b}. The work of separation $G_I$ is estimated by the macroscopic values of the fracture toughness $K_{IC}=3$ $\mathrm{MPa}\sqrt{\mathrm{m}}$ in mode I and in the case of plain strain, of the Young modulus $E=448$ GPa and of the Poisson's ratio $\nu=0.168$ of SiC by means of the relation
\begin{equation}
G_I=\frac{1-\nu^2}{E}K_{IC}^2.
\end{equation}
The inter-granular cohesive properties are assumed to be: $T_\mathrm{max}=500$ MPa; $G_I/G_{II}=1$; $\alpha=1$; $\beta=\sqrt{2}$. The critical displacements $\delta u_n^{cr}$ and $\delta u_s^{cr}$ are then computed using Eqs.(\ref{eq-Ch3:interface equations - work of separation}a) and (\ref{eq-Ch3:interface equations - work of separation}b), respectively.

Differently from what it has been done in the homogenisation tests, regularised \emph{prismatic} and \emph{non-periodic} microstructures are used in the cracking computations. This is due to the circumstance that the simple implementation of Eqs.(\ref{eq-Ch3:periodic boundary conditions}), as periodic boundary conditions for a \emph{non-prismatic} unit cell, without suitable modifications to take into account the presence of possible displacement jumps at the RVE's external boundary, would over-constrain the micro-cracking, preventing the cracks from reaching the external walls of the unit cell. To avoid this, for simplicity, prismatic unit cells subjected to non-periodic kinematical boundary conditions are used here. This is not a limitation of the formulation: there is no problem in using periodic boundary conditions in a micro-cracking simulation, as shown in Ref.\ \cite{benedetti2015}. The only difficulty arises if non-prismatic RVEs are used without suitably modifying Eqs.(\ref{eq-Ch3:periodic boundary conditions}). On the other hand, general kinematic or static boundary conditions can be enforced on the non-prismatic cells. The results presented here serve thus the purpose of testing the use of \emph{regularised} morphologies in conjunction with \emph{optimised} meshes within the scope of microcracking simulations: it has been actually found that the introduction of such enhancements has a remarkable effect in cutting down the microcracking simulation time to several \emph{hours}, for morphologies with few hundred grains, which represents an outstanding gain with respect to the computation times previously reported in Ref.\ \cite{benedetti2013b}, where similar computations required several \emph{days} on similar machines.

Another points of interest concerns the mesh size of the grain-boundary elements. For the basis formulation, this point has been carefully discussed in Ref.\ \cite{benedetti2013b} where it has been highlighted that the characteristic length of the grain-boundary elements $l_e$ should be fine enough to resolve the strain and stress distributions inside the cohesive zone \cite{rice1968,espinosa2003b,tomar2004}. This condition, which ensures the reproducibility of results, is always respected in the present formulation and implementation that, as a matter of fact, smooths the representation of the inter-granular fields without altering the reference mesh size.

In the following, prismatic microstructures whose edges are aligned with the global reference system are tested. Figure (\ref{fig-Ch3:mc bcs}) schematises the boundary conditions that will be used during the numerical tests: for each face, the boundary conditions are given as prescribed displacements along the normal of the face and written as $u_kn_k=\bar{u}_n$ where $n_k$ is the outward unit normal of the face and $\bar{u}_n$ is the prescribed value of displacement. Along the remaining directions, the external faces of the aggregate are traction-free.

\begin{figure}
\centering
\includegraphics[width=0.5\textwidth]{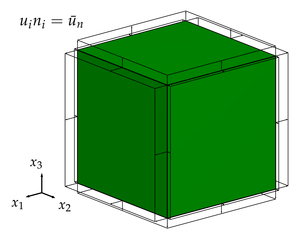}
\caption{Scheme of the boundary conditions used to investigate the micro-cracking of polycrystalline aggregates. The boundary conditions are given as prescribed displacement along the direction defined by the normal of each face.}
\label{fig-Ch3:mc bcs}
\end{figure}

A first set of tests reports the micro-cracking behaviour of aggregates with $N_g=100$ and different ASTM $G$ values, namely ASTM $G=10,\,12,\,14$ \cite{standard1996} under tensile loading in displacement control. The boundary conditions consist of prescribed displacements over the top and bottom faces of the aggregates given by $u_in_i=\lambda/2$, whereas the lateral faces of the aggregates are subject $u_in_i=0$. The remaining loading directions are traction-free. Figure (\ref{fig-Ch3:mc different ASMT G}a) shows three 100-grain morphologies with ASTM $G$ values $G=10,\,12,\,14$ to scale, whereas Figure (\ref{fig-Ch3:mc different ASMT G}b) shows the mesh of the ASTM $G=10$ aggregate. Figure (\ref{fig-Ch3:mcG100}a) reports the value of the macro-stress $\Sigma_{33}$ corresponding to different values of enforced nominal strain $\Gamma_{33}$ computed as $\Gamma_{33}=\lambda/H$ being $H$ the height of the aggregate. It is worth noting as different ASTM grain sizes induce different macro-behaviours in terms of brittleness: aggregates with bigger grains (ASTM $G=10$) are generally more brittle with respect to morphologies with smaller grains, if the inter-granular cohesive strength and toughness are kept constant; this is consistent with the model physics. Figures (\ref{fig-Ch3:mcG100}b), (\ref{fig-Ch3:mcG100}c) and (\ref{fig-Ch3:mcG100}d) show the microcracking patterns and the damage distribution for the above three morphologies with $N_g=100$ and average ASTM grain size $G=10,\,12,\,14$, respectively, at the last computed snapshot reported in Figure (\ref{fig-Ch3:mcG100}a).

\begin{figure}
\centering
	\begin{subfigure}{0.49\textwidth}
	\centering
	\includegraphics[width=\textwidth]{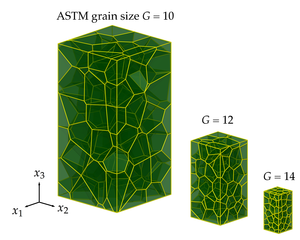}
	\caption{}
	\end{subfigure}
	\
	\begin{subfigure}{0.49\textwidth}
	\centering
	\includegraphics[width=\textwidth]{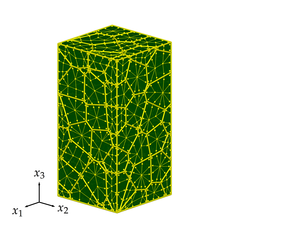}
	\caption{}
	\end{subfigure}
\caption{(\emph{a}) Relative size of three different morphologies with $N_g=100$ and different ASTM grain size $G=10,\,12,\,14$. (\emph{b}) Mesh of the ASTM $G=10$ aggregate and particular of the mesh of a grain.}
\label{fig-Ch3:mc different ASMT G}
\end{figure}

\begin{figure}
\centering
	\begin{subfigure}{0.49\textwidth}
	\centering
	\includegraphics[width=\textwidth]{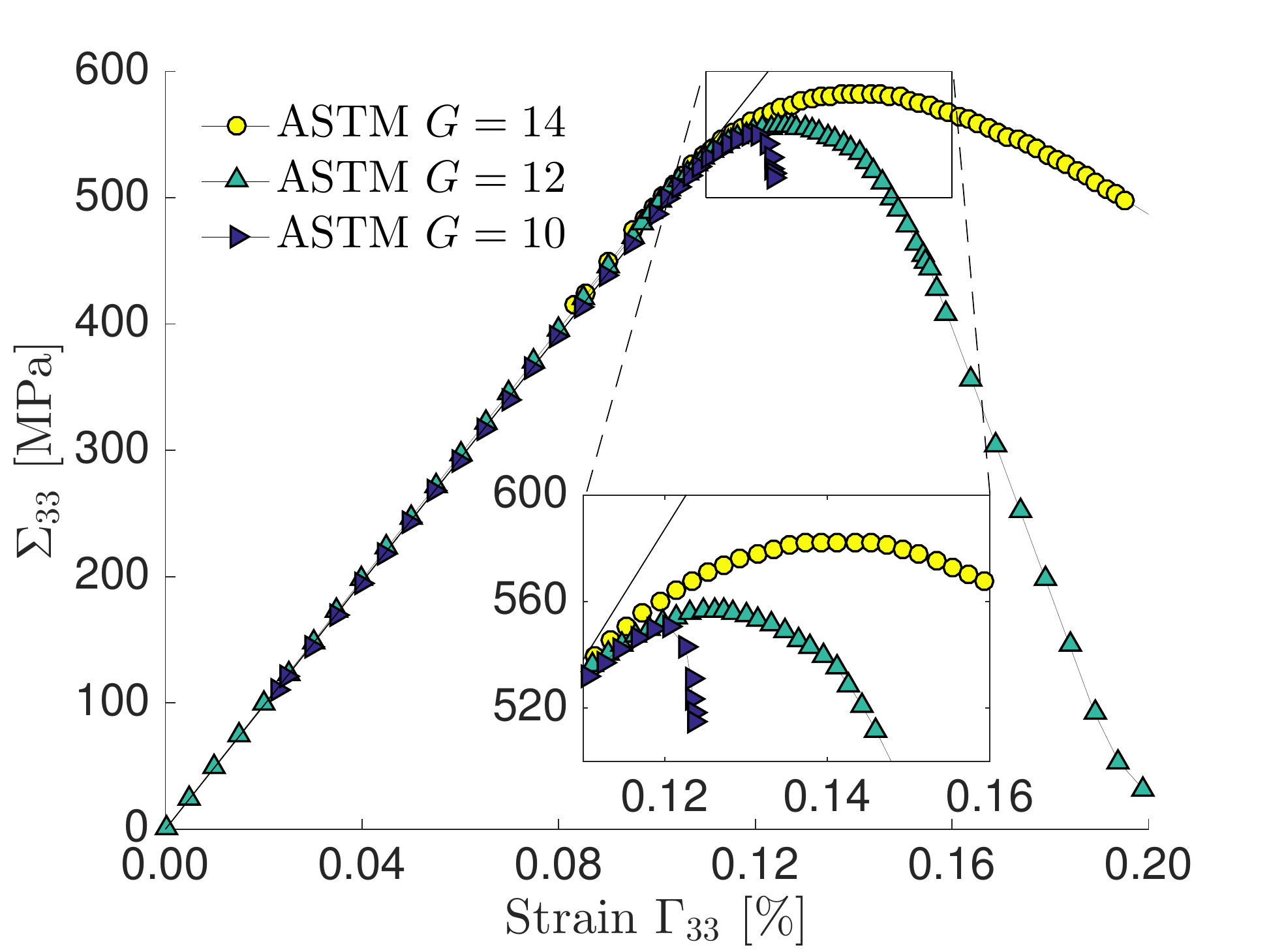}
	\caption{}
	\end{subfigure}
	\
	\begin{subfigure}{0.49\textwidth}
	\centering
	\includegraphics[width=\textwidth]{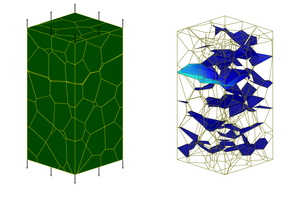}
	\caption{ASTM $G=10$}
	\end{subfigure}
	\
	\begin{subfigure}{0.49\textwidth}
	\centering
	\includegraphics[width=\textwidth]{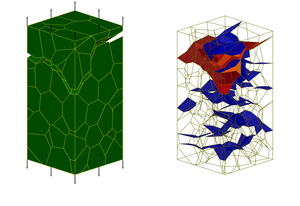}
	\caption{ASTM $G=12$}
	\end{subfigure}
	\
	\begin{subfigure}{0.49\textwidth}
	\centering
	\includegraphics[width=\textwidth]{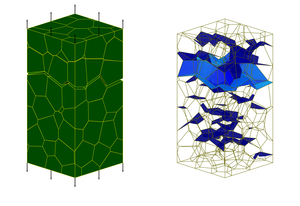}
	\caption{ASTM $G=14$}
	\end{subfigure}
	\\
	\vspace{10pt}
	\begin{subfigure}{0.49\textwidth}
	\centering
	\includegraphics[width=\textwidth]{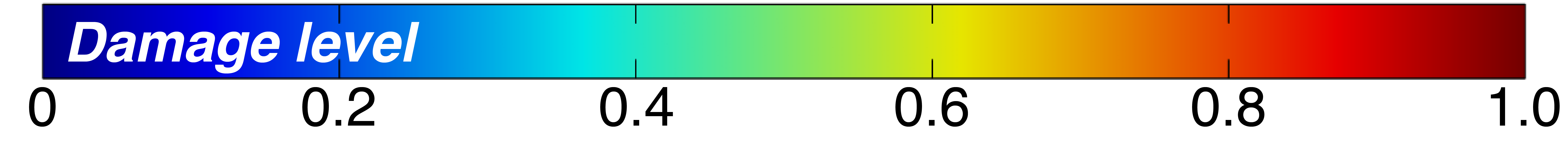}
	\caption{}
	\end{subfigure}
\caption{(\emph{a}) Volume average stress component $\Sigma_{33}$ versus nominal applied strain $\Gamma_{33}$. (\emph{b}-\emph{d}) Microcracking patterns and damage distribution for three morphologies with $N_g=100$ and average ASTM grain size (\emph{b}) $G=10$, (\emph{c}) $G=12$ and (\emph{d}) $G=14$. (\emph{e}) Colormap of the damage.}
\label{fig-Ch3:mcG100}
\end{figure}

The second sets of tests investigates the effect of the number of grains on the micro-cracking response of the aggregates. Similarly to the previous set of tests, the top and bottom faces of the aggregates are subject to $u_in_i=\lambda/2$, the lateral faces are subject to $u_in_i=0$ and the remaining directions are traction-free. Figure (\ref{fig-Ch3:mc different Ng}a) shows three morphologies with average ASTM grain size $G=12$ and number of grains $N_g=50,\,100,\,200$ and Figure (\ref{fig-Ch3:mc different Ng}b) shows the boundary mesh of the internal grains of the 200-grain aggregate, i.e.\ those grains that non are affected by the presence of the prismatic box. The micro-cracking stress-strain behaviour is reported in Figure (\ref{fig-Ch3:mcASTM12}a) in which it is interesting to note that the morphologies with $N_g=100$ and $N_g=200$ have similar response in terms of averaged stress component, so that is appears reasonable to assume a representative behaviour. Figures (\ref{fig-Ch3:mcASTM12}b), (\ref{fig-Ch3:mcASTM12}c) and (\ref{fig-Ch3:mcASTM12}d) show the microcracking patterns and the damage distribution for the morphologies with $N_g=50$, $N_g=100$ and $N_g=200$, respectively, at the last computed snapshot reported in Figure (\ref{fig-Ch3:mcASTM12}a).

\begin{figure}
\centering
	\begin{subfigure}{0.49\textwidth}
	\centering
	\includegraphics[width=\textwidth]{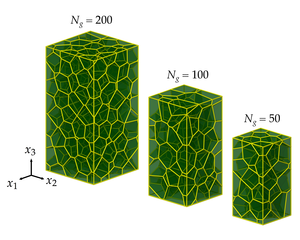}
	\caption{}
	\end{subfigure}
	\
	\begin{subfigure}{0.49\textwidth}
	\centering
	\includegraphics[width=0.6667\textwidth]{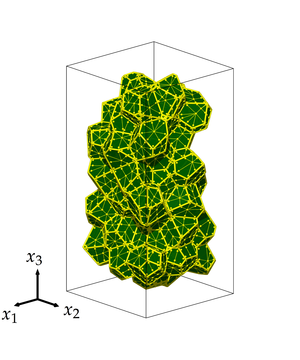}
	\caption{}
	\end{subfigure}
\caption{(\emph{a}) Relative size of three different morphologies with ASTM grain size $G=12$ and different number of grains $N_g=200,\,100,\,50$. (\emph{b}) Grain boundary mesh of those grain not affected by the presence of the prismatic box (represented by black lines) of the 200-grain morphology.}
\label{fig-Ch3:mc different Ng}
\end{figure}

\begin{figure}
\centering
	\begin{subfigure}{0.49\textwidth}
	\centering
	\includegraphics[width=\textwidth]{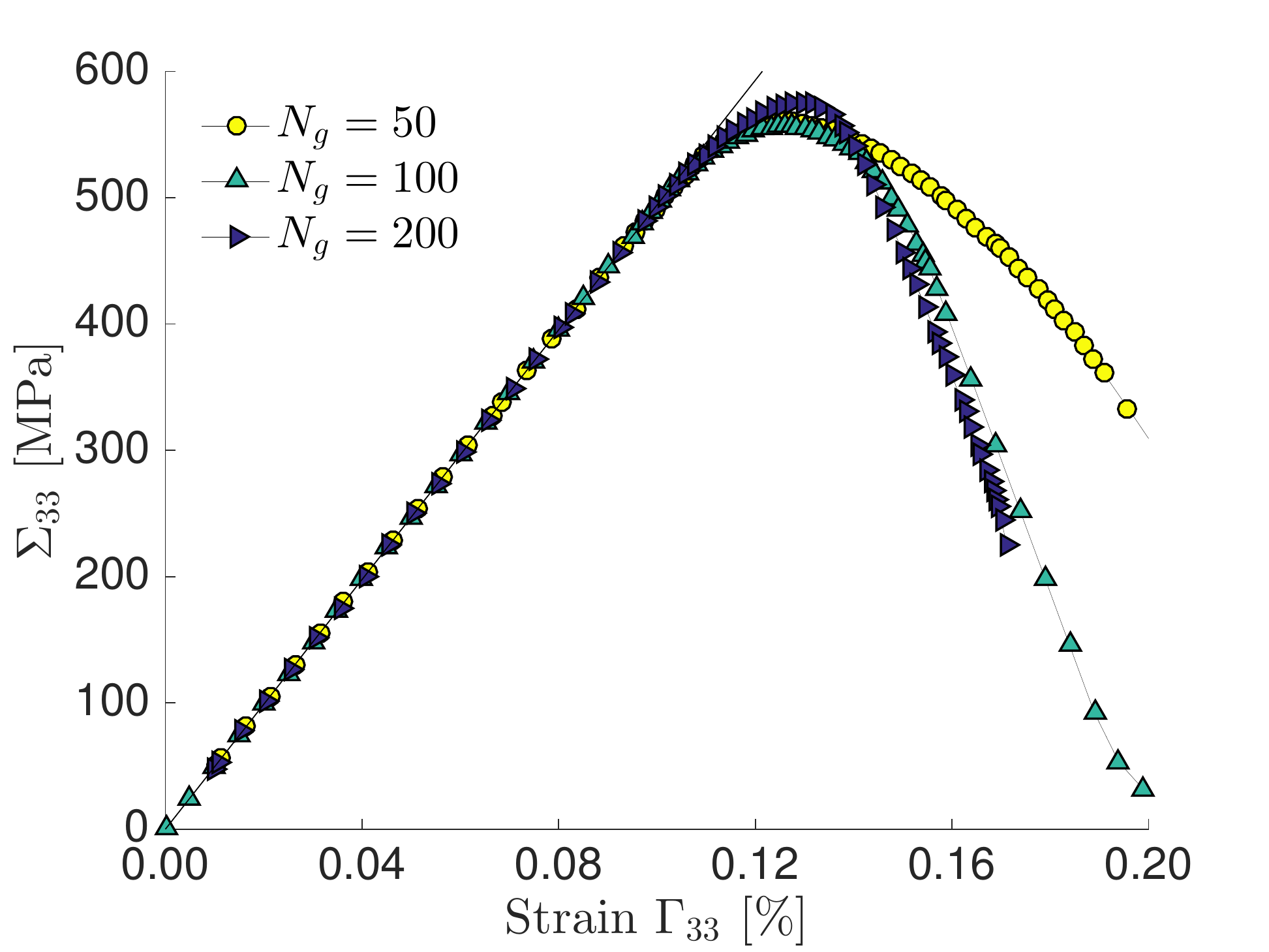}
	\caption{}
	\end{subfigure}
	\
	\begin{subfigure}{0.49\textwidth}
	\centering
	\includegraphics[width=\textwidth]{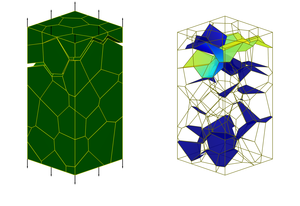}
	\caption{$N_g=50$}
	\end{subfigure}
	\
	\begin{subfigure}{0.49\textwidth}
	\centering
	\includegraphics[width=\textwidth]{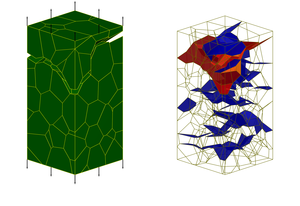}
	\caption{$N_g=100$}
	\end{subfigure}
	\
	\begin{subfigure}{0.49\textwidth}
	\centering
	\includegraphics[width=\textwidth]{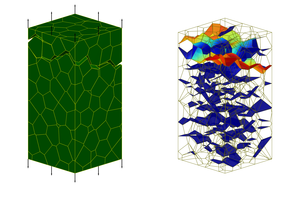}
	\caption{$N_g=200$}
	\end{subfigure}
	\\
	\vspace{10pt}
	\begin{subfigure}{0.49\textwidth}
	\centering
	\includegraphics[width=\textwidth]{./Ch3/images/mc/colorbar.png}
	\caption{}
	\end{subfigure}
\caption{(\emph{a}) Volume average stress component $\Sigma_{33}$ versus nominal applied strain $\Gamma_{33}$. (\emph{b}-\emph{d}) Microcracking patterns and damage distribution for three morphologies with ASTM grain size $G=12$ and number of grains (\emph{b}) $N_g=50$, (\emph{c}) $N_g=100$ and (\emph{d}) $N_g=200$. (\emph{e}) Colormap of the damage.}
\label{fig-Ch3:mcASTM12}
\end{figure}

A third set of tests investigates the micro-cracking behaviour of a polycrystalline aggregate subjected to loads in different directions, so to investigate the isotropic behavior of the macro-aggregate also in micro-cracking simulations. The investigated cubic $200$-grain aggregate, with ASTM $G=12$ grain-size, and the corresponding mesh are shown in Figures (\ref{fig-Ch3:G200ASTM12}a) and (\ref{fig-Ch3:G200ASTM12}b), respectively. Five different sets of boundary conditions are prescribed on the aggregate, which is subject to uniaxial tensile loading along the directions $x_1$, $x_2$ and $x_3$, to biaxial tensile loading on the plane $x_1$-$x_2$ and to triaxial tensile loading.

Similarly to the two previous sets of tests, the uniaxial boundary conditions along the direction $x_k$ are enforced as follows
\begin{align}
u_in_i=
\begin{cases}
\lambda/2, &\quad \mathrm{if} \quad n_k=\pm e_{pk}\\
0, &\quad       \mathrm{otherwise}
\end{cases},
\end{align}
where $n_i$ is the unit normal of the aggregate's faces and $e_{pk}$ is the unit vector of the $k$-th loading direction. It is worth noting that when the aggregate is subject to uniaxial boundary conditions along the $x_k$ direction, the corresponding prescribed strain is given by $\bar{\Gamma}=\Gamma_{kk}=\lambda/H$ being $H$ the side of the aggregate. Figures (\ref{fig-Ch3:mcG200-uniax}a) and (\ref{fig-Ch3:mcG200-uniax-2}a) shows the values of the macro-stress components $\Sigma_{ijkl}$ versus the load factor $\lambda$. In the notation of $\Sigma_{ijkl}$, the first two subscripts $ij$ indicate the considered homogenized stress component, whereas the subscript $k$ indicates the loaded face, i.e.\ the face with $e_{pk}$ as unit normal, and the subscript $l$ indicates the loading direction. In particular, Figure (\ref{fig-Ch3:mcG200-uniax}a) shows the values of the \emph{direct} macro-stress components $\Sigma_{ijkl}$, i.e.\ the components of the macro-stress $\Sigma_{ii}$ when a prescribed displacement in the direction $i$ is applied on the face $i$, which corresponds to a prescribed macro-strain component $\Gamma_{ii}$. Similarly, Figure (\ref{fig-Ch3:mcG200-uniax-2}a) shows the values of the \emph{cross} macro-stress components $\Sigma_{ijkl}$, that is the components of the macro-stress $\Sigma_{ii}$ when a prescribed displacement in the direction $j$ is imposed on the face $j$, with $j\neq i$. The insets in Figures (\ref{fig-Ch3:mcG200-uniax}a) and (\ref{fig-Ch3:mcG200-uniax-2}a) are close-ups of the curves when the stresses reach their maximum values. Figures (\ref{fig-Ch3:mcG200-uniax}b-d) show the cracked morphology when subject to the three uniaxial loading conditions, whereas Figures (\ref{fig-Ch3:mcG200-uniax-2}b-d) show the corresponding damage distribution. It is worth noting the correspondence between the crack pattern,the distribution of damage and the homogenised curves. More specifically, when the morphology is loaded along the direction $x_2$ and $x_3$, it shows an almost well-defined unique crack perpendicular to the direction of loading and the corresponding homogenised curves overlap reasonably well. On the other hand, when the morphology is loaded along the $x_1$ direction, it shows a more tortuous crack, which corresponds to a slightly higher values of the homogenised stresses. However, in all three tests the isotropic behaviour of the macro direct and cross macro components is verified. Unlike Figures (\ref{fig-Ch3:mcG100}b-d) and (\ref{fig-Ch3:mcASTM12}b-d) that show all the damaged interfaces in the aggregates, in order to have a clearer representation of the cracks, Figures (\ref{fig-Ch3:mcG200-uniax-2}b-d) display the interface whose damage is higher than 5\%.

Then, a tensile biaxial and a tensile triaxial test have been performed on the morphology shown in Figure (\ref{fig-Ch3:G200ASTM12}). The biaxial has been simulated enforcing the following boundary conditions: $u_in_i=\lambda/2$ on the lateral faces of the morphology and $u_in_i=0$ on the top and bottom faces. On the other hand, the triaxial has been simulated enforcing $u_in_i=\lambda/2$ on all the faces of the morphology. Figure (\ref{fig-Ch3:mcG200-biax-triax}a) shows the macro-stress components $\Sigma_{11}$, $\Sigma_{22}$ and $\Sigma_{33}$ versus $\lambda$, when the morphology is subject to a biaxial tensile load. For comparison, also the macro-stress components $\Sigma_{1111}$ and $\Sigma_{1122}$, taken from corresponding the uniaxial tests are reported in the figure. Similarly, Figure (\ref{fig-Ch3:mcG200-biax-triax}b) shows the macro-stress components $\Sigma_{11}$, $\Sigma_{22}$ and $\Sigma_{33}$ versus $\lambda$, when the morphology of is subject to a triaxial tensile load. The macro-stress components $\Sigma_{1111}$ and $\Sigma_{1122}$ taken from the corresponding uniaxial test and the macro-stress $\Sigma_{11}$ taken from the biaxial test are reported in the figure. The crack patterns for the biaxial and triaxial loading conditions, at the last computed values of $\lambda$, are also reported in Figures (\ref{fig-Ch3:mcG200-biax-triax}c) and (\ref{fig-Ch3:mcG200-biax-triax}d) and the corresponding damage distribution are reported in Figures (\ref{fig-Ch3:mcG200-biax-triax}e) and (\ref{fig-Ch3:mcG200-biax-triax}f). It is interesting to note that for these last two tests, and in particular for the triaxial case, multiple cracks show during the loading history. Again, in Figures (\ref{fig-Ch3:mcG200-biax-triax}e,f) the interfaces whose damage is higher than 5\% are reported.

\begin{figure}
\centering
	\begin{subfigure}{0.49\textwidth}
	\centering
	\includegraphics[width=\textwidth]{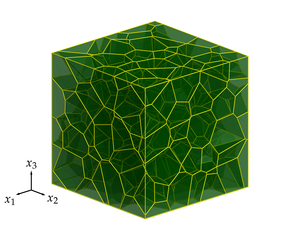}
	\caption{}
	\end{subfigure}
	\
	\begin{subfigure}{0.49\textwidth}
	\centering
	\includegraphics[width=\textwidth]{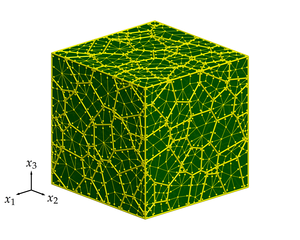}
	\caption{}
	\end{subfigure}
\caption{(\emph{a}) 200-grain morphology with ASTM grain size $G=12$ and (\emph{b}) its boundary mesh.}
\label{fig-Ch3:G200ASTM12}
\end{figure}

\begin{figure}
\centering
	\begin{subfigure}{0.49\textwidth}
	\centering
	\includegraphics[width=\textwidth]{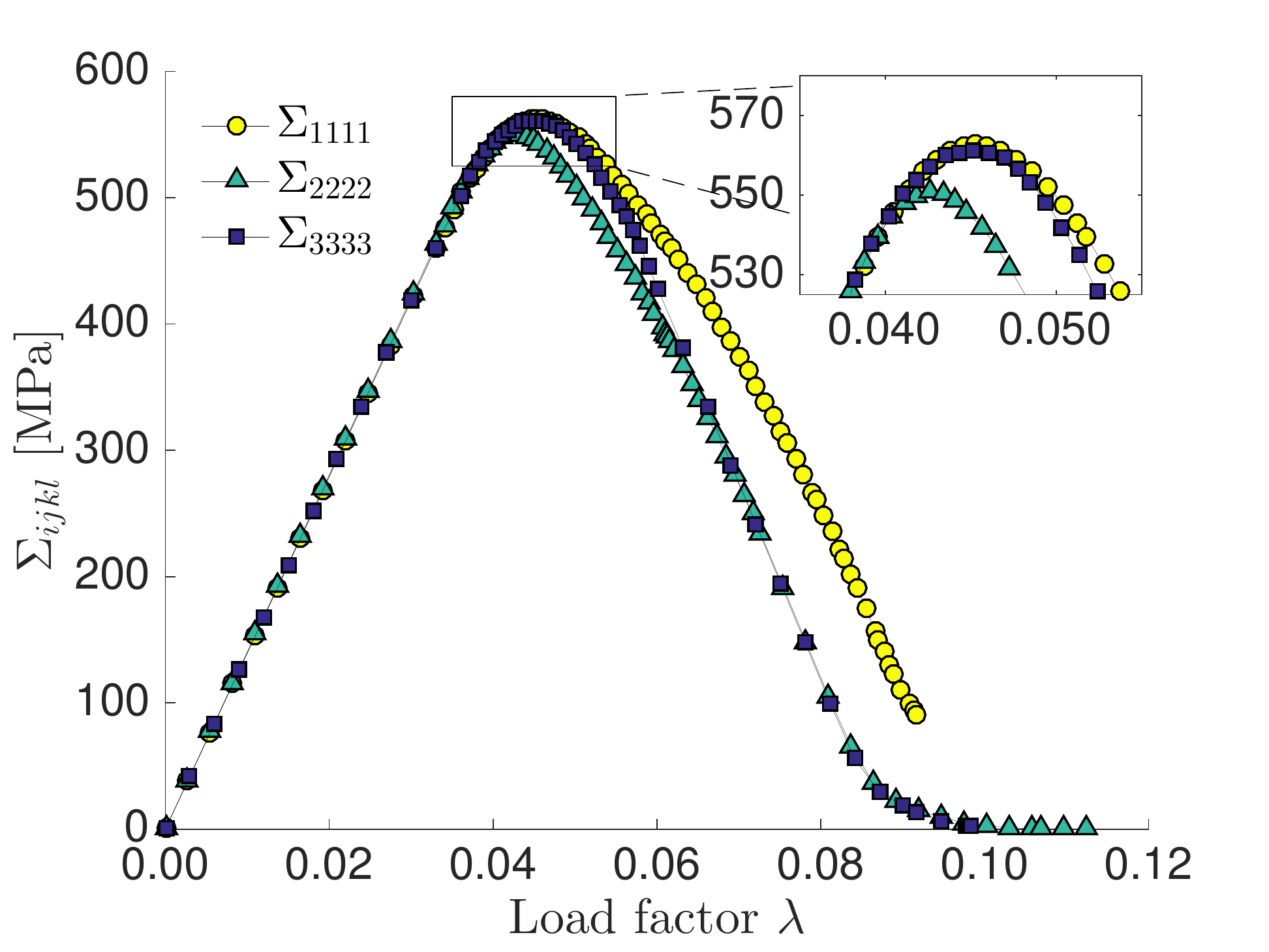}
	\caption{}
	\end{subfigure}
	\
	\begin{subfigure}{0.49\textwidth}
	\centering
	\includegraphics[width=\textwidth]{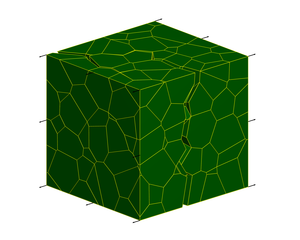}
	\caption{$\bar{\Gamma}=\Gamma_{11}$}
	\end{subfigure}
	\
	\begin{subfigure}{0.49\textwidth}
	\centering
	\includegraphics[width=\textwidth]{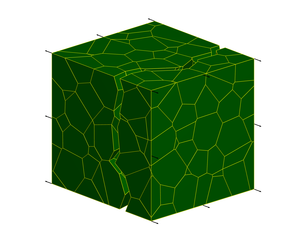}
	\caption{$\bar{\Gamma}=\Gamma_{22}$}
	\end{subfigure}
	\
	\begin{subfigure}{0.49\textwidth}
	\centering
	\includegraphics[width=\textwidth]{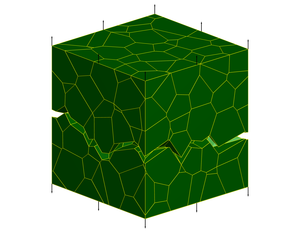}
	\caption{$\bar{\Gamma}=\Gamma_{33}$}
	\end{subfigure}
\caption{(\emph{a}) Volume average of the direct stress component $\Sigma_{ijkl}$ versus applied load factor $\lambda$ for the three uniaxial loading configurations. (\emph{b}-\emph{d}) Microcracking patterns for the three uniaxial loading configurations denoted by $\bar{\Gamma}=\Gamma_{kk}$ where $k$ is the loading direction.}
\label{fig-Ch3:mcG200-uniax}
\end{figure}

\begin{figure}
\centering
	\begin{subfigure}{0.49\textwidth}
	\centering
	\includegraphics[width=\textwidth]{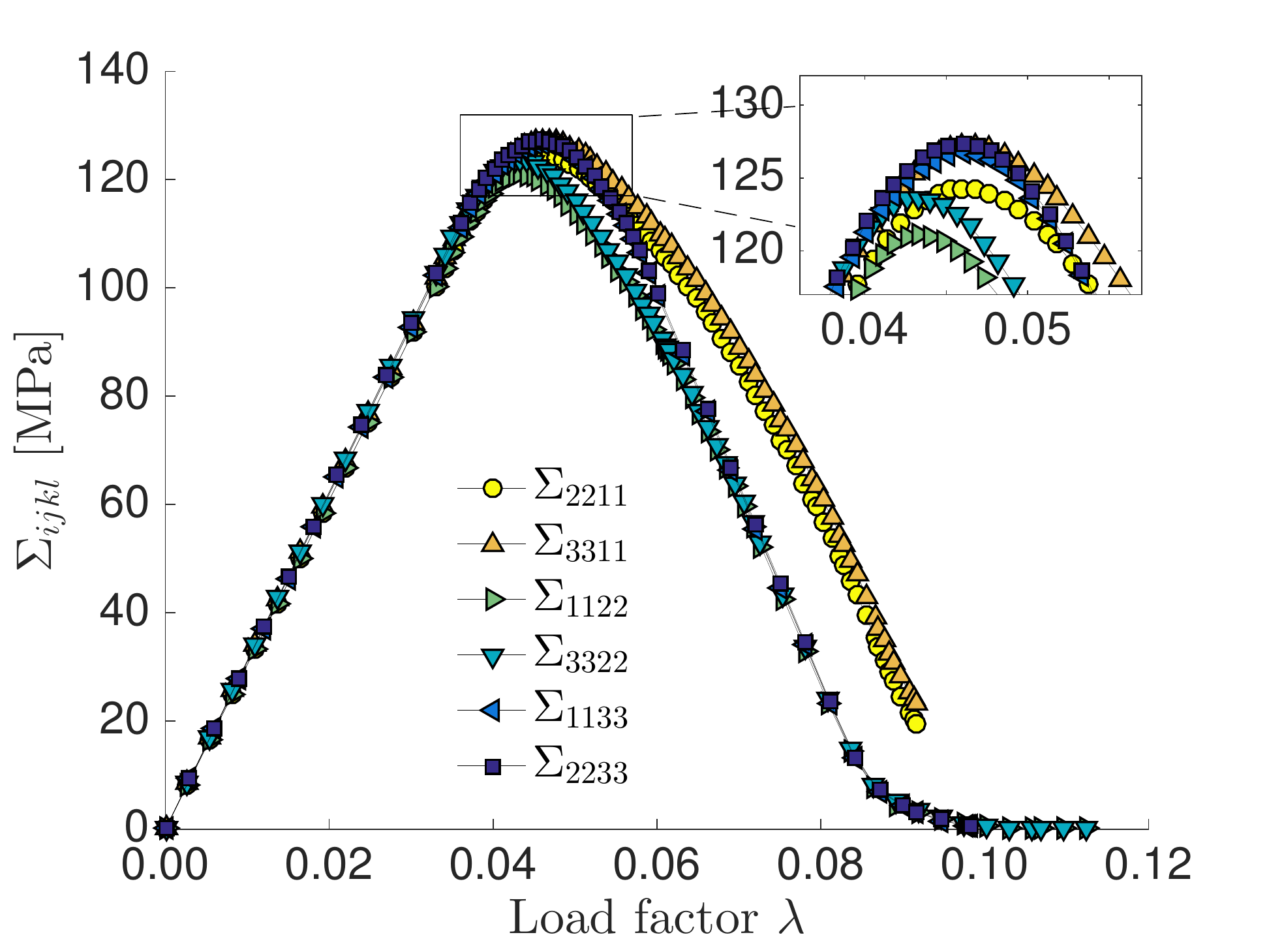}
	\caption{}
	\end{subfigure}
	\
	\begin{subfigure}{0.49\textwidth}
	\centering
	\includegraphics[width=\textwidth]{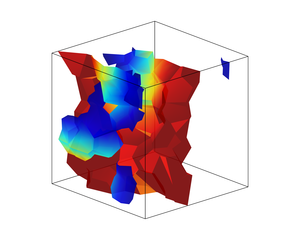}
	\caption{$\bar{\Gamma}=\Gamma_{11}$}
	\end{subfigure}
	\
	\begin{subfigure}{0.49\textwidth}
	\centering
	\includegraphics[width=\textwidth]{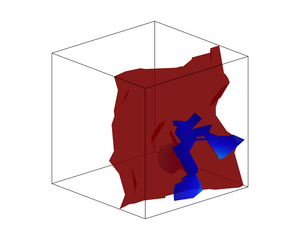}
	\caption{$\bar{\Gamma}=\Gamma_{22}$}
	\end{subfigure}
	\
	\begin{subfigure}{0.49\textwidth}
	\centering
	\includegraphics[width=\textwidth]{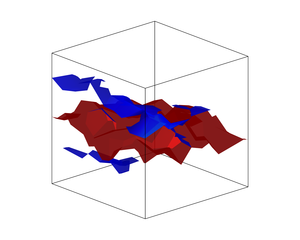}
	\caption{$\bar{\Gamma}=\Gamma_{33}$}
	\end{subfigure}
	\\
	\vspace{10pt}
	\begin{subfigure}{0.49\textwidth}
	\centering
	\includegraphics[width=\textwidth]{./Ch3/images/mc/colorbar.png}
	\caption{}
	\end{subfigure}
\caption{(\emph{a}) Volume average of the cross stress component $\Sigma_{ijkl}$ versus applied load factor $\lambda$ for the three uniaxial loading configurations. (\emph{b}-\emph{d}) Damage distribution for the three uniaxial loading configurations denoted by $\bar{\Gamma}=\Gamma_{kk}$ where $k$ is the loading direction. (\emph{e}) Colormap of the damage.}
\label{fig-Ch3:mcG200-uniax-2}
\end{figure}

\begin{figure}
\centering
	\begin{subfigure}{0.49\textwidth}
	\centering
	\includegraphics[width=\textwidth]{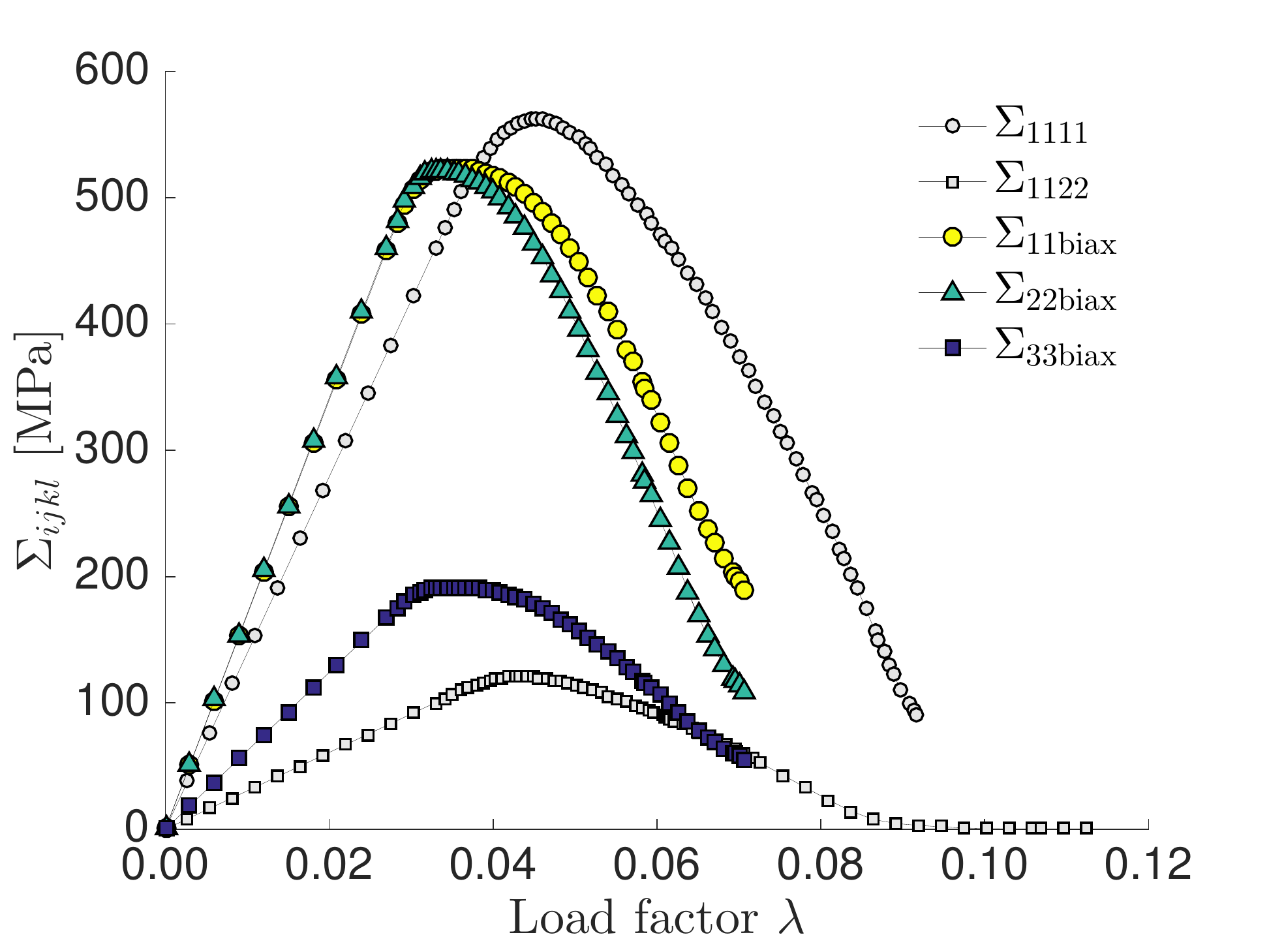}
	\caption{}
	\end{subfigure}
	\
	\begin{subfigure}{0.49\textwidth}
	\centering
	\includegraphics[width=\textwidth]{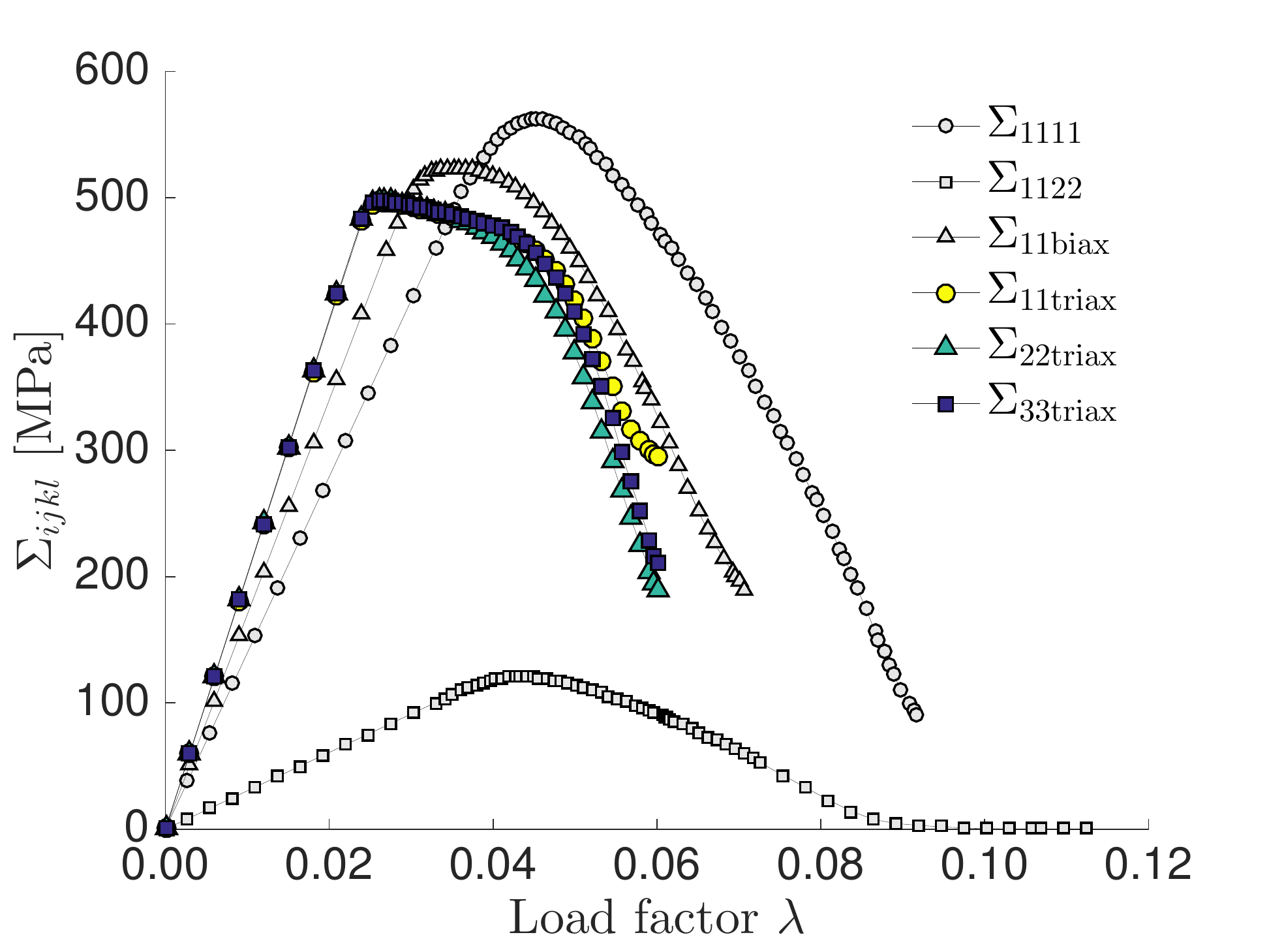}
	\caption{}
	\end{subfigure}
	\
	\begin{subfigure}{0.45\textwidth}
	\centering
	\includegraphics[width=\textwidth]{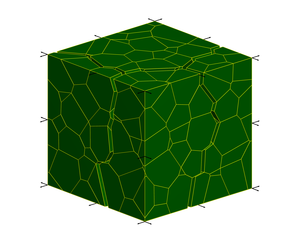}
	\caption{}
	\end{subfigure}
	\
	\begin{subfigure}{0.45\textwidth}
	\centering
	\includegraphics[width=\textwidth]{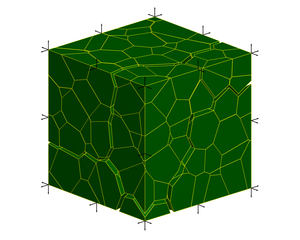}
	\caption{}
	\end{subfigure}
	\
	\begin{subfigure}{0.45\textwidth}
	\centering
	\includegraphics[width=\textwidth]{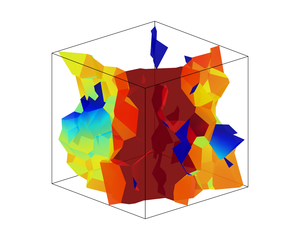}
	\caption{}
	\end{subfigure}
	\
	\begin{subfigure}{0.45\textwidth}
	\centering
	\includegraphics[width=\textwidth]{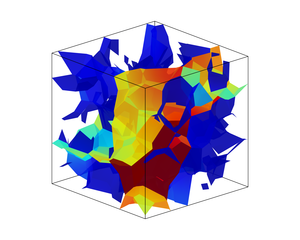}
	\caption{}
	\end{subfigure}
	\\
	\vspace{10pt}
	\begin{subfigure}{0.49\textwidth}
	\centering
	\includegraphics[width=\textwidth]{./Ch3/images/mc/colorbar.png}
	\caption{}
	\end{subfigure}
\caption{Micro-cracking results of (\emph{a},\emph{c},\emph{e}) biaxial test and (\emph{b},\emph{d},\emph{f}) triaxial test: (\emph{a},\emph{b}) Volume averages of the stress components $\Sigma_{11}$, $\Sigma_{22}$ and $\Sigma_{33}$ versus applied load factor $\lambda$; (\emph{c},\emph{d}) micro-crack pattern; (\emph{e},\emph{f}) Damage distribution. (\emph{g}) Colormap of the damage.}
\label{fig-Ch3:mcG200-biax-triax}
\end{figure}

Finally, the microcracking response of a 1000-grain cubic polycrystalline aggregates with ASTM grain size $G=12$ is considered. The 1000-grain morphology is generated in a cubic box by the distribution of the weighted seeds shown in Figure (\ref{fig-Ch3:G1000ASTM12}a). The resulting tessellation is reported in Figure (\ref{fig-Ch3:G1000ASTM12}b). The aggregate is subject to uniaxial tensile boundary conditions along the $x_3$, i.e.\ $u_in_i=\lambda/2$ on the top and bottom faces and $u_in_i=0$ on the lateral faces. Figure (\ref{fig-Ch3:mcG1000}a) show the volume stress averages $\Sigma_{11}$, $\Sigma_{22}$ and $\Sigma_{33}$ as a function of the load factor $\lambda$. The crack pattern at the last computed step of the loading history is shown in Figure (\ref{fig-Ch3:mcG1000}b) where it is possible to see the presence of two fully developed micro-cracks, one originating from the bottom-right corner and the second one originating from the center of leftmost edge of the aggregate. The cracks configuration will be clearer by means of the following figures. Figures (\ref{fig-Ch3:mcG1000-steps}a,c,e) show the evolution of the micro-crack pattern at the different values of the load factor $\lambda$. At the same values of $\lambda$, Figures (\ref{fig-Ch3:mcG1000-steps}b,d,f) show the evolutions of the damage distribution within the aggregate. In this case, all the damaged interface are displayed and it is interesting to note how at $\lambda=0.077$, corresponding the load factor at which the softening initiates, the damage is almost uniform throughout the morphology. During the next load steps, the aggregate undergoes softening and the cracks localize. In order to better show the shape and configuration of the cracks within the aggregate, Figure (\ref{fig-Ch3:mcG1000-steps-up-down}) shows from two different views the interfaces of the aggregate whose damage is higher than 5\% at the three load steps $\lambda=0.077$, $\lambda=0.079$ and $\lambda=0.086$. By looking at the figures, it is possible to recognise the coalescence of failing interfaces into the two fully developed cracks, one close to the bottom face and the second one at half the height of the aggregate. Also, it is clear that the cracks initiate at the external faces and tend to grow towards the interior of the aggregate.

\begin{figure}
\centering
	\begin{subfigure}{0.49\textwidth}
	\centering
	\includegraphics[width=\textwidth]{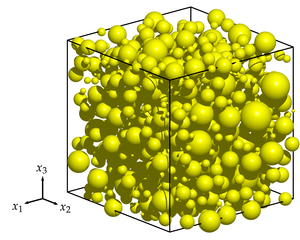}
	\caption{}
	\end{subfigure}
	\
	\begin{subfigure}{0.49\textwidth}
	\centering
	\includegraphics[width=\textwidth]{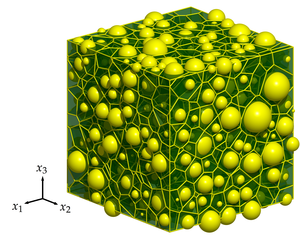}
	\caption{}
	\end{subfigure}
\caption{1000-grain morphology with ASTM grain size $G=12$: (\emph{a}) seeds and weights distribution inside the cubic box; (\emph{b}) resulting polycrystalline aggregate.}
\label{fig-Ch3:G1000ASTM12}
\end{figure}

\begin{figure}
\centering
	\begin{subfigure}{0.49\textwidth}
	\centering
	\includegraphics[width=\textwidth]{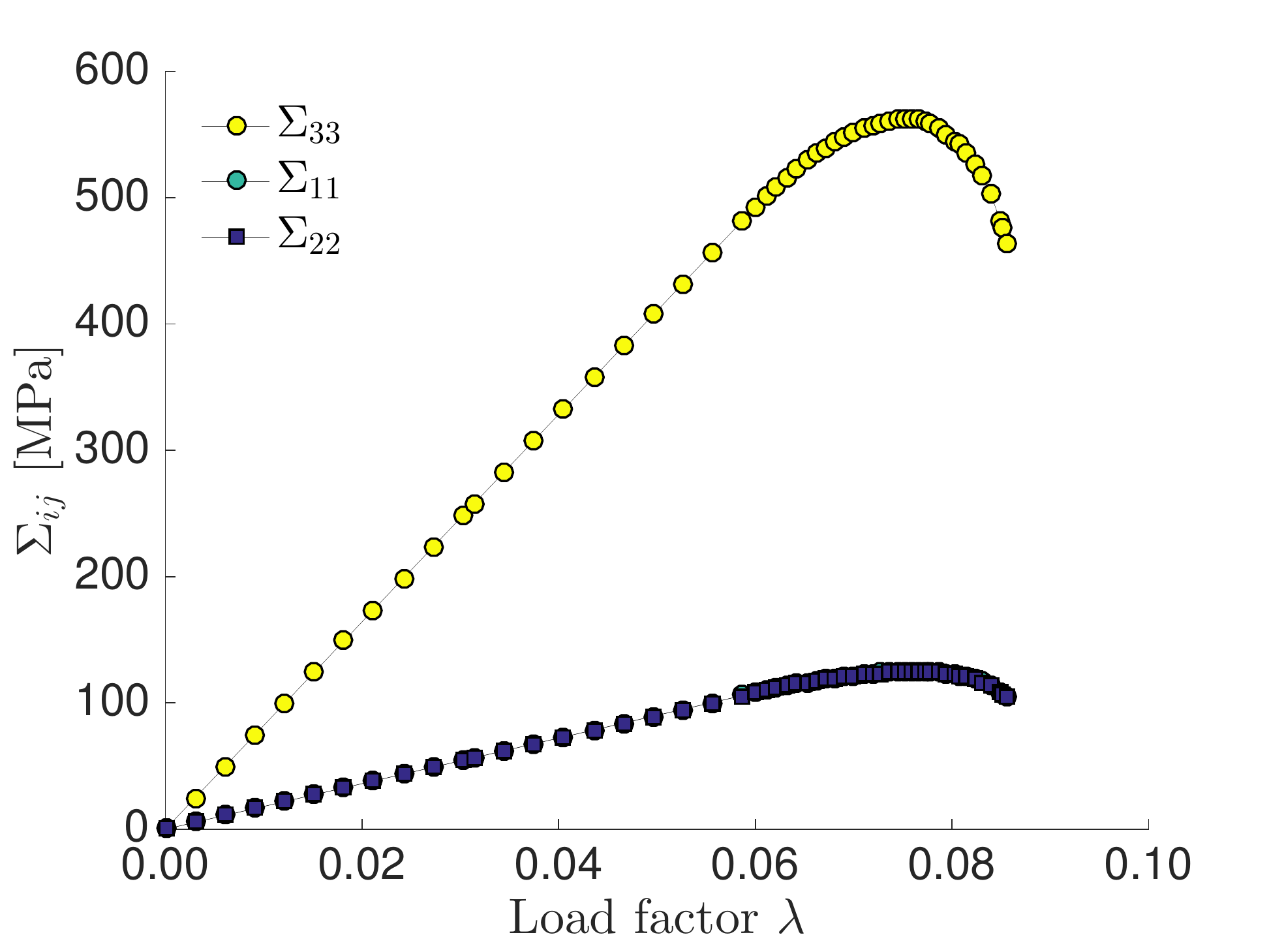}
	\caption{}
	\end{subfigure}
	\
	\begin{subfigure}{0.49\textwidth}
	\centering
	\includegraphics[width=\textwidth]{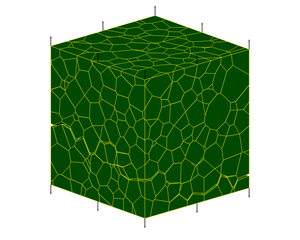}
	\caption{}
	\end{subfigure}
\caption{(\emph{a}) Volume stress averages $\Sigma_{11}$, $\Sigma_{22}$ and $\Sigma_{33}$ as a function of the load factor $\lambda$. (\emph{b}) Micro-crack pattern of the 1000-grain morphology subject to tensile load at the last computer step.}
\label{fig-Ch3:mcG1000}
\end{figure}

\begin{figure}
\centering
	\begin{subfigure}{0.45\textwidth}
	\centering
	\includegraphics[width=\textwidth]{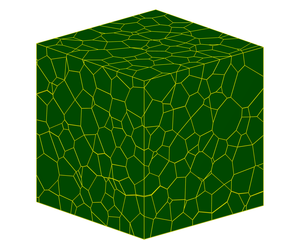}
	\caption{$\lambda=0.077$}
	\end{subfigure}
	\
	\begin{subfigure}{0.45\textwidth}
	\centering
	\includegraphics[width=\textwidth]{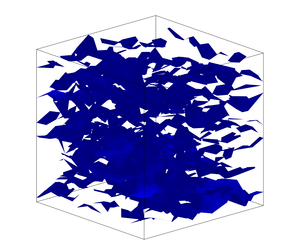}
	\caption{$\lambda=0.077$}
	\end{subfigure}
	\
	\begin{subfigure}{0.45\textwidth}
	\centering
	\includegraphics[width=\textwidth]{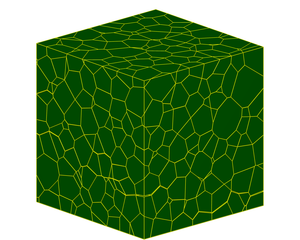}
	\caption{$\lambda=0.079$}
	\end{subfigure}
	\
	\begin{subfigure}{0.45\textwidth}
	\centering
	\includegraphics[width=\textwidth]{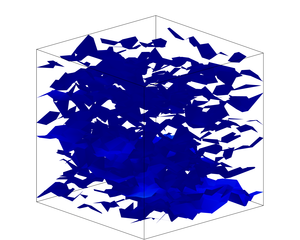}
	\caption{$\lambda=0.079$}
	\end{subfigure}
	\
	\begin{subfigure}{0.45\textwidth}
	\centering
	\includegraphics[width=\textwidth]{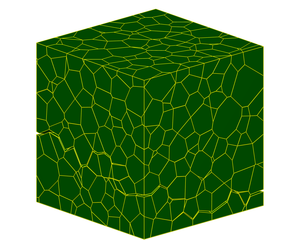}
	\caption{$\lambda=0.086$}
	\end{subfigure}
	\
	\begin{subfigure}{0.45\textwidth}
	\centering
	\includegraphics[width=\textwidth]{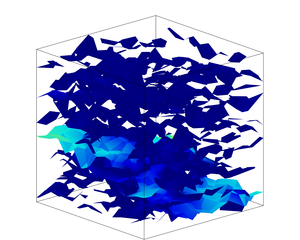}
	\caption{$\lambda=0.086$}
	\end{subfigure}
	\\
	\vspace{5pt}
	\begin{subfigure}{0.49\textwidth}
	\centering
	\includegraphics[width=\textwidth]{./Ch3/images/mc/colorbar.png}
	\caption{}
	\end{subfigure}
\caption{Microcracking snapshots at the loading steps (\emph{a},\emph{b}) $\lambda=0.077$, (\emph{c},\emph{d}) $\lambda=0.079$ and (\emph{e},\emph{f}) $\lambda=0.086$. (\emph{a},\emph{c},\emph{e}) Cracked aggregate. (\emph{b},\emph{d},\emph{f}) Damage distribution. (\emph{g}) Colormap of the damage.}
\label{fig-Ch3:mcG1000-steps}
\end{figure}

\begin{figure}
\centering
	\begin{subfigure}{0.45\textwidth}
	\centering
	\includegraphics[width=\textwidth]{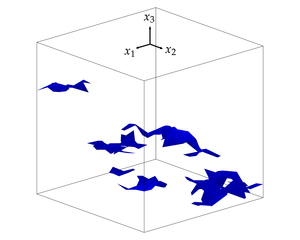}
	\caption{$\lambda=0.077$}
	\end{subfigure}
	\
	\begin{subfigure}{0.45\textwidth}
	\centering
	\includegraphics[width=\textwidth]{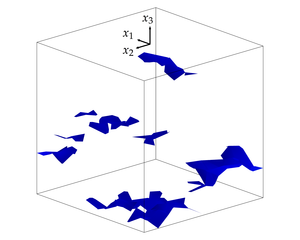}
	\caption{$\lambda=0.077$}
	\end{subfigure}
	\
	\begin{subfigure}{0.45\textwidth}
	\centering
	\includegraphics[width=\textwidth]{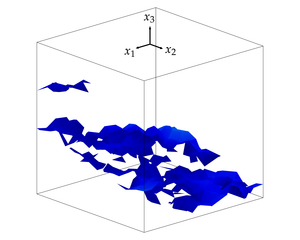}
	\caption{$\lambda=0.079$}
	\end{subfigure}
	\
	\begin{subfigure}{0.45\textwidth}
	\centering
	\includegraphics[width=\textwidth]{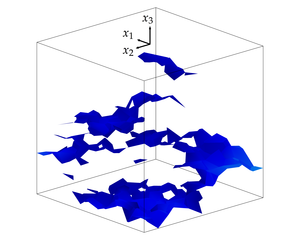}
	\caption{$\lambda=0.079$}
	\end{subfigure}
	\
	\begin{subfigure}{0.45\textwidth}
	\centering
	\includegraphics[width=\textwidth]{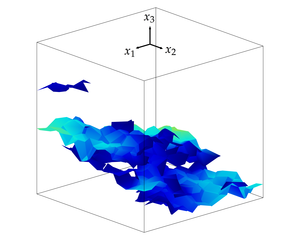}
	\caption{$\lambda=0.086$}
	\end{subfigure}
	\
	\begin{subfigure}{0.45\textwidth}
	\centering
	\includegraphics[width=\textwidth]{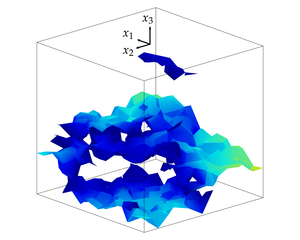}
	\caption{$\lambda=0.086$}
	\end{subfigure}
	\\
	\vspace{5pt}
	\begin{subfigure}{0.49\textwidth}
	\centering
	\includegraphics[width=\textwidth]{./Ch3/images/mc/colorbar.png}
	\caption{}
	\end{subfigure}
\caption{Two different views of the microcracks pattern at the loading steps (\emph{a},\emph{b}) $\lambda=0.077$, (\emph{c},\emph{d}) $\lambda=0.079$ and (\emph{e},\emph{f}) $\lambda=0.086$. (\emph{g}) Colormap of the damage.}
\label{fig-Ch3:mcG1000-steps-up-down}
\end{figure}

\clearpage

\section{Discussion}\label{sec-Ch3: discussion}
In this Chapter of the present thesis, an effective computational tool for the microstructural analysis of polycrystalline materials, with applications in computational homogenisation and inter-granular micro-cracking simulations, has been presented. The key contribution of the present work relies in the use of the grain-boundary formulation in conjunction with regularized Laguerre morphologies and a specific grain-boundary meshing strategy. In particular:
\begin{itemize}
\item{
The employment of the tessellation regularization \cite{quey2011} has allowed to remove the sources of unnecessary mesh refinements in the original morphologies, thus reducing the overall number of degrees of freedom necessary to model the crystal aggregate. The effect of this regularization is particularly appreciated in microcracking simulations: in the original scheme \cite{benedetti2013b}, the occurrence of many small grain-boundary elements hindered the computational performance of the technique and could represent an issue for numerical convergence, due to the concentration of collocation nodes on excessively small grain faces;
}
\item{
The employment of a specific grain boundary meshing strategy, using continuous and semi-discontinuous boundary elements, has simultaneously allowed: \emph{a}) to smooth the representation of the inter-granular displacement and traction fields, with advantages in terms of numerical convergence of the non-linear solution scheme; \emph{b}) to reduce the number of inter-granular unknowns, i.e.\ degrees of freedom; \emph{c}) to avoid the typical complexities arising, within a boundary element framework, when grid nodes are collocated over edges or corners, where a unit normal cannot be uniquely defined \cite{gray1990}.
}
\end{itemize}
Additionally, for computational homogenization applications, non-prismatic RVEs have been adopted, with the subsequent further reduction of meshing complexities potentially induced by the cutting operations usually performed to obtain periodic prismatic morphologies. The above enhancements have induced a remarkable acceleration in the numerical solution strategy, retaining the accuracy of the original formulation \cite{benedetti2013b}, which is particularly welcome if the technique is to be used in multiscale perspective \cite{benedetti2015}.

However, some numerical/computational aspects deserve further study. First of all, although the reduction of the system order, in terms of DoFs, undoubtedly represents a benefit in any numerical technique, it should be noted that it is not the only factor affecting the computational effectiveness. In this sense, the features of the resolving system, e.g.\ the sparsity pattern, the symmetry and the definiteness of the coefficient matrices, play a crucial role. The coefficient matrix produced by the proposed formulation is not symmetric. For such reasons, the general purpose parallel direct sparse solver \texttt{PARDISO} has been used in the present study, as described in Chapter (\ref{ch-intro}). However, the development of \emph{specific iterative Krylov solvers} for sparse systems having the structure of the system described in Section (\ref{ssec-Ch1: system solution}), \cite{araujo2010,araujo2011}, in conjunction with hierarchical matrices \cite{benedetti2008,benedetti2009}, would probably produce computational savings, both in terms of storage memory and solution time. To improve the numerical convergence under general loading conditions, the implementation of arc-length procedures \cite{crisfield1981,mallardo2004} could be of interest.

\clearpage

\chapter{Modelling inter-granular and trans-granular cracking in polycrystalline materials}\label{ch-TG}
In this Chapter, a novel grain boundary formulation for inter-granular and trans-granular cracking of three-dimensional polycrystalline materials is presented. The formulation is based on the displacement and stress boundary integral equations introduced in Chapter (\ref{ch-intro}) and still retains the advantage of expressing the considered polycrystalline problem in terms of grain boundary variables only. Trans-granular cracking is assumed to occur along specific cleavage planes, whose direction depends on the crystallographic orientation of the grains of the aggregate. Both inter- and trans-granular cracking are modelled using the cohesive zone approach. However, unlike grain boundary cracking for which the failure interface is well defined, trans-granular cracks are introduced only if a suitably defined effective stress over the potential cleavage plane overcomes the threshold strength for that plane. During the load history, whenever a trans-granular crack needs to be introduced, the polycrystalline morphology is remeshed accordingly and a cohesive law, whose parameters characterise the behaviour of trans-granular failure, is associated to the newly introduced interface.

Some preliminary results show that the method is able to capture the competition between inter-granular and trans-granular cracking.

\section{Introduction}\label{sec-Ch3b:intro}
The present Chapter presents a model for studying the competition between inter-granular and trans-granular cracking modes, which represent two of the main mechanisms of failure in brittle polycrystalline materials. Inter-granular cracking denotes the failure of the interfaces between adjacent grains whereas trans-granular cracking denotes the failure, often along specific crystallographic planes, of the bulk grains.
The occurrence of the two cracking modes is influenced by several factors such as crystallographic lattice, temperature and aggressive environment \cite{anderson2005,crocker2005}.
The crystallographic lattice plays a key role in determining to which deformation mechanism a polycrystalline materials is susceptible. 
As an example, body-centered cubic (BCC) and hexagonal close packed (HCP) crystals usually show ductile-to-brittle transition \cite{smith2002,hughes2005,hughes2007} with decreasing temperature and inter-granular and trans-granular cracking phenomena are the result of the limited number of slip systems in these crystal lattices at low temperature.
On the other hand, although face-centered cubic (FCC) lattices generally favour ductile deformation over a wide range of temperatures as a consequence of the large number of slip systems, aggressive environments are well-known to induce grain boundary embrittlement and therefore favour inter-granular and trans-granular fracture in naturally ductile materials \cite{mcmahon2001,mori2005,raja2011}.
At room temperature, HPC ceramics such as 6H silicon carbide (SiC) also exhibit inter- and trans-granular brittle fracture \cite{gandhi1979}.

Several studies, both experimental and numerical, have been devoted to understanding the complex interaction between inter- and trans-granular cracking and their relationship with the morphological, chemical and physical properties of the micro-structure of polycrystalline materials.
Sukumar et al.\ \cite{sukumar2003} presented a extended finite element (XFEM) approach to model the inter- and trans-granular behaviour of 2D brittle polycrystalline microstructures. In their work, the authors assumed a linear elastic isotropic behaviour of the crystals and studied the micro-cracking response of the polycrystalline morphologies and the corresponding crack patterns as a function of the ratio between the fracture toughness of the grain boundaries and the fracture toughness of the grain interiors. Within the XFEM framework, Wang et al.\ \cite{wang2013b} investigated the effect of the average grain diameter, of a secondary phase and of the grain boundary fraction on the the fracture toughness of 2D polycrystalline aggregates, and recently, Prakash et al.\ \cite{prakash2016} studied the effect of the grain boundary strength on the crack propagation in polycrystalline tungsten (W) using XFEM.

Verhoosel and Guti\'errez \cite{verhoosel2009} modelled the inter-granular and trans-granular crack propagation in 2D piezoelectric polycrystals. The model was developed in a finite element framework and crack propagation was obtained using interface elements and two different cohesive laws for modelling the grain boundary inter-granular failure and the bulk crystals trans-granular cracking.

Musienko and Cailletaud \cite{musienko2009} used a rate-dependent finite element framework to model inter- and trans-granular stress corrosion cracking of 2D and 3D Zircaloy polycrystalline morphologies in an iodine environment. However, trans-granular crack propagation was allowed only for 2D morphologies. For inter-granular crack propagation, the authors used grain boundary finite elements with a specific constitutive relations involving a damage parameter whose evolution was governed by a rate-dependent power law flow rule. Trans-granular cracking was allowed for a specific crystallographic plane of the Zircaloy crystal, namely the basal plane, which failed to cleavage when the normal stress reached a threshold value.

Kraft and Molinari \cite{kraft2008} developed a 2D finite element model to study the effect of grain boundaries distribution on the inter- and trans-granular fracture of polycrystalline aluminum oxide ($\mathrm{Al}_2\mathrm{O}_3$). The authors employed a cohesive zone approach to model tensile and shear separation of grain boundaries as well as cleavage cracks. Unlike the aforementioned studies, trans-granular fracture was considered in the model by inserting cohesive interfaces along specific crystallographic planes within the crystals and by remeshing the morphology accordingly. New trans-granular planes were introduced during the analyses as soon as the values of the resolved normal and shear stress over those plane overcame their corresponding critical values, which were estimated from the fracture toughness of the trans-grnular interface.

Inter- and trans-granular dynamic cracking in 2D polycrystalline materials were recently investigated by Mousavi et al.\ \cite{mousavi2014,mousavi2015} and by Lin et al.\ \cite{lin2015} using the cohesive zone approach within the finite element framework.

More recently, models based on different approaches such as peridynamics or nonlocal lattice particle method have been presented.
Ghajari et al.\ \cite{ghajari2014} proposed a peridynamic model to study the fracture propagation in 2D $\mathrm{Al}_2\mathrm{O}_3$ polycrystalline aggregates accounting for the random orientation and therefore anisotropy of the microstructure. They reported the crack patterns as a function of the ratio between the fracture toughness of the grains boundaries and the grain interiors showing a transition from inter-granular to trans-granular behaviour corresponding to weaker to tougher grain boundaries.
Chen et al.\ \cite{chen2015} studied the fracture behaviour of 2D polycrystalline morphologies using a nonlocal lattice particle framework. In their model, all the crystallites belonging to a grain were rotated to match the crystallographic orientation of the grain and grain boundaries were naturally generated at the location where two grains meet. Crack propagation was obtained by considering a spring-based model among the particles and inter-granular to trans-granular cracking mode was realised by changing the strength of the grain boundary springs parameters.

While several two-dimensional models accounting for the interplay or competition between inter- and trans-granular polycrystalline cracking mechanisms are present in the literature, few 3D models have been developed, due to increased geometrical and mechanical complexity and much higher computational requirements. However, the combined effect of inter- and trans-granular cracking is a naturally three-dimensional phenomenon due to the crucial role of the 3D crystallographic orientation and therefore of the potential cleavage planes within polycrystalline aggregates. Furthermore, as pointed out by some researchers \cite{smith2002,hughes2005,hughes2007}, 2D model may not be able to fully capture the role of grain boundary fracture in accommodating trans-granular crack propagation through misaligned adjacent grains.

Three-dimensional polycrystalline morphologies were studied by Clayton and Knap \cite{clayton2015} by means of a phase field approach. The authors studied silicon carbide (SiC) and zinc (Zn) polycrystalline aggregates taking into account the anisotropic linear elastic behaviour of the crystals as well as the fracture behaviour of the trans-granular failure, which occurred along the basal planes of the considered HCP crystals. By varying the ratio between the elastic stiffness and the surface energy of the grains with respect to the grain boundaries, they studied inter-granular and trans-granular crack patterns within the considered morphologies. Recently \cite{clayton2016}, they also incorporated deformation twinning in their phase field model. Abdollahi and Arias \cite{abdollahi2012,abdollahi2014} employed the phase field model to study the fracture and toughening mechanisms in ferroelectric polycrystalline microstructures. They studied 2D \cite{abdollahi2012} and 3D \cite{abdollahi2014} morphologies and modelled the inter-granular and trans-granular fracture behaviour by considering a specific ratio of the grain interior and grain boundaries fracture toughnesses. Within the phase-field framework, Shanthraj et al.\ \cite{shanthraj2016} developed an elasto-viscoplastic phase field model to capture anisotropic cleavage failure in 2D polycrystalline aggregates.
Three-dimensional modelling of cleavage in polycrystals has been also proposed by Shterenlikht and Margetts \cite{shterenlikht2015} and Di Caprio et al.\ \cite{di2016} using a cellular automata approach.

In this Chapter, inter- and trans-granular crack evolution in three-dimensional polycrystalline morphologies is modelled using a boundary element and a cohesive zone approach. The Chapter is organised as follows: Section (\ref{sec-Ch3b:formulation}) briefly recalls the grain boundary formulation of Chapter (\ref{ch-intro}) focusing on the displacements and stress boundary integral equations and the cohesive zone modelling of inter- and trans-granular failure; Section (\ref{sec-Ch3b:discrete system}) describes the numerical discretisation of the displacement boundary integral equations as well as the stress boundary integral equations and presents the numerical algorithm accounting for the two considered failure mechanisms and involving the on-the-fly remeshing of the polycrystalline morphology; Section (\ref{sec-Ch3b:numerical tests}) shows some preliminary results of the developed formulation. 

\section{Grain boundary formulation}\label{sec-Ch3b:formulation}
In this Section, the grain boundary formulation described in Chapter (\ref{ch-intro}) is briefly recalled.

\subsection{Boundary integral equations}\label{ssec-Ch3b:BIE}
To model inter-granular and trans-granular cracking in 3D polycrystalline materials, the displacement boundary integral equations (\ref{eq-Ch1:DBIE-2 local RS}) and the stress boundary integral equations (\ref{eq-Ch1:SBIE}) are simultaneously used during the analysis. They are recalled here for the sake of completeness. The displacement boundary integral equations are used to express the behaviour of the generic grain $g$ in terms of the grain boundary values of displacements $\wtilde{u}_i^g(\mathbf{x})$ and tractions $\wtilde{t}_i^g(\mathbf{x})$ and read
\begin{equation}\label{eq-Ch3b:DBIE}
\wtilde{c}_{pi}(\mathbf{y})\wtilde{u}_i^g(\mathbf{y})+
\dashint_{S^g} \wtilde{T}_{pi}^g(\mathbf{x},\mathbf{y})\wtilde{u}_i^g(\mathbf{x})\dd S(\mathbf{x})=
\int_{S^g} \wtilde{U}_{pi}^g(\mathbf{x},\mathbf{y})\wtilde{t}_i^g(\mathbf{x})\dd S(\mathbf{x})
\end{equation}
where $\mathbf{x}$ and $\mathbf{y}$ are the integration and collocation points, respectively, $S^g$ is the boundary of the grain $g$, $\wtilde{c}_{pi}(\mathbf{y})\wtilde{u}_i^g(\mathbf{y})$ are the free-terms, $\wtilde{U}_{pi}^g(\mathbf{x},\mathbf{y})$ and $\wtilde{T}_{pi}^g(\mathbf{x},\mathbf{y})$ are boundary equations kernels and it is recalled that the symbol $\wtilde{\cdot}$ denotes a quantity expressed in the boundary local reference system.

On the other hand, the stress boundary integral equations, as given in Eq.(\ref{eq-Ch1:SBIE}), allow to compute the stress tensor $\sigma_{mn}^g(\mathbf{y})$ at any internal point $\mathbf{y}\in V^g$ of the grain $g$ in terms of the grain boundary displacements and tractions as follows
\begin{equation}\label{eq-Ch3b:SBIE}
\sigma_{mn}^g(\mathbf{y})+
\int_{S^g} T_{mni}^{\sigma,g}(\mathbf{x},\mathbf{y})u_i^g(\mathbf{x})\dd S(\mathbf{x})=
\int_{S^g} U_{mni}^{\sigma,g}(\mathbf{x},\mathbf{y})t_i^g(\mathbf{x})\dd S(\mathbf{x}),
\end{equation}
where once again $\mathbf{x}$ is the integration point and $U_{mni}^{\sigma,g}(\mathbf{x},\mathbf{y})$ and $T_{mni}^{\sigma,g}(\mathbf{x},\mathbf{y})$ are the kernels of the integral equations. In this case, there is no need to express the equation in the local boundary reference system since the stress boundary integral equations are used in a post-processing stage, that is after the solution has been obtained. The reader is referred to Section (\ref{sec-Ch1:grain boundary integral equations}) or Chapter (\ref{ch-intro}) for a more detailed description of terms in Eqs.(\ref{eq-Ch3b:DBIE}) and (\ref{eq-Ch3b:SBIE}).

\subsection{Boundary conditions and interface equations}
To close the polycrystalline problem, a proper set of boundary conditions and interface conditions must be enforced. The boundary conditions are enforced on the external faces of the aggregate whereas the interface conditions express the relation between the inter-granular displacements and tractions as described in Sections (\ref{ssec-Ch1:BCs}) and (\ref{ssec-Ch1:ICs}).

In Chapter (\ref{ch-EF}), inter-granular cracking was modelled using a suitably defined traction-separation cohesive law, which accounted for the degradation and failure of the grain boundaries. In this work, although in a different way with respect to inter-granular failure that will be described in Section (\ref{sec-Ch3b:fracture modes}), trans-granular cracking is also modelled using the cohesive-zone approach. First, the cohesive zone model is recalled.

As discussed in Section (\ref{ssec-Ch3: cohesive law}), damage arises over certain interfaces, which for the purpose of this study may represent inter-granular regions as well as cleavage planes, when a suitably defined \emph{effective traction} $\tau_e$ overcomes the \emph{interface strength} $T_\mathrm{max}$. When such criterion is fulfilled for a point of the generic couple of adjacent grains $g$ and $h$ sharing the interface $\mathscr{I}^{gh}$, an \emph{extrinsic cohesive law} of the form 
\begin{subequations}\label{eq-Ch3b:interface equations - cohesive law}
\begin{align}
&C_s(d^*)\delta\wtilde{u}_i^{gh}-\wtilde{t}_i^{gh}=0,\quad (i=1,2)\\
&C_n(d^*)\delta\wtilde{u}_3^{gh}-\wtilde{t}_3^{gh}=0,
\end{align}
\end{subequations}
is introduced to link the boundary traction components $\wtilde{t}_i^{gh}$ with the boundary displacements jump $\delta\wtilde{u}_i^{gh}$. It is worth noting that if the interface $\mathscr{I}^{gh}$ represent a trans-granular crack, the grains $g$ and $h$ denote those grains originating from the trans-granular failure of a single grain. The cohesive law (\ref{eq-Ch3b:interface equations - cohesive law}) is given in terms of the constitutive constants $C_s(d^*)$ and $C_n(d^*)$, which are function of the \emph{irreversible damage parameter} $d^*$ defined as maximum value of the \emph{effective opening displacement} $d$ during the loading history $\mathscr{H}_d$, see Eqs.(\ref{eq-Ch3:interface equations - cohesive law constants}), (\ref{eq-Ch3:irreversible damage}) and (\ref{eq-Ch3:effective opening}).

Subsequently, upon assuming the work of separation $G_I$, the interface strength $T_\mathrm{max}$, and the relative contribution between opening and sliding failure modes represented by the parameters $\alpha$, $\beta$ and the ratio $G_{II}/G_I$, it is possible to completely determine the cohesive law.

To account for inter-granular and trans-granular failures, two different cohesive laws are employed in the present study and their parameters are set to represent the different properties of the two types of interface as well as the competition between the two mechanisms. Although inter- and trans-granular processes are modelled using the same cohesive approach, it is worth underlining their different nature and the consequence in terms of numerical implementation: inter-granular failures occur along the interfaces defined by the grain boundaries, which are natural sites of crack initiation and are well-defined when the polycrystalline morphology is generated; on the other hand, at a generic point \emph{within} the bulk crystallographic lattice of a crystal, trans-granular failure may potentially occur along different planes, which are not \emph{a priori} known and may or may not activate depending on the fulfilment of a threshold condition involving the comparison between an actual local stress acting over the plane and the corresponding strength. As an example, Figure (\ref{fig-Ch3b:grain boundary and cleavage plane}a) shows a grain boundary interface between two adjacent grains, whereas Figure (\ref{fig-Ch3b:grain boundary and cleavage plane}b) shows an example of a potential trans-granular interface identified by a crystallographic plane described by the unit normal vector $m_i$ and passing through the point $\mathbf{y}$ within a bulk grain. In terms of numerical implementation, the trans-granular failure is addressed by considering the grain as a bulk homogeneous solid and introducing a cohesive interface along the activated cleavage plane whenever the trans-granular threshold condition is met.

\begin{figure}
\centering
	\begin{subfigure}{0.49\textwidth}
	\centering
	\includegraphics[width=\textwidth]{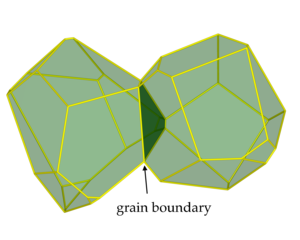}
	\caption{}
	\end{subfigure}
	\
	\begin{subfigure}{0.49\textwidth}
	\centering
	\includegraphics[width=\textwidth]{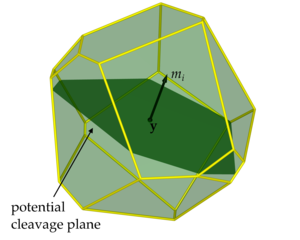}
	\caption{}
	\end{subfigure}
\caption{(\emph{a}) Grain boundary (in darker green) between two adjacent grains and (\emph{b}) potential cleavage plane (in darker green) defined by the normal vector $m_i$ and passing through the point $\mathbf{y}$.}
\label{fig-Ch3b:grain boundary and cleavage plane}
\end{figure}

\subsection{Inter- and trans-granular fracture modes}\label{sec-Ch3b:fracture modes}
In this Section, the parameters entering the cohesive laws employed to describe the two failure mechanisms are discussed. As regards the notation of this Section, the grain boundaries quantities referring to the interface between the adjacent grains $g$ and $h$ are denoted by the superscript $gh$ since they in fact characterise the behaviour of the interface $\mathscr{I}^{gh}$ between the grains $g$ and $h$ and can in general vary from interface to interface. On the other hand, the quantities characterising the behaviour of the cleavage planes of the grain $g$ are denoted by the superscript $g$ only since they refer to the properties of the bulk grain $g$.

In Section (\ref{ssec-Ch3: cohesive law}), it was shown that, at the interfaces between adjacent grains, the effective traction of the grain boundaries at the interface $\mathscr{I}^{gh}$, denoted by $\tau_{e}^{gh}$, is computed in terms of the local grain boundary tractions as
\begin{equation}\label{eq-Ch3b:gb effective traction}
\tau_{e}^{gh}=\sqrt{\left\langle \tau_{n}^{gh}\right\rangle^2+\left(\frac{\beta^{gh}}{\alpha^{gh}}\tau_{s}^{gh}\right)^2},
\end{equation}
where $\beta^{gh}$ and $\alpha^{gh}$ are the values of the cohesive law coefficients introduced in Section (\ref{ssec-Ch3: cohesive law}) and characterising the cohesive behaviour of the grain boundary. $\tau_s^{gh}$ denotes the traction along the sliding direction that is computed as  $\tau_{s}^{gh}=\sqrt{(\wtilde{t}_1^{gh})^2+(\wtilde{t}_2^{gh})^2}$; $\tau_n$ denotes the traction along the normal direction and is simply $\tau_n=\wtilde{t}_3^{gh}$. It is recalled that the tractions $\wtilde{t}_i^{gh}=\wtilde{t}_i^{g}=\wtilde{t}_i^{h}$ are obtained as solution of the displacements BIE (\ref{eq-Ch3b:DBIE}) written for each grain of the aggregate and coupled to the interface equations, whose numerical implementation is described in Section (\ref{sec-Ch3b:discrete system}).

In this case, it is highlighted that the interfaces between different grains are known when the polycrystalline tessellation is generated and are retained during the loading history even if the grain boundaries do not undergo inter-granular cracking. In fact, if the grain boundary between the generic grains $g$ and $h$ is in a pristine state, which holds as long as effective traction $\tau_e^{gh}$ is smaller the grain boundary strength $T_\mathrm{max}^{gh}$, the interface equations simply enforce a zero displacements jump, i.e.\ $\delta \wtilde{u}_i^{gh}=0$.

Similarly to the inter-granular fracture mode, an effective traction $\tau_e^g$ of the trans-granular fracture mode can be defined as follows
\begin{equation}\label{eq:cl effective traction}
\tau_e^g=\sqrt{\left\langle \tau_n^g\right\rangle^2+\left(\frac{\beta^g}{\alpha^g}\tau_s^g\right)^2}.
\end{equation}
However, unlike the inter-granular case where the interface is already well-defined, given a point within the considered grain, the sliding traction $\tau_s^g$ and the normal traction $\tau_n^g$ must be computed as a function of the stress tensor $\sigma_{ij}^g$ and all the potential cleavage planes of the grain $g$ at that point. It is clear that the crystallographic lattice of the grains plays a crucial role in the determination of the orientation of the trans-granular cracks. Using the stress BIE (\ref{eq-Ch3b:SBIE}), the stress tensor $\sigma_{ij}^g$ at any internal point of the grain $g$ can be computed as a function of the boundary displacements and tractions. Then, given the unit normal $m_i^g$ representing the potential cleavage plane of the grain $g$ and two mutually orthogonal directions $p_i^g$ and $q_i^g$ on the plane, the normal traction $\tau_n^g$ is computed as $\tau_n^g=m_i^g\sigma_{ij}^gm_j^g$, whereas the sliding traction $\tau_s$ is computed as $\tau_s^g=\sqrt{(\tau_p^g)^2+(\tau_q^g)^2}$ being $\tau_p^g=p_i^g\sigma_{ij}^gm_j^g$ and $\tau_q^g=q_i^g\sigma_{ij}^gm_j^g$. During the loading history, the values of the effective stress $\tau_e^g$ is computed for each potential cleavage plane. If $\tau_e^g$ overcomes the \emph{cleavage plane strength} $T_\mathrm{max}^g$ of the considered plane, a trans-granular cohesive interface is introduced into the grain as a flat surface extending up to the boundaries of the grain. This assumption is justified by experimental observations of trans-granular cracking in several classes of metallics \cite{riedle1996,kumar2007,hughes2007,joo2012} and ceramics materials \cite{ii2005}.

\subsubsection{Inter- and trans-granular fracture competition}
The competition between the inter- and trans-granular modes of failure in polycrystalline materials is modelled by considering suitable sets of parameters entering the corresponding cohesive laws. Here, it is assumed that the coefficients $\alpha$, $\beta$ and $G_{II}/G_{I}$ do not differ between the two mechanisms, i.e.\ $\alpha^{gh}=\alpha^{g}=\alpha$, $\beta^{gh}=\beta^{g}=\beta$ and $G_{II}^{gh}/G_{I}^{gh}=G_{II}^{g}/G_{I}^{g}$ $\forall g=1,\dots,N_g$ and $\forall gh=1,\dots,N_i$, being $N_g$ the number of grains and $N_i$ the number of grain boundary interfaces. On the other hand, different ratios between the fracture energy $G_I^{gh}$ in mode I of the grain boundaries and the fracture energy $G_I^{g}$ in mode I of the cleavage planes are considered. More specifically, considering the relation $G_I=T_\mathrm{max}\delta u_n^{cr}/2$ (see Eq.(\ref{eq-Ch3:interface equations - work of separation}a)), it is easy to see that the ratio $\gamma_G\equiv{G_I^{g}}/{G_I^{gb}}$ between the two aforementioned fracture energies can be modified by changing $T_\mathrm{max}$ and/or $\delta u_n^{cr}$. In this thesis, to scale the fracture energy $G_I^{g}$ of the factor $\gamma_G$, i.e.\ $G_I^{g}=\gamma_G G_{I}^{gh}$, it is assumed that both the interface strength $T_\mathrm{max}$ and the critical displacement $\delta u_n^{cr}$ in mode I are scaled by a same quantity, i.e.\ $T_\mathrm{max}^{g}=\sqrt{\gamma_G} T_\mathrm{max}^{gh}$ and $\delta u_n^{cr,g}=\sqrt{\gamma_G} \delta u_n^{cr,gh}$. By the same fashion, it is easy to see that the the critical displacement $\delta u_s^{cr}$ in mode II is scaled by $\sqrt{\gamma_G}$, i.e.\ $\delta u_s^{cr,g}=\sqrt{\gamma_G} \delta u_s^{cr,gh}$.

The effect of the parameter $\gamma_G$ on the cohesive law is represented in Figures (\ref{fig-Ch3b:cohesive law}). Figure (\ref{fig-Ch3b:cohesive law}a) represents the tangential component $\tau_s$ as a function of $(\beta\delta u_s/\delta u_s^{cr})$ and $(\delta u_n/\delta u_n^{cr})$ according to Eq.(\ref{eq-Ch3:interface equations - cohesive law 2}) whereas Figure (\ref{fig-Ch3b:cohesive law}b) represents the normal component $\tau_n$ as a function of $(\beta\delta u_s/\delta u_s^{cr})$ and $(\delta u_n/\delta u_n^{cr})$ according to Eq.(\ref{eq-Ch3:interface equations - cohesive law 2}). In the figures, the red surfaces represent a reference cohesive law, e.g.\ associated to the behaviour of the grain boundaries, whereas the blue surfaces represent a scaled cohesive law with $\gamma_G < 1$.

\begin{figure}
\centering
	\begin{subfigure}{0.49\textwidth}
	\centering
	\includegraphics[width=\textwidth]{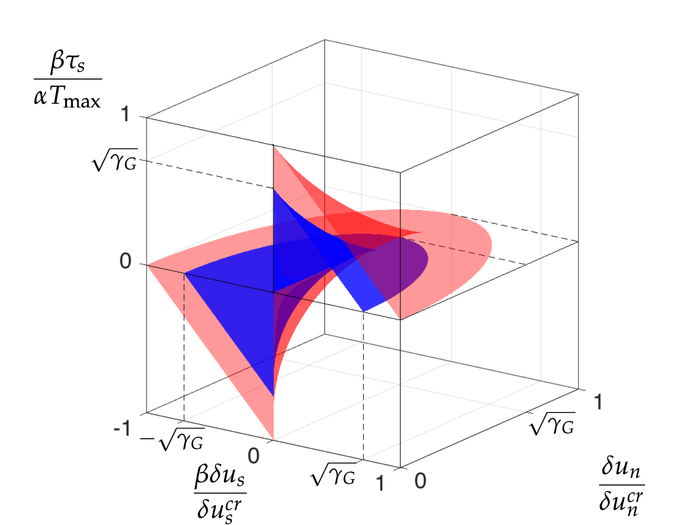}
	\caption{}
	\end{subfigure}
	\
	\begin{subfigure}{0.49\textwidth}
	\centering
	\includegraphics[width=\textwidth]{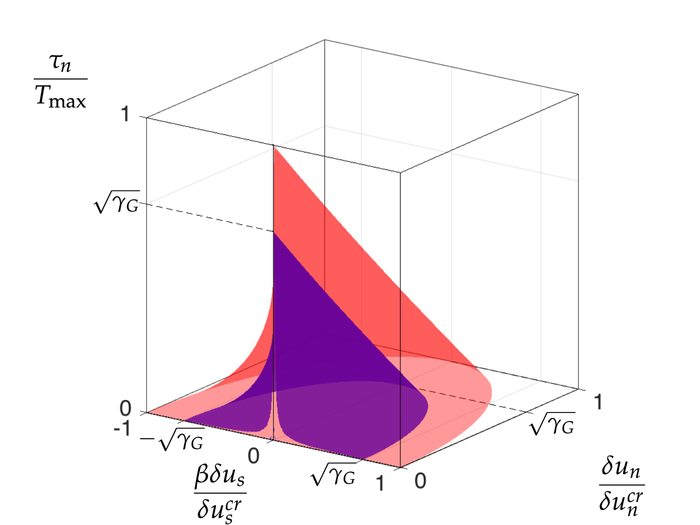}
	\caption{}
	\end{subfigure}
\caption{Schematic representation of (\emph{a}) the tangential and (\emph{b}) normal components of the two cohesive laws used in this work model inter- and trans-granular cracking. The red surfaces represent a reference cohesive law whereas the blu surfaces represent a scaled cohesive law with $\gamma_G < 1$.}
\label{fig-Ch3b:cohesive law}
\end{figure}

\section{Numerical discretisation and solution}\label{sec-Ch3b:discrete system}
The solution of the polycrystalline problems of inter- and trans-granular microcracking is obtained by numerically discretising Eqs.(\ref{eq-Ch3b:DBIE}) and (\ref{eq-Ch3b:SBIE}). The boundary $S^g$ of the generic grain $g$ is divided into non-overlapping elements according to the meshing strategy developed in Chapter (\ref{ch-EF}), where combined continuous and semi-discontinuous triangular and quadrangular elements were used to reduce the computational cost of the polycrystalline problems. The mesh size of the surface mesh is chosen such that the average element length $l_e$ is much smaller than characteristic length $L_{cz}$ of the cohesive zone, which can be estimated in terms of the material fracture toughness and the interface strength \cite{rice1968,espinosa2003b,tomar2004}.

The discretised version of the displacement boundary integral equations (\ref{eq-Ch3b:DBIE}) is computed using the discretisation technique and the meshing strategy combining triangular and quadrangular elements described in Section (\ref{sec-Ch3:meshing}). For the generic grain $g$, a linear system of equations of the form $\mathbf{H}^g\mathbf{U}^g=\mathbf{G}^g\mathbf{T}^g$ is obtained. The equations of the entire polycrystalline aggregate are then retrieved by writing such equations for each grain of the aggregate and by coupling them with the interface equations as discussed in Section (\ref{sec-Ch1: polycrystalline system of equations}). A system of equations of the form $\mathbf{M}(\mathbf{X},\lambda)=\mathbf{0}$ is obtained, where $\mathbf{X}$ collects the unknowns of the system and $\lambda$ is the load factor.

Unlike the displacement BIE, which are evaluated on the boundary of the grains, the stress boundary integral equations (\ref{eq-Ch3b:SBIE}) are evaluated at selected \emph{control} points within the grain. The discretised version of Eq.(\ref{eq-Ch3b:SBIE}) is then obtained as follows
\begin{equation}\label{eq-Ch3b:SBIE discrete}
\boldsymbol{\Sigma}^{g} =
\mathbf{G}^{\sigma,g}\mathbf{T}^{g}-
\mathbf{H}^{\sigma,g}\mathbf{U}^{g},
\end{equation}
where $\boldsymbol{\Sigma}^{g}$ contain the components of the stress tensor evaluated at the internal control points and the matrices $\mathbf{H}^{\sigma,g}$ and $\mathbf{G}^{\sigma,g}$ stem from the integration of the kernels $T_{mni}^{\sigma,g}(\mathbf{x},\mathbf{y})$ and $U_{mni}^{\sigma,g}(\mathbf{x},\mathbf{y})$, respectively, over the boundary $S^g$. It is worth noting that, if $n_v^g$ is the number of control points for the grain $g$ and the stress tensor is represented in Voigt notation, the matrices $\mathbf{H}^{\sigma,g}$ and $\mathbf{G}^{\sigma,g}$ have $6n_c^g$ rows and $3n_s^g$ columns, being $n_s^g$ the number of collocation points of grain $g$. As already mentioned, Eq.(\ref{eq-Ch3b:SBIE discrete}) are used in a post-processing stage and therefore the reordering due the boundary conditions enforced on the aggregate is not introduced.

In the next Section, the use of the stress equations (\ref{eq-Ch3b:SBIE discrete}) in combination with the solution of the polycrystalline system of equations is represented by a flow chart describing the algorithm for inter- and trans-granular cracking.

\subsection{Algorithm for inter- and trans-granular cracking}\label{ssec-Ch3b:algo}
Figure (\ref{tg-algo}) schematises the flow chart of the algorithm used in the present work to address inter- and trans-granular cracking in polycrystalline materials. At the beginning of the analysis, the load factor $\lambda$ is initialised to $0$, the load step counter $n$ is initialised to $1$ and the \emph{remesh} boolean variable is initialised to \texttt{.false.}. The polycrystalline morphology is then generated and the properties of the constituent grains and grain boundaries are loaded. The algorithm then enters the incremental loop that is governed by the counter $n$ and can be described by the following steps:
\begin{enumerate}
\item{
At the beginning of the load step $n$, the load factor is incremented by the load factor increment $\Delta \lambda_n$ chosen based on the convergence speed at the previous load step \cite{benedetti2013b}.
}
\item{
If the analysis has just started, i.e.\ $n=1$, or the remesh is \texttt{.true.}, the aggregate is discretised, the boundary element matrices are computed, the overall element system is assembled and the remesh variable is set to \texttt{.false.}.
}
\item{
At the $n$-th load step, the system of equation $\mathbf{M}(\mathbf{X}_n,\lambda_n)=\mathbf{0}$ is to be solved. The equilibrium solution $\mathbf{X}_n$ is obtained by employing the Newton-Raphson algorithm and the solver \texttt{PARDISO} \cite{pardiso1,pardiso2,pardiso3} as described in Section (\ref{sec-Ch1: polycrystalline system of equations}). During the Newton-Raphson search for the solution, the consistency of the grains interfaces is checked according to the procedure described in Ref.\ \cite{benedetti2013b}
}
\item{
The values of the stress tensor $\sigma_{mn}^g$ at the control points of each grain of the aggregate is computed by means of Eq.(\ref{eq-Ch3b:SBIE discrete}) and the possible occurrence of trans-granular cracking is then checked by comparing the local effective cleavage stress $\tau_e^g$ with the local threshold $T^{g}_\mathrm{max}$.
}
\item{
If the cleavage threshold is overcome, new cohesive interfaces are introduced along the cleavage planes of those grains undergoing trans-granular cracking. The cohesive properties of the newly introduced interfaces are those corresponding to trans-granular cracks and are in general different from the cohesive properties of the grain boundaries. At this point, the morphology has been modified and the boolean variable remesh is set to \texttt{.true.}. As an example, Figure (\ref{fig-Ch3b:split and remesh of a grain}a) shows the boundary mesh of a grain whose threshold condition for trans-granular cracking is fulfilled, Figure (\ref{fig-Ch3b:split and remesh of a grain}b) shows the cleavage plane (in darker green) that needs to be introduced into the grain and Figure (\ref{fig-Ch3b:split and remesh of a grain}c) shows the remesh of the two child grains originated from the cut of the grain in Figure (\ref{fig-Ch3b:split and remesh of a grain}a). The flow goes back to step $2$ in order to find the equilibrium solution of the new aggregate for the same load factor $\lambda_n$.
}
\item{
If no trans-granular cracks are introduced, the current load factor $\lambda_n$ is compared to the final load factor $\lambda_f$ and either the next load step is considered or the analysis is ended.
}
\end{enumerate}

\begin{figure}
\centering
	\begin{subfigure}{0.32\textwidth}
	\centering
	\includegraphics[width=\textwidth]{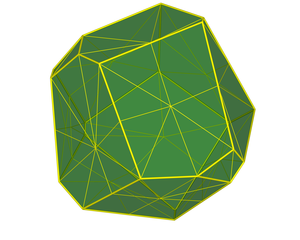}
	\caption{}
	\end{subfigure}
	\
	\begin{subfigure}{0.32\textwidth}
	\centering
	\includegraphics[width=\textwidth]{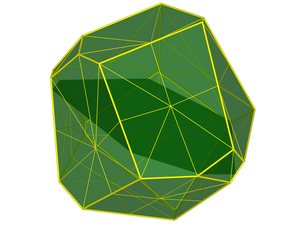}
	\caption{}
	\end{subfigure}
	\
	\begin{subfigure}{0.32\textwidth}
	\centering
	\includegraphics[width=\textwidth]{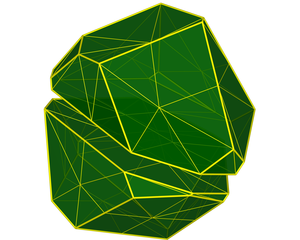}
	\caption{}
	\end{subfigure}
\caption{(\emph{a}) Boundary mesh of a grain at the point of undergoing cleavage cracking; (\emph{b}) cleavage plane (in darker green) that needs to be introduced into the grain; (\emph{c}) mesh of the two child grains originated from the grain in figure (\emph{a}).}
\label{fig-Ch3b:split and remesh of a grain}
\end{figure}

\clearpage
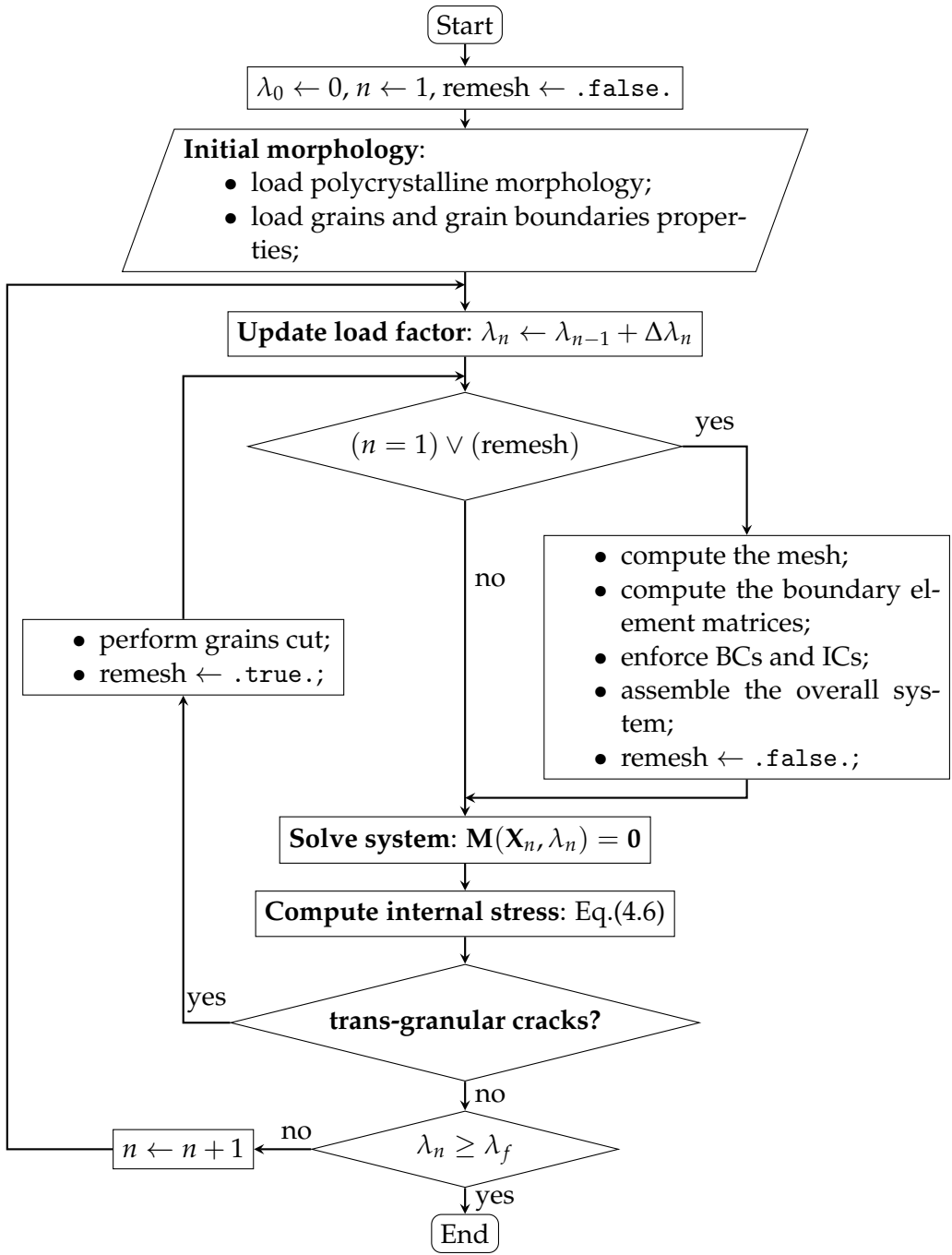
\begin{figure}
\begin{center}
\begin{tikzpicture}[node distance=1.5cm]
\node (start) at (0,0) [rectangle, rounded corners,text centered, draw=black] {Start};

\node (tstart) at (0,-0.9) [rectangle, text centered, draw=black] {$\lambda_0\gets0$, $n\gets1$, $\mathrm{remesh}\gets\texttt{.false.}$};

\node (ingeo) at (0,-2.5) [trapezium, trapezium left angle=70, trapezium right angle=110, text centered, draw=black]{
\begin{minipage}{8cm}
\textbf{Initial morphology}:
\begin{itemize}[noitemsep]
\item load polycrystalline morphology;
\item load grains and grain boundaries properties;
\end{itemize} 
\end{minipage}
};

\node (load_factor) at (0,-4.4) [rectangle, text centered, draw=black]{
\textbf{Update load factor}: $\lambda_{n}\gets\lambda_{n-1}+\Delta \lambda_{n}$
};

\node (n1_or_remesh) at (0,-6.0) [diamond,  aspect=4, text centered, draw=black]{
$(n=1)\lor(\mathrm{remesh})$
};

\node (sys) at (4, -9.0) [rectangle, text centered, draw=black] {
\begin{minipage}{5.5cm}
\begin{itemize}[noitemsep]
\item compute the mesh;
\item compute the boundary element matrices;
\item enforce BCs and ICs;
\item assemble the overall system;
\item $\mathrm{remesh}\gets\texttt{.false.}$;
\end{itemize} 
\end{minipage}
};

\node (sol) at (0,-11.6) [rectangle, text centered, draw=black]{
\textbf{Solve system}: $\mathbf{M}(\mathbf{X}_{n},\lambda_{n})=\mathbf{0}$
};

\node (compute_stress) at (0,-12.65) [rectangle, text centered, draw=black]{
\textbf{Compute internal stress}: Eq.(\ref{eq-Ch3b:SBIE discrete})
};

\node (trans_cut) at (0,-14.2) [diamond,  aspect=4, text centered, draw=black]{
\textbf{trans-granular cracks?}
};

\node (cut_remesh) at (-4, -9.0) [rectangle, draw=black] {
\begin{minipage}{4.3cm}
\begin{itemize}[noitemsep]
\item perform grains cut;
\item $\mathrm{remesh}\gets\texttt{.true.}$;
\end{itemize} 
\end{minipage}
};


\node (end_of_analysis) at (0,-16.0) [diamond,  aspect=4, text centered, draw=black]{
\textbf{$\lambda_{n}\ge \lambda_f$}
};

\node (update_n) at (-4,-16.0) [rectangle, text centered, draw=black]{
$n\gets n+1$
};

\node (end) at (0,-17.2) [rectangle, rounded corners,text centered, draw=black] {End};

\draw [arrow] (start) -- (tstart);
\draw [arrow] (tstart) -- (ingeo);
\draw [arrow] (ingeo) -- (load_factor);
\draw [arrow] (load_factor) -- (n1_or_remesh);
\draw [arrow] (n1_or_remesh) -| node [near start, above] {yes} (sys);
\draw [arrow] (n1_or_remesh) -- node [near start, right] {no} (sol);
\draw [arrow] (sys) |-  (0,-11);
\draw [arrow] (sol) -- (compute_stress);
\draw [arrow] (compute_stress) -- (trans_cut);
\draw [arrow] (trans_cut) -| node [near start,above] {yes} (cut_remesh);
\draw [arrow] (cut_remesh) |- (0,-5.);
\draw [arrow] (trans_cut) -- node [right] {no} (end_of_analysis);
\draw [arrow] (end_of_analysis) -- node [near start,above] {no} (update_n);
\draw [arrow] (update_n) -- (-6.5,-16.0) |- (0,-3.7);
\draw [arrow] (end_of_analysis)  -- node [right] {yes} (end);

\end{tikzpicture}
\end{center}
\caption{Flow chart of the algorithm for inter- and trans-granular cracking.}
\label{tg-algo}
\end{figure}

\clearpage
\section{Numerical tests}\label{sec-Ch3b:numerical tests}
In this tests, polycrystalline SiC aggregates with hexagonal crystal lattice are considered. For hexagonal 6H SiC polytypes, the preferred cleavage plane is the basal plane \cite{clayton2015} identified by $(0001)$ Miller indices. The elastic properties of the crystals and the cohesive properties of the grain boundaries and the cleavage planes are listed in Table (\ref{tab-Ch3b:SiC properties}).

In this thesis, only preliminary results of the formulation are presented. Figure (\ref{fig-Ch3b:mcTG stress averages}a) shows a 10-grain polycrystalline aggregate with ASTM grain size $G=12$. In Figure (\ref{fig-Ch3b:mcTG stress averages}a), the small reference systems inside each grain represents the orientation of the crystallographic lattice where the normal to the basal plane is indicated by the longer arrow. For simplicity, in the following tests, the threshold condition for activating the trans-granular cracking is checked at the centroid of each grain of the aggregate.

Consistently with the notation of Section (\ref{ssec-Ch3: micro-cracking}), the aggregate is loaded along the $x_3$ direction by means of uniaxial boundary conditions prescribed as $u_in_i=\lambda/2$ over the top and bottom faces and $u_in_i=0$ over the lateral faces, whereas the remaining loading directions are kept traction-free. Two values of the ratio $\gamma_G={G_I^{g}}/{G_I^{gb}}$ are considered for these analyses, namely $\gamma_G=1$ and $\gamma_G=1/4$. Figure (\ref{fig-Ch3b:mcTG stress averages}b) shows the macro-stress response $\Sigma_{33}$ as a function of the load factor $\lambda$ corresponding to the two values of $\gamma_G$. As expected, the response corresponding to a lower fracture energy of the cleavage planes, that is $\gamma_G=1/4$, is characterised by a smaller value of the maximum macro-stress reached during the loading history.

The effect of the ratio $\gamma_G$ on the micro-cracking pattern and on the damage distribution is reported in Figures (\ref{fig-Ch3:mcTG-def}) and (\ref{fig-Ch3:mcTG-dam}), respectively. Figures (\ref{fig-Ch3:mcTG-def}\emph{a}), (\ref{fig-Ch3:mcTG-def}\emph{c}) and (\ref{fig-Ch3:mcTG-def}\emph{e}) report the micro-cracking pattern at the load steps $\lambda=0.009$, $\lambda=0.011$ and $\lambda=0.025$, respectively, corresponding to $\gamma_G=1$; in this case, since the two mechanisms have the same fracture energy and the cleavage planes are not favourably oriented with respect to the loading direction, the crack path occur along the grain boundaries of the aggregate.
On the other hand, as shown in Figures (\ref{fig-Ch3:mcTG-def}\emph{b}), (\ref{fig-Ch3:mcTG-def}\emph{d}) and (\ref{fig-Ch3:mcTG-def}\emph{f}), by reducing the fracture energy of the cleavage planes, that is by choosing a value $\gamma_G=1/4$, the cleavage planes start to fail in correspondence to a lower value of the load factor, for which the stress on the grain boundaries is not high enough to initiate the damage. However, the three-dimensional random orientation of the grains does not allow a perfect propagation of the trans-granular crack within the aggregate and, as a consequence, partial grain boundaries failure is necessary to accommodate the crack propagation and the final failure of the aggregate. Similarly, Figures (\ref{fig-Ch3:mcTG-dam}\emph{a}), (\ref{fig-Ch3:mcTG-dam}\emph{c}) and (\ref{fig-Ch3:mcTG-dam}\emph{e}) show the inter-granular damage distribution at the load steps $\lambda=0.009$, $\lambda=0.011$ and $\lambda=0.025$, respectively, corresponding to $\gamma_G=1$. Figures (\ref{fig-Ch3:mcTG-dam}\emph{b}), (\ref{fig-Ch3:mcTG-dam}\emph{d}) and (\ref{fig-Ch3:mcTG-dam}\emph{f}) show the damage distribution corresponding to $\gamma_G=1/4$ at the same load steps. By looking at Figures (\ref{fig-Ch3:mcTG-dam}\emph{a}) and (\ref{fig-Ch3:mcTG-dam}\emph{b}), it is clear that the damage over a cleavage plane initiates in correspondence of a much lower value of the load factor, and, due to the local softening of failing cleavage plane, the corresponding stress redistribution and the three-dimensional random orientation of the grains, the crack path for this choice of $\gamma_G$ is characterised by inter- and trans-granular failure.

\begin{table}[ht]
\begin{center}
\caption{Elastic and cohesive properties of SiC polycrystalline aggregates.}
\label{tab-Ch3b:SiC properties}
\begin{tabular}{llll}
\hline
\hline
domain&property&component&value\\
\hline
Bulk crystals&elastic constants&$c_{1111}$, $c_{2222}$&502\\
&[$10^{9}\,\mathrm{N}/\mathrm{m}^{2}$]&$c_{3333}$&565\\
&&$c_{1122}$&95\\
&&$c_{1133}$, $c_{2233}$&96\\
&&$c_{2323}$, $c_{1313}$&169\\
&&$c_{1212}$&$(c_{1111}-c_{1122})/2$\\
\hline
Grain&interface strength [MPa]&$T_\mathrm{max}^{gh}$&500\\
boundaries&\multirow{2}{*}{cohesive law constants [-]}&$\alpha^{gh}$&1\\
&&$\beta^{gh}$&$\sqrt{2}$\\
&critical displacements&$\delta u_n^{cr,gh}$&$7.8089\cdot10^{-2}$\\
&jumps [$\mu\mathrm{m}$]&$\delta u_s^{cr,gh}$&$1.5618\cdot10^{-1}$\\
\hline
Cleavage&interface strength [MPa]&$T_\mathrm{max}^{g}$&$\sqrt{\gamma_G}\cdot T_\mathrm{max}^{gh}$\\
planes&\multirow{2}{*}{cohesive law constants [-]}&$\alpha^{g}$&$\alpha^{gh}$\\
&&$\beta^{g}$&$\beta^{gh}$\\
&critical displacements&$\delta u_{n}^{cr,g}$&$\sqrt{\gamma_G}\cdot\delta u_{n}^{cr,gh}$\\
&jumps [$\mu\mathrm{m}$]&$\delta u_{s}^{cr,g}$&$\sqrt{\gamma_G}\cdot\delta u_{s}^{cr,gh}$\\
\hline
\hline
\end{tabular}
\end{center}
\end{table}

\begin{figure}
\centering
	\begin{subfigure}{0.49\textwidth}
	\centering
	\includegraphics[width=\textwidth]{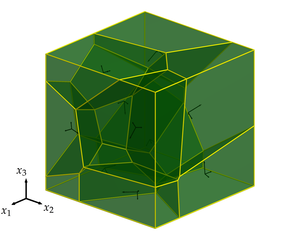}
	\caption{}
	\end{subfigure}
	\
	\begin{subfigure}{0.49\textwidth}
	\centering
	\includegraphics[width=\textwidth]{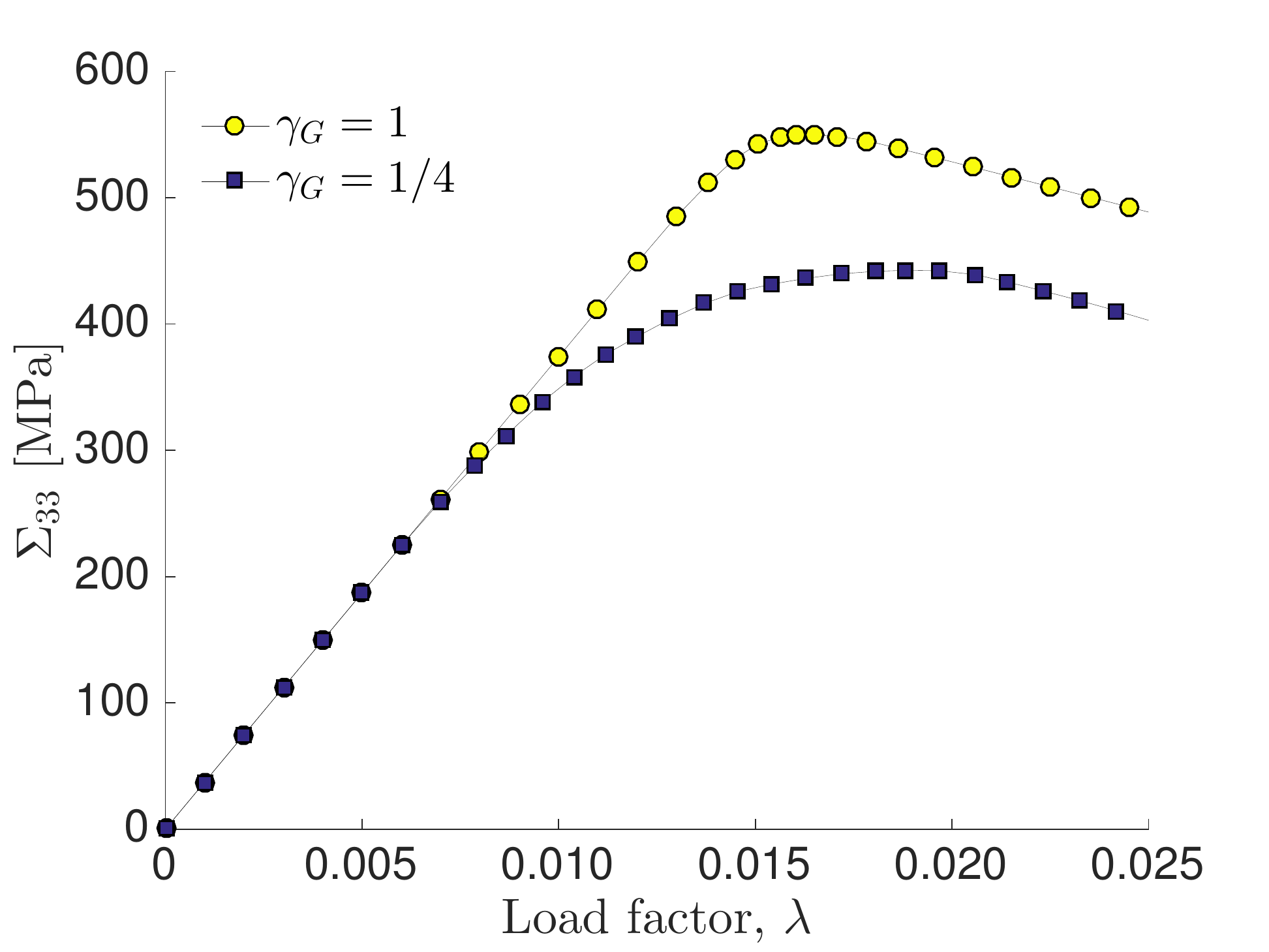}
	\caption{}
	\end{subfigure}
\caption{(\emph{a}) 10-grain polycrystalline morphology with ASTM grain size $G=12$; (\emph{b}) Volume stress averages $\Sigma_{33}$ as a function of the load factor $\lambda$ for the two values of the considered ratio $\gamma_G=G_{I}^g/G_{I}^{gh}$ between the fracture energy $G_{I}^g$ of the cleavage planes and the fracture energy $G_{I}^{gh}$ of the grain boundaries.}
\label{fig-Ch3b:mcTG stress averages}
\end{figure}

\begin{figure}
\centering
	\begin{subfigure}{0.45\textwidth}
	\centering
	\includegraphics[width=\textwidth]{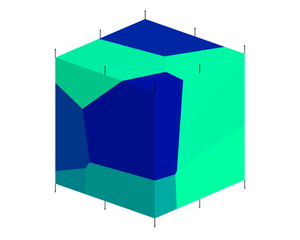}
	\caption{$\lambda=0.009$}
	\end{subfigure}
	\
	\begin{subfigure}{0.45\textwidth}
	\centering
	\includegraphics[width=\textwidth]{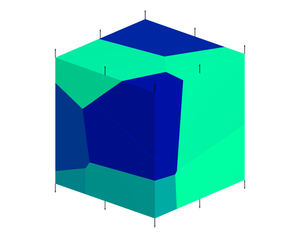}
	\caption{$\lambda=0.009$}
	\end{subfigure}
	\
	\begin{subfigure}{0.45\textwidth}
	\centering
	\includegraphics[width=\textwidth]{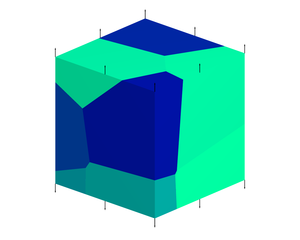}
	\caption{$\lambda=0.011$}
	\end{subfigure}
	\
	\begin{subfigure}{0.45\textwidth}
	\centering
	\includegraphics[width=\textwidth]{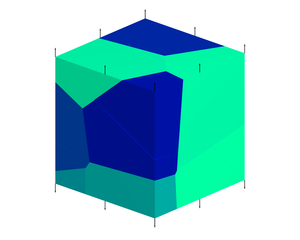}
	\caption{$\lambda=0.011$}
	\end{subfigure}
	\
	\begin{subfigure}{0.45\textwidth}
	\centering
	\includegraphics[width=\textwidth]{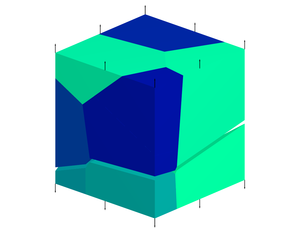}
	\caption{$\lambda=0.025$}
	\end{subfigure}
	\
	\begin{subfigure}{0.45\textwidth}
	\centering
	\includegraphics[width=\textwidth]{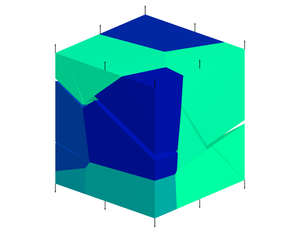}
	\caption{$\lambda=0.025$}
	\end{subfigure}
\caption{Micro-cracking snapshots at the loading steps (\emph{a},\emph{b}) $\lambda=0.009$, (\emph{c},\emph{d}) $\lambda=0.011$ and (\emph{e},\emph{f}) $\lambda=0.025$. (\emph{a},\emph{c},\emph{e}) $\gamma_G=1$. (\emph{b},\emph{d},\emph{f}) $\gamma_G=1/4$.}
\label{fig-Ch3:mcTG-def}
\end{figure}

\begin{figure}
\centering
	\begin{subfigure}{0.45\textwidth}
	\centering
	\includegraphics[width=\textwidth]{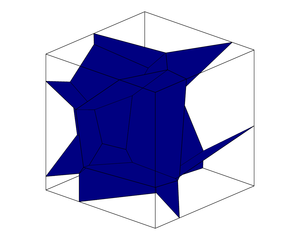}
	\caption{$\lambda=0.009$}
	\end{subfigure}
	\
	\begin{subfigure}{0.45\textwidth}
	\centering
	\includegraphics[width=\textwidth]{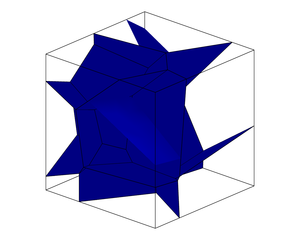}
	\caption{$\lambda=0.009$}
	\end{subfigure}
	\
	\begin{subfigure}{0.45\textwidth}
	\centering
	\includegraphics[width=\textwidth]{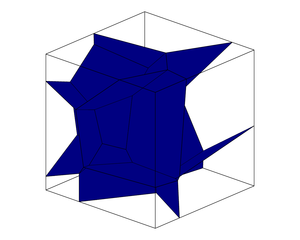}
	\caption{$\lambda=0.011$}
	\end{subfigure}
	\
	\begin{subfigure}{0.45\textwidth}
	\centering
	\includegraphics[width=\textwidth]{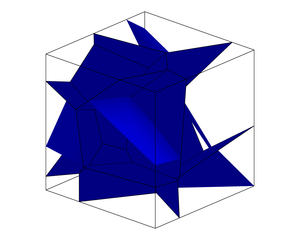}
	\caption{$\lambda=0.011$}
	\end{subfigure}
	\
	\begin{subfigure}{0.45\textwidth}
	\centering
	\includegraphics[width=\textwidth]{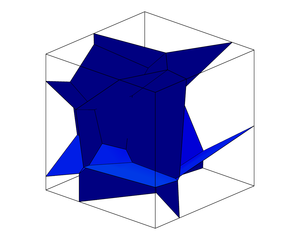}
	\caption{$\lambda=0.025$}
	\end{subfigure}
	\
	\begin{subfigure}{0.45\textwidth}
	\centering
	\includegraphics[width=\textwidth]{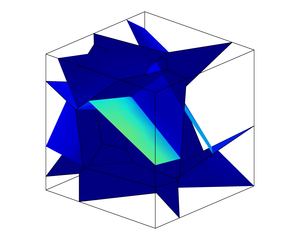}
	\caption{$\lambda=0.025$}
	\end{subfigure}
	\\
	\vspace{5pt}
	\begin{subfigure}{0.49\textwidth}
	\centering
	\includegraphics[width=\textwidth]{./Ch3/images/mc/colorbar.png}
	\caption{}
	\end{subfigure}
\caption{Micro-cracking snapshots of the damage distribution at the loading steps (\emph{a},\emph{b}) $\lambda=0.009$, (\emph{c},\emph{d}) $\lambda=0.011$ and (\emph{e},\emph{f}) $\lambda=0.025$. (\emph{a},\emph{c},\emph{e}) $\gamma_G=1$. (\emph{b},\emph{d},\emph{f}) $\gamma_G=1/4$. (\emph{g}) Colormap of the damage.}
\label{fig-Ch3:mcTG-dam}
\end{figure}

\section{Discussion and future developments}\label{sec-Ch3b:discussion}
In this Chapter, a numerical scheme for addressing the competition between inter- and trans-granular crack propagation has been developed and tested. Although the numerical tests involved a polycrystalline aggregate with a rather small number of constituent grains and subject to one type of boundary conditions only, the results showed that the developed formulation is able to capture the behaviour of polycrystalline aggregates when the two mechanisms are active. The next step of this formulation is to consider polycrystalline morphologies with a much larger number of grains in order to avoid the boundary effects of the finiteness of the container and to investigate the coupling between the statistical variability of polycrystalline micro-structure and the competition between the two mechanisms. Furthermore, different sets of boundary conditions can also be tested to study the polycrystalline response along different directions and eventually how the existence of a representative volume element can be affected by the simultaneous presence of the two mechanisms.

\clearpage

\chapter{Grain boundary formulation for crystal plasticity}\label{ch-CP}
In this Chapter, a three-dimensional grain-boundary formulation for small-strain crystal plasticity in polycrystalline aggregates is presented. The method is developed starting from single grain crystal plasticity equations and implemented in polycrystalline aggregates \cite{benedetti2016,benedetti2016b,gulizzi2016c}. It is based on the use of a suitable set of boundary integral equations for modelling the individual grains, which are represented as anisotropic elasto-plastic domains. In the boundary integral framework, crystal plasticity is modelled resorting to an initial strains approach, and specific aspects related to the integration of strongly singular volume integrals in the anisotropic elasto-plastic grain-boundary equations are discussed and suitably addressed. In the polycrystalline case, Laguerre-type micro-morphologies are discretised using robust non-structured boundary and volume meshes. A general grain-boundary incremental/iterative algorithm, embedding rate-dependent flow and hardening rules for crystal plasticity, is developed and discussed. The method has been assessed through several numerical simulations, for both single and polycrystalline aggregates, which confirm its robustness and accuracy and suggest directions for further developments. The formulation retains the key feature of the grain boundary formulation described in Chapter (\ref{ch-intro}), that is the expression of the micro-mechanical problem in terms of grain-boundary variables only, namely inter-granular displacements and tractions, which results in a reduction of the total number of degrees of freedom and may be appealing in a multi-scale framework.

\section{Introduction}\label{sec-Ch4:intro}
Crystal plasticity represents one of the main deformation mechanisms active in grains of a polycrystal and consists of plastic slip over well defined crystallographic planes. The phenomenon is characterised by its inherent anisotropy, induced by the structure of the crystalline lattice itself. Several numerical and computational methods have been proposed in the literature for the analysis of plasticity in single crystals and crystal aggregates.
A detailed comparison among three different theoretical frameworks, i.e.\ conventional continuum plasticity, discrete dislocation plasticity and nonlocal plasticity, is reported by Needleman \cite{needleman2000}, who strongly underlines the importance of computational studies.
The importance of computational approaches addressing multiple scales and involving the explicit representation of the micro-structural details characterising the crystal plasticity phenomenon is highlighted in the work of Dawson \cite{dawson2000}.
A very complete review of Crystal Plasticity Finite Element Methods (CPFEMs) can be found in \cite{roters2010} where a wide range of continuum-based techniques and contributions, with applications in fields as diverse as texture evolution, creep, nano indentation, forming and others, is discussed.

Single crystal plastic slip studies can be traced back to the works of Taylor et al.\ \cite{taylor1923,taylor1925,taylor1932}.
Classical contributions in the field are those by Mandel \cite{mandel1965}, Hill \cite{hill1966}, Rice \cite{rice1971}, Hill and Rice \cite{hill1972}, Asaro and Rice \cite{asaro1977}, Asaro \cite{asaro1983} (also addressing the mechanics of polycrystals), Peirce et al.\ \cite{peirce1982,peirce1983} and Bassani et al.\ \cite{wu1991,bassani1991}. In Ref.\ \cite{cuitino1993}, Cuitino and Ortiz reviewed some computational models for single crystal plasticity and proposed a statistical mechanical model based on dislocation mechanics.
Both rate-dependent (RD) \cite{rice1971,peirce1983,ling2005} and rate-independent (RI) \cite{anand1996} models have been proposed and compared \cite{miehe2001}. Much research has also been focused on the development of crystal plasticity models for polycrystalline aggregates \cite{roters2010}. Several 3D CPFEMs were initially developed to assess texture development in polycrystals. In several studies, the aggregate was represented with simple morphologies and coarse meshes, with individual grains often modelled by a single finite element. Despite such approximations, the models were quite accurate in predicting texture evolution. In \cite{wu1996}, a Taylor-type polycrystal plasticity model was tested in large strain reverse torsion tests, with the different slip hardening laws presented by Asaro and Needleman \cite{asaro1985}, by Harren et al.\ \cite{harren1989}, by Bassani and Wu \cite{bassani1991} and by Kalidindi et al.\ \cite{kalidindi1992}.

Both 2D \cite{kovavc2005} and fully 3D \cite{sarma1998} polycrystalline models, with the explicit representation of individual grains, have been developed, using a rate-dependent CPFEM formulations. Quilici and Cailletaud \cite{quilici1999} developed a CPFEM framework that was subsequently used for the analysis of stress/strain fields within 3D polycrystalline aggregates in the small strains regime \cite{barbe2001a,barbe2001b}. Voronoi tessellations have been used to represent the microstructure and a visco-plastic crystallographic constitutive model, with isotropic and kinematic hardening, was employed. Aggregates up to 200 grains were analysed and the FETI (Finite Element Tearing and Interconnecting) method \cite{farhat1994} was employed, in conjunction with High Performance Computing, to tackle the high computational requirements of 3D computations. The same strategy was later employed by Diard et al.\ \cite{diard2005}, who analysed micro-plasticity in hexagonal close-packed materials, with as realistic as possible microstructures. Sarma and Radhakrishnan \cite{sarma2004} applied CPFEM for studying the hot deformation of grains in polycrystalline aluminum. They report that, due to the time-consuming nature of such simulations, it was possible to consider only a limited number of specially constructed microstructures. Zeghadi et al.\ \cite{zeghadi2007} performed, in a CPFEM framework, a statistical analysis on several different 3D morphologies generated starting from the same 2D surface. Aggregates with around $40$ grains were considered and the plastic strain fields over the reference 2D surface, induced by different internal 3D morphologies of the assessed specimen, were investigated. It was found that large fluctuations in the equivalent plastic slip level are induced by different internal 3D morphologies, under the same external loading conditions, thus advocating the need of fully 3D studies over simplified pseudo 3D (or 2D columnar) studies.

Similar studies, focused on the comparison between experimental and computational results, with the aim of identifying possible sources of error and better calibrating microstructural constitutive parameters were performed by H\'{e}ripr\'{e} et al.\ \cite{heripre2007}, Musienko et al.\ \cite{musienko2007}, Zhang et al.\ \cite{zhang2015} and Pinna et al.\ \cite{pinna2015}. Musienko et al.\ \cite{musienko2007} analyzed an experimentally reconstructed copper specimen with approximately 100 grains. CPFEM was used and the results of fully 3D and columnar 2D (pseudo-3D) simulations were compared to assess the bias introduced by the simplified morphologies. It was found that the computational results of the fully 3D mesh were in good agreement with the experimental data, while only some traces of the experimental behaviour were retained in the extended 2D simulations. Zhang et al.\ \cite{zhang2015} compared, in terms of stress-strain curves and lattice reorientation, the experimental results obtained from a tensile test of a Ti-5Al-2.5Sn specimen with those obtained from numerical CPFEM simulations using 2D columnar and 3D microstructures. The stress-strain curve was used to calibrate the parameters of the crystal plasticity constitutive model. Using the lattice reorientation maps, the authors showed that fully 3D microstructures better reproduced the experimental observations with respect to the pseudo 3D microstructures. Pinna et al.\ \cite{pinna2015} simulated the deformation of a FCC aluminium alloy AA5052 at high temperatures using a structured mesh where each cubic volume element had its own crystallographic orientation, and a Fe-30wt$\%$Ni alloy using a two dimensional mesh with plane strain assumption. Besides the qualitative agreement between the numerical results and the experimental measurements in terms of stress-strain curves, the authors reported the limitations of simplified CPFEM models to capture the intra-granular strain distribution and suggested the use of fully three-dimensional models. Barbe and Quey \cite{barbe2011} presented a CPFEM model for 3D polycrystalline microstructures, including polycrystal-to-polycrystal diffusive transformations. Schneider et al.\ \cite{schneider2010,schneider2013} investigated the mechanical behaviour of $\alpha$Fe-Cu composites under large plastic deformations in simple tension and compression, using 3D FEM simulations (ABAQUS) with an elasto-visco-plastic material model. Crack initiation and evolution in polycrystalline aggregates is addressed in \cite{simonovski2015}, where the cohesive-zone type contact is implemented in a CPFEM framework employing Bassani-Wu hardening laws. In \cite{robert2015}, a comparison between the self-consistent, finite element and spectral methods is carried out in the context of homogenisation of polycrystalline aggregates with an elastic-viscoplastic behaviour. Zhang et al.\ \cite{zhang2015b} studied the anisotropic mechanical response of a rolled AA1050 aluminium plate with respect to the loading direction using five different crystal plasticity models, showing the ability of CPFEM to qualitatively reproduce the experimental findings. Zeng et al.\ \cite{zeng2015} developed a smoothed crystal plasticity finite element framework for modelling two-dimensional single crystals and polycrystals. Kim and Yoon \cite{kim2015} simulated the necking behaviour of AA 6022-T4 sheet using a CPFEM formulation involving four different continuum damage mechanics models, discussing the performance of such damage models when compared to experimental tests.

More advanced crystal plasticity phenomenological models, able to capture different deformation mechanisms besides slip, such as twinning in HCP crystals or non-Schmid effects in BCC crystals, are still under investigation.As a few recent examples,  Abdolvand et al.\ \cite{abdolvand2011} implemented in a finite element formulation a crystal plasticity model comprising twinning as well as slip in HCP materials, in order to simulate the deformation of a rolled Zircaloy-2 element. Wang et al.\ \cite{wang2013} proposed a physically-based crystal plasticity model including also de-twinning for HCP crystals and used it in a self-consistent model to simulate cyclic loading in a magnesium alloy. Lim et al.\ \cite{lim2013} developed a BCC crystal plasticity finite element model and employed it to simulate the deformation of tantalum oligo-crystals \cite{lim2014}. The authors compared numerical and experimental results finding good agreement in terms of both in-plane and out-of-plane strain fields and lattice rotations. An interesting point raised by the authors was the lower agreement between predicted and measured fields at the grain boundaries, which prompted the authors to recommend more detailed grain boundary models in future studies.

The short review given above confirms the maturity of the CPFEM for the investigation of plasticity in crystalline solids and also the current engineering interest toward the development of new or more enriched models. On the other hand, while several boundary element formulations for elasto-plastic isotropic problems are present in the literature, see e.g.\ Refs.\ \cite{brebbia2012,banerjee1981,aliabadi2002,bonnet1996,cisilino1998,benallal2002,mallardo2004,mallardo2009}, the present Chapter represents the first boundary element application to anisotropic or crystal plasticity in polycrystalline materials. A boundary element approach may in fact result in accurate stress/strain predictions at reduced discretisation effort.

In the present Chapter, a novel three-dimensional grain-boundary formulation for small strains crystal plasticity in single crystals and polycrystalline materials is presented for the first time. The boundary integral equations given in Chapter (\ref{ch-intro}) are suitably modified to account for general anisotropic elasto-plasticity and are given in Section (\ref{sec-Ch4:formulation}). The presence of plastic flow is taken into account through an \emph{initial strains} approach, which is commonly adopted in boundary element formulations for addressing elasto-plastic problems \cite{banerjee1981,aliabadi2002,brebbia2012}. The free terms involved in the formulation are explicitly given, Section (\ref{sec-Ch4:BIE}), and the strategy for numerical integration of the  strongly singular volume integrals is discussed in Section (\ref{sec-Ch4:integration}). The adopted phenomenological continuum-based crystal plasticity description is discussed in Section (\ref{sec-Ch4:cp modelling}).

The grain-boundary \emph{polycrystalline} implementation for crystal plasticity, which is based on the formulation given in Chapter (\ref{ch-intro}), is given in Section (\ref{sec-Ch4:cp algorithm + poly}) and the incremental/iterative grain-boundary algorithm for crystal plasticity is presented in a rate-dependent framework in Section (\ref{ssec-Ch4:polycrystalline iia}). The salient feature of the formulation is the expression of the problem in terms of inter-granular variables only, namely grain-boundary displacements and tractions, also in presence of an interior plastic flow, which allows a remarkable reduction in the order of the numerical system, especially for polycrystalline analysis. In Section (\ref{sec-Ch4:numerical tests}), numerical tests are performed both for single crystals and crystal aggregates, and they confirm the potential of the technique. Directions of further investigation are discussed in Section (\ref{sec-Ch4:discussion}).
\section{Anisotropic elasto-plastic boundary element formulation}\label{sec-Ch4:formulation}
In the present Section, the different ingredients of the boundary integral formulation of the crystal plasticity problem are presented. To maintain a unified notation, the basic equations are presented in rate form. This is a common procedure for time-dependent problems, such as creep or visco-plasticity. Rate-independent crystal plasticity problems, on the other hand, could be formulated in purely incremental terms, due to the lack of time-dependent effects. However, also rate-independent problems can be generally associated to a time-like parameter that plays the role of an ordering parameter, such as the load factor $\lambda$. Keeping this in mind, it is possible to retain a unique rate notation for both classes of problems.

Considered a generic domain $V^g \subset \mathbb{R}^3$ representing a single grain with specific symmetry undergoing elasto-plastic deformations, the proposed formulation is based on the use of the set of boundary integral equations representing the displacements and stresses of the points of the domain in terms of boundary integrals of the displacements and tractions of points on the surface $S^g$ of the domain itself, see Section (\ref{sec-Ch1:grain boundary integral equations}) of Chapter (\ref{ch-intro}). When a region $V^{p,g}\subseteq V^g$ undergoes plastic slip, such equations are modified by the presence of some volume integral terms. The salient feature of the method is that only the boundary points carry degrees of freedom, irrespective of the presence of internal plastic regions, allowing a considerable reduction in the size of the solved system, which may be appealing in computational terms.

In the following Sections, the previous outline is further developed and attention is devoted to the fundamental boundary integral equations, to the expressions of the integral kernels appearing in the equations and their numerical integration, and to the form of the discrete algebraic systems.

\subsection{Boundary integral equations}\label{sec-Ch4:BIE}
Let us consider the generic grain $V^g$ and its boundary $S^g=\partial V^g$. When plastic deformations are considered within the context of the \emph{small strains theory}, the \emph{total} strain rate $\dot{\eps}_{ij}^{g}(\mathbf{x})$ can be expressed as the sum of the \emph{elastic} strain rate $\dot{\eps}_{ij}^{e,g}(\mathbf{x})$ and \emph{plastic} strain rate $\dot{\eps}_{ij}^{p,g}(\mathbf{x})$, i.e.\
\begin{equation}\label{eq-Ch4:total strain}
\dot{\eps}_{ij}^g(\mathbf{x})=\frac{1}{2}\left[\frac{\pd \dot{u}_i^g}{\pd x_j}(\mathbf{x})+\frac{\pd \dot{u}_j^g}{\pd x_i}(\mathbf{x})\right]=
\dot{\eps}_{ij}^{e,g}(\mathbf{x})+\dot{\eps}_{ij}^{p,g}(\mathbf{x}),
\end{equation}
where it is recalled that $u_{i}$ denotes the displacements components. The presence of plastic strains can be addressed in boundary element formulations by adopting an \emph{initial strain} or \emph{initial stress} approach \cite{banerjee1981,aliabadi2002,brebbia2012}: in the present work, the initial strain formulation is used, as it appears particularly suitable to the form of the crystal plasticity constitutive equations, see Section (\ref{sec-Ch4:cp modelling}).

When some initial \textit{plastic strain} $\eps_{ij}^{p,g}$ is present in $V^{p,g}\subseteq V^g$, the \emph{displacement boundary integral equations} (\ref{eq-Ch1:DBIE-2}) are modified and rewritten, in rate form, as
\begin{multline}\label{eq-Ch4:DBIE-2}
c_{pi}(\mathbf{y})\dot{u}_i^g(\mathbf{y})+
\dashint_{S^g} T_{pi}^g(\mathbf{x},\mathbf{y})\dot{u}_i^g(\mathbf{x})\dd S(\mathbf{x})=\\
\int_{S^g} U_{pi}^g(\mathbf{x},\mathbf{y})\dot{t}_i^g(\mathbf{x})\dd S(\mathbf{x})+
\int_{V^{p,g}} \Sigma_{pij}^g(\mathbf{x},\mathbf{y})\dot{\eps}_{ij}^{p,g}(\mathbf{x})\dd V(\mathbf{x})
\end{multline}
where, consistently with Eq.(\ref{eq-Ch1:DBIE-2}), $\mathbf{y}$ denote the \emph{collocation} point, $\mathbf{x}$ denotes the boundary and volume integration point and $u_i(\mathbf{x})$ and $t_i(\mathbf{x})$ are the components of boundary displacements and tractions, respectively. The kernels appearing in Eq.(\ref{eq-Ch4:DBIE-2}) are given in Eq.(\ref{eq-Ch1:DBIE kernels}) of Chapter (\ref{ch-intro}). $\dashint$ denotes the Cauchy principal value integral and $c_{pi}(\mathbf{y})\dot{u}_i(\mathbf{y})$ are free terms. The volume integral in Eq.(\ref{eq-Ch4:DBIE-2}) is performed over the interior region $V^{p,g}\subseteq V^g$ undergoing plastic deformation, i.e.\ where $\dot{\eps}_{ij}^{p,g}\ne 0$. Similarly to the integral containing the kernel $U_{pi}^{g}(\mathbf{x},\mathbf{y})$, the volume integral, although containing a singular kernel of order $r^{-2}$, is a weakly singular integral. In fact, upon considering a reference system centered at $\mathbf{y}$ and taking the limit $\mathbf{x}\rightarrow\mathbf{y}$, the order of the volume element $\dd V(\mathbf{x})$ is $r^2\dd r$ whereas the kernel $\Sigma_{pij}^g(\mathbf{x},\mathbf{y})$ is of the order $r^{-2}$, being $r$ is the distance between the point $\mathbf{x}$ and $\mathbf{y}$.

By considering an interior point $\mathbf{y}\in V^g$, the strain field inside the grain $g$ is computed by suitably taking the derivatives of Eq.(\ref{eq-Ch4:DBIE-2}) with respect to the coordinates $y_i$ of the collocation point $\mathbf{y}$. In rate form, the strain boundary integral equations (\ref{eq-Ch1:EBIE}) are modified as
\begin{multline}\label{eq-Ch4:EBIE}
\dot{\eps}_{pq}^g(\mathbf{y})+
\int_{S^g} T_{pqi}^{\eps,g}(\mathbf{x},\mathbf{y})\dot{u}_i^g(\mathbf{x})\dd S(\mathbf{x})=\\
\int_{S^g} U_{pqi}^{\eps,g}(\mathbf{x},\mathbf{y})\dot{t}_i^g(\mathbf{x})\dd S(\mathbf{x})+
\dashint_{V^{p,g}} \Sigma_{pqij}^{\eps,g}(\mathbf{x},\mathbf{y})\dot{\eps}_{ij}^{p,g}(\mathbf{x})\dd V(\mathbf{x})+
f_{pqij}^\eps\dot{\eps}_{ij}^{p,g}(\mathbf{y}),
\end{multline}
where the kernels $U_{pqi}^{\eps,g}(\mathbf{x},\mathbf{y})$, $\Sigma_{pqij}^g(\mathbf{x},\mathbf{y})$ and $T_{pqi}^{\eps,g}(\mathbf{x},\mathbf{y})$ are given in Eq.(\ref{eq-Ch1:EBIE kernels}). In this case, unlike the DBIE (\ref{eq-Ch4:DBIE-2}), the volume integral must be evaluated in the Cauchy principal value sense since, as $\mathbf{x}\rightarrow\mathbf{y}$, the order of the volume element $\dd V(\mathbf{x})$ is $r^2\dd r \dd S$ and the order of the kernel $\Sigma_{pqij}^{\eps,g}(\mathbf{x},\mathbf{y})$ is $r^{-3}$. Consequently, after the limiting process, the free terms $f_{pqij}^\eps\dot{\eps}_{ij}^{g,p}(\mathbf{y})$ remain and the coefficients $f_{pqij}^\eps$ are computed as
\begin{equation}\label{eq-Ch4:EBIE - free terms}
f_{pqij}^{\eps}=-\frac{1}{2}\int_{S_1}[\Sigma_{pij}(\uv{r})n_q(\uv{r})+\Sigma_{qij}n_p(\uv{r})]\dd S(\uv{r}),
\end{equation}
where $S_1$ is the three dimensional unit sphere and $\uv{r}$ is the unit vector centered at $\mathbf{y}$. The numerical treatment of the volume integrals in Eq.(\ref{eq-Ch4:EBIE}) is addressed in Section \ref{sec-Ch4:integration}.

The boundary integral representation for the stress rate $\dot{\sigma}_{mn}(\mathbf{y})$ at the generic interior point $\mathbf{y}\in V^g$ is obtained from Eq.(\ref{eq-Ch4:EBIE}) using the constitutive relationships $\dot{\sigma}_{mn}(\mathbf{y})=c_{mnpq}\dot{\eps}_{pq}^{e,g}(\mathbf{y})=c_{mnpq}[\dot{\eps}_{pq}^g(\mathbf{y})-\dot{\eps}_{pq}^{p,g}(\mathbf{y})]$. In rate form, the stress boundary integral representation reads
\begin{multline}\label{eq-Ch4:SBIE}
\dot{\sigma}_{mn}^g(\mathbf{y})+
\int_{S^g} T_{mni}^{\sigma,g}(\mathbf{x},\mathbf{y})\dot{u}_i^g(\mathbf{x})\dd S(\mathbf{x})=\\
\int_{S^g} U_{mni}^{\sigma,g}(\mathbf{x},\mathbf{y})\dot{t}_i^g(\mathbf{x})\dd S(\mathbf{x})+
\dashint_{V^{p,g}} \Sigma_{mnij}^{\sigma,g}(\mathbf{x},\mathbf{y})\dot{\eps}_{ij}^{p,g}(\mathbf{x})\dd V(\mathbf{x})+
f_{mnij}^\sigma\dot{\eps}_{ij}^{p,g}(\mathbf{y}),
\end{multline}
where the kernels $U_{mni}^{\sigma,g}(\mathbf{x},\mathbf{y})$, $T_{mni}^{\sigma,g}(\mathbf{x},\mathbf{y})$ and $\Sigma_{mnij}^{\sigma,g}(\mathbf{x},\mathbf{y})$ are given in Eq.(\ref{eq-Ch1:SBIE kernels}) and the coefficients of the free terms are
\begin{equation}\label{eq-Ch4:SBIE - free terms}
f_{mnij}^\sigma=-c_{mnpq}\left[\int_{S_1}\Sigma_{pij}(\uv{r})n_q(\uv{r})\dd S(\uv{r})+\delta_{pi}\delta_{qj}\right].
\end{equation}

\subsection{Numerical discretisation}\label{sec-Ch4:discretisation}
In the initial strains approach \cite{brebbia2012,aliabadi2002}, the crystal plasticity problem is closed considering the displacement and stress boundary integral equations, Eq.(\ref{eq-Ch4:DBIE-2}) and Eq.(\ref{eq-Ch4:SBIE}), together with the crystal plasticity constitutive equations given in Section (\ref{sec-Ch4:cp modelling}). In order to write the discrete counterpart of the boundary integral equations, the following steps are followed:
\begin{itemize}
\item{
The boundary $S^g$ and the volume $V^g$ of each grain domain are subdivided into sets of non-overlapping surface and volume elements:
	\begin{itemize}
	\item{
	the surface is discretised using the strategy developed in Chapter (\ref{ch-EF}), where a mesh combining continuous/semi-discontinuous triangular and quadrangular two-dimensional elements is implemented in order to tackle and reduce the high computational costs of the polycrystalline problem
	}
	\item{
	the volume of the grain is divided into linear tetrahedral elements and the plastic strain field is assumed constant and equal to the value at the centroid of each element. Such elements are usually referred to as \emph{sub-parametric} elements\footnote{In \emph{sub-parametric} elements the order of interpolation of the unknown fields is lower than the order of interpolation of the geometry.};
	}
	\end{itemize}
}
\item{
The unknown fields, namely displacements and tractions on the surface and plastic strains in the volume, are expressed in terms of nodal values and suitable shape functions;
}
\item{
Eq.(\ref{eq-Ch4:DBIE-2}) is written for every boundary discretisation node and it is then numerically integrated;
}
\item{
Eq.(\ref{eq-Ch4:SBIE}) is written at the centroids of each tetrahedral volume element and it is then numerically integrated.
}
\end{itemize}

The discrete version of the DBIE written in rate form coincide with Eq.(\ref{eq-Ch1:DBIE discrete}) plus an additional term accounting for the plastic deformation of the volume as follows
\begin{equation}\label{eq-Ch4:DBIE discrete}
\mathbf{H}^{g}\dot{\mathbf{U}}^{g} = \mathbf{G}^{g}\dot{\mathbf{T}}^{g}+\mathbf{D}^{g}\dot{\boldsymbol{\mathcal{E}}}^{p,g},
\end{equation}
where $\dot{\mathbf{U}}^{g}$ and $\dot{\mathbf{T}}^{g}$ contain the components of the nodal values of the boundary displacements and tractions, the matrices $\mathbf{H}^{g}$ and $\mathbf{G}^{g}$ stem from the integration over $S^g$ of the kernels $\wtilde{U}_{pi}^g(\mathbf{x},\mathbf{y})$ and $\wtilde{T}_{pi}^g(\mathbf{x},\mathbf{y})$ in the local boundary reference system (see Eq.(\ref{eq-Ch1:DBIE-2 local RS}) in Section(\ref{sec-Ch1:grain boundary integral equations})), respectively, $\dot{\boldsymbol{\mathcal{E}}}^{p,g}$ contain the components of the values of plastic strains at the tetrahedral elements' centroids and the matrix $\mathbf{D}^{g}$ is generated from the numerical integration of $\Sigma_{pij}^g(\mathbf{x},\mathbf{y})$ over the interior volume of the grain $g$ undergoing plastic flow.

To solve the crystal plasticity problem, Eq.(\ref{eq-Ch4:DBIE discrete}) has to be used with the discretised version of the stress integral equations (\ref{eq-Ch4:SBIE}) that, written for each internal point, read
\begin{equation}\label{eq-Ch4:SBIE discrete}
\dot{\boldsymbol{\Sigma}}^{g} =
\mathbf{G}^{\sigma,g}\dot{\mathbf{T}}^{g}-
\mathbf{H}^{\sigma,g}\dot{\mathbf{U}}^{g}+
\mathbf{D}^{\sigma,g}\dot{\boldsymbol{\mathcal{E}}}^{p,g},
\end{equation}
where $\dot{\boldsymbol{\Sigma}}^{g}$ contain the components of the nodal values of stress at the centroids of the tetrahedra, the matrices $\mathbf{H}^{\sigma,g}$ and $\mathbf{G}^{\sigma,g}$ stem from the integration of the kernels $T_{mni}^{\sigma,g}(\mathbf{x},\mathbf{y})$ and $U_{mni}^{\sigma,g}(\mathbf{x},\mathbf{y})$, respectively, over the boundary $S^g$ and $\mathbf{D}^{\sigma,g}$ stems from the numerical integration of $\Sigma_{mnij}^{\sigma,g}(\mathbf{x},\mathbf{y})$ over the interior volume of the grain $g$ undergoing plastic flow.

Eqs.(\ref{eq-Ch4:DBIE discrete}) and (\ref{eq-Ch4:SBIE discrete}) must be used in conjunction with the suitable boundary conditions and the constitutive crystal plasticity equations for the solution of the grain-boundary elasto-plastic problem, as will be described in Section (\ref{sec-Ch4:cp algorithm + poly}). The interested reader is referred to references \cite{banerjee1981,aliabadi2002,brebbia2012} for further details about the boundary element method.

\subsection{Numerical integration and singular integrals}\label{sec-Ch4:integration}
To obtain the discrete equations introduced above, it is necessary to accurately evaluate the \emph{singular} integrals appearing in Eqs.(\ref{eq-Ch4:DBIE-2}) and (\ref{eq-Ch4:SBIE}). While the singular integrals appearing in the displacement equations, Eq.(\ref{eq-Ch4:DBIE-2}), have been widely considered in the literature, the integration of the singular kernels $\Sigma_{pqij}^{\eps,g}(\mathbf{x},\mathbf{y})$ and $\Sigma_{mnij}^{\sigma,g}(\mathbf{x},\mathbf{y})$ \emph{for the anisotropic case} was recently presented in \cite{benedetti2015}. In this Section, the procedures used for the evaluation of the singular integrals are briefly recalled, and the strategy used for treating the integrals involving $\Sigma_{mnij}^{\sigma,g}(\mathbf{x},\mathbf{y})$ is described. The same approach applies to the integrals involving $\Sigma_{pqij}^{\eps,g}(\mathbf{x},\mathbf{y})$.

Considering Eq.(\ref{eq-Ch4:DBIE-2}), the weak singularity induced on the boundary by $U_{ij}^g(\mathbf{x},\mathbf{y})$ is treated subdividing the singular elements into triangles and representing each sub-triangle as a quadrangle collapsed into the singular point; the strong singularity induced by $T_{ij}^g(\mathbf{x},\mathbf{y})$, treated as a Cauchy singular value, is tackled using rigid body considerations \cite{banerjee1981,aliabadi2002}; the volume terms appearing in Eq.(\ref{eq-Ch4:DBIE-2}) may give rise to weak singularities only when the plastic region reaches the boundary: in this case, the interested volume cells are subdivided into sub-tetrahedra, which afterwards are represented as 8-node cubes collapsed into the singular point. In this way, the Jacobian of the integration cancels the singularity.

In the strain and stress integral equations, Eqs.(\ref{eq-Ch4:EBIE}) and (\ref{eq-Ch4:SBIE}), since constant volume cells are used in the implementation, the collocation points $\mathbf{y}$ always fall in the \emph{interior} domain, being the centroids of the volume cells. For such a reason, the boundary integrals may become at most nearly singular, and this happens when the plastic region reaches the boundary. Nearly singular integrals are simply evaluated through boundary element subdivision. The \emph{volume} integrals, on the contrary, must be treated in the Cauchy singular value sense, being the kernels $\Sigma_{pqij}^{\eps,g}(\mathbf{x},\mathbf{y})$ and $\Sigma_{mnij}^{\sigma,g}(\mathbf{x},\mathbf{y})$ strongly singular in the volume.

In the present work, the starting point for the evaluation of the volume Cauchy singular values is the work by Gao and Davies \cite{gao2000}, who, for isotropic materials, have shown that the integral of $\Sigma_{mnij}^{\sigma,g}(\mathbf{x},\mathbf{y})$ over the singular volume cell $V^c$ can be expressed as
\begin{equation}\label{eq-Ch4:gao and davies}
\dashint_{V^c}\Sigma_{mnij}^{\sigma,g}(\mathbf{x},\mathbf{y})\dd V(\mathbf{x})=
\int_{S^c}\Sigma_{mnij}^{\sigma,g}(\mathbf{x},\mathbf{y})(x_p-y_p) n_p(\mathbf{x}) \log(r) \dd S(\mathbf{x})
\end{equation}
as long as the condition
\begin{equation}\label{eq-Ch4:gao and davies condition}
\int_{S_1}\Sigma_{mnij}^{\sigma,g}(\uv{r})\dd S(\uv{r})=0,
\end{equation}
where $S_1$ is the unit sphere centered at $\mathbf{y}$, is satisfied. In Eq.(\ref{eq-Ch4:gao and davies}), $S^c=\pd V^c$ is the external boundary of the singular volume cell. Gao and Davies \cite{gao2000} proved Eq.(\ref{eq-Ch4:gao and davies condition}) for isotropic materials, taking into account the explicit analytical form of the isotropic fundamental solutions. Using the spherical harmonics expansion developed in Chapter (\ref{ch-FS}), the fact that the integral of the spherical harmonics $Y_\ell^m(\uv{r})$ over the unit sphere $\int_{S_1}Y^\ell_m(\uv{r})\dd S(\uv{r})=0$ for $m,\ell\ne0$ and that $P^2_0(0)=0$ ($P_\ell^m(0)$ is the associated Legendre function), it is possible to demonstrate that Eq.(\ref{eq-Ch4:gao and davies condition}) also holds for general anisotropic materials.

In the next Section, the crystal plasticity constitutive model linking Eqs.(\ref{eq-Ch4:DBIE discrete}) and (\ref{eq-Ch4:SBIE discrete}) and employed in this thesis is discussed.

\section{Crystal plasticity phenomenological modelling}\label{sec-Ch4:cp modelling}
In crystallography, a \emph{slip system} family describes a set of crystallographically equivalent \emph{slip planes} associated with a family of crystallographically equivalent \emph{slip directions} for which dislocation motion can occur, producing plastic deformation. Different slip systems are associated with different classes of crystal lattice. Generally, slip planes are those with the highest surface density of atoms, i.e.\ close-packed planes, while slip directions are those corresponding to highest linear density of atoms, i.e.\ close-packed directions. Once a specific slip plane and a specific slip direction over it are identified, a certain resolved shear stress is needed to trigger the \emph{slip}, which is an important deformation mechanism in crystals.

Crystal plasticity can be generally modeled through \emph{phenomenological} or \emph{physically-based} approaches \cite{roters2010}. Moreover, the crystallographic slip within single crystals can be idealised as either \emph{rate-dependent} or \emph{rate-independent}. In the present work, a phenomenological rate-dependent description is adopted. In the present Section, an introductory functional description of crystal slip is briefly recalled.

The slip over a specific slip plane $\alpha$ identified by the unit normal vector $n_i^\alpha$, and along a specific direction identified by the unit vector $s_i^\alpha$, is \emph{activated} by the \emph{Schmid resolved shear stress}
\begin{equation}\label{eq-Ch4:resolved shear}
\tau^{\alpha}=s^{\alpha}_i\sigma_{ij}n^{\alpha}_j,
\end{equation}
which triggers the corresponding \emph{slip} or \emph{shear rate}
\begin{equation}\label{eq-Ch4:flow rule}
\dot{\gamma}^{\alpha}=\Psi^{\alpha}\left(\tau^{\alpha},\tau^{\alpha}_c\right),
\end{equation}
where $\Psi^{\alpha}$ expresses a specific \emph{flow rule} and $\tau^{\alpha}_c$ is a certain \emph{critical} resolved shear stress, which plays the role of a \emph{state variable} for the slip system $\alpha$.
$\tau^{\alpha}_c$ generally depends on the \emph{accumulated slips} $\gamma^\beta={\int_{t}{|\dot{\gamma}^{\beta}|\dd t}}$ and on the slip rates $\dot{\gamma}^{\beta}$ on \emph{all} the $N_s$ slip systems and its evolution can be expressed as
\begin{equation}\label{eq-Ch4:hardening law}
\dot{\tau}^{\alpha}_c=\Xi^{\alpha}\left(\gamma^\beta,\dot{\gamma}^{\beta}\right)
\end{equation}
where $\Xi^{\alpha}$ expresses some specific \emph{hardening law} and $\beta=1,\dots,N_g$, meaning that the evolution of $\tau_c^\alpha$ of the generic slip system $\alpha$ is a function of slip rate and accumulated slip of all the slip systems. To avoid possible confusion, the sum of the accumulated slips $\gamma^\alpha$ over all the $N_s$ slip systems, which is defined as $\gamma\equiv\sum_{\alpha}\gamma^\alpha$,  will be referred to as the \emph{overall cumulative slip}. Eqs.(\ref{eq-Ch4:resolved shear}-\ref{eq-Ch4:hardening law}) are valid $\forall \alpha=1,\cdots,N_s$.

Once the slip rates are known, the plastic strains rate at a point in the crystal can be expressed as
\begin{equation}\label{eq-Ch4:plasticstrain}
\dot{\eps}_{ij}^{p}=\frac{1}{2}\sum_{\alpha=1}^{N_s}\dot{\gamma}^{\alpha}\left({n_i^{\alpha}\:s_j^{\alpha}+s_i^{\alpha}\:n_j^{\alpha}}\right).
\end{equation}
The developed formulation can be coupled with different types of flow and hardening rules. The flow and hardening rules used in this Chapter are reported in Section (\ref{app-Ch4:flow N hardening laws}).

\section{The grain-boundary incremental-iterative algorithm for crystal plasticity in polycrystalline aggregates}\label{sec-Ch4:cp algorithm + poly}
Following Refs.\ \cite{telles1981,brebbia2012}, in the framework of the initial strain formulation the elasto-plastic grain-boundary problem can be expressed in terms of \emph{accumulated} values of displacements $\mathbf{U}^g$, tractions $\mathbf{T}^g$ and plastic strains $\boldsymbol{\eps}^{p,g}$. Eq.(\ref{eq-Ch4:DBIE discrete}) can then be rewritten as
\begin{equation}\label{eq-Ch4:DBIE discrete + BCs}
\mathbf{H}^{g}\mathbf{U}^{g} = \mathbf{G}^{g}\mathbf{T}^{g}+\mathbf{D}^{g}\boldsymbol{\mathcal{E}}^{p,g}
\quad\xrightarrow{\mathrm{BCs}}\quad
\mathbf{A}^{g}\mathbf{X}^g=\mathbf{C}^g\mathbf{Y}^g(\lambda)+\mathbf{D}^{g}\boldsymbol{\mathcal{E}}^{p,g}
\end{equation}
where the column reordering induced by the enforcement of the boundary conditions (BCs) is indicated. In Eq.(\ref{eq-Ch4:DBIE discrete + BCs}), the matrix $\mathbf{A}^g$ contains a selection of columns from $\mathbf{H}^g$ and $\textbf{G}^g$ corresponding to unknown components of boundary displacements and tractions, collected in the unknowns vector $\mathbf{X}^g$; the matrix $\mathbf{C}^g$ is a linear combinations of the columns from $\mathbf{H}^g$ and $\mathbf{G}^g$ multiplying the known components of boundary displacements and tractions, collected in the vector $\mathbf{Y}^g$, and whose value is governed by the load factor $\lambda$. Similarly, Eq.(\ref{eq-Ch4:SBIE discrete}) can be rewritten, in terms of accumulated values, as
\begin{equation}\label{eq-Ch4:SBIE discrete accumulated}
\boldsymbol{\Sigma}^{g} =
\mathbf{A}^{\sigma,g}\mathbf{X}^g-
\mathbf{C}^{\sigma,g}\mathbf{Y}^g(\lambda)+
\mathbf{D}^{\sigma,g}\boldsymbol{\mathcal{E}}^{p,g},
\end{equation}
where, similarly to the matrices $\mathbf{A}^g$ and $\mathbf{C}^g$, the matrices $\mathbf{A}^{\sigma,g}$ and $\mathbf{C}^{\sigma,g}$ contains a suitable combination of columns from $\mathbf{H}^{\sigma,g}$ and $\textbf{G}^{\sigma,g}$ corresponding to unknown and known, respectively, components of boundary displacements and tractions. However, it is recalled that these equations are generally used in a \emph{post-processing} stage and the distinction between known and unknown values of boundary fields is adopted for the sake of notation.

\subsection{Numerical discretisation of polycrystalline aggregates}\label{ssec-Ch4: polycrystalline system of equations}
Similarly to the assemblage procedure of the polycrystalline aggregate described in Section (\ref{sec-Ch1: polycrystalline system of equations}) of Chapter (\ref{ch-intro}), Eq.(\ref{eq-Ch4:DBIE discrete + BCs}) is written for each grain of the aggregate, i.e.\ for $g=1,\dots,N_g$, and coupled to an appropriate set of interface equations. The system of equations of the whole aggregate is then written as
\begin{equation}\label{eq-Ch4:polycrystalline DBIE discrete}
\mathbf{M}(\mathbf{X},\lambda,\boldsymbol{\mathcal{E}}^p)=\left\{
\begin{array}{c}
\mathbf{A}\mathbf{X}-\mathbf{B}(\lambda)-\mathbf{D}\boldsymbol{\mathcal{E}}^p\\
\mathbf{I}\mathbf{X}
\end{array}
\right\}=\mathbf{0}
\end{equation}
where the matrix $\mathbf{A}$ and the vector $\mathbf{B}$ are given in Eq.(\ref{eq-Ch1:polycrystalline DBIE matrix A b X}). Unlike the system of the equations (\ref{eq-Ch1:polycrystalline DBIE discrete}), in which non-linear behaviours within the grains are not considered, in Eq.(\ref{eq-Ch4:polycrystalline DBIE discrete}), the presence of plastic slip within the polycrystalline aggregate is represented by the term $\mathbf{D}\boldsymbol{\mathcal{E}}^p$ where
\begin{equation}\label{eq-Ch4:polycrystalline DBIE matrix D eps}
\mathbf{D}=
\left[\begin{array}{cccc}
\mathbf{D}^{1}& \mathbf{0}&\cdots&\mathbf{0}\\
\mathbf{0}& \mathbf{D}^2&\cdots&\mathbf{0}\\
\vdots&\vdots&\ddots&\vdots\\
\mathbf{0}&\mathbf{0}&\cdots&\mathbf{D}^{N_g}
\end{array}\right]
\quad\mathrm{and}\quad
\boldsymbol{\mathcal{E}}^p=
\left[\begin{array}{c}
\boldsymbol{\mathcal{E}}^{p,1}\\
\boldsymbol{\mathcal{E}}^{p,2}\\
\vdots\\
\boldsymbol{\mathcal{E}}^{p,N_g}
\end{array}\right].
\end{equation}
In the system of equations (\ref{eq-Ch4:polycrystalline DBIE discrete}), the interface equations implement perfect bonding between contiguous grains and are given by the relation $\mathbf{I}\mathbf{X}=\mathbf{0}$, where the matrix $\mathbf{I}$ contains just zeros and $\pm1$'s enforcing the interface conditions $\delta\wtilde{u}_i^{gh}=0$ and $\wtilde{t}_i^g=\wtilde{t}_i^h$ for pairs of interface points belonging to the generic adjacent grains $g$ and $h$.

In the solution algorithm described in Section (\ref{ssec-Ch4:polycrystalline iia}), the system of equations (\ref{eq-Ch4:polycrystalline DBIE discrete}) is used in conjunction with the stress equations (\ref{eq-Ch4:SBIE discrete accumulated}) which are used to evaluate the stress at each interior test point within each grain.
It is worth noting that, being Eq.(\ref{eq-Ch4:SBIE discrete accumulated}) used after the solution of Eq.(\ref{eq-Ch4:polycrystalline DBIE discrete}), in a post-process stage within each iteration, it can be used for each grain separately, as $\mathbf{U}^g$ and $\mathbf{T}^g$ are known. However, for the sake of notation, the computation of the stress at the interior points of each grain of the polycrystalline aggregate is indicated as
\begin{equation}\label{eq-Ch4:polycrystalline SBIE discrete}
\boldsymbol{\Sigma}(\mathbf{X},\lambda,\boldsymbol{\mathcal{E}}^{p}) =
\mathbf{A}^{\sigma}\mathbf{X}-
\mathbf{B}^{\sigma}(\lambda)+
\mathbf{D}^{\sigma}\boldsymbol{\mathcal{E}}^{p},
\end{equation}
where
\begin{equation*}
\boldsymbol{\Sigma}=
\left[\begin{array}{c}
\boldsymbol{\Sigma}^{1}\\
\boldsymbol{\Sigma}^{2}\\
\vdots\\
\boldsymbol{\Sigma}^{N_g}
\end{array}\right],\
\mathbf{A}^{\sigma}\mathbf{X}=
\left[\begin{array}{c}
\mathbf{A}^{\sigma,1}\mathbf{X}^1\\
\mathbf{A}^{\sigma,2}\mathbf{X}^2\\
\vdots\\
\mathbf{A}^{\sigma,N_g}\mathbf{X}^{N_g}
\end{array}\right],
\end{equation*}
\begin{equation*}
\mathbf{B}^{\sigma}=
\left[\begin{array}{c}
\mathbf{C}^{\sigma,1}\mathbf{Y}^1\\
\mathbf{C}^{\sigma,2}\mathbf{Y}^2\\
\vdots\\
\mathbf{C}^{\sigma,N_g}\mathbf{Y}^{N_g}
\end{array}\right],\
\mathbf{D}^{\sigma}\boldsymbol{\mathcal{E}}^p=
\left[\begin{array}{c}
\mathbf{D}^{\sigma,1}\boldsymbol{\mathcal{E}}^{p,1}\\
\mathbf{D}^{\sigma,2}\boldsymbol{\mathcal{E}}^{p,2}\\
\vdots\\
\mathbf{D}^{\sigma,N_g}\boldsymbol{\mathcal{E}}^{p,N_g}
\end{array}\right].
\end{equation*}

The incremental-iterative algorithm for the solution of the crystal plasticity problem is described in the next Section.

\subsection{Polycrystalline incremental-iterative system solution}\label{ssec-Ch4:polycrystalline iia}
The incremental-iterative algorithm used to solve the crystal plasticity problem is described in Algorithm (\ref{alg-Ch4:CPBEM}), which is detailed in the following. In the Algorithm, the subscript $n$ refers to the $n$-th load step whereas the superscript $k$ refers to the $k$-th iterative step of the considered load step. Moreover, the superscript $g$ refers to the $g$-th grain of the aggregate, $\mathbf{y}_e$ is the position of the centroid of $e$-th volume cell and $\alpha$ denotes the $\alpha$-th slip system of the grain.

The Algorithm starts at Step (\ref{step:initiate}) at which the load factor is initialised to 0 and the index $n$ is initialised to 1. The Algorithm runs until the load factor $\lambda$ reaches the given final load factor $\lambda_f$. Once the solution at the load step $n-1$ corresponding to the load factor $\lambda_{n-1}$ has been found, the load increment $\dot{\lambda}\Delta t_n$ is introduced, being $\dot{\lambda}$ the \emph{loading rate} and $\Delta t_n$ the time step, and the new load factor $\lambda_n$ is computed. At the beginning of the analysis, the plastic strain $\dot{\boldsymbol{\mathcal{E}}}^{p}_{n=0}$ throughout the whole polycrystalline aggregate as well as the plastic slip $\gamma_{n=0}^{\alpha}(\mathbf{y}_e^g)$ for each slip system of the volume cells of each grain are set to zero, i.e.\ $\dot{\boldsymbol{\mathcal{E}}}^{p}_{n=0}=\mathbf{0}$ and $\gamma_{n=0}^{\alpha}(\mathbf{y}_e^g)=0$. On the other hand, the initial value of the critical resolved shear stress $\tau_{c,n=0}^{\alpha}(\mathbf{y}_e^g)$ is set to an initial value depending on the considered material.

At Step (\ref{step:iterative algorithm}), the iterative cycle starts and the solution of the system of equations (\ref{eq-Ch4:polycrystalline DBIE discrete}) and (\ref{eq-Ch4:polycrystalline SBIE discrete}) coupled to the constitutive relations (\ref{eq-Ch4:resolved shear}), (\ref{eq-Ch4:flow rule}) and (\ref{eq-Ch4:hardening law}) is computed. At Step (\ref{step:solve DBIE}), Eq.(\ref{eq-Ch4:polycrystalline DBIE discrete}) is solved for the current $(k+1)$-th iteration as a function of $\lambda_n$ and of the value of the plastic strain at the $k$-th iteration, which is obtained as the sum of the last converged value $\boldsymbol{\mathcal{E}}^p_{n-1}$ plus the increment at the $k$-th step given by the product of the time step $\Delta t_n$ and the last computed value of the plastic strain rate $\dot{\boldsymbol{\mathcal{E}}}^{p,k}_{n}$. It is worth noting that in the Algorithm, $\boldsymbol{\mathcal{E}}^p_{n-1}$ collects the converged plastic strain components up to the load level $\lambda_{n-1}$, but not including the last increment $\dot{\lambda}\Delta t_n$.

The updated values of boundary displacements and tractions, given by $\mathbf{X}_n^{k+1}$ and by the prescribed values governed by $\lambda_n$, are then used in Step (\ref{step:compute stress}) to compute the internal stress components at the volume cells' centroids collected in $\boldsymbol{\Sigma}_n^{k+1}$, which in turn are used to obtain the resolved shear stress $\tau^{\alpha,k+1}_n(\mathbf{y}_e^g)$ for each slip system $\alpha$ of the centroids of all the volume cells of each grain (Step (\ref{step:resolved shear stress})).

At Step (\ref{step:slip rate}), Eq.(\ref{eq-Ch4:flow rule}) is used to compute the slips rate $\dot{\gamma}_n^{\alpha,k+1}(\mathbf{y}_e^g)$ and the accumulated slips $\gamma_n^{\alpha,k+1}(\mathbf{y}_e^g)$ for each slip system of the volume cells' centroids of each grain. Similarly, at Step (\ref{step:hardening law}) the critical shear stress rate $\dot{\tau}_{c,n}^{\alpha,k+1}(\mathbf{y}_e^g)$ is computed using Eq.(\ref{eq-Ch4:hardening law}) and the previously computed values of the slips rate and the accumulated slips. At Step (\ref{step:plasticstrain}), the rate of the plastic strain components $\dot{\eps}_{ij,n}^{p,k+1}(\mathbf{y}_e^g)$ is computed using Eq.(\ref{eq-Ch4:plasticstrain}).

At this point, the overall cumulative slip $\gamma_n^{k+1}(\mathbf{y}_e^g)$ is computed for the volume cells of each grain and the convergence condition is checked. The convergence conditions is given in terms of the difference between two consecutive cumulative slip between $\gamma_n^{k}(\mathbf{y}_e^g)$ compared to a tolerance value $\epsilon_{conv}$.
If the convergence is met then the accumulated slips, the critical shear stresses and the accumulated plastic strains are updated and stored and another load increment can be introduced and computed if needed. If the convergence is not met, the iteration counter $k$ is updated and another iteration is started.

\clearpage
\begin{algorithm}
\caption{CPBEM algorithm}
\label{alg-Ch4:CPBEM}
\begin{algorithmic}[1]
\Statex{\textbf{START}}
\State{$\lambda_0 \leftarrow 0$, $n \leftarrow 1$;}\label{step:initiate}

\While{($\lambda_n < \lambda_f$)}

	\State{$\lambda_n \leftarrow \lambda_{n-1}+\dot{\lambda}\Delta t_n$;}
	\State{$k \leftarrow 0$;}
	\State{$\mathrm{converged} \leftarrow \texttt{.false.}$;}

	\While{(\texttt{.not.}$\mathrm{converged}$)}\label{step:iterative algorithm}
	\State{
	Solve:
	}\label{step:solve DBIE}
	\NoNumber{
	$\mathbf{M}(\mathbf{X}_n^{k+1},\lambda_n,\boldsymbol{\mathcal{E}}^p_{n-1}+\dot{\boldsymbol{\mathcal{E}}}^{p,k}_{n}\Delta t_n)=\mathbf{0}$;
	\Comment{Eq.(\ref{eq-Ch4:polycrystalline DBIE discrete})}
	}
	\State{
	Compute the internal stress field:
	}\label{step:compute stress}
	\NoNumber{
	$\boldsymbol{\Sigma}_n^{k+1}=\boldsymbol{\Sigma}(\mathbf{X}_n^{k+1},\lambda_n,\boldsymbol{\mathcal{E}}^p_{n-1}+\dot{\boldsymbol{\mathcal{E}}}^{p,k}_{n}\Delta t_n)$;
	\Comment{Eq.(\ref{eq-Ch4:polycrystalline SBIE discrete})}
	}
	\NoNumber{}
	\State{
	$\forall g,e,\alpha$ compute the resolved shear stress:
	}\label{step:resolved shear stress}
	\NoNumber{
	$\tau^{\alpha,k+1}_n(\mathbf{y}_e^g)=s_i^\alpha(\mathbf{y}_e^g) \sigma_{ij,n}^{g,k+1}(\mathbf{y}_e^g)n_j^\alpha(\mathbf{y}_e^g)$;
	\Comment{Eq.(\ref{eq-Ch4:resolved shear})}
	}
	\NoNumber{}
	\State{
	$\forall g,e,\alpha$ compute the slips \emph{rate} and the \emph{accumulated} slips:
	}\label{step:slip rate}
	\NoNumber{
	$\dot{\gamma}_n^{\alpha,k+1}(\mathbf{y}_e^g)=\Psi^\alpha[\tau^{\alpha,k+1}_n(\mathbf{y}_e^g),\tau^{\alpha}_{c,n-1}(\mathbf{y}_e^g)+\dot{\tau}^{\alpha,k}_{c,n}(\mathbf{y}_e^g)\Delta t_n]$;
	\Comment{Eq.(\ref{eq-Ch4:flow rule})}
	}
	\NoNumber{}
	\NoNumber{
	$\gamma_n^{\alpha,k+1}(\mathbf{y}_e^g)=\gamma_{n-1}^{\alpha}(\mathbf{y}_e^g)+|\dot{\gamma}_n^{\alpha,k+1}(\mathbf{y}_e^g)|\Delta t_n$;
	}
	\NoNumber{}
	\State{
	$\forall g,e,\alpha$ compute the critical shear \emph{rate}:
	}\label{step:hardening law}
	\NoNumber{
	$\dot{\tau}_{c,n}^{\alpha,k+1}(\mathbf{y}_e^g)=\Xi^\alpha[\gamma^{\beta,k+1}_n(\mathbf{y}_e^g),\dot{\gamma}^{\beta,k+1}_n(\mathbf{y}_e^g)]$;
	\Comment{Eq.(\ref{eq-Ch4:hardening law})}
	}
	\NoNumber{}
	\State{
    $\forall g,e$ compute the plastic strain \emph{rate}: \Comment{Eq.(\ref{eq-Ch4:plasticstrain})}
    }\label{step:plasticstrain}
    \NoNumber{
	$\dot{\eps}_{ij,n}^{p,k+1}(\mathbf{y}_e^g)=\frac{1}{2}\sum_{\alpha}\dot{\gamma}_n^{\alpha,k+1}(\mathbf{y}_e^g)[n_i^\alpha(\mathbf{y}_e^g) s_j^\alpha(\mathbf{y}_e^g)+n_j^\alpha(\mathbf{y}_e^g) s_i^\alpha(\mathbf{y}_e^g)]$
	}
	
	\NoNumber{}
	\State{
	$\forall g,e$ compute the \emph{cumulative} slip $\gamma_n^{k+1}(\mathbf{y}_e^g)=\sum_{\alpha}\gamma_n^{\alpha,k+1}(\mathbf{y}_e^g)$;
	}
	
	\If{($\forall g,e$\,$\left|\gamma_n^{k+1}(\mathbf{y}_e^g)-\gamma_n^k(\mathbf{y}_e^g)\right|<\epsilon_{conv}$)}
	\State{$\mathrm{converged} \leftarrow \texttt{.true.}$}
	\Else
	\State{$k \leftarrow k+1$}
	\EndIf
	
	\EndWhile
	
	\State{$\forall g,e,\alpha$ update and store
	$\begin{cases}
	\gamma_n^{\alpha}(\mathbf{y}_e^g) \leftarrow \gamma_{n-1}^{\alpha}(\mathbf{y}_e^g)+|\dot{\gamma}_n^{\alpha,k+1}(\mathbf{y}_e^g)|\Delta t_n\\
	\tau_{c,n}^{\alpha}(\mathbf{y}_e^g) \leftarrow \tau_{c,n-1}^{\alpha}(\mathbf{y}_e^g)+\dot{\tau}_{c,n}^{\alpha,k+1}(\mathbf{y}_e^g)\Delta t_n\\
	\boldsymbol{\mathcal{E}}_{n}^{p} \leftarrow \boldsymbol{\mathcal{E}}_{n-1}^{p}+\dot{\boldsymbol{\mathcal{E}}}_{n}^{p,k+1}\Delta t_n
	\end{cases}$}
     
    \State{$n \leftarrow n+1$;}

\EndWhile

\Statex{\textbf{EXIT}}
\end{algorithmic}
\end{algorithm}

\clearpage
\section{Numerical tests}\label{sec-Ch4:numerical tests}
In this Section, the developed and implemented formulation is numerically and computationally assessed. The phenomenological framework by Bassani and Wu \cite{bassani1991} is used in the first series of tests performed in the present work, considering FCC Copper (Cu) as benchmark material. The explicit expressions of the flow rule and the hardening law of the Bassani and Wu crystal plasticity framework are given in Section (\ref{app-Ch4:flow N hardening laws}). The single crystal properties as well as the slip systems numbering used in the present study coincide with those in Refs.\ \cite{wu1991,bassani1991}. The elastic constants and the crystal plasticity parameters for FCC copper single crystals are summarised in Table (\ref{tab-Ch4:CP properties of Cu}).

\begin{table}[ht]
\begin{center}
\caption{Elastic and crystal plasticity properties for FCC Copper from \cite{bassani1991}. To interpret the symbols, the reader is referred to Section (\ref{ssec-Ch4:flow N hardening laws - BW}).}
\label{tab-Ch4:CP properties of Cu}
\begin{tabular}{lll}
\hline
\hline
property&component&value\\
\hline
elastic constants [$10^{9}\,\mathrm{N}/\mathrm{m}^{2}$]&$c_{1111}$, $c_{2222}$, $c_{3333}$&170\\
&$c_{1122}$, $c_{1133}$, $c_{2233}$&124\\
&$c_{1212}$, $c_{1313}$, $c_{2323}$&64.5\\
initial critical resolved&\multirow{2}{*}{$\tau_0^\alpha\equiv\tau_0$, $\alpha=1,\cdots,N_s$}&\multirow{2}{*}{17}\\
shear stress [$10^{6}\,\mathrm{N}/\mathrm{m}^{2}$]&&\\
reference shear rate [$1/\mathrm{s}$]&$\dot{\gamma}_0^\alpha$, $\alpha=1,\cdots,N_s$&0.001\\
rate sensitivity [-]&$n$&50\\
stage I reference stress [$10^{6}\,\mathrm{N}/\mathrm{m}^{2}$]&$\tau_s$&$1.3\cdot\tau_0$\\
initial hardening modulus [$10^{6}\,\mathrm{N}/\mathrm{m}^{2}$]&$h_0$&$90\cdot\tau_0$\\
hardening during easy glide [$10^{6}\,\mathrm{N}/\mathrm{m}^{2}$]&$h_s$&$1.5\cdot\tau_0$\\
interaction peak slip [-]&$\gamma_p$&0.001\\
latent hardening factor [-]&$q$&0\\
\multirow{5}{*}{cross-hardening moduli $f_{\alpha\beta}$ [-]}&no junction&8.0\\
&Hirth lock&8.0\\
&Coplanar interaction&8.0\\
&Glissile junction&15.0\\
&Sessile junction&20.0\\
\hline
\hline
\end{tabular}
\end{center}
\end{table}

\subsection{Single crystal tests}\label{ssec-Ch4: single crystal}
To assess the accuracy and numerical performance of the formulation, tests on a single crystal are performed first. The analysed domain is a cubical crystal, with the edges aligned with the reference system and subjected to uniaxial stress acting along the $x_3$ direction.

\subsubsection{Single crystal test: algorithm test}
In Figure (\ref{fig-Ch4:RDnConv}), the behaviour of the implemented scheme is tested with respect to the value of the rate sensitivity $n$ appearing in Eq.(\ref{eq-Ch4:BW flow rule}), for the three different crystallographic loading directions $[001]$, $[111]$ and $[632]$, chosen as in Ref.\ \cite{bassani1991} for comparative purposes. In all tests, the loading rate $\dot{\lambda}$ is always set as $\dot{\lambda}=10^{-3}\textit{s}^{-1}$. The curves plot the stress component $\sigma_{33}$ at the centroid of the cube versus the strain component $\eps_{33}$ at the same point. It is worth noting that nominal components of strains are considered, being the developed formulation valid for small strains. However, strains up to $10\%$ are plotted to investigate when the small strains assumption breaks down: it is interesting to note that, up to this value of strains, the converged curves match reasonably well with those reported in \cite{bassani1991}, which have been replicated in the figure using a simple rate-independent scheme and represented with dashed lines. It is confirmed that, as $n$ rises, the rate-dependent scheme simulates a rate-independent process, although the convergence is not the same for all directions, being noticeably slower (with respect to $n$) for the direction $[632]$. Also, it is interesting to note that for the $[001]$ and $[111]$ loading directions, the rate-dependent scheme predicts a lower bound for the stress, whereas for the $[632]$ loading direction it predicts an upper bound.

\begin{figure}
\centering
	\begin{subfigure}{0.49\textwidth}
	\centering
	\includegraphics[width=\textwidth]{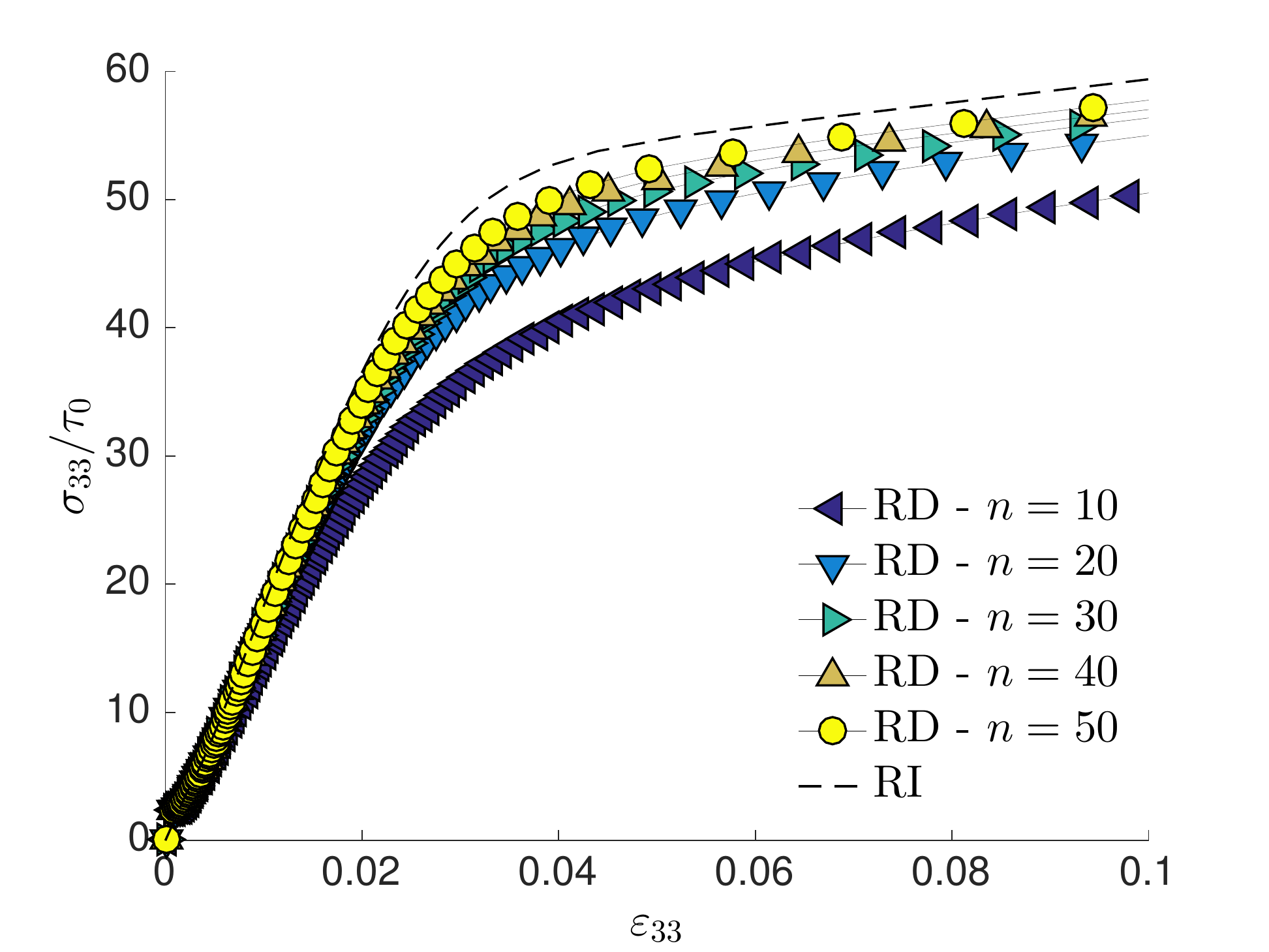}
	\caption{Loading direction [001]}
	\end{subfigure}
\
	\begin{subfigure}{0.49\textwidth}
	\centering
	\includegraphics[width=\textwidth]{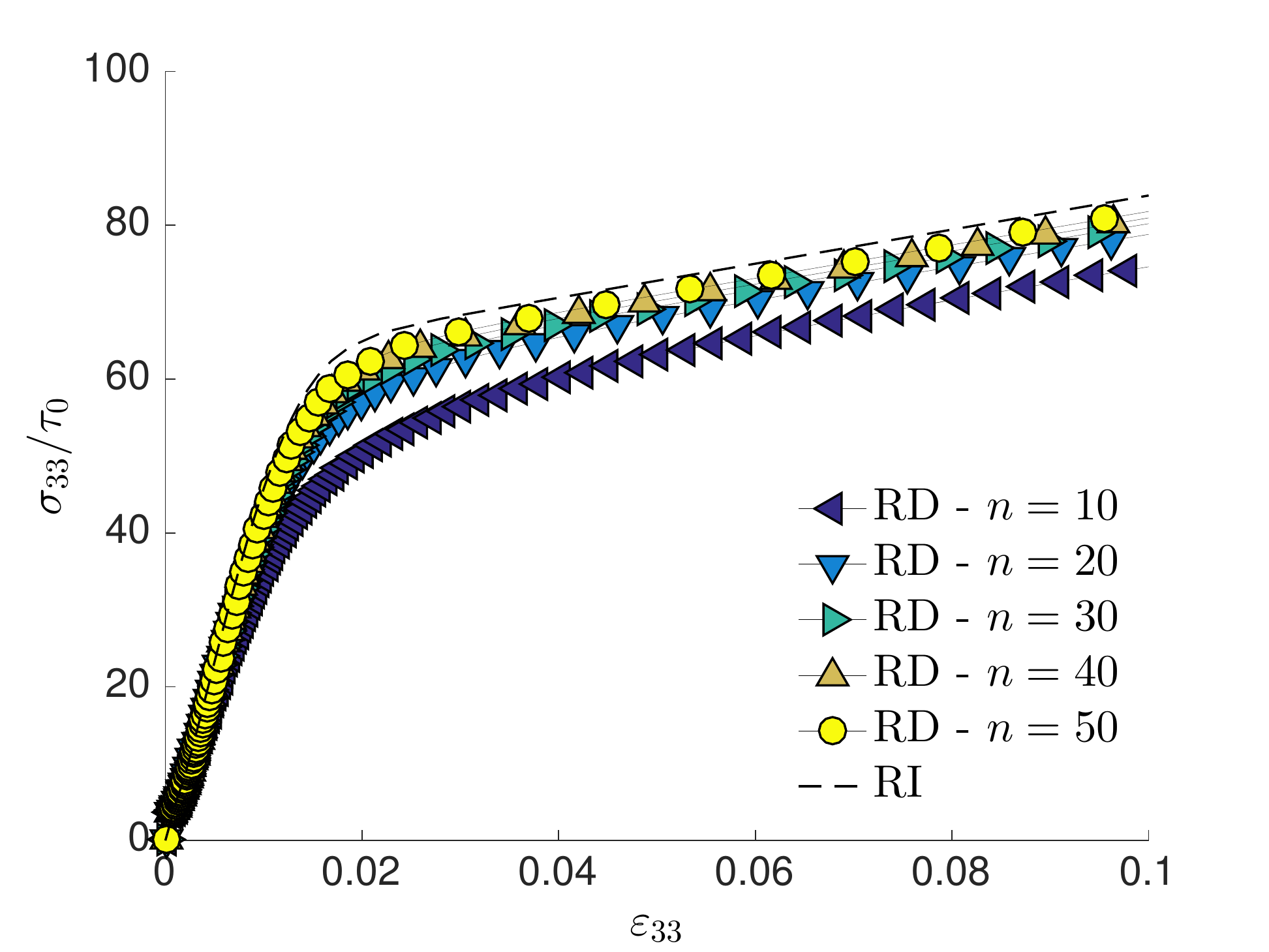}
	\caption{Loading direction [111]}
	\end{subfigure}
\
	\begin{subfigure}{0.49\textwidth}
	\centering
	\includegraphics[width=\textwidth]{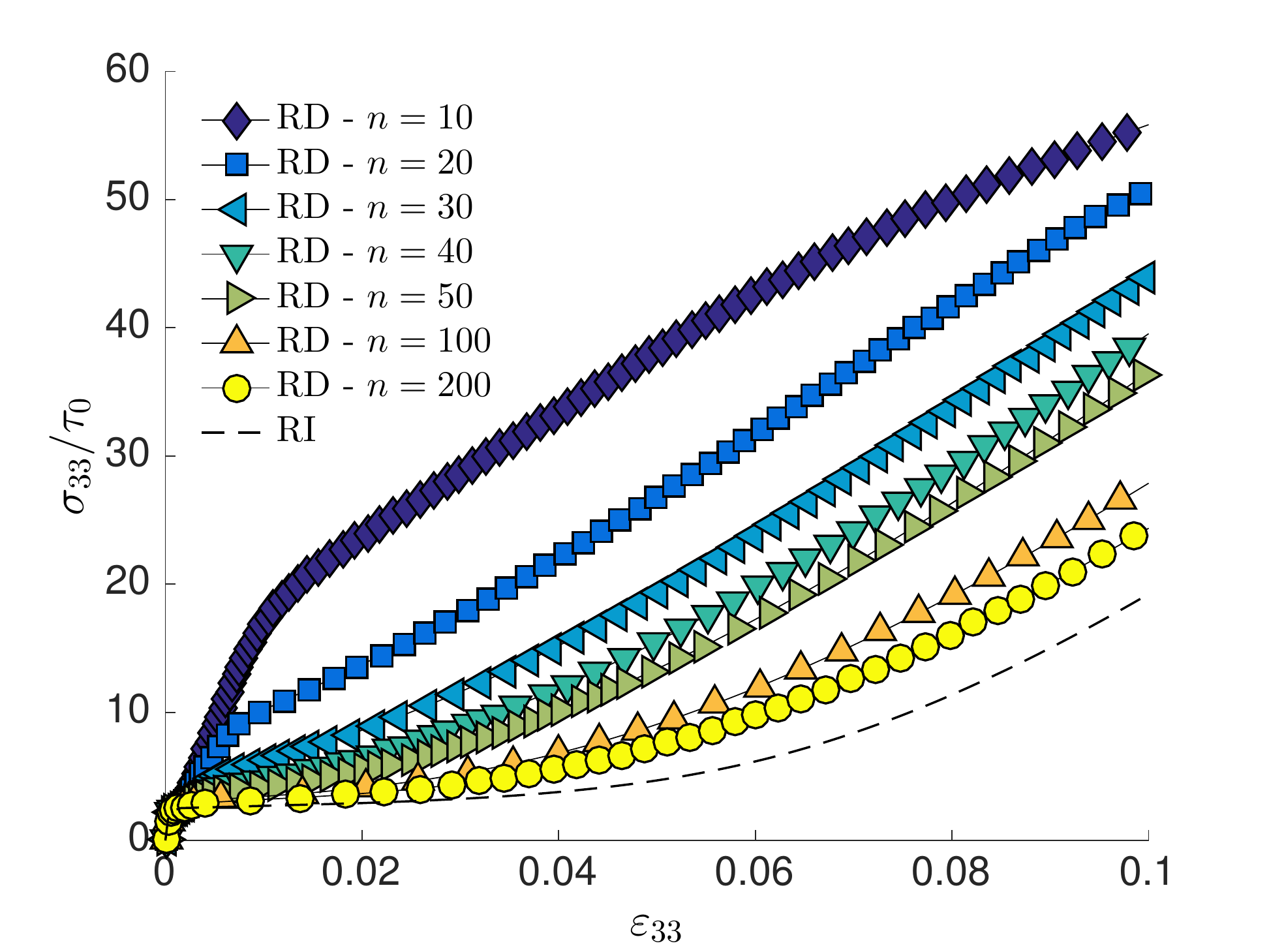}
	\caption{Loading direction [632]}
	\end{subfigure}
\
	\begin{subfigure}{0.49\textwidth}
	\centering
	\includegraphics[width=\textwidth]{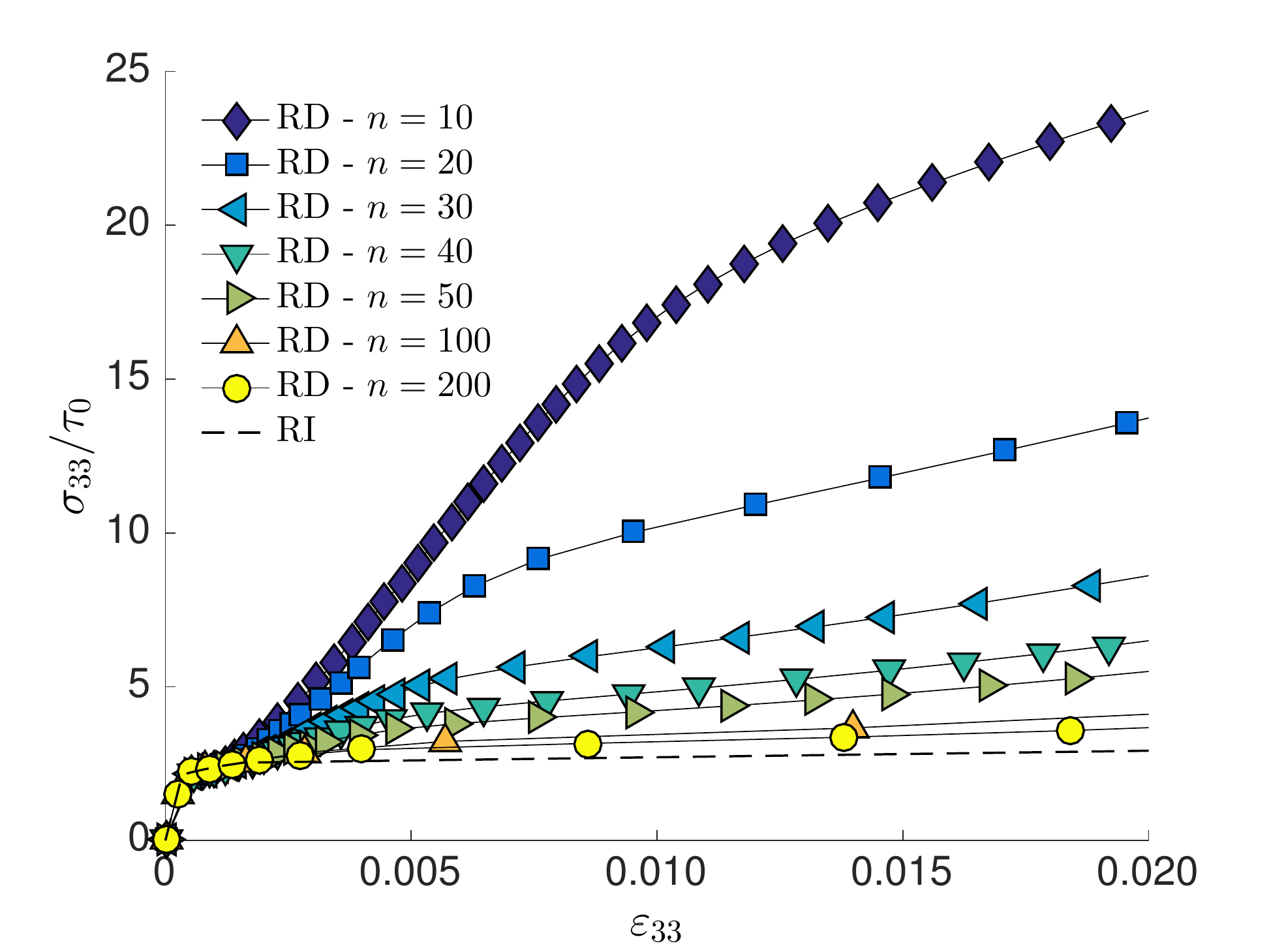}
	\caption{Loading direction [632]}
	\end{subfigure}
\
\caption{Effect of the rate sensitivity parameter $n$ for three different loading crystallographic directions. Plot (\emph{d}) shows a zoom from plot (\emph{c}). It is apparent as the rate-dependent results converge to the rate-independent ones with increasing values of $n$.}
\label{fig-Ch4:RDnConv}
\end{figure}

\subsubsection{Single crystal test: mesh convergence}
In the following series of tests, the single crystal response is tested against different domain volume meshes, in order to assess the robustness of the computational homogenisation scheme that will be used in the polycrystalline tests. Unlike the curves reported in Figs.(\ref{fig-Ch4:RDnConv}), which are obtained considering the values of stress and strain at the centroid of the cubic single crystal domain, the stress-strain curves in the subsequent part of this work are obtained as averaged results of the stress and the strain fields over the crystalline or polycrystalline domain. Strain and stress volume averages are given by
\begin{equation}\label{eq-Ch4:strain_average}
\Gamma_{ij}=\frac{1}{V}\int_V\gamma_{ij}(\mathbf{x})dV(\mathbf{x})=\frac{1}{2V}\int_{S}[u_i(\mathbf{x})n_j(\mathbf{x})+u_j(\mathbf{x})n_i(\mathbf{x})]dS(\mathbf{x})
\end{equation}
\begin{equation}\label{eq-Ch4:stress_average}
\Sigma_{ij}=\frac{1}{V}\int_V\sigma_{ij}(\mathbf{x})dV(\mathbf{x})=\frac{1}{V}\int_{S}x_i(\mathbf{x})t_j(\mathbf{x})dS(\mathbf{x})
\end{equation}
where $S$ and $V$ represents the boundary and the volume of the domain, respectively. The terms involving the surface integrals allow to write the volume averages in terms of boundary displacements and tractions, which are the primary variables of the model. The terms involving the volume integrals, on the other hand, are obtained as the weighed sum of the strains and stresses computed at the centroids of the tetrahedra, i.e. $\int_V(\cdot)dV=\sum_{e}(\cdot)_eV_e$ being $V_e$ the volume of the $e$-th volume element. The comparison between the volume averages computed using the volume and the surface integrals can be used as a tool to assess the accuracy in the integration of the kernels in the displacement and stress boundary integral equations.

Figure (\ref{fig-Ch4:C1_meshes}) shows four surface and volume meshes of a cubic domain. Following the procedure described in Chapter (\ref{ch-EF}) and used in Refs.\ \cite{benedetti2013a,benedetti2013b,gulizzi2015}, the surface and volume meshes of the crystalline domain are built considering an as uniform as possible mesh size, which is computed as the average length of the edges of the considered crystals divided by a mesh density parameter $d_m$. In the considered case, the average edge length is clearly the cube's edge.

Figures (\ref{fig-Ch4:C1_meshes}a,e), (\ref{fig-Ch4:C1_meshes}b,f), (\ref{fig-Ch4:C1_meshes}c,g), (\ref{fig-Ch4:C1_meshes}e,h) correspond to $d_m=2,3,4,5$, respectively. Although they may not represent the best meshing choice for a cubic domain, the combination of triangular and quadrangular elements in the surface meshes and the tetrahedra in the volume meshes is the same used to adapt to the geometrical variability of general Laguerre-based polycrystalline morphologies. Some statistics about the meshing algorithm are summarised in Table (\ref{tab-Ch4:meshes-stats}), which reports the number of degrees of freedom and the number of surface and volume elements of the single crystal meshes involved in the present test, as well as those for the polycrystalline meshes considered in the next Section.

The cube in Figure (\ref{fig-Ch4:C1_meshes}) is loaded in uniaxial stress, with uniform tractions acting over the top and bottom faces, whereas the lateral faces are traction-free. To avoid the rigid body motion, the technique proposed by Lutz et al.\ \cite{lutz1998} is used. Figure (\ref{fig-Ch4:C1_632}a) shows the stress-strain curves of a single crystal cubic domain loaded along the $[632]$ direction for the four considered meshes. The curves are plotted in the strain range 0-2\% considering a slip rate sensitivity $n=50$ and are compared to that obtained for the single crystal centroid. The four meshes replicate the results obtained in Figure (\ref{fig-Ch4:RDnConv}d) as uniform strain and stress fields are expected for this configuration. In Figure (\ref{fig-Ch4:C1_632}a), the close-up view shows a comparison between the curves obtained using the surface and the volume integrals of Eqs.(\ref{eq-Ch4:strain_average}) and (\ref{eq-Ch4:stress_average}). A good match is found among the curves showing a satisfactory accuracy in the integration of the kernels of the boundary integral equations. Finally, Figure (\ref{fig-Ch4:C1_632}b) shows the same response in a smaller strain range, i.e.\ 0-0.2\%.

\begin{figure}
\centering
	\begin{subfigure}{0.2\textwidth}
	\centering
	\includegraphics[width=\textwidth]{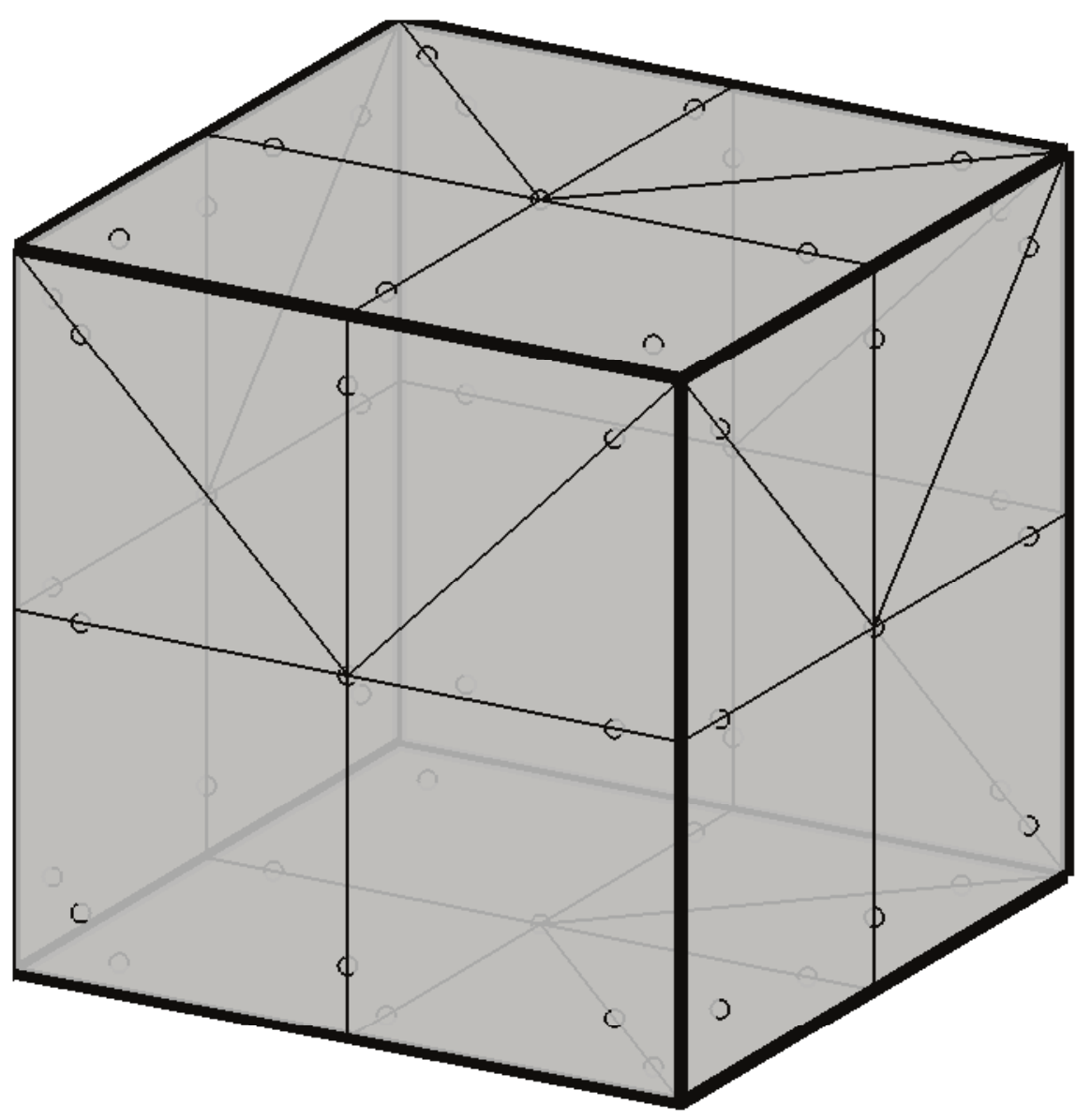}
	\caption{}
	\end{subfigure}
\
	\begin{subfigure}{0.2\textwidth}
	\centering
	\includegraphics[width=\textwidth]{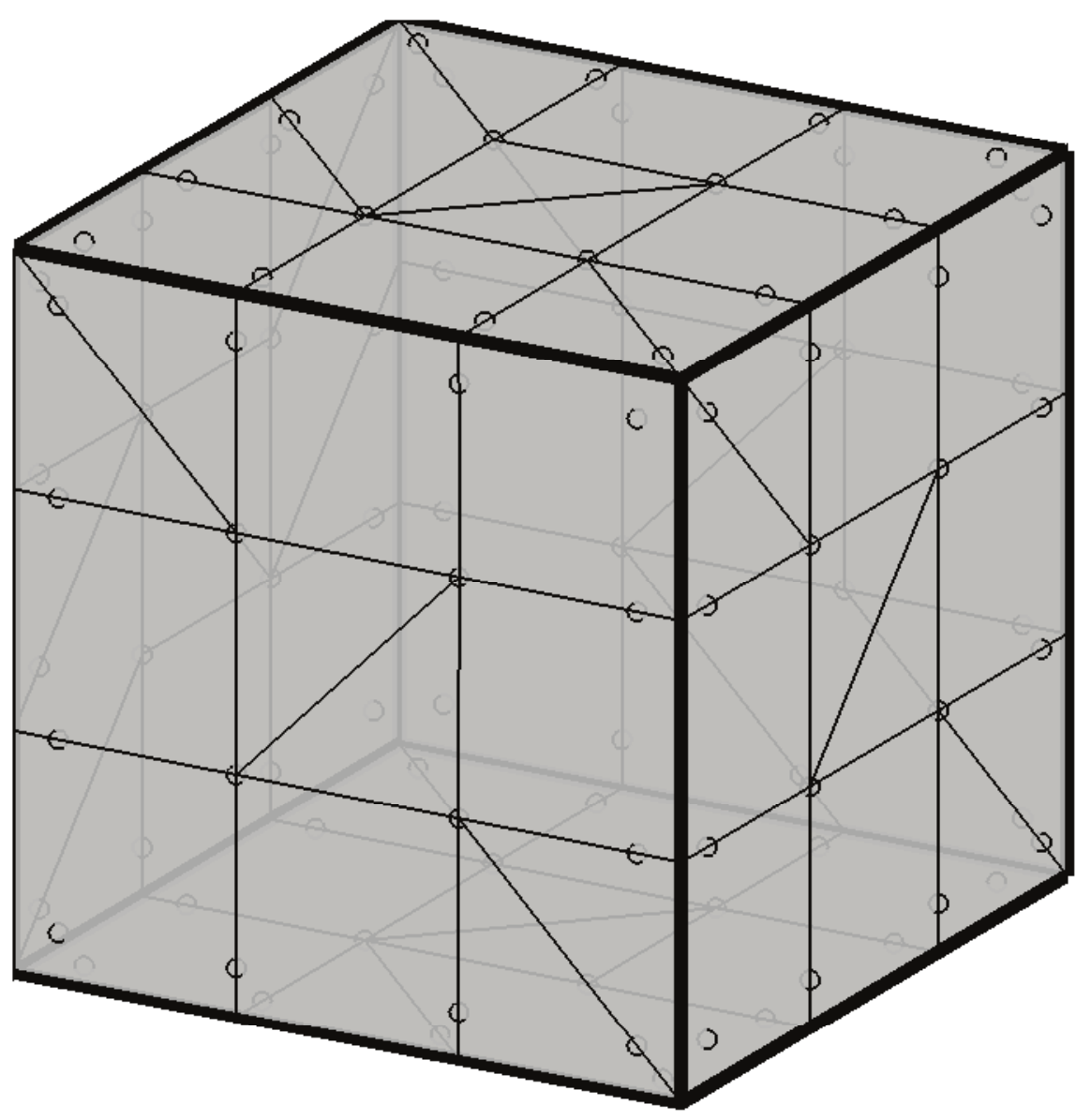}
	\caption{}
	\end{subfigure}
\
	\begin{subfigure}{0.2\textwidth}
	\centering
	\includegraphics[width=\textwidth]{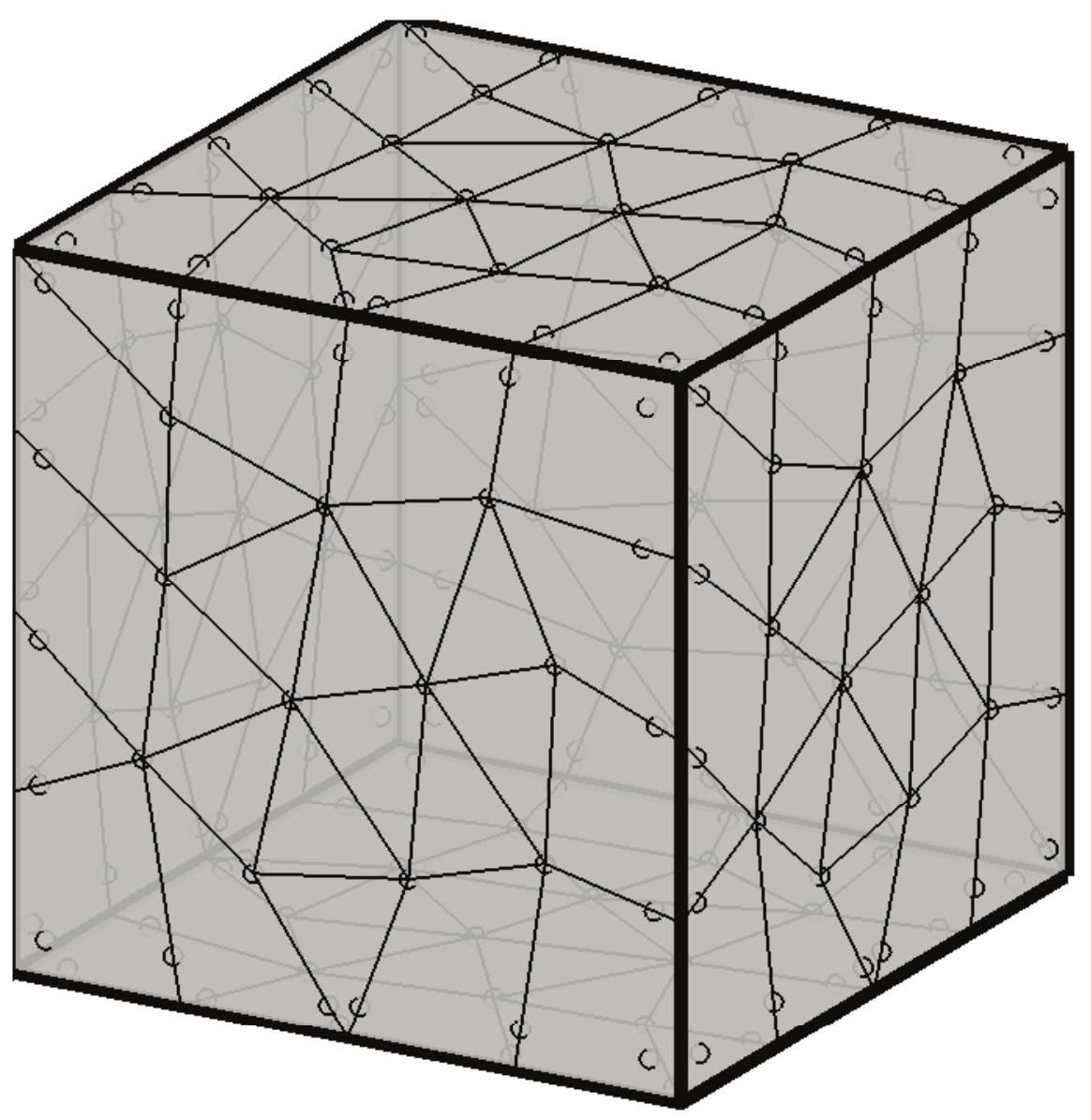}
	\caption{}
	\end{subfigure}
\
	\begin{subfigure}{0.2\textwidth}
	\centering
	\includegraphics[width=\textwidth]{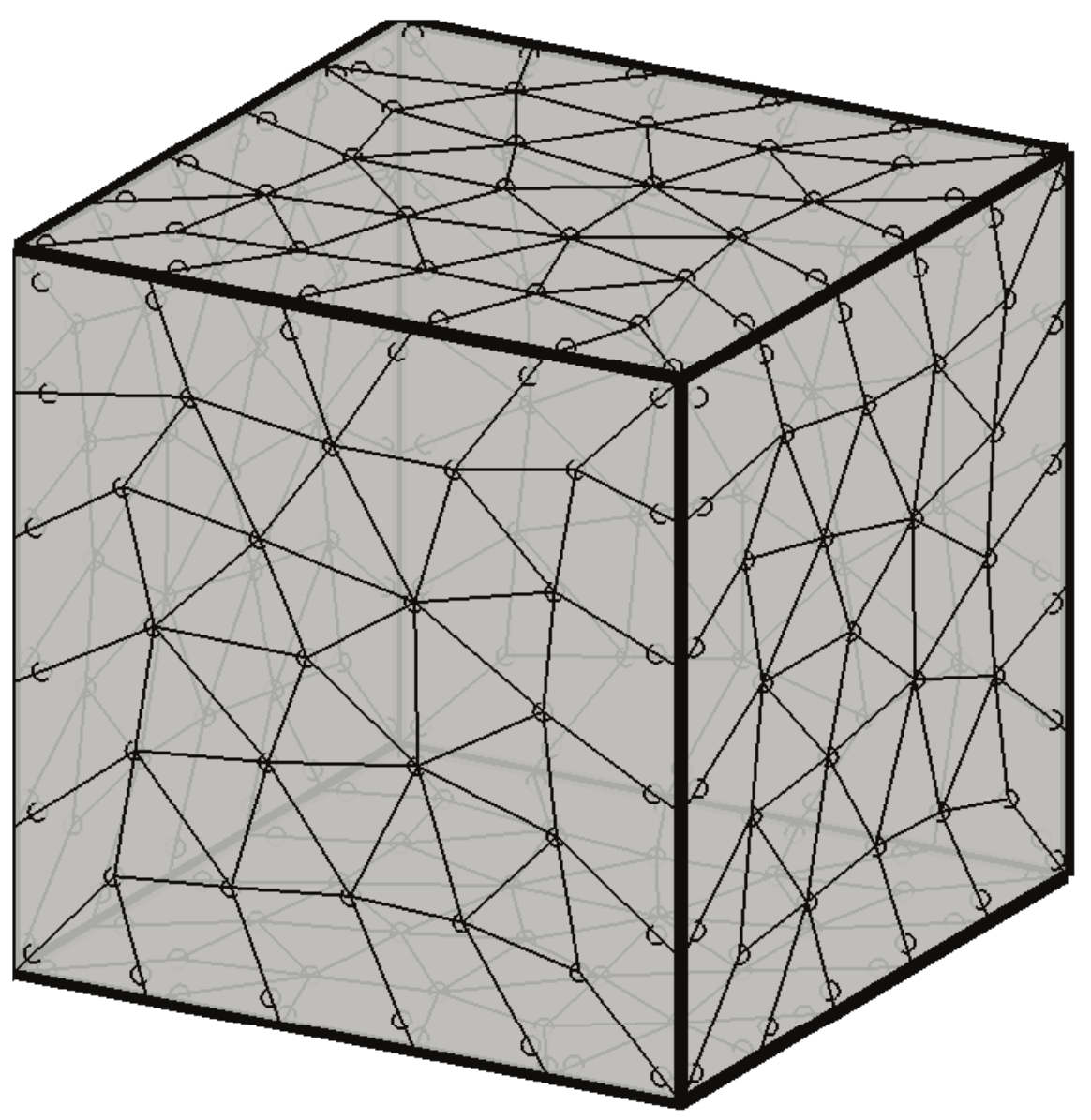}
	\caption{}
	\end{subfigure}
\\
	\begin{subfigure}{0.2\textwidth}
	\centering
	\includegraphics[width=\textwidth]{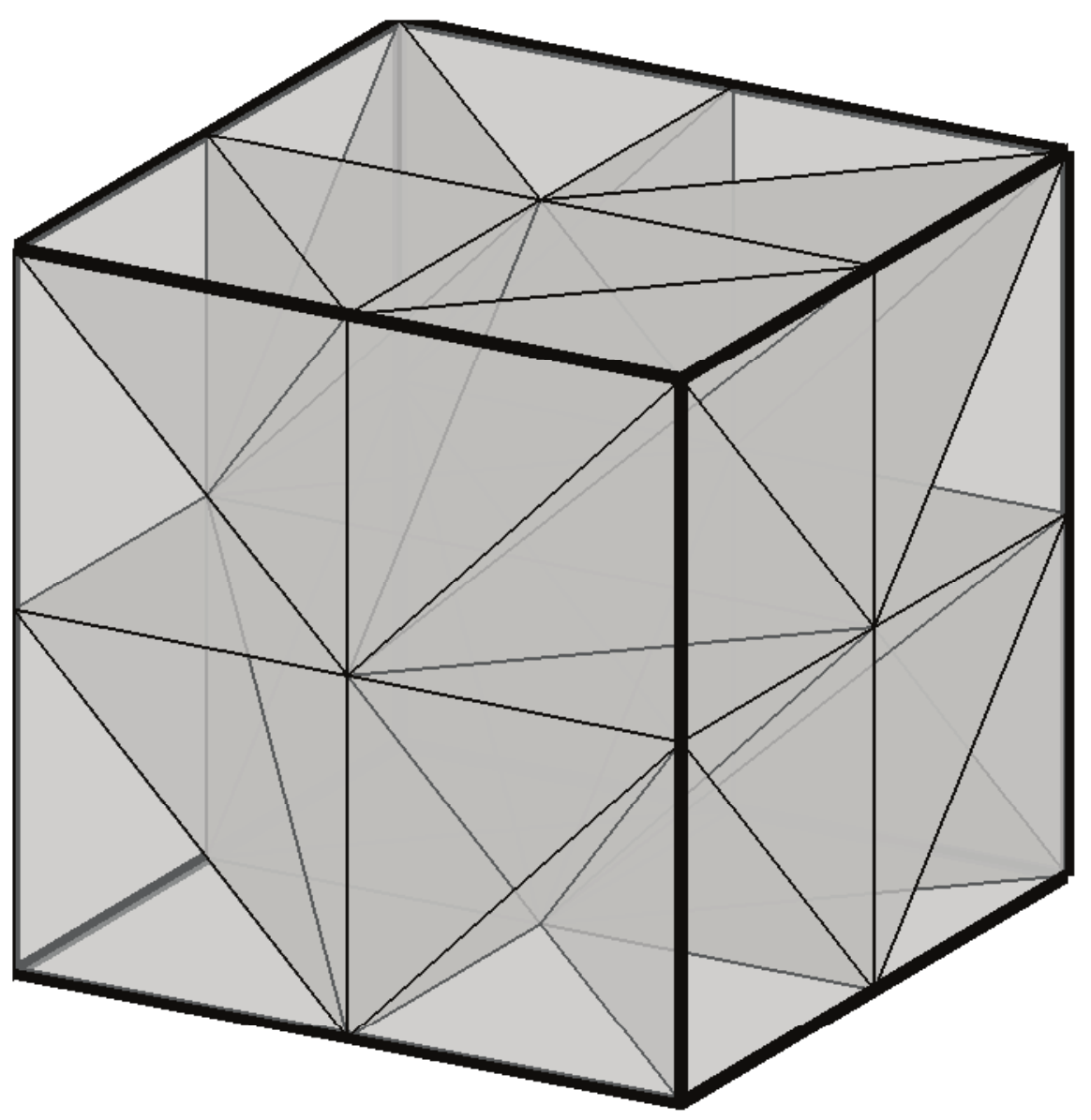}
	\caption{}
	\end{subfigure}
\
	\begin{subfigure}{0.2\textwidth}
	\centering
	\includegraphics[width=\textwidth]{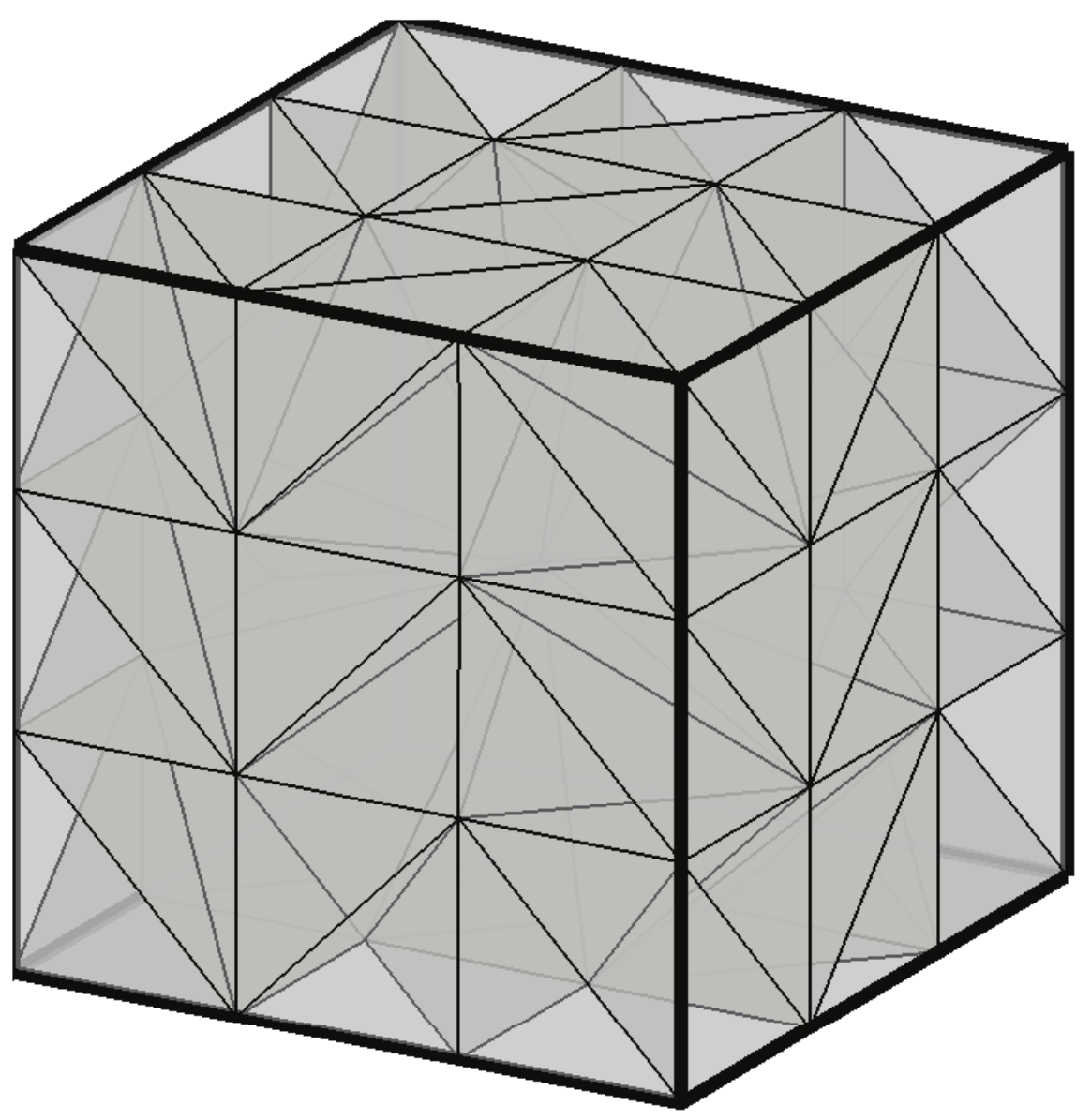}
	\caption{}
	\end{subfigure}
\
	\begin{subfigure}{0.2\textwidth}
	\centering
	\includegraphics[width=\textwidth]{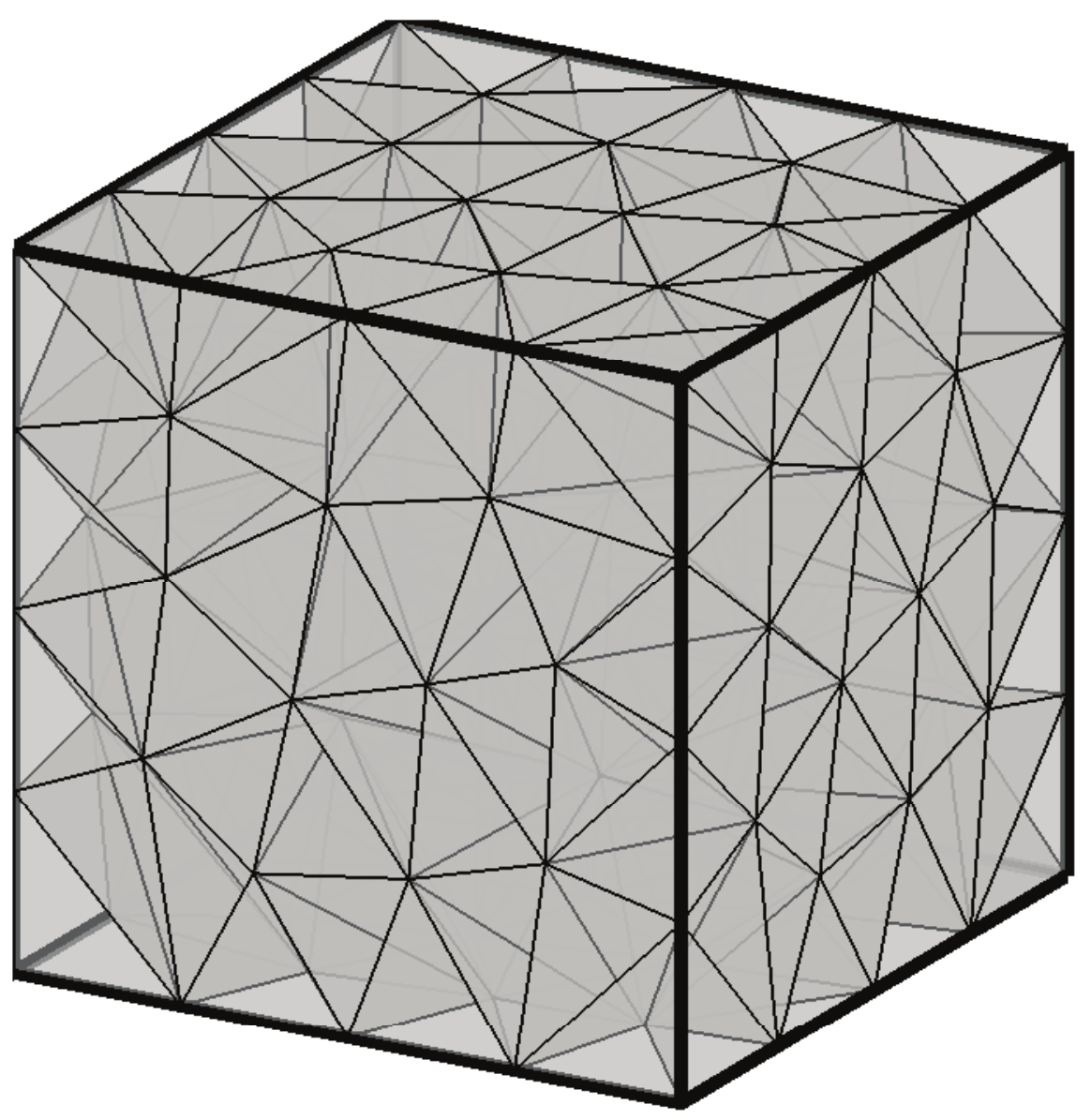}
	\caption{}
	\end{subfigure}
\
	\begin{subfigure}{0.2\textwidth}
	\centering
	\includegraphics[width=\textwidth]{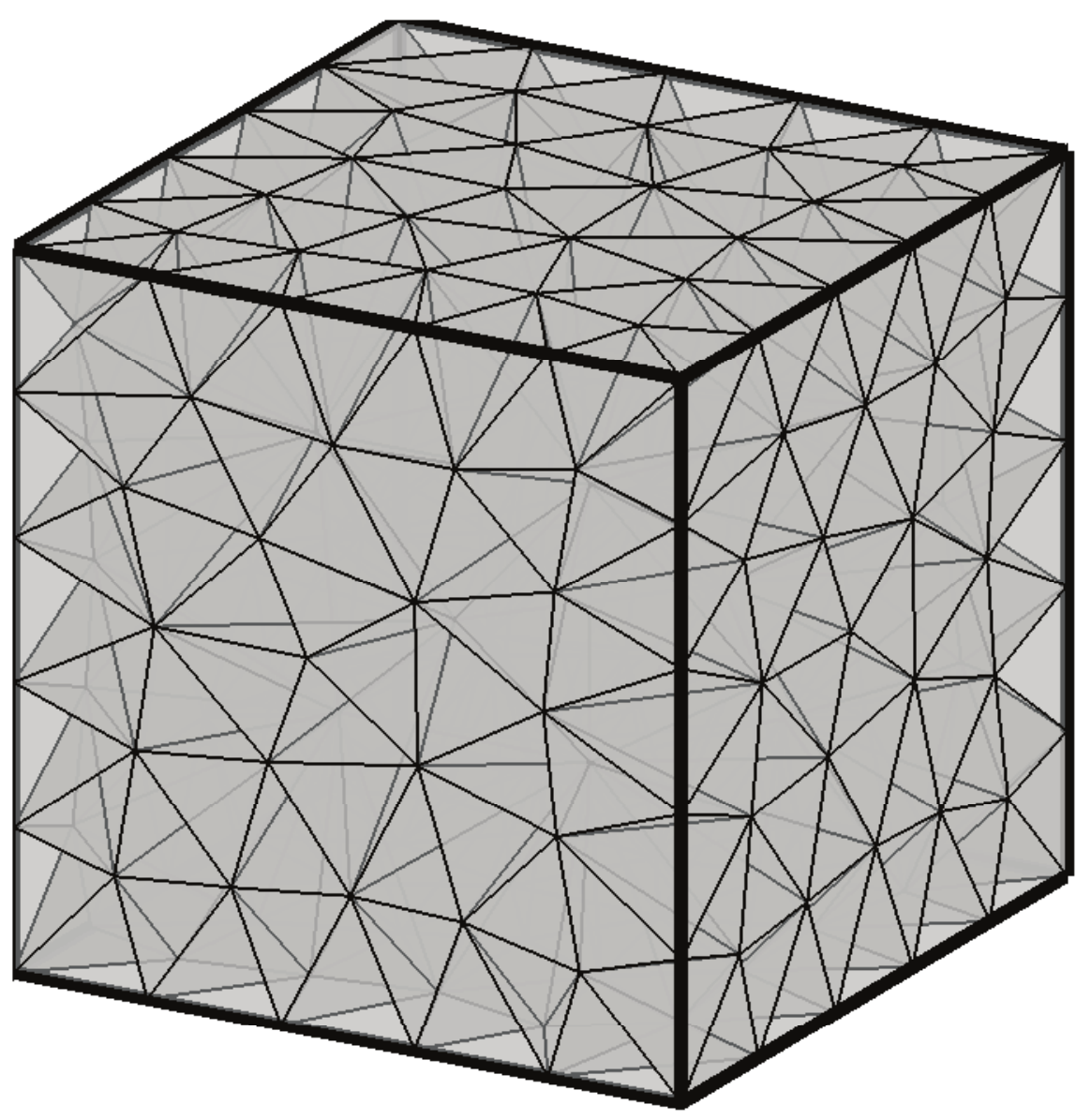}
	\caption{}
	\end{subfigure}
\
\caption{Single crystal surface (\emph{a}-\emph{d}) and volume (\emph{e}-\emph{h}) meshes: (\emph{a},\emph{e}) $d_m=2$, (\emph{b},\emph{f}) $d_m=3$, (\emph{c},\emph{g}) $d_m=4$, (\emph{d},\emph{h}) $d_m=5$. The small circles in the surface meshes represent the collocation points.}
\label{fig-Ch4:C1_meshes}
\end{figure}

\begin{table}[h]
\small
\begin{center}
\caption{Statistics about the meshes of the single crystal and polycrystalline domains: $d_m$ is the mesh density parameter; $N_e^s$ is the number of surface elements; $DoFs$ is the number of degrees of freedom for the system in Eq.(\ref{eq-Ch4:polycrystalline DBIE discrete}); $N_e^v$ is the number of volume elements. The last column reports the number of volume elements per grain, which is considered as reference to capture the volume field distribution.}\label{tab-Ch4:meshes-stats}
\begin{tabular}{c|c|cccc}
\hline
\hline
morphology&$d_m$&$N_e^s$&$DoFs$&$N_e^v$&$N_e^v/N_g$\\
\hline
\multirow{4}{*}{single crystal}&2&36&162&46&46\\
&3&72&288&161&161\\
&4&150&504&436&436\\
&5&255&729&836&836\\
\hline
\multirow{5}{*}{polycrystalline, 10 grains}&1&430&3594&562&56\\
&2&895&6285&1576&158\\
&3&1811&10131&4097&410\\
&4&3291&15669&8791&897\\
&5&5242&22143&16119&1612\\
\hline
polycrystalline, 100 grains (I)&2.5&19902&132369&45704&457\\
\hline
polycrystalline, 100 grains (II)&2.5&18075&122166&40181&402\\
\hline
\hline
\end{tabular}
\end{center}
\end{table}

\begin{figure}
\centering
	\begin{subfigure}{0.49\textwidth}
	\centering
	\includegraphics[width=\textwidth]{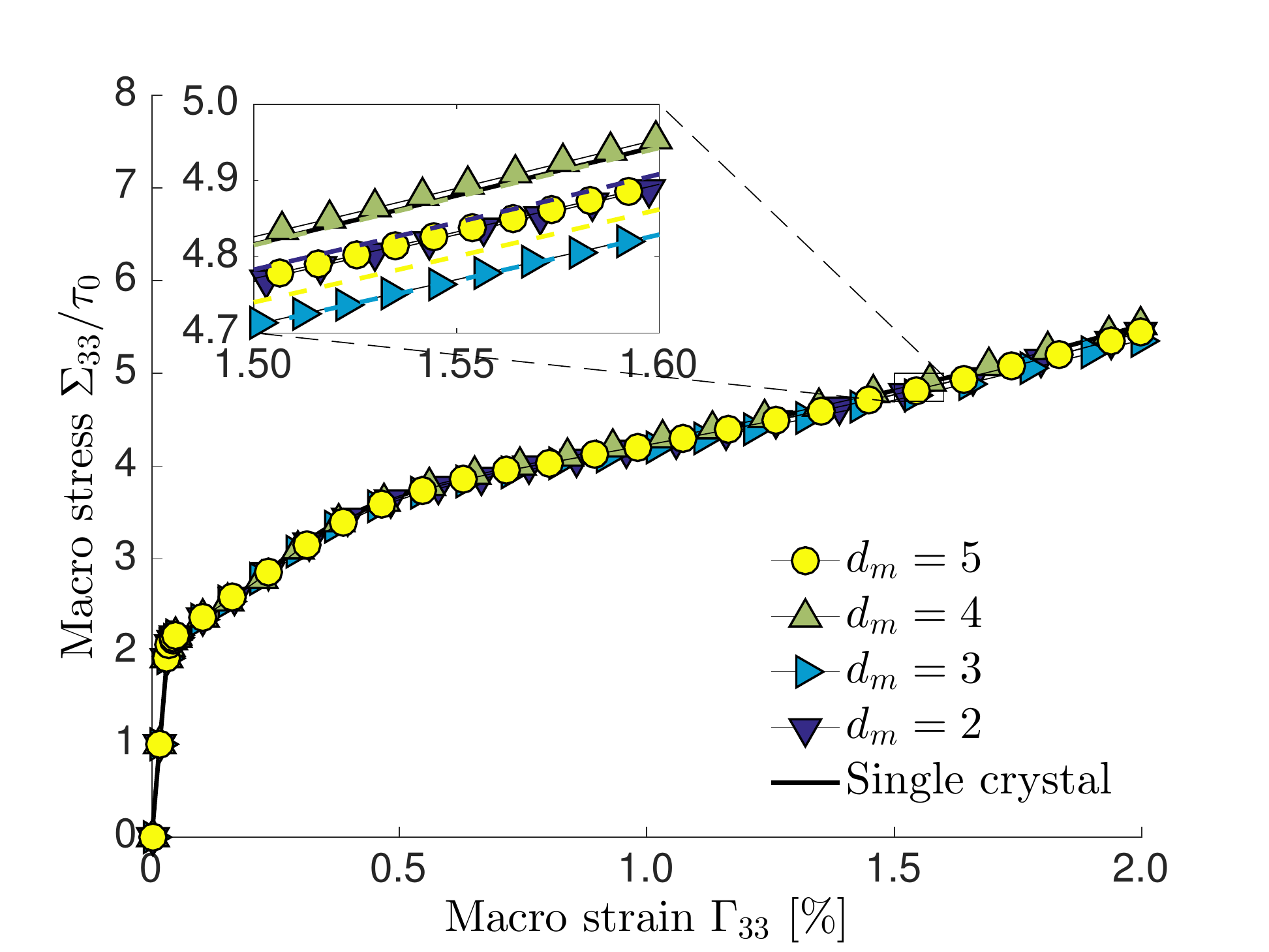}
	\caption{}
	\end{subfigure}
\
	\begin{subfigure}{0.49\textwidth}
	\centering
	\includegraphics[width=\textwidth]{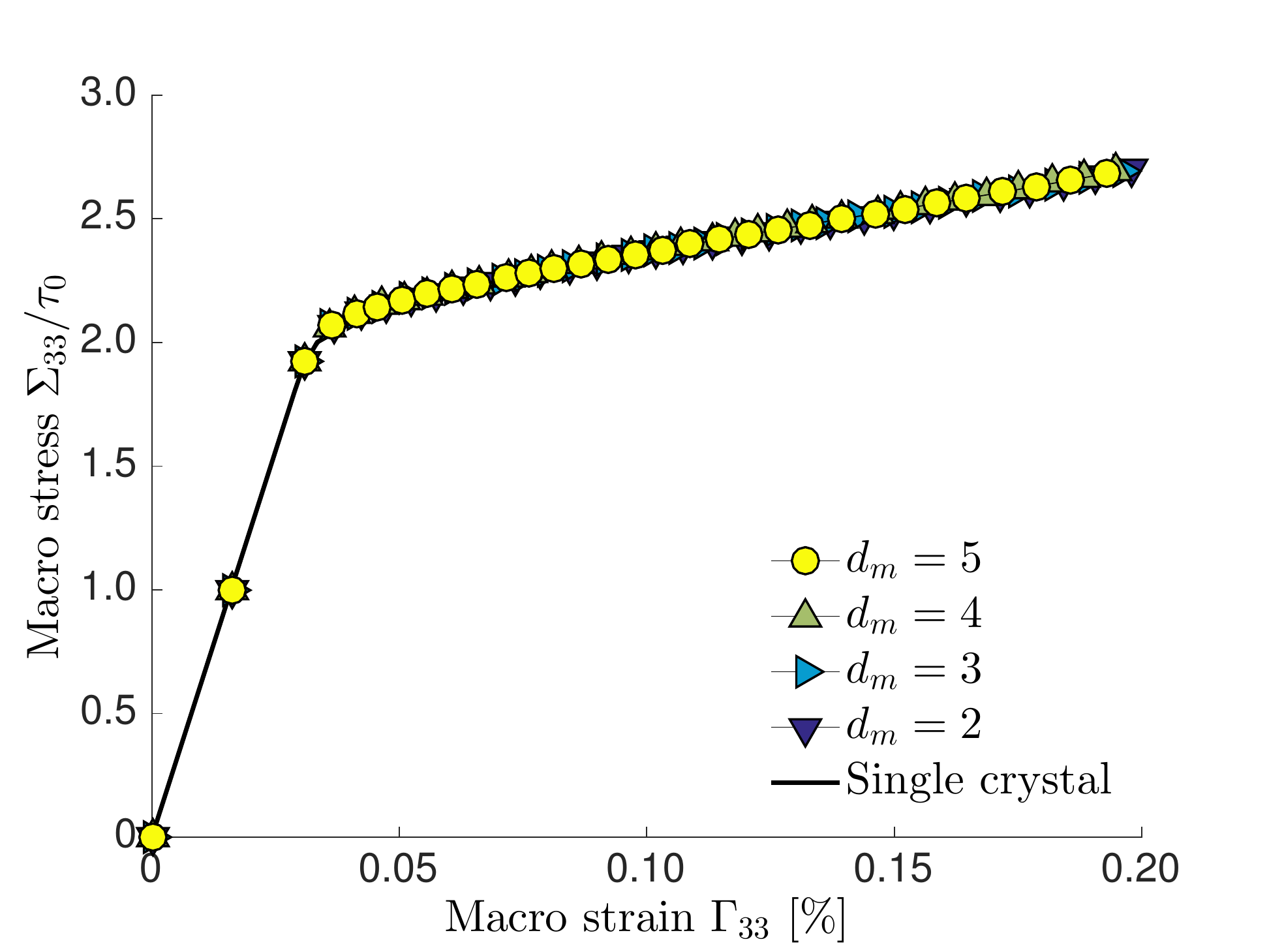}
	\caption{}
	\end{subfigure}
\
\caption{Volume stress average $\Sigma_{33}$ plotted versus volume strain average $\Gamma_{33}$ for the four considered meshes of the single crystal loaded along the [632] direction. (\emph{a}) stress-strain response in the strain range 0-2.0\%; the close-up shows the values obtained using the surface integral (solid lines) compared to those obtained using the volume integral (dashed lines). (\emph{b}) stress-strain response in the range strain 0-0.2\%.}
\label{fig-Ch4:C1_632}
\end{figure}

\clearpage

\begin{figure}[h]
\centering
	\begin{subfigure}{0.18\textwidth}
	\centering
	\includegraphics[width=\textwidth]{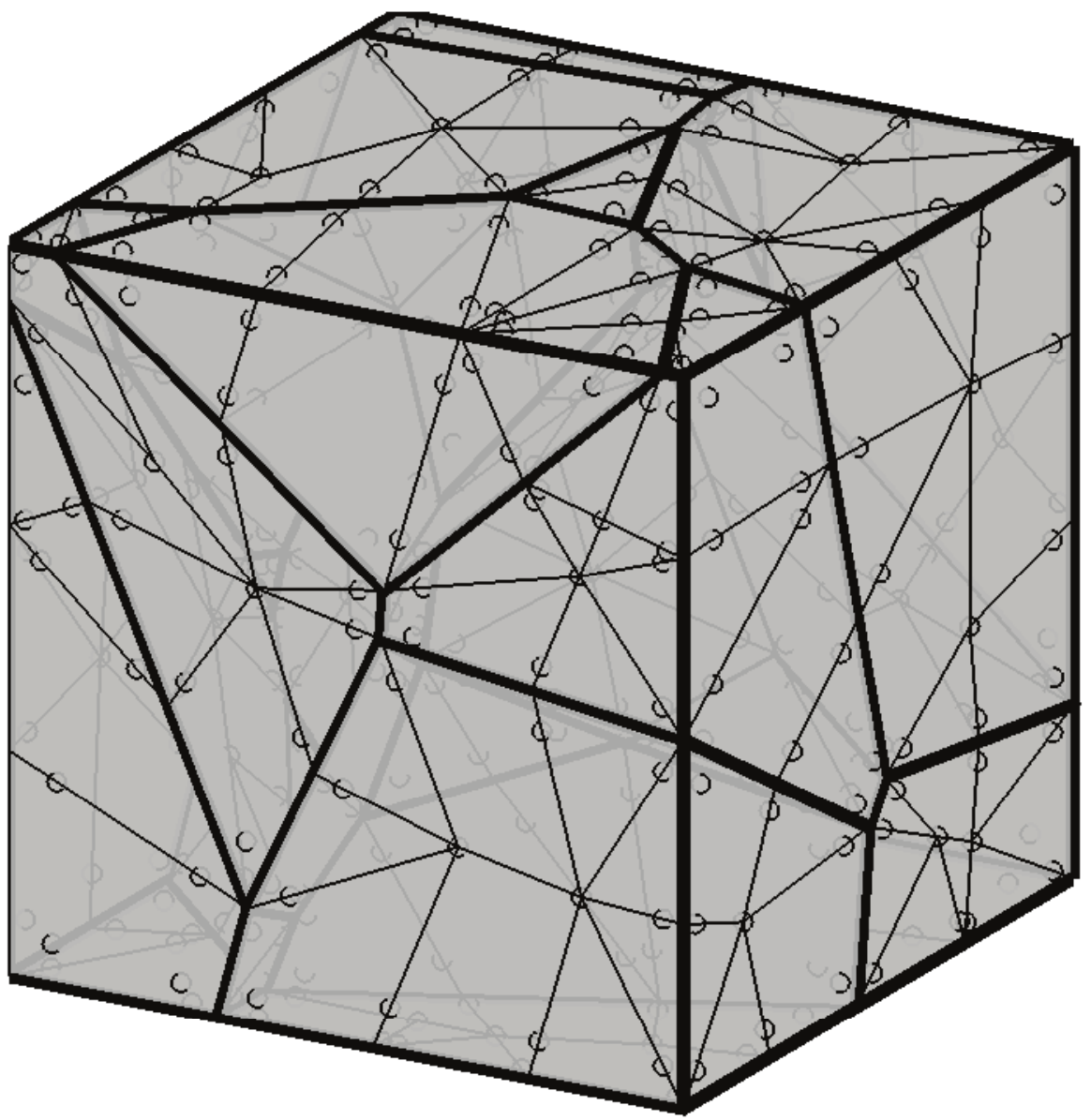}
	\caption{}
	\end{subfigure}
\
	\begin{subfigure}{0.18\textwidth}
	\centering
	\includegraphics[width=\textwidth]{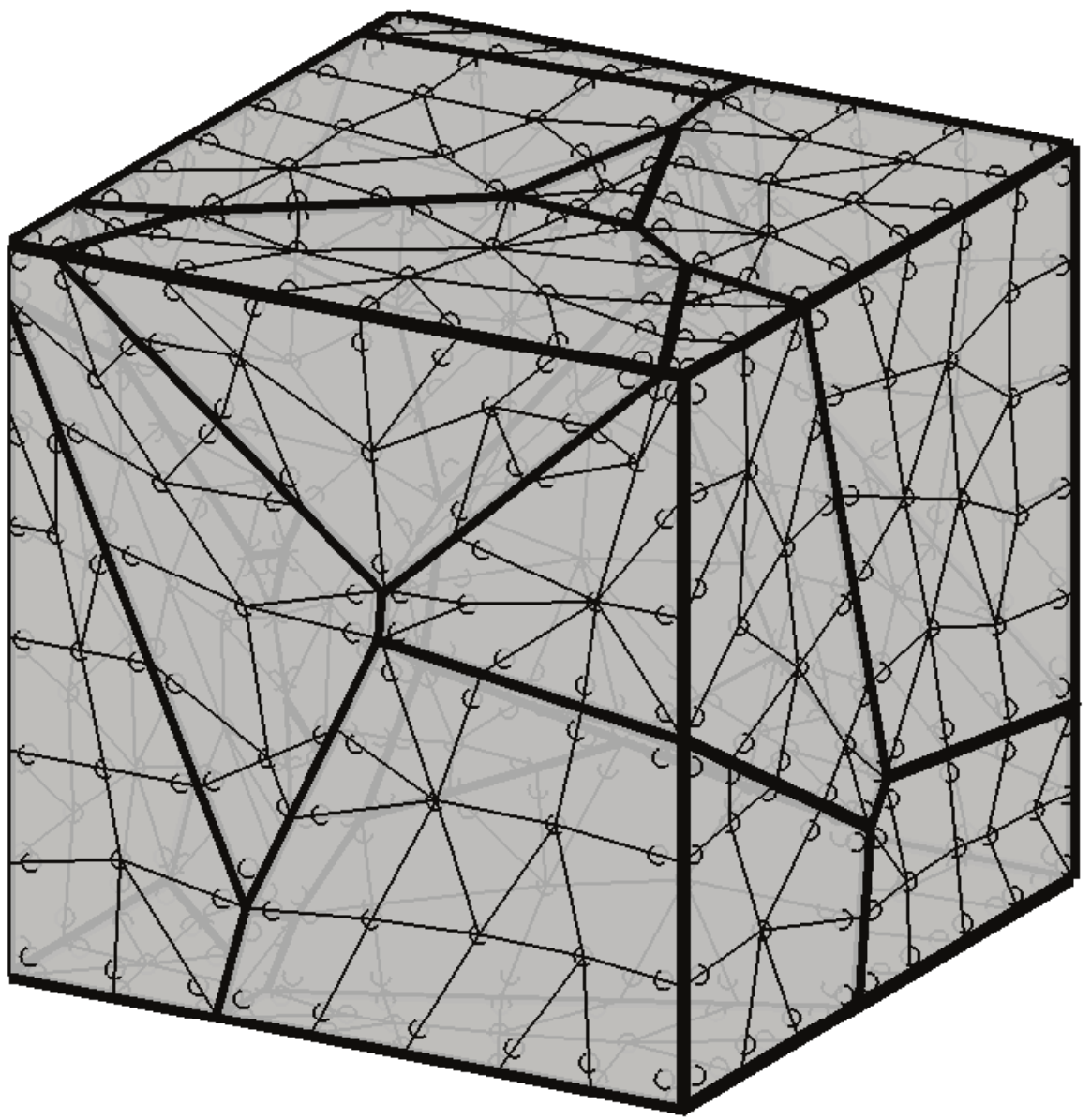}
	\caption{}
	\end{subfigure}
\
	\begin{subfigure}{0.18\textwidth}
	\centering
	\includegraphics[width=\textwidth]{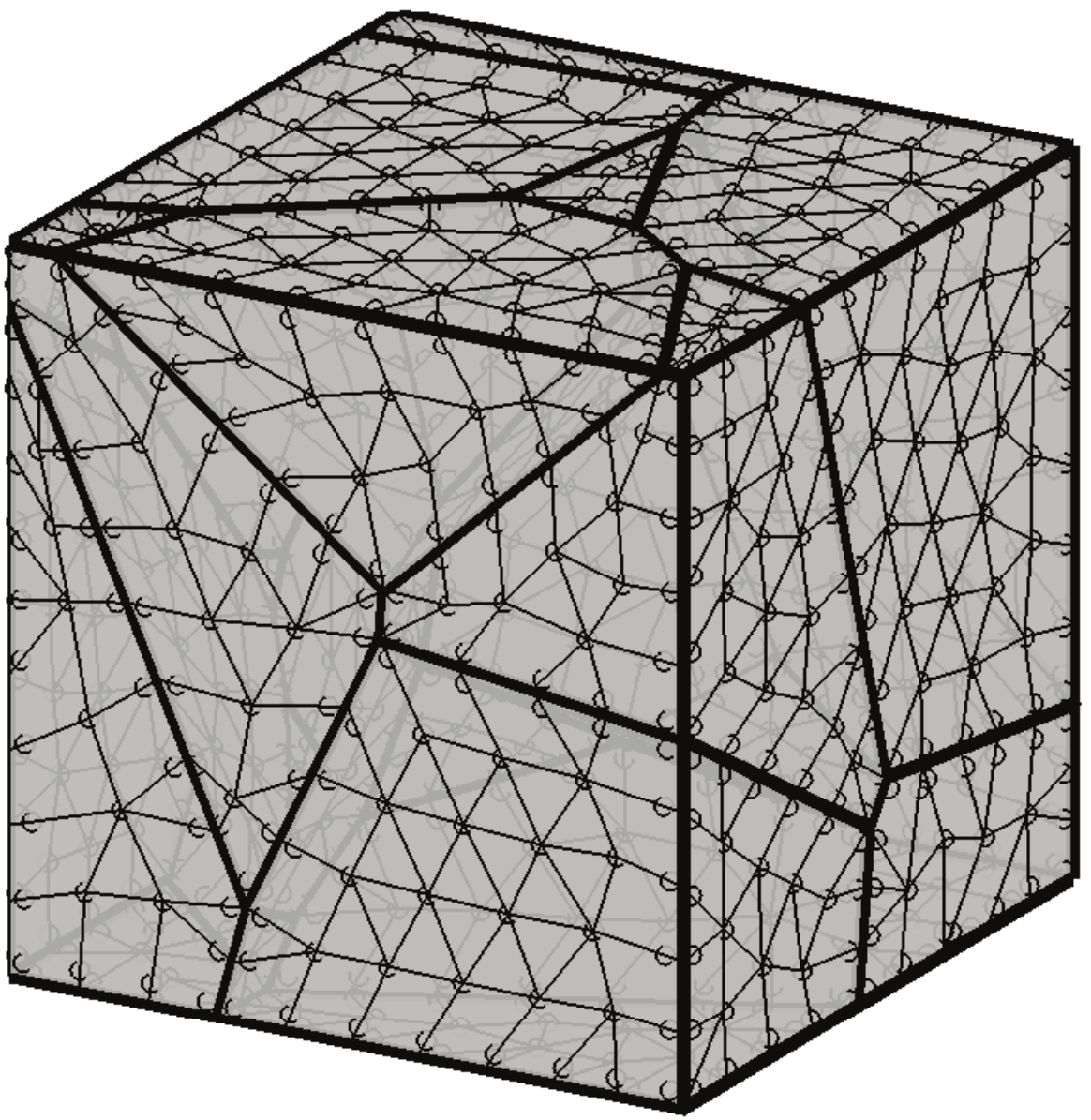}
	\caption{}
	\end{subfigure}
\
	\begin{subfigure}{0.18\textwidth}
	\centering
	\includegraphics[width=\textwidth]{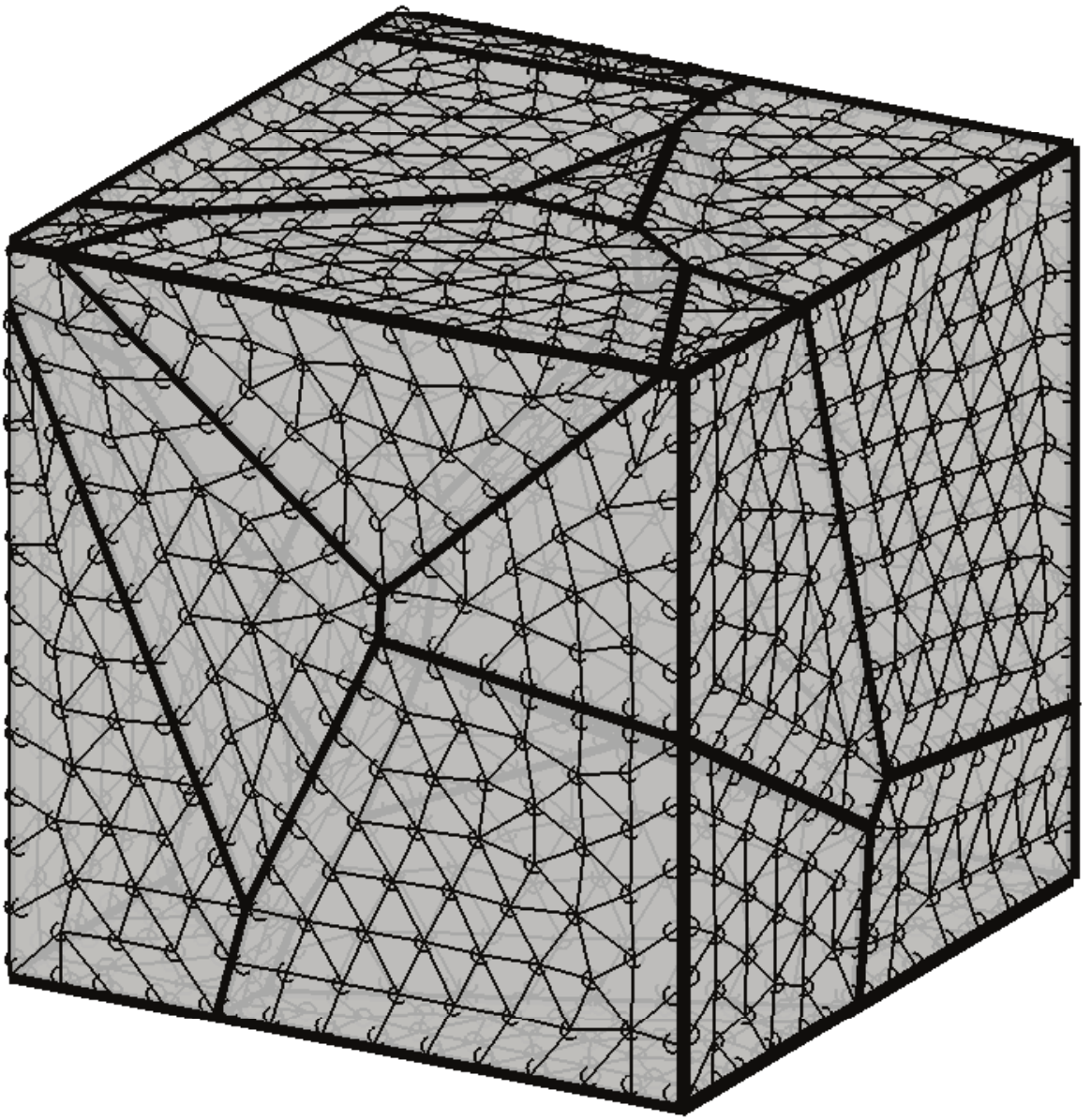}
	\caption{}
	\end{subfigure}
\
	\begin{subfigure}{0.18\textwidth}
	\centering
	\includegraphics[width=\textwidth]{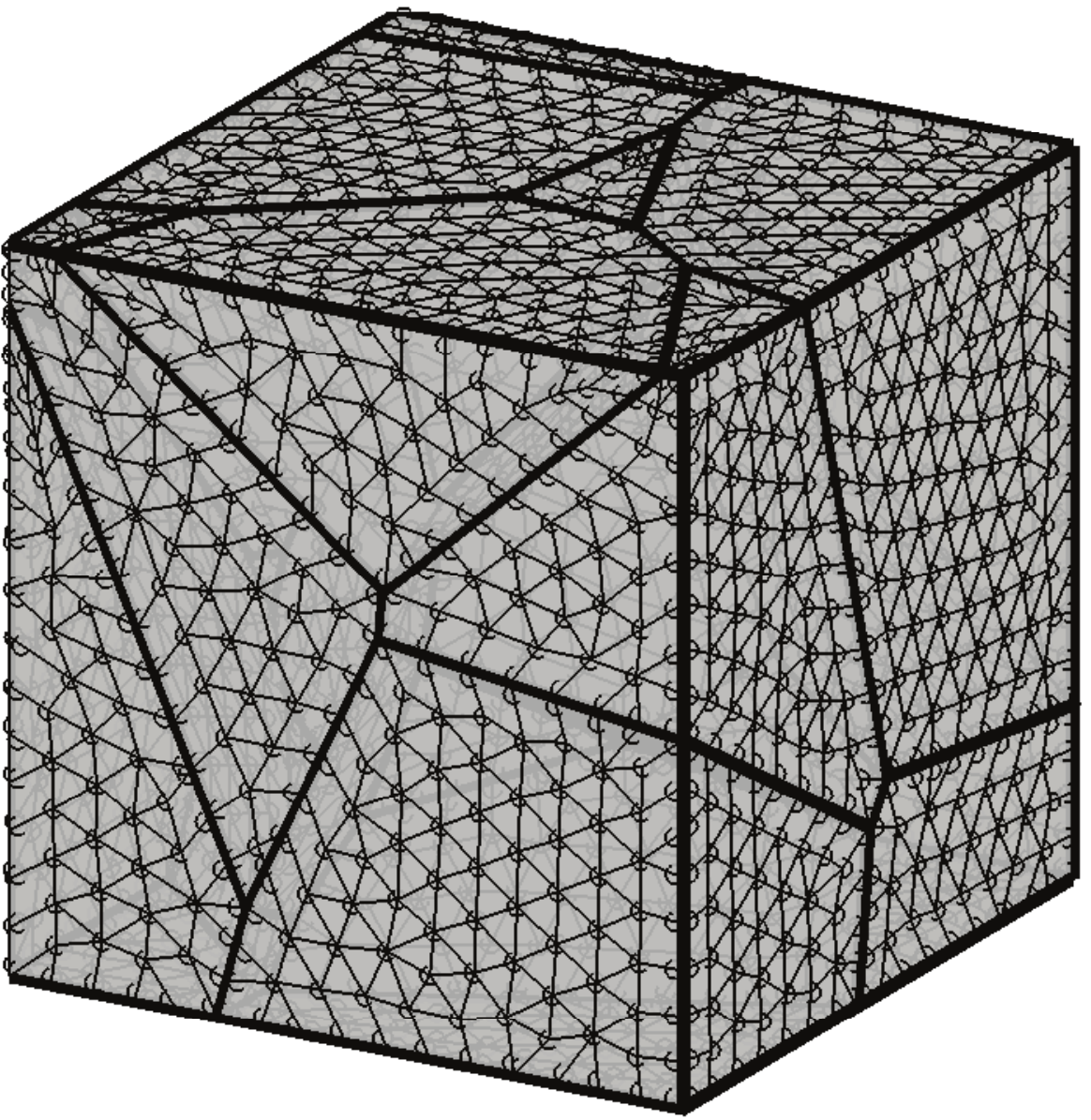}
	\caption{}
	\end{subfigure}
\\
	\begin{subfigure}{0.18\textwidth}
	\centering
	\includegraphics[width=\textwidth]{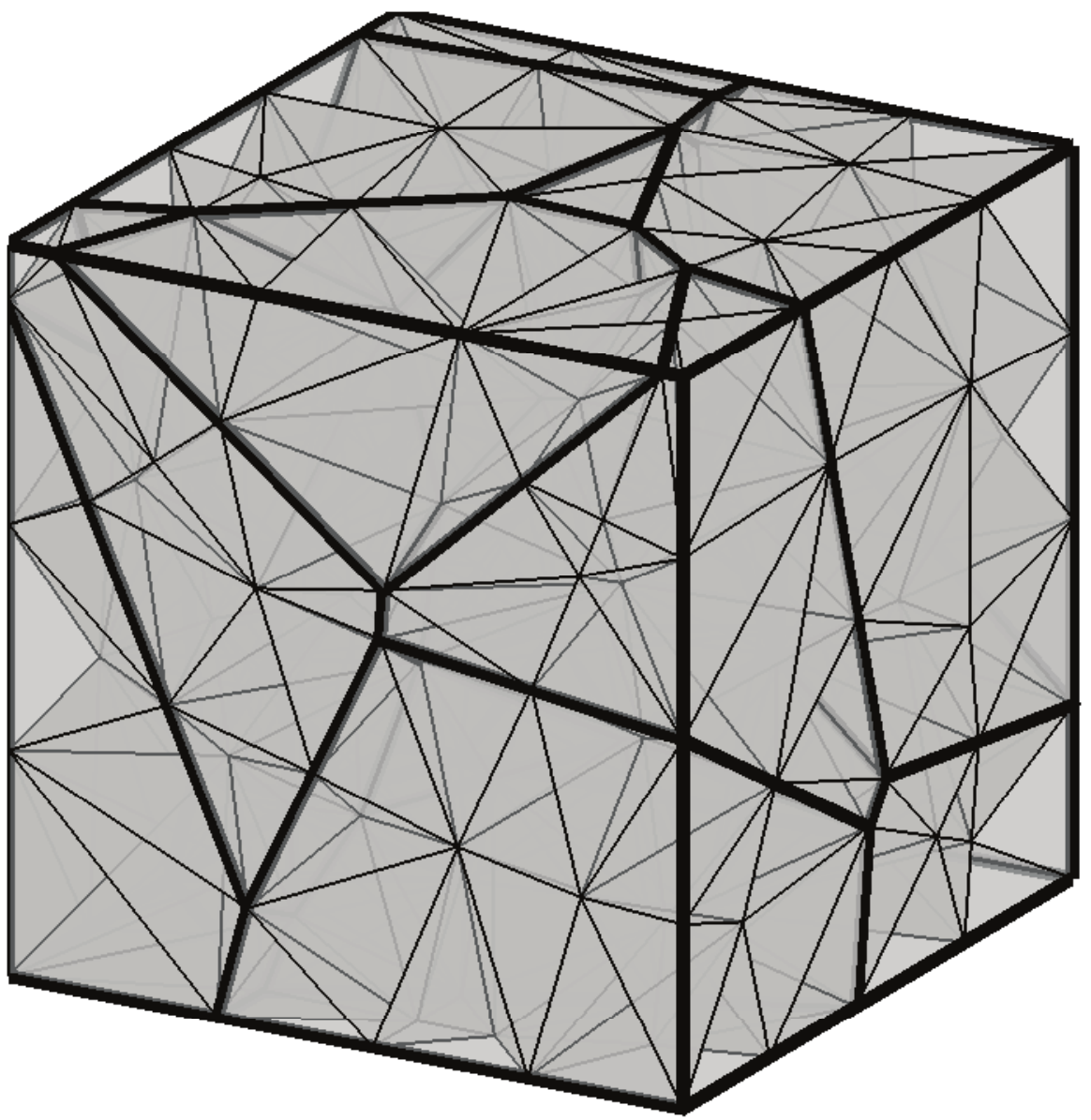}
	\caption{}
	\end{subfigure}
\
	\begin{subfigure}{0.18\textwidth}
	\centering
	\includegraphics[width=\textwidth]{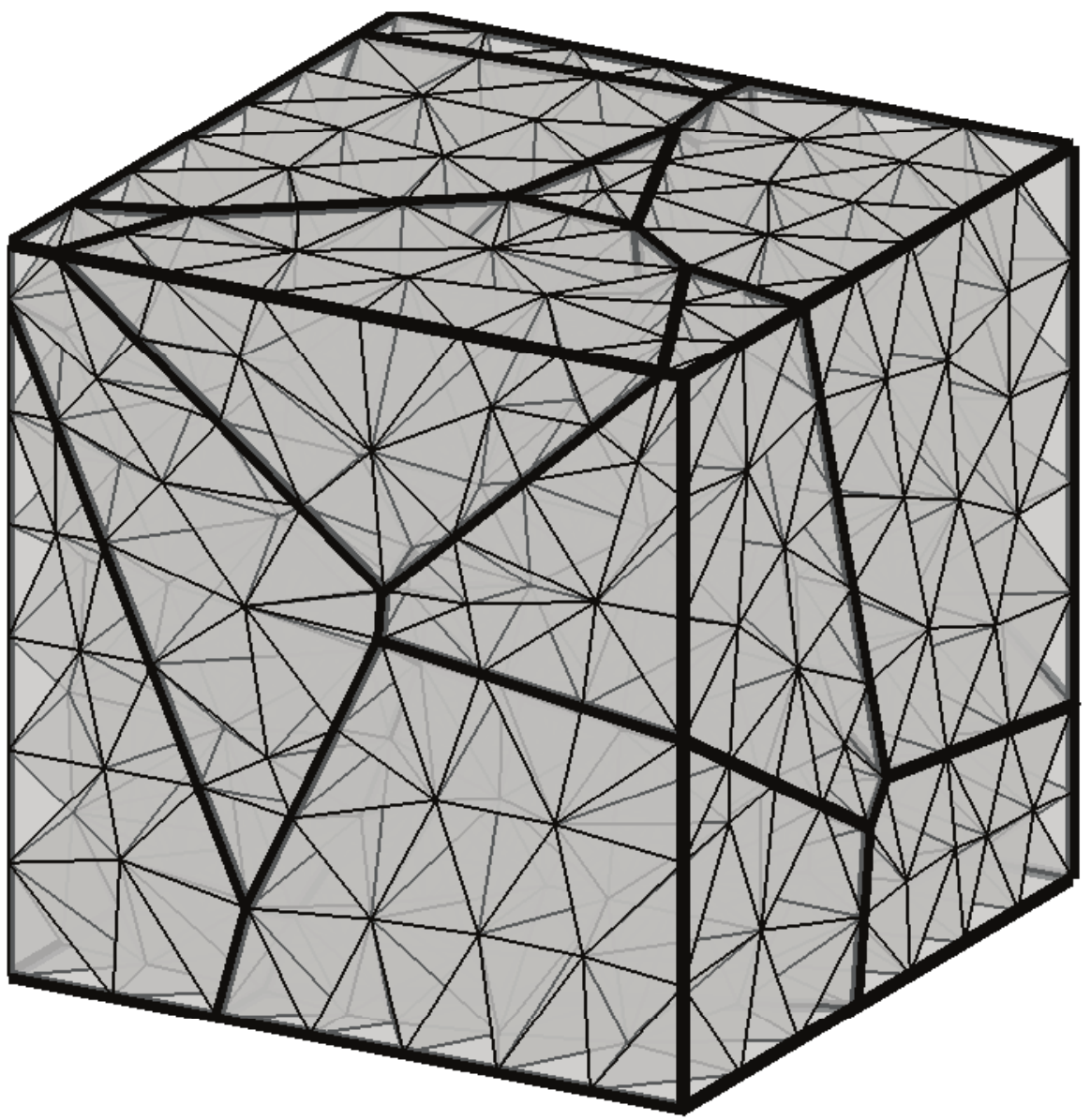}
	\caption{}
	\end{subfigure}
\
	\begin{subfigure}{0.18\textwidth}
	\centering
	\includegraphics[width=\textwidth]{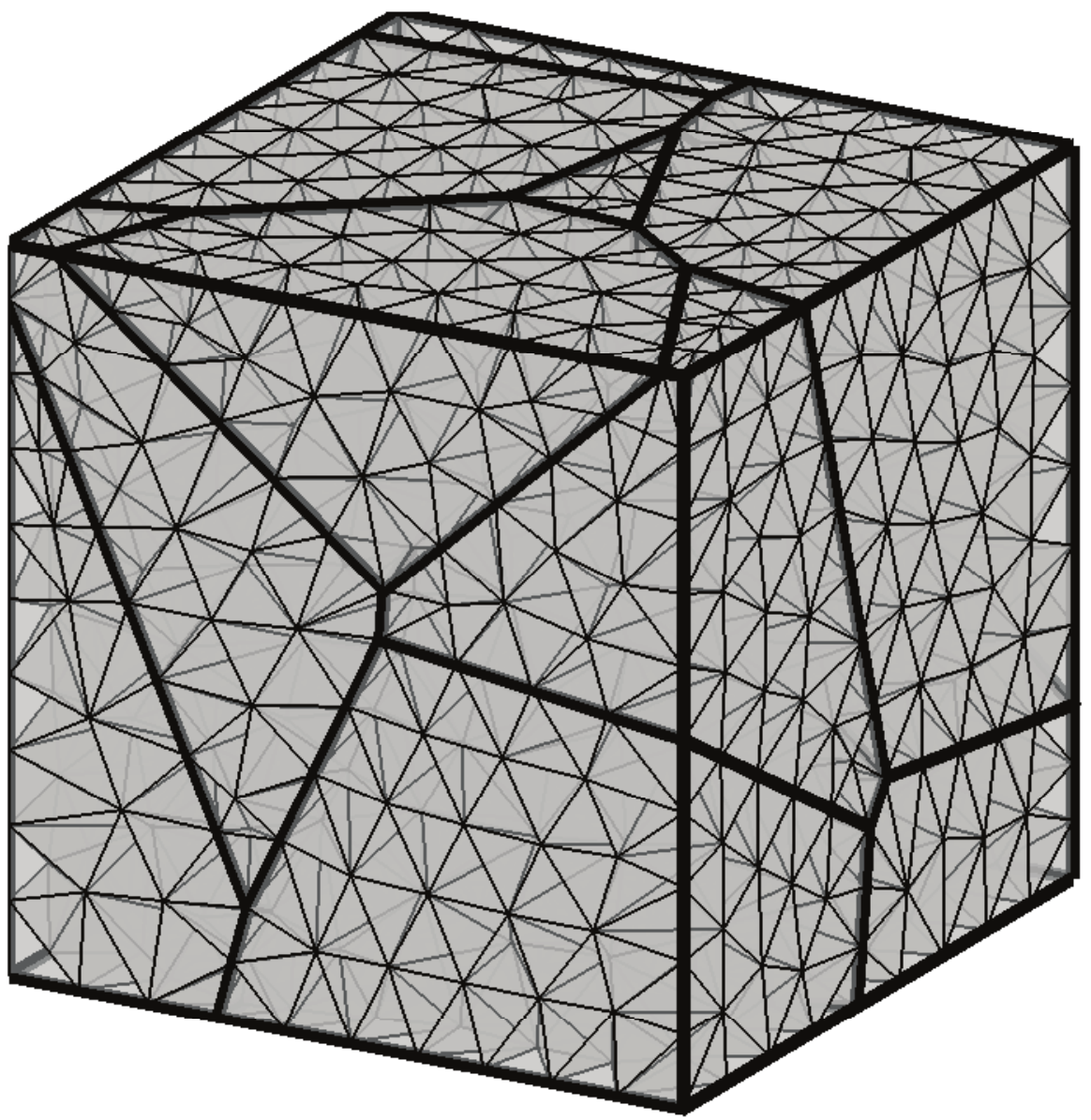}
	\caption{}
	\end{subfigure}
\
	\begin{subfigure}{0.18\textwidth}
	\centering
	\includegraphics[width=\textwidth]{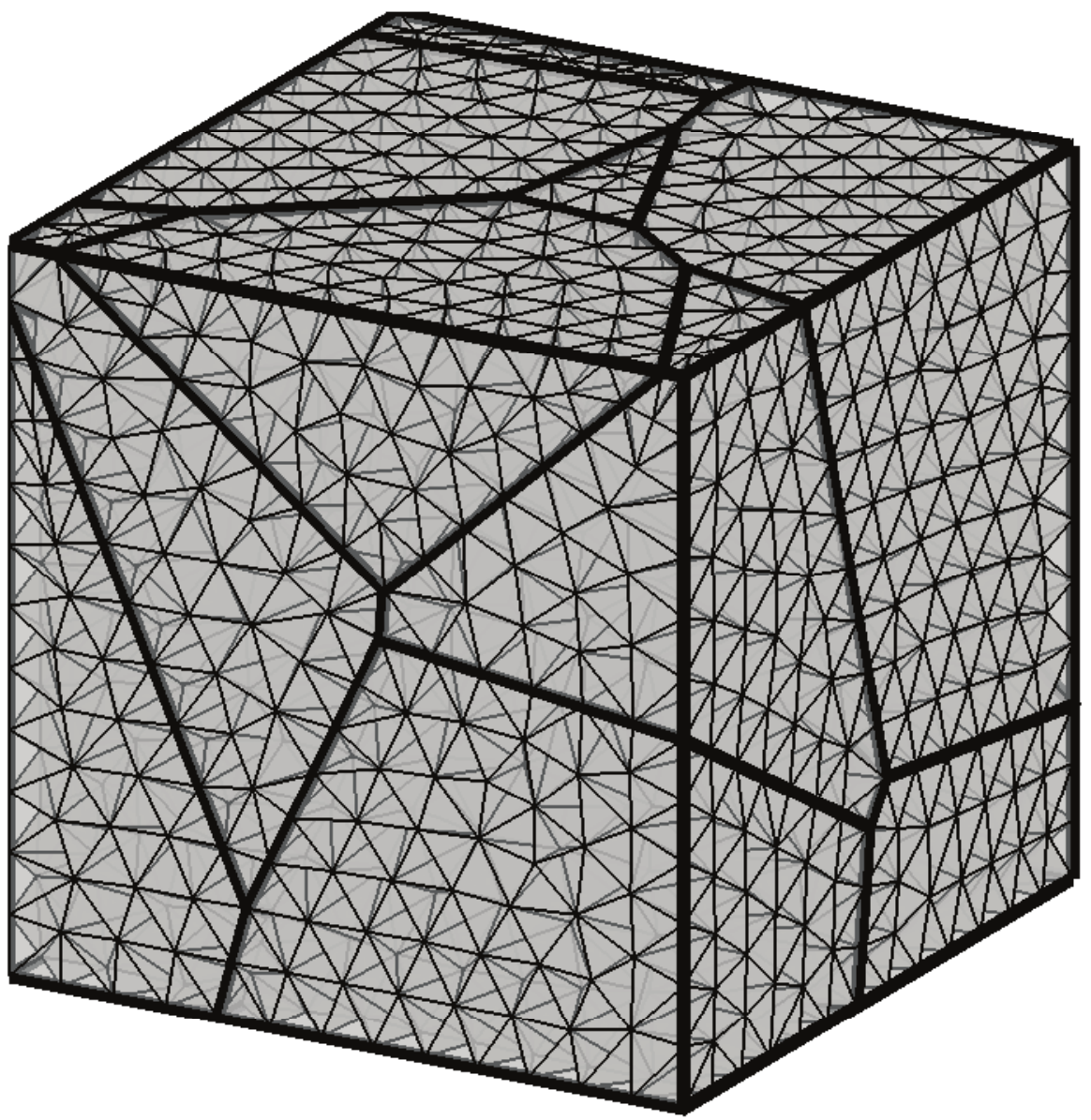}
	\caption{}
	\end{subfigure}
\
	\begin{subfigure}{0.18\textwidth}
	\centering
	\includegraphics[width=\textwidth]{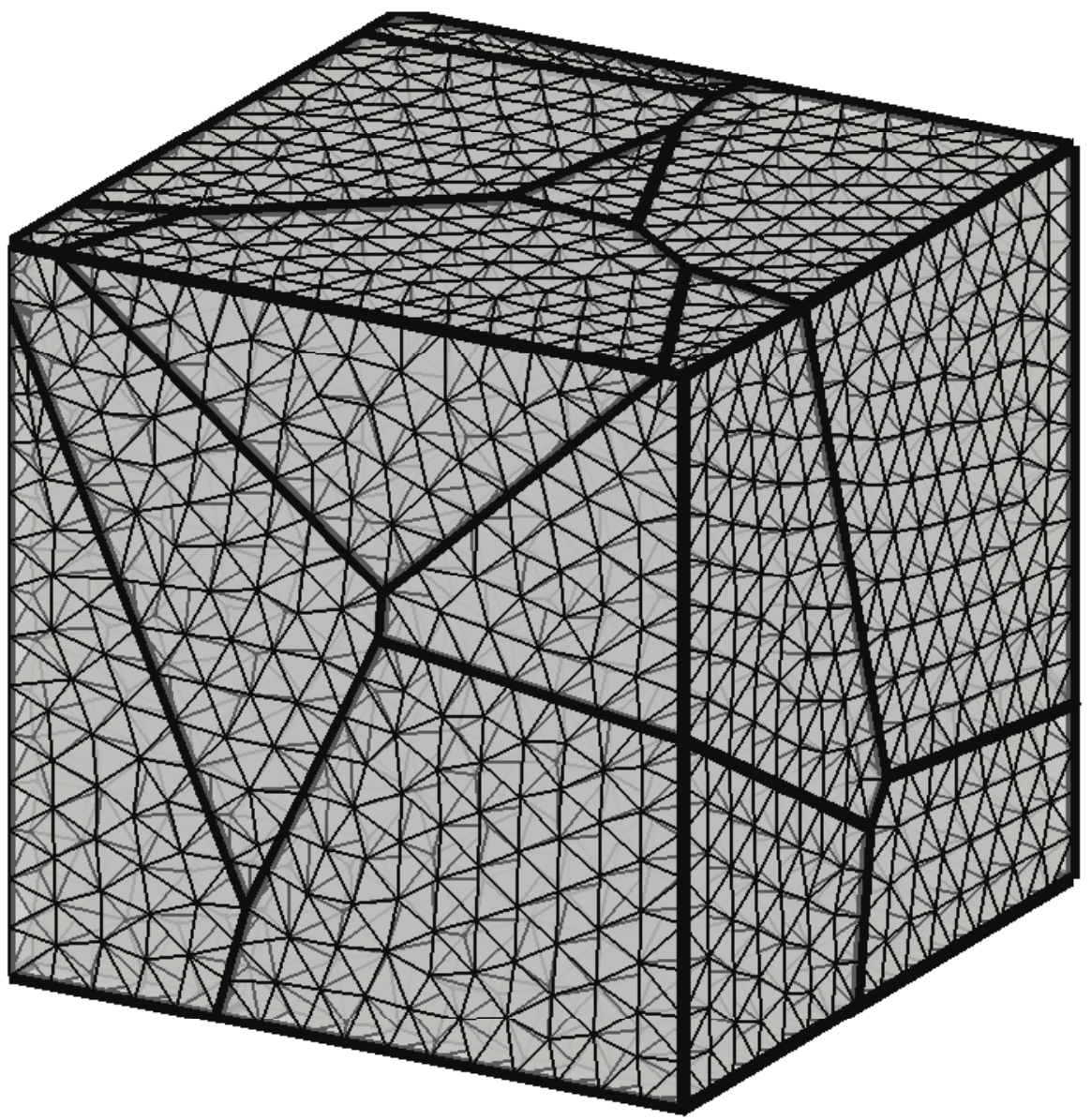}
	\caption{}
	\end{subfigure}
\
\caption{Surface (\emph{a}-\emph{e}) and volume (\emph{f}-\emph{j}) meshes of a 10-grain aggregate: (\emph{a},\emph{f}) $d_m=1$, (\emph{b},\emph{g}) $d_m=2$, (\emph{c},\emph{h}) $d_m=3$, (\emph{d},\emph{i}) $d_m=4$, (\emph{e},\emph{j}) $d_m=5$. The small circles in the surface meshes represent the collocation points.}
\label{fig-Ch4:G10_meshes}
\end{figure}

\subsection{Polycrystalline tests}\label{ssec-Ch4: poly crystal}
In the present Section, the proposed model is used to simulate the plastic behaviour of Cu polycrystalline aggregates. To generate the polycrystalline morphologies, the Neper package \cite{quey2011} is used for its capability of eliminating pathologically small geometrical entities typical of domains subdivided by the Voronoi or Laguerre algorithm, which may induce unnecessarily heavy meshes.

\subsubsection{10-grain aggregate}
First, a convergence analysis is performed to obtain a proper mesh size ensuring appropriate grains boundary and volume fields resolution and mesh independency of the results.
Figure (\ref{fig-Ch4:G10_meshes}) shows five different meshes ($d_m=1,2,...,5$) of a 10-grain aggregate. The statistics about the number of DoFs and the number of surface and volume elements are summarised in Table (\ref{tab-Ch4:meshes-stats}). The aggregate is loaded in uniaxial tension, with uniform tractions acting over the top and bottom faces, while the lateral faces are traction-free. In this case, to avoid rigid body motion, the elimination of rigid body modes technique is applied to one grain of the aggregate.

\begin{figure}[h]
\centering
	\begin{subfigure}{0.49\textwidth}
	\centering
	\includegraphics[width=\textwidth]{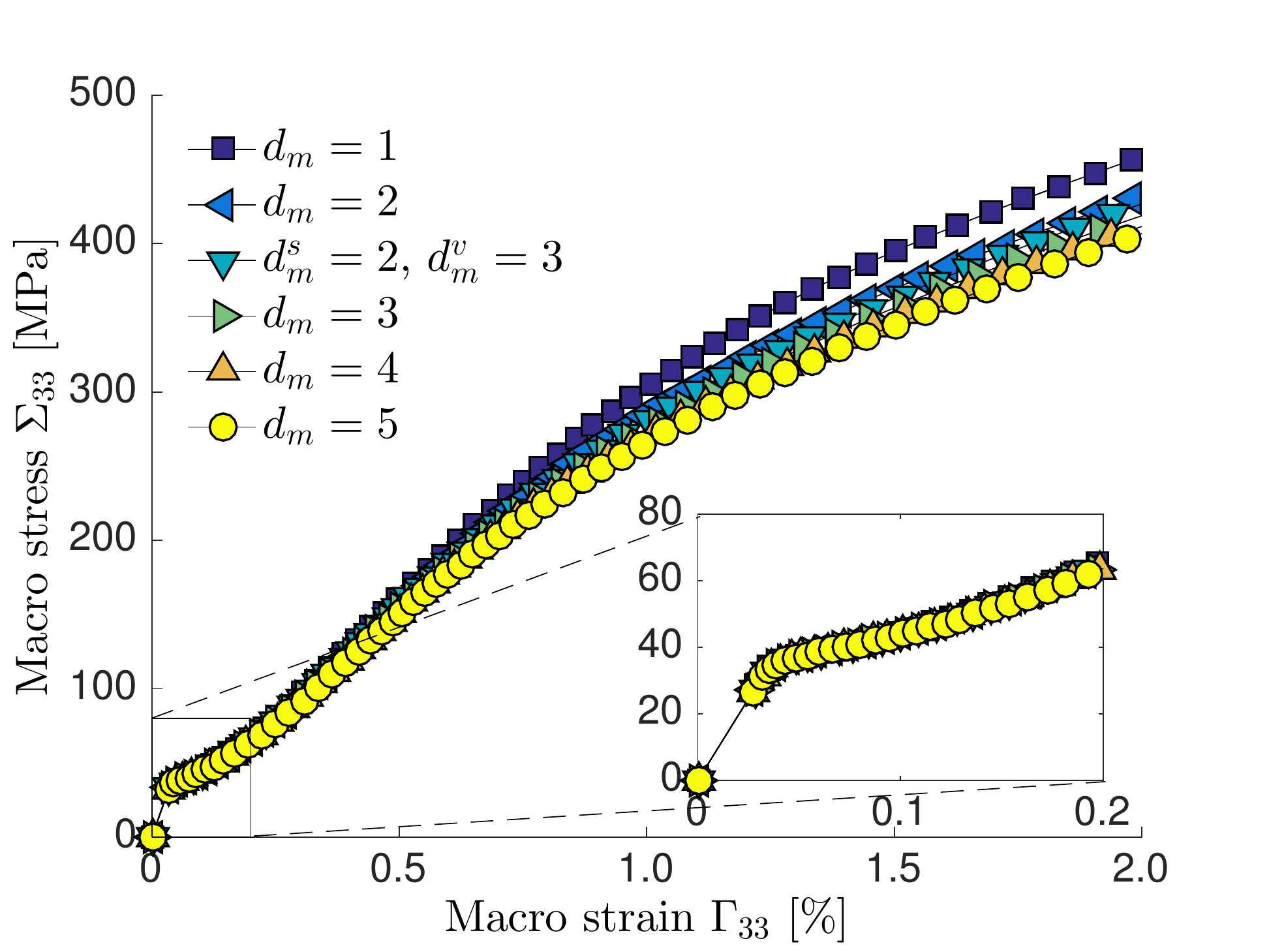}
	\caption{}
	\end{subfigure}
\
	\begin{subfigure}{0.49\textwidth}
	\centering
	\includegraphics[width=\textwidth]{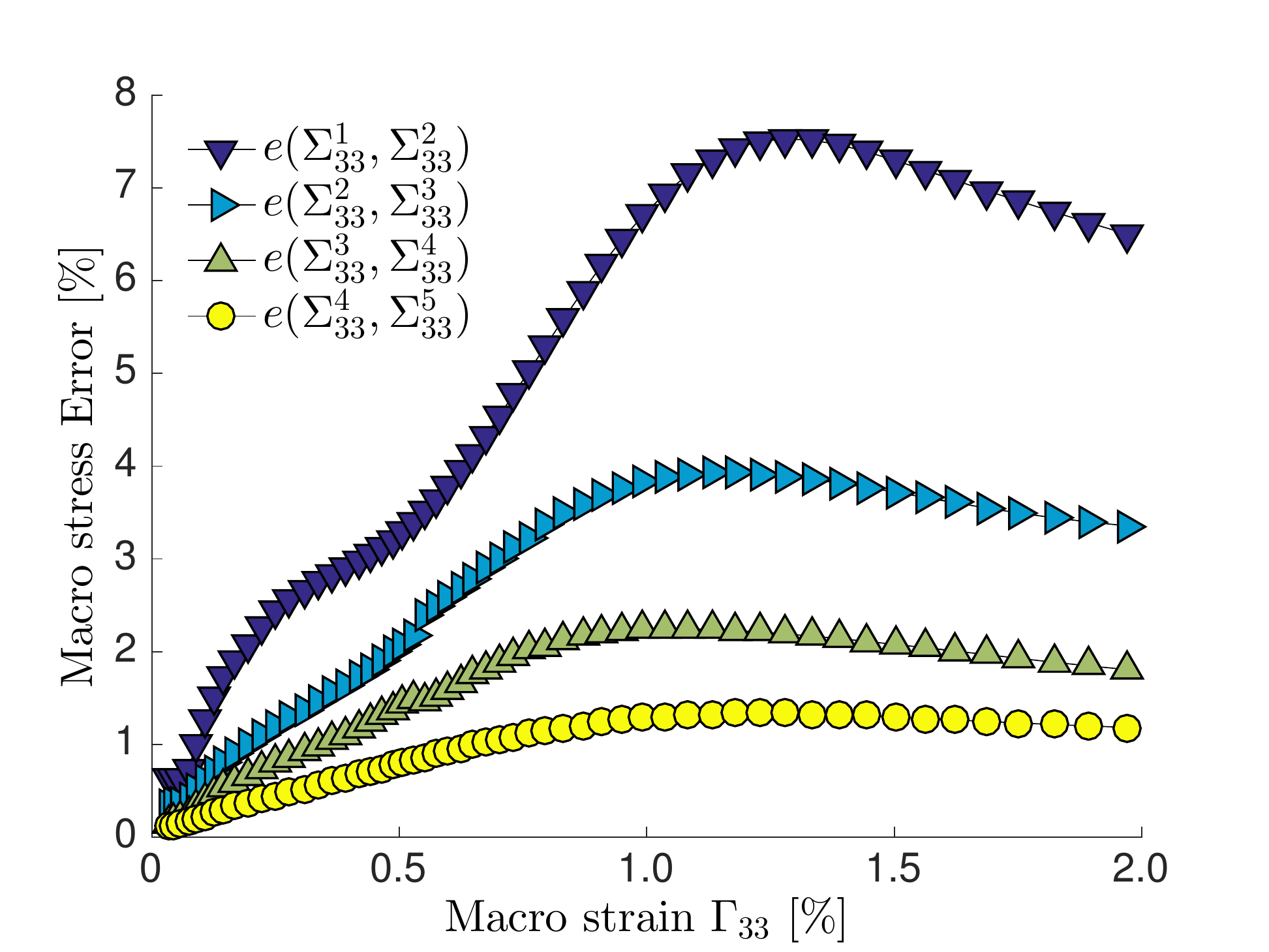}
	\caption{}
	\end{subfigure}
\
\caption{Volume stress average $\Sigma_{33}$ versus volume strain average $\Gamma_{33}$ for the considered meshes of the 10-grain aggregate. (\emph{a}) Stress-strain response in the strain range 0-2.0\%; the close-up shows the values obtained in the strain range 0-0.2\%. (\emph{b}) Error between two consecutive mesh refinements.}
\label{fig-Ch4:G10_meshconv}
\end{figure}

Figure (\ref{fig-Ch4:G10_meshconv}a) shows the convergence of the macro stress-strain response of the Cu 10-grain aggregate with respect to the considered meshes. As the model allows to build the surface and the volume meshes separately, Figure (\ref{fig-Ch4:G10_meshconv}a) shows also the macro response obtained combining a surface mesh with a mesh density parameter $d_m=2$ and a volume mesh with a mesh density parameter $d_m=3$, which corresponds to the curve labelled with $d_m^s=2$, $d_m^v=3$. Figure (\ref{fig-Ch4:G10_meshconv}b) reports the macro-stress error during the loading history between two consecutive mesh refinements defined as:
\begin{equation}
e(\Sigma_{33}^i,\Sigma_{33}^{i+1})=100\cdot\frac{|\Sigma_{33}^{i} - \Sigma_{33}^{i+1}|}{\Sigma_{33}^5}
\end{equation}
where $\Sigma_{33}^i$ corresponds to the macro stress computed for $d_m=i$.

Figure (\ref{fig-Ch4:G10_meshes_def}) shows the deformed configurations of the aggregate for the meshes of Figure (\ref{fig-Ch4:G10_meshes}) at the last computed step of the load history, i.e.\ for $\Gamma_{33}=2\%$. A good convergence is obtained in terms of the grain boundary displacements starting from $d_m=2$. Figure (\ref{fig-Ch4:G10_meshes_vonmises}) and Figure (\ref{fig-Ch4:G10_meshes_cumgamma}) report the contour plots of the Von Mises stress and cumulative slip fields, respectively. Also in this case, the convergence in terms of the overall behaviour and stress and slip concentration is satisfactorily reached through mesh refinement. It is however noted that a $d_m=3$ mesh is required to appreciate the fields distribution, likely due to the use of constant volume elements. Considering Table (\ref{tab-Ch4:meshes-stats}), an estimate of the number of volume elements per grain to obtain satisfactory fields distributions within the aggregate appears to be around 400 elements per grain, at least for the present load condition.

\begin{figure}[H]
\centering
	\begin{subfigure}{0.18\textwidth}
	\centering
	\includegraphics[width=\textwidth]{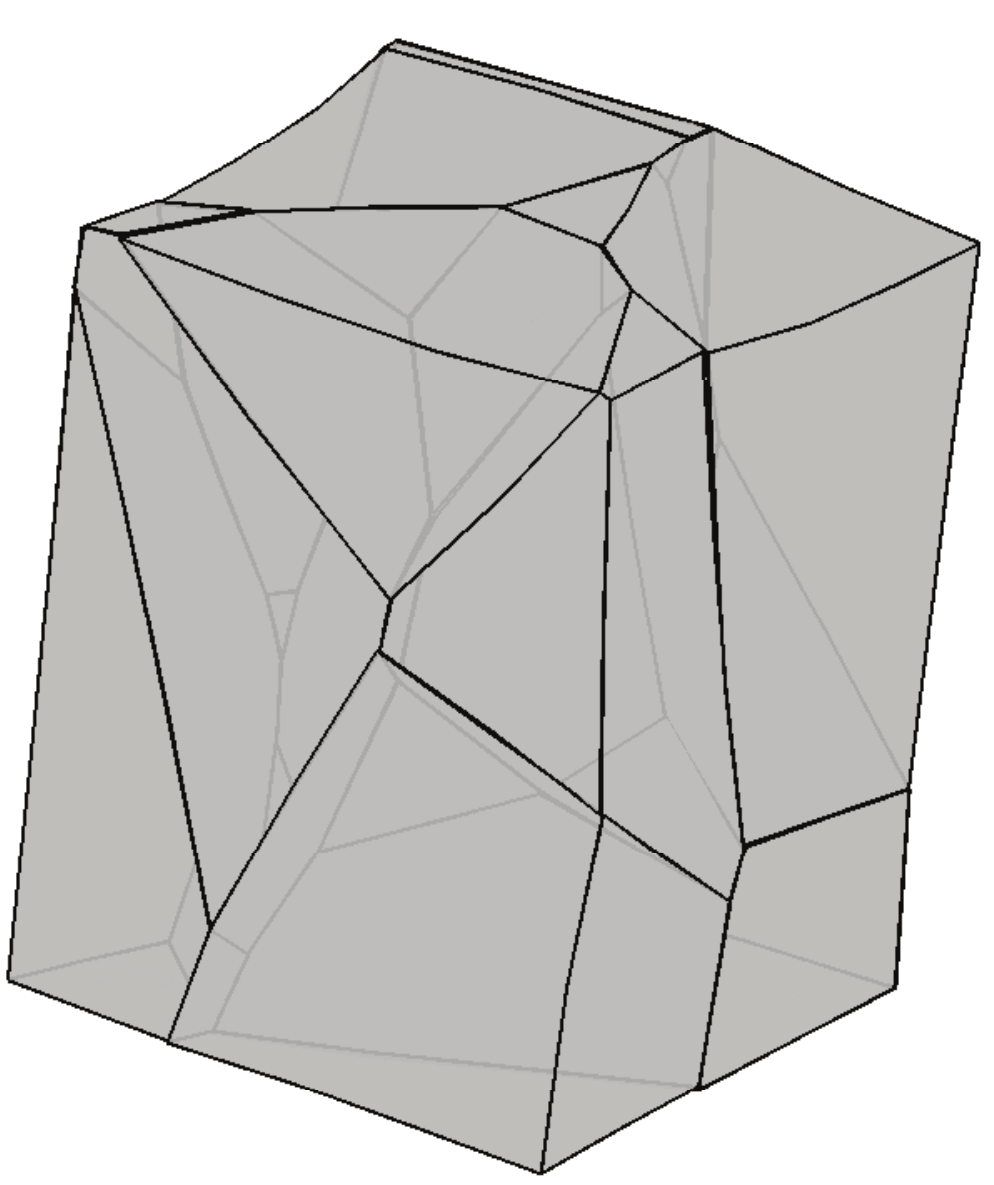}
	\caption{}
	\end{subfigure}
\
	\begin{subfigure}{0.18\textwidth}
	\centering
	\includegraphics[width=\textwidth]{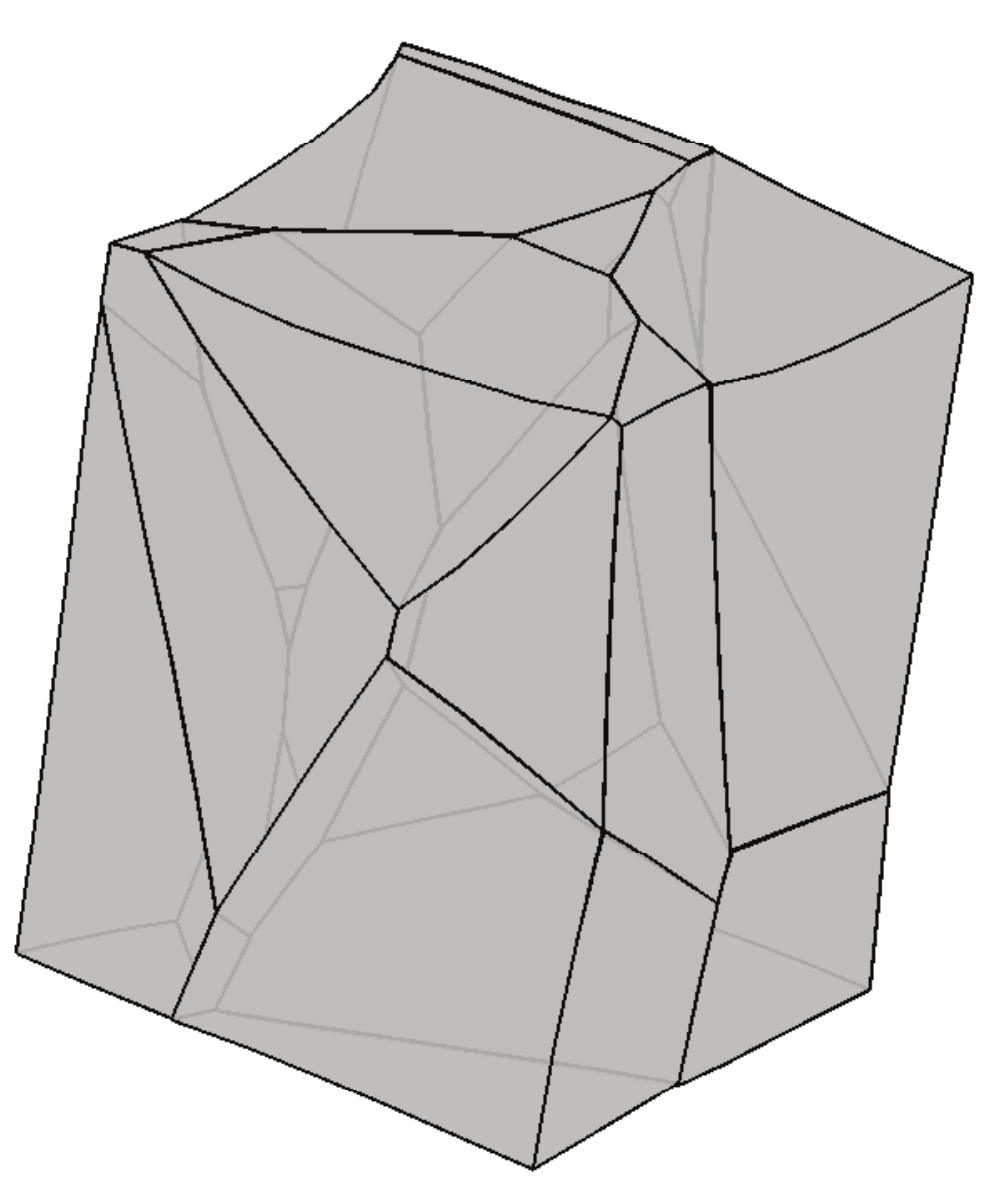}
	\caption{}
	\end{subfigure}
\
	\begin{subfigure}{0.18\textwidth}
	\centering
	\includegraphics[width=\textwidth]{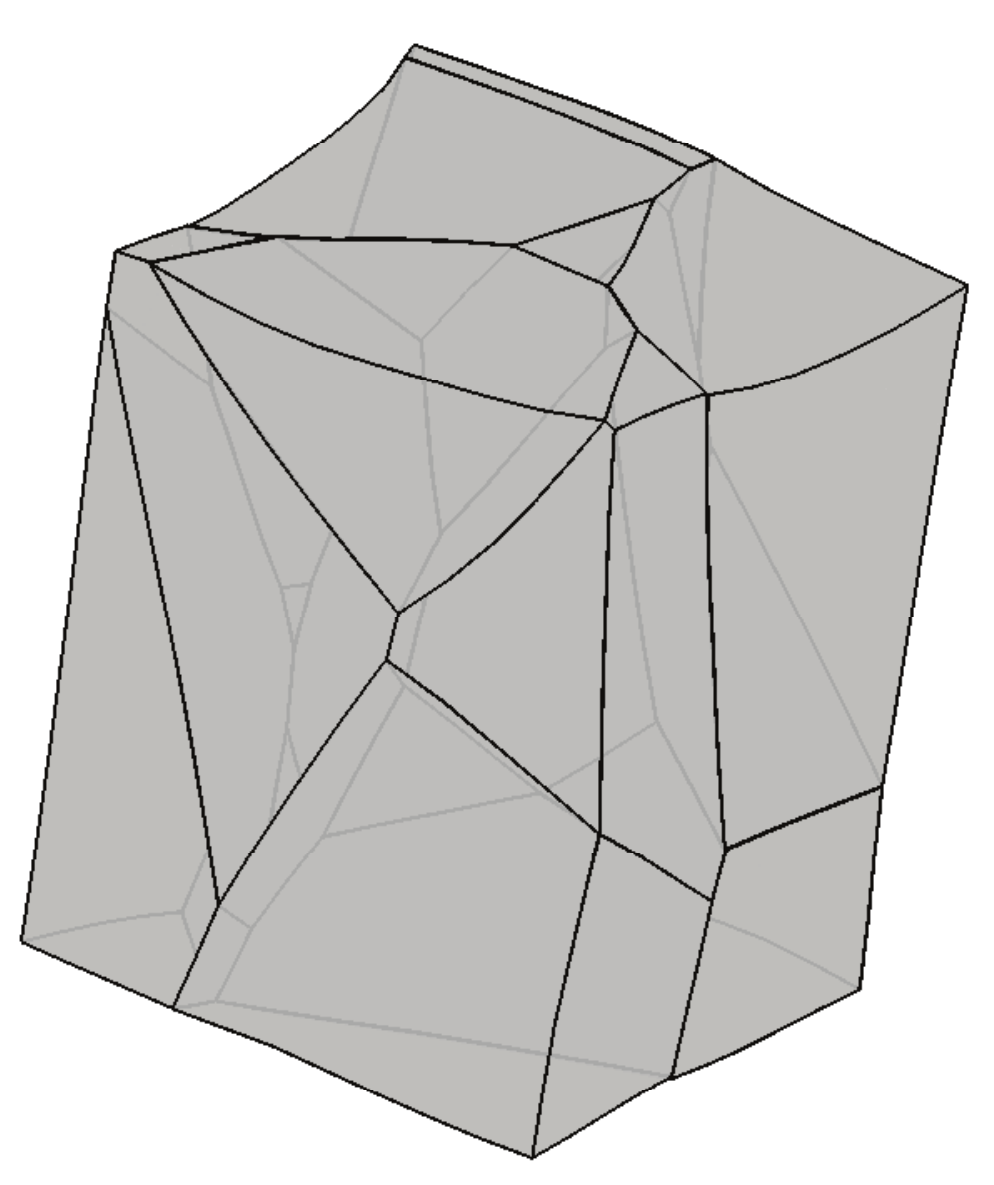}
	\caption{}
	\end{subfigure}
\
	\begin{subfigure}{0.18\textwidth}
	\centering
	\includegraphics[width=\textwidth]{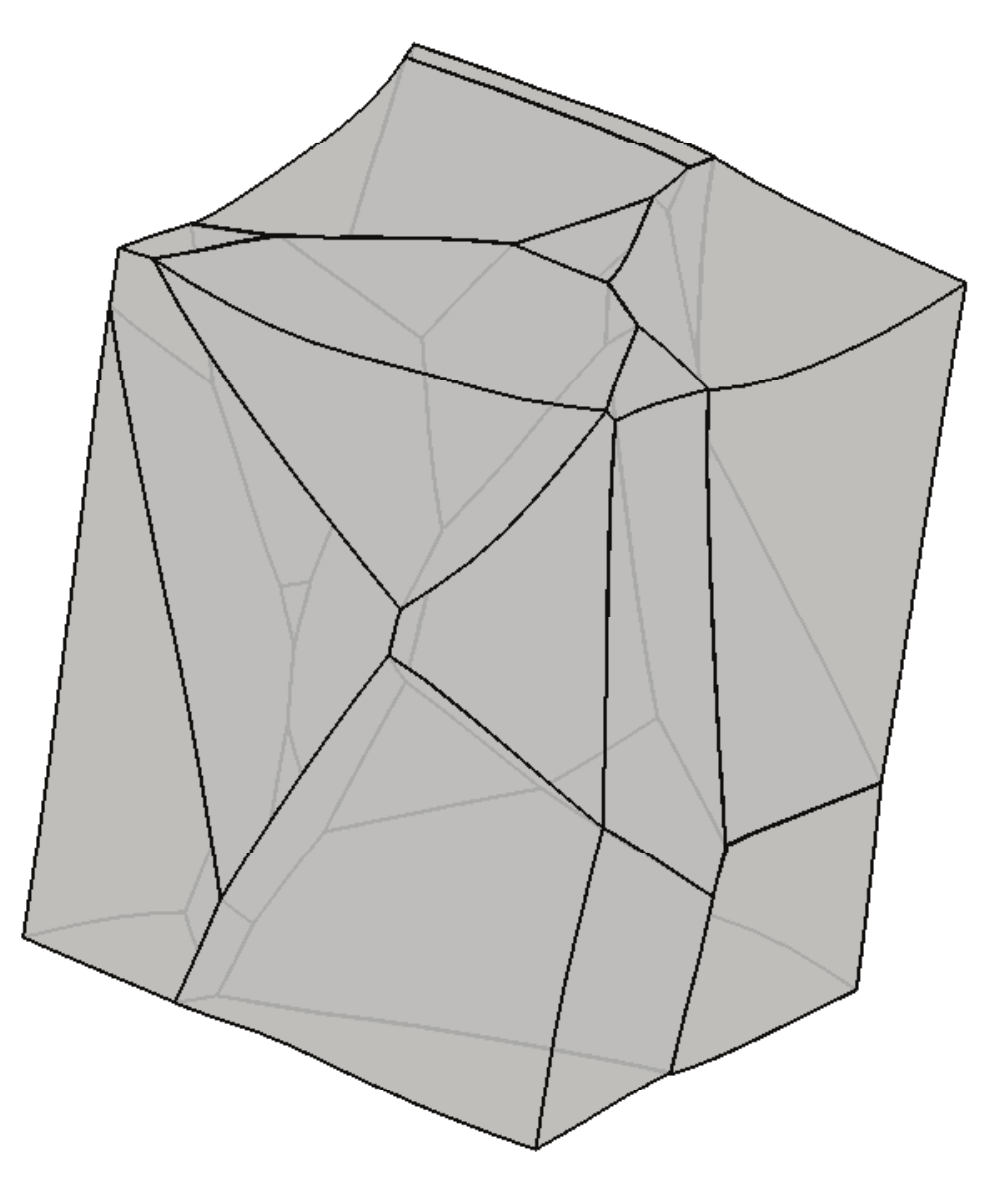}
	\caption{}
	\end{subfigure}
\
	\begin{subfigure}{0.18\textwidth}
	\centering
	\includegraphics[width=\textwidth]{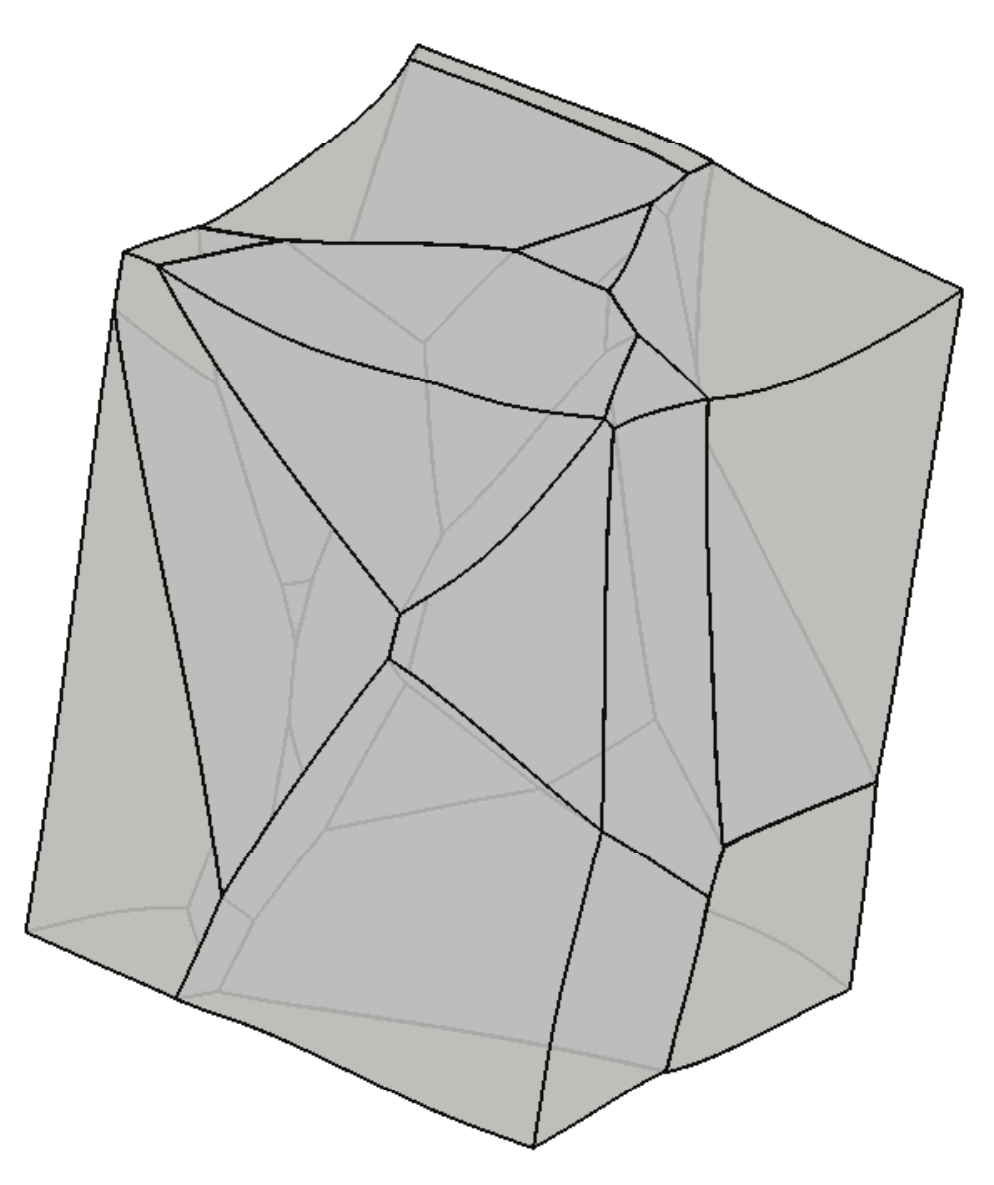}
	\caption{}
	\end{subfigure}
\caption{Deformed shape for the five considered meshes of the 10-grain polycrystalline copper aggregate subjected to tensile load at the last computed step ($\Gamma_{33}=2\%$). Figs.(\emph{a}) to (\emph{e}) correspond to the (\emph{a}) to (\emph{e}) surface meshes and (\emph{f}) to (\emph{j}) volume meshes of Figure (\ref{fig-Ch4:G10_meshes}), respectively.}
\label{fig-Ch4:G10_meshes_def}
\end{figure}

\begin{figure}[H]
\centering
	\begin{subfigure}{0.18\textwidth}
	\centering
	\includegraphics[width=\textwidth]{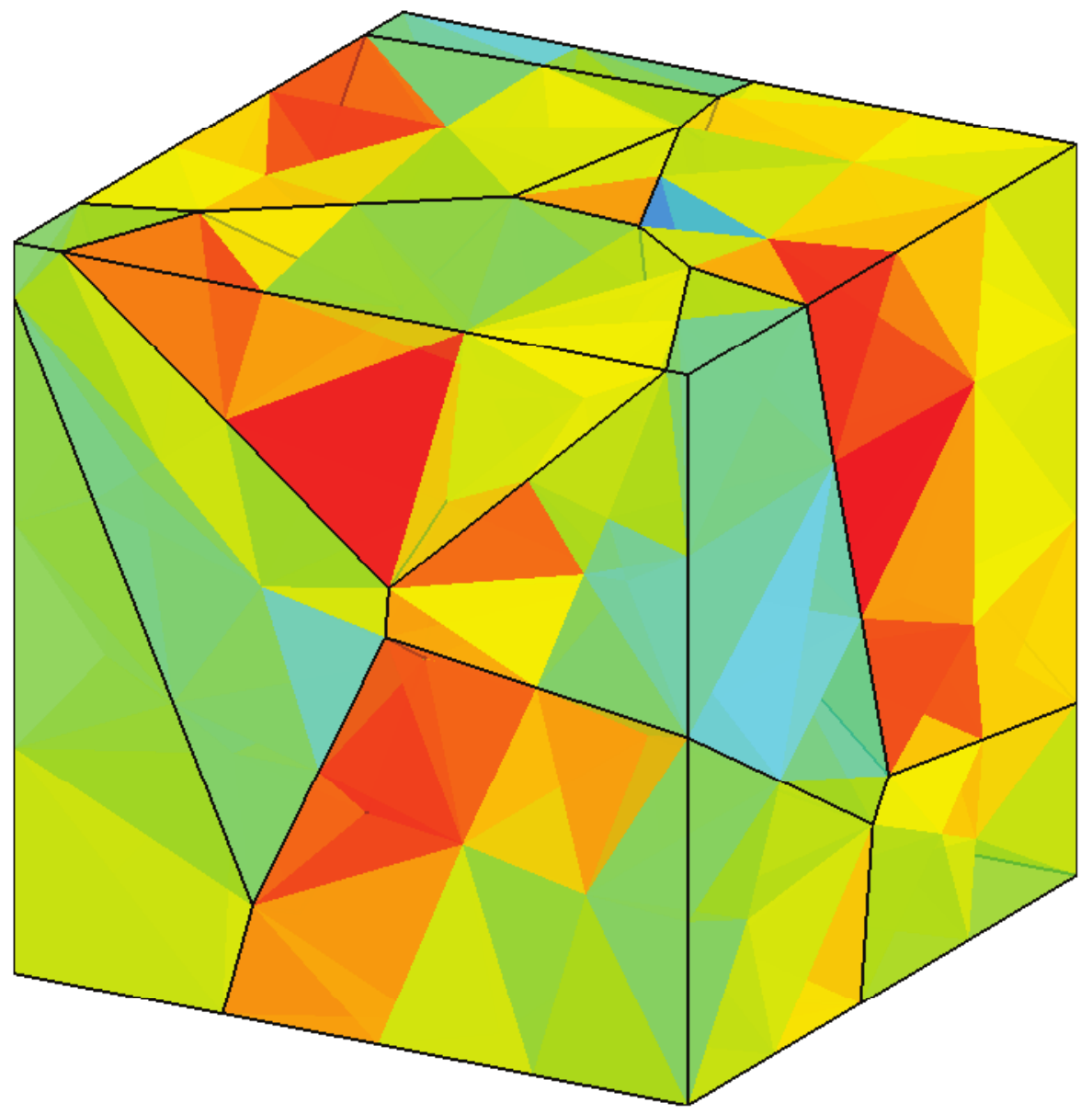}
	\caption{}
	\end{subfigure}
\
	\begin{subfigure}{0.18\textwidth}
	\centering
	\includegraphics[width=\textwidth]{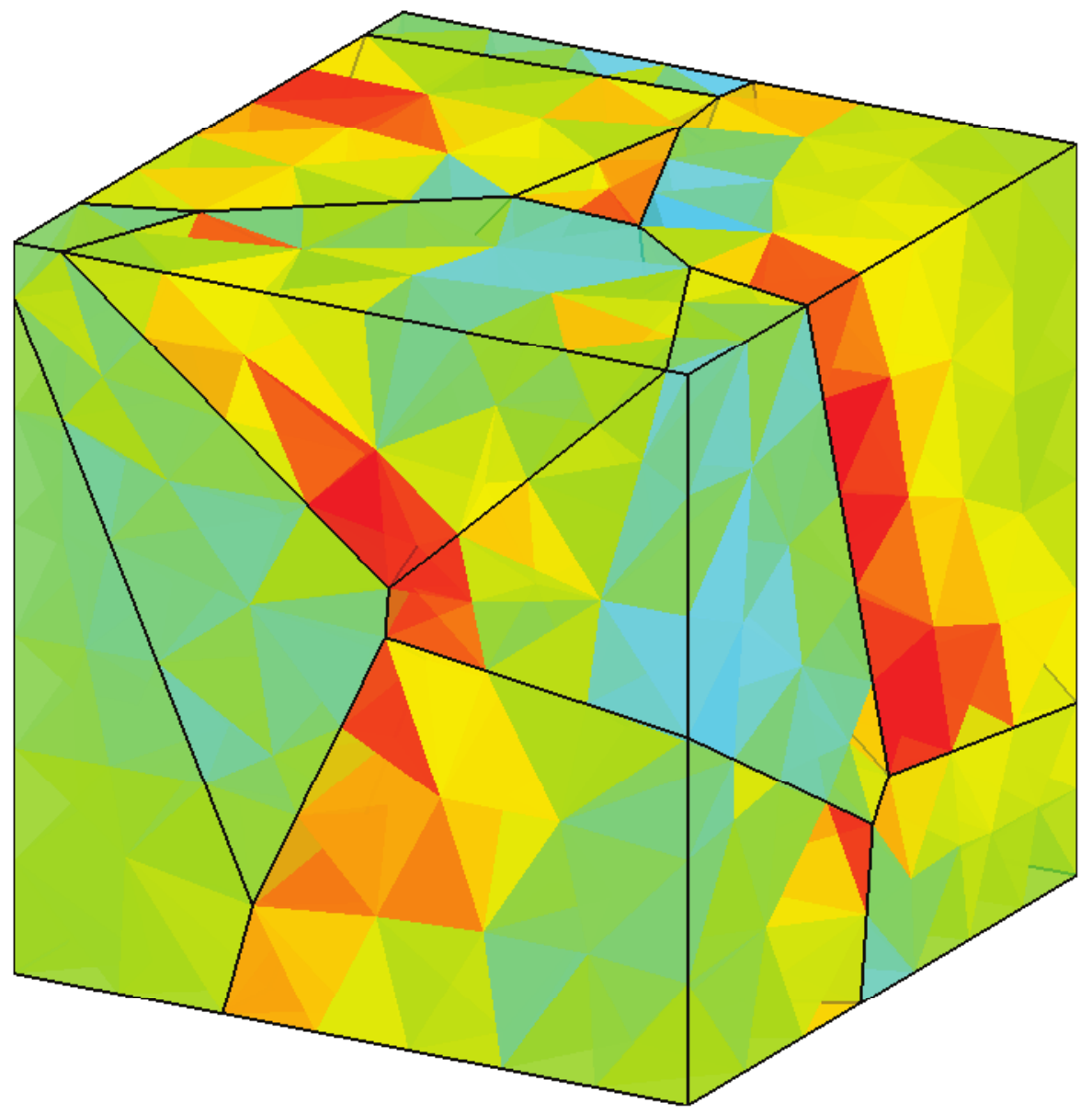}
	\caption{}
	\end{subfigure}
\
	\begin{subfigure}{0.18\textwidth}
	\centering
	\includegraphics[width=\textwidth]{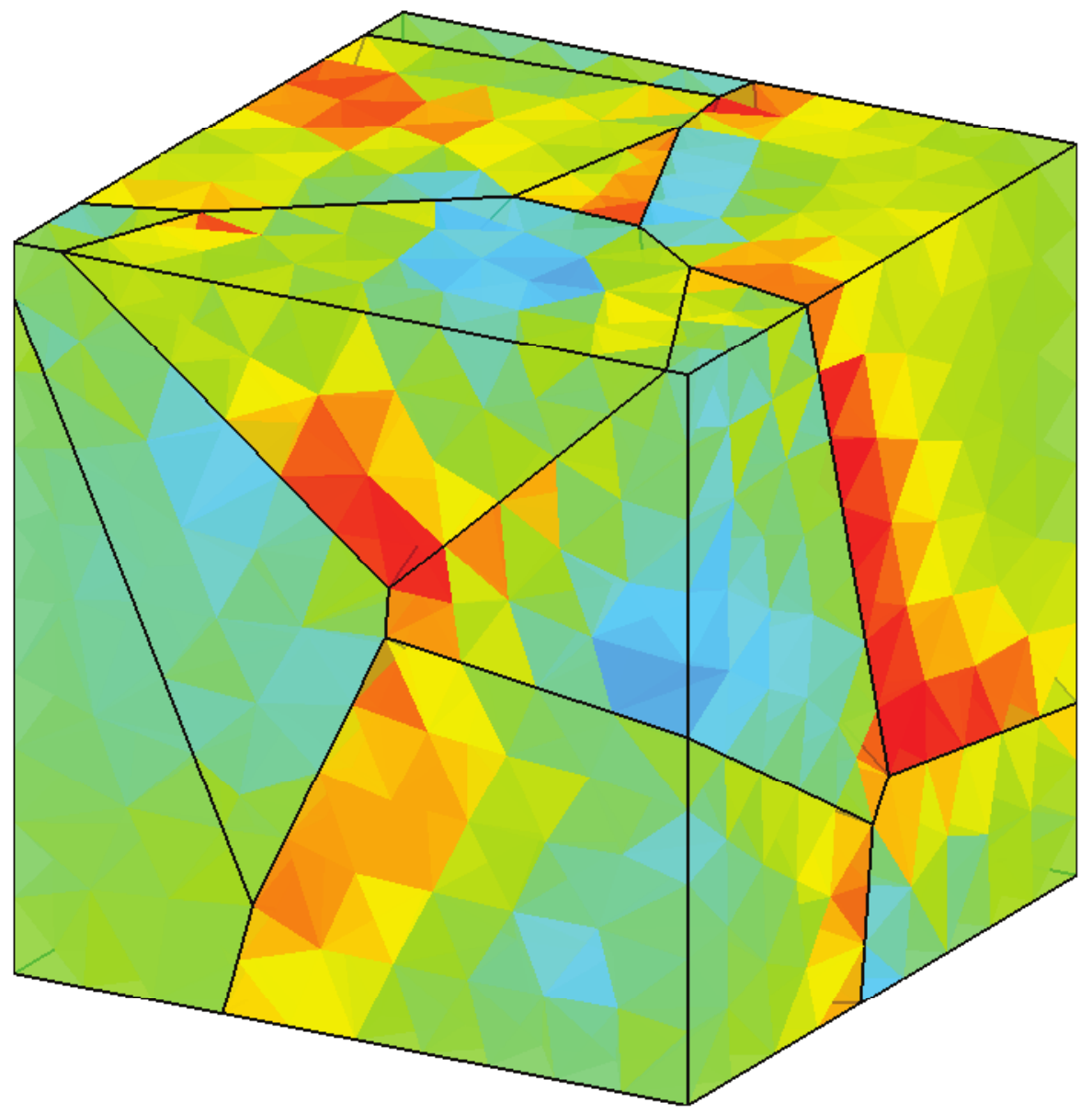}
	\caption{}
	\end{subfigure}
\
	\begin{subfigure}{0.18\textwidth}
	\centering
	\includegraphics[width=\textwidth]{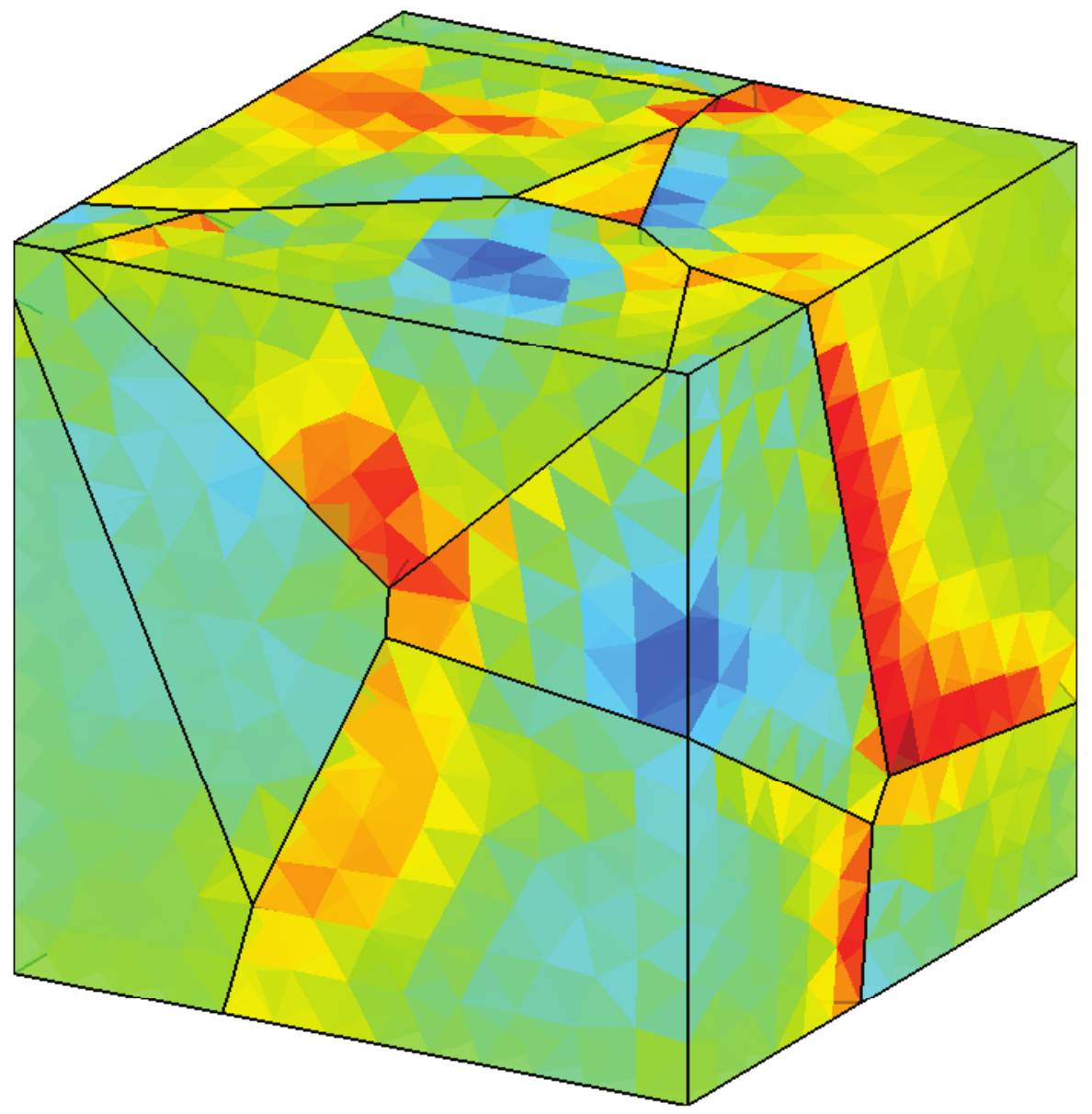}
	\caption{}
	\end{subfigure}
\
	\begin{subfigure}{0.18\textwidth}
	\centering
	\includegraphics[width=\textwidth]{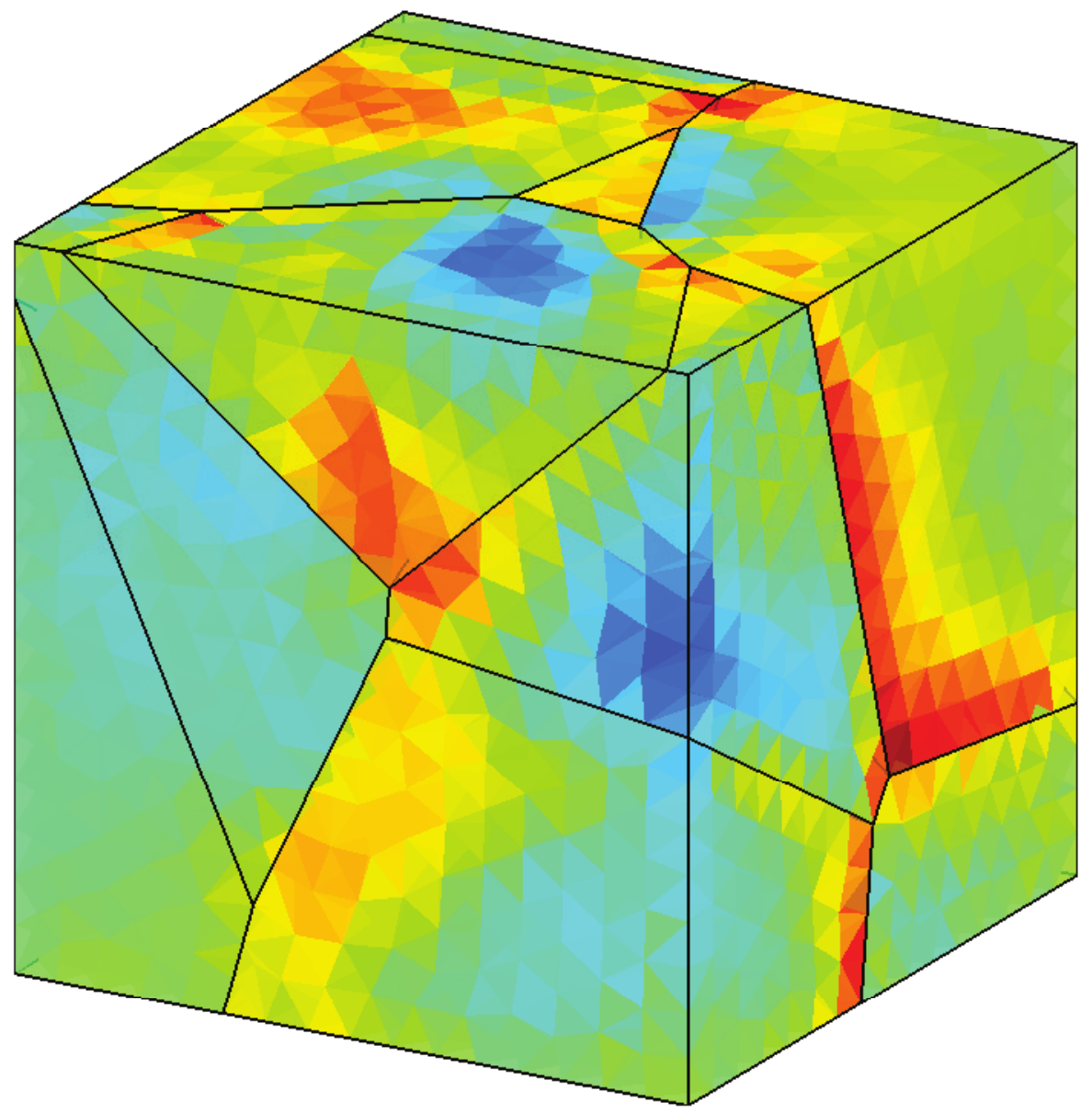}
	\caption{}
	\end{subfigure}
\\
	\begin{subfigure}{0.18\textwidth}
	\centering
	\includegraphics[width=\textwidth]{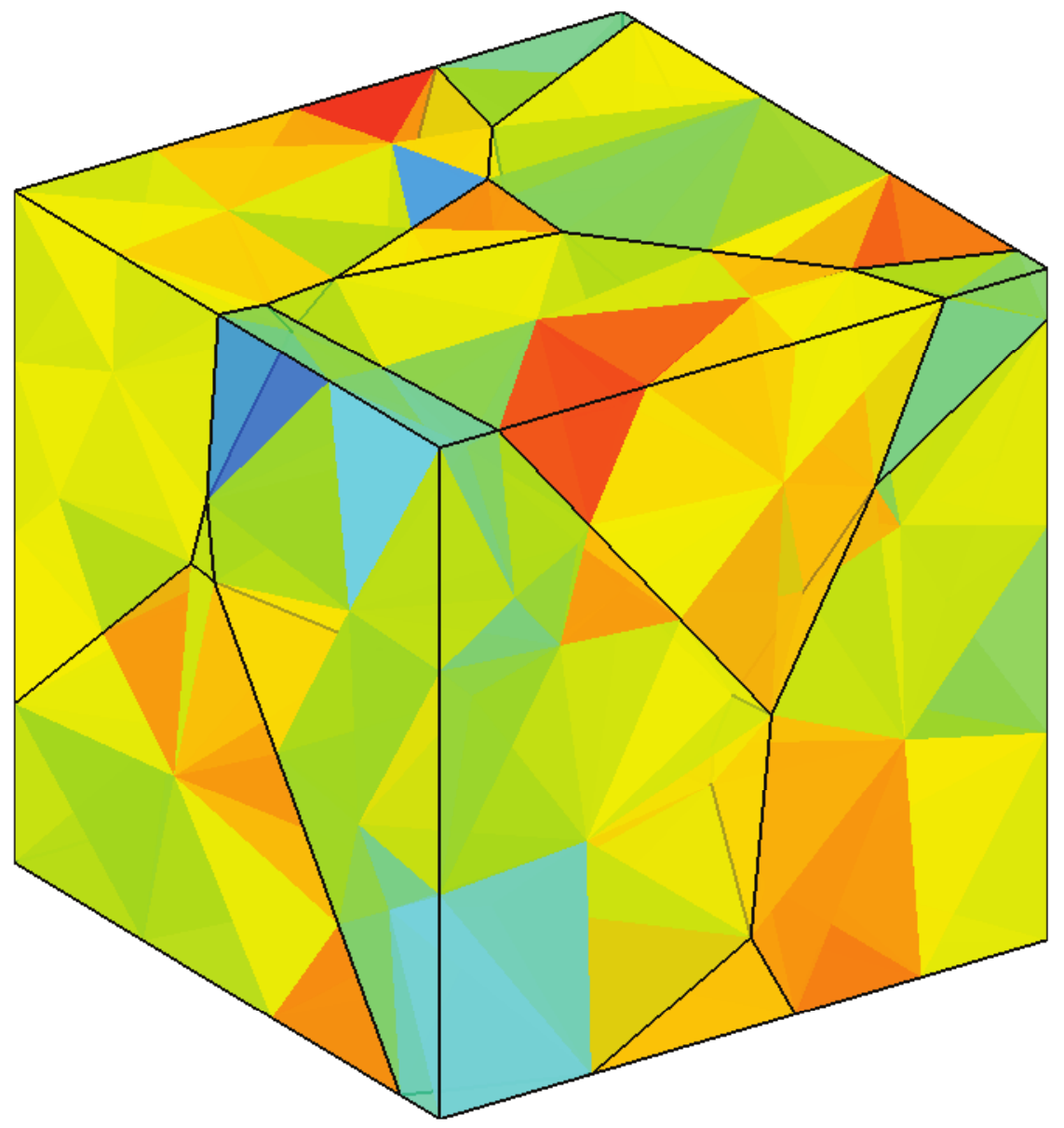}
	\caption{}
	\end{subfigure}
\
	\begin{subfigure}{0.18\textwidth}
	\centering
	\includegraphics[width=\textwidth]{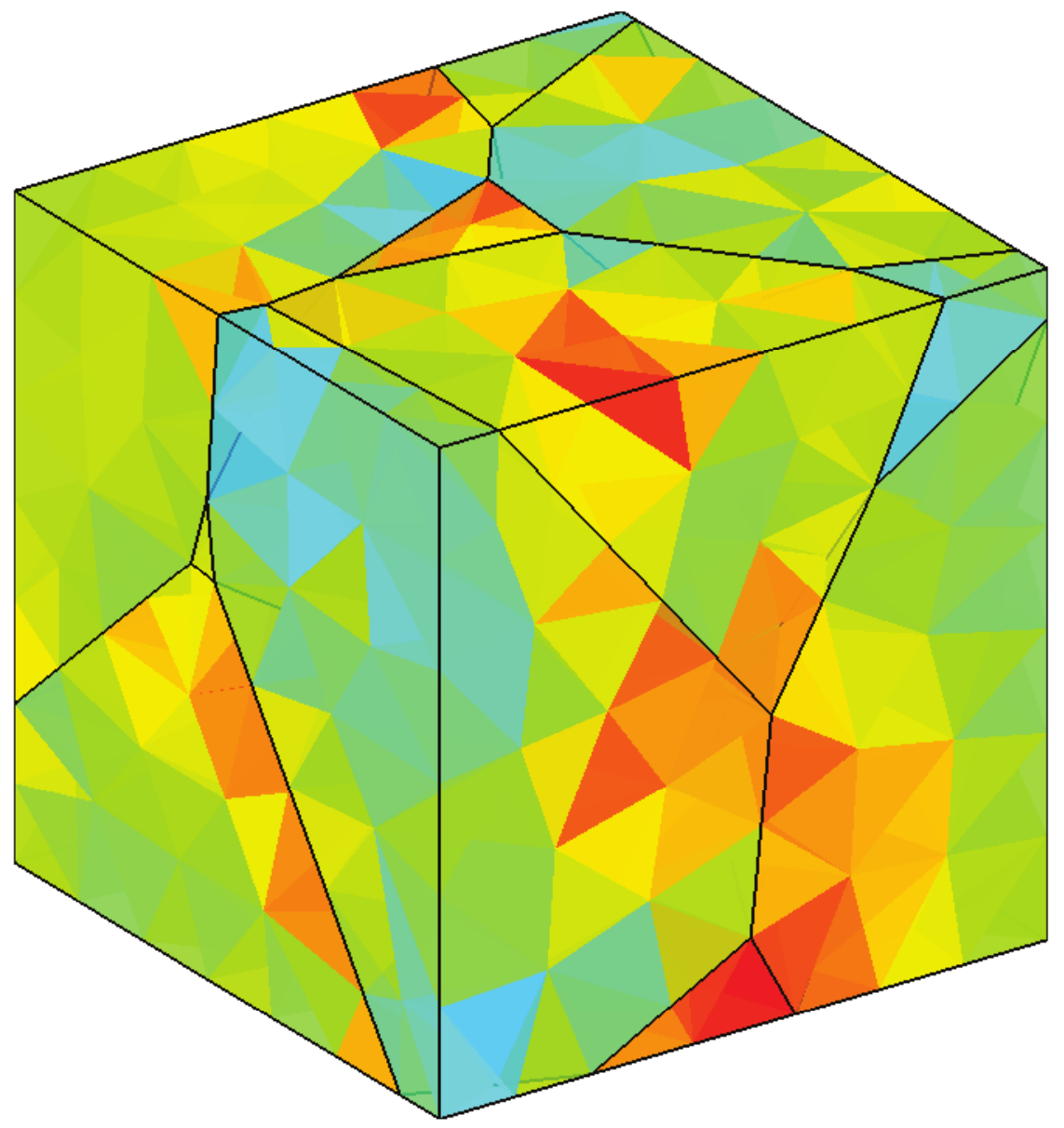}
	\caption{}
	\end{subfigure}
\
	\begin{subfigure}{0.18\textwidth}
	\centering
	\includegraphics[width=\textwidth]{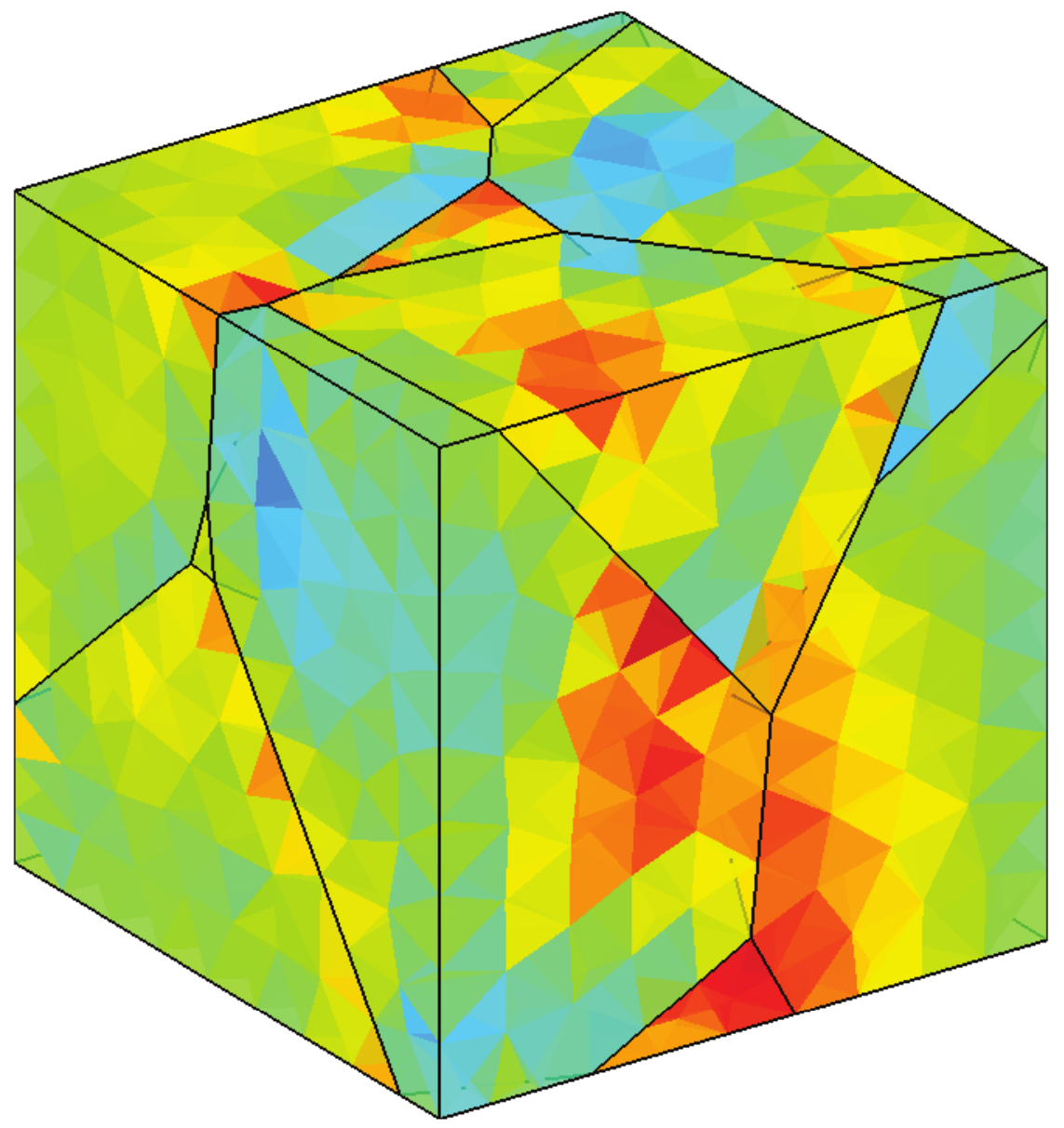}
	\caption{}
	\end{subfigure}
\
	\begin{subfigure}{0.18\textwidth}
	\centering
	\includegraphics[width=\textwidth]{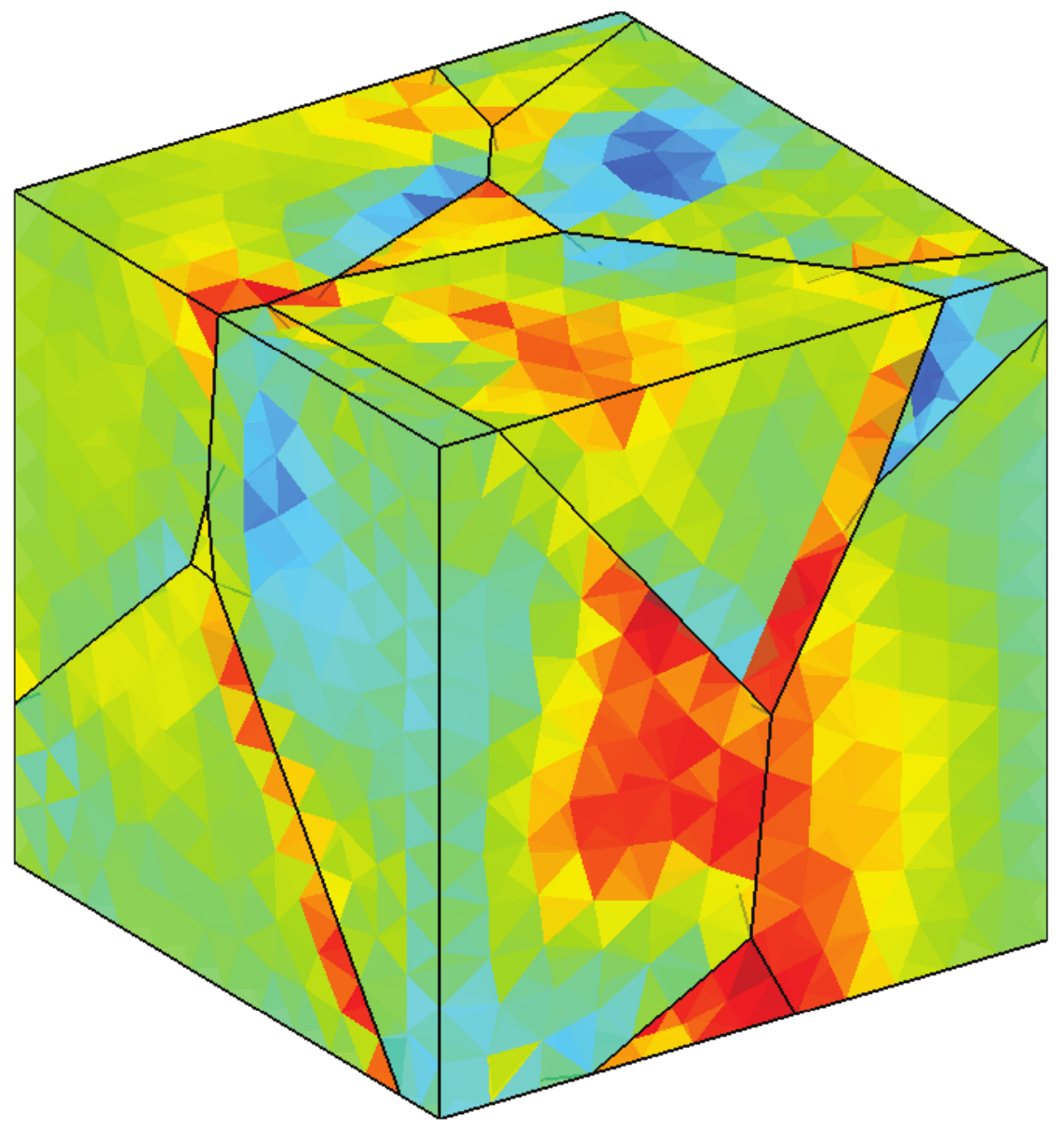}
	\caption{}
	\end{subfigure}
\
	\begin{subfigure}{0.18\textwidth}
	\centering
	\includegraphics[width=\textwidth]{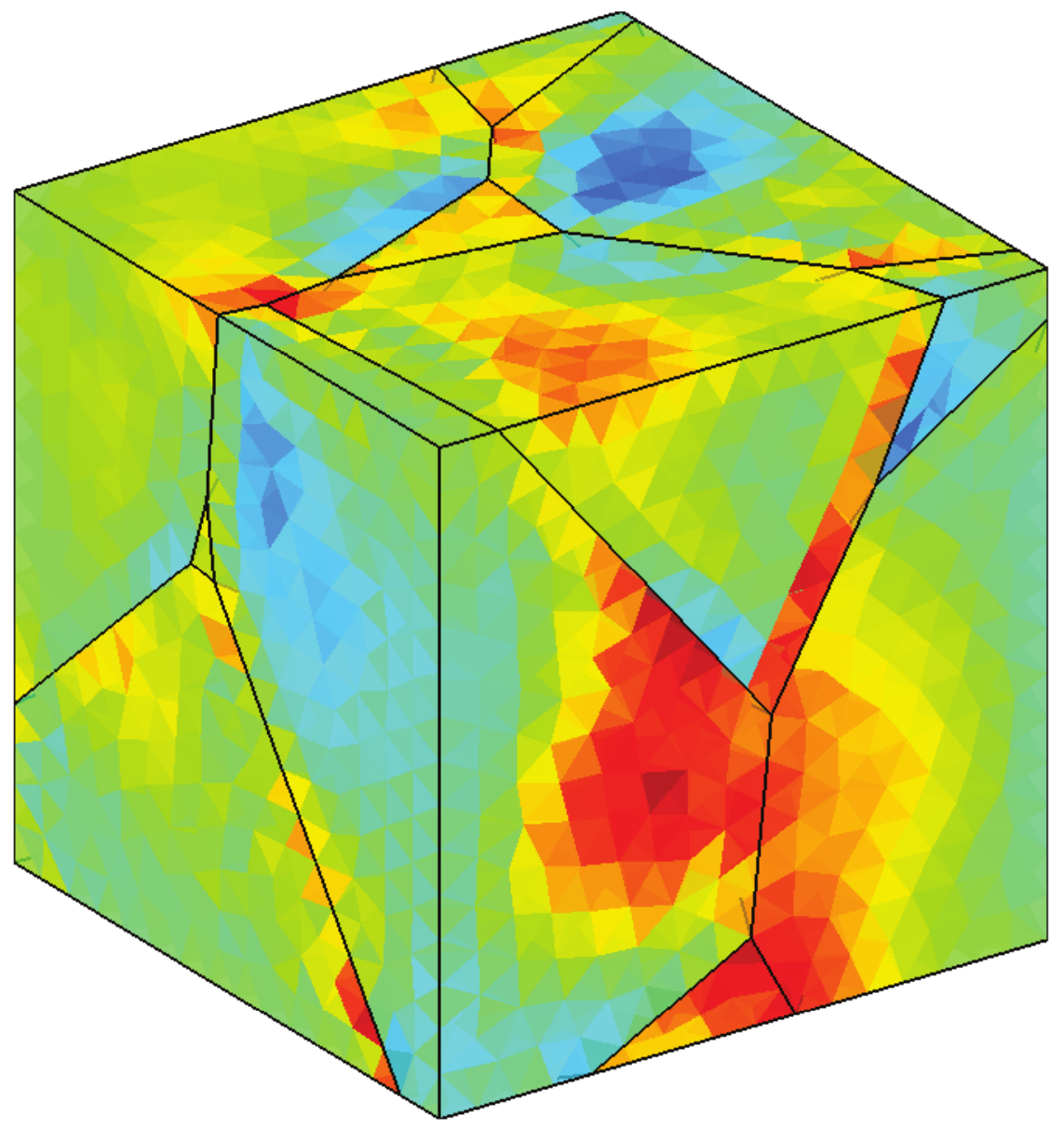}
	\caption{}
	\end{subfigure}
\\
	\begin{subfigure}{0.6\textwidth}
	\centering
	\includegraphics[width=\textwidth]{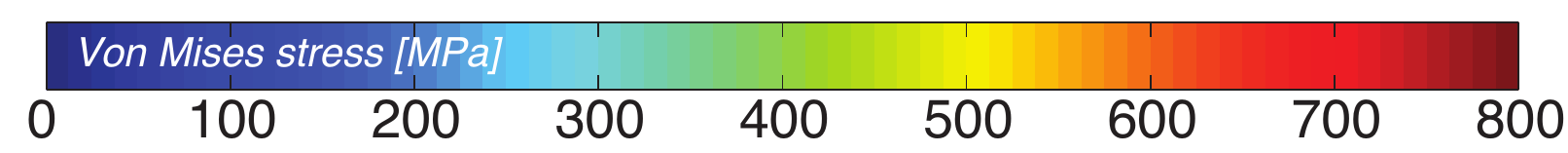}
	\caption{}
	\end{subfigure}
\caption{Von Mises stress contour plots for the 10-grain polycrystalline copper aggregate subjected to tensile load at the last computed step ($\Gamma_{33}=2\%$). Figs.(\emph{a}-\emph{e}) and Figs.(\emph{h}-\emph{j}) correspond to two different views of the aggregate. (\emph{k}) Colormap.}
\label{fig-Ch4:G10_meshes_vonmises}
\end{figure}

\begin{figure}[H]
\centering
	\begin{subfigure}{0.18\textwidth}
	\centering
	\includegraphics[width=\textwidth]{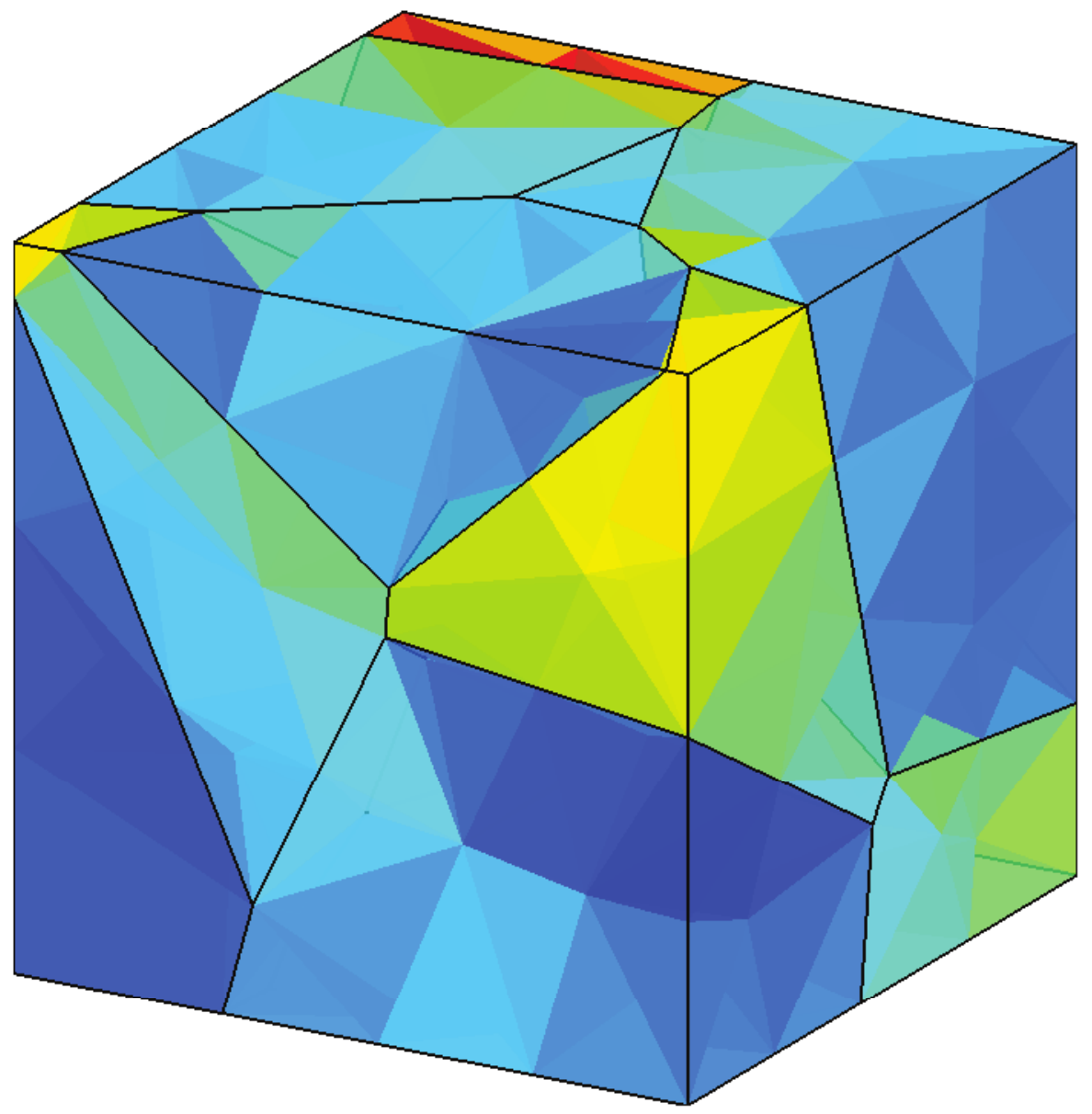}
	\caption{}
	\end{subfigure}
\
	\begin{subfigure}{0.18\textwidth}
	\centering
	\includegraphics[width=\textwidth]{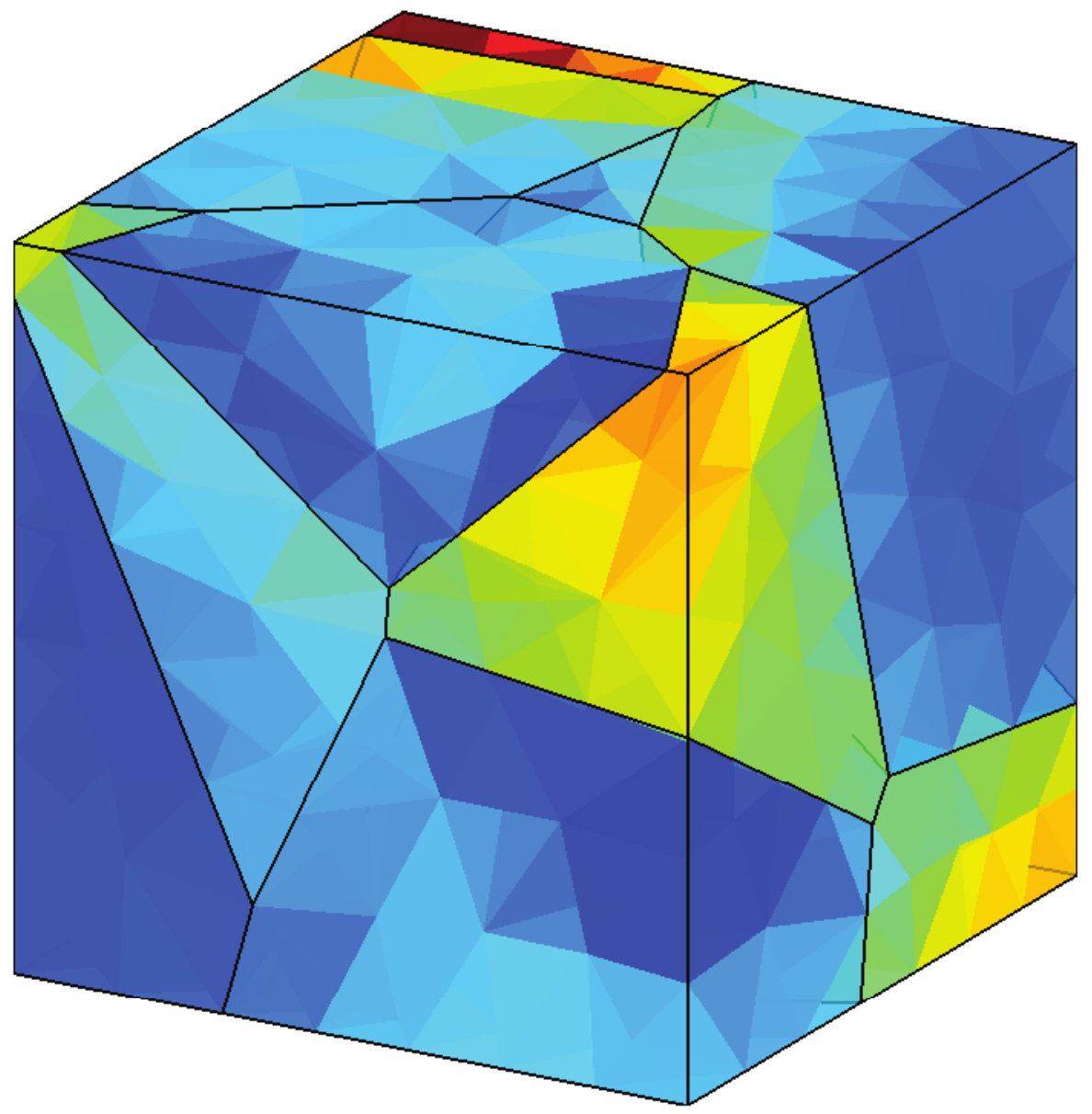}
	\caption{}
	\end{subfigure}
\
	\begin{subfigure}{0.18\textwidth}
	\centering
	\includegraphics[width=\textwidth]{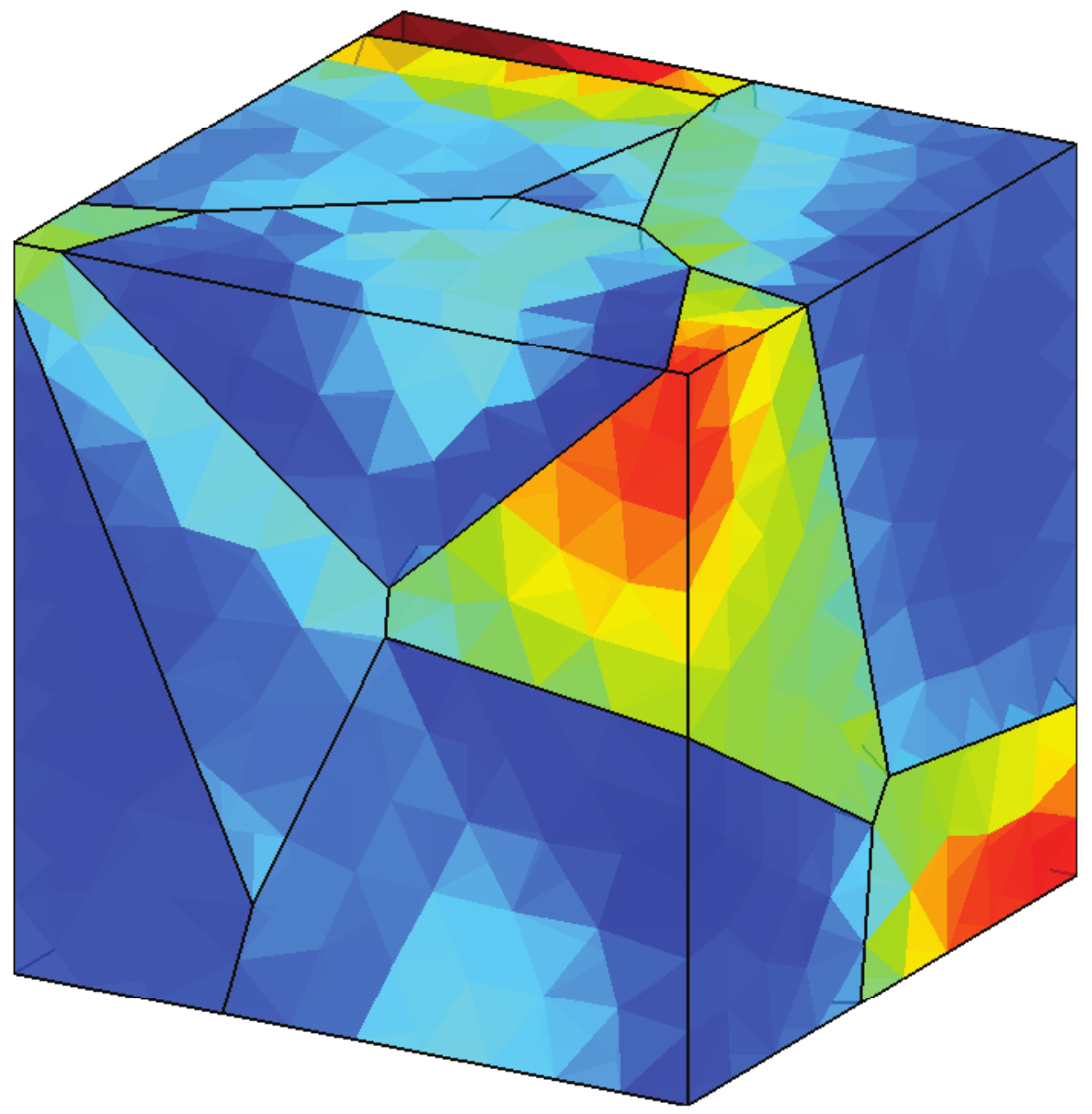}
	\caption{}
	\end{subfigure}
\
	\begin{subfigure}{0.18\textwidth}
	\centering
	\includegraphics[width=\textwidth]{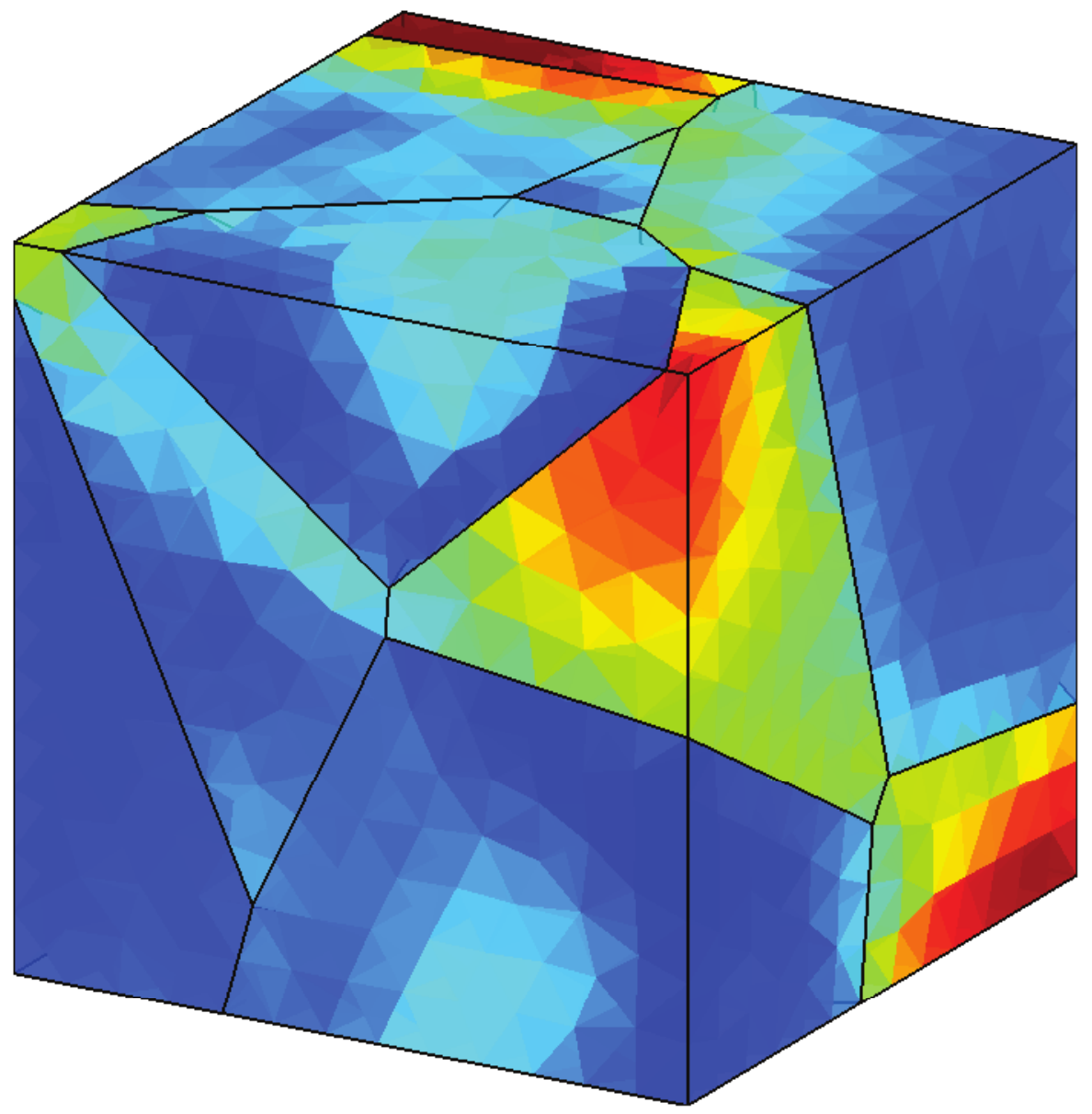}
	\caption{}
	\end{subfigure}
\
	\begin{subfigure}{0.18\textwidth}
	\centering
	\includegraphics[width=\textwidth]{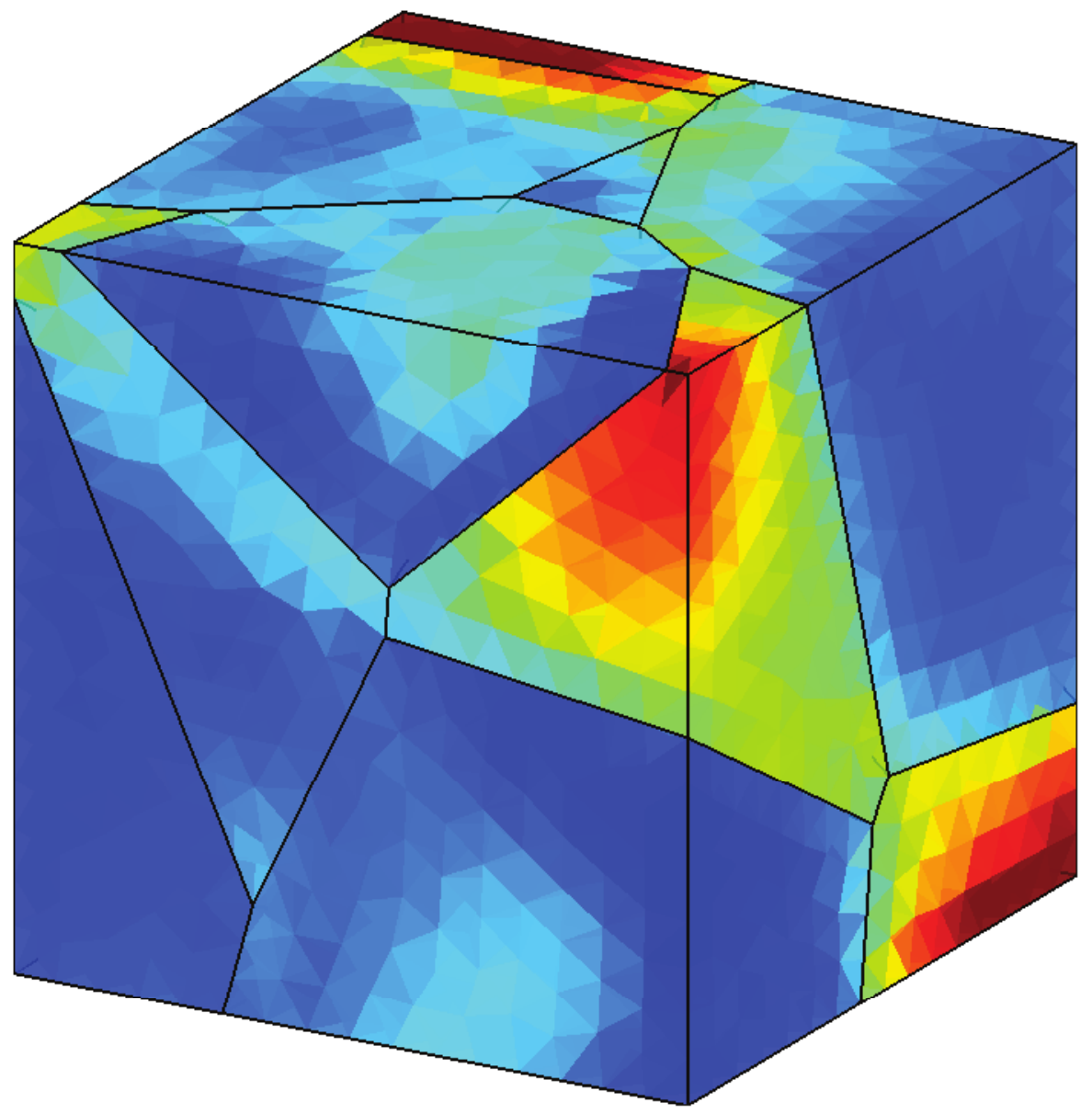}
	\caption{}
	\end{subfigure}
\\
	\begin{subfigure}{0.18\textwidth}
	\centering
	\includegraphics[width=\textwidth]{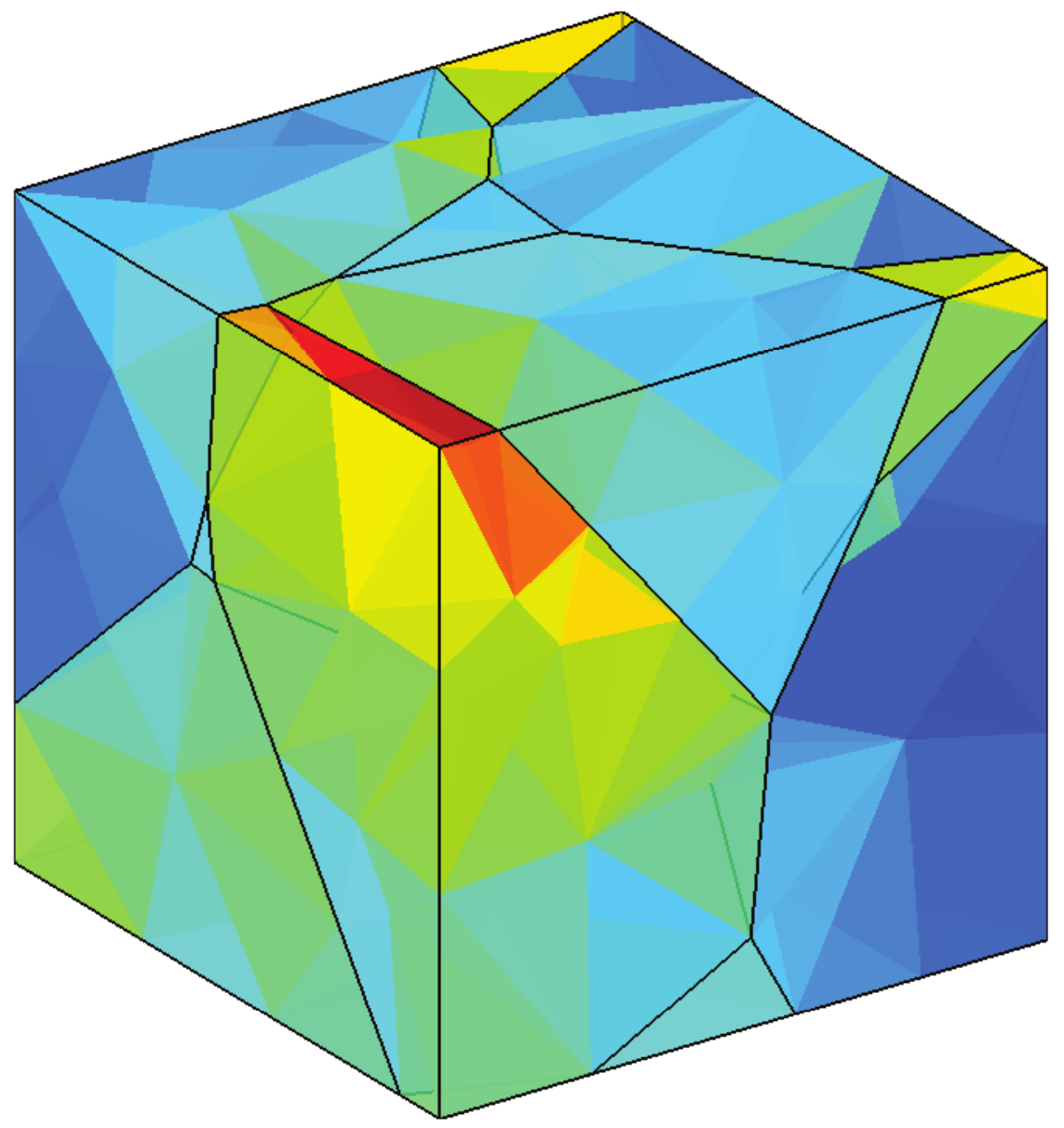}
	\caption{}
	\end{subfigure}
\
	\begin{subfigure}{0.18\textwidth}
	\centering
	\includegraphics[width=\textwidth]{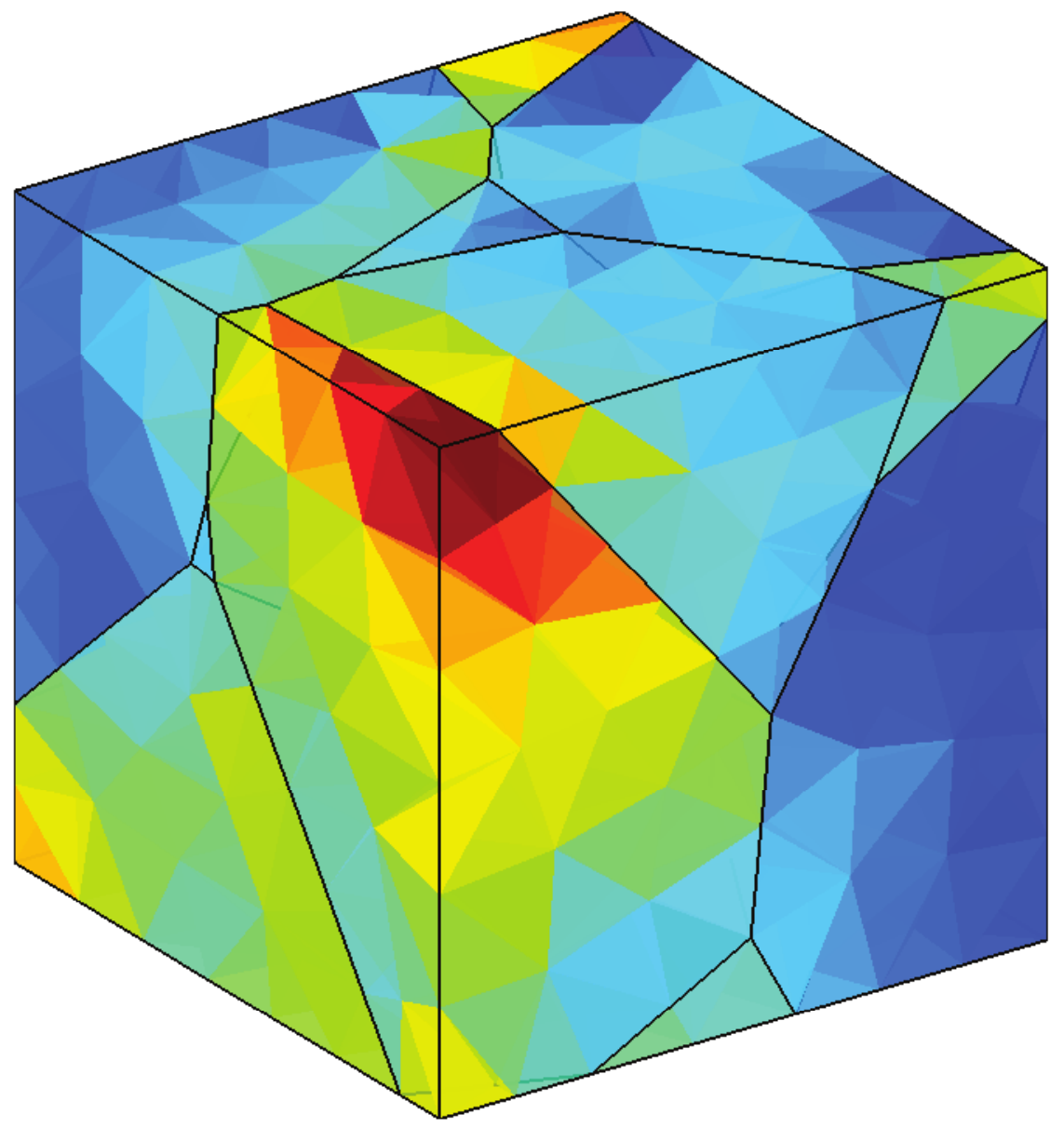}
	\caption{}
	\end{subfigure}
\
	\begin{subfigure}{0.18\textwidth}
	\centering
	\includegraphics[width=\textwidth]{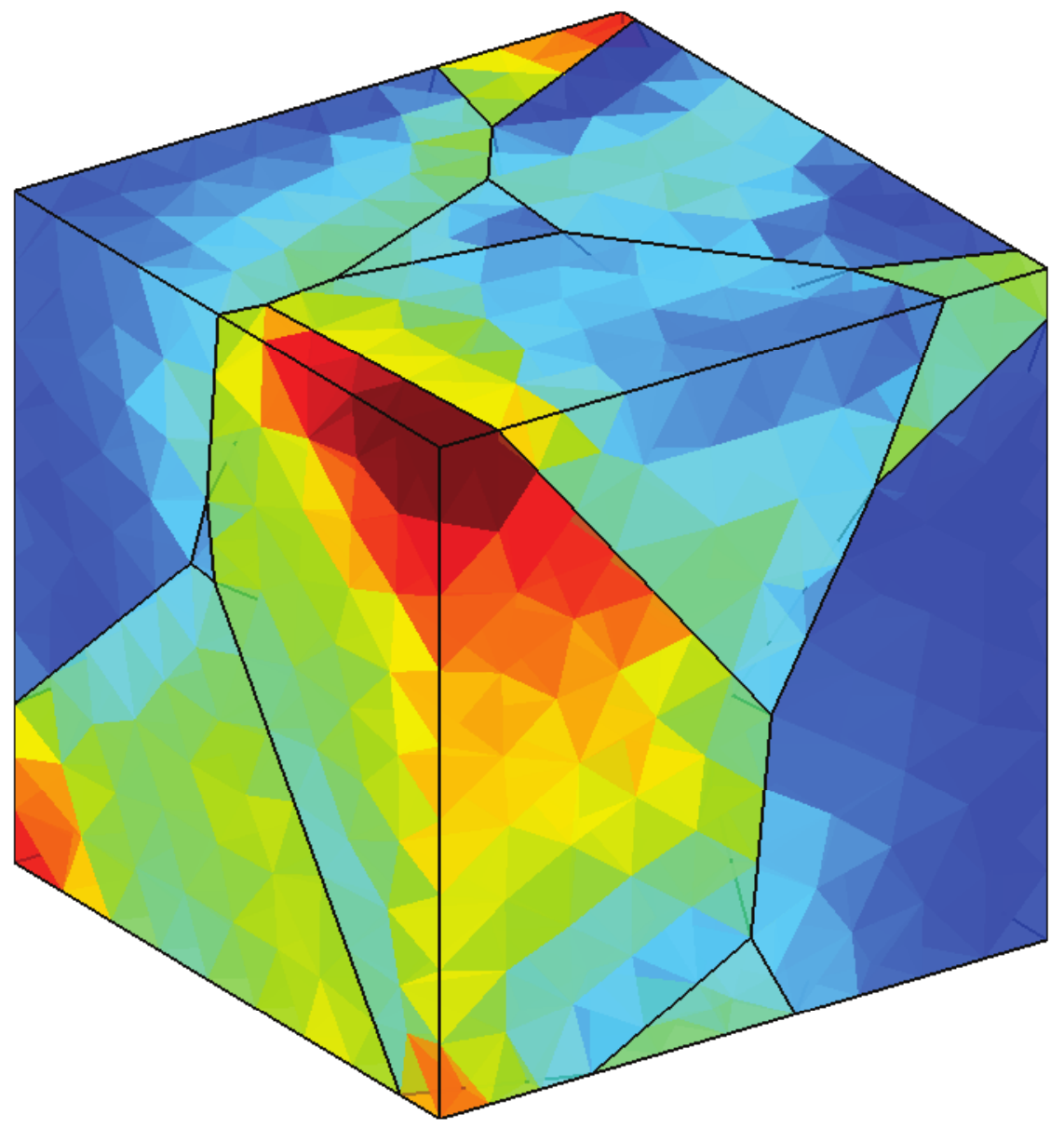}
	\caption{}
	\end{subfigure}
\
	\begin{subfigure}{0.18\textwidth}
	\centering
	\includegraphics[width=\textwidth]{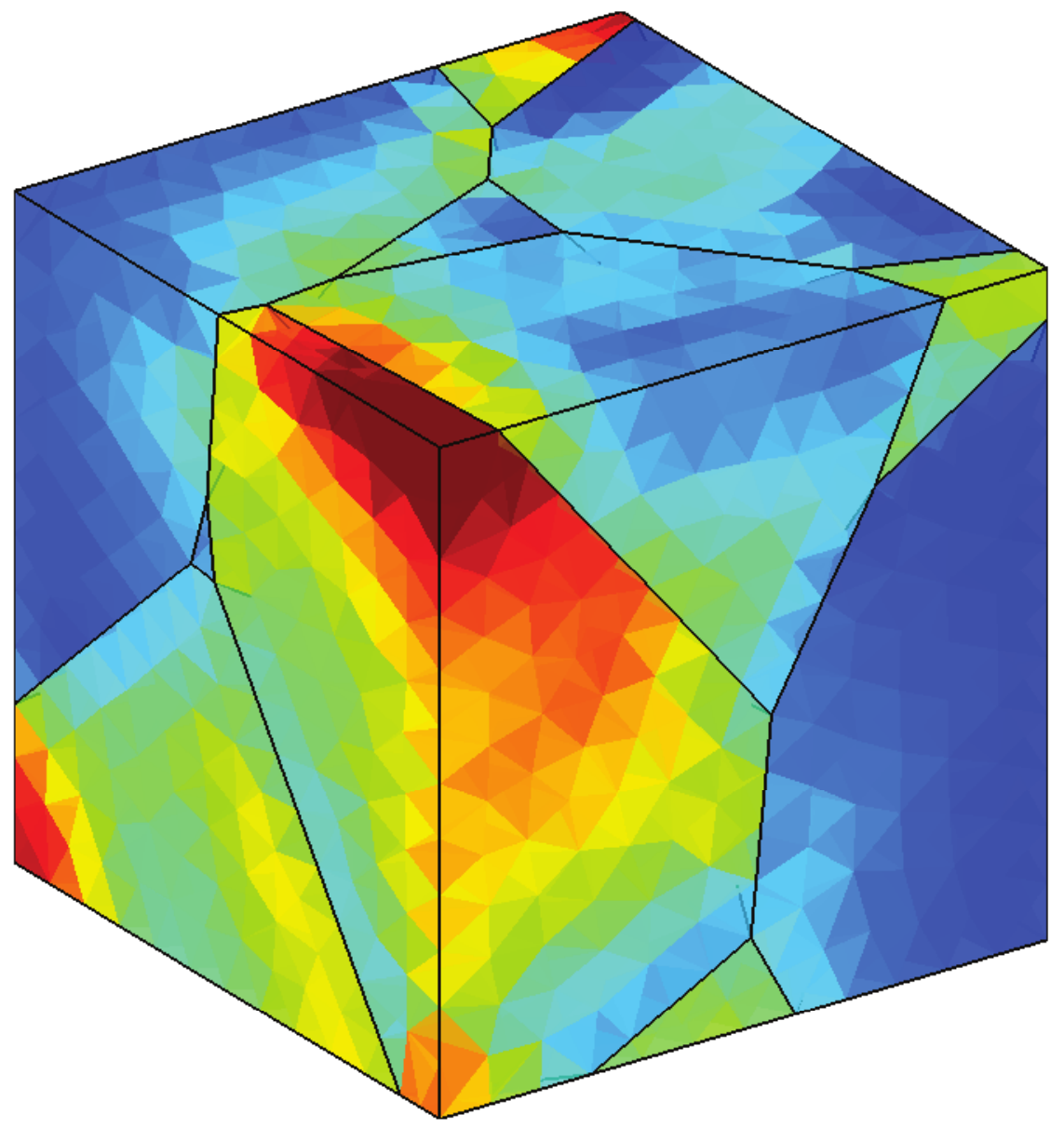}
	\caption{}
	\end{subfigure}
\
	\begin{subfigure}{0.18\textwidth}
	\centering
	\includegraphics[width=\textwidth]{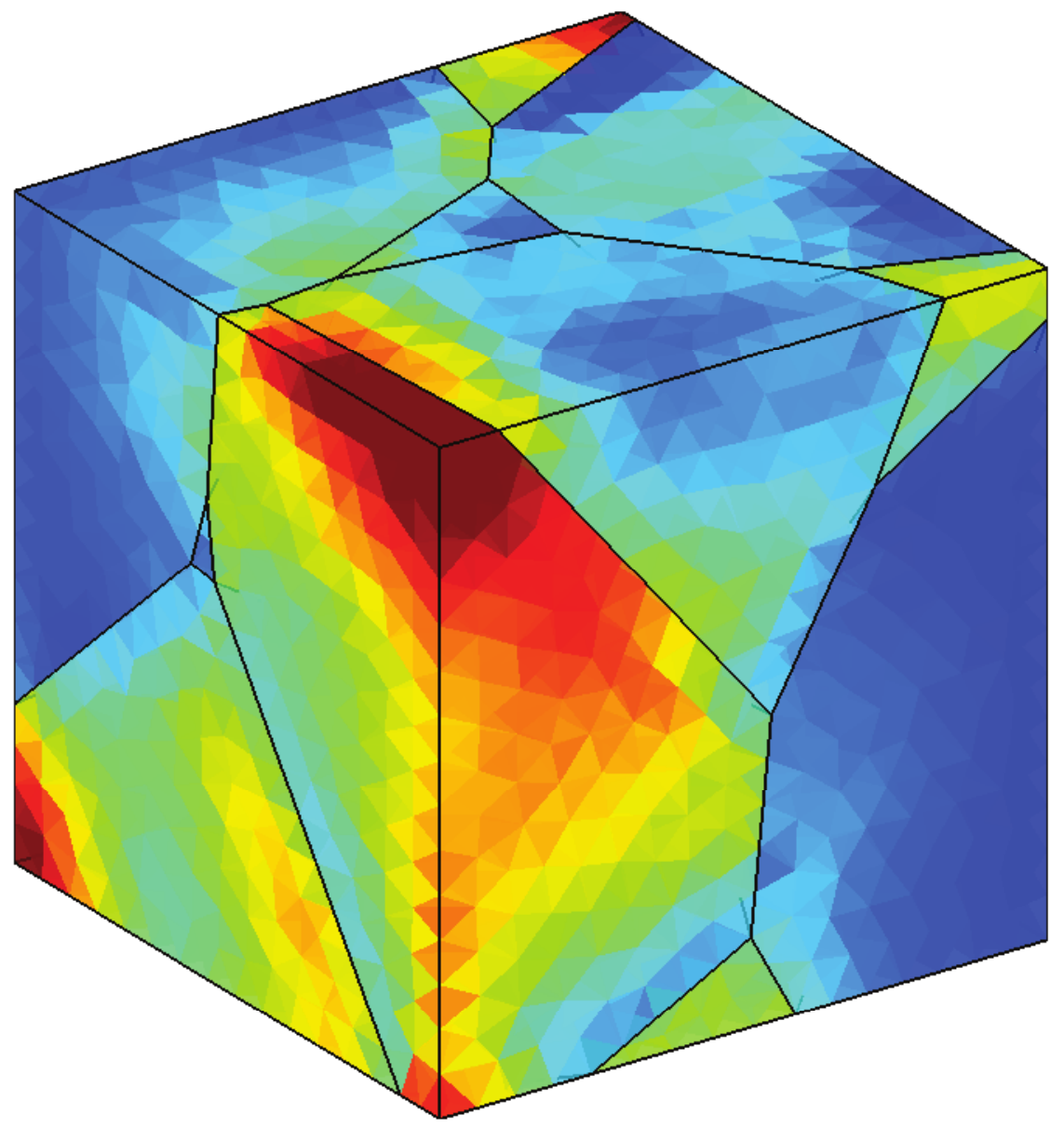}
	\caption{}
	\end{subfigure}
\\
	\begin{subfigure}{0.6\textwidth}
	\centering
	\includegraphics[width=\textwidth]{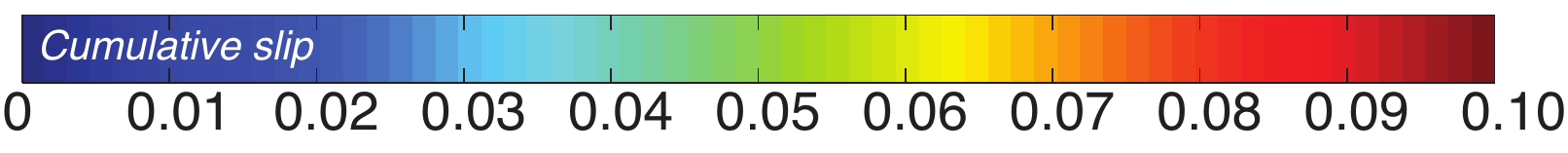}
	\caption{}
	\end{subfigure}
\caption{Cumulative slip contour plots of the 10-grain polycrystalline copper aggregate subjected to tensile load at the last computed step ($\Gamma_{33}=2\%$). Figs.(\emph{a}-\emph{e}) and Figs.(\emph{h}-\emph{j}) correspond to two different views of the aggregate. (\emph{k}) Colormap.}
\label{fig-Ch4:G10_meshes_cumgamma}
\end{figure}

\begin{figure}[H]
\centering
	\begin{subfigure}{0.49\textwidth}
	\centering
	\includegraphics[width=\textwidth]{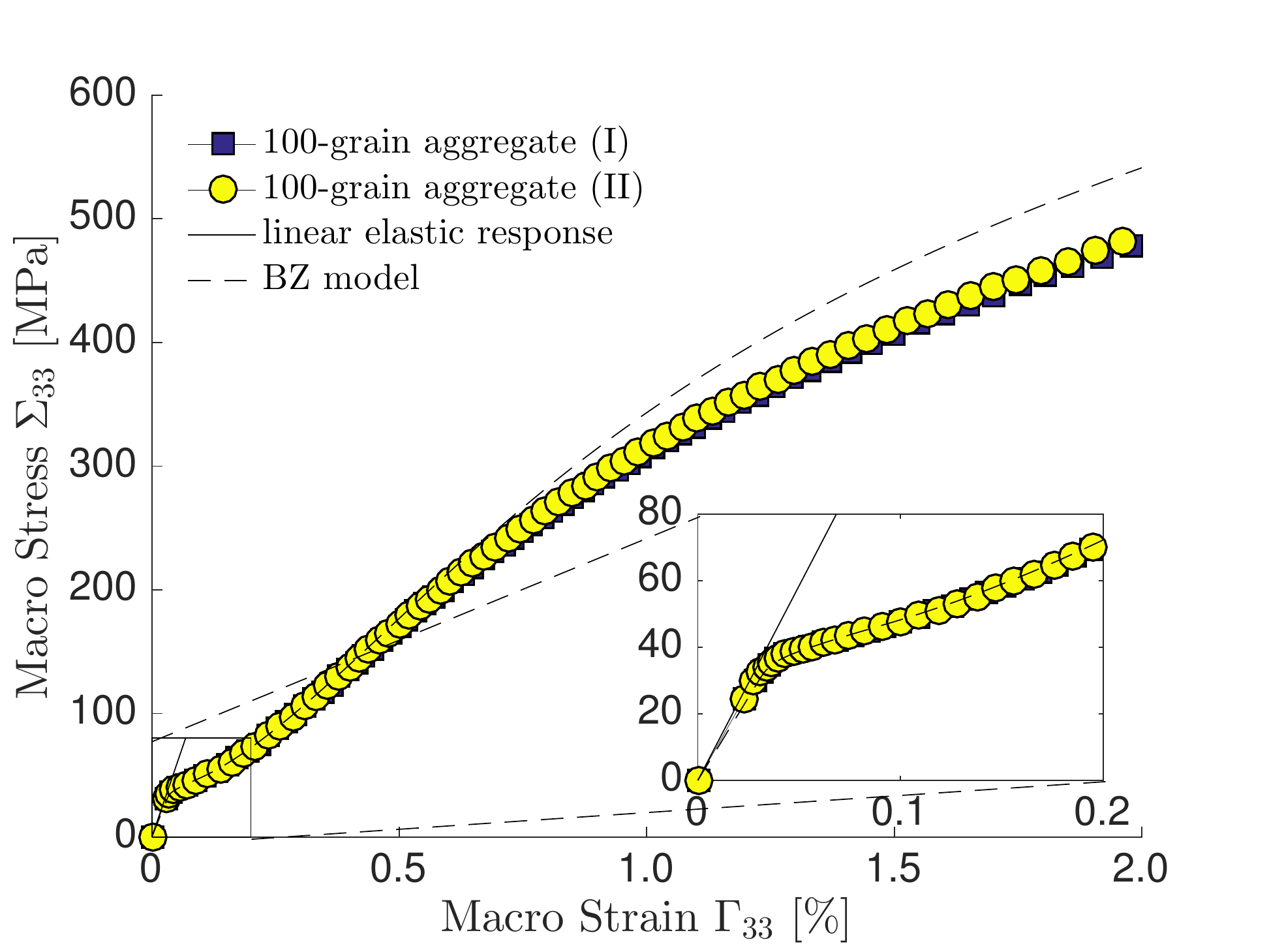}
	\caption{}
	\end{subfigure}
\
	\begin{subfigure}{0.49\textwidth}
	\centering
	\includegraphics[width=\textwidth]{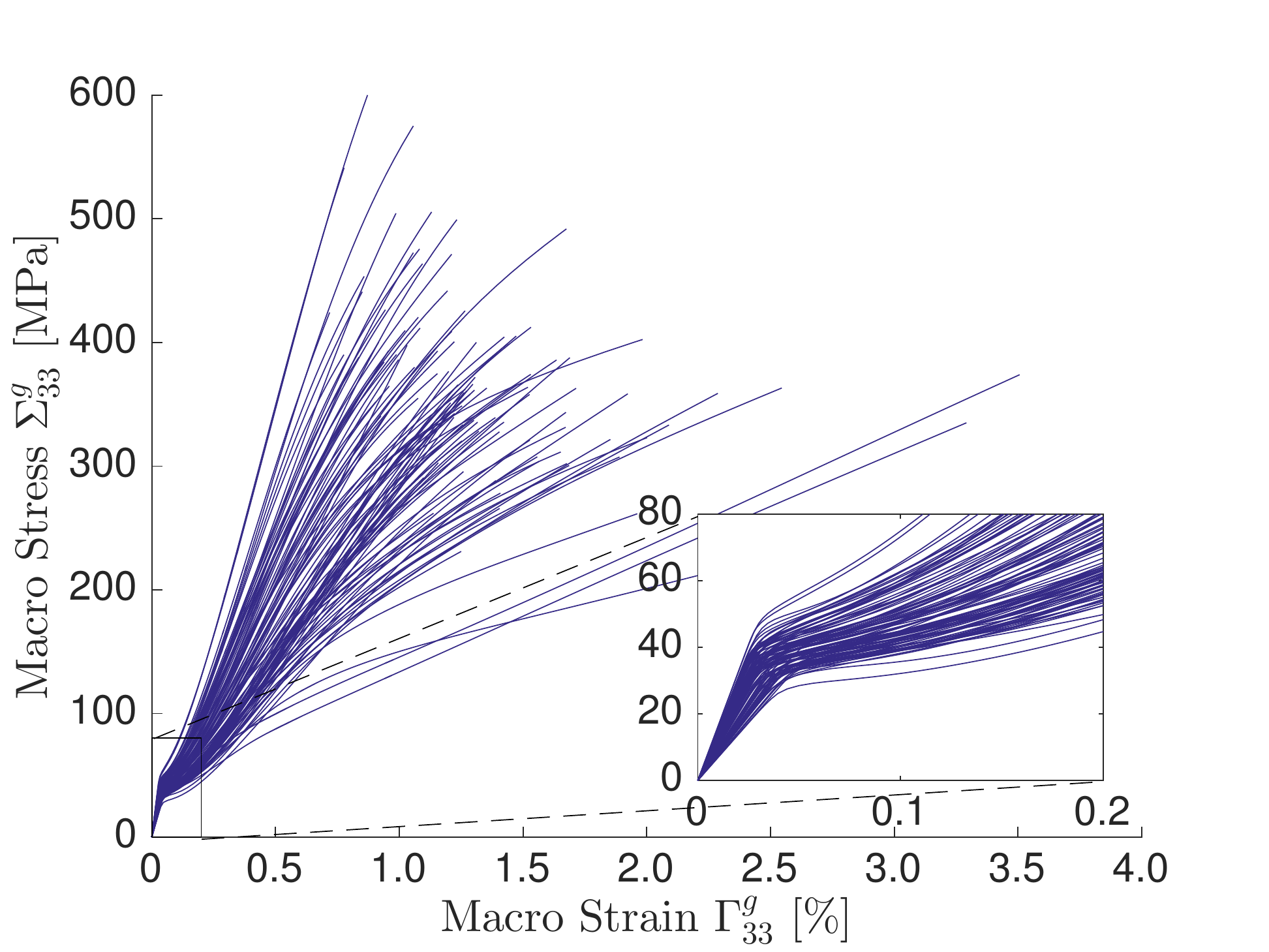}
	\caption{}
	\end{subfigure}
\
\caption{(\emph{a}) Volume stress average $\Sigma_{33}$ versus volume strain average $\Gamma_{33}$ for the two 100-grain aggregates; (\emph{b}) Averaged stress-strain behaviour for \emph{each} constituent grain of the aggregate (I).}
\label{fig-Ch4:G100_macro-response}
\end{figure}

\subsubsection{100-grain aggregate}
The same tensile test is performed on two 100-grain Cu aggregates, in order to evaluate the plastic response of more complex and representative Cu microstructures. Figure (\ref{fig-Ch4:G100_macro-response}a) shows the macro stress-strain behaviour of the two aggregates, which are labelled as aggregate (I) and (II) and reported in Figures (\ref{fig-Ch4:G100_microstructures}a) and (\ref{fig-Ch4:G100_microstructures}d), respectively. Figures (\ref{fig-Ch4:G100_microstructures}b,e) show the corresponding surface meshes, whereas Figure (\ref{fig-Ch4:G100_microstructures}c,f) show the volume meshes. The meshes are built using a mesh density parameter that ensures an average number of 400 volume elements per grain, see Table (\ref{tab-Ch4:meshes-stats}), which represents a good tradeoff between solution accuracy and number of DoFs.

The macro stress-strain behaviour reported in Figure (\ref{fig-Ch4:G100_macro-response}a) is qualitatively similar to that obtained for the 10-grain aggregate. The insert of the figure reports a close-up view of the macro response in the range $0$-$0.2\%$, showing the transition from the linear to the plastic slip behaviour.
By comparing the response of the two microstructures, it can be noted that an aggregate with 100 grains is satisfactorily representative for polycrystalline Cu undergoing small strains crystal plasticity. The solid line in Figure (\ref{fig-Ch4:G100_macro-response}a) represents the linear response of the microstructure whose averaged stiffness is calculated using the Hashin-Shtrikman bounds \cite{hashin1962a,hashin1962b}, whereas the dashed line reports the response of the aggregate using the Berveiller and Zaoui (BZ) model \cite{berveiller1978}. It can be noted that the BZ model and the present model agree very well with respect to the elastic to plastic behavior transition. On the other hand, the BZ model provides values of stress higher than the present model, but this appears consistent with Barbe et al.\ \cite{barbe2001a}, who reported an overestimate with respect to FEM as well. Figure (\ref{fig-Ch4:G100_macro-response}b) shows the averaged stress-strain behavior of each of the 100 grains of the aggregates, obtained using Eqs.(\ref{eq-Ch4:strain_average}) and (\ref{eq-Ch4:stress_average}) where the integration is performed over the boundary of \emph{each} grain.

Figures (\ref{fig-Ch4:G100_vonmises}a,c,e) show the contour plot of the Von Mises stress over the external boundary of the aggregate (I) at the macro-strain steps $\Gamma_{33}=0.3\%$, $1.0\%$ and $2.0\%$, respectively, whereas Figures (\ref{fig-Ch4:G100_vonmises}b,d,f) report the same contour plot at the same macro-strain steps for a section of the aggregate giving an insight of the internal volume fields distribution. Similarly, Figures (\ref{fig-Ch4:G100_cumgamma}a,c,e) show the contour plot of the cumulative slip at the macro-strain steps $\Gamma_{33}=0.3\%$, $1.0\%$ and $2.0\%$, respectively, and Figures (\ref{fig-Ch4:G100_cumgamma}b,d,f) report the same contour plot at the same macro-strain steps for a section of the aggregate.

\begin{figure}[h]
\centering
	\begin{subfigure}{0.49\textwidth}
	\centering
	\includegraphics[width=0.7\textwidth]{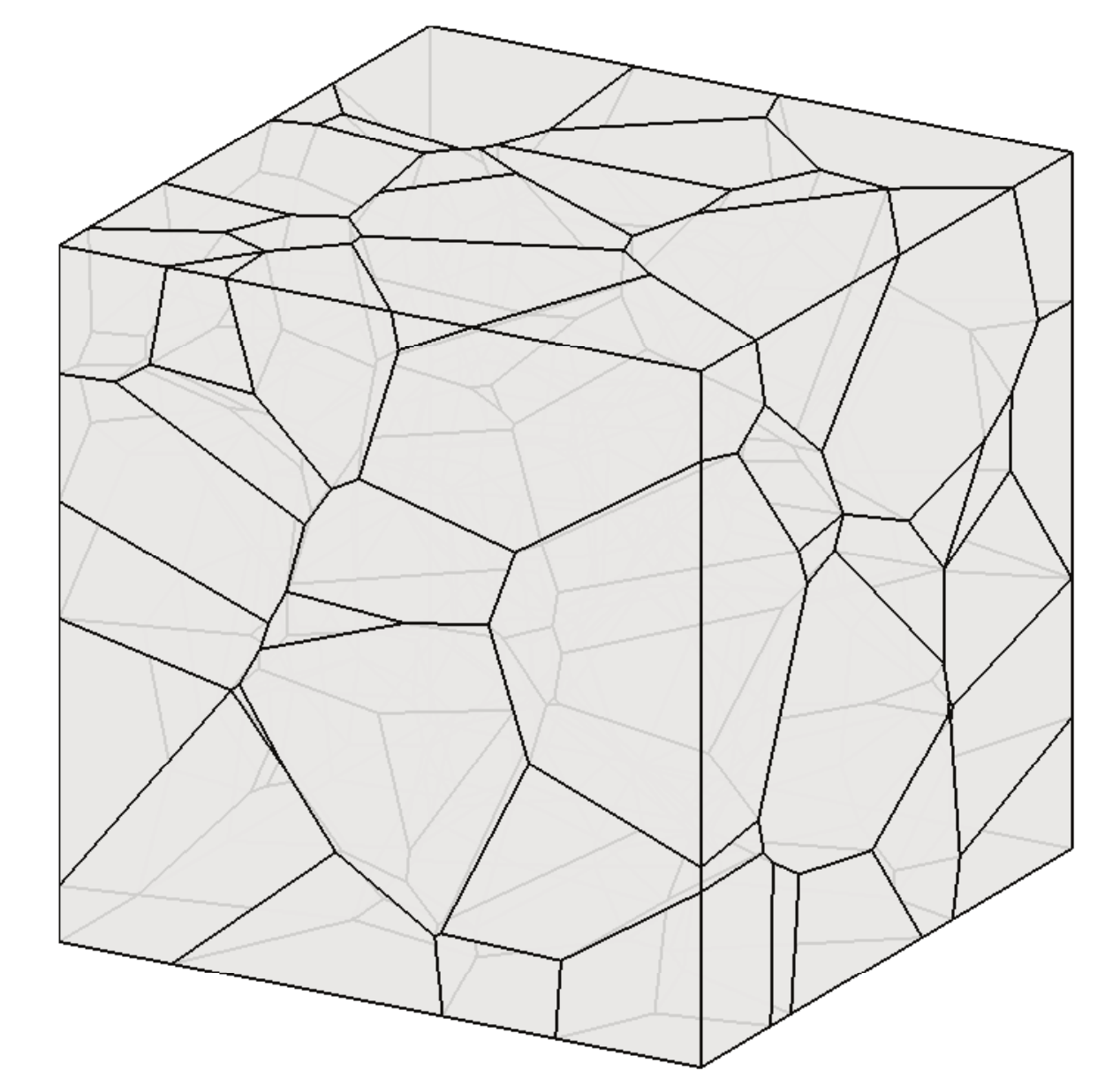}
	\caption{}
	\end{subfigure}
\
	\begin{subfigure}{0.49\textwidth}
	\centering
	\includegraphics[width=0.7\textwidth]{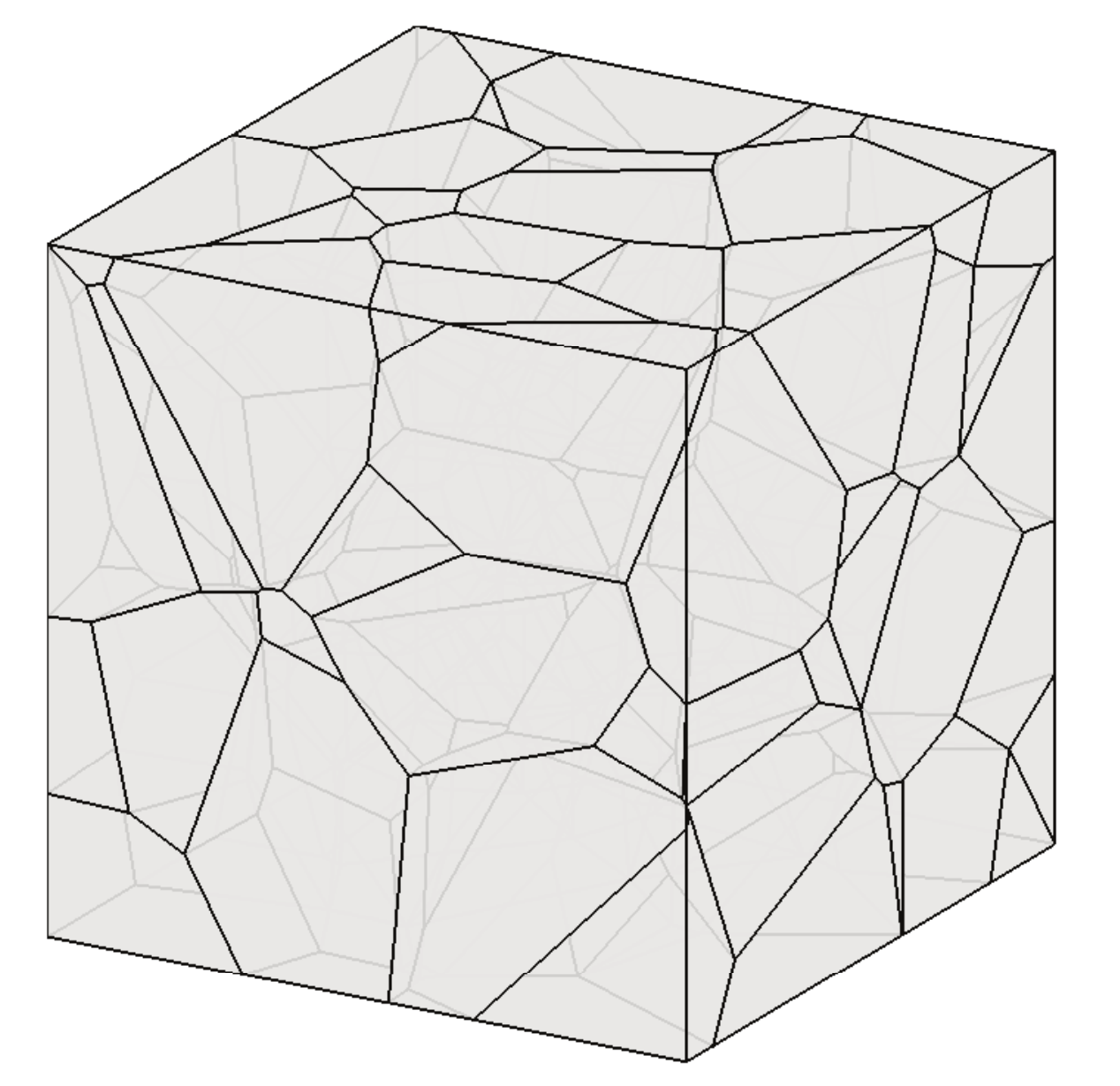}
	\caption{}
	\end{subfigure}
\\
	\begin{subfigure}{0.49\textwidth}
	\centering
	\includegraphics[width=0.7\textwidth]{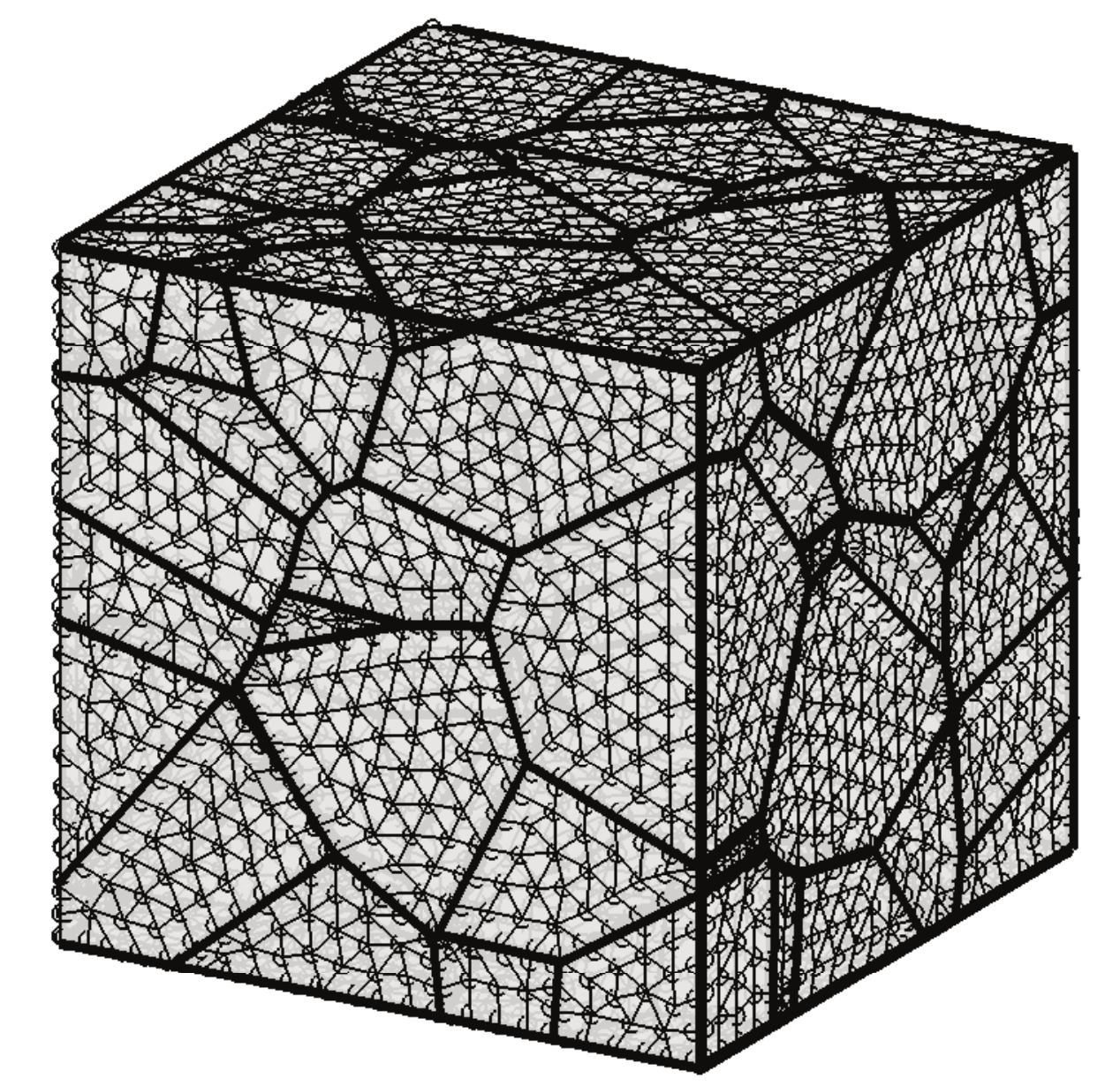}
	\caption{}
	\end{subfigure}
\
	\begin{subfigure}{0.49\textwidth}
	\centering
	\includegraphics[width=0.7\textwidth]{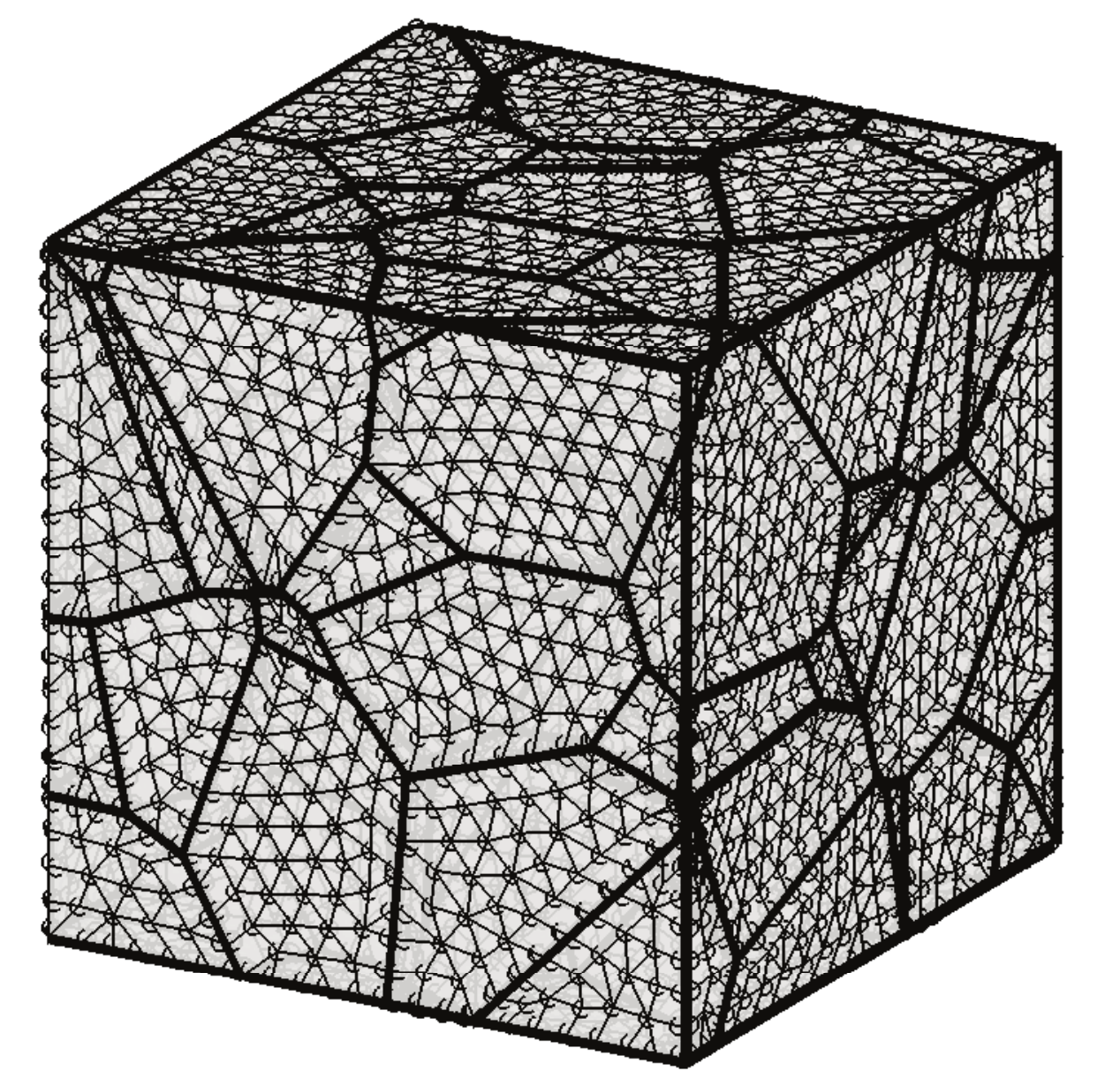}
	\caption{}
	\end{subfigure}
\\
	\begin{subfigure}{0.49\textwidth}
	\centering
	\includegraphics[width=0.7\textwidth]{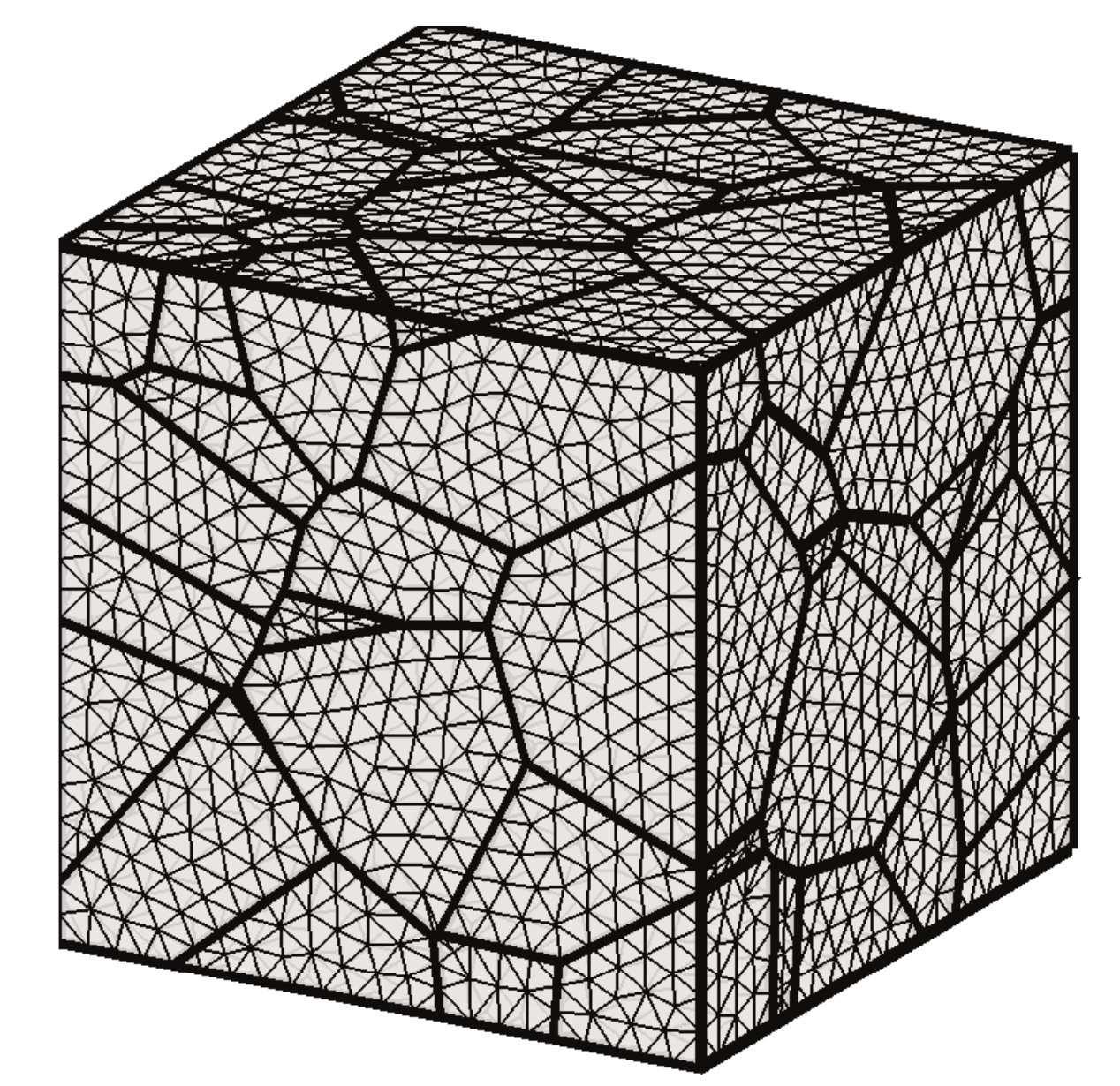}
	\caption{}
	\end{subfigure}
\
	\begin{subfigure}{0.49\textwidth}
	\centering
	\includegraphics[width=0.7\textwidth]{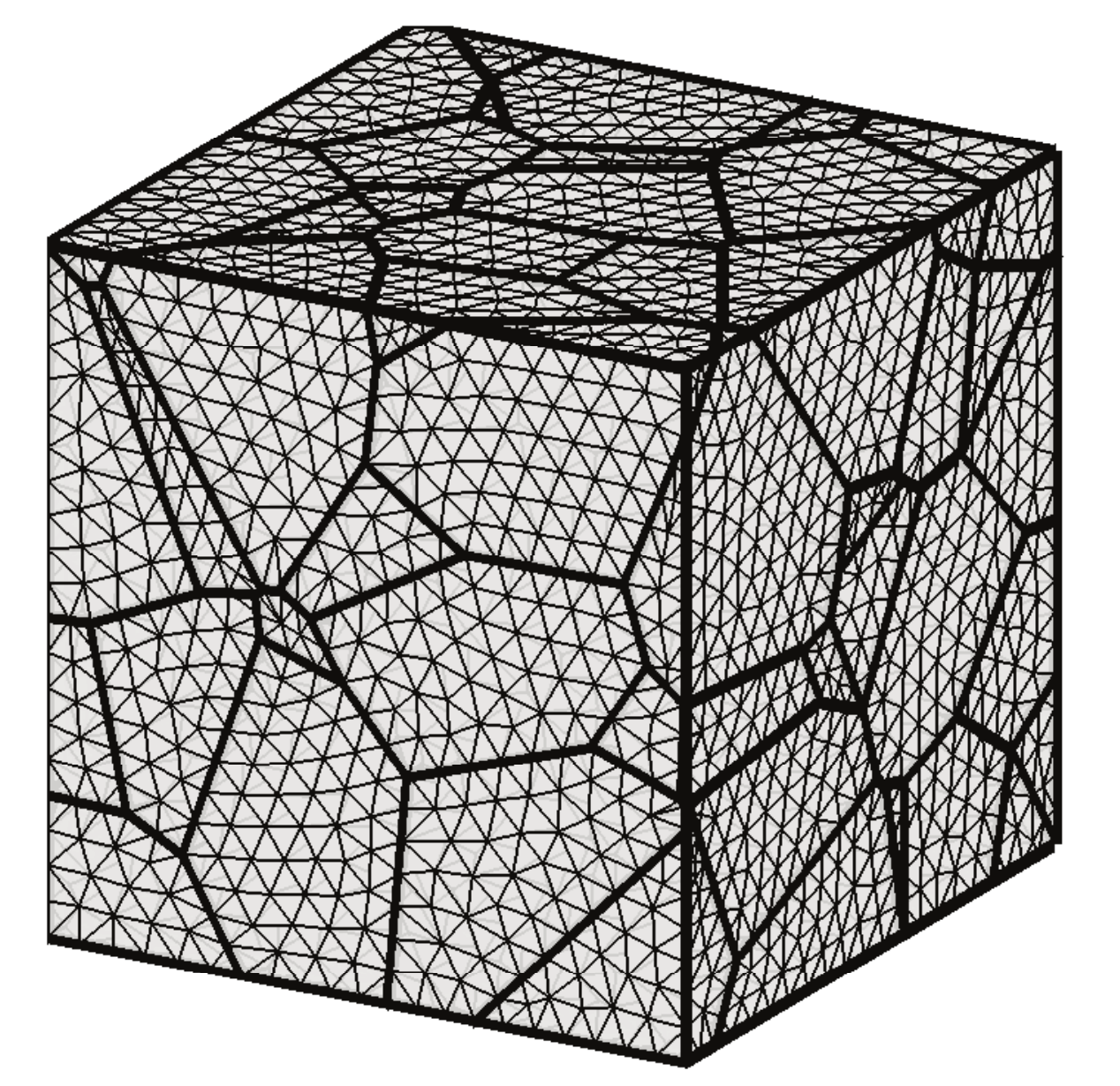}
	\caption{}
	\end{subfigure}
\caption{(\emph{a},\emph{b}) 100-grain polycrystalline microstructure (I) and (II), respectively; corresponding (\emph{c},\emph{d}) surface and (\emph{e},\emph{f}) volume meshes.}
\label{fig-Ch4:G100_microstructures}
\end{figure}

\begin{figure}[h]
\centering
	\begin{subfigure}{0.49\textwidth}
	\centering
	\includegraphics[width=0.7\textwidth]{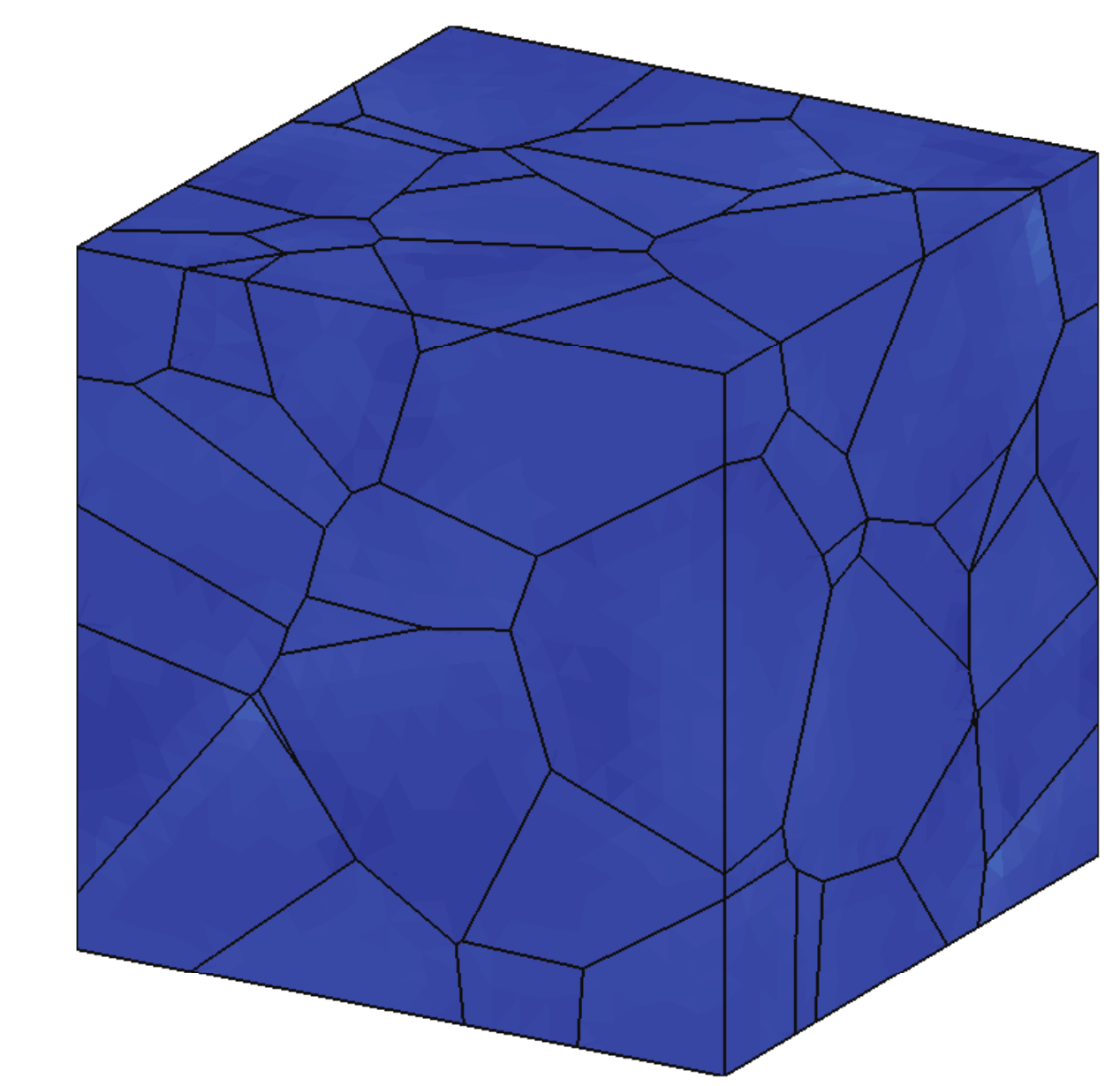}
	\caption{$\Gamma_{33}=0.3\%$}
	\end{subfigure}
\
	\begin{subfigure}{0.49\textwidth}
	\centering
	\includegraphics[width=0.7\textwidth]{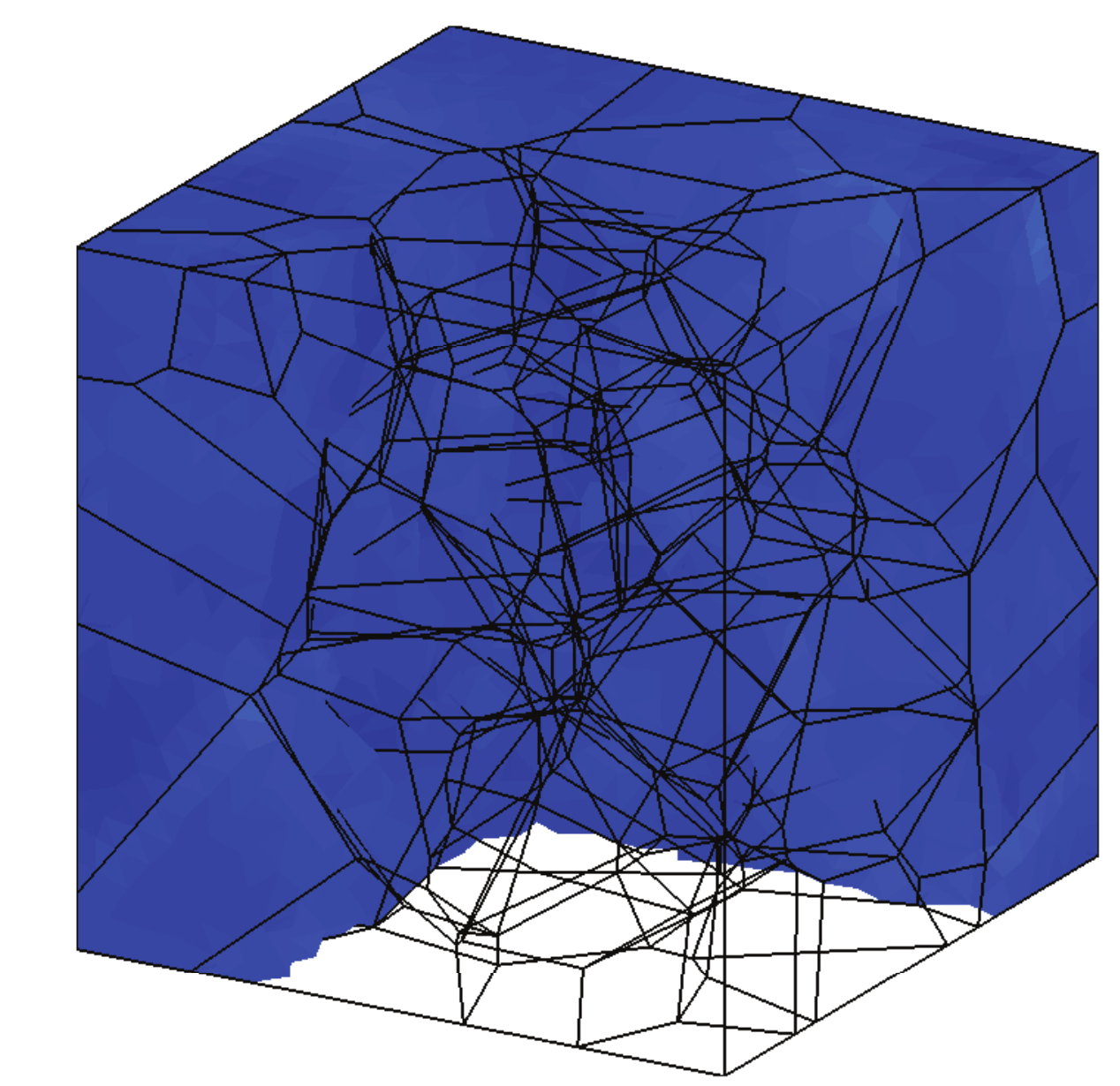}
	\caption{$\Gamma_{33}=0.3\%$}
	\end{subfigure}
\\
	\begin{subfigure}{0.49\textwidth}
	\centering
	\includegraphics[width=0.7\textwidth]{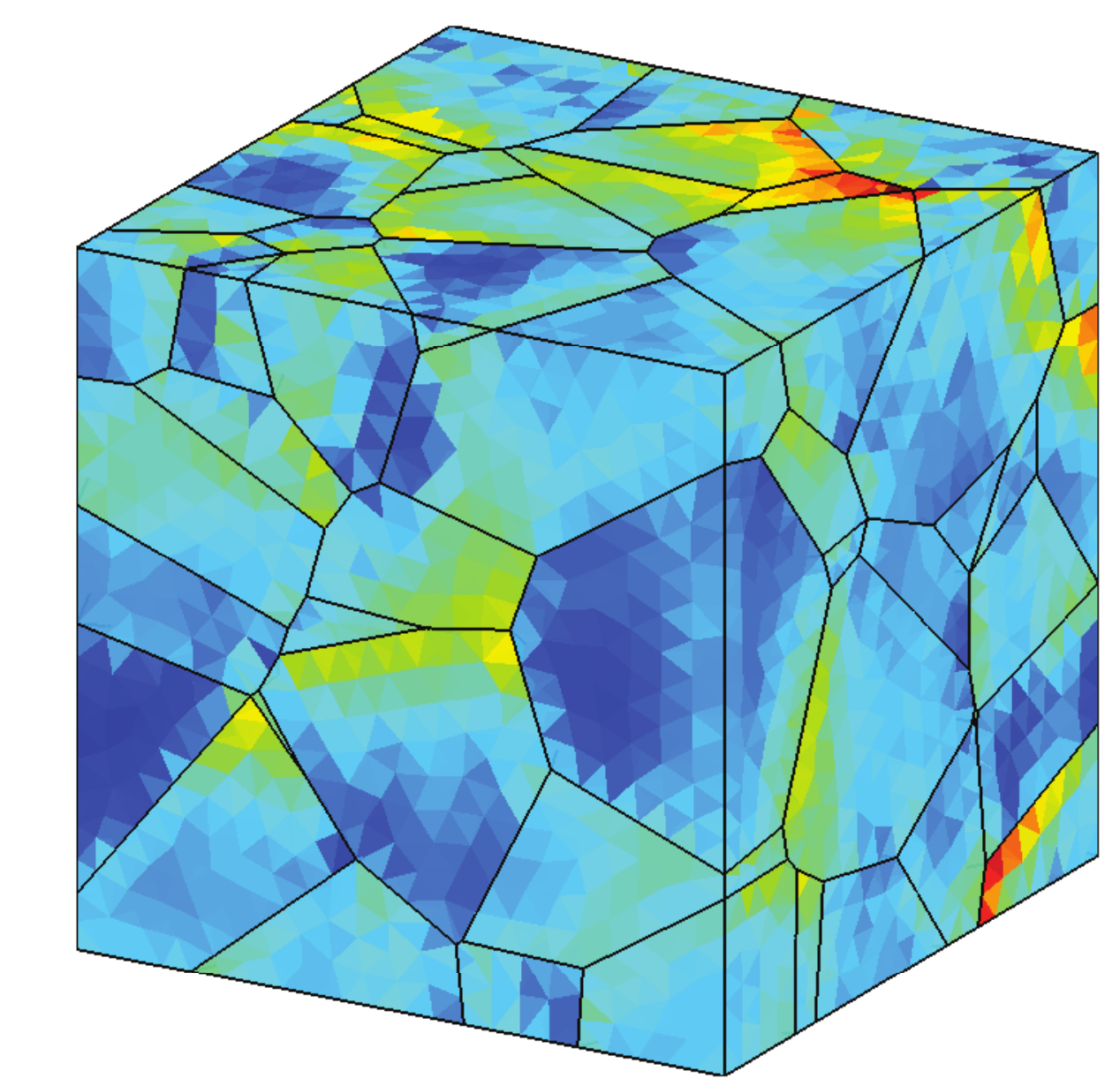}
	\caption{$\Gamma_{33}=1.0\%$}
	\end{subfigure}
\
	\begin{subfigure}{0.49\textwidth}
	\centering
	\includegraphics[width=0.7\textwidth]{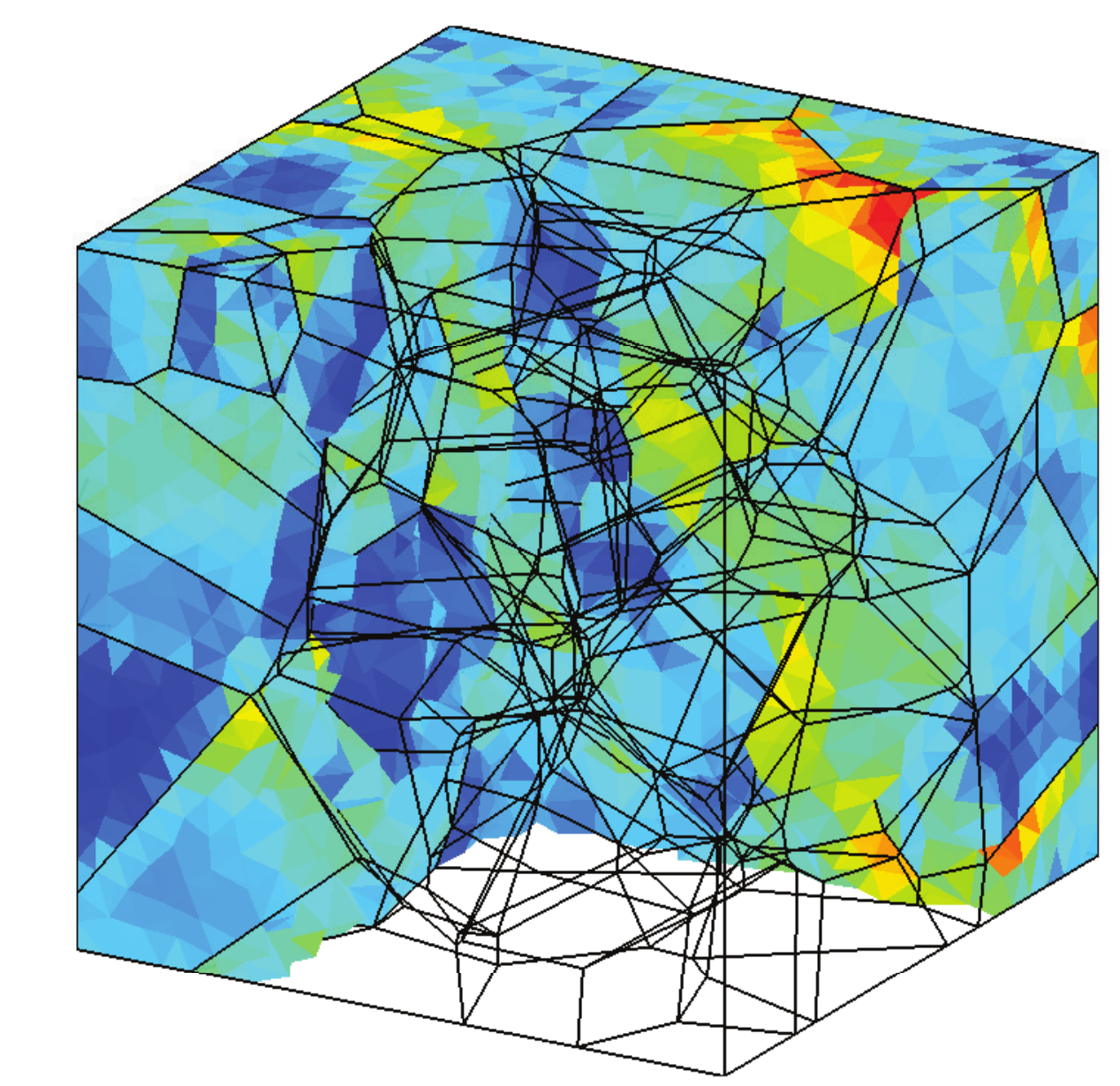}
	\caption{$\Gamma_{33}=1.0\%$}
	\end{subfigure}
\\
	\begin{subfigure}{0.49\textwidth}
	\centering
	\includegraphics[width=0.7\textwidth]{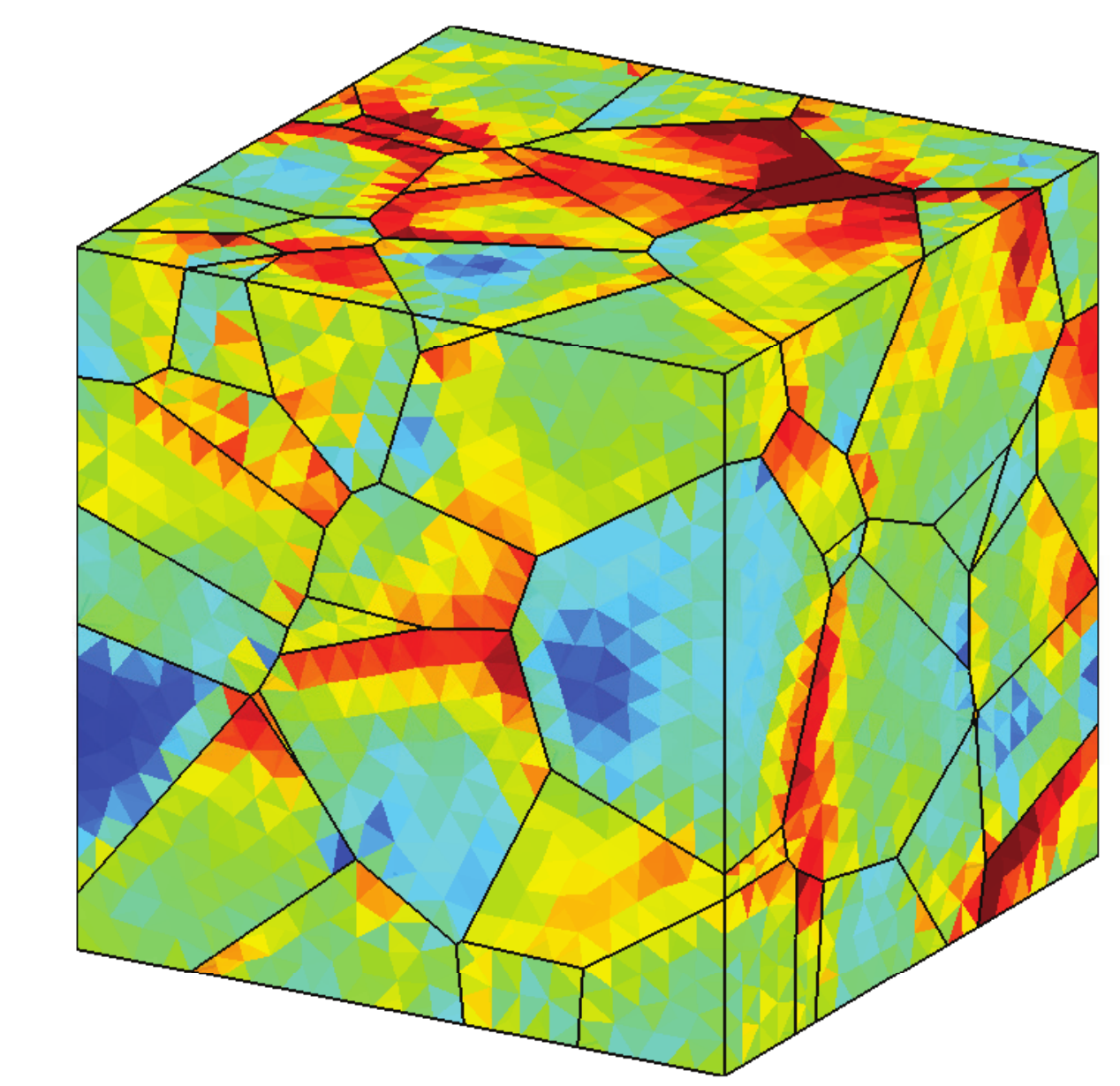}
	\caption{$\Gamma_{33}=2.0\%$}
	\end{subfigure}
\
	\begin{subfigure}{0.49\textwidth}
	\centering
	\includegraphics[width=0.7\textwidth]{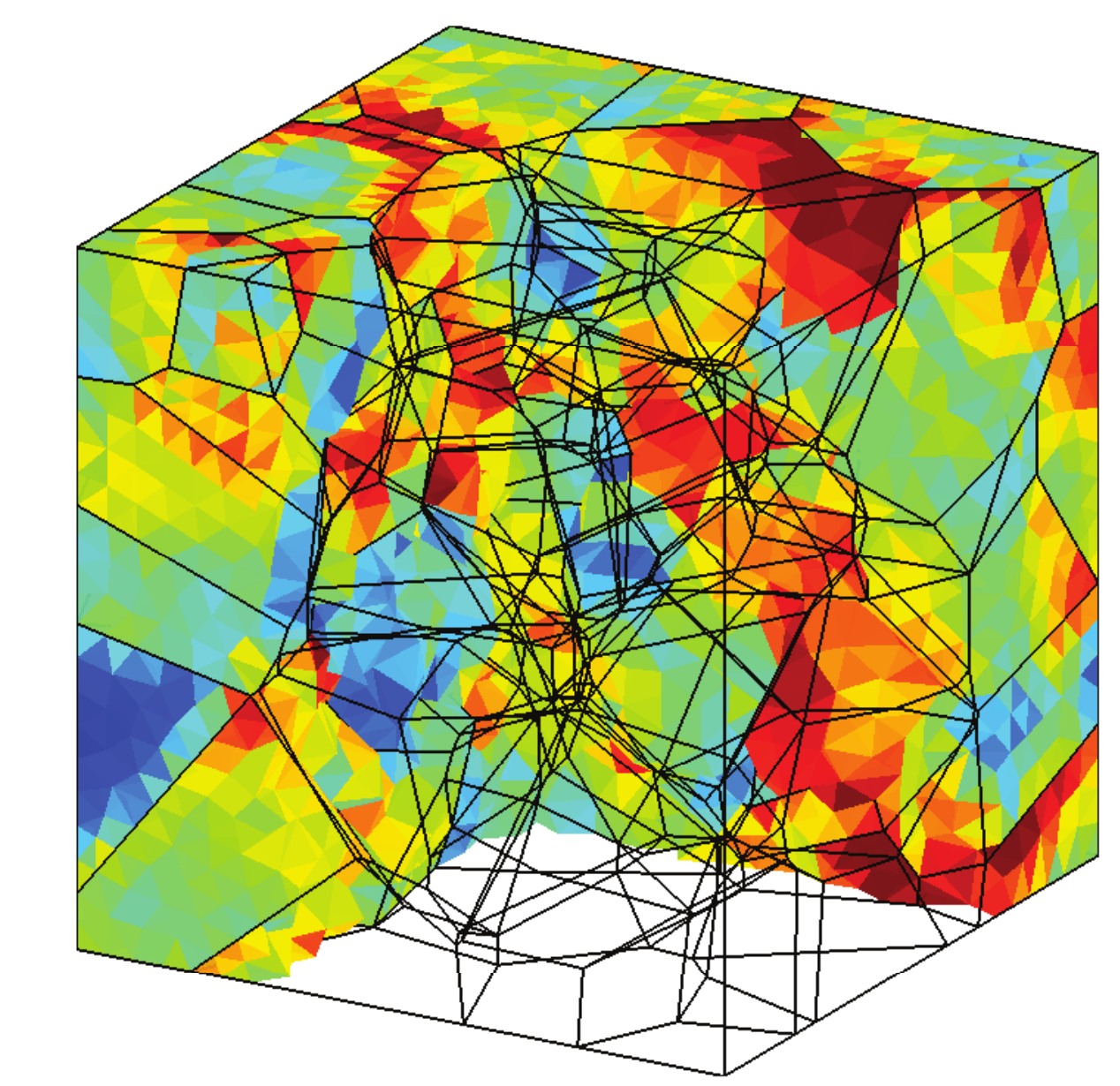}
	\caption{$\Gamma_{33}=2.0\%$}
	\end{subfigure}
\\
	\vspace{5pt}
	\begin{subfigure}{0.6\textwidth}
	\centering
	\includegraphics[width=\textwidth]{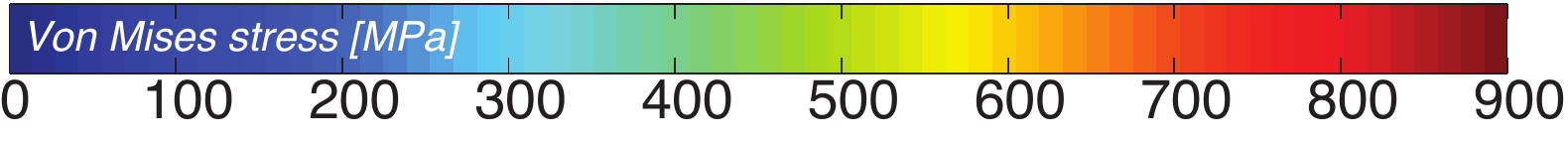}
	\caption{}
	\end{subfigure}
\caption{Von Mises stress contour plots for the 100-grain polycrystalline copper aggregate (I) at the macro-strain steps (\emph{a},\emph{b}) $\Gamma_{33}=0.3\%$, (\emph{c},\emph{d}) $1.0\%$ and (\emph{e},\emph{f}) $2.0\%$; (\emph{g}) Colormap.}
\label{fig-Ch4:G100_vonmises}
\end{figure}

\begin{figure}[h]
\centering
	\begin{subfigure}{0.49\textwidth}
	\centering
	\includegraphics[width=0.7\textwidth]{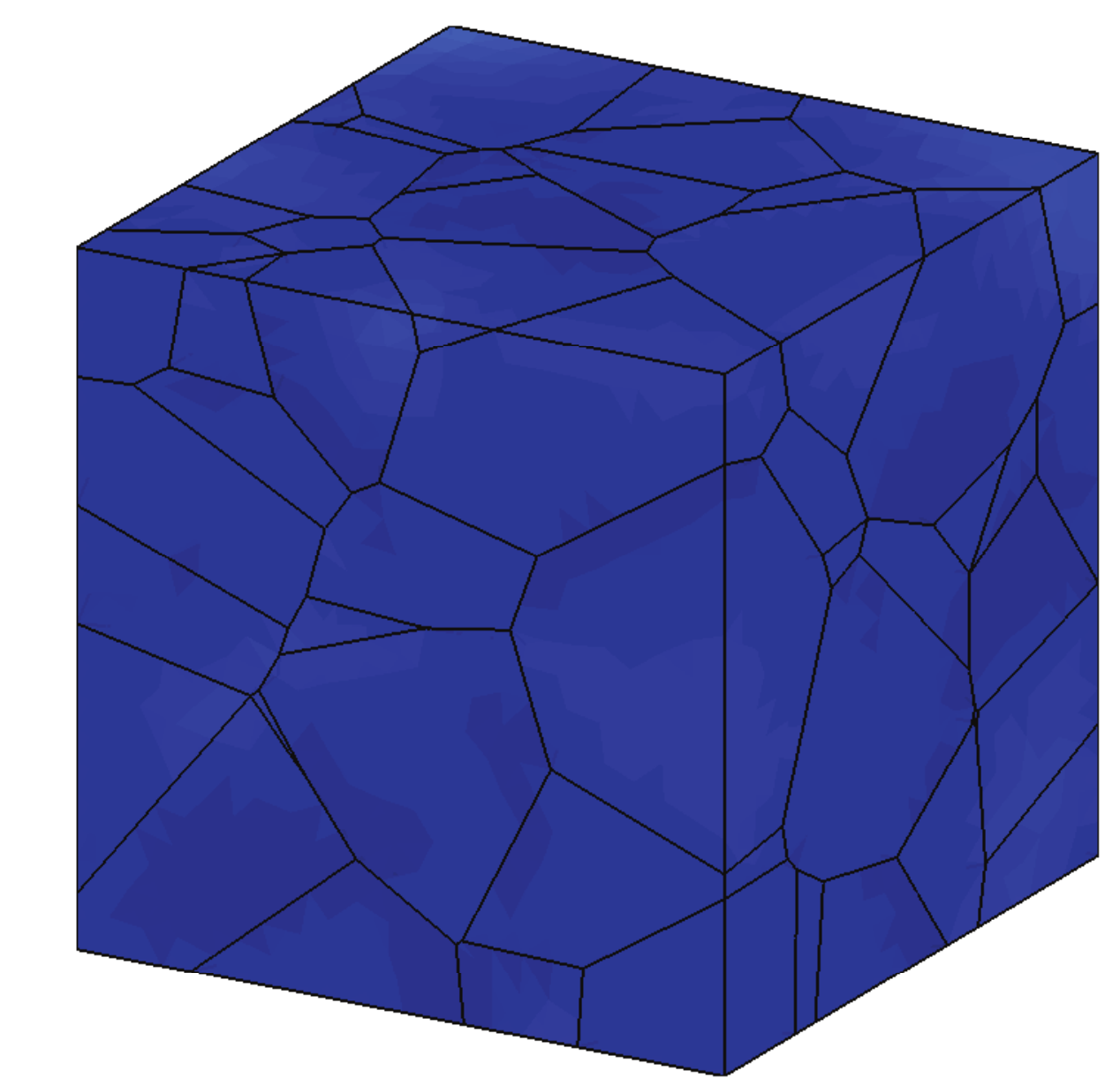}
	\caption{$\Gamma_{33}=0.3\%$}
	\end{subfigure}
\
	\begin{subfigure}{0.49\textwidth}
	\centering
	\includegraphics[width=0.7\textwidth]{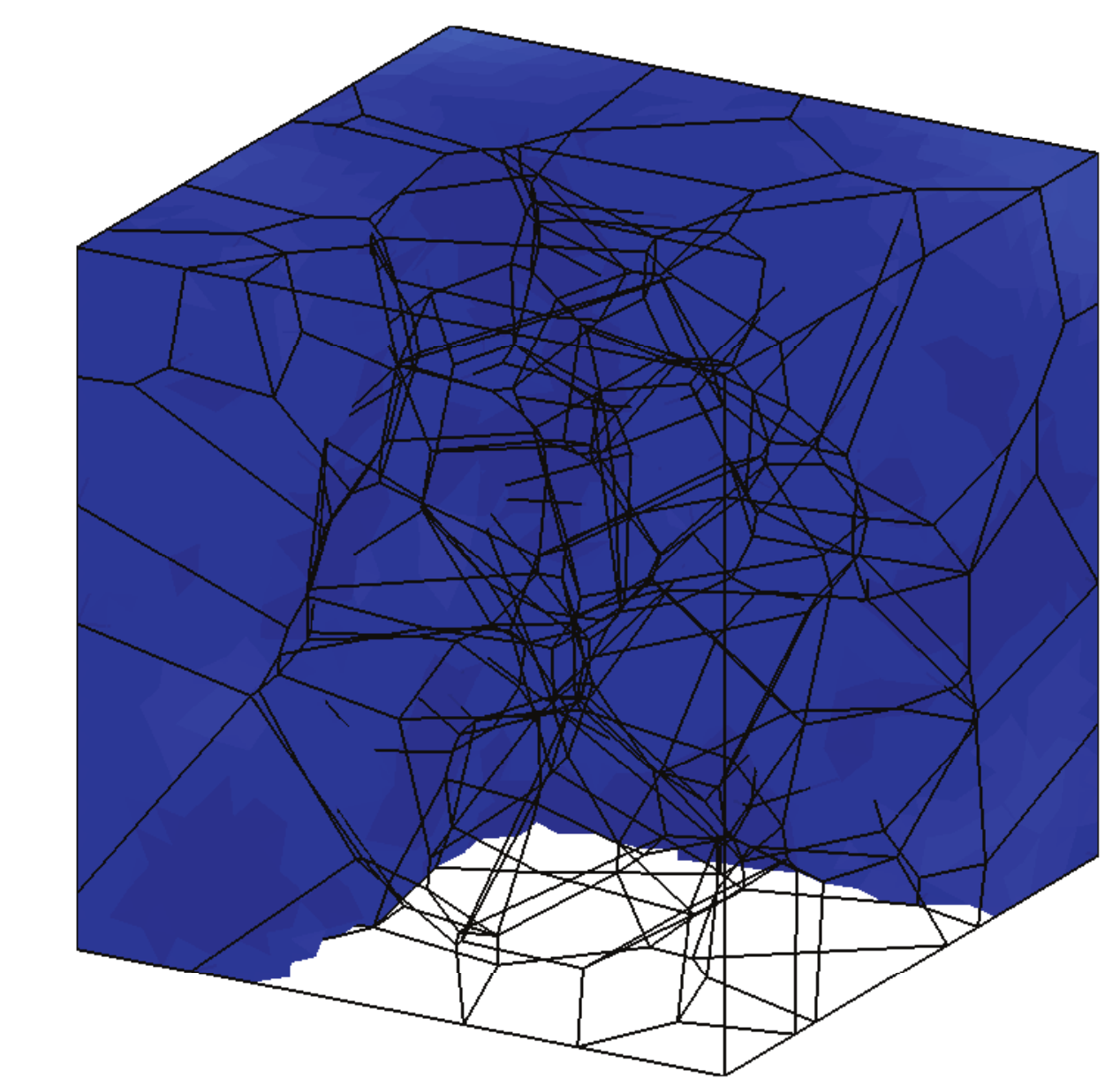}
	\caption{$\Gamma_{33}=0.3\%$}
	\end{subfigure}
\\
	\begin{subfigure}{0.49\textwidth}
	\centering
	\includegraphics[width=0.7\textwidth]{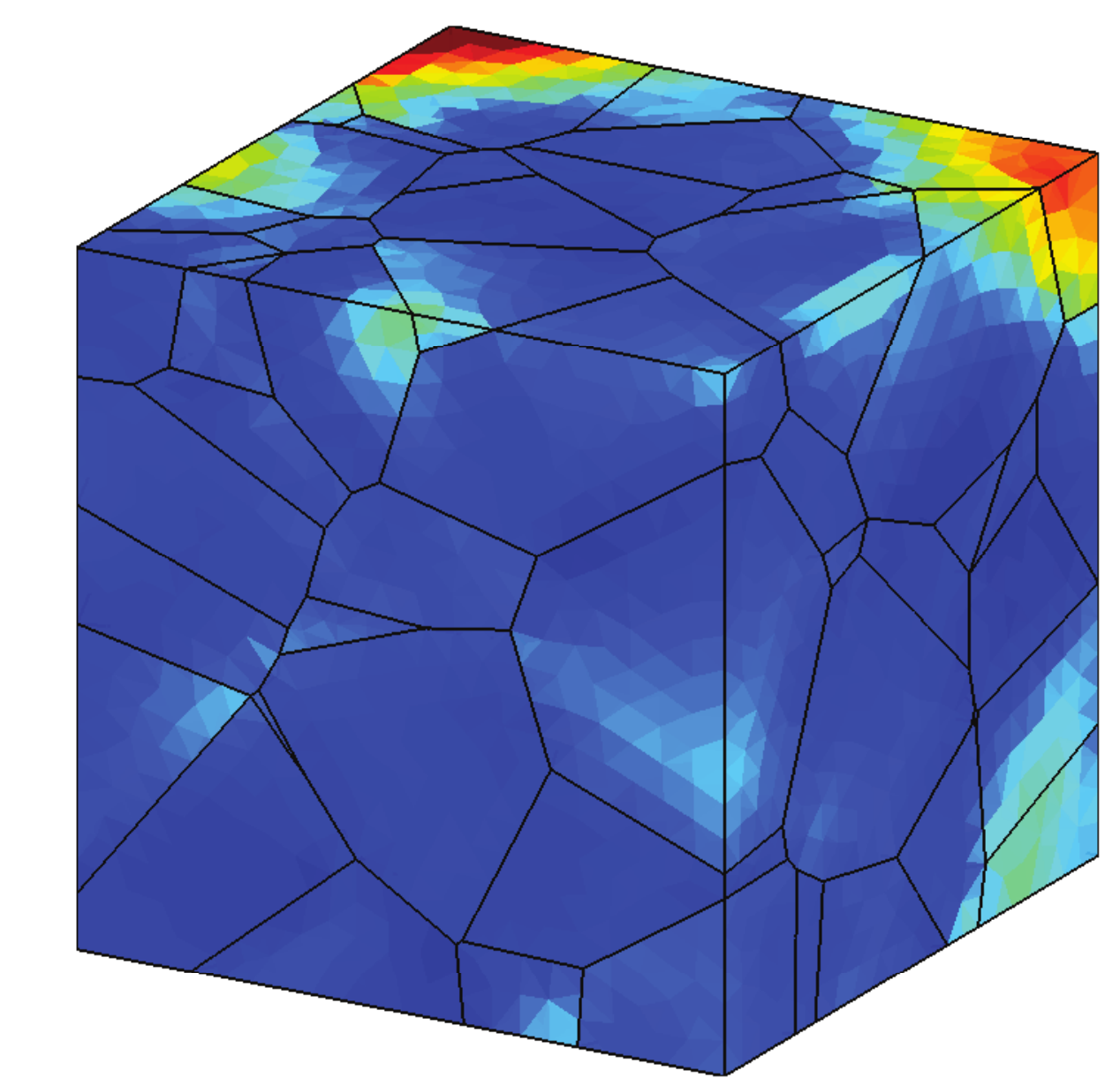}
	\caption{$\Gamma_{33}=1.0\%$}
	\end{subfigure}
\
	\begin{subfigure}{0.49\textwidth}
	\centering
	\includegraphics[width=0.7\textwidth]{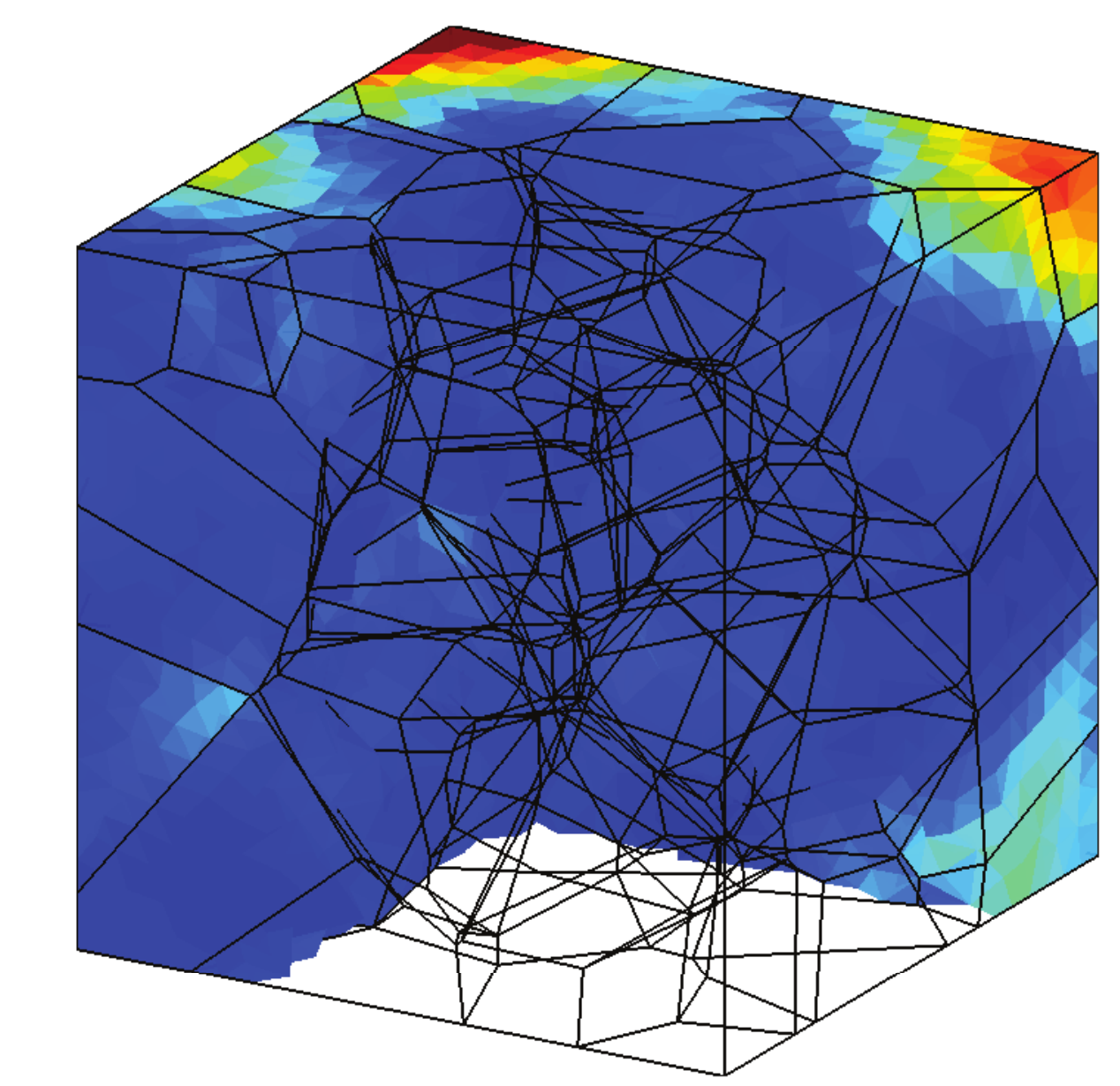}
	\caption{$\Gamma_{33}=1.0\%$}
	\end{subfigure}
\\
	\begin{subfigure}{0.49\textwidth}
	\centering
	\includegraphics[width=0.7\textwidth]{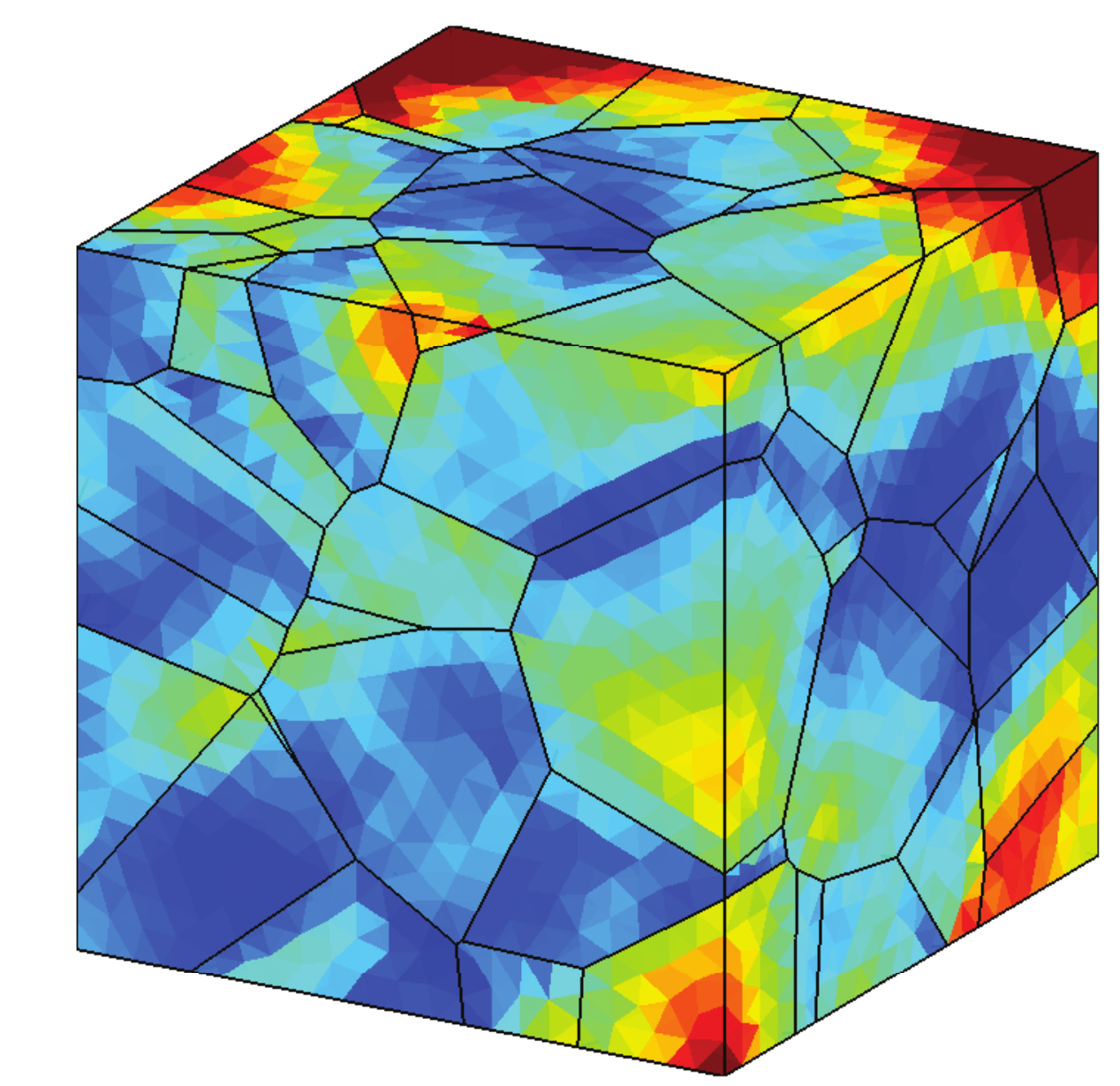}
	\caption{$\Gamma_{33}=2.0\%$}
	\end{subfigure}
\
	\begin{subfigure}{0.49\textwidth}
	\centering
	\includegraphics[width=0.7\textwidth]{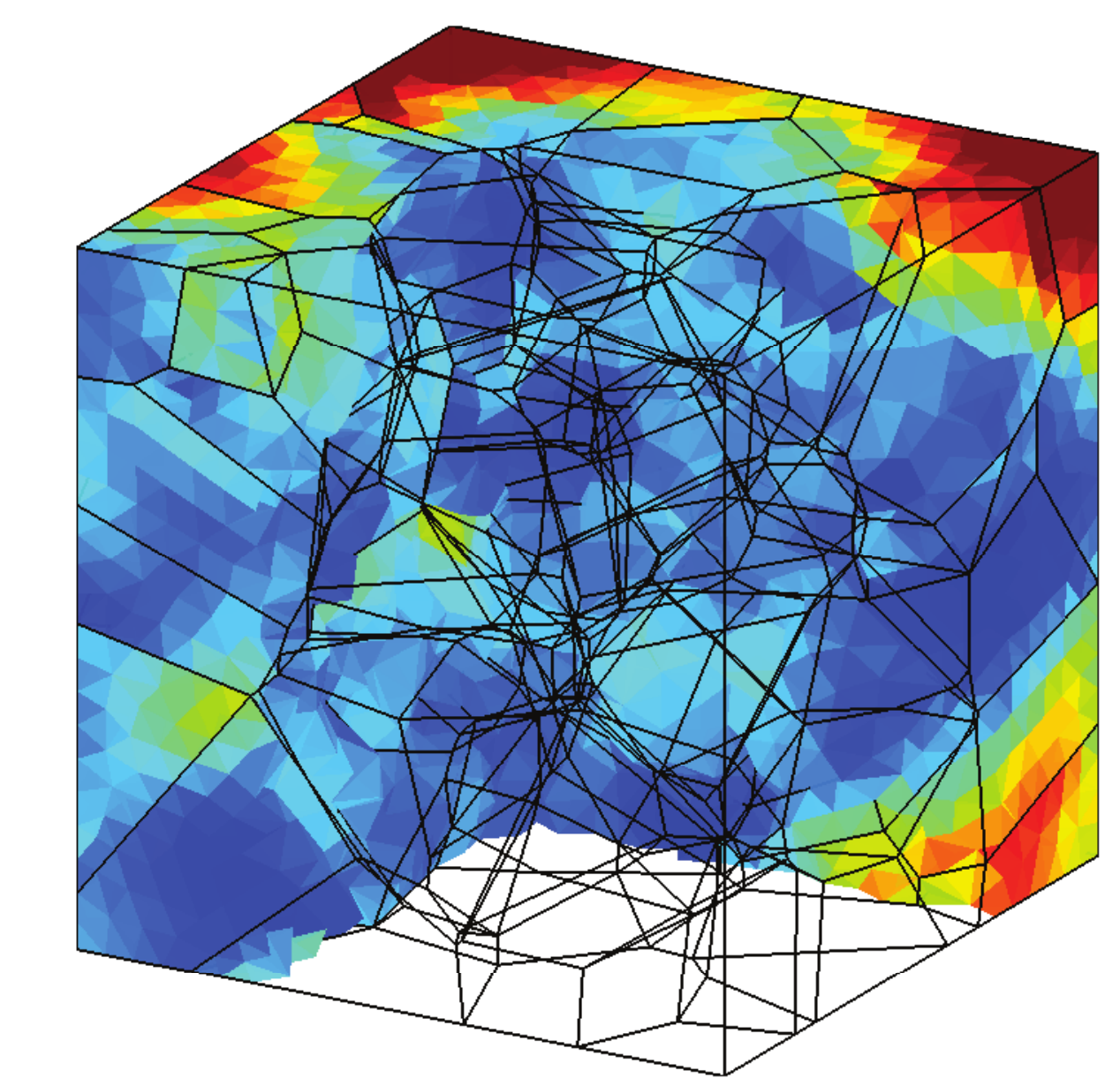}
	\caption{$\Gamma_{33}=2.0\%$}
	\end{subfigure}
\\
	\vspace{5pt}
	\begin{subfigure}{0.6\textwidth}
	\centering
	\includegraphics[width=\textwidth]{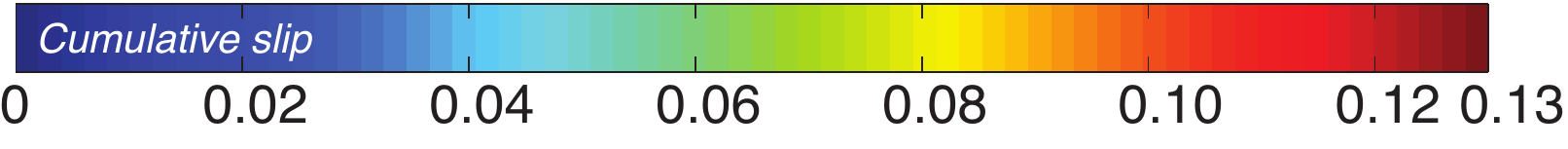}
	\caption{}
	\end{subfigure}
\caption{Cumulative slip contour plots for the 100-grain polycrystalline copper aggregate (I) at the macro-strain steps (\emph{a},\emph{b}) $\Gamma_{33}=0.3\%$, (\emph{c},\emph{d}) $1.0\%$ and (\emph{e},\emph{f}) $2.0\%$; (\emph{g}) Colormap.}
\label{fig-Ch4:G100_cumgamma}
\end{figure}

\clearpage

In conclusion, a comparison with the homogenisation results obtained using the crystal plasticity finite element method is presented.
The reference is the work by Barbe et al.\ \cite{barbe2001a}, where the response of polycrystalline aggregates at small strains is studied using FEM.
In their work, the authors employ the model proposed by M{\'e}ric and Cailletaud \cite{meric1991}, which is briefly
recalled in Section (\ref{app-Ch4:flow N hardening laws}) for the sake of completeness.

In \cite{barbe2001a}, the FEM aggregate consists of 200 grains with random crystallographic orientations, discretized using a structured mesh and subject to iso-volumic boundary conditions. The BEM results are instead obtained with the polycrystalline aggregate (I) whose morphology, surface and volumes meshes are reported in Figures (\ref{fig-Ch4:G100_microstructures}a,c,e), respectively. As in Ref.\ \cite{barbe2001a}, the BEM polycrystalline aggregate is subjected to iso-volumic boundary conditions and is loaded along the $x_3$ direction, so that the axial stress and strain correspond to the averaged macro-stress $\Sigma_{33}$ and macro-strain $\Gamma_{33}$, respectively. The lateral stresses correspond to the averaged macro-stresses $\Sigma_{11}$ and $\Sigma_{22}$. The elastic, slip rate and hardening constants are taken from \cite{barbe2001a}, whereas the interaction hardening moduli from the work of G\'erard et al.\ \cite{gerard2013}, who adopted such coefficients based on the work on dislocation dynamic simulations for FCC crystals of Devincre et al.\ \cite{devincre2006}. All the constants are reported in Table (\ref{tab-Ch4:CP properties of Cu for MC model}). It is worth noting that in the elastic regime, the grains are considered isotropic, as in \cite{barbe2001a}, and this is why their elastic behaviour is given in terms of Young's modulus $E$ and Poisson's ratio $\nu$.

Figure (\ref{fig-Ch4:G100_barbe}) shows the comparison between the results obtained using the FEM model in \cite{barbe2001a} and those obtained from the present formulation. In Figure (\ref{fig-Ch4:G100_barbe}), the FEM points are denoted by small circles whereas the BEM points are denoted by small squares. As shown in the figures, for the strain range $0$-$0.6\%$, it is possible to assess the capabilities of the present formulation, as the plastic behaviour of the polycrystalline aggregate is well developed. The presented results highlight a satisfactory agreement between the results of the present formulation and those reported in the literature.

Eventually, Figures (\ref{fig-Ch4:G100_MC}a,c,e) show the contour plot of the Von Mises stress over the external boundary of the aggregate (I) at the macro-strain steps $\Gamma_{33}=0.14\%$, $0.30\%$ and $0.59\%$, respectively, using the present grain boundary formulation and the M{\'e}ric and Cailletaud model \cite{meric1991}. Similarly, Figures (\ref{fig-Ch4:G100_MC}b,d,f) show the contour plot of the cumulative slip at the same macro-strain steps.

\begin{table}[ht]
\begin{center}
\caption{Material parameters for FEM-BEM comparison. The elastic, slip rate and hardening constants are taken from \cite{barbe2001a}, whereas the interaction hardening moduli from \cite{gerard2013}. To interpret the symbols, the reader is referred to Section (\ref{ssec-Ch4:flow N hardening laws - MC}).}
\label{tab-Ch4:CP properties of Cu for MC model}
\begin{tabular}{lll}
\hline
\hline
property&component&value\\
\hline
Young's modulus [$10^{9}\,\mathrm{N}/\mathrm{m}^{2}$]&$E$&169\\
Poisson's ratio [-]&$\nu$&0.3\\
slip rate constant [$10^6\, \mathrm{N}\cdot \mathrm{s}^{1/n}$]&$K$&10\\
rate sensitivity [-]&$n$&25\\
initial critical resolved&\multirow{2}{*}{$\tau_0$}&\multirow{2}{*}{111}\\
shear stress [$10^{6}\,\mathrm{N}/\mathrm{m}^{2}$]&&\\
hardening constant [$10^{6}\,\mathrm{N}/\mathrm{m}^{2}$]&$Q$&$35$\\
hardening constant [-]&$b$&7\\
\multirow{6}{*}{hardening moduli [-]}&self hardening&8.0\\
&Hirth lock&0.2\\
&Coplanar interaction&1.0\\
&Glissile junction&3.0\\
&Collinear interaction&90.0\\
&Lomer lock&2.5\\
\hline
\hline
\end{tabular}
\end{center}
\end{table}

\begin{figure}
\centering
	\begin{subfigure}{0.49\textwidth}
	\centering
	\includegraphics[width=\textwidth]{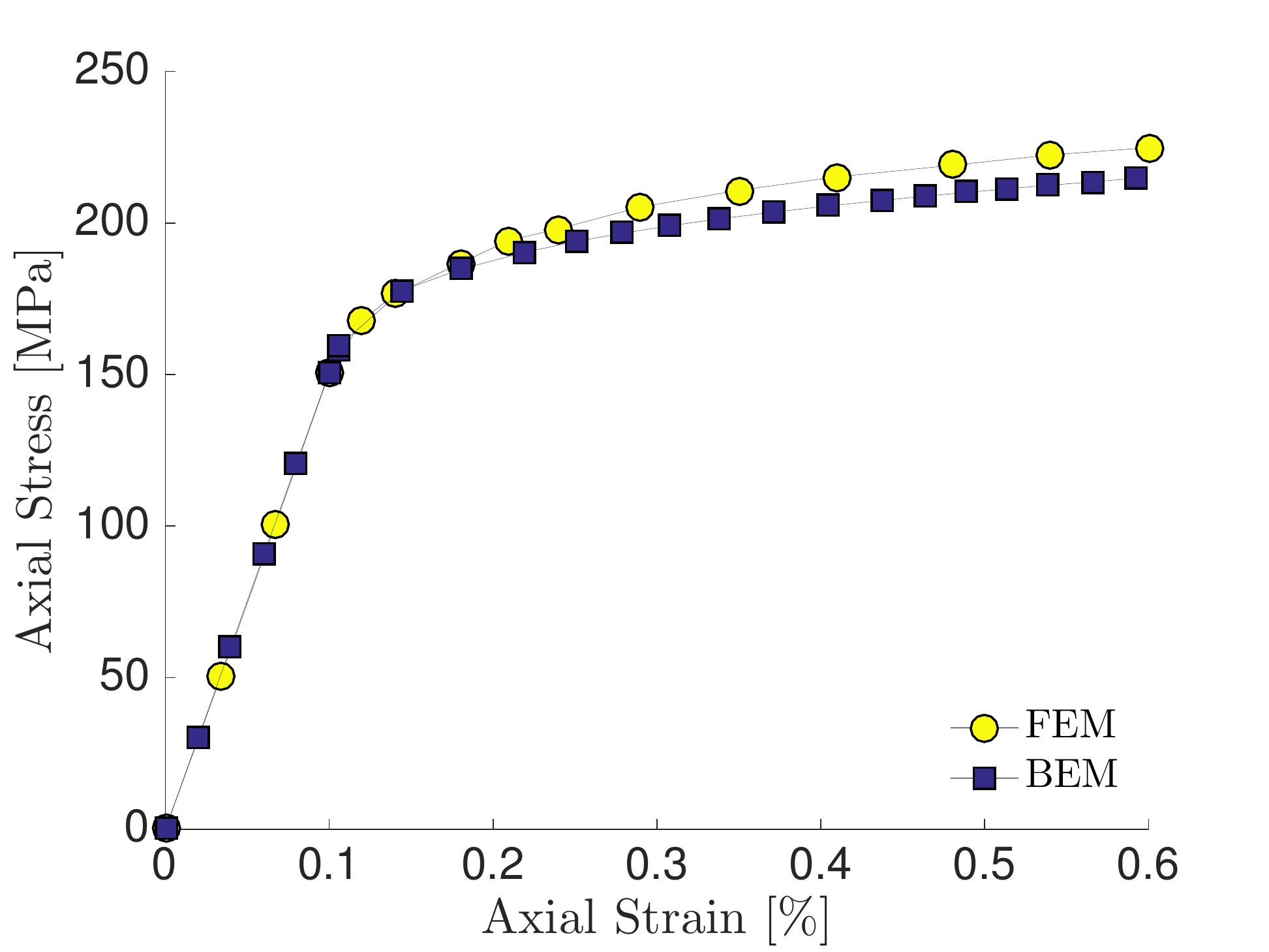}
	\caption{}
	\end{subfigure}
\	
	\begin{subfigure}{0.49\textwidth}
	\centering
	\includegraphics[width=\textwidth]{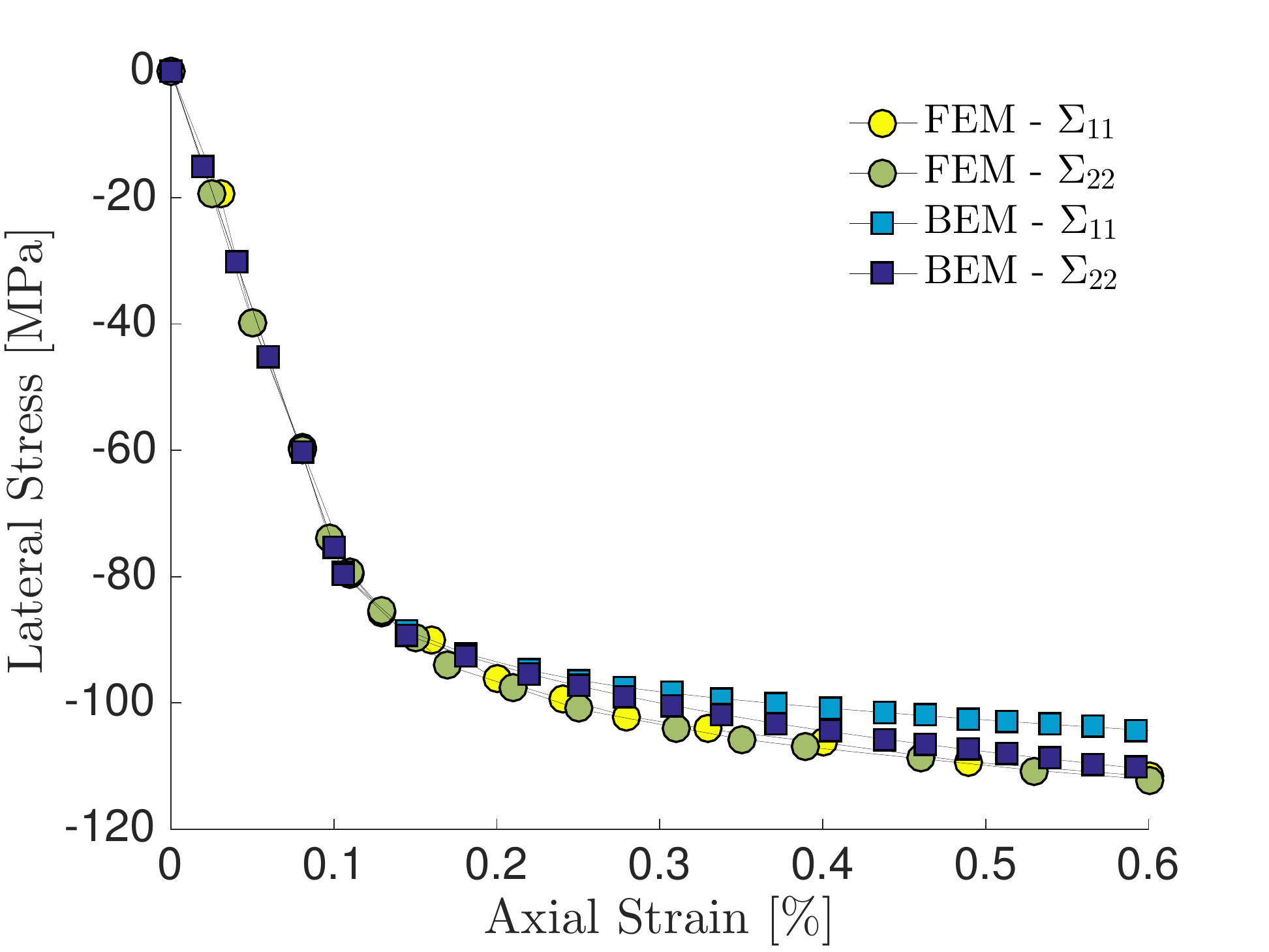}
	\caption{}
	\end{subfigure}
\caption{Comparison between FEM results \cite{barbe2001a} and the present BEM formulation. (\emph{a}) Macro stress-strain behaviour of a polycrystalline aggregate subjected to iso-volumic boundary conditions; (\emph{b}) Lateral macro-stresses versus axial macro-strain.}
\label{fig-Ch4:G100_barbe}
\end{figure}

\begin{figure}[h]
\centering
	\begin{subfigure}{0.49\textwidth}
	\centering
	\includegraphics[width=0.7\textwidth]{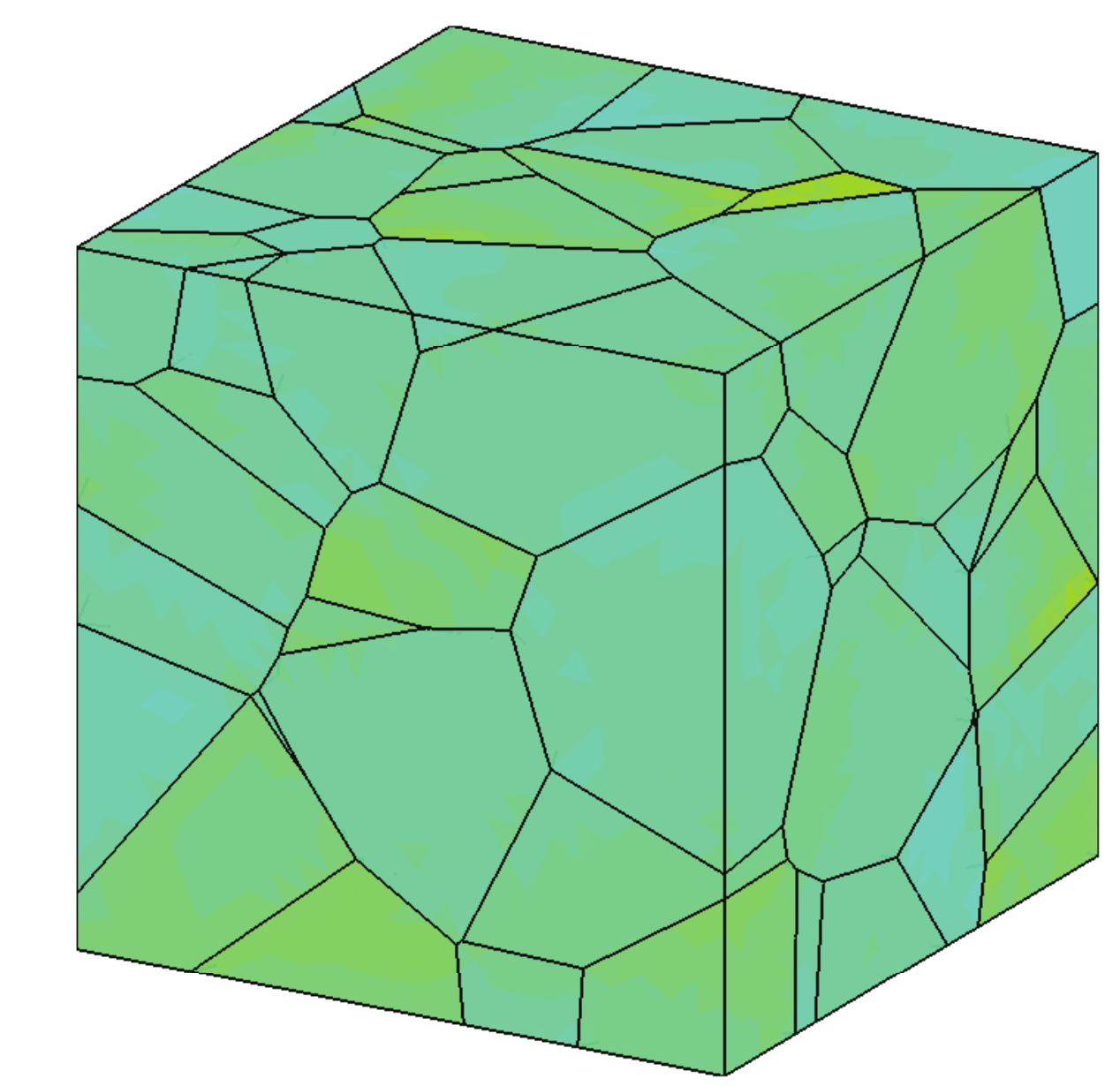}
	\caption{$\Gamma_{33}=0.14\%$}
	\end{subfigure}
\
	\begin{subfigure}{0.49\textwidth}
	\centering
	\includegraphics[width=0.7\textwidth]{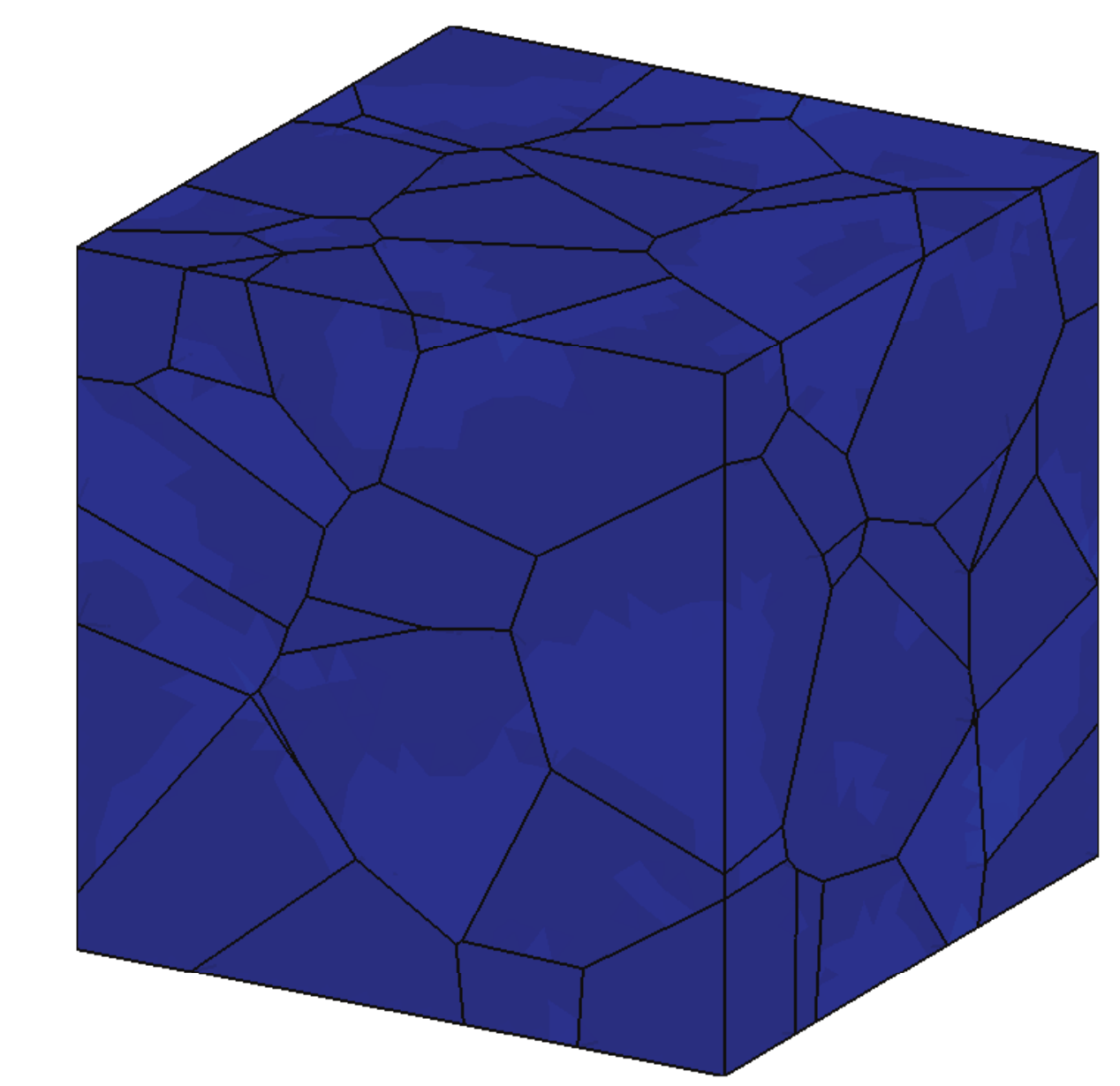}
	\caption{$\Gamma_{33}=0.14\%$}
	\end{subfigure}
\\
	\begin{subfigure}{0.49\textwidth}
	\centering
	\includegraphics[width=0.7\textwidth]{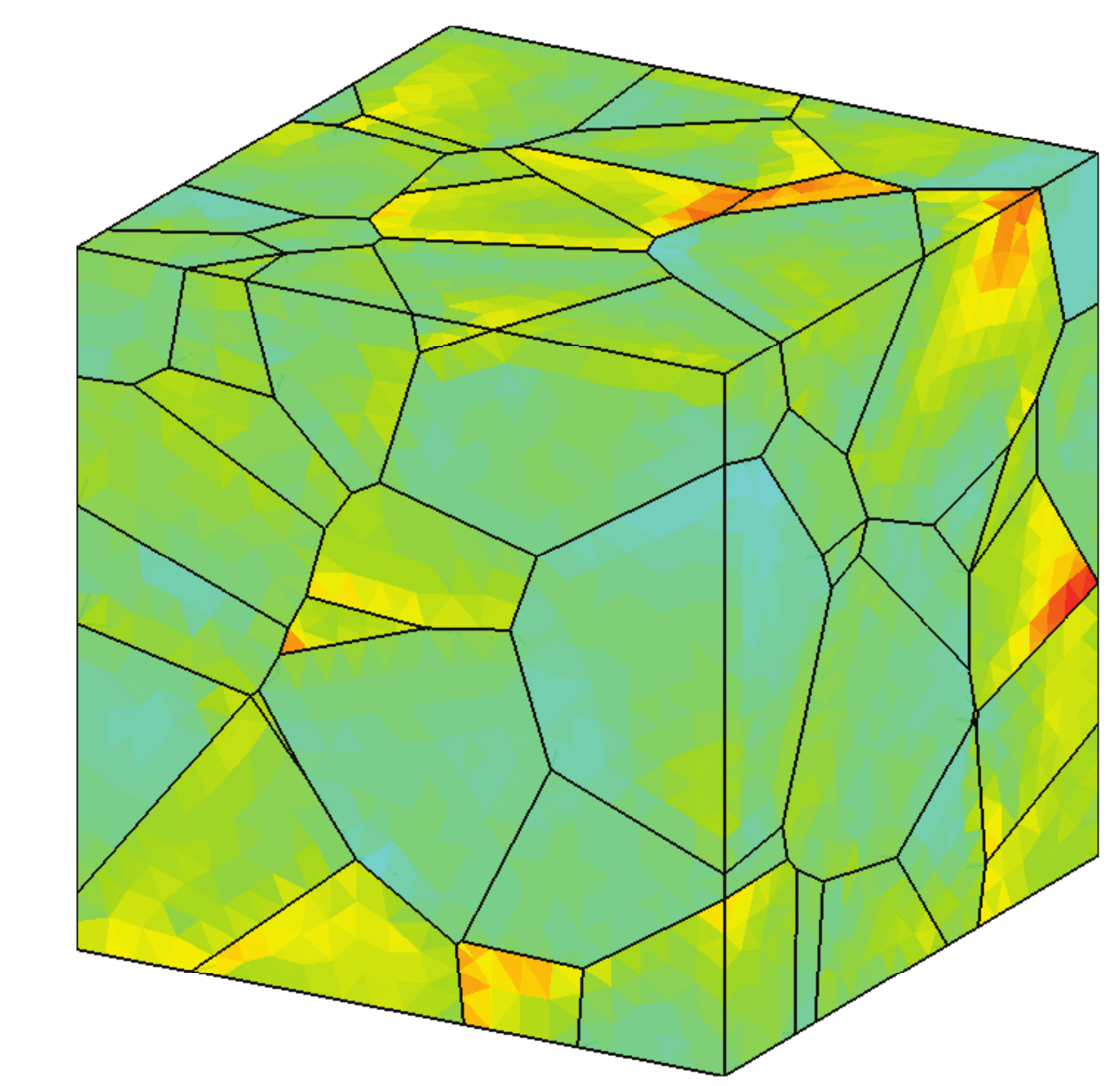}
	\caption{$\Gamma_{33}=0.30\%$}
	\end{subfigure}
\
	\begin{subfigure}{0.49\textwidth}
	\centering
	\includegraphics[width=0.7\textwidth]{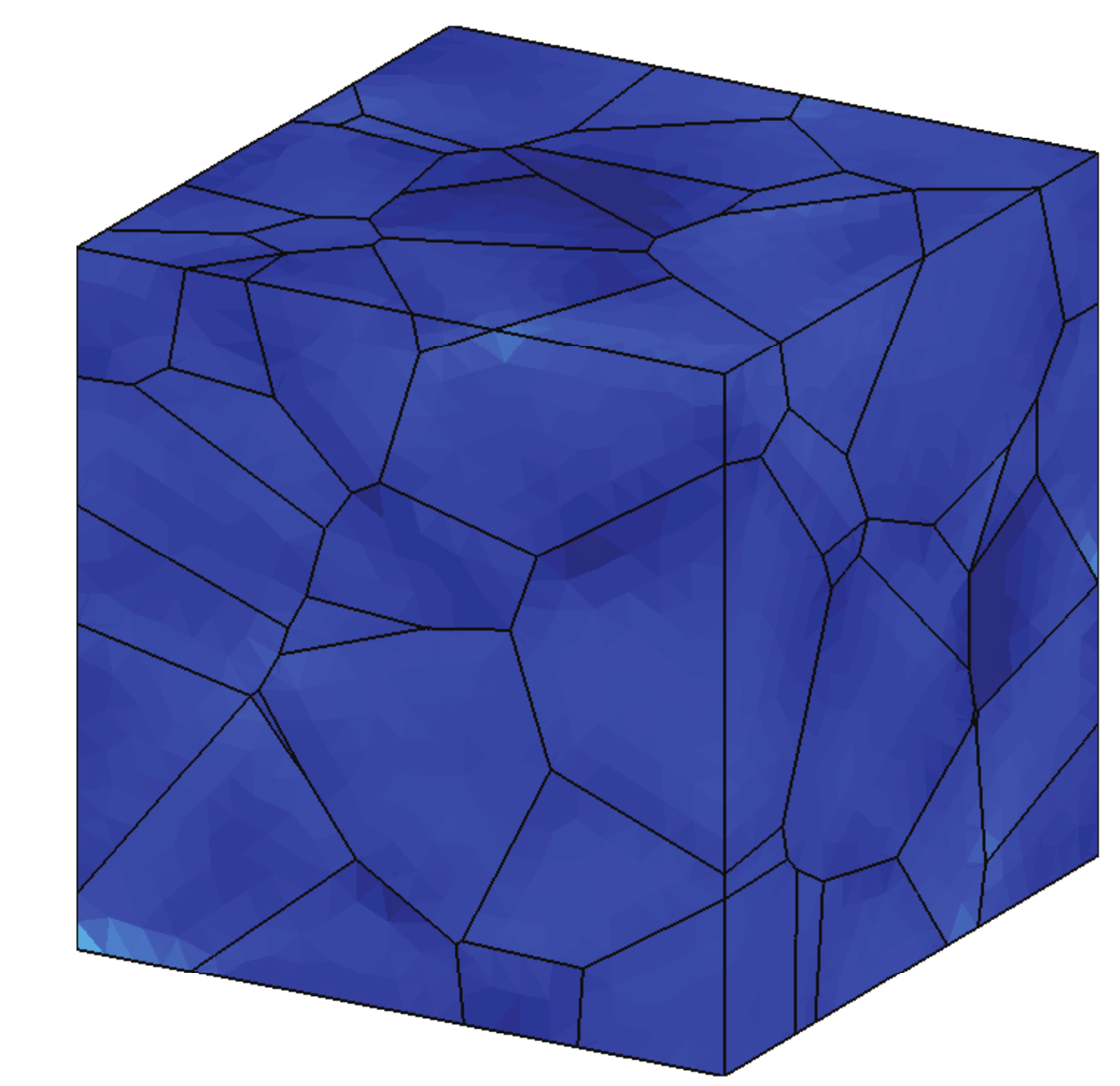}
	\caption{$\Gamma_{33}=0.30\%$}
	\end{subfigure}
\\
	\begin{subfigure}{0.49\textwidth}
	\centering
	\includegraphics[width=0.7\textwidth]{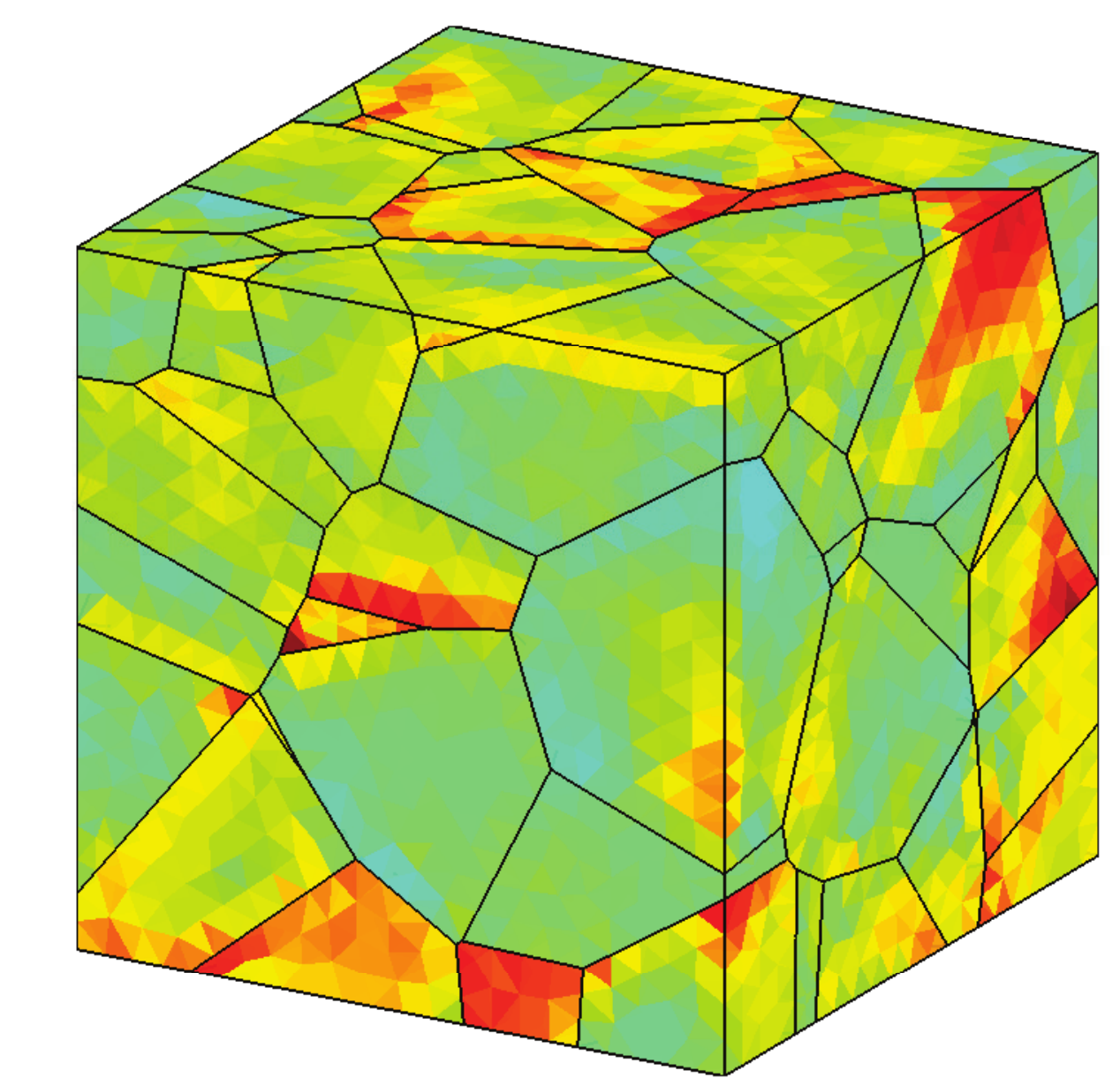}
	\caption{$\Gamma_{33}=0.59\%$}
	\end{subfigure}
\
	\begin{subfigure}{0.49\textwidth}
	\centering
	\includegraphics[width=0.7\textwidth]{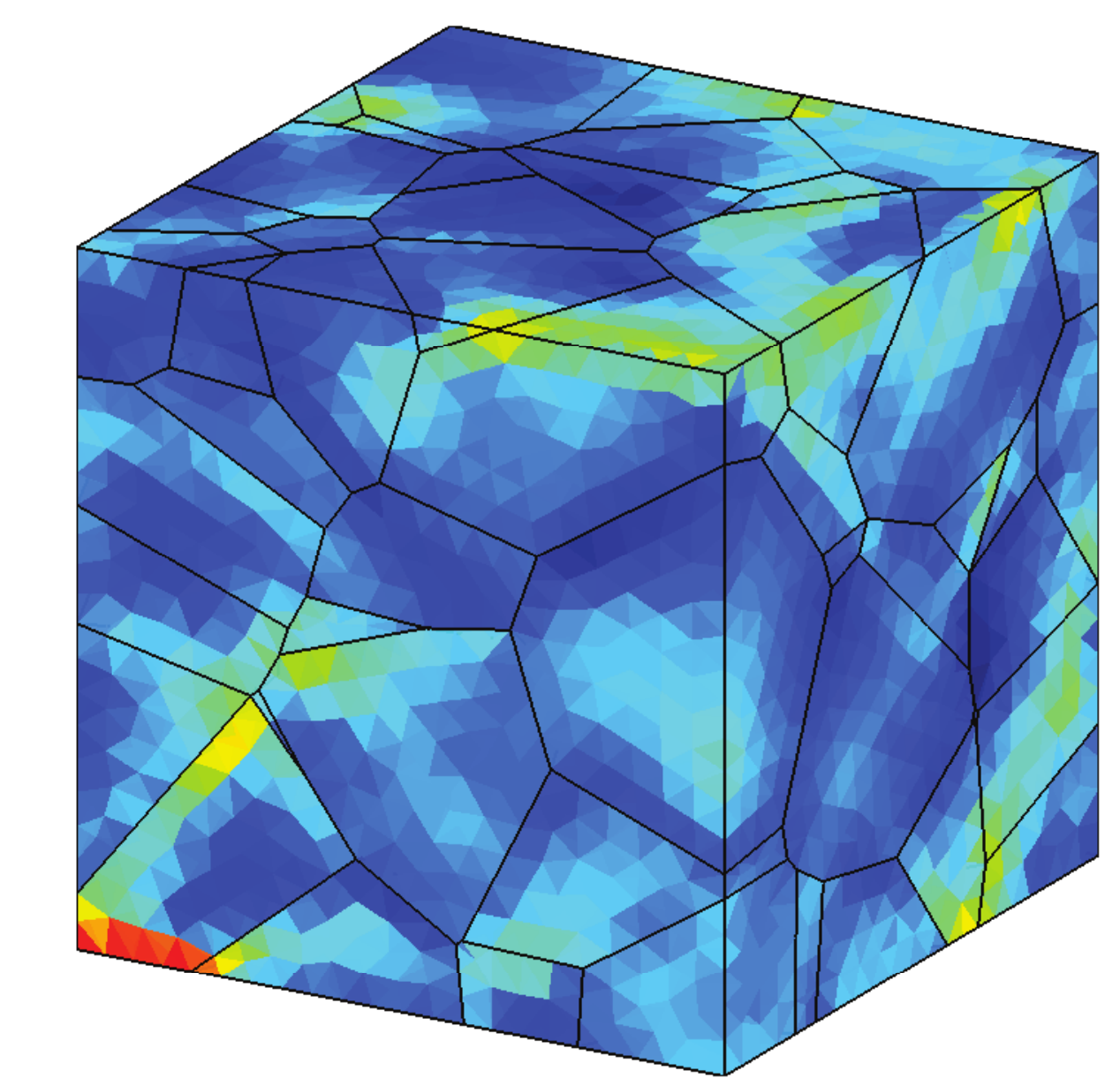}
	\caption{$\Gamma_{33}=0.59\%$}
	\end{subfigure}
\\
	\vspace{5pt}
	
	\begin{subfigure}{0.49\textwidth}
	\centering
	\includegraphics[width=\textwidth]{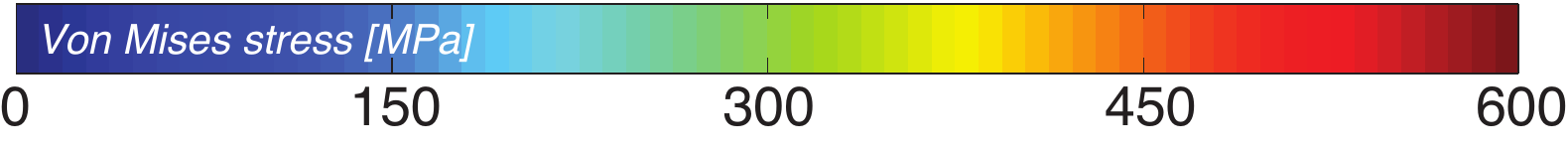}
	\caption{}
	\end{subfigure}
\
	\begin{subfigure}{0.49\textwidth}
	\centering
	\includegraphics[width=\textwidth]{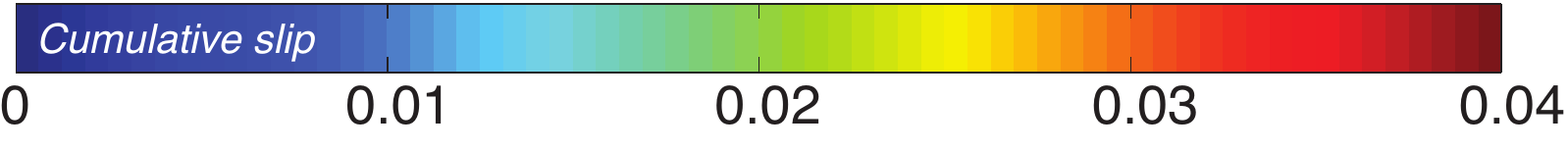}
	\caption{}
	\end{subfigure}
\caption{(\emph{a},\emph{c},\emph{e}) Von Mises stress contour plots for the polycrystalline aggregate at the macro-strain steps $\Gamma_{33}=0.14\%$, $0.30\%$ and $0.59\%$, respectively. (\emph{b},\emph{d},\emph{f}) Cumulative slip contour plots for the polycrystalline aggregate at the macro-strain steps $\Gamma_{33}=0.14\%$, $0.30\%$ and $0.59\%$, respectively. (\emph{g},\emph{h}) Colormaps of (\emph{g}) Von Mises stress and (\emph{h}) cumulative slip.}
\label{fig-Ch4:G100_MC}
\end{figure}

\clearpage

\section{Discussion and possible developments}\label{sec-Ch4:discussion}
The presented formulation addresses crystal plasticity in a boundary element framework, both for single crystals and crystal aggregates. From this point of view, it may be interesting to note that, while the possibility of using boundary elements for crystal plasticity is mentioned in the very exhaustive review \cite{roters2010}, no reference is provided to crystal plasticity boundary elements papers.

The salient feature of the method is its expression in terms of grain-boundary variables only, namely in terms of inter-granular displacements and tractions. In previous applications, see e.g.\ Chapters (\ref{ch-intro}), (\ref{ch-EF}) and (\ref{ch-TG}) and \cite{benedetti2013a,benedetti2013b,gulizzi2015}, this aspect ensured \emph{simplification} in data preparation, as only meshing of the inter-granular interfaces was required, and a \emph{reduction} in the number of DoFs for the analysis, particularly appreciated in polycrystalline problems. In crystal plasticity applications, the data preparation simplification with respect to other techniques is lost, due to the need of meshing the grains interior volume; however, the system order reduction is maintained, as \emph{the volume mesh does not carry additional degrees of freedom}. The volume mesh is in fact only used to compute the plastic contribution on the right-hand side of Eq.(\ref{eq-Ch4:polycrystalline SBIE discrete}), while \emph{the number of degrees of freedom with respect to the purely elastic case remains unchanged}. Additionally, it is worth noting that the plastic accumulation just modifies the right-hand side of Eq.(\ref{eq-Ch4:polycrystalline SBIE discrete}): such circumstance is particularly advantageous in terms of numerical solution, as the factorisation of the coefficient matrix in the left-hand side of Eq.(\ref{eq-Ch4:polycrystalline SBIE discrete}) is performed just once and used throughout the analysis, thus ensuring numerical effectiveness.

Although the proposed scheme allows a considerable reduction in the number of DoFs with respect to other volume discretisation schemes, the solution of system (\ref{eq-Ch4:polycrystalline SBIE discrete}) still represents a formidable computational task, especially in the polycrystalline case, due to the fact that the numerical blocks associated to each grain in the boundary integral scheme are fully populated. For such reason, at least in the present implementation, the typical computation time is longer than the typical times with CPFEMs or Fast Fourier Transform based formulations, see e.g.\ Ref.\ \cite{lebensohn2012}.
An interesting direction of further investigation, from this point of view, could be related to the use of fast multipoles \cite{liu2009} or hierarchical matrices \cite{bebendorf2008,benedetti2008,benedetti2009,benedetti2010,milazzo2012} in conjunction with iterative solvers to reduce the computational time and further compress the storage memory requirements  of the proposed framework. 
On the other hand, the implementation of effective rate-independent schemes \cite{anand1996,miehe1999,bettaieb2012}, could speed up the iterative convergence of the crystal plastic analysis, avoiding the numerical \emph{stiffness}, and the associated computational costs, associated to high values of the rate sensitivity $n$.

In the developed model, a phenomenological description of crystal plasticity has been adopted. However, the model offers flexibility and could be coupled with more sophisticated crystal plasticity laws, included physically based approaches.

Another natural extension could be the introduction of different models of grain interfaces, which at the moment are modelled as perfect and remain intact during the loading history. Sliding, separation and contact could be included within the formulation by suitably redefining the interface equations in (\ref{eq-Ch4:polycrystalline SBIE discrete}).

The presented formulation has also some limitations, which provide input for further investigation. First of all, the formulation is confined within the framework of \emph{small strains}, so that \emph{finite} lattice rotations are not taken into account. For this reason, applications to \emph{texture evolution} problems are not currently possible. In fact, the present formulation is based on an \emph{elasto-plastic} analysis, while crystal plastic finite element formulations for texture evolution are often based on \emph{rigid-plastic} approaches, where the initial elastic behaviour is neglected \cite{bassani1991,huang1991}. The extension of the developed grain-boundary framework to finite strains and the inclusion of finite rotations is not a trivial task and poses considerable research challenges that are left as future task.

\section{Flow rules and hardening laws}\label{app-Ch4:flow N hardening laws}

In the present work two different crystal plasticity phenomenological frameworks have been used in the performed numerical tests: the model by Bassani and Wu \cite{bassani1991} and the model by M\'eric and Cailletaud \cite{meric1991}.

\subsection{Bassani and Wu model}\label{ssec-Ch4:flow N hardening laws - BW}
Bassani and Wu used as flow rule a power law of the form
\begin{equation}\label{eq-Ch4:BW flow rule}
\dot{\gamma}^{\alpha}=\dot{\gamma}^{\alpha}_0\left|\frac{\tau^{\alpha}}{\tau^{\alpha}_c}\right|^{n}\,sgn\left(\tau^{\alpha}\right),\quad \forall \alpha=1,\cdots,N_s,
\end{equation}
where $\dot{\gamma}^{\alpha}_0$ and $n$ are material parameters known as \textit{reference shear rate} and \textit{rate sensitivity of slip} respectively, while $\tau^{\alpha}_c$ is the evolving critical stress. Eq.(\ref{eq-Ch4:BW flow rule}) describes, in general, a rate-dependent phenomenon in which all the slip systems are always \emph{active} with a specific slip rate $\dot{\gamma}^{\alpha}$, as long as the corresponding resolved shear stress is non-zero, so that no distinction is made between \emph{active} and \emph{inactive} slip systems. It can be demonstrated that when $n\rightarrow\infty$ the formulation becomes rate-independent.

They also expressed the \emph{hardening laws}, i.e.\ the evolution of $\tau^{\alpha}_c$ with respect to the material state and evolution, as
\begin{equation}\label{eq-Ch4:BW hardening rule}
\dot{\tau}^{\alpha}_c=\sum_{\beta=1}^{N_{s}}h_{\alpha\beta}\:\left|\dot{\gamma}^{\beta}\right|,\quad\forall \alpha=1,\cdots,N_s,
\end{equation}
where the \emph{self hardening} moduli $h_{\alpha\alpha}$ and the \emph{latent hardening} moduli $h_{\alpha\beta}$ were expressed as
\begin{subequations}\label{eq-Ch4:BW hardening moduli}
\begin{equation}
h_{\alpha\alpha}=\left[h_s+\left(h_0-h_s\right)\mathrm{sech}^2\left(\frac{h_0-h_s}{\tau_s-\tau^{\alpha}_{0}}\cdot\gamma^{\alpha}\right) \right]\cdot \left[1+\sum_{\beta\neq\alpha}f_{\alpha\beta}\tanh\left(\frac{\gamma^{\beta}}{\gamma_p}\right)\right]
\end{equation}
\begin{equation}
h_{\alpha\beta}=q\cdot h_{\alpha\alpha},\quad \beta\neq\alpha,
\end{equation}
\end{subequations}
where: $h_0$ is the initial hardening modulus; $h_s$ is the hardening modulus during easy glide within stage I hardening; $\tau_s$ is the stage I reference stress, where large plastic flow initiates; $\tau^{\alpha}_{0}=\tau^{\alpha}_{c}(0)$ is the initial critical resolved shear stress on the $\alpha$-th slip system; $\gamma^{\alpha}=\int_{t}{|\dot{\gamma}^{\alpha}|dt}$; $\gamma_p$ is the amount of slip after which the interaction between slip systems reaches the peak strength; $f_{\alpha\beta}$ is a specific slip interaction strength; $q$ is a constant latent hardening factor. The model by Bassani and Wu has been implemented in several FEM formulations, also for polycrystalline micromechanics \cite{simonovski2011b,simonovski2012}, often using the implementation developed by Huang \cite{huang1991}.

\subsection{M\'eric and Cailletaud model}\label{ssec-Ch4:flow N hardening laws - MC}
On the other hand, neglecting the kinematic hardening, M\'eric and Cailletaud represented the slip rate $\dot{\gamma}^\alpha$ of the $\alpha$-th slip system using the following flow rule
\begin{equation}\label{eq-Ch4:MC flow rule}
\dot{\gamma}^\alpha=\left\langle\frac{|\tau^\alpha|-\tau_c^\alpha}{K}\right\rangle^n\,sgn\left(\tau^{\alpha}\right), \quad \forall \alpha=1,\cdots,N_s,
\end{equation}
where $\langle\cdot\rangle=\max(0,\cdot)$ and $K$ and $n$ are material parameters. In addition, in their model, the hardening law, which comprises self and latent hardening through the hardening moduli $h_{\alpha\beta}$, has the form
\begin{equation}\label{eq-Ch4:MC hardening rule}
\tau_c^\alpha=\tau_0+Q\sum_{\beta=1}^{N_s}h_{\alpha\beta}\left[1-\exp(-b|\gamma^{\beta}|)\right],\quad\forall \alpha=1,\cdots,N_s
\end{equation}
where $\tau_0$ is the initial critical resolved shear stress; $Q$ is a material parameter that, multiplied by the hardening modulus $h_{\alpha\beta}$, represents the maximal increase of the critical shear stress on the $\alpha$-th slip system due to the slip in the $\beta$-th system; $\tau_0+Q\sum_{\beta}h_{\alpha\beta}$ represents the maximum, or \emph{saturated}, value that the critical shear stress on the $\alpha$-th slip system can reach; eventually, $b$ is a material parameter that governs the exponential saturation law.

\clearpage

\chapter*{Conclusions}
\addcontentsline{toc}{chapter}{Conclusions}
In this thesis, a computational framework for micro-structural modelling of polycrystalline materials with damage and failure has been developed and presented. The framework is based on a non-linear multi-domain grain boundary formulation for generally anisotropic materials and has been employed for modelling the deformation, damage and failure mechanisms of polycrystalline aggregates at the scale of the constituent grains.

In the thesis, several aspects of the polycrystalline problem within a boundary element formulation have been investigated and addressed.

In Chapter (\ref{ch-FS}), a novel unified compact scheme to compute the fundamental solutions of generally anisotropic multi-field materials, as well as their derivatives up to the desired order, has been presented. The scheme is based on the relationship between the Rayleigh expansion and the three-dimensional Fourier representation of a second-order homogenous partial differential operator, which allows to obtain a convergent series representation of the fundamental solutions and their derivatives in terms of spherical harmonics. The presented expression is easy to implement as it does not require any term-by-term differentiation to obtain the derivatives of the fundamental solution. Furthermore, the coefficients of the series depend on the material properties only and it has been shown that the technique does not suffer from material degeneracies, thus making it very attractive for efficient boundary element implementations.
Two isotropic operators, namely Laplace and isotropic elasticity operators, have been analysed as particular cases and it has been shown that, in such cases, the exact solutions are easily retrieved as the series contain a finite number of terms. Numerical tests have been presented to assess the accuracy, convergence and robustness of the proposed scheme when isotropic elastic, generally anisotropic elastic, transversely isotropic and generally anisotropic piezo-electric and magneto-electro-elastic materials are considered.

In Chapter (\ref{ch-EF}), a computational effective framework for micro-structural analysis of polycrystalline materials, with micro damage and cracks, has been presented.
The framework, based on the grain-boundary formulation of Chapter (\ref{ch-intro}), has been developed for computational effectiveness, with an aim towards affordable three-dimensional modelling. The sources of high computational costs in the original strategy have been identified and suitably treated. First, the original Voronoi or Laguerre morphologies have been regularised, to remove the small entities usually inducing excessively refined meshes. Then, special periodic non-prismatic unit cells have been considered, to remove the pathological grains often induced by the operations used to generate prismatic periodic RVEs. Eventually, a specific grain meshing strategy has been adopted, to conjugate mesh effectiveness with formulation simplicity. The combined effects of such enhancements has been demonstrated to lead to outstanding reductions in terms of overall number of degrees of freedom, with no compromise over accuracy for both homogenisation and cracking analyses. In turn, this is reflected in remarkable savings in terms of solution time, especially for micro-damage and micro-cracking simulations. In conclusion, micro-cracking simulations involving few hundred grains, typically requiring several \emph{days} with the original scheme, are now performed in several \emph{hours}.

In Chapter (\ref{ch-TG}), a numerical formulation for inter- and trans-granular cracking in three-dimensional polycrystalline materials has been presented. The formulation still retains the advantages of expressing the polycrystalline problem involving trans-granular cracking in terms of grain boundary variables thus reducing the computational cost of polycrystalline simulations. Trans-granular cracking has been modelled by computing the stress in the interior of the grains, by introducing cohesive interfaces in correspondence of the failing cleavage planes according to a threshold condition and by remeshing the polycrystalline morphology accordingly. The competition between inter- and trans-granular cracking has been investigated by changing the parameters of the cohesive laws used to model the two mechanisms. Some preliminary results of the formulation have been presented showing that, by changing the ratio between the fracture energies of the two mechanisms, the failure behaviour switch from inter-granular to trans-granular cracking. Future improvements of the presented formulation will include analyses of polycrystalline morphologies with a higher number of grains and different loading conditions to investigate the effect of the statistical variability of polycrystalline morphologies under different boundary conditions on the competition between the two considered mechanisms.

In Chapter (\ref{ch-CP}), a boundary element formulation for small strains crystal plasticity in polycrystalline aggregates has been presented. The method, based on the use of grain-boundary integral equations for the anisotropic elasto-plastic problem, is formulated in terms of boundary displacements and tractions, which play the role of primary variables, thus allowing a considerable reduction in the number of DoFs with respect to other volume discretisation techniques. The scheme has been implemented both for single crystals and polycrystalline aggregates and a general rate-dependent scheme has been used in the proposed incremental/iterative grain-boundary crystal plasticity algorithm. Several numerical tests have been performed to assess the accuracy and robustness of the method and the results confirm the potential of the developed tool. Although the still high computational costs associated with the analysis of general plastic polycrystalline problems call for further developments and refinements, the presented framework provides an alternative tool for the analysis of this class of problems.

To conclude, it is worth highlighting that, since the developed framework is capable of dealing with non-linear behaviours of bulk anisotropic domains (as in the crystal plasticity study), with non-linear behaviours of the inter-domain interfaces (as in the study of inter-granular failure), with on-the-fly remeshing (as in the trans-granular cracking study), it can be easily extended to investigate the micro-mechanisms of damage and failure of other classes of materials whose macroscopic behaviour is strongly influenced by their micro-structure, such as fibre-reinforced composites. Furthermore, it was shown that a broader class of physical behaviours can be modelled using the unified representation of the fundamental solutions obtained in Chapter (\ref{ch-FS}) and therefore the same framework can be also employed to model the damage and failure mechanisms of materials showing multi-field coupling.

\backmatter

\bibliographystyle{abbrv}
\bibliography{mybib}

\end{document}